\documentclass{tADP2e}

\usepackage{graphicx}         
\usepackage{amsmath,amssymb,mathrsfs,pifont,array,makecell}  
\usepackage{mathtools}
\usepackage{verbatim}          
\usepackage{enumerate,esint,dcolumn}
\usepackage{breakurl}
\usepackage[normalem]{ulem} 
\usepackage{tabularx,multirow,booktabs}
\usepackage[usenames,dvipsnames]{color}
\usepackage{bbold}
\usepackage{times}
\usepackage{setspace}

\usepackage{amsthm}
\theoremstyle{plain}
\newcommand{\cmark}{\ding{51}}%
\newcommand{\xmark}{\ding{55}}%

\newcommand{\eqn}[1]{
\begin{eqnarray}
	#1
\end{eqnarray}
}

\theoremstyle{plain}
\newtheorem{theorem}{Theorem}[section]
\theoremstyle{definition}
\newtheorem{example}{Example}[section]
\newcommand{\thm}[1]{
\begin{theorem}
	#1
\end{theorem}
}
\newcommand{\exmp}[1]{
\begin{example}
	#1
\end{example}
}

\begin{document}

\articletype{{\fontsize{13pt}{0pt}\selectfont Review}}

\title{{\fontsize{13pt}{0pt}\selectfont Non-Hermitian Physics}}

\author{\fontsize{12pt}{0pt}\selectfont Yuto Ashida$^{\rm a}$$^{\ast}$\thanks{\fontsize{10pt}{0pt}\selectfont$^\ast$Email: ashida@ap.t.u-tokyo.ac.jp}, Zongping Gong$^{\rm b}$, and Masahito Ueda$^{\rm b,c}$
\\\vspace{12pt}
$^{a}${{\fontsize{10pt}{0pt}\em{Department of Applied Physics, University of Tokyo, 7-3-1 Hongo, Bunkyo-ku, Tokyo 113-8656, Japan}}}\\
$^{b}${{\fontsize{10pt}{0pt}\em{Department of Physics, University of Tokyo, 7-3-1 Hongo, Bunkyo-ku, Tokyo 113-0033, Japan}}}\\
$^{c}${{\fontsize{10pt}{0pt}\em{RIKEN Center for Emergent Matter Science (CEMS), Wako, Saitama 351-0198, Japan}}}
}

\maketitle

\begin{abstract} 
{\begin{spacing}{1.2}
\fontsize{11pt}{11pt}\selectfont
\noindent A review is given on the foundations and applications of non-Hermitian classical and quantum physics. First, key theorems and central concepts in non-Hermitian linear algebra, including Jordan normal form, biorthogonality, exceptional points, pseudo-Hermiticity and parity-time symmetry, are delineated in a pedagogical and mathematically coherent manner. Building on these, we provide an overview of how diverse classical systems, ranging from photonics, mechanics, electrical circuits, acoustics to active matter, can be used to simulate non-Hermitian wave physics. In particular, we discuss rich and unique phenomena found therein, such as unidirectional invisibility, enhanced sensitivity, topological energy transfer, coherent perfect absorption, single-mode lasing, and robust biological transport. We then explain in detail how non-Hermitian operators emerge as an effective description of open quantum systems on the basis of the Feshbach projection approach and the quantum trajectory approach. We discuss their applications to physical systems relevant to a variety of fields, including atomic, molecular and optical physics, mesoscopic physics, and nuclear physics with emphasis on prominent phenomena/subjects in quantum regimes, such as quantum resonances, superradiance, continuous quantum Zeno effect, quantum critical phenomena, Dirac spectra in quantum chromodynamics, and nonunitary conformal field theories. Finally, we introduce the notion of band topology in complex spectra of non-Hermitian systems and present their classifications by providing the proof, firstly given by this review in a complete manner, as well as a number of instructive examples. Other topics related to non-Hermitian physics, including nonreciprocal transport, speed limits, nonunitary quantum walk, are also reviewed.
\end{spacing}}
\vspace{11pt}
{\begin{spacing}{1.2}\fontsize{11pt}{11pt}\selectfont
\end{spacing}
}

\newpage

\newpage
\centerline{\bfseries Contents}\medskip
\hbox to \textwidth{\hsize\textwidth\vbox{\hsize48pc

\hspace*{-13pt} {1.  Introduction}\\
{2. Mathematical foundations of non-Hermitian physics}\\
\hspace*{8pt} {2.1. Spectral decomposition}\\
\hspace*{8pt} {2.2. Singular value decomposition and polar decomposition}\\
\hspace*{8pt} {2.3. Spectrum of non-Hermitian matrices}\\
\hspace*{8pt} {2.4. Eigenvectors of non-Hermitian matrices}\\
\hspace*{25pt} {2.4.1. Resolvent and perturbation formula of eigenprojector}\\
\hspace*{25pt} {2.4.2. Petermann factor}\\
\hspace*{8pt} {2.5. Pseudo Hermiticity and quasi Hermiticity}\\
\hspace*{8pt} {2.6. Exceptional points}\\
\hspace*{25pt} {2.6.1. Definition and basic properties}\\ 
\hspace*{25pt} {2.6.2. Physical applications}\\
\hspace*{25pt} {2.6.3. Topological properties}\\
{3. Non-Hermitian classical physics}\\
\hspace*{8pt} {3.1. Photonics}\\
\hspace*{25pt} {3.1.1. Optical wave propagation}\\
\hspace*{25pt} {3.1.2. Light scattering in complex media}\\
\hspace*{8pt} {3.2. Mechanics}\\
\hspace*{8pt} {3.3. Electrical circuits}\\
\hspace*{8pt} {3.4. Biological physics, transport phenomena, and neural networks}\\
\hspace*{25pt} {3.4.1. Master equation and transport in biological physics}\\
\hspace*{25pt} {3.4.2. Random matrices in population evolution and machine learning}\\
\hspace*{8pt} {3.5. Optomechanics and optomagnonics}\\
\hspace*{8pt} {3.6. Hydrodynamics}\\
\hspace*{25pt} {3.6.1. Non-Hermitian acoustics in fluids, metamaterials, and active matter}\\
\hspace*{25pt} {3.6.2. Exciton polaritons and plasmonics}\\
{4. Non-Hermitian quantum physics}\\
\hspace*{8pt} {4.1. Feshbach projection approach}\\
\hspace*{25pt} {4.1.1. Non-Hermitian operator}\\
\hspace*{25pt} {4.1.2. Quantum resonances}\\
\hspace*{25pt} {4.1.3. Superradiance}\\
\hspace*{25pt} {4.1.4. Physical applications}\\
\hspace*{8pt} {4.2. Quantum optical approach}\\
\hspace*{25pt} {4.2.1. Indirect measurement and quantum trajectory}\\
\hspace*{25pt} {4.2.2. Role of conditional dynamics}\\
\hspace*{8pt} {4.3. Quantum many-body physics}\\
\hspace*{25pt} {4.3.1. Criticality, dynamics, and chaos}\\
\hspace*{25pt} {4.3.2. Physical systems}\\
\hspace*{25pt} {4.3.3. Beyond the Markovian regimes}\\
\hspace*{8pt} {4.4. Quadratic problems}\\
\hspace*{8pt} {4.5. Nonunitary conformal field theory}\\
\hspace*{8pt} {4.6. Non-Hermitian analysis of Hermitian systems}\\
{5. Band topology in non-Hermitian systems}\\
\hspace*{8pt} {5.1. Brief review of band topology in Hermitian systems}\\
\hspace*{25pt} {5.1.1. Definition of band topology}\\
\hspace*{25pt} {5.1.2. Prototypical systems}\\
\hspace*{25pt} {5.1.3. Periodic table for Altland-Zirnbauer classes}\\
\hspace*{8pt} {5.2. Complex energy gaps}\\
\hspace*{8pt} {5.3. Prototypical examples and topological invariants}\\
\hspace*{8pt} {5.4. Bulk-edge correspondence}\\
\hspace*{8pt} {5.5. Topological classifications}\\
\hspace*{25pt} {5.5.1. Bernard-LeClair classes}\\
\hspace*{25pt} {5.5.2. Periodic tables}\\
\hspace*{25pt} {5.5.3. Non-Hermitian-Hermitian correspondence}\\
{6. Miscellaneous subjects}\\
\hspace*{8pt} {6.1. Nonreciprocal transport}\\
\hspace*{8pt} {6.2. Speed limits, shortcuts to adiabacity, and quantum thermodynamics}\\
\hspace*{8pt} {6.3. Miscellaneous topics on non-Hermitian topological systems}\\
{7. Summary and outlook}\\
\\
{Appendix A. Details on the Jordan normal form and the proofs}\\
{Appendix B. General description of quadratic Hamiltonians}\\
{Appendix C. Bound on correlations in matrix-product states}\\
{Appendix D. Continuous Hermitianization of line-gapped Bloch Hamiltonians}\\
{Appendix E. Topological classifications of the Bernard-LeClair classes}\\
}}
\end{abstract}

\newpage
\section{Introduction\label{sec1}}

Hermiticity of a Hamiltonian is one of the key postulates in quantum mechanics. It ensures the conservation of probability in an isolated quantum system and guarantees the real-valuedness of an expectation value of energy with respect to a quantum state. However, it is ubiquitous in nature that the probability effectively becomes nonconserving due to the presence of flows of energy, particles, and information to external degrees of freedom that are out of the Hilbert space of our interest. Historically, studies of such {\it open} systems date back to the early works by Gamow \cite{GG28}, Siegert \cite{SAJ39},  Majorana \cite{ME06}, and Feshbach \cite{HF58,HF62}. They considered the radiative decay in reactive nucleus, which has been analyzed by an effective non-Hermitian Hamiltonian associated with the decay of the norm of a quantum state, indicating the presence of nonzero probability flow to the outside of nucleus. A theoretical approach developed along this line, known as the Feshbach or Cohen-Tannoudji projection approach \cite{HF58,HF62,CCT68}, has found its applications in numerous subsequent studies in mesoscopic physics as well as atomic and molecular physics. 

Meanwhile, with the advances in controlling quantum coherence, yet another theoretical approach to open quantum systems has been developed in the field of quantum optics, where researchers explore few-body regimes with a highly controllable setup \cite{CCT98}. In contrast, a phenomenon of great interest in quantum physics is the collective behavior that can emerge only when a large number of constituent particles interact with each other \cite{APW72}. Recent advances in atomic, molecular and optical (AMO) physics have indeed enabled one to study open systems in such {\it many-body} regimes. From a broad perspective, these developments shed new light on the earlier studies that have once been considered as purely academic interest, such as the Yang-Lee  singularity \cite{LTD52} in nonunitary quantum field theory \cite{MEF78}.

On another front, a non-Hermitian description has also been applied to a variety of nonconservative {\it classical} systems, which provide a versatile platform for exploring unconventional wave phenomena among many subfields, including optics, photonics, electronic circuits, mechanical systems, optomechanics, biological transport, acoustics, and fluids. This is made possible by a formal equivalence between single-particle quantum mechanics and the classical wave equation as first pointed out by Schr{\"o}dinger \cite{SE26}. Owing to such an equivalence and the wave nature, band theory originally developed in solid-state physics can straightforwardly be applied to classical systems when they possess periodic structures.  Recently, mainly motivated by this formal relevance of non-Hermitian wave physics to classical systems, the concept of band topology has been extended to non-Hermitian regimes, leading to rich phenomena going beyond the conventional Hermitian band theory in condensed matter physics.   

\begin{table*}[b]
\caption{\label{table1} A wide variety of classical and quantum systems described by non-Hermitian matrices/operators together with their physical origins of non-Hermiticity, presented in order of appearance in the present review. }
\footnotesize
\begin{tabular}{p{3.5cm}p{5cm}p{5cm}l}
\midrule \midrule
Systems / Processes & Physical origin of non-Hermiticity& Theoretical methods \\ \midrule 
Photonics & Gain and loss of photons & 
Maxwell equations \cite{AR05,REG07}
\\
Mechanics   & Friction & Newton equation \cite{KCL14,HSD16}
 \\
Electrical circuits &  Joule heating &  Circuit equation \cite{YNJ09}
\\
Stochastic processes & Nonreciprocity of state transitions & 
Fokker-Planck equation \cite{RH96,GCW04}
\\
Soft matter and fluid & Nonlinear instability & 
Linearized hydrodynamics \cite{MMC13,CI13,JXZ16}
\\
Nuclear reactions & Radiative decays & Projection methods \cite{HF58,HF62,CCT68}
\\
Mesoscopic systems& Finite lifetimes of resonances & Scattering theory \cite{CDT92,NM98}
\\
Open quantum systems & Dissipation& Master equation \cite{GV76,LG76}
\\
Quantum measurement & Measurement backaction& Quantum trajectory approach \cite{MU89,UM90,DJ92,DR92,GCH92,HC93}
\\
\midrule \midrule
\end{tabular}
\end{table*}

The main goal of the present review is to discuss physics of open systems in both quantum and classical regimes in a coherent way from a common perspective of non-Hermitian physics (see Table~\ref{table1}). While there have certainly been more studies than we can cover here, we have attempted our best to clarify essential aspects of non-Hermitian physics in a self-contained manner by including key contributions as many as possible. Each section is self-contained and can be read by its own. A common physical mechanism behind seemingly different phenomena is elucidated and stressed wherever appropriate. It is our hope that this review article will be a useful guide for researchers who are interested in this burgeoning field. Going beyond the works included in this review, interested readers may refer to other excellent reviews that are more specialized in particular areas/topics, such as optics \cite{MH17,EG18,Mirieaar7709}, acoustics \cite{Mae1501595,CSA16,FZN19}, parity-time-symmetric systems \cite{CMB07rev,PD07,SKO19},    mesoscopic physics and quantum resonances \cite{NM98,JGM04,CH15,RS17}, 
ultracold atoms \cite{DAJ14,KS16}, driven-dissipative systems  \cite{DH10,MM12,RH13,Sieberer_2016,WH19}, optomechanics \cite{AM14}, exciton-polariton condensates \cite{CI13,Schneider_2016}, nonlinear phenomena \cite{KVV16}, biological transport \cite{Chou_2011},  active matter \cite{Marchetti2013}, and random matrices and disordered systems \cite{BCWJ97,Fyodorov_2003,MGE10}.

It is our hope that this review article serves to bridge an apparent gap between different branches of physics so far mainly discussed in a separate manner, such as open quantum systems, classical optics, biophysics, statistical physics, and topological physics. For this purpose, below we summarize and address questions that are frequently asked by newcomers in the field by referring to the contents presented in the review article for further details. 
\vspace{7pt}
\begin{center}
\textit{ \textbf{Frequently asked questions}}
\end{center}
\vspace{10pt}
\noindent{\it I have no experience about non-Hermitian matrices. Could you remind me the basic mathematical properties in the latter?}
\\
\\
In Sec.~\ref{sec2}, we summarize important theorems in non-Hermitian linear algebra in a mathematically coherent and pedagogical manner. 
There, we also introduce and explain the central concepts such as exceptional points, biorthogonality of eigenstates, pseudo-Hermiticity and parity-time symmetry, which will play major roles in our understanding of non-Hermitian classical and quantum physics in the subsequent sections. 
\\
\\
{\it I often encounter parity-time (PT) symmetric systems in non-Hermitian physics, but what is advantage/significance of considering this class of systems?}
\\
\\ 
Historically, PT-symmetric non-Hermitian systems have originally been introduced and investigated with purely academic interests as they can permit entirely real spectra, despite being non-Hermitian. Their physical significance was later clarified mainly together with experimental advances in controlling gain and loss in photonics to simulate PT-symmetric wave phenomena. There, rich and unconventional physical properties are found to typically occur in the vicinity of the spectral transition in PT-symmetric non-Hermitian systems. While  related phenomena can be realized in non-Hermitian systems with other antilinear symmetries, an advantage of the PT symmetry is its simplicity that allows one to implement the symmetry by spatially engineering gain-loss structures. In Sec.~\ref{secphqh}, we introduce and explain the notion of PT symmetry and its generalization,  pseudo-Hermiticity, by providing a number of examples relevant to experiments. 
\\
\\
{\it I suppose that a non-Hermitian description is approximate and can be effective only when certain conditions are satisfied. Under what type of physical situations can non-Hermitian descriptions be useful?}
\\
\\
In classical regimes, a non-Hermitian description can typically be justified on the basis of a formal equivalence between a linearized wave equation and the one-body Schr{\"o}dinger equation. Thus, the range of validity of a non-Hermitian description hinges on which physical system one uses to realize nonconservative setups. In Sec.~\ref{sec3}, we explain how and when a non-Hermitian description can be applied to analyze each of a diverse range of such classical systems. In contrast, in quantum regimes there are mainly two common frameworks, in which the non-Hermitian description can be relevant, known as the Feshbach projection approach and the quantum trajectory approach.  In Sec.~\ref{sec4}, we clarify in detail under what physical situations these approaches can be justified and applied to understand rich physics of various open quantum systems. We also review several attempts to go beyond these frameworks in Sec.~\ref{Sec:BMR}.
\\
\\
{\it What  are interesting aspects of non-Hermitian systems from a phenomenological perspective of physics research? Please provide prominent examples.}
\\
\\
Physical phenomena that are especially interesting from an application point of view include unconventional wave transport known as unidirectional invisibility, enhanced transmission and enhanced sensitivity owing to spectral singularity known as exceptional points, chiral and topological phenomena protected by the Riemann surface structures, coherent perfect absorption and single-mode lasing, robust biological transport, and even efficient learning of deep neural networks; a detailed explanation and further examples can be found in Secs.~\ref{secepphys} and \ref{sec3}. 
While these phenomena mainly emerge in classical regimes, there also exist a number of prominent physical phenomena occurring in quantum regimes. Examples include quantum resonances, superradiance, continuous quantum Zeno effect, quantum critical phenomena, the Dirac spectra in quantum chromodynamics, and nonunitary conformal field theories having peculiar algebraic structures; a detailed explanation and further examples can be found in Secs.~\ref{FP}, \ref{Sec:QMBP}, and \ref{nonunitary_cft}.  While these studies are often motivated from purely academic interest, many of them are indeed relevant to state-of-the-art experimental systems as discussed in Secs.~\ref{Sec:4pa}  and \ref{Sec:PS}. 
Finally, in view of a surge of interest in topological physics, non-Hermitian systems may find applications to realizing novel edge modes. From a phenomenological perspective, the most prominent example in this context is lasing of a topological edge mode. The so-called skin effect might also be of interest; however, at the present time it is not obvious whether or not this phenomenon is something more than other known phenomena such as convective flow in fluids or atmospheric pressure change due to gravity. A more detailed account of these phenomena can be found in Secs.~\ref{sechydro} and \ref{sec:bec}. 
\\
\\
{\it I get interested in non-Hermitian extensions of topological bands. However, the fundamental postulate in the field of topological physics, namely the bulk-edge correspondence, appears to be violated in non-Hermitian regimes. Then, how can one trust topological invariants/periodic tables as indicators for inferring the presence of edge modes in the first place?}
\\
\\ 
It is true that the bulk-edge correspondence in non-Hermitian regimes is very subtle and complicated, and that it is an on-going subject. However, there exist several efforts on restoring this correspondence by, e.g., modifying the definition of edge modes or introducing modified topological invariants appropriate for open boundary conditions. Further discussions can be found in Sec.~\ref{sec:bec}. In Sec.~\ref{ceg}, we  introduce possible extensions of band topology to complex spectra of non-Hermitian systems. In Secs.~\ref{Sec:5peti}, \ref{Sec:5topcl} and Appendices~\ref{app4},\ref{app5}, we present their classifications in a  constructive manner by providing the proof (firstly given by the present review in a complete way) as well as a number of prototypical examples. For readers who are not familiar with topological physics, we provide a concise summary of key results on the band topology in Hermitian systems in Sec.~\ref{Sec:5rev}. 
\\
\\
{\it Are there any significant other developments in non-Hermitian physics?}
\\
\\
Other developments of particular interest, including nonreciprocal transport, speed limits, shortcuts to adiabacity, quantum thermodynamics, higher-order topology, nonunitary quantum walk, are reviewed in Sec.~\ref{sec6}.

\section{Mathematical foundations of non-Hermitian physics\label{sec2}}
We give a general and mathematically coherent description of non-Hermitian matrices. We present important theorems including the Jordan normal form, the singular value decomposition and the spectral sensibility. Key notions such as exceptional points, biorthogonality of eigenstates, pseudo-Hermiticity are delineated in relation to the parity-time symmetry. We illustrate these concepts in prototypical examples that find applications in physics.

\subsection{Spectral decomposition\label{secspecdec}}
\noindent{\it Eigenvalue and Eigenvector}

\vspace{3pt}
\noindent
We focus our attention on generic square matrices with complex entries, i.e, $M\in\mathbb{C}^{n\times n}$. Compared with $c$-numbers, the calculation and analysis of matrices are much more complicated, especially when the dimension $n$ is very large. Nevertheless, if we consider the action of $M$ on some \emph{nonzero} vector $\bold{v}\in\mathbb{C}^n\backslash\{\bold{0}\}$ that satisfies 
\begin{equation}
M\bold{v}=\lambda \bold{v},\;\;\;\; \lambda\in\mathbb{C},
\label{Mvl}
\end{equation}
then $M$ behaves simply as a multiplier of $\lambda$. In particular, for an arbitrary polynomial $p(x)=\sum^N_{l=1}c_lx^l$, we have
\begin{equation}
p(M)\bold{v}\equiv\sum^N_{l=1} c_l M^l\bold{v}=\sum^N_{l=1}c_l\lambda^l\bold{v}=p(\lambda)\bold{v}.
\end{equation}
Whenever a pair $(\lambda,\bold{v})$ satisfies Eq.~(\ref{Mvl}), we call $\lambda$ and $\bold{v}$ an \emph{eigenvalue} and an \emph{eigenvector} of $M$, respectively \cite{CDM00}. 

We first discuss how to determine the eigenvalues of a matrix. Denoting $I$ as the identity matrix in $\mathbb{C}^{n\times n}$, we can rewrite Eq.~(\ref{Mvl}) into 
\begin{equation}
(\lambda I-M)\bold{v}=\bold{0}, 
\end{equation}
implying that $\lambda I-M$ \emph{cannot} be invertible (otherwise $\bold{v}=\bold{0}$). Therefore, all the possible eigenvalues $\lambda$'s can be determined from
\begin{equation}
p_M(\lambda)\equiv {\rm det}(\lambda I-M)=0,
\label{pM}
\end{equation}
where $p_M(\lambda)$ is called the \emph{characteristic polynomial} of $M$. According to the fundamental theorem of algebra \cite{LVA79}, we can generally decompose $p_M(\lambda)$ into
\begin{equation}
p_M(\lambda)=\prod^J_{j=1}(\lambda-\lambda_j)^{m^{\rm a}_j},
\end{equation}
where $\lambda_{j}\neq\lambda_{j'}$ for $\forall j\neq j'$ and $m^{\rm a}_j$'s are called \emph{algebraic multiplicities}, which are positive integers satisfying $\sum^J_{j=1}m^{\rm a}_j=n$. The {\emph{spectrum}} of $M$, denoted as $\Lambda(M)$, is defined by the union of all the eigenvalues with their (algebraic) multiplicities taken into account:
\begin{equation}
\Lambda(M)\equiv\bigcup^J_{j=1}\{\lambda_j\}^{\cup m^{\rm a}_j}.
\end{equation}
Note that each element in the \emph{multiset} $\Lambda(M)$ is real for a Hermitian matrix $M$, since 
\begin{equation}
\lambda=\frac{\bold{v}^\dag M\bold{v}}{\bold{v}^\dag\bold{v}}=\frac{\bold{v}^\dag M^\dag\bold{v}}{\bold{v}^\dag\bold{v}}=\lambda^*
\label{lamreal}
\end{equation}
for any pair of eigenvalue $\lambda$ and eigenvector $\bold{v}$. Here $\dag$ is the Hermitian conjugation (i.e., transposition $\rm T$ combined with complex conjugation $*$). This is no longer the case for a non-Hermitian matrix, whose spectrum is complex in general.
\\
\\
{\it Jordan normal form}

\vspace{3pt}
\noindent
Given an eigenvalue $\lambda_j$, we can find one or several eigenvectors $\bold{v}_j$. We define the \emph{eigenspace} associated with $\lambda_{j}$ as
\begin{equation}
\textsf{V}_M(\lambda_j)\equiv{\rm Ker}(M-\lambda_jI)\equiv {\rm span}\{\bold{v}_j:M\bold{v}_j=\lambda_j\bold{v}_j,\bold{v}_j\in\mathbb{C}^n\}. 
\label{VM}
\end{equation}
Then its dimension $m^{\rm g}_j\equiv{\rm dim}\textsf{V}_M(\lambda_j)$ must be a positive integer, which is called \emph{geometric multiplicity}. For any eigenvalue $\lambda_j$, we have
\begin{equation}
m^{\rm g}_j\le m^{\rm a}_j
\end{equation}
because $\textsf{V}_M(\lambda_j)$ is an \emph{invariant subspace} of $M$ (i.e., for any vector $\bold{v}$ in the subspace, $M\bold{v}$ still lies in the same subspace), implying that $p_M(\lambda)$ must contain a factor $(\lambda-\lambda_j)^{m^{\rm g}_j}$.

If $M$ is Hermitian, we have $m^{\rm g}_j=m^{\rm a}_j$. This is because whenever we find at least one normalized eigenvector $\bold{v}_j$ with eigenvalue $\lambda_j$, we can construct a Hermitian projector $P_j=\bold{v}\bold{v}^\dag$ such that $P_jM=MP_j=\lambda_j P_j$. This implies that $M$ can be block diagonalized into
\begin{equation}
M=\lambda_j P_j+(1-P_j)M(1-P_j).
\label{MP1P}
\end{equation}
The characteristic polynomial for $(1-P_j)M(1-P_j)$ should be $(\lambda-\lambda_j)^{-1}p_M(\lambda)$, so we can repeat the above procedure as long as $m^{\rm a}_j-1>0$. Finally, we can find $m^{\rm a}_j$ orthogonal eigenvectors, which constitute a complete basis of $\textsf{V}_M(\lambda_j)$. 

If $M$ is non-Hermitian, things are much more complicated in general. For instance, even if we find an eigenvector $\bold{v}_j$, it is not ensured that $M$ can be block diagonalized on $\bold{v}_j$ and $\mathbb{C}^n\backslash\bold{v}_j$. We illustrate below the simplest example of this.

\exmp{\label{minimalnond}(Nondiagonalizable $2\times 2$ matrix).
We consider
\begin{equation}
M=\begin{bmatrix} \;\lambda\; & \;1\;\; \\ \;0\; & \;\lambda\;\; \end{bmatrix},
\label{Mlam}
\end{equation}
which has the eigenvector $\bold{v}=(1,0)^{\rm T}$ with a doubly degenerate eigenvalue $\lambda$. However, the decomposition in Eq.~(\ref{MP1P}) cannot be applied because there exists a nonzero off-diagonal component. In this case, we have $m^{\rm a}=2>m^{\rm g}=1$.
} 

\begin{figure}[!t]
\begin{center}
\includegraphics[width=5cm]{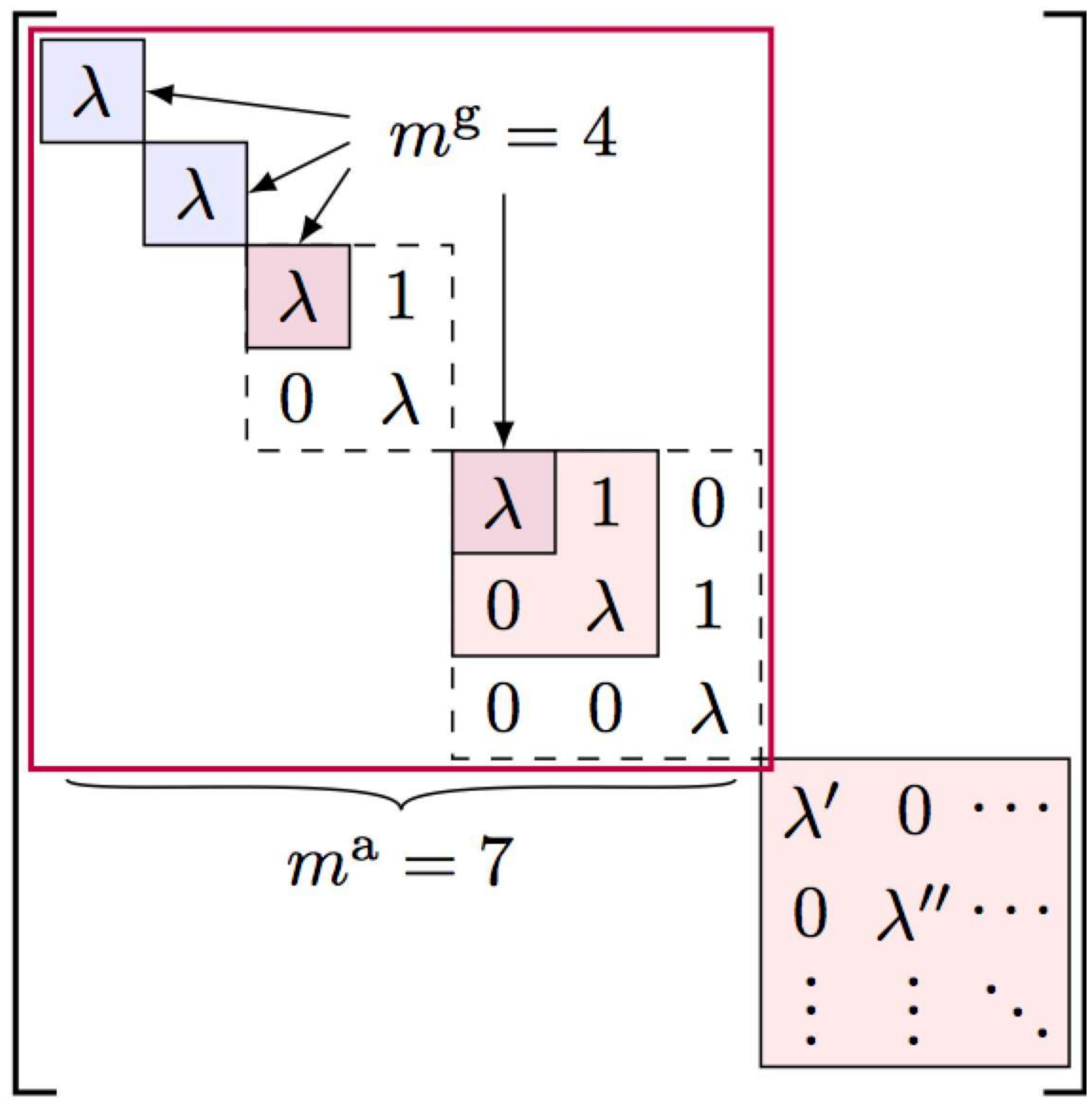}
\end{center}
\caption{Example of the Jordan normal form of a non-Hermitian matrix,  where {for eigenvalue $\lambda$} its geometric multiplicity $m^{\rm g}=4$ is smaller than the algebraic multiplicity $m^{\rm a}=7$ due to the existence of size-$2$  and  size-$3$ Jordan blocks.  Equation~\eqref{jnfrel} is satisfied as $1+1+2+3=7$. 
}
\label{fig:Jordan}
\end{figure}

Nevertheless, we can still simplify an arbitrary matrix $M$ to the {\it Jordan normal form} \cite{CDM00}, which can be considered as the generalization of the diagonalized form for Hermitian matrices to general non-Hermitian matrices as follows (see Fig.~\ref{fig:Jordan} for an illustrative example and Appendix~\ref{app1} for the proof).
\thm{[Jordan normal form]
\label{Thm:JNF}
For any square matrix $M\in\mathbb{C}^{n\times n}$, we can always find an invertible matrix $V\in\mathbb{C}^{n\times n}$ (not unique) such that $M$ is related to a direct sum of Jordan blocks via the similarity transformation by $V$, i.e.,
\begin{equation}
M=V\left[\bigoplus^J_{j=1}\bigoplus^{m^{\rm g}_j}_{\alpha=1} J_{n_{j\alpha}}(\lambda_j)\right]V^{-1},
\label{JNF}
\end{equation}
where $\lambda_j$ is an eigenvalue of $M$ with geometric and algebraic multiplicities $m^{\rm g}_j$ and $m^{\rm a}_j$, and $J_{n}(\lambda_j)$ is the size-$n$ Jordan block with eigenvalue $\lambda_j$ defined by
\begin{equation}
J_{n}(\lambda_j)\equiv
\begin{bmatrix}  
\;\lambda_j\; & \;1\; & \;0\; & \cdots & \;0\; & \;0\; \\ 
\;0\; & \;\lambda_j\; & \;1\; & \cdots & \;0\; & \;0\; \\ 
\;0\; & \;0\; & \;\lambda_j\; & \cdots & \;0\; & \;0\; \\ 
\vdots & \vdots & \vdots & \ddots & \vdots & \vdots \\ 
0 & 0 & 0 & \cdots & \;\lambda_j\; & \;1\; \\
0 & 0 & 0 & \cdots & 0 & \;\lambda_j\; 
\end{bmatrix}_{n\times n}.\label{njblock}
\end{equation}
The positive integer $n_{j\alpha}$, which is the size of the $\alpha$th Jordan block with eigenvalue $\lambda_j$, satisfies the relation
\begin{equation}
\sum^{m^{\rm g}_j}_{\alpha=1}n_{j\alpha}=m^{\rm a}_j,\;\;\;\;
\forall j=1,2,...,J.\label{jnfrel}
\end{equation}
}

\vspace{5pt}

\noindent{\it Spectral decomposition}

\vspace{3pt}
\noindent
If we decompose each Jordan block in Eq.~(\ref{JNF}) into a diagonal part and an off-diagonal part, i.e.,
\begin{equation}
J_{n_{j\alpha}}(\lambda_j)=\lambda_j I_{n_{j\alpha}}+J_{n_{j\alpha}},
\end{equation} 
where $[I_{n_{j\alpha}}]_{ab}=\delta_{ab}$ is the identity matrix and $[J_{n_{j\alpha}}]_{ab}=\delta_{a,b-1}$ ($a,b=1,2,...,n_{j\alpha}$), we obtain the \emph{spectral decomposition} as \cite{TK80}
\begin{equation}
\begin{split}
M&=D+N,\\
D&=\sum^J_{j=1}\sum^{m^{\rm g}_j}_{\alpha=1} \lambda_jP_{j\alpha},\;\;\;\;N=\sum^J_{j=1}\sum^{m^{\rm g}_j}_{\alpha=1} N_{j\alpha}.
\end{split}
\label{spedec}
\end{equation}
Here $D$ is the diagonal part and 
$P_{j\alpha}$ is a rank-$n_{j\alpha}$ projector given by
\begin{equation}\label{2_projector}
P_{j\alpha}=VI_{j\alpha}V^{-1},\;\;\;\;
I_{j\alpha}\equiv \bigoplus^J_{j'=1}\bigoplus^{m^{\rm g}_{j'}}_{\alpha'=1} \delta_{j'j}\delta_{\alpha'\alpha}I_{n_{j'\alpha'}}.
\end{equation}
We note that $P_{j\alpha}$'s are generally non-Hermitian, yet they are orthogonal and complete: \begin{equation}
P_{j\alpha}P_{j'\alpha'}=\delta_{j'j}\delta_{\alpha'\alpha}P_{j\alpha},\;\;\;\;
\sum^J_{j=1}\sum^{m^{\rm g}_j}_{\alpha=1}P_{j\alpha}=I.
\label{onp}
\end{equation}
As for the off-diagonal part $N$, each component $N_{j\alpha}$ with $n_{j\alpha}\ge2$ (otherwise $N_{j\alpha}=0$) is a \emph{nilpotent} given by 
\begin{equation}\label{2_nilpotent}
N_{j\alpha}=VJ_{j\alpha}V^{-1},\;\;\;\;
J_{j\alpha}\equiv \bigoplus^J_{j'=1}\bigoplus^{m^{\rm g}_{j'}}_{\alpha'=1} \delta_{j'j}\delta_{\alpha'\alpha}J_{n_{j'\alpha'}}.
\end{equation}
This is a nilpotent since
\begin{equation}
N_{j\alpha}^{n_{j\alpha}-1}\neq0,\;\;\;\;
N_{j\alpha}^{n_{j\alpha}}=0.
\label{nip}
\end{equation}
We can also easily check that 
\begin{equation}
N_{j\alpha}N_{j'\alpha'}=\delta_{j'j}\delta_{\alpha'\alpha}N_{j\alpha}^2,\;\;\;\;
P_{j'\alpha'}N_{j\alpha}=N_{j\alpha}P_{j'\alpha'}=\delta_{j'j}\delta_{\alpha'\alpha}N_{j\alpha}. 
\label{pn}
\end{equation} 
We say that $M$ is \emph{diagonalizable} if and only if $N=0$ in Eq.~(\ref{spedec}), i.e., there is only the diagonal part $D$. This occurs when $n_{j\alpha}=1$ for $\forall j=1,2,...,J$ and $\forall\alpha=1,2,...,m^{\rm g}_j$ $\Leftrightarrow m^{\rm g}_j=m^{\rm a}_j$ for $\forall j=1,2,...,J$, as is always the case for Hermitian matrices. 
The notion of Jordan normal form and diagonalizability will play a central role in discussing \emph{exceptional points} \cite{TK80} for a continuous family of matrices (see Sec.~\ref{Sec:EP}), which are one of the most important concepts in non-Hermitian physics \cite{WDH12}. 
\\
\\
{\it Function of a matrix}

\vspace{3pt}
\noindent
As already mentioned in the beginning of this section, finding eigenvalues and eigenvectors of matrices can greatly simplify their algebraic calculations. This is because, when acting on an eigenvector, a matrix behaves just like a $c$-number, i.e., the eigenvalue. It turns out that all the eigenvectors, although linearly independent, generally \emph{do not} span the entire linear space. Nevertheless, we can find a complete basis on which the matrix action becomes a direct sum of Jordan blocks (see Theorem~\ref{Thm:JNF}). Thanks to Eqs.~(\ref{nip}) and (\ref{pn}), we will see that the calculations of Jordan blocks are only slightly more complicated than those of $c$-numbers.  

Specifically, we consider $f(M)$ for a general complex analytic function $f(z)$ on a domain ${\rm D}\subseteq\mathbb{C}$ containing $z_0$. By assumption, we can apply for $\forall z\in{\rm D}$ the Taylor expansion \cite{LVA79} 
\begin{equation}
f(z)=\sum^\infty_{l=0}c_l(z-z_0)^l. 
\end{equation}
Without loss of generality, we set $z_0=0$ since otherwise we can consider $f(z-z_0)$ and $M+z_0 I$, the latter of which shares the same $P_{j\alpha}$'s and $N_{j\alpha}$'s as $M$. Given the general spectral decomposition Eq.~(\ref{spedec}) and provided that the full spectrum falls into ${\rm D}$, we have (see Appendix~\ref{app1} for details)
\begin{equation}
f(M)=\sum^J_{j=1}\left[f(\lambda_j)P_j+\sum^{m^{\rm g}_j}_{\alpha=1}\sum^{n_{j\alpha}-1}_{p=1}\frac{1}{p!}f^{(p)}(\lambda_j)N_{j\alpha}^p\right],
\label{fM}
\end{equation}
where $P_j\equiv\sum^{m^{\rm g}_j}_{\alpha=1}P_{j\alpha}$ and $f^{(p)}$ is the $p$th derivative of $f$. Note that for diagonalizable matrices, especially Hermitian matrices, we simply have
\begin{equation}
f(M)=\sum^J_{j=1}f(\lambda_j)P_j.
\end{equation}
Moreover, for an arbitrary $M$, we can always get rid of the nilpotent part by taking the trace:
\begin{equation}
{\rm Tr}[f(M)]=\sum^J_{j=1}m^{\rm a}_j f(\lambda_j).
\end{equation}

As an important application of Eq.~(\ref{fM}), we consider the exponential of $M$, in which case we take $f(z)=e^{zt}$ with $t$ being a parameter. Note that $f(M)=e^{Mt}$ is nothing but the propagator of a general linear differential system with constant coefficients \cite{EAC55}: 
\begin{equation}
\frac{d\bold{v}_t}{dt}=M\bold{v}_t\;\;\;\;\Rightarrow\;\;\;\;
\bold{v}_t=e^{Mt}\bold{v}_0.
\end{equation}
Substituting $f^{(p)}(z)=t^pe^{zt}$ for $\forall p\in\mathbb{Z}^+\equiv\{1,2,3,\ldots\}$ into Eq.~(\ref{fM}), we obtain
\begin{equation}\label{jorlinear}
e^{Mt}=\sum^J_{j=1}e^{\lambda_jt}\left(P_j+\sum^{m^{\rm g}_j}_{\alpha=1}\sum^{n_{j\alpha}-1}_{p=1}\frac{t^p}{p!}N_{j\alpha}^p\right).
\end{equation}
In later sections, we will see that this result is relevant to, e.g., the linear-time growth at exceptional points \cite{Heiss2010}  and polynomial corrections to exponential decays in spatial correlations of matrix-product states \cite{FV08}.

\subsection{Singular value decomposition and polar decomposition\label{secsvd}}
We recall that the spectral decomposition (\ref{spedec}) is algebraically motivated in an attempt to simplify the algebraic calculations of a matrix into those of $c$-numbers. On the other hand, a matrix also has a clear \emph{geometric} meaning as a linear transformation of vectors. From this viewpoint, it is natural to ask whether a matrix can be decomposed into some fundamental geometric operations such as rotations and scalings. This is answered in the affirmative by the \emph{singular value decomposition}, which we define in the following.

For an arbitrary matrix $M\in\mathbb{C}^{n\times n}$, we can always perform the singular value decomposition as \cite{CDM00}
\begin{equation}
M=W\Sigma V^\dag,
\label{SVD}
\end{equation}
where $W$ and $V$ are both \emph{unitary} matrices in $\mathbb{C}^{n\times n}$ and $\Sigma={\rm diag}\{\sigma_j\}^n_{j=1}$ is a \emph{diagonal matrix of non-negative values} that constitue the \emph{singular-value spectrum} $\Sigma(M)$. The existence of such a decomposition implies that to obtain a general linear transformation, it is sufficient to combine just three fundamental operations --- two rotations and one scaling. Such a decomposition is particularly useful when we consider how the length of a vector, which is quantified by the vector norm $\|\bold{v}\|\equiv\sqrt{\bold{v}^\dag\bold{v}}$, changes upon a linear transformation. Indeed, only the scaling part $\Sigma$ contributes to the change of the vector norm. This implies that the \emph{operator norm} $\|M\|$ induced by the vector norm is nothing but the maximal singular value:
\begin{equation}
\|M\|\equiv\max_{\|\bold{v}\|=1}\|M\bold{v}\|=\max_{\|\bold{v}\|=1}\|W\Sigma V^\dag\bold{v}\|=\max_{\|\bold{v}\|=1}\|\Sigma\bold{v}\|=\max_{1\le j\le n}\sigma_j,
\label{Mnorm}
\end{equation}
where we have used the unitary invariance of the vector norm and the fact that $V^\dag\bold{v}$ goes over all the normalized vectors as $\bold{v}$ does.

To demonstrate the singular value decomposition, we first note that $M^\dag M$ is always Hermitian so that we can diagonalize it through a unitary transformation, i.e.,
\begin{equation}
V^\dag M^\dag MV = \Sigma^2,
\label{VMMV}
\end{equation}
for some unitary $V$ and real diagonal matrix $\Sigma\ge0$. If $M$ is invertible (i.e., $\mathrm{det}M\neq0\Leftrightarrow$ all the eigenvalues of $M$ are nonzero), then $\Sigma>0$ and we can introduce
\begin{equation}
W=MV\Sigma^{-1},
\label{WMVS}
\end{equation}
which turns out to be unitary since $W^\dag W=\Sigma^{-1}V^\dag M^\dag MV\Sigma^{-1}=I$ according to Eq.~(\ref{VMMV}). The singular value decomposition (\ref{SVD}) immediately follows from Eq.~(\ref{WMVS}). The argument can readily be generalized to the case when $M$ is noninvertible  (see Appendix~\ref{app1}). Note that Eq.~\eqref{VMMV} makes it clear that the singular value spectrum is nothing but the root of the spectrum of $M^\dag M$.

We introduce a closely related concept called as \emph{polar decomposition}, which takes the following form \cite{BCH15}:
\begin{equation}
M=UQ,
\label{PD}
\end{equation}
where $U$ is a unitary matrix and $Q$ is a \emph{positive semi-definite} Hermitian matrix. In fact, given a singular value decomposition as Eq.~(\ref{SVD}), we can immediately obtain a polar decomposition with 
\begin{equation}
U=WV^\dag,\;\;\;\;Q=V\Sigma V^\dag\equiv \sqrt{M^\dag M}. 
\label{UWV}
\end{equation}
Note that the singular value spectrum of $M$ is nothing but the spectrum of $Q$. Such a polar decomposition is \emph{unique} if $M$ is invertible. To see this, suppose there is another polar decomposition $M=U'Q'$, then at least 
\begin{equation}
M^\dag M=Q^2=Q'^2\;\;\;\;\Rightarrow\;\;\;\; (Q+Q')(Q-Q')=0.
\end{equation}
By assumption, both $Q$ and $Q'$ are positive definite, and thus $Q+Q'$ is also positive definite and invertible, implying $Q=Q'$. Accordingly, $U'=MQ'^{-1}=MQ^{-1}=U$. In contrast, the singular value decomposition is \emph{not} unique even if $M$ is invertible. This is because we can always perform a gauge transformation
\begin{equation}
V\to V\Theta,\;\;\;\; W\to W\Theta,
\label{VWgauge}
\end{equation} 
where $\Theta$ can be an arbitrary unitary matrix that commutes with $\Sigma$, such as $\Theta={\rm diag}\{e^{i\theta_j}\}^n_{j=1}$ for arbitrary real $\theta_j$'s. On the other hand, the uniqueness of polar decomposition together with Eq.~(\ref{UWV}) implies that $WV^\dag$ is unique, or equivalently, $W$ and $V$ are unique up to the gauge transformation given in Eq.~(\ref{VWgauge}).

Several remarks are in order. First, we note that if $M$ is normal, i.e.,
\eqn{M^\dagger M=MM^\dagger\Leftrightarrow M=VDV^\dagger\label{2normalmat}}
with $D$ and $V$ being the diagonal and unitary matrix, respectively, significant simplifications occur, namely, the diagonalization and the singular value decomposition (as well as the polar decomposition) essentially become equivalent. To see this, we can decompose the diagonal matrix $D=\Theta_D|D|$ into its phase $\Theta_D$ and absolute value $|D|$, leading to the following identifications:
\eqn{
M=\underbrace{V\Theta_D}_{W}\underbrace{|D|}_{\Sigma}V^\dagger=\underbrace{V\Theta_DV^\dagger}_{U}\underbrace{V|D|V^\dagger}_{Q},
}
where the first (second) equality can immediately be identified as the singular value decomposition (polar decomposition).
In this case, the singular value spectrum $\Sigma$ is simply the absolute value of the eigenspectrum $|D|$. 
In particular, these simplifications apply to any Hermitian or unitary matrices that are normal.

Second, however, there is in general no such a concrete relation between the singular values and the eigenvalues of a non-Hermitian matrix (or more precisely, a non-normal matrix).  
Nevertheless, defining the \emph{spectral radius} $\rho(M)$ of a matrix $M$ as
\begin{equation}
\rho(M)\equiv\max_{1\le j\le J}|\lambda_j|,
\end{equation}
we have \cite{CDM00}
\begin{equation}
\rho(M)\le\|M\|,\;\;\;\;\rho(M)=\lim_{k\to\infty}\|M^k\|^{\frac{1}{k}}.
\label{rhonorm}
\end{equation}
The left inequality can be understood from the definition of operator norm (\ref{Mnorm}), while the proof of the right identity is given in Appendix~\ref{app1}. If we regard $M$ as a transfer matrix that generates a series of vectors, the right identity in Eq.~(\ref{rhonorm}) means that the large-scale exponential decay or amplification is dominated by the largest eigenvalue.  

Finally, we remark that the singular value decomposition or polar decomposition cannot be possible in general in the case of {\it infinite} dimensions. To show this, it is sufficient to focus on the class of \emph{Fredholm operators}, which have finite-dimensional kernels and cokernels. Such an assumption enables us to define the \emph{index} as \cite{TK80}
\begin{equation}
{\rm ind}M\equiv{\rm dim\;Ker}M-{\rm codim\;Im}M.
\end{equation}
We can easily check that ${\rm ind}M=-{\rm ind}M^\dag$, so the index of a Hermitian operator is always zero. According to the additivity of $\rm ind$ for operator associations (matrix multiplications) and the fact that any unitary operator has a zero index, the possibility of singular value decomposition or polar decomposition necessarily leads to a zero index. Taking the contraposition, we find that the singular value decomposition or polar decomposition cannot be carried out for any Fredholm operator with a nonzero index. 

A remarkable implication of this fact in physics is the absence of the Hermitian phase operator for a boson mode. To see this, we only have to take $M=a$ to be the annihilation operator \cite{FK95}. This operator has index $1$ since $a$ annihilates the zero-particle state (vacuum) while each Fock state $|n\rangle$ can be obtained from $a|n+1\rangle$ followed by an appropriate normalization. That is, we have ${\rm dim\;Ker}a=1$, ${\rm codim\;Im}a=0$ and thus ${\rm ind}a=1$.

\subsection{Spectrum of non-Hermitian matrices\label{secspecnh}}
In this section, we discuss the stabilities of eigenvalues and singular values of general matrices upon some perturbations. That is, given $M,E\in\mathbb{C}^{n\times n}$, we would like to ask how much the spectrum or the singular value spectrum of $M$ can differ from those of $M+E$, especially when $\|E\|$ is small. It turns out that the singular value spectrum is always stable, but the eigenvalue spectra of non-Hermitian matrices can change dramatically.
\\
\\
{\it Spectral stability of Hermitian matrices}

\vspace{3pt}
\noindent
Before discussing general non-Hermitian matrices, it is instructive to first analyze the case of Hermitian matrices, whose singular value spectra are simply the absolute values of the spectra. We will see that the spectra of Hermitian matrices are stable in the sense that the spectral shift is rigorously bounded by the norm of perturbation. This result actually implies that the singular value spectra of general matrices are stable.

We first consider the case in which \emph{both}  the original matrix $M$ and the perturbation $E$ are Hermitian. In this case, the spectral stability is ensured by  \emph{Weyl's perturbation theorem}  \cite{RB97} (see Appendix~\ref{app1} for the proof): 
\begin{theorem}[Weyl's perturbation theorem]
For two arbitrary $n\times n$ Hermitian matrices $M$ and $E$, denoting the $j$th largest eigenvalue in $\Lambda(M)$ and $\Lambda(M+E)$ as $\lambda_j$ and $\lambda'_j$, respectively, we have
\begin{equation}
|\lambda'_j-\lambda_j|\le \|E\|,\;\;\;\;\forall j=1,2,...,n.
\label{Weylpert}
\end{equation}
\end{theorem}
We emphasize that while the title of the theorem involves ``perturbation", this theorem is valid no matter how large $\|E\|$ is. The equality can be achieved by choosing $E\propto I$. Even if $[M,E]=0$, this inequality is already nontrivial since the perturbation may change the order of $\lambda_j$'s. Of course, the inequality holds true even if $M$ and $E$ do not commute.

As an important application of  Weyl's perturbation theorem, let us demonstrate that the singular-value spectrum is always stable for general matrices and perturbations. To this end, we first note that the singular value spectrum of an arbitrary $M\in\mathbb{C}^{n\times n}$ is the non-negative part in the eigenspectrum of the
\emph{Hermitianized} matrix \cite{JF97}
\begin{equation}
H_M=\begin{bmatrix} \;0\; & \;\;M\;\; \\ \;M^\dag\; & \;\;0\;\; \end{bmatrix}. 
\label{HM}
\end{equation}
This is because, according to the singular value decomposition $M=W\Sigma V^\dag$, we can diagonalize $H_M$ as
\begin{equation}
H_M=U \begin{bmatrix} \;\Sigma\; & \;0\; \\ \;\;0\;\; & \;\;-\Sigma\;\; \end{bmatrix}U^\dag,\;\;\;\; 
U=\frac{1}{\sqrt{2}}\begin{bmatrix} \;W\; & \;\;W\;\; \\ \;V\; & \;\;-V\;\; \end{bmatrix},
\end{equation}
from which it is clear that the spectrum of $H_M$ is the combination of the singular value of $M$ with its minus copy. Such an inversion invariance of the spectrum of  $H_M$ is actually due to an underlying \emph{chiral symmetry} $\Gamma$ that anti-communtes with $H_M$: 
\begin{equation}
\Gamma H_M \Gamma^{-1}=-H_M,\;\;\;\;
\Gamma\equiv\begin{bmatrix} \;I\; & \;\;0\;\; \\ \;0\; & \;\;-I\;\; \end{bmatrix}.
\label{HMchiral}
\end{equation}
We mention that such a Hermitianization technique (\ref{HM}), which was originally developed for dealing with non-Hermitian random matrices \cite{JF97}, will play a crucial role in the classification of non-Hermitian Bloch Hamiltonians \cite{ZG18}. 
Now we apply Weyl's perturbation theorem to $H_M$ and $H_{M+E}$, obtaining the following theorem. 
\begin{theorem}[Stability of singular value spectra]
For two arbitrary $n\times n$ matrices $M$ and $E$, denoting the $j$th largest singular value of $M$ and that of $M+E$ as $\sigma_j$ and $\sigma'_j$, respectively, we have 
\begin{equation}
|\sigma'_j-\sigma_j|\le\|E\|,\;\;\;\;\forall j=1,2,...,n.
\end{equation}
\end{theorem} 
\leftline{Here we have used the fact that $\|H_E\|=\|E\|$.}

We then consider a general non-Hermitian perturbation to a Hermitian matrix. In this case, we generally cannot define an order relation for the \emph{complex} spectrum of $M+E$. Nevertheless, the following theorem holds. 
\begin{theorem}[Stability against general perturbations]
\label{HNH}
For an arbitrary $n\times n$ Hermitian matrix $M$ and an arbitrary $n\times n$ matrix $E$, denoting the $j$th largest eigenvalue in $\Lambda(M)$ as $\lambda_j$, we have
\begin{equation}
\min_j|\lambda'-\lambda_j|\le \|E\|,\;\;\;\;\forall \lambda'\in\Lambda(M+E).
\label{NHpert}
\end{equation}
\end{theorem}
\noindent This result implies that $\Lambda(M+E)$ is completely covered by
\begin{equation}
\bigcup^n_{j=1}\{z\in\mathbb{C}:|z-\lambda_j|\le\|E\|\},
\end{equation}
which means that the spectra of Hermitian matrices are also robust against non-Hermitian perturbations. We do not provide the proof of Theorem~\ref{HNH} here since this is actually a corollary of a more general theorem on the spectral shift for non-Hermitian matrices perturbed by non-Hermitian matrices (see Theorem~\ref{NHSS} below).
\\
\\
{\it Spectral sensibility of non-Hermitian matrices}

\vspace{3pt}
\noindent
In stark contrast to Hermitian matrices, if a non-Hermitian matrix $M$ is perturbed by an arbitrary matrix $E$, the spectral shift can no longer be bounded by $\|E\|$ in general. 

\exmp{\label{specsens}(Boundary sensibility of the Hatano-Nelson model). 
We illustrate the spectral sensibility of non-Hermitian matrices by taking an unperturbed matrix $M$ to be a size-$n$ Jordan block $J_n(\lambda)$ with eigenvalue $\lambda$ whose spectrum is $\Lambda(M)=\{\lambda\}^{\cup n}$ (cf. Eq.~\eqref{njblock}). If we perturb $M$ by the non-Hermitian matrix $[E]_{mm'}=\epsilon\delta_{mn}\delta_{m'1}$, we find that the spectrum becomes
\begin{equation}
\Lambda(M+E)=\{\lambda+\epsilon^\frac{1}{n}e^{i\frac{2\pi m}{n}}:m=1,2,...,n\},
\label{MEep}
\end{equation}
which lies on a circle centered at $\lambda$ with radius $|\epsilon|^{\frac{1}{n}}$. The spectral shift is thus $|\epsilon|^{\frac{1}{n}}$, which can be much larger than $\|E\|=|\epsilon|$ for a small $|\epsilon|$. In other words, if we want to achieve a spectral shift with order $\epsilon$, then $\|E\|$ only has to be \emph{exponentially} (in terms of $n$) small as $\epsilon^n$. We mention that $M=J_n(\lambda)$ can actually be considered as the maximally asymmetric limit of the \emph{Hatano-Nelson model} \cite{HN96} with a constant on-site potential and under the open boundary condition. The perturbation corresponds to a slight modification of the boundary condition, which, however, leads to a dramatic change of the spectrum. In the thermodynamic limit (i.e., $n\to\infty$), the spectral shift is always one for an arbitrarily small but nonzero $\epsilon$.
}

In general, we can show the following theorem on the spectral sensibility of non-Hermitian matrices \cite{RB97} (see Appendix~\ref{app1} for the proof):
\begin{theorem}[Spectral shift]
\label{NHSS}
Consider a diagonalizable $n\times n$ matrix $M=VDV^{-1}$, where $V$ is invertible and $D$ is diagonal, and an arbitrary $n\times n$ matrix $E$. Denoting the spectrum of $M$ and $M+E$ as $\{\lambda_j\}^n_{j=1}$ and $\{\lambda'_j\}^n_{j=1}$, respectively, we have
\begin{equation}
\max_j\min_{j'}|\lambda'_j-\lambda_{j'}|\le{\rm cond}(V)\|E\|,
\label{SpeS}
\end{equation}
where ${\rm cond}(V)\equiv \|V\|\|V^{-1}\|\ge1$ is the condition number\footnote{The condition number of an invertible matrix $A$ measures how much the error in the solution $\bold{x}$ to a linear equation $A\bold{x}=\bold{b}$ can be amplified by the error in $\bold{b}$ \cite{CDM00}. Denoting the latter as $\bold{e}$, the ratio of the relative error is upper bounded by the condition number: $(\|A^{-1}\bold{e}\|/\|A^{-1}\bold{b}\|)/(\|\bold{e}\|/\|\bold{b}\|)=(\|A^{-1}\bold{e}\|/\|\bold{e}\|)(\|AA^{-1}\bold{b}\|/\|A^{-1}\bold{b}\|)\le \|A^{-1}\|\|A\|={\rm cond}(A)$.} of $V$. 
\end{theorem}

We note that the condition number is unity if $M$ is Hermitian, for which we have Thoerem~\ref{HNH}. In fact, we can apply Thoerem~\ref{HNH} to any \emph{normal} matrix with $[M,M^\dag]=0$, such as unitary matrices. On the other hand, the condition number can be very large if $M$ is close to a nondiagonalizable point (dubbed as an \emph{exceptional point}, see Sec.~\ref{Sec:EP}), at which the condition number diverges. In particular, ${\rm cond}(V)$ may be of the order of $\|E\|^{-\alpha}$ ($0<\alpha<1$) if $M+E$ is nondiagonalizable. 

The single Jordan block $J_n(\lambda)$ in Example~\ref{specsens} gives a nontrivial realization of the \emph{equality} in Eq.~(\ref{SpeS}). To see this, we consider equivalently $M=J_n(\lambda)+E$ perturbed by $-E$, so that $M$ is diagonalizable and Theorem~\ref{NHSS} applies. The eigenvalues of $M$ have been given in Eq.~(\ref{MEep}) and, for the invertible matrix $V$ in $M=VDV^{-1}$, we can check that
\begin{equation}
V=D_\epsilon F,\;\;\;\;
D_\epsilon\equiv{\rm diag}\{\epsilon^{\frac{j-1}{n}}\}^n_{j=1},\;\;
[F]_{mm'}\equiv\frac{1}{\sqrt{n}}e^{i\frac{2\pi mm'}{n}},
\end{equation}
where $D_\epsilon$ is diagonal and $F$ is the unitary matrix of the discrete Fourier transformation. Therefore, the condition number of $V$ is given by
\begin{equation}
{\rm cond}(V)=\|D_\epsilon\|\|D_\epsilon^{-1}\|=|\epsilon|^{\frac{1}{n}-1}.
\end{equation}
Indeed, the spectral shift saturates ${\rm cond}(V)|\epsilon|=|\epsilon|^{\frac{1}{n}}$.

To directly characterize the spectral sensitivity of general non-Hermitian matrices, mathematicians introduce the notion of the \emph{$\epsilon$-pseudospectrum} \cite{VJM79,Reichel1992}, which is a \emph{set} (instead of multiset) defined as
\begin{equation}
\Lambda_\epsilon(M)\equiv\bigcup_{E\in \mathbb{C}^{n\times n}: \|E\|\le\epsilon}\Lambda(M+E).\label{Lamep}
\end{equation}
For Hermitian matrices, according to Theorem~\ref{HNH}, this quantity simply reduces to
\begin{equation}
\Lambda_\epsilon(M)=\bigcup_{\lambda\in\Lambda(M)}\{z\in\mathbb{C}:|z-\lambda|\le\epsilon\}.
\end{equation}
As demonstrated by the above argument for Example~\ref{specsens}, the pseudospectrum $\Lambda_\epsilon(M)$ can differ from the original one $\Lambda(M)$ by $\delta<1$ even for an exponentially small perturbation $\epsilon\sim\delta^n$.  
This indicates that, if we consider a series of $M_n$'s with increasing size $n$ and a bounded spectrum, we generally have 
\begin{equation}
\lim_{n\to\infty}\lim_{\epsilon\to0}\Lambda_\epsilon(M_n)\neq\lim_{\epsilon\to0}\lim_{n\to\infty}\Lambda_\epsilon(M_n).
\label{dlim}
\end{equation}
For example, taking $M_n=J_n(\lambda)$ we have
\begin{equation} 
\lim_{n\to\infty}\lim_{\epsilon\to0}\Lambda_\epsilon(J_n(\lambda))=\{\lambda\},\;\;\;\;
\lim_{\epsilon\to0}\lim_{n\to\infty}\Lambda_\epsilon(J_n(\lambda))=\{z\in\mathbb{C}:|z-\lambda|<1\}.
\end{equation}
The inequivalence~(\ref{dlim}) is reminiscent of spontaneous symmetry breaking, to obtain which we have to first take the thermodynamic limit and then the zero-perturbation limit \cite{HW15}. Such an inequivalence has important implications on the physics of non-Hermitian tight-binding models, which will be discussed in later sections.

\subsection{Eigenvectors of non-Hermitian matrices\label{seceigv}}
In this section, we focus on the fundamental properties of eigenvectors, or eigenstates in physics, of general non-Hermitian matrices. We introduce the notions of left and right eigenvectors and highlight their biorthogonal relationship, which makes a crucial distinction compared with the orthogonal relationship between eigenvectors of Hermitian matrices. We also introduce a powerful tool for the analysis of the properties of eigenvectors.
\\ \\ {\it Nonorthogonality}

\vspace{3pt}
\noindent
In the beginning of this section, we have proved that the spectrum of a Hermitian matrix is always real (see Eq.~(\ref{lamreal})). Yet another well-known property for Hermitian matrices is the orthogonality between different eigenvectors. Given $\bold{v}_1$ and $\bold{v}_2$ as two eigenvectors of a Hermitian matrix $M$ with different eigenvalues $\lambda_1\neq\lambda_2$, then $\bold{v}_1^\dag\bold{v}_2=0$. This is because
\begin{equation}
\bold{v}_1^\dag M\bold{v}_2=\lambda_2\bold{v}_1^\dag\bold{v}_2=(M\bold{v}_1)^\dag\bold{v}_2=\lambda_1\bold{v}_1^\dag\bold{v}_2,\;\;\;\;
\Rightarrow\;\;\;\;(\lambda_1-\lambda_2)\bold{v}_1^\dag\bold{v}_2=0,
\end{equation}
where we have used the Hermiticity of $M$ to obtain $\bold{v}_1^\dag M=(M^\dag \bold{v}_1)^\dag=(M \bold{v}_1)^\dag$. 
More generally, eigenvectors can be chosen to be orthogonal to each other, or said differently, a matrix can be diagonalized by a unitary matrix if and only if the matrix is {\it normal}, i.e., $MM^\dagger=M^\dagger M$ (cf. Eq.~\eqref{2normalmat}). 

To generalize the orthogonality to non-Hermitian (or more precisely, non-normal) matrices, we should require $\bold{v}_1$ to be an eigenvector of $M^\dag$ with eigenvalue $\lambda_1^*$, i.e.,
\begin{equation}
M^\dag\bold{v}_1=\lambda_1^*\bold{v}_1\;\;\;\;\Leftrightarrow\;\;\;\;\bold{v}_1^\dag M=\lambda_1\bold{v}_1^\dag,
\label{lev}
\end{equation}
so that we again have
\begin{equation}
\bold{v}_1^\dag M\bold{v}_2=(M^\dag\bold{v}_1)^\dag\bold{v}_2=\lambda_1\bold{v}_1^\dag\bold{v}_2.
\end{equation}
We call an eigenvector of $M^\dag$ like $\bold{v}_1$ in Eq.~(\ref{lev}) as a \emph{left eigenvector} of $M$, while the conventional one a \emph{right eigenvector}. It follows that there is only an orthogonal relation between left and right eigenvectors, which is called \emph{biorthogonality} \cite{BS87}. The notion of biorhogonality is particularly important when we consider singular behaviors close to exceptional points in the later sections.

In fact, the biorthogonal relation has essentially been given in Eq.~(\ref{onp}). More generally, an arbitrary  non-Hermitian $M$ can be decomposed into (cf. Eq.~\eqref{spedec})
\begin{equation}
M=\sum^J_{j=1}\sum^{m^{\rm g}_j}_{\alpha=1}\left(\sum^{n_{j\alpha}}_{p=1}\lambda_j\bold{r}_{j\alpha p}\bold{l}_{j\alpha p}^\dag+\sum^{n_{j\alpha}-1}_{p=1}\bold{r}_{j\alpha, p+1} \bold{l}_{j\alpha p}^\dag\right),
\label{Mrl}
\end{equation}
where $\bold{r}_{j\alpha p}$ and $\bold{l}_{j\alpha p}$ are vectors satisfying the biorthogonal relation:
\begin{equation}
\bold{l}_{j'\alpha' p'}^\dag\bold{r}_{j\alpha p}=\delta_{j'j}\delta_{\alpha'\alpha}\delta_{p'p}.
\label{bio}
\end{equation}
Here, $\bold{r}_{j\alpha p}$ ($\bold{l}_{j\alpha p}$) is a usual right (left) eigenvector for $p=n_{j\alpha}$ or $n_{j\alpha}=1$ and otherwise called as the $p$th {\it generalized}  eigenvector in the $\alpha$th Jordan block with eigenvalue $\lambda_j$. 
A notable feature of non-Hermitian matrices is that $\bold{r}_{j\alpha p}$'s ($\bold{l}_{j\alpha p}$'s) themselves are generally nonorthogonal
\begin{equation}\label{nonorthorr}
\bold{r}_{j'\alpha' p'}^\dag\bold{r}_{j\alpha p}\neq\delta_{j'j}\delta_{\alpha'\alpha}\delta_{p'p},\;\;\;\;
\bold{l}_{j'\alpha' p'}^\dag\bold{l}_{j\alpha p}\neq\delta_{j'j}\delta_{\alpha'\alpha}\delta_{p'p}.
\end{equation}
Note that both Eqs.~(\ref{Mrl}) and (\ref{bio}) are invariant under $\bold{r}_{j\alpha p}\to c_{j\alpha p} \bold{r}_{j\alpha p}$ and $\bold{l}_{j\alpha p}\to \bold{l}_{j\alpha p}/c_{j\alpha p}^*$ for $\forall c_{j\alpha p}\in\mathbb{C}\backslash\{0\}$, so that we can always further impose 
\eqn{\label{rrnorm}
\bold{r}_{j\alpha p}^{\dagger}\bold{r}_{j\alpha p}=1.
} 
We emphasize that such a normalization condition generally does not imply\footnote{Conversely, we can also impose the normalization condition to the left eigenvectors, but again the right eigenvectors are not normalized in general.} $\bold{l}_{j\alpha p}^{\dagger}\bold{l}_{j\alpha p}=1$. 

If $M$ is diagonalizable, we have $m^{\rm a}_j=m^{\rm g}_j$, $n_{j\alpha}=1$ and Eqs.~(\ref{Mrl}), (\ref{bio}), and \eqref{rrnorm} can be simplified into
\begin{equation}
M=\sum^J_{j=1}\sum^{m^{\rm a}_j}_{\alpha=1}\lambda_j \bold{r}_{j\alpha}\bold{l}_{j\alpha}^\dag,\;\;\;\;
\bold{l}_{j'\alpha'}^\dag\bold{r}_{j\alpha}=\delta_{j'j}\delta_{\alpha'\alpha},\;\;\;\;\bold{r}_{j\alpha}^{\dagger}\bold{r}_{j\alpha}=1.
\label{mathdec}
\end{equation}
When $M$ is an operator acting on the Hilbert space, a more familiar form of Eq.~(\ref{mathdec}) in physics is
\begin{equation}
M=\sum_j \lambda_j| r_j\rangle\langle l_j|,\;\;\;\; \langle l_{j'} | r_j\rangle=\delta_{j'j},\;\;\;\; \langle r_{j}|r_j\rangle=1,
\label{NHdec}
\end{equation}
where $\lambda_j$'s with different labels are not necessarily the same. At the fundamental level, the nonorthogonality $\langle r_{j'}|r_j\rangle\neq\delta_{j'j}$ originates from effective self-interactions mediated via environmental degrees of freedom integrated out upon obtaining non-Hermitian Hamiltonians as shown later. The nonorthogonality has profound physical consequences such as power oscillations in optics \cite{KGM08,KS082,RCE10}, propagation of correlations beyond the conventional bound in many-particle systems \cite{YA18} and even enhanced expressivity of the recurrent neural networks \cite{KG19}. Further, it is noteworthy that non-orthogonality is in general restricted by the complex spectrum \cite{WJ19}. 
Specifically, for a non-Hermitian operator $M$ that satisfies $i(M-M^\dag)$ is either positive or negative semi-definite, we can upper bound the non-orthogonality by \cite{LTD57,LTD65,WJ19} (see Appendix~\ref{app1} for the proof)
\begin{equation}
|\langle r_j| r_{j'}\rangle|^2\le\frac{|\lambda_j-\lambda_j^*||\lambda_{j'}-\lambda_{j'}^*|}{|\lambda_{j'}^*-\lambda_j|^2}=\frac{4\gamma_j\gamma_{j'}}{\Delta_{jj'}^2+(\gamma_j+\gamma_{j'})^2},
\label{rjrjp}
\end{equation}
where $\gamma_j={\rm Im}\lambda_j$ is the imaginary part and $\Delta_{jj'}={\rm Re}(\lambda_j-\lambda_{j'})$ is the real-part difference. Therefore, the maximal non-orthogonality $|\langle r_j|r_{j'}\rangle|=1$ can be achieved only if $\Delta_{jj'}=0$ and $\gamma_j=\gamma_{j'}$, which means $\lambda_j=\lambda_{j'}$. In other words, the coalesce of two eigenvalues is a \emph{necessary condition} for the emergence of maximal non-orthogonality. Note that this statement holds true for a general $M$ since we can shift $M$ by $icI$ ($c\in\mathbb{R}$) with a sufficiently large $|c|$ to make $i(M-M^\dag)$ either positive or negative semi-definite. Then we can apply Eq.~(\ref{rjrjp}) with all the eigenstates staying unchanged.  This intimate relation between non-Hermitian degeneracy and strong non-orthogonality  lies at the heart of physical phenomena associated with the exceptional points as detailed in Sec.~\ref{secepphys}.

\subsubsection{Resolvent and perturbation formula of eigenprojector\label{sec2resolvent}}
In Sec.~\ref{secspecnh}, we have analyzed the stability of spectrum against perturbations. It is natural to ask how about the eigenvectors. To take into account the possible existence of degeneracy and difference between left and right eigenvectors, we focus on the change of \emph{eigenprojectors}, i.e., 
\begin{equation}
P_j\equiv\sum^{m^{\rm g}_j}_{\alpha=1}\sum^{n_{j\alpha}}_{p=1} \bold{r}_{j\alpha p}\bold{l}_{j\alpha p}^\dag.
\label{eproj}
\end{equation}
Concretely speaking, we focus on an \emph{isolated} eigenvalue $\lambda_j$ of $M$ and its associated $P_j$ together with its perturbed counterpart $P'_j$ of $M'=M+E$. We allow the lift of degeneracy due to $E$ so that $P'_j$ is generally a sum of several eigenprojectors. On the other hand, we assume that the eigenvalues of $M'$ arising from $\lambda_j$, which are called the \emph{$\lambda$-group} (and $P'_j$ is called the \emph{total projection} of the $\lambda$-group) \cite{TK80}, stay separated from other eigenvalues.\footnote{This is always possible for a sufficiently small $\|E\|$ even if $M$ is not diagonalizable, since the eigenvalues are continuous functions in terms of the parameters of a matrix \cite{TK80}, although their derivatives can be singular.} Our purpose is to find a bound on $\|P'_j-P_j\|$ in terms of $\|E\|$ and other parameters.

To this end, we introduce a useful quantity for studying 
eigenprojectors of a general matrix $M$ called \emph{resolvent}. The resolvent of $M$ is a function $\mathbb{C}\to\mathbb{C}^{n\times n}$ defined as \cite{TK80}
\begin{equation}
R(z)\equiv(M-zI)^{-1}.
\end{equation} 
We can show that the resolvent is analytic on a region that does not contain any eigenvalues of $M$, where we have the following expression: 
\begin{equation}
R(z)=-\sum^J_{j=1}\left[\frac{P_j}{z-\lambda_j}+\sum^{m^{\rm g}_j}_{\alpha=1}\sum^{n_{j\alpha}-1}_{p=1}\frac{N^p_{j\alpha}}{(z-\lambda_j)^{p+1}}\right].
\end{equation}
This formula can be derived by taking $f(\zeta)=(\zeta-z)^{-1}$ in Eq.~(\ref{fM}). Suppose that $\lambda_j$ is an isolated eigenvalue with an arbitrary degeneracy. 
Then the projector can be related to the resolvent through 
\begin{equation}
P_j=-\oint_{C_j}\frac{dz}{2\pi i}R(z),
\label{PjRz}
\end{equation}
where $C_j$ can be an arbitrary contour that separates $\lambda_j$ from the remaining eigenvalues. Similarly, we can determine the nilpotent from the resolvent via\footnote{One can readily  check the relation $N_j=(M-\lambda_j I) P_j$ from this relation.}
\begin{equation}
N_j\equiv\sum^{m^{\rm g}_j}_{\alpha=1}N_{j\alpha}=-\oint_{C_j}\frac{dz}{2\pi i}(z-\lambda_j)R(z).
\label{NjRz}
\end{equation}
We will discuss an interesting application of the projector formula (\ref{PjRz}) to proving exponential decay of correlations from band gaps for quadratic fermionic systems in Sec.~\ref{Sec:NHH}.

We now employ the resolvent approach to analyze how eigenprojectors change upon perturbations. By assumption, we have a common contour $C_j$ that separates $\lambda_j$ from the remaining eigenvalues of $M$ and also the corresponding $\lambda$-group from the remaining eigenvalues of $M'$. According to Eq.~(\ref{PjRz}), we have
\begin{equation}
P'_j-P_j=\oint_{C_j}\frac{dz}{2\pi i}\frac{1}{M-zI}E\frac{1}{M+E-zI}.
\end{equation}
Defining $\Delta_{C_j}\equiv(\max_{z\in C_j}\|R(z)\|)^{-1}$, $l_{C_j}\equiv\oint_{C_j}|dz|$ and assuming that $\|E\|<\Delta_{C_j}$, we can expand $(M+E-zI)^{-1}$ as $(M-zI)^{-1}\sum^\infty_{m=0}(E\frac{1}{zI-M})^m$ and obtain
\begin{equation}
P'_j-P_j=\sum^\infty_{m=1}(-)^{m+1}\oint_{C_j}\frac{dz}{2\pi i} R(z)[ER(z)]^m.
\label{Pjpert}
\end{equation}
This result is important on its own right as a general perturbation formula for eigenprojectors.  As for our current purpose to bound the shift of eigenprojectors, we take the norm for both sides of Eq.~(\ref{Pjpert}) to obtain
\begin{equation}
\|P'_j-P_j\|\le\frac{l_{C_j}\|E\|}{2\pi\Delta_{C_j}(\Delta_{C_j}-\|E\|)}.
\label{Pjshift}
\end{equation}
The bound~(\ref{Pjshift}) is at least $\frac{\|E\|}{\Delta_{C_j}-\|E\|}$ since $l_{C_j}\ge 2\pi\min_{z\in C_j}|z-\lambda_j|\ge 2\pi\Delta_{C_j}$\footnote{The second inequality results from the fact that $\Delta_{C_j}\le[\max_{z\in C_j}\rho(\frac{1}{M-zI})]^{-1}\le(\max_{z\in C_j}|z-\lambda_j|^{-1})^{-1}=\min_{z\in C_j}|z-\lambda_j|$, where we have used the inequality in Eq.~(\ref{rhonorm}).}. This can indeed be achieved by Hermitian matrices, for which $\Delta_{C_j}$ is a measure of energy gap and the bound is essentially $\frac{\|E\|}{\Delta_{C_j}}$ provided that $\Delta_{C_j}=\min_{z\in C_j}|z-\lambda_j|\gg\|E\|$. However, for non-Hermitian matrices, the bound in Eq.~(\ref{Pjshift}) can in general be \emph{of order of one} since $\Delta_{C_j}$ can be much smaller than $l_{C_j}$ (i.e., the energy gap $\min_{z\in C_j}|z-\lambda_j|$) even if $\Delta_{C_j}\gg\|E\|$. 

\exmp{(Eigenvector sensibility in $2\times 2$ matrix). 
\label{Ex:ev}
To demonstrate the sensibility of non-Hermitian eigenvectors against a small perturbation, we consider a $2\times 2$ matrix
\begin{equation}
M=\begin{bmatrix} \;0\; & \;1\; \\ \;\;\kappa\;\; & \;\;0\;\; \end{bmatrix},
\label{Mkappa}
\end{equation}
which has eigenvalues $\pm\sqrt{\kappa}$ and the resolvent is given by
\begin{equation}
R(z)=\frac{1}{\kappa - z^2} \begin{bmatrix} \;z\; & \;\;1\;\; \\ \;\kappa\; & \;\;z\;\; \end{bmatrix}.
\end{equation}
The eigenprojectors corresponding to $\pm\sqrt{\kappa}$ can thus be determined from Eq.~(\ref{PjRz}) as
\begin{equation}
P_\pm=-\oint_{C_\pm} \frac{dz}{2\pi i}R(z)=\frac{1}{2}\begin{bmatrix} \;1\; & \;\;\pm\kappa^{-\frac{1}{2}}\;\; \\ \;\pm\kappa^{\frac{1}{2}}\; & \;\;1\;\; \end{bmatrix},
\label{Ppm}
\end{equation}
where $C_\pm=\{z:|z\pm\sqrt{\kappa}|=\sqrt{|\kappa|}\}$. If we change $\kappa$ in Eq.~(\ref{Mkappa}) into $\kappa'$ and denote the corresponding projectors as $P'_\pm$, we have
\begin{equation}
\|P'_\pm-P_\pm\|=\left|\frac{\kappa'-\kappa}{2\sqrt{\kappa'\kappa}(\sqrt{\kappa'}+\sqrt{\kappa})}\right|,
\end{equation}
provided that $|\kappa|\sim|\kappa'|\ll1$ (in fact, $|\kappa'\kappa|<1$ is enough). One can achieve an order-one $\|P'_\pm-P_\pm\|$ by taking $\kappa'-\kappa\sim\kappa^{\frac{3}{2}}$.  In this case, $\Delta_{C_\pm}$, which is of the order of $\kappa$, is much smaller than the energy gap with order $\kappa^{\frac{1}{2}}$ even if  $\Delta_{C_\pm}\gg\|E\|\sim\kappa^{\frac{3}{2}}$.  This extreme sensibility of eigenvectors is ultimately related to the singular behavior at the exceptional point $\kappa=0$ as we discuss in Sec.~\ref{Sec:EP}. 
}

\subsubsection{Petermann factor\label{sec2peter}}
\begin{figure}
\begin{center}
\includegraphics[width=7cm]{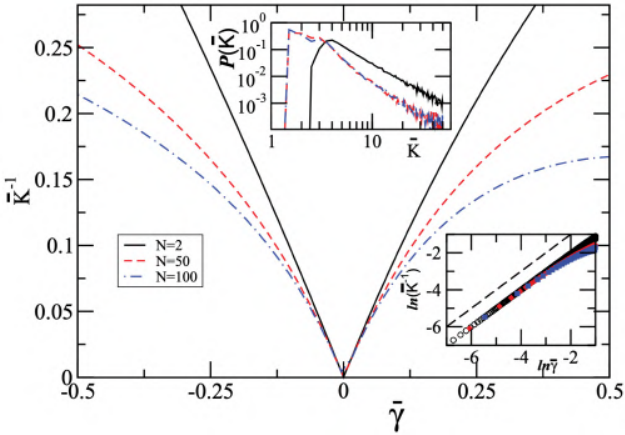}
\end{center}
\caption{Inverse mean Petermann factor $1/\bar{K}_N$ versus the deviation $\bar{\gamma}=(\gamma-\gamma_\mathrm{EP}){N}$ of the gain/loss parameter $\gamma$ from its value at the exceptional point $\gamma_\mathrm{EP}$ multiplied by the length of disordered dimer chains ${N}$ \cite{ZMC10}. The Petermann factor diverges at the EP as $\bar{K}\propto\bar{\gamma}^{-1}$. The upper inset shows the distribution of the Petermann factors around the EP. The lower inset demonstrates the scaling $\bar{K}\propto \bar{\gamma}^{-1}$ in the vicinity of the EP, where the dashed line shows the slope with the exponent $-1$. Adapted from Ref.~\cite{ZMC10}. Copyright \copyright\,  2010 by the American Physical Society.
}
\label{fig:2petermann}
\end{figure}
Another useful way to characterize the nonorthogonality of non-Hermitian eigenstates is to use the Petermann factor \cite{KP79}. It quantifies the deviation of non-Hermitian eigenstates from the standard orthonormalization condition and is defined as
\eqn{
K_{j'j}\equiv({\bold l}_{j'}^{\dagger}{\bold l}_{j})({\bold r}_{j}^{\dagger}{\bold r}_{j'}),
}
where we assume that the non-Hermitian matrix is diagonalizable and eigenvalues with different labels $j\neq j'$ are not necessarily the same.  It is also useful to consider the mean of the diagonal Petermann factor \cite{ZMC10}
\eqn{
\bar{K}_n=\frac{1}{n}\sum_{j=1}^{n}K_{jj},
}
whose deviation from the unity corresponds to the nonorthogonality of eigenstates.
The divergence of the Petermann factor indicates the presence of exceptional points at which the orthonormalization condition can be maximally violated.

As a simple illustration, we consider the matrix in Example~\ref{Ex:ev}, whose $\bar K_n$ reads
\eqn{
\bar{K}_2=\frac{1}{2}(\|P_+\|^2_2+\|P_-\|^2_2)=\frac{1}{2}+\frac{1}{4}\left(|\kappa|+\frac{1}{|\kappa|}\right),
}
where $\|P_\pm\|_2\equiv\sqrt{{\rm Tr}[P_\pm^\dag P_\pm]}$ is the Schatten-2 norm (Hilbert-Schmidt norm). As expected, $\bar K_2$ reaches its minimum  at the Hermitian point $\kappa=1$ and diverges at the nondiagonalizable point $\kappa=0$. See also Fig.~\ref{fig:2petermann} for another illustrative example of the coupled dimers with gain and loss \cite{ZMC10}.

One of the earliest studies on physical applications of the Petermann factor is the explanation of linewidth broadening \cite{HWA90} in laser beyond the Schawlow-Townes linewidth \cite{SAL58}. The broadening originates from the non-Hermicity in laser oscillation by resonant cavities and becomes most pronounced near exceptional points \cite{MVB03}. For this reason, this factor has traditionally been known also as the Petermann {\it excess noise} factor, which can be regarded as a generalized Purcell factor \cite{LZ16}.   

\subsection{Pseudo Hermiticity and quasi Hermiticity\label{secphqh}}
The spectrum is trivially real in a Hermitian matrix.  However, the Hermiticity is not a necessary condition for eigenvalues to be real. In physics, to our knowledge this has firstly been pointed out in the context of studies on the hard-core Bose gas using the Fermi's pseudopotential in 1959 \cite{WTT59}. Also, the spectrum of a non-Hermitian variant of the Toda lattice has been found to be entirely real \cite{TH92}.  Later, it has been noticed that there exist certain classes of matrices whose spectra can be entirely real, despite being non-Hermitian. Among the most important is the class satisfying the \emph{parity-time (PT) symmetry} \cite{BCM98}. While eigenvalues can also be real for other types of antilinear symmetries \cite{CMB02}, one distinctive advantage of the PT symmetry is its simplicity that enables one to physically implement the symmetry by spatially engineering gain-loss structures, stimulating a rich interplay between theory and experiment \cite{EG18}. Meanwhile, it is the notion of \emph{pseudo Hermiticity} that provides a general perspective on characterizing a class of non-Hermitian matrices with purely real spectra \cite{AM10}.

Historically, the notion of pseudo Hermiticy has been first introduced by Dirac and Pauli in the context of indefinite metric of quantum field theories due to negative norms \cite{PW43}. Later, motivated by studies of the Yang-Lee edge singularity based on the nonunitary quantum field theory \cite{MEF78,JC85}, Bessis and Zinn-Justin have conjectured that the corresponding single-particle Hamiltonian with a cubic imaginary potential should exhibit the entirely real spectrum \cite{PD07}. This conjecture has been numerically verified by Bender and B{\"o}ttcher \cite{BCM98} and later rigorously proven by Dorey, Dunning and Tateo \cite{PD01} based on the Bethe ansatz equations.  It has been first argued in Ref.~\cite{BCM98} that the PT symmetry in the underlying Hamiltonian plays a crucial role for its spectrum to be real. This idea was further developed by Mostafazadeh \cite{AM02,AM022} who employed the concept of pseudo Hermiticity to identify general conditions for the spectrum to be real. 

In this section, we assume that $M$ is a diagonalizable linear operator acting on the Hilbert space and has a complete biorthonormal eigenbasis and a discrete spectrum:
\eqn{
M=\sum_{j}\lambda_{j}|r_j\rangle\langle l_j|,\;\;\;\;\langle l_{j'}|r_j\rangle=\delta_{j'j},\;\;\;\;\sum_{j}|r_j\rangle\langle l_j|=1,
}  
where $\lambda_j$'s with different labels are not necessarily the same. 
To formulate the notion of pseudo Hermiticity, we define that an operator $M$ is said to be $\eta$-{\it pseudo-Hermitian} if it satisfies 
\eqn{\label{etapseudo}
M^\dagger=\eta M \eta^{-1},
}
where $\eta=\eta^\dagger$ is a Hermitian invertible operator. We note that if we choose the identity operator as $\eta=I$, $\eta$-pseudo Hermiticity reduces to an ordinary Hermiticity $M=M^\dagger$. More generally, an operator $M$ is said to be {\it pseudo-Hermitian} if there exists a Hermitian invertible matrix $\eta$ such that $M$ is $\eta$-pseudo Hermitian.  

The following theorem characterizes the necessary and sufficient conditions for the pseudo Hermiticity \cite{AM02}:
\begin{theorem}[Pseudo Hermiticity]
\label{pseudoh} 
A linear operator $M$ acting on the Hilbert space with a complete biorthonormal eigenbasis and a discrete spectrum is pseudo-Hermitian if and only if one of the following conditions hold:
\begin{itemize}
\item The spectrum of $M$ is entirely real.
\item The eigenvalues appear in complex conjugate pairs and the degeneracy of complex conjugate eigenvalues are the same.
\end{itemize}  
\end{theorem}

It is worthwhile to note that studies of this property can be traced back to an observation made by Wigner \cite{WE60}, who pointed out that operators satisfying  antilinear symmetry have either real eigenvalues or eigenvalues with complex conjugate pairs depending on whether or not the corresponding eigenstates  respect the symmetry.

\exmp{(Quantum particle in a one-dimensional PT-symmetric potential). Consider a single quantum particle subject to a one-dimensional PT-symmetric complex potential. It is described by a non-Hermitian Hamiltonian
\eqn{
H=p^2+V_{\rm r}(x)+iV_{\rm i}(x),
}
where $p$ and $x$ are the momentum and position operators, respectively. The potentials $V_{\rm r}$ and $V_{\rm i}$ are real and satisfy $V_{\rm r}(x)=V_{\rm r}(-x)$ and $V_{\rm i}(x)=-V_{\rm i}(-x)$ to be consistent with the PT symmetry, where
\eqn{
PxP^{-1}=-x,\;\;PpP^{-1}=-p,\;\;TxT^{-1}&=&x,\;\;TpT^{-1}=-p,\;\;TiT^{-1}=-i,\\
PT H (PT)^{-1}&=&H.
}
Since the time-reversal operator $T$ acts as complex conjugation, we arrive at
\eqn{
PHP^{-1}=H^{\dagger},
}
which shows that the parity operator $P$ is nothing but the $\eta$ operator in Eq.~\eqref{etapseudo} \cite{Mostafazadeh_2005}. Thus, the Hamiltonian $H$ is $P$-pseudo-Hermitian. From Theorem~\ref{pseudoh}, its eigenvalues must either be real or form complex conjugate pairs.
}

The above example demonstrates that  PT-symmetric operators provide a particularly simple subclass of pseudo Hermitian operators. PT symmetry is said to be {\it unbroken} if every  eigenstate of a PT-symmetric non-Hermitian operator satisfies  PT symmetry; then, the entire spectrum is real even though the operator of interest is not Hermitian. PT symmetry is said to be  spontaneously {\it broken} if some  eigenstates are not the eigenstates of the PT operator; then, some pairs of eigenvalues become complex conjugate to each other. This real-to-complex spectral transition is often called the PT transition. As detailed in Sec.~\ref{Sec:EP}, the transition is typically accompanied by the coalescence of  eigenstates and that of the corresponding eigenvalues at  an exceptional point  \cite{TK80} in the discrete spectrum or the spectral singularity  \cite{AM09} in the continuum spectrum. 

While PT symmetry, or more generally, pseudo Hermiticity implies that the spectrum can be entirely real despite being non-Hermitian, they do not characterize it in a conclusive manner. In fact, they alone are neither necessary nor sufficient conditions for an operator to have a real spectrum \cite{AM02}.  Instead, the following theorem provides a necessary and sufficient condition for this statement in terms of the positivity of  $\eta$ for a pseudo Hermitian operator \cite{AM022}.
\begin{theorem}[Characterization of real spectrum]
\label{realspectrum} A linear operator $M$ acting on the Hilbert space with a complete biorthonormal eigenbasis and a discrete spectrum has real spectrum if and only if there is an invertible linear operator $O$ such that $M$ is $OO^\dagger$-pseudo Hermitian. 
\end{theorem}
This theorem indicates the existence of a Hermitian operator $h$ whose spectrum coincides with that of $OO^\dagger$-pseudo Hermitian operator $M$. This can be understood from the fact that $h$ and $M$ can be related to each other via the similarity transformation with $O$:
\eqn{\label{similar}
h=O^{-1}MO=O^{\dagger}M^{\dagger}{(O^\dagger)^{-1}}.
}
We note that $O$ is not unique in general. Since the transformation in Eq.~\eqref{similar} is not a unitary transformation, eigenvectors of $h$ and $M$ can be qualitatively different while the eigenvalues are the same. Specifically, eigenstates of the Hermitian operator $h$ are orthogonal while those of non-Hermitian operator $M$ are not in general.  This point can lead to a number of physical phenomena unique to nonconservative systems as we detail in later sections.

We note that the concept of pseudo Hermiticity is different from that of the so-called \emph{quasi-Hermiticity}. There is a subtle, yet important difference which is often overlooked in literature. An operator $M$ is said to be {\it quasi-Hermitian} if there exists a Hermitian and positive definite operator $\xi$ such that  $\xi M=M^\dagger \xi$ \cite{WJP69}. Note that $\xi$ is \emph{not} necessarily invertible (i.e., $\xi^{-1}$ can be unbounded) in contrast to an invertible $\eta$ operator in the pseudo-Hermitian case; we will see a concrete example for this (cf. Eq.~\eqref{quasiherexmp}). It was shown that, in the quasi-Hermitian case, every eigenvalue is real \cite{WJP69,FGS92}. Moreover, a quasi-Hermitian operator can always be regarded as a Hermitian operator on the modified Hilbert space whose inner product is defined by using the operator $\xi$. In contrast, a pseudo-Hermitian operator associates with an invertible Hermitian (but not necessarily positive definite) operator $\eta$. In this case, the existence of the modified metric (for which an operator can be regarded as a Hermitian one) is not necessarily guaranteed in  contrast to the quasi-Hermitian case. The linear space associated with an indefinite metric is known as the Krein space or the Pontrjagin space for a finite-dimensional case \cite{TYA89}.

{In the following, let us provide some prototypical examples of pseudo-Hermitian matrices that are relevant to physics,  covering few-level, single-particle and many-body systems.}
\exmp{\label{ph22}(Pseudo-Hermitian $2\times 2$ matrices).
Consider $2\times 2$ non-Hermitian matrices. This class of problems is particularly important as it finds direct applications to a multitude of experimental systems as reviewed in later sections. A two-level system also plays a crucial role when we discuss exceptional points Sec.~\ref{Sec:EP}. Exceptional points emerge when (more than) two eigenvalues and eigenvectors coalesce. The original problem with a (possibly) large dimension can thus be typically reduced to the analysis of two-level matrices describing two merging levels in the vicinity of exceptional points. Also, the two-dimensional problem is important in view of studies of band topology based on the $2\times 2$ Bloch Hamiltonian as reviewed in Sec.~\ref{sec5}.

We here provide a complete characterization of pseudo-Hermitian $2\times 2$ matrices.  Theorem~\ref{pseudoh} indicates that the pseudo Hermiticity imposes two constraints on $2\times 2$ matrices. Thus, pseudo-Hermitian $2\times 2$ matrices are parameterized by 6 variables while Hermitian matrices have 4 parameters. A general expression for $2\times 2$ pseudo-Hermitian matrices is  given by \cite{WCT10}
\eqn{\label{pseudo2b2}
M=\epsilon\sigma_0+(\gamma\,{\boldsymbol n}_{1}+i\rho\sin\alpha\,{\boldsymbol n}_2+i\rho\cos\alpha\,{\boldsymbol n}_3)\cdot{\boldsymbol \sigma},
}
where $\epsilon,\gamma,\alpha,\rho$ are real parameters, ${\boldsymbol n}_1=(\sin\theta\cos\phi,\sin\theta\sin\phi,\cos\theta)$, ${\boldsymbol n}_2=(\cos\theta\cos\phi,\cos\theta\sin\phi,-\sin\theta)$, and ${\boldsymbol n}_3=(-\sin\phi,\cos\phi,0)$ are orthogonal unit vectors with real parameters $\theta,\phi$, and $\sigma_0,{\boldsymbol \sigma}=(\sigma^x,\sigma^y,\sigma^z)^{\rm T}$ are the identity matrix and Pauli matrices, respectively,
\eqn{
\sigma_0=
\begin{bmatrix}
\;1\;&\;\;0\;\;\\
\;0\;&\;\;1\;\;\\
\end{bmatrix},\;\;\;
\sigma^x=
\begin{bmatrix}
\;0\;&\;\;1\;\;\\
\;1\;&\;\;0\;\;\\
\end{bmatrix},\;\;\;
\sigma^y=
\begin{bmatrix}
\;0\;&\;-i\;\\
\;i\;&\;0\;\\
\end{bmatrix},\;\;\;
\sigma^z=
\begin{bmatrix}
\;1\;&\;0\;\\
\;0\;&\;-1\;\\
\end{bmatrix}.
}
One can check that matrix $M$ in Eq.~\eqref{pseudo2b2} is ${\boldsymbol n}_1\cdot{\boldsymbol \sigma}$-pseudo Hermitian. According to Theorem~\ref{pseudoh},  its eigenvalues are thus either real or pairwise complex conjugate. Note that all $2\times 2$ Hermitian matrices are included as special cases with $\rho=0$ in Eq.~\eqref{pseudo2b2}. The eigenvalues of $M$ are
\eqn{
\lambda_{1,2}=\epsilon\pm\sqrt{\gamma^2-\rho^2}.
}
The matrix $M$ is diagonalizable and its spectrum is real when $\gamma^2-\rho^2>0$. In this case, one can find a positive definite operator $\xi$ such that $\xi M=M^{\dagger}\xi$ (this $\xi$ corresponds to $OO^\dagger$ in  Theorem~\ref{realspectrum}). Its general expression is
\eqn{
\xi=u\left[\gamma\sigma_0+(v{\boldsymbol n}_1+\rho\cos\alpha{\boldsymbol n}_2-\rho\sin\alpha{\boldsymbol n}_3)\cdot{\boldsymbol \sigma}\right],
}
where $u$ and $v$ are arbitrary real constants satisfying the conditions $u\gamma>0$ and $v^2<\gamma^2-\rho^2$. Matrix $\xi$ has positive eigenvalues $u(\gamma\pm\sqrt{\rho^2+v^2})$ and its square root will provide an operator $O$ in Theorem~\ref{realspectrum} for the present finite-dimensional problem. As shown later, the noninvertibility of $\xi$ (and $M$) in the limit of $\gamma^2-\rho^2\to 0$ indicates the presence of exceptional point at $\gamma^2=\rho^2>0$. At this point, matrix $M$ is nondiagonalizable but can be transformed to the Jordan normal form  via the similarity transformation (cf.  Eq.~\eqref{JNF}).
It is worthwhile to note that a set of pseudo-Hermitian matrices is equivalent to that of PT-symmetric matrices in the two-level case. This can be shown by explicitly 
constructing a parity operator $P$ such that ${\boldsymbol n}_1\cdot{\boldsymbol \sigma}$-pseudo Hermitian matrices are PT-symmetric.
}

\exmp{\label{phparticle}(Quantum particle in a one-dimensional complex potential).
\begin{figure}
\begin{center}
\includegraphics[width=6cm]{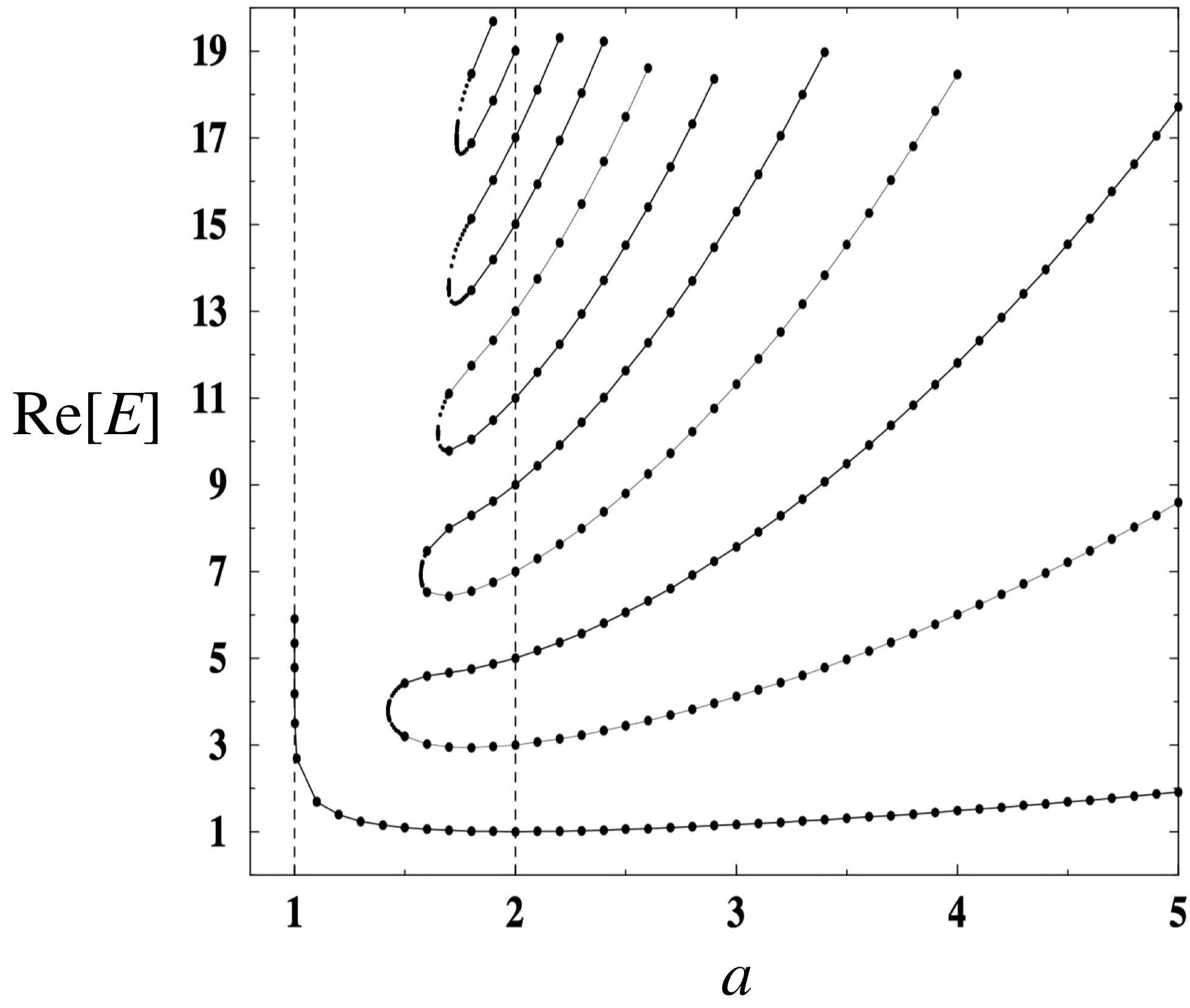}
\end{center}
\caption{Real part of the eigenspectrum of the PT-symmetric non-Hermitian Hamiltonian $H_{a}=p^2-(ix)^a$ as a function of $a\in{\mathbb R}$. For $a\geq 2$, the PT symmetry is unbroken and eigenvalues are all real and positive. For $a< 2$, the PT symmetry is broken and an infinite number of complex conjugate pairs of eigenvalues appear in  excited states. Note that only real parts of eigenvalues are plotted in this figure. Adapted from Ref.~\cite{BCM98}. Copyright \copyright\,  1998 by the American Physical Society.
\label{fig:2bender}
}
\end{figure}
Historically, studies of non-Hermitian operators with real spectra have been stimulated by the analyses of single-particle Hamiltonians with one-dimensional complex potentials. The original example is
\eqn{\label{benderham}
H_{a}=p^2-(ix)^a,\;\;\;a\in {\mathbb R},
}
whose spectrum is real and positive for $a\geq 2$. This real-valuedness has been first numerically found in Ref.~\cite{BCM98} (see Fig.~\ref{fig:2bender}) and later rigorously proven by Ref.~\cite{PD01}. Another important example is a single particle in a periodic complex potential 
\eqn{\label{singlecossin}
H_{V}=p^2+\cos(2x)+iV\sin(2x),\;\;\;V\in {\mathbb R},
}
which found an application to optics \cite{KGM08}. 
Its spectrum is real if $|V|<1$ in which one can find a similarity transformation $O=\exp[(p/2)\tanh^{-1}V]$ that maps $H_V$ into the Hermitian Hamiltonian $h_V$ having the same real spectrum $h_V=O^{-1}HO=p^2+\sqrt{1-V^2}\cos(2x)$ \cite{BM10}. Note that $O$ is invertible if and only if  $|V|<1$. At $|V|=1$, the Hamiltonian $H_V$ is nondiagonalizable and exhibits the spectral singularity \cite{AM09}, where anomalous wave propagation such as the unidirectional invisibility can appear \cite{LZ11}. This case is also interesting in terms of its relation to the Liouville theory having a complex potential $e^{\pm2ix}$ as reviewed in Sec.~\ref{nonunitary_cft} (cf. Eq.~\eqref{livcft}). 

We note that the Hamiltonians~\eqref{benderham} and \eqref{singlecossin} satisfy the PT symmetry with respect to a spatial parity operator $P$ and thus are $P$-pseudo Hermitian. However, the PT symmetry is not a necessary condition for non-Hermitian operators to have real spectra. One possible class of non-Hermitian operators that can have a real spectrum without satisfying the PT symmetry is \cite{NS16}
\eqn{\label{quasiherexmp}
H_{g(\cdot)}=p^2-\left[g^2(x)+ig'(x)\right],
} 
where $g(x)$ is an arbitrary real function and $g'(x)\equiv dg/dx$. It satisfies $\zeta H_{g(\cdot)}=H^{\dagger}_{g(\cdot)}\zeta$ with $\zeta=i[p+g(x)]$ being an operator that is not necessarily invertible  \cite{NS16}. A real-to-complex spectral transition can occur also in this class of non-Hermitian Hamiltonians while its transition point depends on a specific choice of $g(x)$.
}

\exmp{\label{phmany}(Quantum many-body systems).
Yang and Lee studied the distribution of the zeros of the partition function of Ising models in the complex plane of a magnetic field $h_{\rm mag}$ \cite{LTD52} (see also Sec.~\ref{Sec:NHH}). They found that, in the thermodynamic limit, the zeros become dense and form edge singularities at $h_{\rm mag}=\pm i\kappa_c$ with $\kappa_c\in{\mathbb R}$. Michael Fisher analyzed this Yang-Lee edge singularity in terms of field theory with an imaginary field $-i(\kappa-\kappa_c)\phi$ and an imaginary cubic interaction $ig\phi^3$ in the case of dimension $d=6$ \cite{MEF78} (cf. Eq.~\eqref{iphi3}). Such an action is PT-symmetric, i.e., it is invariant upon $\phi\to-\phi$ and $i\to -i$. Cardy discussed its two-dimensional version by using the techniques of conformal field theory (CFT) and pointed out that the field theory corresponds to the ${\cal M}_{5,2}$-model in the minimal CFTs \cite{JC85} having the central charge $c=-22/5$ (see  Sec.~\ref{nonunitary_cft} for further reviews on the nonuntiary CFT).

A lattice model that demonstrates the Yang-Lee edge singularity, which has been first analyzed by von Gehlen \cite{GvG91}, is the Ising quantum spin chain in the presence of a magnetic field in the $z$-direction as well as a longitudinal imaginary field:
\eqn{
H_{\lambda,\kappa}=-\frac{1}{2}\sum_{j=1}^N(\sigma_j^z+\lambda\sigma_j^x\sigma_{j+1}^x+i\kappa\sigma_j^x),\;\;\;\;\lambda,\kappa\in{\mathbb R}.
}
This model is of interest as its effective field theory can be regarded as the above mentioned ${\cal M}_{5,2}$ nonunitary CFT \cite{JC85}. It satisfies the PT symmetry with respect to the spin parity operator $P=\prod_{j=1}^N \sigma_j^z$. It has been numerically verified that, for $\lambda<1$, the real-to-complex spectral transition occurs at finite threshold $\kappa_c$, which does not vanish in the thermodynamic limit \cite{GvG91}. The perturbative analysis with respect to $\kappa$ has also been performed in Ref.~\cite{OAC09}. 

Another well-known family of non-Hermitian many-body Hamiltonians is the integrable XXZ spin chain with imaginary boundary magnetic fields \cite{VP90}:
\eqn{\label{xxzps}
H_{r}=\frac{1}{2}\sum_{j=1}^{N}\left[\sigma_j^x\sigma_{j+1}^x+\sigma_j^y\sigma_{j+1}^y+\cos\left(\frac{\pi}{r}\right)(\sigma_j^z\sigma_{j+1}^z-1)\right]+i\sin\left(\frac{\pi}{r}\right)\frac{\sigma_1^z-\sigma_{N+1}^z}{2},
}
where $r\in{\mathbb R}$. It is believed that its effective field theory corresponds to a nonunitary CFT with central charge $c=1-6/[r(r-1)]$ \cite{FCA87,VP90}. This model is of interest also because of its quantum group invariance and its connection with the Temperley-Lieb algebras. It satisfies the PT symmetry with respect to the spatial parity operator $P: \sigma_{i}\to\sigma_{N+1-i}$ \cite{KC072}. The spectrum is real and eigenvalues can be exactly obtained from the Bethe-ansatz analysis. 

\begin{figure}
\begin{center}
\includegraphics[width=12cm]{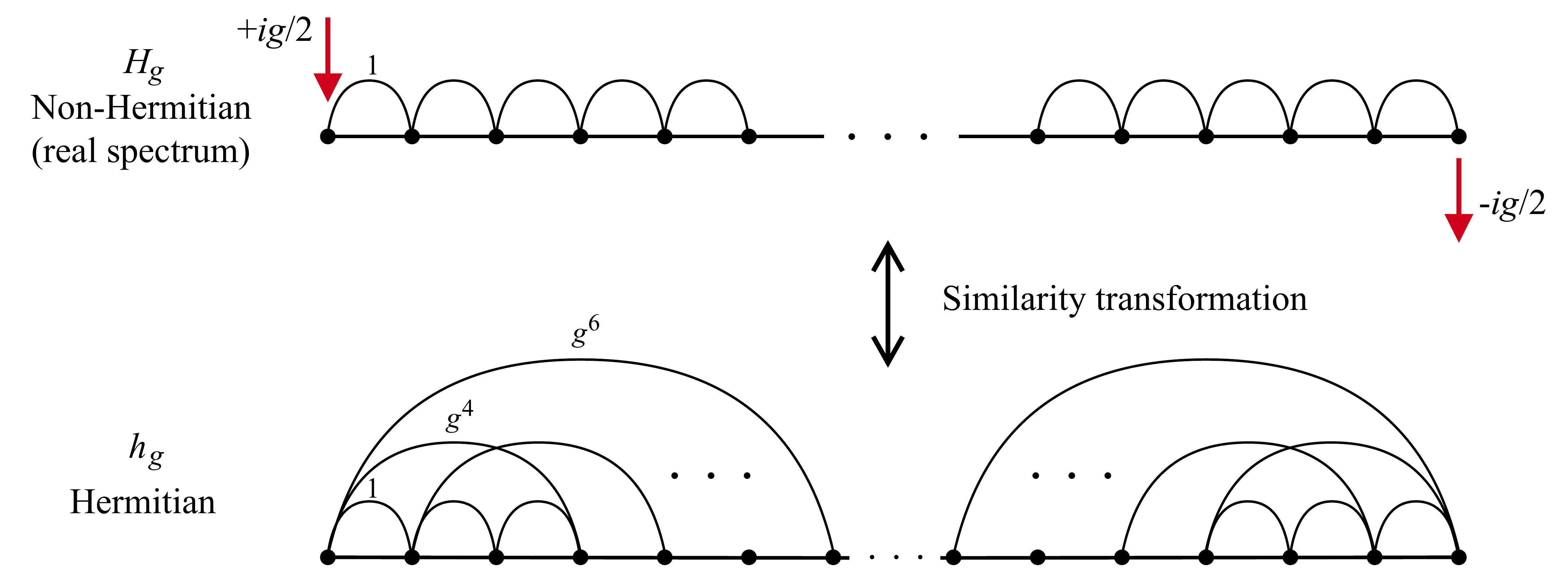}
\end{center}
\caption{Schematic  illustration of the relation between the non-Hermitian Hamiltonian $H_g$ in Eq.~\eqref{korff1} whose spectrum is real for $g\leq1$ and the Hermitian Hamiltonian $h_g$  in Eq.~\eqref{korff2} that has the same spectrum \cite{CK08}. In the non-Hermitian model $H_g$, all the terms are local including imaginary magnetic fields $\pm ig/2$ at the boundaries. After a similarity transformation $h_g=O^{-1}H_gO$, the Hermitian model acquires long-range hopping terms, which become increasingly long-ranged as $g$ approaches 1 at which the real-to-complex spectral transition occurs.   
}
\label{fig:2korff}
\end{figure}

Korff studied a noninteracting variant of this model \cite{CK08}
\eqn{\label{korff1}
H_g=\frac{1}{2}\sum_{j=1}^{N}\left[\sigma_j^x\sigma_{j+1}^x+\sigma_j^y\sigma_{j+1}^y\right]+ig\frac{\sigma_1^z-\sigma_{N+1}^z}{2}.
}
Note that, at $g=1$, this model can be regarded as a special case of Eq.~\eqref{xxzps}, i.e., $H_{g=1}=H_{r=2}$. For $0\leq g<1$, its spectrum is also real and its effective field theory corresponds to the CFT with $c=1$ while the correspondence changes discontinuously at $g=1$ for which the effective field theory is the nonunitary CFT  with $c=-2$. According to Theorem~\ref{realspectrum},  for $0\leq g<1$, there exists a Hermitian counterpart $h_g=O^{-1}H_gO$ that shares the same real spectrum with $H_g$ (cf. Eq.~\eqref{similar}). From the perturbative analysis, such a Hermitian Hamiltonian $h_g$ has been constructed as follows \cite{CK08}:
\eqn{\label{korff2}
h_{g}=-\sum_{n>0}\sum_{j=1}^{N+1-n}p_{j}^{(n)}(g^2)(c^{\dagger}_{j}c_{j+n}+{\rm H.c.}),
}
where $c_{j}^\dagger$ and $c_{j}$ are fermionic creation and annihilation operators at site $j$. As $g$ is increased from zero, the hopping coefficients $p_{j}^{(n)}$ become more long-ranged and can eventually cover the entire system in the limit of $g\to 1$ (see Fig.~\ref{fig:2korff}). This example clearly demonstrates that the Hermitian counterpart having the same real spectrum can be highly nonlocal even if the original non-Hermitian Hamiltonian satisfies the locality. This fact can be understood from the strong skewness of the similarity transformation $O$ in Eq.~\eqref{similar}, or said differently, the strong nonorthogonality of eigenstates of the non-Hermitian Hamiltonian.

It is in general not easy to exactly find a Hermitian counterpart for a non-Hermitian Hamiltonian with a real spectrum especially for interacting many-body problems. One notable exception is the generalized sine-Gordon model \cite{CMB05,YA17nc}:
\eqn{\label{sG}
H_{\alpha_{\rm r},\alpha_{\rm i}}=\int dx\left\{\frac{\hbar v}{2\pi}\left[K(\partial_{x}{\theta})^2+\frac{1}{K}(\partial_{x}{\phi})^2\right]+\frac{\alpha_{\rm r}}{\pi}\cos(2{\phi})+\frac{i \alpha_{\rm i}}{\pi}\sin(2{\phi})\right\},
}
where $v$ is the sound velocity, $K$ is the Tomonaga-Luttinger liquid parameter, and $\alpha_{\rm r,i}$ represent the depths of real and imaginary potentials. The scalar fields $\phi,\theta$ satisfy the commutation relation $[\phi(x),\partial_x\theta(x')]=-i\pi\delta(x-x')$. This model satisfies the PT symmetry with respect to the parity operator acting on the field operator $\phi$ as $P\phi P^{-1}=-\phi$ and exhibits the real spectrum if $\alpha_{\rm r}>\alpha_{\rm i}$. In this case, one can exactly find its Hermitian counterpart $h_{\alpha_{\rm r},\alpha_{\rm i}}$ that has the same real spectrum as in $H_{\alpha_{\rm r},\alpha_{\rm i}}$. To see this, we introduce an operator $O=\exp[(\theta_{0}/2)\tanh^{-1}(\alpha_{\rm i}/\alpha_{\rm r})]$ with $\theta_0$ being a constant part of the field $\theta$, which generates the shift of the conjugate field as $\phi\to\phi+i\tanh^{-1}(\alpha_{\rm i}/\alpha_{\rm r})$.  
Using the similarity transformation $O^{-1}H_{\alpha_{\rm r},\alpha_{\rm i}}O=h_{\alpha_{\rm r},\alpha_{\rm i}}$, we obtain the Hermitian counterpart as
\eqn{
h_{\alpha_{\rm r},\alpha_{\rm i}}=\int dx\left\{\frac{\hbar v}{2\pi}\left[K(\partial_{x}{\theta})^2+\frac{1}{K}(\partial_{x}{\phi})^2\right]+\frac{\sqrt{\alpha_{\rm r}^2-\alpha_{\rm i}^2}}{\pi}\cos(2{\phi})\right\}.
}
The field theory~\eqref{sG} is of interest because of its anomalous quantum critical behavior violating the $c$-theorem \cite{YA17nc} and its relation to the quantum Liouville field theory at $\alpha_{\rm r}=\alpha_{\rm i}$ \cite{IY16,NS90} (cf. Eq.~\eqref{livcft} and Sec.~\ref{nonunitary_cft}).  One may view this model as a many-body generalization of the single-particle Hamiltonian $H_V$ in Eq.~\eqref{singlecossin}. Figure~\ref{fig:2ashida_NC}(a) illustrates a possible microscopic realization of the generalized sine-Gordon model \cite{YA17nc},  and Fig.~\ref{fig:2ashida_NC}(b) plots its low-energy many-body spectrum. In the latter, the real-to-complex spectral transition occurs at the first EP encountered with increasing the non-Hermitian term $\gamma$. Above this finite threshold $\gamma_{\rm EP}>0$ (not vanishing in the thermodynamic limit),  many excited energies coalesce and a proliferation of EPs occurs. 
}

\begin{figure}
\begin{center}
\includegraphics[width=12cm]{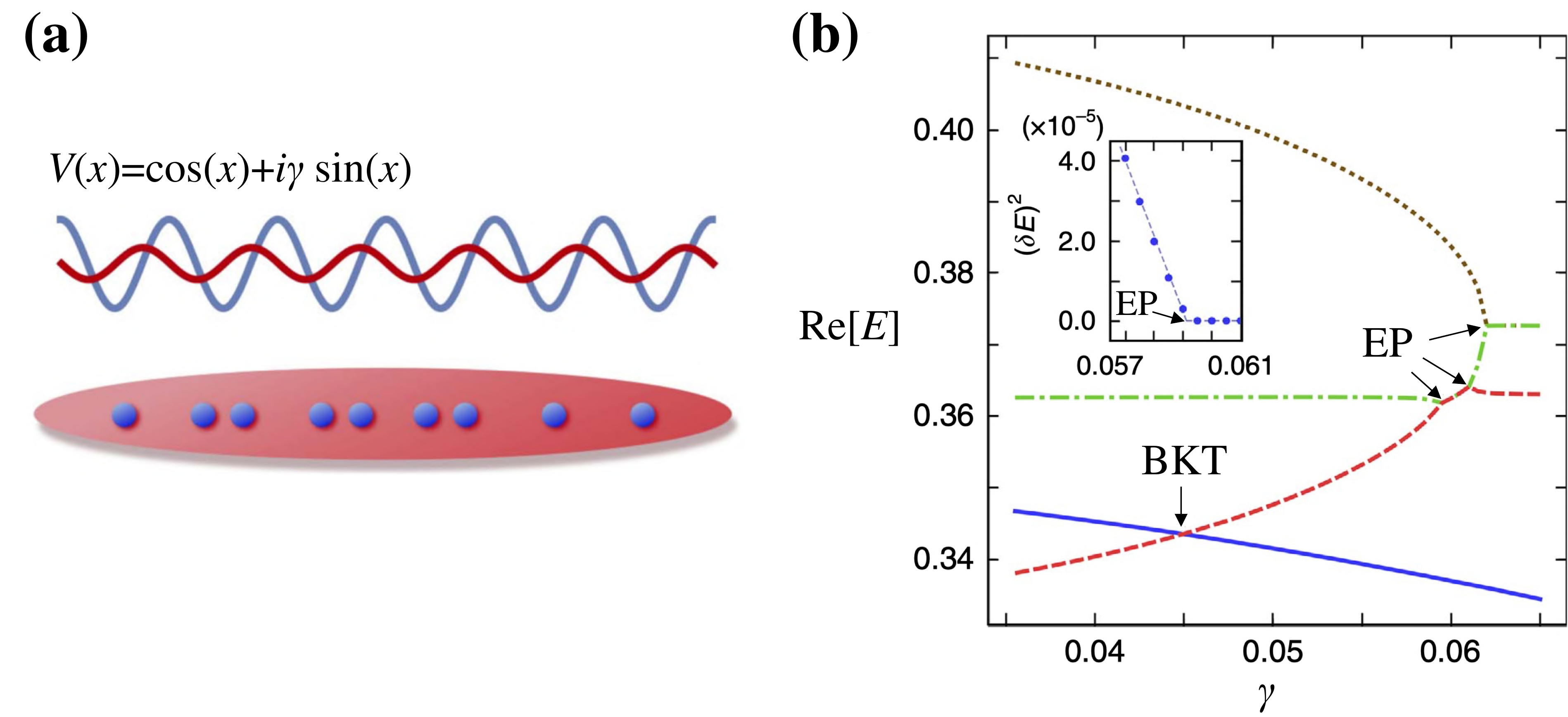}
\end{center}
\caption{(a) Schematic figure illustrating a microscopic realization of the generalized sine-Gordon model~\eqref{sG}, where a one-dimensional strongly correlated system is subject to a dissipative periodic potential $V(x)=\cos(x)+i\gamma\sin(x)$. (b) Low-energy many-body spectrum of the lattice model for the generalized sine-Gordon model. Real parts of eigenvalues are plotted against a non-Hermitian term $\gamma$. With increasing $\gamma$, the exceptional points (EPs) occur at finite $\gamma$ and the corresponding eigenvalues coalesce with the square-root scaling (inset). The crossing point between red and blue eigenvalues represents the Berezinskii-Kosterlitz-Thouless transition point  \cite{YA17nc}. See Sec.~\ref{Sec:QMBP} for further discussions.}
\label{fig:2ashida_NC}
\end{figure}

Examples~\ref{ph22}, \ref{phparticle}, and \ref{phmany} demonstrate that if a non-Hermitian operator satisfies the pseudo Hermiticity, there exists a certain parameter region for which the spectrum remains to be entirely real, indicating that the real-valuedness of the spectrum is robust against perturbations. In fact, one can show the following theorem \cite{EC04}:

\begin{theorem}[Robustness of the real-valuedness of spectrum of a pseudo-Hermitian operator]
Consider a Hermitian positive operator $M_1$ on the Hilbert space with discrete spectrum $\{0\leq\lambda_0<\lambda_1<\ldots\}$. Consider a (not necessarily Hermitian) operator $M_2$. Let $\eta$ be an invertible Hermitian operator satisfying $\eta^2=1$ and let $M_1$ and $M_2$ satisfy the pseudo Hermiticity as $\eta M_1\eta^{-1}=M_1^\dagger$ and $\eta M_2\eta^{-1}=M_2^\dagger$. The spectrum of $M_1+\epsilon M_2$ is then real if $|\epsilon|<{\rm inf}_{j\geq0}[\lambda_{j+1}-\lambda_j]/(2\|M_2\|)$.
\end{theorem}
Note that the above condition on $\epsilon$ is sufficient, but not necessary for the real-valuedness of the spectrum. Indeed, as discussed in Example~\ref{phmany}, a certain class of many-body models can exhibit real spectra in nonvanishing parameter regions even though their level spacings are in general exponentially small and the condition on $\epsilon$ in the above theorem can be easily violated (see e.g., Fig.~\ref{fig:2ashida_NC}(b)). In such a many-body case, the robustness of the real-valuedness is much more nontrivial than single-particle models. 
 
\subsection{Exceptional points\label{Sec:EP}}
Many problems in non-Hermitian physics deal with a \emph{continuous family} of matrices, i.e., a continuous map from some parameter space to the matrix space. Since any matrix can be written into the Jordan normal form (cf. Eq.~\eqref{JNF}), it is natural to ask whether, and if yes, how will the Jordan normal form change with the parameter. For the case of matrices being parametrized by a complex variable, it turns out that the Jordan normal form is stable except for singularities known as \emph{exceptional points}, which form measure-zero set in the parameter space. Two or more eigenvectors can coalesce at the exceptional point, and due to such  skewness of the vector space, a non-Hermitian system behaves as if it loses its dimensionality near the exceptional point. In this section, we explain fundamental aspects of exceptional points and review a number of physical phenomena originating from the singular behavior around exceptional points. 

\subsubsection{Definition and basic properties\label{secepdef}}
We first present the mathematical definitions for two different types of  \emph{exceptional points (EPs)}, which have been firstly introduced by Toshio Kato over half a century ago \cite{TK80}. We note that these definitions are more general than the notion of the physical EPs usually discussed in a majority of the literature in physics as detailed below. 
\\
\\
{\it First-type EPs}

\vspace{3pt}
\noindent 
We consider a family of $n\times n$ matrix $M(\kappa): \mathbb{C}\to\mathbb{C}^{n\times n}$, of which each entry is a complex analytic function of $\kappa$. Without loss of generality, we assume that these analytic functions are well-defined on a neighborhood ${\rm D}_0$ of the origin, so that we can expand $M$ as
\begin{equation}
M(\kappa)=\sum^\infty_{p=0} \kappa^p M_p,\;\;\;\;\forall\kappa\in{\rm D}_0.
\end{equation}
Let the number of different eigenvalues $\lambda_j(\kappa)$ of $M(\kappa)$ be $J$. Then $J$ should be a constant integer on ${\rm D}_0$ except for several discrete points, which are defined as the \emph{first-type EPs}. We denote the set of all the first-type EPs as $\mathscr{E}_1$ at which spectral degeneracies must occur.  

Two remarks are in order. First, we should explain why $J$ basically remains invariant so that the first-type EPs are indeed well-defined. The reason is that all the eigenvalues are the solutions to an order-$n$ algebraic equation ${\rm det}(\lambda-M(\kappa))=0$ with all the coefficients being analytic functions. This means that each $\lambda(\kappa)$, which is always continuous, has at most a finite number of \emph{algebraic singularities}\footnote{These are the points around which $\lambda(\kappa)$ behaves asymptotically like $(\kappa-\kappa_0)^\alpha$, with $\alpha$ being a non-integer rational number. Since the eigenvalues should be bounded, we actually have $\alpha>0$.} if any. Thus, $\lambda(\kappa)$'s are almost everywhere analytic functions, and two solutions should coincide entirely or at a finite number of points in ${\rm D}_0$ (denoted by $\mathscr{E}_1$), leading to a constant $J$ in ${\rm D_0}\backslash\mathscr{E}_1$. Second, it is possible that $J<n$ almost everywhere, a situation called \emph{permanently degenerate}. Nevertheless, the degeneracy will only increase but never decrease at the first-type EPs. Otherwise, it would contradict the continuity of $\lambda(\kappa)$'s.\footnote{If the degree of degeneracy at $\kappa_0$ decreases due to the lift of degeneracy between $\lambda_1(\kappa)$ and $\lambda_2(\kappa)$, we have $\lambda_{12}(\kappa)\equiv\lambda_1(\kappa)-\lambda_2(\kappa)\neq0$ at $\kappa_0$ but $\lambda_{12}(\kappa)=0$ near $\kappa_0$. This is impossible for a continuous $\lambda_{12}(\kappa)$.} 
\\
\\
{\it Second-type EPs}

\vspace{3pt}
\noindent 
Excluding the first-type EPs, the projectors $P_j(\kappa)$ (cf. Eq.~\eqref{eproj}) share the same dimensions for $\kappa\in{\rm D_0}\backslash\mathscr{E}_1$ and it can be proved that, within an arbitrary simply connected domain ${\rm D}\subset{\rm D_0}\backslash\mathscr{E}_1$, there exists a family of \emph{analytic} and invertible matrix $U_j(\kappa)$ such that $P_j(\kappa)=U_j(\kappa)P_j(\kappa_0)U_j(\kappa)^{-1}$ for some $\kappa_0\in{\rm D}$. On the other hand, we can only find a family of \emph{rational} matrix $V_j(\kappa)$ such that the nilpotent part (cf. Eq~\eqref{2_nilpotent}) satisfies $N_j(\kappa)=V_j(\kappa)N_j(\kappa_0)V_j(\kappa)^{-1}$. By rational (instead of analytic), we mean that $V_j(\kappa)$ or $V_j(\kappa)^{-1}$ may have a finite number of \emph{poles} in ${\rm D}$, which are defined as the \emph{second-type  EPs}. We denote the set of all the second-type EPs  
as $\mathscr{E}_2$. It is clear that $M(\kappa)$ has a stable Jordan normal form \eqref{JNF} except for $\kappa\in\mathscr{E}_1\cup\mathscr{E}_2$. 
\\
\\
{\it Physical EPs and diabolic points}

\vspace{3pt}
\noindent 
These definitions are slightly different from the ones found in the literature of physics, where the EPs are usually considered as the special points in the parameter space such that a parametrized matrix, which is almost everywhere diagonalizable, becomes nondiagonalizable. Accordingly, here we define the {\it physical EP} as the point at which two or more eigenvalues and their corresponding eigenvectors coalesce \cite{Berry2004,Heiss_2004}. Equivalently, this definition characterizes the physical EPs as the singularities in the parameter space, at which {\it both} projector and nilpotent in the Jordan normal form exhibit  discontinuous changes. It then follows that the spectrum must exhibit algebraic singularities around the physical EPs.

While the above physical definition of EPs is a special case of a more general notion of the first-type EPs introduced above, such a definition is useful in practice because, in real physical systems without parameter fine-tuning, the matrix describing the system should typically be diagonalizable almost everywhere in the parameter space. 
The simplest example of a first-type EP as a physical EP has already been given in Example~\ref{Ex:ev}, where the two eigenvalues $\lambda(\kappa)=\pm\sqrt{\kappa}$ and the corresponding eigenvectors coalesce at $\kappa=0$, around which algebraic singularities with the square-root scaling emerge.

Meanwhile, the notion of the first-type EP also includes another set of singular points, where some of eigenvalues coalesce (and thus spectral degeneracies occur), while the corresponding eigenvectors can  still be chosen to be orthogonal and thus algebraic singularities are absent. Such a kind of points are called as the {\it diabolic points} \cite{YDR96} and not usually considered as a physical EP in   literature, since it can occur also in Hermitian cases. The following minimal example illustrates the notion of diabolic points.

\exmp{(The first-type EP as a diabolic point in $2\times 2$ matrices). 
\label{Ex:1EP}
Consider a family of $2\times 2$ matrices: 
\begin{equation}
M(\kappa)=\begin{bmatrix} \;\kappa\; & \;0\; \\ \;0\; & \;-\kappa\; \end{bmatrix}, 
\label{EPsz}
\end{equation}
where $\kappa=0$ is the diabolic point. The eigenvalues obviously stay analytic everywhere, i.e., ${\rm D}_0=\mathbb{C}$, and thus no algebraic singularities exist. 
}
\noindent From now on, among the first-type EPs, we shall call only the ones accompanying algebraic singularities, which are also known as \emph{branchpoints} \cite{HD16},  simply as the {\it EPs}. This definition is  consistent with the practical definition that is usually considered in the literature. The notion of EPs can readily be generalized to multiple variables,  leading to higher dimensional exceptional objects such as exceptional rings and surfaces.

Nevertheless, we mention that this physical definition still does not capture a second-type EP, which is not characterized by the coalescence of eigenvalues and can even be an atypical \emph{diagonalizable} point as illustrated by Example~\ref{Ex:2EP} below. This situation has been overlooked in the majority of the literature. 

\exmp{(The second-type EP in $2\times 2$ matrices).
\label{Ex:2EP}
Let us consider a minimal example of the second-type EP realized by a family of $2\times2$ matrices:
\begin{equation}
M(\kappa)=\begin{bmatrix}  \;0\; &  \;\;\kappa\;\; \\ \;0\; & \;\;0\;\; \end{bmatrix}.
\label{exstep}
\end{equation}
This is a family of permanently degenerate matrices without any first-type EP. The eigenprojector is always the identity and we can choose $U(\kappa)=I$. On the other hand, the Jordan normal form of $M(\kappa)$ reads
\begin{equation}
M(\kappa)=V(\kappa)\begin{bmatrix} \;0\; & \;\;1\;\; \\ \;0\; & \;\;0\;\; \end{bmatrix} V(\kappa)^{-1},\;\;\;\;
V(\kappa)=\begin{bmatrix} \;1\; & \;\;b\;\; \\ \;0\; & \;\;\kappa^{-1}\;\; \end{bmatrix},
\end{equation}
where $b$ can be an arbitrary complex number. Obviously, the nilpotent part exhibits the singularity at $\kappa=0$ and thus this point is the second-type EP.
}

\subsubsection{Physical applications\label{secepphys}}
The singular behavior around EPs in the complex parameter space  often leads to dramatic effects in a wide range of fields in physics such as optics, mechanics, atomic and molecular physics, quantum phase transitions, and even quantum chaos.
Historically, the physical significance of EPs has been pointed out in early theoretical works \cite{Pancharatnam1958,Berry_1998,Heiss1999} and their branchpoint singularities have been studied in the context of quantum resonances in atomic and nuclear physics \cite{NM98}. 
Unambiguous experimental observations of EPs and their topological aspects have been firstly achieved by using microwave cavities \cite{PE00,DC01,DC03}.  For quantum resonances in scattering problems, the EP emerges as a branchpoint singularity at a pole of the $S$-matrix \cite{NRG02}. Physically, this nonanalytic feature results in unconventional behavior of line widths in resonances, which goes beyond the standard expectations from Fermi's golden rule \cite{RI00}, and also leads to the enhanced transmission through quantum dots below EPs \cite{AKM05,LYA00}. EPs for resonances in the non-Hermitian chaotic Sinai billiard have been experimentally observed in exciton-polariton systems \cite{GT15}. Recently, the enhanced laser linewidth at an EP \cite{KP79} has been experimentally observed in a phonon laser created by optomechanical systems \cite{ZJ18}.

Experimental developments in optics have in turn enabled one to manipulate gain and loss of photons in a controlled way, especially in a spatially engineered manner.
Together with a surge of studies of PT-symmetric non-Hermitian systems \cite{BCM98,AZ01,AZ012,PD07}, photonics has become one of the most ideal experimental platforms to study the physics of EPs \cite{REG07,KGM08,KS082,LSS10,DK15,ZB15,CA16}. As demonstrated in Example~\ref{ptband} below, an increase of the strength of gain and loss in PT-symmetric photonic crystals typically leads to gap closing of band dispersions at which eigenvectors coalesce, thus generating EPs. This feature allows for realizing peculiar types of EPs that were difficult to access in previous setups of quantum resonances and microwave cavities, in which a band structure is absent since they have no spatial invariance. An illustrative example is a ring of EPs in two-dimensional photonic crystals, which is induced by the band merging of two dispersions around the (gapless) Dirac point \cite{ZB15}. 
 
The counterpart of EPs in the continuum spectrum is known as a {\it spectral singularity} and usually introduced as a singularity at which the completeness of the basis for the continuum spectrum is lost \cite{N60}. In one-body regimes, spectral singularity can appear as a merging point in the band spectrum at which two dispersions and the corresponding eigenvectors coalesce at the gapless point embedded in the continuum spectrum. There, anomalous wave propagation and reflection have been predicted \cite{AM09,LSS10,MA15}. In many-body regimes, a spectral singularity can occur at a quantum phase transition, where the non-Hermiticity induces gap closing in the many-body spectrum that is continuum in the thermodynamic limit. Unconventional renormalization-group fixed points \cite{YA17nc} and critical behavior \cite{IY16} have been found at such many-body spectral singularities. Proliferation of spectral singularities in the chaotic many-body system has also been numerically predicted \cite{LDJ19}. 

Below we review the notion of exceptional points in more detail with introducing illustrative examples and reviewing their key applications to physical systems.

\exmp{\label{ep2b2}(EPs in $2\times 2$  matrices and unidirectional invisibility). 
For EPs of physical interest, one usually encounters a branchpoint singularity, where two repelling levels are connected by a square-root scaling. We illustrate it by discussing the simplest example  of $2\times 2$ matrices. To be concrete, we focus on the non-Hermitian matrices considered in Eq.~\eqref{pseudo2b2} but with  $\theta=\phi=0$ and $\gamma\in\mathbb{C}$:
\eqn{\label{sec22b2}
M(\gamma)=\begin{bmatrix}
\;\epsilon+\gamma\; & \;\;\rho e^{i\alpha}\;\; \\
\;-\rho e^{-i\alpha}\; & \;\;\epsilon-\gamma\;\;
\end{bmatrix},\;\;\;\epsilon,\rho,\alpha\in\mathbb{R},
} 
where the parameters $\epsilon,\alpha$ and $\rho\neq0$ are assumed to be fixed and $\gamma\in\mathbb{C}$ parametrizes a continuous family of matrices.
Its eigenvalues are given by $\lambda_{1,2}=\epsilon\pm\sqrt{\gamma^2-\rho^2}$ and the corresponding right eigenvectors are (besides a constant factor)
\eqn{
{\bf r}_{1,2}\propto\begin{bmatrix}
\;\gamma\pm\sqrt{\gamma^2-\rho^2}\;\;\\
\;-\rho e^{-i\alpha}\;\;
\end{bmatrix}.
}
It is evident that EPs occur at $\gamma=\pm|\rho|$ at which the two eigenvalues $\lambda_{1,2}$ and the corresponding eigenvectors ${\bf r}_{1,2}$ coalesce with the nonanalytic, square-root scaling $\propto \sqrt{\gamma-\rho}$. 
Despite the simplicity, the EPs in $2\times 2$ matrices find many physical applications. The reason is that the original problem with a (possibly) large dimension can often reduce to the analysis of two-level matrices describing two merging levels in the vicinity of EPs.  Physical systems relevant to the EPs in $2\times 2$ matrices include coupled waveguides and cavities \cite{AY73,HAH91,PE00,DC01,DC03}, polarization modes and parametric amplification in optical waveguides \cite{Gordon00,RS82}, quantum resonances in atomic physics \cite{Magunov_2000}, dynamical instability in exciton-polariton systems \cite{CI13} 
and optomechanical systems \cite{AM14}. In the infinite-dimensional case, an infinite sequence of EPs can be found in both one-body and many-body spectra  \cite{BCM68,BCM69}. 

\begin{figure}
\begin{center}
\includegraphics[width=10.5cm]{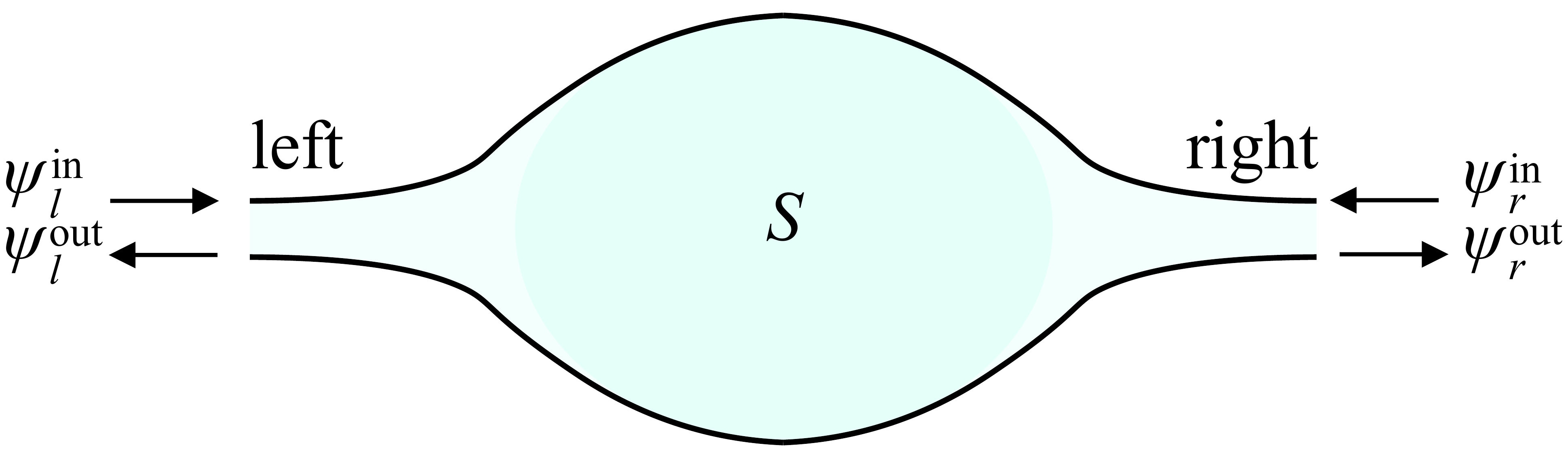}
\end{center}
\caption{Schematic figure illustrating the input-output scattering formalism for the two channel case. The scattering media has left ($l$) and right ($r$) optical channels and $\psi_{l,r}^{\rm out (in)}$ represent the electromagnetic fields of the outgoing (incoming) modes at each optical channel. Scattering and transport of these modes are characterized by the scattering matrix $S$ via Eq.~\eqref{epsca}. If the media is conservative, the matrix $S$ must be unitary, whereas the unitarity can be violated if it is nonconservative due to the presence of gain or loss.}
\label{fig:2scattering}
\end{figure}

A simple, yet notable physical phenomenon at EPs in $2\times 2$ non-Hermitian matrices is unidirectional and reflectionless light propagation through non-Hermitian waveguides \cite{MK05,AM09,LZ11,Longhi_2011,MMA12,CG132}. The non-Hermiticity can originate from alternating layers with different strengths of gain or loss in optical gratings.
In the simplest case, the scattering property through the waveguide can be understood from the following $2\times 2$ scattering matrix $S$ (see Sec.~\ref{secplight} for a general formalism):
\eqn{\label{epsca}
\begin{bmatrix}
\;\psi_{r}^{\rm out}\;\;\\
\;\psi_{l}^{\rm out}\;\;
\end{bmatrix}
=S
\begin{bmatrix}
\;\psi_{l}^{\rm in}\;\;\\
\;\psi_{r}^{\rm in}\;\;
\end{bmatrix}
\equiv
\begin{bmatrix}
\;T_l\; & \;\;R_{r}\;\; \\
\;R_{l}\; & \;\;T_r\;\; \\
\end{bmatrix}
\begin{bmatrix}
\;\psi_{l}^{\rm in}\;\;\\
\;\psi_{r}^{\rm in}\;\;
\end{bmatrix},
}
where $\psi_{l,r}^{\rm out (in)}$ represent the electromagnetic fields of the outgoing (incoming) modes at the left ($l$) and right ($r$) optical channels, respectively (see Fig.~\ref{fig:2scattering}), $T_{l,r}\in\mathbb{C}$ and $R_{l,r}\in\mathbb{C}$ are the complex transmission and reflection coefficients at the left and right channels. We assume $T_{l}=T_r\equiv T$ as satisfied in the physical examples discussed below (see Sec.~\ref{sec:6np} for general arguments). The eigenvalues and the corresponding eigenvectors with $R_r\neq 0$ are then given by
\eqn{
\lambda_{\pm}=T\pm\sqrt{R_{l}R_{r}},\;\;\;\;{\bf r}_{\pm}\propto
\begin{bmatrix}
\;1\;\;\\
\;\pm\sqrt{\frac{R_l}{R_r}}\;\;
\end{bmatrix}.
}
In a Hermitian case, equal transmission coefficients always lead to equal reflection coefficients, $R_l=R_r$, due to the unitarity of the $S$ matrix. However, in a non-Hermitian case, this unitarity can be violated, allowing one to achieve different reflection coefficients $R_l\neq R_r$, despite the equal transmissions. In the most extreme case, one can realize the condition $R_l=0$ with $R_r\neq 0$ at the EP and the eigenstates coalesce into $\boldsymbol{\psi}\propto[1,0]^{\rm T}$. Due to this dimensional reduction, the light incoming from the left transmits through the waveguide without reflections, thus leading to the unidirectional and reflectionless wave propagation at the EP. 

Physical examples for realizing the EPs in the scattering matrix~\eqref{epsca} include a Fabry-Perot cavity subject to PT-symmetric gain and loss acting as a laser and a perfect absorber at the EPs \cite{LS102,CYD11}, and an EP at the wafer-scale optical structure \cite{LF142}. A peculiar wave propagation has  also been  experimentally observed in the semiconductor waveguide gratings using Ge/Cr and Si structures whose permittivity is manipulated to ensure the PT symmetry \cite{LF13}. There, the asymmetry in the light reflection at EPs is induced by the destructive interference of the coalescing light modes. The unidirectionality and the related phenomena  have also been observed in photonic lattices \cite{RA12pt}, organic composite films \cite{YY142}, acoustic sensors \cite{FR15}, micro resonators with definite angular momentum modes \cite{MP16} and chiral rotating modes \cite{WJ11,MK14,PB16}. 
Another example is a scattering in a single cavity fabricated by semiconductor structures with gain and loss \cite{WZ16}.

Despite the unidirectionality described above, we emphasize that the reciprocity (i.e., $T_l=T_r$) in the light transmission still holds true if the time-reversal symmetry, ${\cal T}H{\cal T}^{-1}=H^\dagger$, is satisfied for a non-Hermitian Hamiltonian $H$. In fact, in linear regimes, this fact indicates that the reciprocity cannot be violated by merely realizing a complex permittivity in medium  \cite{JGM04,AZ01}. Thus, to realize nonreciprocal transmission, one must attain at least one of the following conditions: (i) include nonlinear effects \cite{BP14}, (ii) break the time-reversal symmetry (in the above sense) \cite{BM19}, or (iii) go beyond the two-channel setup \cite{Fleury516}  (see Sec.~\ref{sec:6np} for further reviews on nonreciprocal transport). 
}

\exmp{(EPs with higher-order singularities and enhanced sensitivities). 
\begin{figure}
\begin{center}
\includegraphics[width=13.5cm]{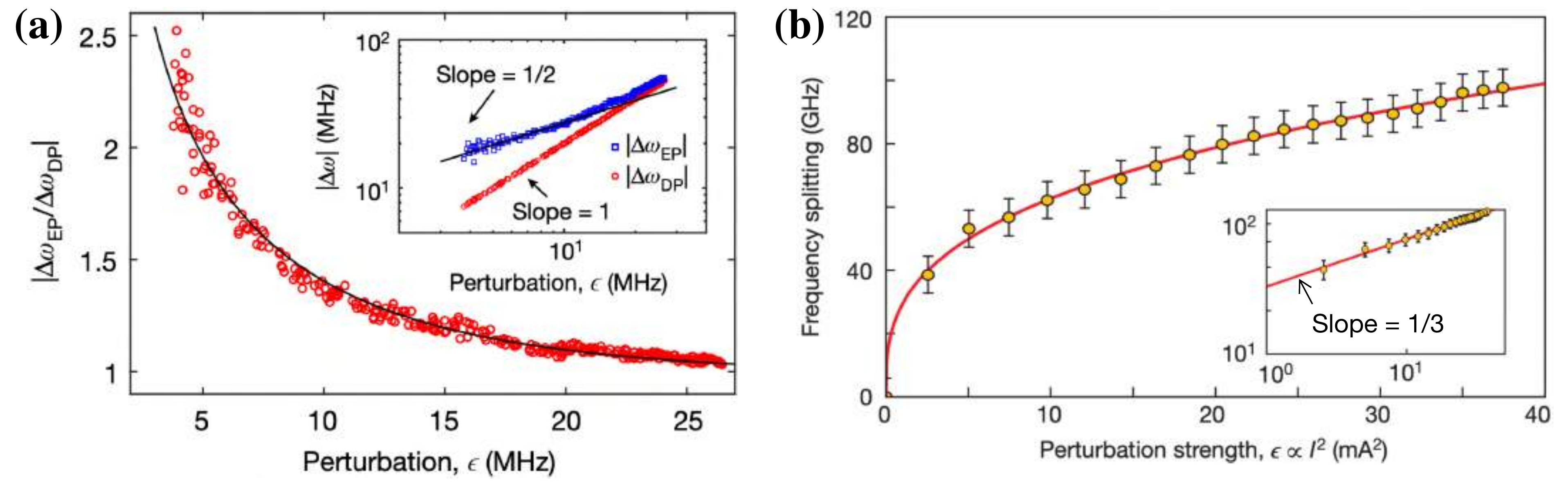}
\end{center}
\caption{Enhanced sensitivity at exceptional points (EPs). (a) Experimental results on the enhanced sensitivity of a nanorparticle sensor operating at a two-fold EP. As shown in inset, the eigenvalue gap $\Delta\omega_{\rm EP}$ around the EP is consistent with the scaling $\propto\sqrt{\epsilon}$ with $\epsilon$ characterizing the deviation from the EP. In contrast, around the diabolic point (DP), it linearly scales as $\Delta\omega_{\rm DP}\propto \epsilon$. Adapted from Ref.~\cite{CW17}. Copyright \copyright\,   2017 by Springer Nature. (b) Experimental results on the enhanced sensitivity of a refractive index sensor operating at a three-fold EP. The inset confirms the scaling $(\epsilon)^{1/3}$ around this higher-order EP.  
Adapted from Ref.~\cite{HH17}. Copyright \copyright\,   2017 by Springer Nature.
}
\label{fig:2enhanced}
\end{figure}
When  two or more eigenstates coalesce, the eigenvalue difference around an EP exhibits a branchpoint singularity. While the singularity typically associates with the square-root scaling (see Fig.~\ref{fig:2enhanced}(a)), higher-order singularity is also possible when more eigenstates coalesce at an EP.  To demonstrate this, let us consider the $3\times 3$ non-Hermitian matrix \cite{Demange_2011}:
\eqn{\label{3b3nh}
M(\gamma)=
\begin{bmatrix}
\;\gamma+i\; & \;\frac{1}{\sqrt{2}}\; & \;0\; \\
\;\frac{1}{\sqrt{2}}\; & \;0\; & \;\frac{1}{\sqrt{2}}\; \\
\;0\; & \;\frac{1}{\sqrt{2}}\; & \;\gamma-i\; 
\end{bmatrix},
} 
where $\gamma\in\mathbb{C}$  is a variable characterizing a continuous family of matrices. All the three eigenvalues and eigenvectors coalesce at $\gamma=0$.  
In the vicinity of this threefold EP, i.e., for a very small $|\gamma|$, a gap $\delta\lambda$ between different eigenvalues closes with the third-root scaling instead of the square-root scaling in the previous example:\footnote{The characteristic polynomial (\ref{pM}) is given by $(x-\gamma)^3+\gamma(x-\gamma)^2+\gamma$, so its zeros are given by $x=y+\gamma$ with $y^3+\gamma y^2+\gamma=0$. When $|\gamma|$ is small, then both $x$ and $y$ are of the order of $O(\gamma^{1/3})$ due to the fact that $\gamma$ and $\gamma y^2$ are higher-order small quantities. }
\eqn{
\delta\lambda\propto \gamma^{1/3}.
}
A non-Hermitian system described by Eq.~\eqref{3b3nh} has been realized by three coupled cavities, where outer two ones have balanced gain and loss, while the middle one has none of them \cite{HH17} (see Fig.~\ref{fig:2enhanced}(b)). 

The argument can be generalized to a larger $n$-level case. To see this, we recall that, in a Hermitian system, one can use the perturbation theory to derive an analytic expression $\delta\lambda=\sum_{j=1}^{\infty}c_j(\gamma-\gamma_{\rm DP})^{j}$ with $c_j$ being some coefficients and  DP denoting a diabolic point; this analytical perturbation indicates the absence of branchpoint singularities. 
In contrast, in non-Hermitian cases, this standard perturbation theory can be hampered because branchpoint singularities associated with EPs limit the convergence radius of the standard perturbation series \cite{THS72}. 
Nevertheless, the following modified series solution can still exist in the vicinity of $n$-fold EP \cite{Heiss_2008}
\eqn{\label{nfoldep}
\delta\lambda=\sum_{j=1}^{\infty}c_j(\gamma-\gamma_{n{\rm EP}})^{\frac{j}{n}},
}
where the $n$-fold EP is assumed to occur at $\gamma=\gamma_{n{\rm EP}}$. This result is a direct consequence of the block structure of the size-$n$ Jordan form, where an $n$-fold coalescence of eigenvalues leads to a branchpoint singularity of $n$th order.

A notable physical consequence of EPs is the singularly high sensitivity to parameter changes.   
More explicitly, in the vicinity of an EP, the derivative of the eigenvalue gap with respect to $\gamma$ becomes singular at the EP as
\eqn{
 \frac{\partial\,\delta\lambda}{\partial\gamma}\propto \frac{1}{(\gamma-\gamma_{n{\rm EP}})^{1-\frac{1}{n}}}\to\infty.
}
This clearly shows a high sensitivity to a varying parameter $\gamma$ away from the EP.
As the order $n$ of an EP is increased, this singular sensitivity becomes stronger. 
Such sensitivity enhancement at EPs compared with Hermitian systems find interesting physical applications to a wide variety of systems including 
enhanced sensing with graphene metasurfaces \cite{CPY16} and microtoroid cavities \cite{CW17}, integrated micro resonators with a three-fold EP \cite{HH17}, 
coupled photonic cavities \cite{ZS16}, 
microcavity sensors for nanoparticle detection \cite{WJ142},  
enhanced sensing for mechanical motion \cite{LZP16}  and electronics \cite{CPY18}.
In contrast to classical systems described above, it has been argued that the exceptionally high sensitivity at EPs does not necessarily lead to enhancement  when effects from quantum noise are taken into account \cite{LW18,ZM19,LHK18}. 
Besides the enhanced sensitivity, an EP can also be used to discriminate multiple laser modes. The most dominant contribution in Eq.~\eqref{nfoldep} has a branchpoint singularity of $n$th order $\delta\lambda_{n}\propto(\gamma-\gamma_{n{\rm EP}})^{1/n}$. As the ability of the mode discrimination at the $n$-fold EP is governed by a factor $(\gamma^n-\gamma_{n{\rm EP}}^n)^{1/n}$ \cite{HH14},  it can considerably be larger than the corresponding linear term $\gamma-\gamma_{\rm DP}$ at diabolic points. This property has found applications to coupled micro resonators \cite{HH14}, rings with embedded gain-loss gratings \cite{LF14},  filtering of  transverse modes in ring resonators \cite{HH162}, and single-mode operation of microring laser \cite{LW17}.
}

\exmp{\label{ptband}(Chiral eigenvectors at spectral singularity in periodic structures). 
\begin{figure}
\begin{center}
\includegraphics[width=12cm]{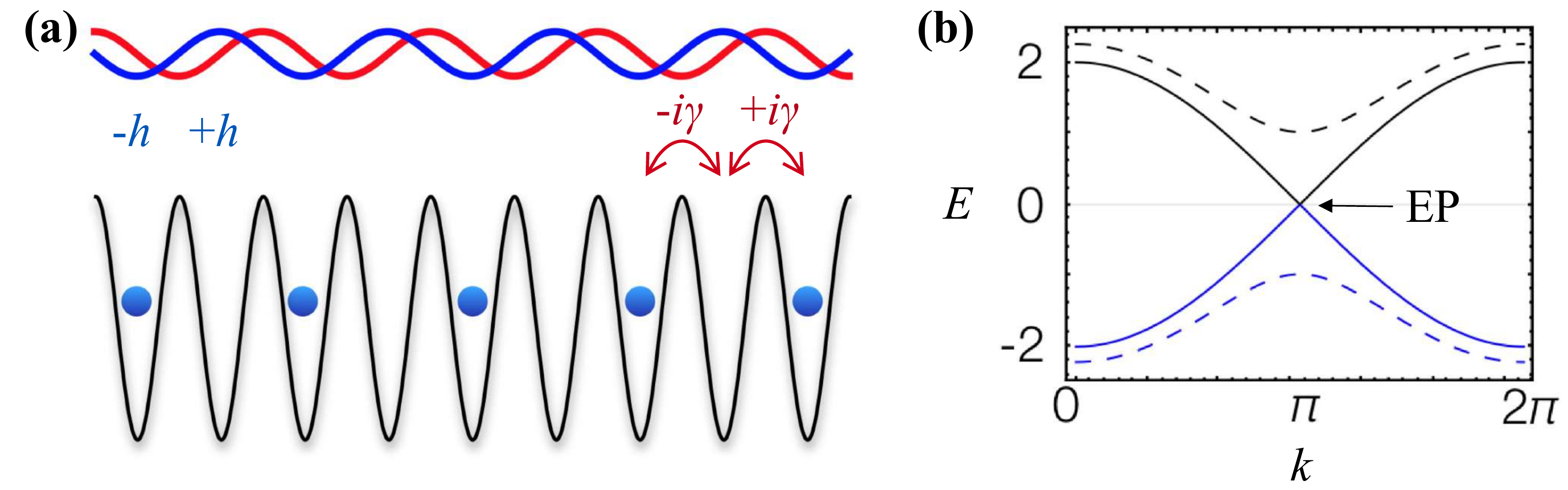}
\end{center}
\caption{(a) Schematic illustration of the tight-binding model in Eq.~\eqref{pt2S}.
In addition to the real (blue curve) and imaginary (red curve) potentials  (cf. the microscopic model in Fig.~\ref{fig:2ashida_NC}),  the deep optical potential (black curve) is superimposed on the system to localize atomic wave packets. The real potential leads to the on-site staggered potential $\pm h$, while the imaginary potential generates the staggered imaginary hopping $\pm i\gamma$ in addition to the real-valued tunneling. (b) Band spectrum of the tight-binding model~\eqref{pt2S} illustrated in (a) as a function of wavenumber $k$. The dashed curves show the gapped, real spectrum for $h=1$ and $\gamma=0$. As the imaginary coupling $\gamma$ is increased, the lower (blue) and upper (black) branches approach to each other and coalesce at $h=2\gamma$ as shown by the solid curves. The gapless coalescence point at $k=\pi$ is the EP \cite{YA18}.}
\label{fig:2epchiral1}
\end{figure}
When one considers a translationally invariant system, eigenvalues form a continuous spectrum as a function of wavenumber $k$, which is known as the Bloch bands. In non-Hermitian systems, the counterpart of EPs in such a continuous spectrum is called the spectral singularity, which can occur when the completeness of the basis is lost. One of the most intriguing aspects at spectral singularities is the strong skewness or the chirality  of eigenvectors in their vicinity. To gain further insights, we consider a concrete example of the periodic tight-binding model \cite{YA18} (see Fig.~\ref{fig:2epchiral1}(a)). Owing to the spatial invariance with period 2, the problem reduces to diagonalizing the following $2\times 2$ non-Hermitian Bloch Hamiltonian  for each wavenumber $k$:
 \eqn{
{H}\!&=&\!\!
\sum_{0\leq k<2\pi}\!\!\left[\begin{array}{cc}
{a}^{\dagger}_k\; & \;{b}^{\dagger}_k \end{array}\right]\!\left[\begin{array}{cc}
-h & -1-i\gamma+(-1+i\gamma)e^{-ik}\\
-1-i\gamma+(-1+i\gamma)e^{ik} & h
\end{array}\right]\!\!\left[\begin{array}{c}
{a}_k\\
{b}_k
\end{array}\right], \label{pt2S}
 }
 where $\gamma$ and $h$ are real parameters, $a_{k}$ and $b_{k}$ denote independent bosonic or fermionic  operators associated with wavenumber $k$ and satisfy the commutation or the anticommutation relation $[a_k,a_{k'}^\dagger]_{\pm}=\delta_{kk'}$ and $[b_k,b_{k'}^\dagger]_{\pm}=\delta_{kk'}$ with $[A,B]_{\pm}\equiv AB\mp BA$.   
It is instructive to express the diagonalized form of $H$ in terms of the normalized second-quantized operators as
\eqn{
{H}=\sum_{k}\sum_{\lambda=\pm}\epsilon_{\lambda}(k){g}_{\lambda k}^{\dagger}{f}_{\lambda k},\;\;\;\epsilon_{\pm}(k)=\pm\sqrt{h^{2}-4\gamma^{2}+2J^2(1+\cos k)},\label{diagS}
}
where $\epsilon_{\lambda}(k)$ are two eigenvalues for each  $k$ with band index $\lambda=\pm$, the operator ${g}_{\lambda k}^{\dagger}$ creates a right eigenvector of a single particle state in ${H}$, i.e., ${H}{g}^{\dagger}_{\lambda k}|0\rangle=\epsilon_{\lambda}(k){g}^{\dagger}_{\lambda k}|0\rangle$,  the operator ${f}_{\lambda k}$ creates the corresponding left eigenvector of ${H}$, i.e., $\langle 0|{f}_{\lambda k}{H}=\langle 0|{f}_{\lambda k} \epsilon_{\lambda}(k)$, and $J=\sqrt{1+\gamma^2}$. In the second-quantized language, the biorthogonal relationship and the nonorthogonality of right eigenvectors can be formulated as (cf. Eqs.~\eqref{bio} and \eqref{nonorthorr}) 
\eqn{
[{f}_{\lambda k},{g}_{\lambda' k'}^{\dagger}]_{\pm}=\delta_{k,k'}\delta_{\lambda,\lambda'},\;\;\;\;[{g}_{\lambda k},{g}^{\dagger}_{\lambda' k'}]_{\pm}=\delta_{k,k'}\Delta_{\lambda\lambda'}(k),
}  where  the $2\times2$ matrices $\Delta_{\lambda\lambda'}(k)$ characterize an unusual commutation relation and their nonzero off-diagonal elements indicate the nonorthogonality between different bands at wavenumber $k$. 
One can further impose the normalization condition for the right eigenvectors as $\Delta_{++}=\Delta_{--}=1$ (cf. Eq.~\eqref{rrnorm}). The formulation described here can readily be generalized to a generic class of noninteracting, quadratic Hamiltonians (cf. Appendix~\ref{app2}). 

Let us now consider increasing $\gamma$ from zero while fixing the other parameters as $J,h>0$. The model satisfies the PT symmetry with respect to the spatial parity, and the real-to-complex spectral transition occurs at the gap closing point $h^2-4\gamma^2=0$. In particular, in the case of wavenumber $k=\pi$, this point corresponds to the second-order EP at which the $2\times 2$ matrix in Eq.~\eqref{pt2S} is nondiagonalizable and the gap in the spectrum closes with the square-root scaling $\delta\epsilon\propto \sqrt{\gamma-\gamma_{\rm EP}}$ with $\gamma_{\rm EP}=h/2$ (see Fig.~\ref{fig:2epchiral1}(b)). As this EP is embedded in the continuum spectrum, technically it should be interpreted as the spectral singularity.

 In the vicinity of the spectral singularity, the skewness or chirality of eigenvectors become singularly strong. To see this, let us first consider the  Hermitian case, i.e., the Hamiltonian with $\gamma=h=0$ in Eq.~(\ref{pt2S}).  There are two band dispersions: one has a positive group velocity ($\epsilon_{>}(k)=2\sin(k/2)$) and the other has a negative group velocity ($\epsilon_{<}(k)=-2\sin(k/2)$). In the vicinity of the gapless point at $k=\pi$, the corresponding eigenvectors with positive ($>$) and negative ($<$) group velocities for the 2$\times$2 Hermitian matrices are given by
\eqn{
{\bf v}_{>}(\delta k)&=&\frac{1}{\sqrt{2}}\left[\begin{array}{c}
-i-\frac{\delta k}{2}\\
1
\end{array}\right]+O\left((\delta k)^2\right),\label{clposi}\\
{\bf v}_{<}(\delta k)&=&\frac{1}{\sqrt{2}}\left[\begin{array}{c}
i+\frac{\delta k}{2}\\
1
\end{array}\right]+O\left((\delta k)^2\right),\label{clnega}
}
where $\delta k=k-\pi$ is the displacement satisfying $|\delta k|\ll 1$. 
At $\delta k=0$, two eigenvectors are orthogonal while eigenvalues coincide $\epsilon_{<}(\pi)=\epsilon_>(\pi)=0$, i.e., this point is a diabolic point.

\begin{figure}
\begin{center}
\includegraphics[width=10.5cm]{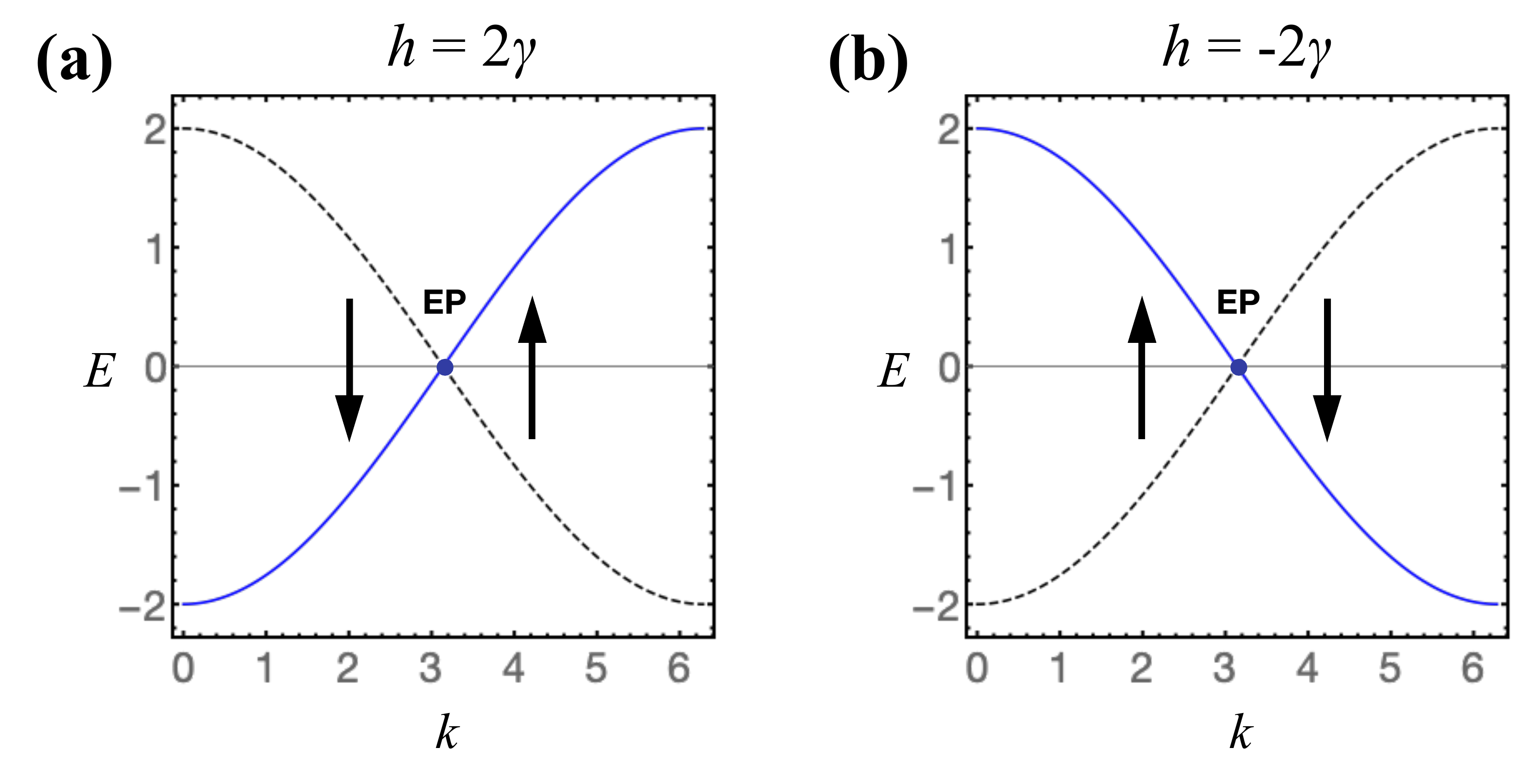}
\end{center}
\caption{Coalescence and chirality of eigenvectors near the EP of the tight-binding model~\eqref{pt2S} (cf. Fig.~\ref{fig:2epchiral1}(b)). (a) When $h=2\gamma$, the gapless point at $k=\pi$ forms an EP around which the two eigenvectors in different bands coalesce into the one associated with the band dispersion having positive group velocities (blue solid curve). This can result in the pronounced propagation in the positive direction. (b) In contrast, if the gain-loss structure is reversed (i.e., if we set $h=-2\gamma$), the eigenvectors in different bands coalesce into the one associated with the band having negative group velocities (blue solid curve), resulting in the pronounced propagation in the negative direction \cite{YA18}.}
\label{fig:2epchiral2}
\end{figure}

We next consider the non-Hermitian case with nonzero $\gamma$ satisfying $\gamma=h/2$. In this case, while the eigenvalue spectrum is exactly the same as in the above Hermitian case aside a constant proportional factor, two eigenvectors of different bands coalesce at the EP for $k=\pi$. This results in the strong skewness of the eigenvectors in the vicinity of the EP as inferred from the following perturbative expressions of right eigenvectors of the 2$\times$2 non-Hermitian matrices:
\eqn{
{\bf r}_{>}(\delta k)&=&\frac{1}{\sqrt{2}}\left[\begin{array}{c}
-i\\
1
\end{array}\right]+\frac{1}{4\sqrt{2}\gamma}\left[\begin{array}{c}
-i(1-2i\gamma-\sqrt{1+\gamma^2})\\
\sqrt{1+\gamma^2}-1
\end{array}\right]\delta k+O\left((\delta k)^2\right),\nonumber
\\
{\bf r}_{<}(\delta k)&=&\frac{1}{\sqrt{2}}\left[\begin{array}{c}
-i\\
1
\end{array}\right]+\frac{1}{4\sqrt{2}\gamma}\left[\begin{array}{c}
-i(1-2i\gamma+\sqrt{1+\gamma^2})\\
-\sqrt{1+\gamma^2}-1
\end{array}\right]\delta k+O\left((\delta k)^2\right),\nonumber
}
which are valid for $|\delta k|\ll \min(\gamma,1)$.  Note that in the limit of $\delta k\to 0$ both of these two eigenvectors coalesce into ${\bf v}_{>}(0)=[-i,1]^{\rm T}/\sqrt{2}$ having a positive group velocity (cf. Eq.~(\ref{clposi})); the degree of freedom associated with the other eigenvector, ${\bf v}_{<}(0)=[i,1]^{\rm T}/\sqrt{2}$, having the opposite ``chirality" is lost (see Fig.~\ref{fig:2epchiral2}(a)).
 Physically, the coalescence of eigenvectors near the spectral singularity thus leads to the chiral propagation in the positive direction \cite{YA18}.  When the two-dimensional vector space can be regarded as the polarization space in the basis of linear polarizations \cite{CA172,BB18}, the different chirality can represent the circular polarization in the opposite direction.  
We remark that if the gain-loss structure is reversed, i.e., if we set $h=-2\gamma$, the two right eigenvectors can be shown to coalesce into one having a negative group velocity given in Eq.~(\ref{clnega})  (see Fig.~\ref{fig:2epchiral2}(b)). 

 The spectral singularity also induces the divergent behavior of the corresponding left eigenvectors as inferred from the following perturbative results:
\eqn{
{\bf l}_{>}(\delta k)&=&\!\frac{2\gamma}{\sqrt{2(1+\gamma^2)}}\left[\begin{array}{c}
i\\
1
\end{array}\right]\frac{1}{\delta k}\!+\!\frac{1}{2\sqrt{2(1+\gamma^2)}}\left[\begin{array}{c}
-i(1+2i\gamma+\sqrt{1+\gamma^2})\\
\sqrt{1+\gamma^2}+1
\end{array}\right]\!+\!O\!\left(\delta k\right),\nonumber
\\
{\bf l}_{<}(\delta k)&=&-\frac{2\gamma}{\sqrt{2(1+\gamma^2)}}\left[\begin{array}{c}
i\\
1
\end{array}\right]\!\frac{1}{\delta k}\!+\!\frac{1}{2\sqrt{2(1+\gamma^2)}}\left[\!\begin{array}{c}
-i(-1-2i\gamma+\sqrt{1+\gamma^2})\\
\sqrt{1+\gamma^2}-1
\end{array}\!\right]\!\!+\!O\!\left(\delta k\right),\nonumber
}
which are valid for $|\delta k|\ll{\rm min}(\gamma,1)$. The divergence of the left eigenvectors in the limit $\delta k\to 0$ originates from the vanishing inner product between the normalized right and left eigenvectors at the EP \cite{HWD01}. 

The skewness or chirality of eigenstates associated with the spectral singularity demonstrated above lies at the heart of a variety of unconventional physical phenomena. Examples include the asymmetric wave propagation in the single-particle model~\eqref{singlecossin} \cite{LSS10}, the supersonic propagation of correlations in the many-particle fermionic models   \cite{YA18,DB19},  the unusual scale invariance of the field theory~\eqref{sG} \cite{SN1990,YA17nc,IY16} and the bifurcation in the Bose-Einstein condensation with attractive $1/r$ interaction \cite{CH08}. The peculiar wave propagation associated with linearly increasing amplitudes in time is also predicted besides the usual exponential behavior \cite{LSS10}. In fact, this linear increase originates from a generic feature of the exponential of the Jordan block form (see Eq.~\eqref{jorlinear}). The spectral singularity in periodic structures was experimentally observed in photonic lattices, where the continuum band of the tight-binding model subject to spatially periodic gain-loss structures leads to the anomalous optical response of short laser pulses \cite{RA12pt,ZB15}. 
}

\subsubsection{Topological properties\label{seceptopo}}
\begin{figure}
\begin{center}
\includegraphics[width=12.5cm]{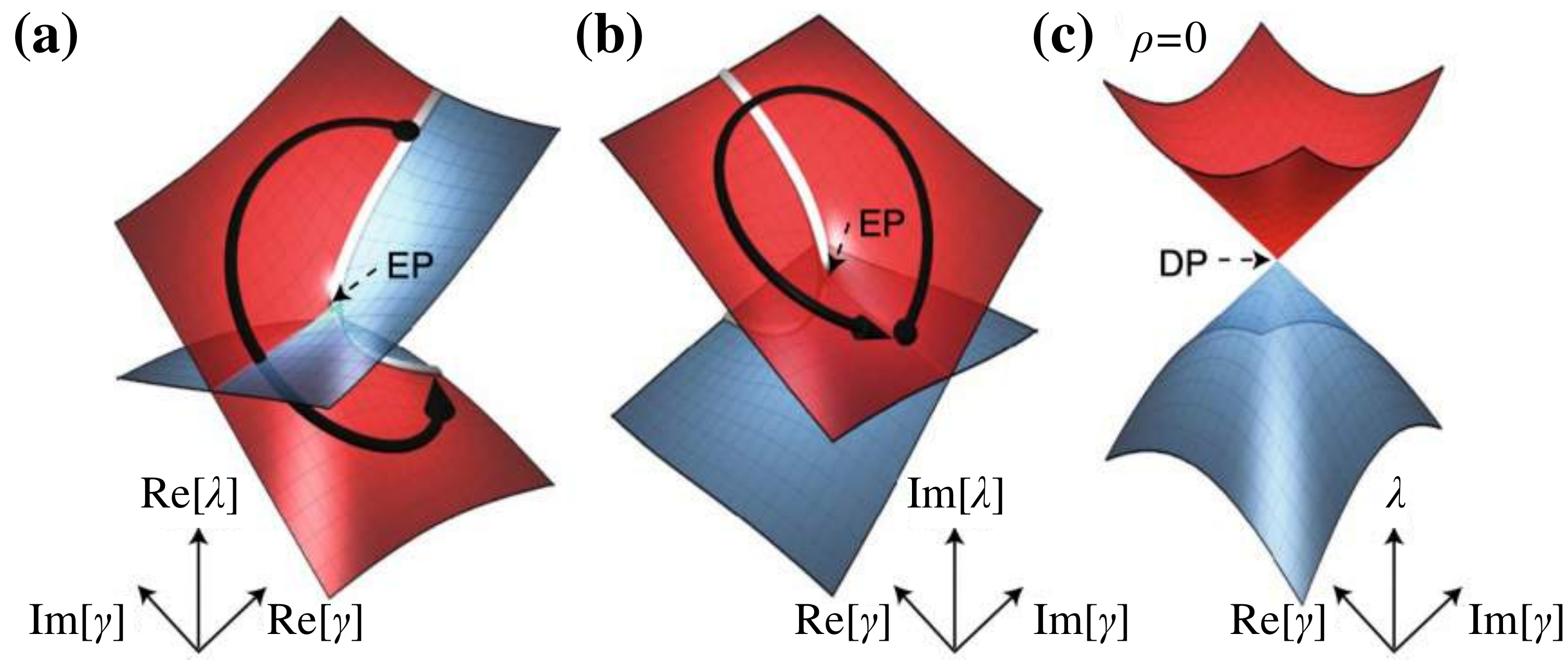}
\end{center}
\caption{ Topological structures of the Riemann surfaces around an EP. The real part (a) and imaginary part (b) of eigenvalues of the $2\times 2$ non-Hermitian matrices $M(\gamma)$ in Eq.~\eqref{sec22b2} are plotted as a function of a complex variable $\gamma$. Two Riemann surfaces intersect with each other  on the real axis with the branch cut terminating at the EP, $\gamma_{\rm EP}=\rho>0$. The black thick arrow indicates that encircling around the EP once leads to the interchange of eigenvectors. The white curves show the square-root scaling $\propto\sqrt{\gamma-\gamma_{\rm EP}}$ around the EP. (c) Eigenvalues for the Hermitian case with $\rho=0$ in Eq.~\eqref{sec22b2}. The degeneracy known as the diabolic point (DP) occurs at $\gamma=0$ around which the eigenspectrum features linear scaling $\propto\gamma$. Adapted from Ref.~\cite{SKO19}. Copyright \copyright\,   2019 by Springer Nature.}
\label{fig:2topo}
\end{figure}
Although an EP is a point of measure zero in the parameter space, it can have strong and singular influences on non-Hermitian systems over (measure nonzero) parameter region around the EP. In the previous subsection, we have demonstrated this feature by revealing the spectral nonanalyticity and the skewness of eigenstates around EPs in some illustrative examples. In this subsection, we provide yet another perspective on the singular influences of EPs based on \emph{topological} aspects of EPs in the parameter space. In a Hermitian case, when one adiabatically changes a parameter along a closed loop without gap closing, an eigenvector returns to the original one aside a possible geometric phase known as the Berry phase \cite{BMV84}. In contrast, in  non-Hermitian systems, encircling an EP swaps  eigenvectors and only a part of them can acquire the geometric phase \cite{DC04,LR09}.

\begin{figure}
\begin{center}
\includegraphics[width=13cm]{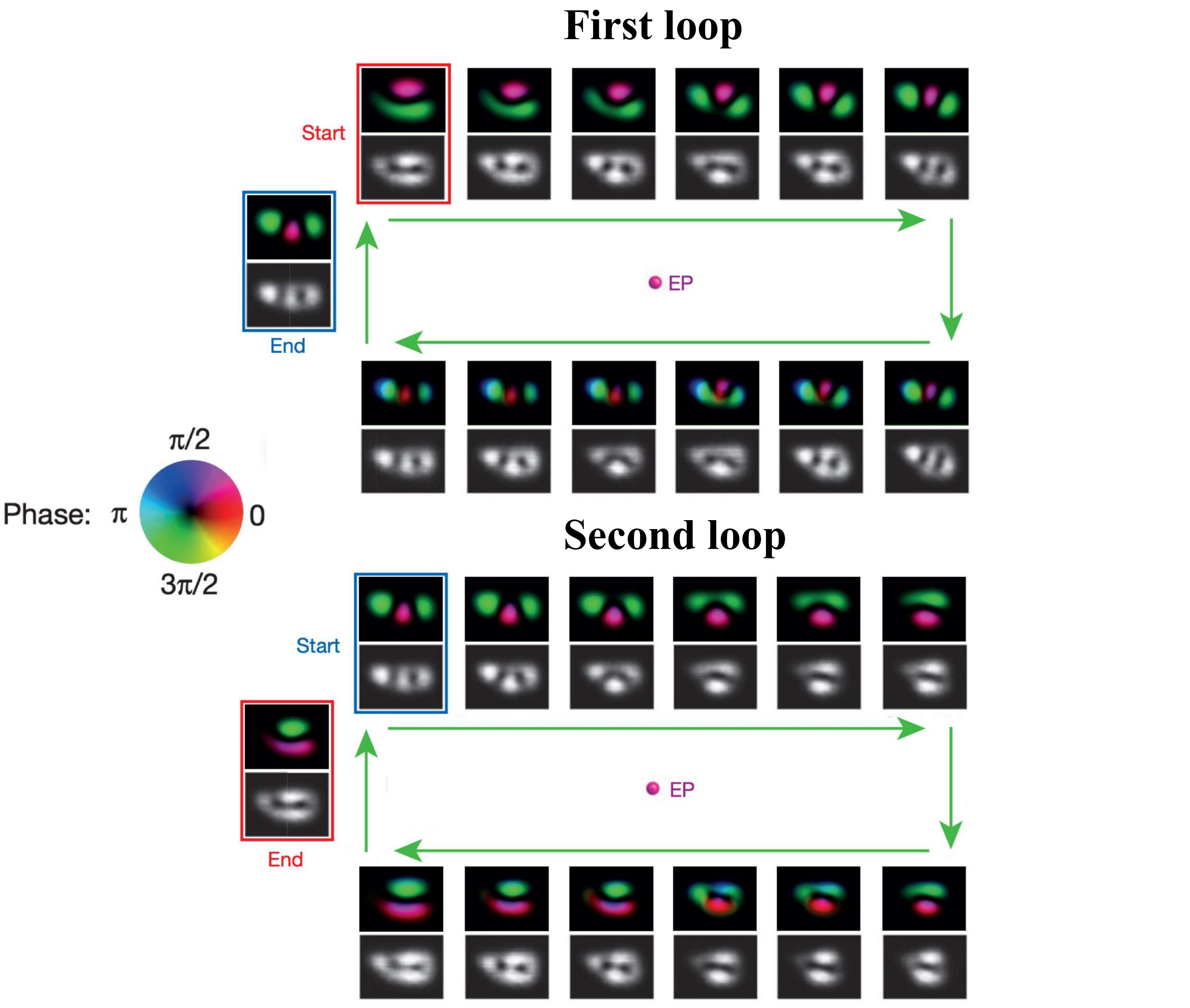}
\end{center}
\caption{  
Experimental demonstration of the acquisition of the Berry phase of $\pi$ after circumnavigating the EP twice.  The circumnavigation is implemented by modulating the spatial configuration of the complex potential in exciton-polariton condensates with gain and loss (see also Fig.~\ref{fig:3excpol}(b) for a schematic illustration of the setup).
Adapted from Ref.~\cite{GT15}. Copyright \copyright\,   2015 by Springer Nature.}
\label{fig:2encircle}
\end{figure}

To be concrete, consider a continuous family of $2\times 2$ non-Hermitian matrices $\{M(\gamma)\}$ parametrized by a complex variable $\gamma$ as discussed in Example~\ref{ep2b2}. There, EPs occur at $\gamma=\pm|\rho|$ and the eigenvalues and eigenvectors exhibit the square-root branch singularities around EPs. 
Let $\cal C$ be a closed loop on the complex plane of $\gamma$ and let us consider varying $\gamma$ along $\cal C$. Whether or not eigenvalues and the corresponding eigenvectors  return to the same ones after going round the loop $\cal C$ is then determined by the topology of the Riemann surfaces along $\cal C$. In the present consideration, there are two Riemann sheets that intersect each other at an EP. If the loop encircles one of the EPs on the $\gamma$-plane, an eigenvalue and eigenvector moves from the initial Riemann sheet to the other one such that the initial state moves around the EP and ends up with a different state in the other sheet \cite{DC01,UR12} (see Fig.~\ref{fig:2topo}(a,b)). One  cannot return to the initial sheet unless going round additional, second loop. We note that if the normalization conditions ${\bf l}^\dagger_{j}{\bf r}_{j'}=\delta_{jj'}$ with $j,j'=1,2$ are imposed in the course of the parameter change, {\it one} of the eigenvectors can change its sign after a one round trip. This results in the acquisition of the Berry phase of $\pi$ for both of the eigenvectors after circumnavigating the loop twice  \cite{DC04,MAA05,LR09,WDH12,GT15} (see Fig.~\ref{fig:2encircle}). Such nontrivial topological structures of Riemann sheets are absent around a DP (see Fig.~\ref{fig:2topo}(c)).

Let us give a simple demonstration ot this based on Example~\ref{Ex:ev}.
\exmp{(Quantization of the Berry phase around an EP). We can choose the biorthogonal left and right eigenvectors of Eq.~(\ref{Mkappa}) as 
\eqn{
\bold{r}_\pm=\frac{1}{\sqrt{2}}\begin{bmatrix} 1 \\ \pm\sqrt{\kappa} \end{bmatrix},\;\;\;\;
\bold{l}_\pm=\frac{1}{\sqrt{2}}\begin{bmatrix} 1 \\ \pm\frac{1}{\sqrt{\kappa^*}} \end{bmatrix}.
}
Therefore, the Berry phase accumulated from encircling the EP at $\kappa=0$ twice is given by
\eqn{
i^{-1}\oint_{2\cal C}d\kappa\bold{l}_\pm^\dag\frac{d}{d\kappa}\bold{r}_\pm =\frac{\pi}{2} \oint_{2\cal C}\frac{d\kappa}{2\pi i}\frac{1}{\kappa}=\pi.
}
} 
In general, if an EP exhibits the $n$-order branchpoint singularity, there are $n$ self-intersecting Riemann sheets and one needs encircling the EP $n$ times to return to the initial sheet. Such topological aspects allow one to introduce a group-theoretical formulation of higher-order EPs \cite{RJW12,JH18,ZQ18}.
These observations indicate that a loop $\cal C$ encircling an EP indeed identifies a topological object in the sense that the above properties do not rely on specific properties of $\cal C$ and are robust against continuous deformation of $\cal C$ unless it crosses the EP.  

When one considers dynamical encircling of an EP by regarding a non-Hermitian matrix as a generator of time evolution \cite{HWD98}, one of the eigenstates will dominate after going round a loop. This is due to the inevitable complexification of eigenvalues around an EP and the subsequent breakdown of the standard adiabatic theorem valid in Hermitian systems \cite{Uzdin_2011,GEM13,MT15}. Interestingly, the dominating final eigenstate depends on the direction of parametric rotation and thus dynamical encircling of an EP can exhibit asymmetric behavior. This feature is often called as a topological transfer or conversion between different states, and has already found applications in energy transfer between multiple modes in an optomechanical system, where coupled mechanical modes of a membrane are coherently manipulated by a laser light \cite{GS16,HX2016}. Similarly, dynamical encircling of an EP is demonstrated also in asymmetric conversions of polarization states \cite{HAU17}, where outgoing light has different polarizations depending on the direction of light propagation, and in asymmetric transmission between modes having different parities \cite{DJ16}, in which the strong attenuation of one of two transverse modes was achieved by asymmetric mode switching.

The chirality of the eigenvectors around EPs, which is illustrated in Example~\ref{ptband},  has its root in topology of the involved Riemann sheets and thus is a generic feature relevant to a variety of non-Hermitian systems. We recall that the analysis of the $2\times 2$ matrices in Examples~\ref{ep2b2} are not specific to a two-level problem but relevant to a larger-level problem since they can be considered as an effective model for two coalescing levels near an EP among many levels.
The chirality manifests itself as the breaking of mirror symmetry between clock-wise and counter-clock-wise propagating modes in cavities \cite{LJM14,PB16,SWR19}.  The chiral property was also experimentally demonstrated in micro resonators \cite{DC04}, a coupled ferromagnetic waveguide \cite{ZXL18}, optical waveguides with integrated  structures \cite{YJW18,ZXL19}, and circularly polarized light \cite{CA172,BB18}. It has been proposed that the concept of the chirality of EPs can be generalized or transferred to different types of physical phenomena such as  the anomalous edge state emerging from the encircling of the EP in the momentum space \cite{LTE16},  transverse zero spin-angular momentum of light \cite{PX18}, surface modes in Maxwell's equations \cite{BKY19}, and the chiral polarization in the relativistic drift effect \cite{KH98,YS16}. The physics of EPs has attracted renewed interest especially in its relation to the band topology, where a higher-dimensional analogue of EPs, namely, exceptional rings can emerge in  the Dirac cones \cite{ZB15}  and the Weyl points \cite{XY17} (see Sec.~\ref{sec5} for further reviews on recent developments in topological aspects of non-Hermitian bands).

\section{Non-Hermitian classical physics\label{sec3}}
In recent years, non-Hermitian physics has attracted considerable interest from a variety of fields in classical physics due partly to the fact that there exists a mathematical equivalence between the one-body Schr{\"o}dinger equation in quantum mechanics and a set of wave equations or linearized equations of motion in classical physics, thus allowing one to simulate non-Hermitian wave physics with  classical systems. 
Studies of such a formal equivalence between single-particle quantum mechanics and classical physics can be traced back to the work by Schr{\"o}dinger \cite{SE26} who has pointed out the analogy between the emergent classical motion of a point particle in quantum mechanics and the ray optics approximation of electromagnetic waves.
The recent proposals  \cite{AR05,REG07,KGM08,MZH08} and realizations \cite{GA09,RCE10} of non-Hermitian optical structures have motivated a considerable amount of subsequent theoretical and experimental studies of anomalous wave propagation and light scattering in photonics. The progress in understanding non-Hermitian wave physics has also been motivated from growing interest in other types of classical systems such as mechanical systems \cite{KB17}, electrical circuits \cite{CHL18}, biological transport \cite{Chou_2011}, neural network and machine learning \cite{GS08,KG19},  optomechanics \cite{AM14,XH18}, acoustics and fluids \cite{CSA16}, thermal atomic gases \cite{CH17} and exciton-polariton condensates \cite{DH10,Schneider_2016}.
In this section, we review non-Hermitian physics in these diverse classical systems on the basis of fundamental properties discussed in Sec.~\ref{sec2}. 

\subsection{Photonics\label{secphoto}}
Though not explicitly mentioned, the non-Hermitian Schr{\"o}dinger-like equation in photonics has already appeared since  the late 1960's \cite{EAJM69,GE72,YC92,Malomed:96} as a theoretical description for the optical couplers, where the non-Hermiticity originates from  gain and loss of power in waveguide couplers. Yet, it was the works in Refs.~\cite{AR05,REG07,KGM08,MZH08} that sparked the subsequent studies on non-Hermitian physics in photonics. They  pointed out that parity-time (PT) symmetric quantum mechanics, which has  originally been studied in the context of mathematical physics (see Sec.~\ref{secphqh}), can be simulated by using optical wave propagation with gain-loss landscapes. Together with experimental progress in fabricating gain and loss in chip-scale nanophotonic systems, non-Hermitian systems have now attracted considerable attention in photonic metamaterials as they can realize unconventional optical properties by implementing nonconservative structures in artificial devices.   These developments have revealed a great potential of non-Hermitian systems for realizing peculiar photonic properties such as unidirectional invisibility \cite{LZ11,YX13,LF13}, coherent perfect absorbers \cite{LS102,CYD11,SY14}, enhanced sensitivity \cite{LZP16,CPY16,HH17,CPY18}, nonreciprocal light propagation \cite{RH10,BP14,CL14}, loss-induced transparency \cite{GA09}, Bragg scattering \cite{Berry_2008,LSS10,MMA12}, Bloch oscillations \cite{LS09,WM15}, nonlinear effects \cite{SAA10,RD11}, orbital angular momentum lasers \cite{MP16}, and  single-mode lasing \cite{PB14,BM14}. Below we review two major formulations for describing non-Hermitian optical systems, namely, the paraxial Helmholtz equation and the scattering formalism, and discuss their applications to understanding the notable wave propagation and light scattering properties observed in experiments.  

\subsubsection{Optical wave propagation\label{secpop}}
A common approach to simulate a one-body Schr{\"o}dinger equation in photonics is to utilize three-dimensional pillar systems, which are homogeneous in the third spatial direction, $z$ that mimics the direction of time, $t$ in quantum mechanics. In practice, this type of systems can be realized, for instance, by aligned optical fibers with a fixed cross section. 
One can categorize the solutions of the Maxwell equations for optical waves propagating along the $z$ direction into two types of solutions, whose electric or magnetic field is fully confined to the $xy$ plane \cite{JJD75}. The solutions are thus called as the transverse electric (TE) or transverse magnetic (TM) modes.  The $z$-component of the magnetic (electric) field $H_z$ ($E_z$) in the TE (TM) solutions can be treated as a scalar variable that acts as the wavefunction in the Schr{\"o}dinger-like equation. Assuming the oscillatory solutions as $ H({\bf r},z)e^{i\kappa z-i\omega t}$ or $E({\bf r},z) e^{i\kappa z-i\omega t}$, where $H({\bf r},z),\,E({\bf r},z)$ are envelop functions with two-dimensional coordinates ${\bf r}=(x,y)$, $\kappa$ is the longitudinal wavenumber, and $\omega$ is the oscillation frequency, and making the paraxial approximation (small diffraction angles of the electromagnetic fields), one arrives at the propagating equation for the optical waves as
\eqn{\label{temode}
i\kappa\frac{\partial}{\partial z}H({\bf r},z)&\!=\!&-\frac{1}{2}\left[n^2({\bf r})\nabla_\perp n^{-2}({\bf r})\cdot\nabla_\perp\!+\!\frac{\omega^2n^2({\bf r})}{c^2}\!-\!\kappa^2\right]\!H({\bf r},z)\;\;\;{\rm (TE\;modes)},\\
i\kappa\frac{\partial}{\partial z}E({\bf r},z)&=&-\frac{1}{2}\left[\nabla^2_\perp+\frac{\omega^2n^2({\bf r})}{c^2}-\kappa^2\right]E({\bf r},z)\;\;\;{\rm (TM\;modes)},
\label{tmmode}
}
where $c$ is the speed of light, $\nabla_\perp$ represents the spatial derivative in the $xy$ plane, and $n({\bf r})$ is a complex refractive index whose positive (negative) imaginary part indicates loss (gain) of optical waves. We note that the description still remains valid even if the refractive index $n$ has spatial dependence also in the $z$-direction as long as the condition $|\partial_{z}n/n|\ll\kappa$ is satisfied. 

The above optical wave equations can be regarded as the time-dependent one-body Schr{\"o}dinger equation, where the propagation distance $z$, the electromagnetic fields $H,\,E$, the longitudinal wavenumber $\kappa$, and the refractive index $n({\bf r})$ correspond to the time, the wavefunction, Planck's constant, and a potential in quantum mechanics, respectively, as pointed out in the book \cite{DD04} in a concrete manner. Later, this idea has found applications to simulate the PT-symmetric non-Hermitian quantum mechanics by making the refractive index complex by introducing gain and loss for optical waves and ensuring the symmetry condition $n({\bf r})=n^*(-{\bf r})$ \cite{AR05,REG07}. 
Below we review a variety of anomalous optical wave phenomena that can be explored in this type of non-Hermitian setups, but not in Hermitian systems.
\begin{figure}
\begin{center}
\includegraphics[width=11cm]{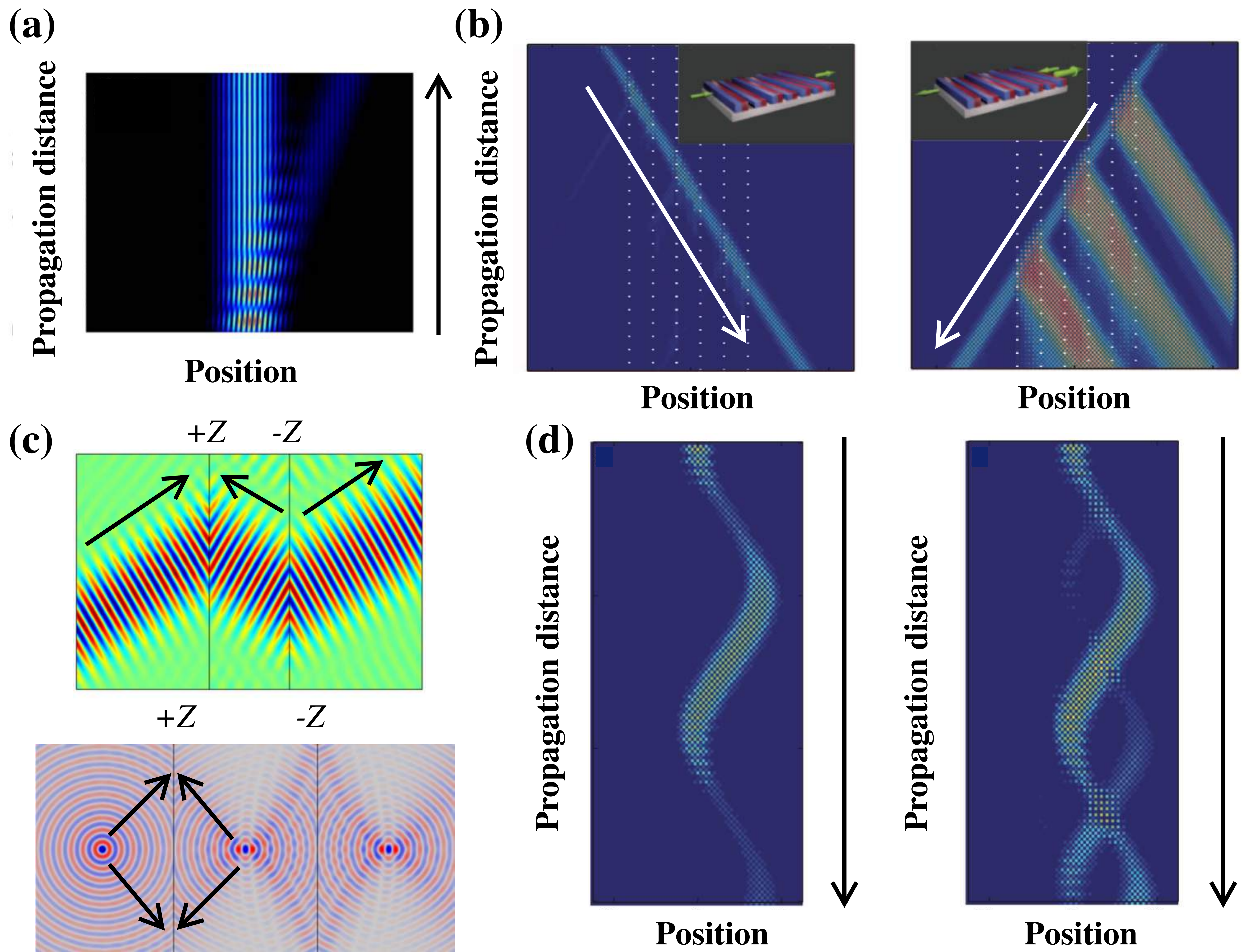}
\end{center}
\caption{
Unconventional optical wave propagation in a non-Hermitian medium.
(a) Numerical results on the secondary emission in the PT-symmetric optical waveguide due to the nonorthogonality of eigenmodes. The propagation of the secondary emission in the right direction originates from the chirality of eigenvectors with particular group velocities (see also Fig.~\ref{fig:2epchiral2} for an illustration of the chirality around an EP). 
Adapted from Ref.~\cite{KGM08}. Copyright \copyright\, 2008 by the American Physical Society. 
(b) Experimental observation of the unidirectional invisibility in the PT-symmetric photonic lattice. Light incoming from the left direction propagates through the medium without reflection (left panel) while light incoming from the right direction undergoes a strong reflection (right panel). 
Adapted from Ref.~\cite{RA12pt}. Copyright \copyright\, 2012 by Springer Nature. 
(c) Numerical results on the negative refraction in the PT-symmetric metasurfaces. The energy of light propagates from the surface with impedance $-Z$ to that with $+Z$, which can be opposite to the incoming direction; such a reverse propagation is unique to nonconservative media. 
Adapted from Ref.~\cite{FR14}. Copyright \copyright\, 2014 by the American Physical Society. 
(d) Experimental observation of the Bloch oscillations in the PT-symmetric photonic lattice. The left panel shows the usual Bloch oscillation on the lattice without gain. The right panel shows the unconventional Bloch oscillation on the lattice with gain, where the nonorthogonality triggers secondary emissions associated with delayed phases. Adapted from Ref.~\cite{RA12pt}. Copyright \copyright\, 2012 by Springer Nature. 
}\label{fig:3optwave}
\end{figure}
\\
\\
{\it Secondary emission and power oscillations in non-Hermitian waves}

\vspace{3pt}
\noindent In general, when the complex potential $n({\bf r})$ has periodic structures,  eigenvalues $\omega$ of the non-Hermitian effective Hamiltonian in Eqs.~\eqref{temode} and \eqref{tmmode} constitute the Bloch bands in the complex plane. The non-Hermiticity can induce an attraction between different bands which, in turn, leads to the band merging (see, e.g., Fig.~\ref{fig:2epchiral1}(b))  or the ring-shaped exceptional point (EP) in photonic crystal slabs \cite{ZB15}.  The tight-binding non-Hermitian chain can be implemented by the synthetic photonic lattice. This was first realized in coupled fiber loops through introduction of PT-symmetric gain-loss spatial profiles \cite{RA12pt}, where the gain is induced by semiconductor optical amplifiers and the loss is caused by acoustic modulators.   

As discussed in Sec.~\ref{seceigv}, one of the most notable features in non-Hermitian systems, which is absent in Hermitian systems, is the nonorthogonality between eigenvectors for different modes. Suppose that one injects an optical wavepacket with a finite width and a well-defined direction of propagation into a non-Hermitian periodic potential. In Hermitian systems, the wavepacket keeps propagating along the initial direction. In contrast, in non-Hermitian cases, the nonorthogonality manifests itself as a generation of a secondary emission whose propagating direction deviates from the initial one \cite{KGM08,MZH08} (Fig.~\ref{fig:3optwave}(a)). This effect originates from the chirality around an EP, i.e., the merging of eigenvectors into a band having a particular group velocity. For instance, in Example~\ref{ptband}, an eigenvector at small $\delta k\equiv k-\pi$ with a positive group velocity becomes very close to that  at $-\delta k$ with a negative group velocity  (cf. Fig.~\ref{fig:2epchiral2}(a) and the related descriptions). When the initial wavepacket has amplitudes of both of these modes, they can contribute to the secondary emission of a chiral wavepacket propagating only in the positive direction \cite{KGM08}. 
    
Another direct physical consequence of the nonorthogonality is power oscillations. In contrast to Hermitian systems, the norm of the wavefunction (i.e., the power in the optical wave) can vary in time (i.e., along the direction of propagation) in non-Hermitian cases. In PT-symmetric systems, the power exhibits oscillations due to interference of nonorthogonal Bloch modes in the PT-unbroken regime while it amplifies or decays exponentially in time in the PT-broken regime \cite{REG07,KS082}. Interestingly, specific initial conditions for the light input allow one to realize power amplification even though the complex lattice is  lossy on average \cite{MKG14}, and also to create a uniform power in an inhomogeneous scattering potential \cite{MKG15,MKG17}. 
Beyond these one-body wave phenomena, the counterpart of the norm oscillations in nonorthogonal many-particle eigenstates can manifest itself as the supersonic propagation of correlations \cite{YA18,DB19} (cf. Fig.~\ref{fig:4nonloc} in  Sec.~\ref{sec4}). 
\\
\\
{\it Unidirectional invisibility and linear power growth at the exceptional point}

\vspace{3pt}
\noindent At the EP, the wave propagation exhibits particularly intriguing features. First of all, the power at the EP increases linearly in time as a direct consequence of the  formula~\eqref{jorlinear} for the Jordan normal form in Sec.~\ref{secspecdec}. 
When the EP is located in the continuum band, this power increase can lead to defect states that exhibit a linear growth of emissions \cite{RAA13}.
Another notable feature is the unidirectional invisibility. Suppose that one injects an optical beam into a one-dimensional complex waveguide at the EP. Then, the input from one direction exhibits no reflection without phase imprinting while the incident light from the opposite direction experiences reflection whose coefficient can exceed one as first observed in a photonic lattice \cite{RA12pt} (Fig.~\ref{fig:3optwave}(b)). This effect results from the absence of the transmitting plane-wave solution for one propagating direction, which is a direct manifestation of the dimensional reduction of non-Hermitian Hamiltonians at EPs (see also discussions in Sec.~\ref{secepphys}). The related phenomena were also observed in other systems such as multilayer Si/SiO$_{2}$ structures \cite{LF142} and silicon nanowires \cite{LF13}.
\\
\\
{\it Negative refraction}

\vspace{3pt}
\noindent 
One can also use non-Hermitian optical systems to realize a robust, loss-immune mechanism for implementing negative refraction.
Suppose that several non-Hermitian elements are placed along the waveguide in a spatially separated manner. When two same PT-symmetric elements are separated by a certain distance but with opposite chirality, it is possible to have the transmitted wave experience a phase advance unlike the  delay in free space, thus effectively realizing the negative refraction \cite{FR14} in a way different from a Hermitian system. Physically, this property originates from the wave components propagating between the two elements in the direction opposite to that of the incident wave (Fig.~\ref{fig:3optwave}(c)). 
From a general perspective, the negative refraction can be considered as a specific case of anisotropic transmission resonances \cite{GL12} and find applications to realize all-angle negative refraction based on the PT-symmetric metasurfaces \cite{MF16}. 
\\
\\
{\it Bloch oscillation}

\vspace{3pt}
\noindent As photonics can simulate one-body quantum mechanical phenomena in non-Hermitian regimes, it is natural to ask how the Bloch oscillation \cite{BF29} should be modified in the presence of non-Hermiticity. In Hermitian systems, the Bloch oscillation is known as the oscillatory motion of a wavepacket on a periodic lattice subject to an external constant force. In contrast, the nonconservative nature in non-Hermitian systems can manifest itself as the complex Berry phases of the Bloch wave functions, leading to unidirectional and amplified Bloch oscillations \cite{LS09}. 
Non-Hermitian Bloch oscillations can also be accompanied by the secondary emission discussed above as observed in the photonic lattice \cite{RA12pt,WM15,XYL16} (see Fig.~\ref{fig:3optwave}(d)), where an external constant force was implemented by a phase gradient \cite{WM15} or a synthetic geometric curvature \cite{XYL16}.
\\
\\
{\it Wave propagation through atomic medium}

\vspace{3pt}
\noindent Analogous to semiconductor optical amplifiers and modulators, one can  utilize the light-induced  coherence of multi-level {\it atomic medium} to implement non-Hermitian optical potentials. It has been pointed out in Refs.~\cite{GBJ95,SJ112} that techniques based on Raman resonances and electromagnetically induced transparency (EIT) can be applied to generate non-Hermitian potentials by modulating the spatial distribution of photon detuning with the Stark shift. Along this line, an anti-PT-symmetric coupling between two optical modes has been realized in a thermal gas of a $\Lambda$-type three-level configuration of $^{87}$Rb atoms. Together with the technique to control diffraction patterns by an optical lattice \cite{JS15}, the EIT near-field diffraction effect (i.e., Talbot effect) has also been demonstrated in a coherently manipulated three-level atomic gas \cite{WJ112}. Introducing an additional pump field, the PT-symmetric Talbot effect has been  realized in an atomic vapor cell \cite{RH12}.
\\
\\
{\it Band topology and other notable aspects}

\vspace{3pt}
\noindent Since the seminal proposal of quantum-Hall edge states in photonic crystals by Raghu and Haldane \cite{HFDM08,RS08} and its experimental realization \cite{WZ08,WZ09}, exploring single-particle band topology in photonic systems has become an emergent field at the interface of optics and condensed matter physics \cite{LL14,KAB17}. Compared with the solid-state counterparts, topological photonic systems have the advantage of the high visibility of topological edge modes and the accessibility to their nonequilibrium dynamics \cite{KF12,MH13,MCR13}. Moreover, in light of the rapid experimental development of non-Hermitian photonics \cite{MH17}, photonic systems have become one of the most ideal platforms for studying the interplay between non-Hermiticity and band topology. In the following, we discuss several prototypical and experimentally relevant examples in non-Hermitian topological photonics. The general recipe used for constructing these examples is simply to introduce non-Hermiticity, typically gain and loss, into well-known Hermitian topological systems. Further recent developments on non-Hermitian band topology will later be presented in Sec.~\ref{sec5}. We also note that some previously known phenomena, such as surface Maxwell waves \cite{KYB19} and plasmon-polaritons \cite{BKY19}, might be reinterpreted in the framework of non-Hermitian band topology.

\begin{figure}
\begin{center}
\includegraphics[width=14.5cm]{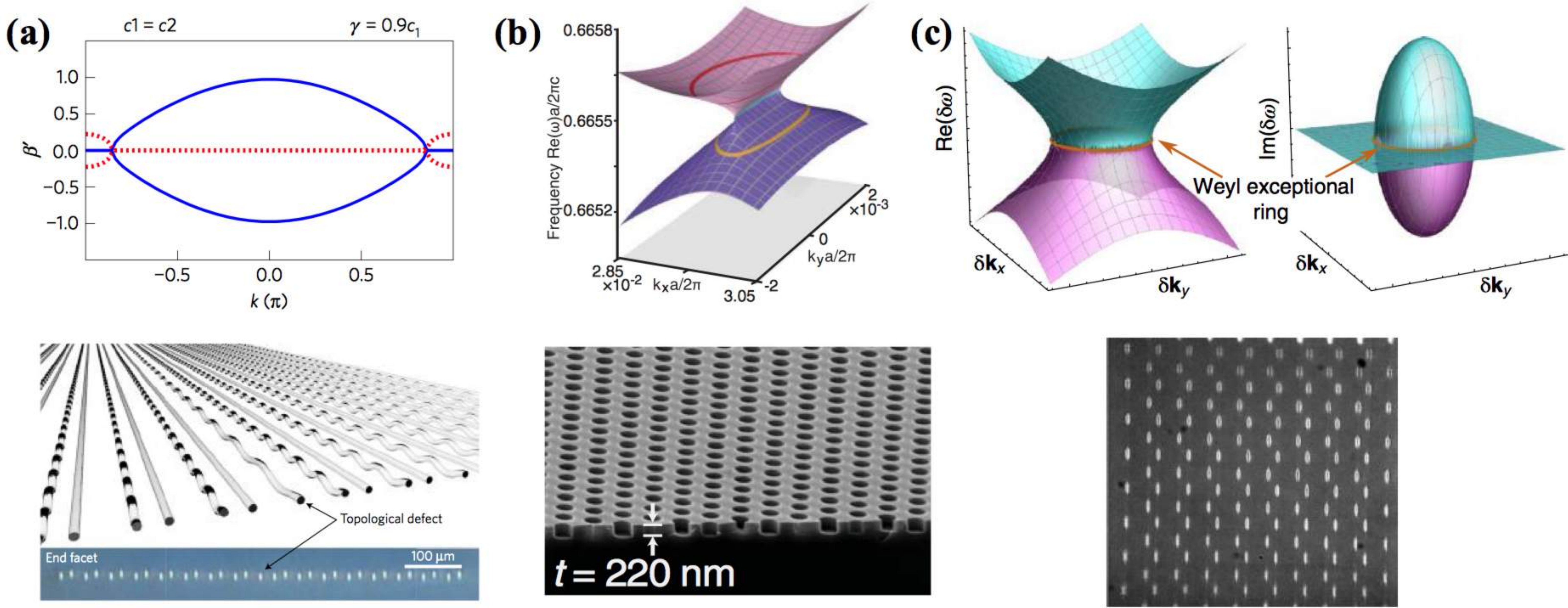}
\end{center}
\caption{(Upper panels) Band structures of (a) a non-Hermitian Su-Schrieffer-Heeger model with broken PT symmetry in the bulk (see Eq.~(\ref{PTSSH})); (b) a lossy two-dimensional lattice with a bulk Fermi arc (see Eq.~(\ref{BFA})); (c) a lossy three-dimensional lattice with a Weyl exceptional ring (see Eq.~(\ref{EWR})). The corresponding photonic-lattice realizations are shown in the lower panels. Figure (a) is adapted from Ref.~\cite{WS17}. Copyright \copyright\,   2016 by Springer Nature. Figure (b) is adapted from Ref.~\cite{ZH18}. Copyright \copyright\,   2018 by American Association for the
Advancement of Science.  Figure (c) is adapted from Ref.~\cite{AC19}. Copyright \copyright\,   2019 by Springer Nature.}
\label{fig:3band}
\end{figure}

The arguably simplest topological insulator in Hermitian physics is the Su-Schrieffer-Heeger (SSH) model \cite{SWP79}, which is a 1D chain with two alternating hopping amplitudes $c_1$ and $c_2$. This model is characterized by an integer winding number protected by the chiral symmetry, and it also exhibits the time-reversal and particle-hole symmetries (class BDI \cite{SAP08}). If we introduce an imaginary staggered potential such that the Bloch Hamiltonian becomes
\begin{equation}
H(k)=-(c_1+c_2\cos k)\sigma^x-c_2\sin k \sigma^y-i\gamma \sigma^z,
\label{PTSSH}
\end{equation} 
then all the above-mentioned symmetries are broken, yet the PT symmetry represented by $\sigma^x\mathcal{K}$ survives with $\mathcal{K}$ being the time-reversal operator. One can show that, at least for gapped two-band systems, the winding number stays well-defined as a homotopy invariant \cite{ZG18}. Starting from the topological Hermitian limit with $0<c_1<c_2$ and subject to the open boundary condition, if we introduce a small $\gamma$, then the bulk spectrum will stay real, while the edge eigenvalues will immediately undergo the PT symmetry breaking \cite{HYC11}. A similar phenomenon was predicted for the interface or the defect between a trivial ($c_1>c_2>0$) and a topological bulk \cite{HS13}. The latter prediction has been experimentally verified in a waveguide \cite{WS17} and dielectric-resonator \cite{CP15} photonic lattices with balanced gain and loss. 
With loss alone, the SSH chain was predict to exhibit a topological transition in the quantized mean displacement of a decaying wave packet \cite{RMS09}. This phenomenon has also been verified in a photonic waveguide array \cite{ZJM15}. We also note that exceptional points can emerge in the bulk when $\gamma$ exceeds $|c_1-c_2|$ (see Fig.~\ref{fig:3band}(a)). In this one-dimensional model, PT symmetry is crucial for the emergence of bulk exceptional points. Otherwise, a single parameter $k$ cannot generally ensure a band touching without fine tuning of other parameters.

In Hermitian physics, quantum Hall and Chern insulators serve as the most important topological systems without the need of symmetry protection \cite{SAP08}. These systems exhibit topologically protected chiral edge modes immune from back scattering by impurities, as have been confirmed in their photonic realizations \cite{WZ08,WZ09}. It is natural to ask whether the edge modes and their Chern-number characterizations stay robust in the presence of non-Hermiticity. The answer turns out to be affirmative, because the Chern number defined from the left or right eigenvectors (or the mixture of left and right eigenvectors) coincide, although the Berry curvature does not \cite{SH18}. Combined with nonlinearity, topological-insulator lasers with high-density and unidirectional edge lasing have been predicted and realized \cite{HG18,BM18}. Robust light steering in topological reconfigurable non-Hermitian junctions has been also realized \cite{ZH192}.  Moreover, in stark contrast to the one-dimensional case, exceptional points can emerge in two-dimensional lattice systems without any parameter fine tuning or symmetry protection. Due to the conservation of total topological charge, the exceptional points can only appear or annihilate in pairs and each pair should be connected by a ``Fermi arc" along which the real energy coincides. Such so-called ``bulk Fermi arc" has been observed in a rhombic photonic lattice that effectively realizes 
\begin{equation}
H(\boldsymbol{k}=\boldsymbol{k}_{\rm D}+\delta\boldsymbol{k})\simeq \omega_{\rm D}-i\gamma_0+(v_g\delta k_x-i\gamma)\sigma^z+v_g\delta k_y\sigma_x
\label{BFA}
\end{equation}
near the original (dissipationless limit, i.e., $\gamma=0$) Dirac point $\boldsymbol{k}_{\rm D}$, where $\gamma\pm\gamma_0$ denote the radiation decay rate of even and odd 
reflection-symmetric modes \cite{ZH18}. By measuring the scattered light from the photonic lattice, which contains the information of density of states, the existence of bulk Fermi arc was confirmed.

Due to the Bott periodicity \cite{AK09}, Hermitian topological insulators in three dimensions again require symmetry protections just like the one-dimensional case. Nevertheless, Weyl semimetals appear as a stable \emph{gapless} topological system without symmetry protection \cite{ANP18}. Recalling the fact that the emergence of an exceptional point requires two (real) parameters, we know that a Weyl point in the three-dimensional Brillouin zone will generally split into a \emph{Weyl exceptional ring}, which is characterized by an additional winding number on top of the Chern number \cite{XY17}. Such a unique Weyl exceptional ring and its associated Fermi-arc edge states have recently been observed in a bipartite waveguide array \cite{AC19}, which can effectively be described by 
\begin{equation}
H(\boldsymbol{k}=\boldsymbol{k}_{\rm W}+\delta\boldsymbol{k})\simeq v_{\perp}(\delta k_x\sigma^x+\delta k_y\sigma^y)+v_z\delta k_z(\sigma_0-|b|\sigma^z)-iv_z\tau(\sigma_0+\sigma^z)
\label{EWR}
\end{equation}
near the original Weyl point $\boldsymbol{k}_{\rm W}$. One can check that the exceptional ring is determined by $\delta k_z=0$ and $|\delta\boldsymbol{k}_\perp|=\sqrt{\delta k_x^2+\delta k_y^2}=\frac{v_z}{v_\perp}\tau$. On the other hand, to our knowledge, gapped non-Hermitian topological phases in 3D have not yet been experimentally realized. Indeed, they have not yet been seriously studied even theoretically, while their classifications and constructions have already been proposed as reviewed in Sec.~\ref{sec5}. 

As mentioned also in the discussion of topological insulator lasers \cite{HG18,BM18}, another notable aspect of wave phenomena in the non-Hermitian medium is \emph{nonlinearity}. While nonlinear effects are inherently small in usual optical setups since photons cannot directly interact with each other, recently there have been several efforts to realize nonlinear and non-Hermitian photonic systems \cite{BP14,WMM15}. 
Theoretically, nonlinear aspects of PT-symmetric non-Hermitian photonics such as  optical solitons and nonlinear Fano resonances \cite{FU61} have been well explored \cite{MZH08,AFK11,MAE11,KVV16}. It merits further study to explore nonlinear effects of non-Hermitian wave phenomena in other physical systems such as Bose-Einstein condensates and active matter, where  strong nonlinearity is inherent (see Sec.~\ref{sechydro}).
Magneto-optical effects in non-Hermitian systems are also of interest; they can be addressed by replacing the spatial derivative $\nabla_\perp$ in Eqs.~\eqref{temode} and \eqref{tmmode} by $\nabla_\perp+i{\bf A}({\bf r})$ with $\bf A$ being an effective gauge field \cite{PH11}. 
While the PT symmetry is commonly used as a guiding principle to design eigenspectra and engineer the EP, recently the supersymmetry has been studied as yet another way to construct non-Hermitian models with entirely real spectra \cite{FC95}. The core idea is to decompose the Schr{\"o}dinger operator into two first-order differential operators and construct a complex potential that shares the same eigenspectrum with a real potential \cite{KA89}. This idea has already found applications to optical structures \cite{MMA13}, mode converters \cite{HM14} and laser arrays \cite{EGR15}.
The other interesting aspects of non-Hermitian physics include all-optical PT-symmetric logic gates \cite{DS15}, PT-symmetric honeycomb lattices \cite{SA112}, amplitude-to-phase converters \cite{Gutierrez:16} and Anderson localization \cite{DMJ12}.

\subsubsection{Light scattering in complex media\label{secplight}}
Non-Hermitian physics in photonics can also be explored without relying on the analogy between the Schr{\"o}dinger equation and the paraxial approximation of Maxwell's equations as discussed in Sec.~\ref{secpop}. Another description of non-Hermitian open optical systems is the scattering formalism \cite{SP01,BCWJ97}, where dynamical properties can be probed by optical scattering and leakage is inherent due to, for example, internal reflection, cavity decay and fiber tips. In this formalism, all the spectral properties are characterized by the scattering matrix $S(\omega)$ that gives the input-output relation for stationary solutions of Maxwell's equations at frequency $\omega$:
\eqn{
{\bf a}^{\rm out}=S(\omega)\,{\bf a}^{\rm in},
}
where ${\bf a}^{\rm in}$ and ${\bf a}^{\rm out}$ are the input and output optical amplitudes, respectively. A common setup is a two-channel open system (cf. Fig.~\ref{fig:2scattering}), in which the amplitudes can be divided into two parts depending on different transmitting channels (denoted as channel $1$ and $2$ here):
\eqn{
{\bf a}^{\rm out}=\begin{bmatrix}{\bf a}^{\rm out}_{2}\\
{\bf a}^{\rm out}_{1}
\end{bmatrix},
\;\;\;
{\bf a}^{\rm in}=\begin{bmatrix}{\bf a}^{\rm in}_{1}\\
{\bf a}^{\rm in}_{2}
\end{bmatrix},
}
where ${\bf a}_{1}^{\rm in,\,out}$ (${\bf a}_{2}^{\rm in,\,out}$) are $N_1$ ($N_2$) dimensional complex vectors with $N_1$ ($N_2$) being the number of scattering states in channel $1$ ($2$). The complex elements in $S(\omega)$ can be divided into four block matrices as
\eqn{\label{scamat}
S=\begin{bmatrix}
\;T_{1}\;&\;\;R_{2}\;\;\\
\;R_{1}\;&\;\;T_{2}\;\;
\end{bmatrix},
}
where $R_{1}$ ($R_2$) is an $N_1\times N_1$ ($N_2\times N_2$) reflection matrix and  $T_{1}$ ($T_2$) is an $N_2\times N_1$ ($N_1\times N_2$) transmission matrix.  
Note that the latter is, in general, a rectangular matrix, while the former is always a square one. The total transmission and reflection coefficients, $t_i$ and $r_i$ for optical waves through channel $i=1,2$ can be expressed as
\eqn{
t_{i}={\rm Tr}[T_i^\dagger T_i],\;\;\; r_{i}={\rm Tr}[R_i^\dagger R_i].
} 

In Hermitian systems, the conservation law of incoming and outgoing fluxes leads to the unitarity of the scattering matrix $SS^\dagger=I$, indicating the relations for the transmission and reflection matrices $T_i^\dagger T_i+R_i^\dagger R_i=I$ and $R_1T_1^\dagger+T_2R_2^\dagger=R_1^\dagger T_2+T_1^\dagger R_2=0$. Provided that $N_1=N_2$, these conditions allow one to use the unitary matrices $W_{1,2},U_{1,2}$ to perform the singular value decomposition (cf. Sec.~\ref{secsvd}) of transmission matrices $T_{1,2}$ and reflection matrices $R_{1,2}$, leading to a useful parameterization of the scattering matrix \cite{PAM88,MTh92}:
\eqn{\label{SVsca}
S=
\begin{bmatrix}
\;W_1\;&\;\;0\;\;\\
\;0\;&\;\;W_2\;\;
\end{bmatrix}
\begin{bmatrix}
\sqrt{\tau}&\sqrt{I-\tau}\\
-\sqrt{I-\tau}&\sqrt{\tau}
\end{bmatrix}
\begin{bmatrix}
\;U_1^\dagger\;&\;\;0\;\;\\
\;0\;&\;\;U_2^\dagger\;\;
\end{bmatrix},
}
Here, $\tau$ is a diagonal matrix that contains the square of the singular values of $T_1$ on its diagonal. This basis fully characterizes transmission eigenmodes (i.e., eigenvectors of matrix $T_i^\dagger T_i$) and the corresponding transmission coefficients while keeping the separation between contributions from different channels unaltered. In addition, when the time-reversal symmetry is satisfied, the scattering becomes reciprocal, i.e.,  $T_1=T_2^{\rm T}$ and $R_i=R_i^{\rm T}$ with $i=1,2$ (i.e., $S=\Sigma_xS\Sigma_x$), which are known as the Onsager reciprocity relations \cite{OL31}. In this case, the decomposed expression~\eqref{SVsca} can be simplified by replacing  $U_{1}^\dagger$ and $U_{2}^\dagger$ with $W_{2}^{\rm T}$ and $W_{1}^{\rm T}$, respectively.

In non-Hermitian systems, the unitarity of the scattering matrix can be violated. Nevertheless, a generalized 
conservation law has been shown in the PT-symmetric gain/loss systems with $N_1=N_2$ for which the following invariance holds \cite{GL12,GL15}:
\eqn{
S(\omega)=\Sigma_x\left[S^{-1}(\omega^{*})\right]^*\Sigma_x\;\;\;{\rm with}\;\;\; \Sigma_x=\sigma_x\otimes I_{N_1}. 
}
The resulting generalized conservation can be read as $|T-1|=\sqrt{R_1R_2}$ for the simplest case of a single-mode reciprocal waveguide with $N_1=N_2=1$ and $T_1=T_2$. We emphasize that, at least in linear regimes discussed here, the reciprocity in the transmission coefficients cannot be violated by merely introducing scalar non-Hermitian terms such as gain/loss  (see Sec.~\ref{sec:6np}). 

Physically, an optical microcavity offers an ideal playground to study non-Hermitian scattering phenomena; it confines light to small volumes by resonant recirculation while it is intrinsically an open system due to inherent loss and gain. Harnessing the non-Hermiticity in optical microcavities found many interesting applications such as coherent perfect absorber \cite{CYD10}, high-sensitive sensing at the EP \cite{LZP16}, single-mode laser \cite{Miri:12}, and wave chaos \cite{YCH11} which will be reviewed below. 
\\ \\ {\it Coherent perfect absorbers and quasi-bound states}

\vspace{3pt}
\noindent
 Without gain and loss, the scattering matrix $S(\omega)$ is unitary and can have poles with $|S|=\infty$ (zeros with $S=0$) only at complex frequencies with negative (positive) imaginary parts.  
 In this Hermitian case, the poles at complex frequencies correspond to solutions with purely outgoing modes (i.e., ${\bf a}^{\rm in}=0$ with finite ${\bf a}^{\rm out}$) having finite lifetimes due to negative imaginary parts, thus known as quasi-bound states or resonances \cite{NM98}. 
In a non-Hermitian case with gain being added, these poles shift up in the complex plane of $\omega$ and the lasing occurs when the poles firstly touch on the real axis, where an infinitesimally small input can lead to a divergent output. In contrast, as loss is added, the poles and zeros shift down (i.e., go down along the imaginary axis) and the coherent perfect absorption (CPA) sets in when the zeros firstly touch on the real axis. With further increasing loss, the CPA occurs everytime the zeros across the real axis \cite{CYD10}. Based on the idea of the PT symmetry, it has been pointed out that both  lasing and CPA  can occur in a single device \cite{LS102,CYD11}. This can be realized by using balanced gain and loss to allow the poles and zeros to occur on the real axis in the same system. 
The CPA in the PT-symmetric system has been experimentally observed in coupled microresonators \cite{SY14}. Lasing and anti-lasing at the same frequency have also been realized in a single PT-symmetric microcavity \cite{WZ16}. Similar ideas have found applications to other systems such as acoustics \cite{RF16} and metasurfaces \cite{MF16} as reviewed later in Sec.~\ref{sechydroaco}.
\begin{figure}
\begin{center}
\includegraphics[width=14.8cm]{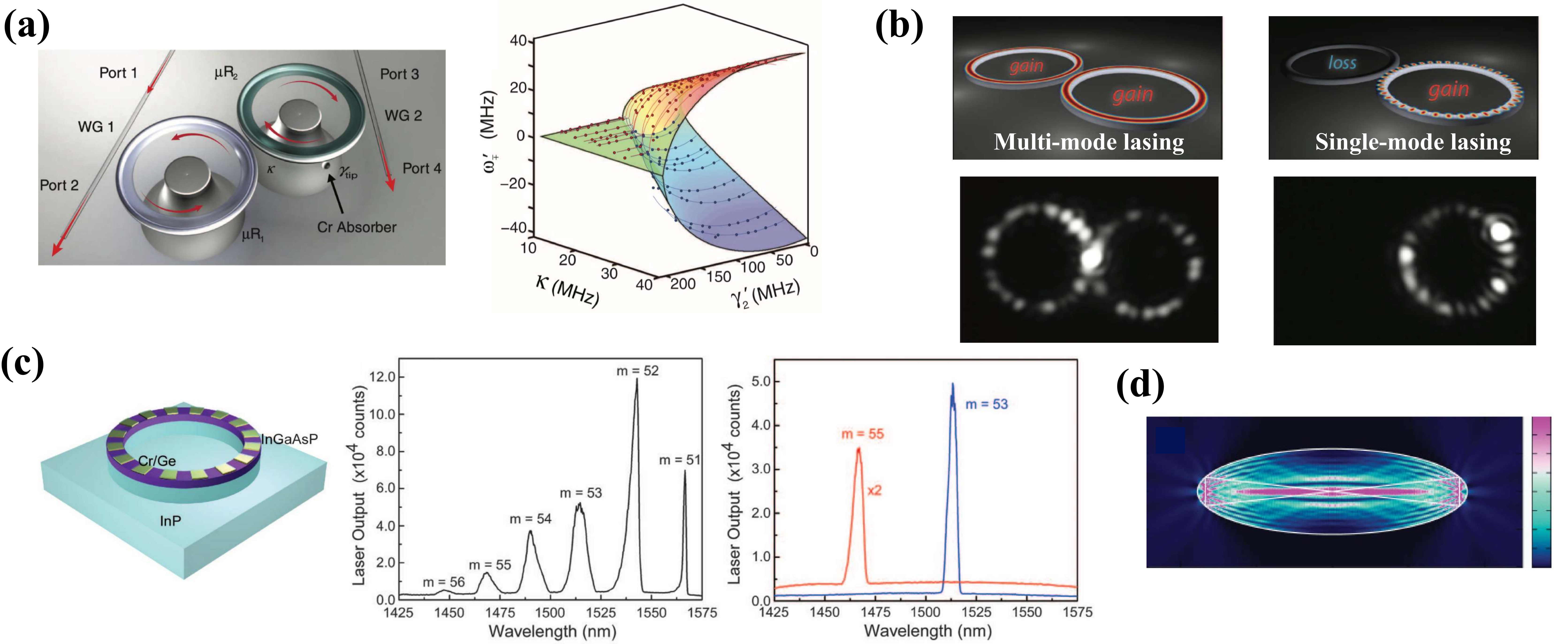}
\end{center}
\caption{
Non-Hermitian physics in optical microcavities. 
(a) Left. Schematic illustration of the experimental system of the coupled whispering-gallery-mode resonators. Right. Experimental results on the real part of  eigenvalues on the complex parameter space around the EP.
Adapted from Ref.~\cite{PB14}. Copyright \copyright\, 2014 by American Association for the Advancement of Science. 
 (b) Single-mode lasing in coupled active-passive microring resonators by breaking the PT symmetry.  Adapted from Ref.~\cite{HH14}. Copyright \copyright\, 2014 by American Association for the Advancement of Science. 
 (c) Left. Schematic illustration of the experimental system of the PT-symmetric microring resonator. Middle. Experimental results on the multi-mode lasing in the resonator. Right. Single-mode lasing of two different modes by breaking the PT symmetry. 
 Adapted from Ref.~\cite{LF14}. Copyright \copyright\, 2014 by American Association for the Advancement of Science. 
 (d) Spatial profile of the scarlike optical mode having the complex eigenvalue in the microcavity. Adapted from Ref.~\cite{YCH11}. Copyright \copyright\,   2011 by the American Physical Society.
}\label{fig:3optcavity}
\end{figure}
\\ \\ {\it Single-mode lasing, width behavior and the exceptional points}

\vspace{3pt}
\noindent
An intriguing possibility in non-Hermitian optical systems is to realize high-quality lasing behavior (see Fig.~\ref{fig:3optcavity}(a)-(c)). For instance, single-mode lasing  requires only one amplifying mode and, for this purpose, one can use a partially broken PT-symmetric phase, i.e., only one pair of eigenstates has complex conjugate eigenvalues, while the other eigenstates have real ones. This idea has been demonstrated in the semiconductor platforms such as coupled microcavities with gain and loss \cite{Miri:12,HH14} (Fig.~\ref{fig:3optcavity}(b)), and a microring cavity with a grating \cite{LF14} (Fig.~\ref{fig:3optcavity}(c)). 
The PT transition can be also used to realize counterintuitive laser manipulations. For instance, increasing the loss in one cavity of coupled microcavities can cause the transition to the PT-broken phase and enhance the gain of a certain mode, thus realizing the loss-induced laser action \cite{PB14} (Fig.~\ref{fig:3optcavity}(a)). The opposite behavior has also been demonstrated by causing the anti-lasing with increasing the pump gain of one microcavity \cite{LM12,BM14}.  
 
Besides the spectral properties, the nonorthogonality of eigenmodes in non-Hermitian systems also leads to unconventional lasing behavior. While the laser linewidth can be described by the Schawlow-Townes one in usual setups \cite{SAL58}, a significant excess of the linewidth from this fundamental limit can occur when the nonorthogonality beomces singularly strong \cite{HWA90}, e.g., in the vicinity of the EP \cite{MVB03}. This width enhancement originates from the excess noise caused by loss-induced coupling in microcavities and can be characterized by the Petermann factor. The divergence of the Petermann factor \cite{KP79} indicates the presence of EPs and has been numerically investigated in the stadium-shaped microcavity \cite{LSY08} (see also related descriptions in Sec.~\ref{seceigv} and Fig.~\ref{fig:2petermann}).
With gain and loss in microcavities, a controllable generation of the EP has also found applications to ultrasensitive measurements \cite{LZP16,CW17,HH17} and unconventional response properties \cite{WJ11,MK14,PB16,MP16}. We refer to Sec.~\ref{Sec:EP} for further explanations on physical aspects of the EPs, including their unique topological aspects, and also on experimental realizations.
\\ \\ {\it Wave chaos and scar modes}

\vspace{3pt}
\noindent
Fabricating the shape of microcavity, one can study wave chaos and the accompanying localization modes known as {\it scar} modes \cite{HEJ84}. A quantum-mechanical particle trapped in a hard-wall potential, whose classical counterpart exhibits chaos, is accompanied by a fraction of eigenstates whose spatial distribution concentrates along unstable periodic orbits of the corresponding classical system. This type of eigenmodes is called the scar mode and, in the context of optics, it manifests itself as the localization of optical wave intensity along unstable periodic ray trajectories. Such scar modes in optics have been experimentally observed in liquid-jet and semiconductor microlasers \cite{LSB02,RNB02,Gmachl:02}. Later, the quasi-scar mode, i.e., the wave localization along marginally stable periodic rays has been studied in the non-Hermitian regimes of the near-field microcavity \cite{WJ06,YCH11} (Fig.~\ref{fig:3optcavity}(d)). The openness of such non-Hermitian systems can be taken into account by including the Fresnel-filtering effect \cite{Tureci:02}, which affects the long-time trajectory dynamics and thus results in the deviation of the ray dynamics from its time-reversed partner of periodic orbitals \cite{Altmann_2008,SQH11,RB12}. Such quasi-scar modes have also been observed in a microring cavity \cite{LSY04}.

\subsection{Mechanics}
\label{sec:mech}
Let us move on to mechanical systems, whose length scales are typically much larger than those of photonic systems discussed in Sec.~\ref{secphoto}. While the study of mechanical wave propagation was initiated decades ago right after Newton's revolutionary works \cite{HMI14}, there is reviving recent interest in light of the rapid development of mechanical metamaterials \cite{KB17}. As a general setup, the equation of motion of 
mechanical modes is given by \cite{SR16}
\eqn{\label{mechanics}
\ddot{\bold{x}}=-{ D}\bold{x}+\Gamma\dot{\bold{x}},
\label{mecheq}
}
where $\bold{x}$ is a displacement vector of mechanical degrees of freedom, and $D$ and $\Gamma$ are both real matrices. In ideal linear mechanical systems with no dissipation, $D$ should be a positive semi-definite symmetric matrix\footnote{Rigorously speaking, we have to assume that all the mass points share the same mass. Nevertheless, even if the diagonal mass matrix $M$ is not proportional to the identity, by redefining $\bold{x}$ as $\sqrt{M}\bold{x}$, we can recover Eq.~(\ref{mecheq}).} so that the zero displacement state is stable. Also, $\Gamma$ is a skew-symmetric matrix (i.e., $\Gamma=-\Gamma^{\rm T}$) and accounts for the frictionless forces that break the reciprocity, such as the Lorentz force and the Coriolis force.\footnote{These forces take the form $\bold{f}=\sum_{j,k,l}\bold{\Omega}\times \dot{\bold{x}}=\epsilon_{jkl}\Omega_j \dot{x}_k\bold{e}_l$, where $\sum_j\epsilon_{jkl}\Omega_j$ indeed reverses its sign upon the interchange of $k$ and $l$.} However, frictions are inevitable in real situations and thus $\Gamma$ is generally not skew-symmetric. This provides a natural mechanism for introducing non-Hermiticity \cite{YT19}. Also, it is possible to explicitly make $D$ non-symmetric in well-designed mechanical metamaterials \cite{BM19}. In the following, we will discuss these two situations separately. In addition, we will see that non-Hermitian topology emerges also in some static mechanical systems instead of wave dynamics.
\\ \\ {\it Non-Hermiticity from friction}

\vspace{3pt}
\noindent
At first glance, the mechanical equation of motion (\ref{mecheq}) seems quite different from the Schr\"odinger equation due to the second derivative. However, by enlarging the variables from $\bold{x}$ to $[\bold{x},\dot{\bold{x}}]^{\rm T}$, Eq.~(\ref{mecheq}) becomes \cite{SR16}
\begin{equation}
\frac{d}{dt}\begin{bmatrix} \bold{x} \\ \dot{\bold{x}} \end{bmatrix} = \begin{bmatrix} \;\;0\;\; & \;\;I\;\; \\ \;\;-D\;\; & \;\;\Gamma\;\; \end{bmatrix} \begin{bmatrix} \bold{x} \\ \dot{\bold{x}} \end{bmatrix},
\label{mechmtrx}
\end{equation}
which is now a first-order differential equation. Recalling that $D\ge0$, we have a well-defined square-root $\sqrt{D}$ and can introduce $\bold{u}\equiv[\sqrt{D}\bold{x},i\dot{\bold{x}}]$ whose dynamics is governed by \cite{KCL14,TCL15,HSD16}
\begin{equation}
i\frac{d}{dt}\bold{u}=\begin{bmatrix} \;\;0\;\; & \;\;\sqrt{D}\;\; \\ \;\;\sqrt{D}\;\; & \;\;i\Gamma\;\; \end{bmatrix}\bold{u},
\label{mechschro}
\end{equation}
which exactly takes the form of a Schr\"odinger equation. Note that Eq.~(\ref{mechschro}) follows straightforwardly from multiplying $\begin{bsmallmatrix} \sqrt{D} & 0 \\ 0 & iI \end{bsmallmatrix}$ to Eq.~(\ref{mechmtrx}) and thus the identity
\begin{equation}
 \begin{bmatrix} \;\;\sqrt{D}\;\; & \;\;0\;\; \\ \;\;0\;\; & \;\;iI\;\; \end{bmatrix} \begin{bmatrix} \;\;0\;\; & \;\;I\;\; \\ \;\;-D\;\; & \;\;\Gamma\;\; \end{bmatrix} = \begin{bmatrix} \;\;0\;\; & \;\;-i\sqrt{D}\;\; \\ \;\;-i\sqrt{D}\;\; & \;\;\Gamma\;\; \end{bmatrix} \begin{bmatrix} \;\;\sqrt{D}\;\; & \;\;0\;\; \\ \;\;0\;\; & \;\;iI\;\; \end{bmatrix},
\end{equation}
does not require $D$ to be invertible. Although the transformation from $\bold{x}$ into $\bold{u}$ discard the information of the components in ${\rm Ker}D$, which are called \emph{floppy modes} \cite{KCL14}, these modes are completely decoupled and do not alter the nontrivial mechanical dynamics of the remaining degrees of freedom. Moreover, the floppy modes also appear as the zero eigenstates of the Hamiltonian in Eq.~(\ref{mechschro}),  so the Schr\"odinger-equation form (\ref{mechschro}) contains information equivalent to the original equation of motion (\ref{mecheq}).

Now suppose that the mass points form a periodic lattice $\Lambda$ such that $D_{\boldsymbol{r}\alpha,\boldsymbol{r}'\beta}=D_{\boldsymbol{r}-\boldsymbol{r}',\alpha\beta}$ and $\Gamma_{\boldsymbol{r}\alpha,\bold{r}'\beta}=\Gamma_{\boldsymbol{r}-\boldsymbol{r}',\alpha\beta}$, where $\boldsymbol{r}$ denotes a lattice site (unit cell) while $\alpha$ labels an internal degree of freedom (e.g., sublattice and direction of displacement). In this case, we can Fourier transform 
the components $x_{\boldsymbol{r}\alpha}$ of $\bold{x}$ into $x_{\boldsymbol{k}\alpha}\equiv|\Lambda|^{-1/2}\sum_{\boldsymbol{r},\alpha}e^{-i\boldsymbol{k}\cdot\boldsymbol{r}}x_{\bold{r}\alpha}$ ($|\Lambda|$: total number of unit cells) such that the equation of motion (\ref{mecheq}) becomes
\begin{equation}
\ddot{\bold{x}}_{\boldsymbol{k}}=-D(\boldsymbol{k})\bold{x}_{\boldsymbol{k}}+\Gamma(\boldsymbol{k})\dot{\bold{x}}_{\boldsymbol{k}},
\end{equation}
where $\bold{x}_{\boldsymbol{k}}=[x_{\boldsymbol{k}\alpha}]^{\rm T}_\alpha$ and $[D(\boldsymbol{k})]_{\alpha\beta}=|\Lambda|^{-1/2}\sum_{\boldsymbol{r}}D_{\boldsymbol{r},\alpha\beta}e^{-i\boldsymbol{k}\cdot\boldsymbol{r}}$ (similar for $\Gamma(\boldsymbol{k})$). One can check from $\Gamma$ and the positive-semidefinite property that $D(\boldsymbol{k})$ is Hermitian and positive-semidefinite (in the absence of frictions, $\Gamma(\boldsymbol{k})$ is anti-Hermitian). The phonon dispersion $\omega_{\boldsymbol{k}}$ thus satisfies \cite{YTW15,KT15}
\begin{equation}
\det[\omega_{\boldsymbol{k}}^2I-i\omega_{\boldsymbol{k}}\Gamma(\boldsymbol{k})-D(\boldsymbol{k})]=0.
\label{mechdet}
\end{equation}
This result can directly be derived from calculating the eigenvalues of the Fourier transformed Hamiltonian in Eq.~(\ref{mechschro}) \cite{SR16,YT19}:  
\begin{equation}
H(\boldsymbol{k})=\begin{bmatrix} \;\;0\;\; & \;\;Q(\boldsymbol{k})\;\; \\ \;\;Q(\boldsymbol{k})\;\; & \;\;i\Gamma(\boldsymbol{k})\;\; \end{bmatrix},
\label{mechHk}
\end{equation}
where $Q(\boldsymbol{k})=\sqrt{D(\boldsymbol{k})}$ is well-defined due to the fact that $D(\boldsymbol{k})$ is positive-semidefinite. 

Since both $D$ and $\Gamma$ are real, the effective Bloch Hamiltonian (\ref{mechHk}) always has a particle-hole symmetry \cite{SR16,YT19}: 
\begin{equation}
\mathcal{C}H(\boldsymbol{k})\mathcal{C}^{-1}=-H(-\boldsymbol{k}),\;\;\;\;\mathcal{C}\equiv(\sigma^z\otimes I)\mathcal{K},
\label{mechPC}
\end{equation}
where $\mathcal{K}$ denotes the complex conjugate. This is a intrinsic property of mechanical systems and does not require any parameter fine tuning. In the absence of the Lorentz force or the Coriolis force, $H(\boldsymbol{k})$ always respects a chiral symmetry:
\begin{equation}
 \Gamma H(\boldsymbol{k})^\dag \Gamma = -H(\boldsymbol{k}),\;\;\;\;\Gamma\equiv\sigma^z\otimes I.
\end{equation}
Note that this differs slightly from Eq.~(\ref{HMchiral}) in that the Hermitian conjugate is taken for $H(\boldsymbol{k})$, which is generally non-Hermitian. The chiral symmetry can support \emph{symmetry-protected exceptional rings} in 2D \cite{YT19}, which should otherwise appear in 3D in the absence of symmetry protection \cite{XY17,CA18}. In contrast, we emphasize that the particle-hole symmetry itself (\ref{mechPC}) cannot protect an exceptional ring in 2D due to the reverse of $\boldsymbol{k}$. Nevertheless, this symmetry can be promoted to the parity-particle-hole (CP) symmetry by an inversion symmetry, which can support exceptional rings in 2D even in the presence of the Lorentz force or the Coriolis force \cite{YT19}.
 
Let us consider a mechanical graphene with friction as a prototypical example with various intrinsic symmetries discussed above. The mechanical graphene was originally introduced to mimic a Chern insulator in classical systems \cite{YTW15,KT15}, and has recently been generalized to the non-Hermitian version \cite{YT19}. Such a model is described by 
\begin{equation}
\begin{split}
D(\boldsymbol{k})&=\frac{3\kappa}{m}\left(1-\frac{\eta}{2}\right)\sigma_0\otimes\sigma_0 +\frac{\kappa}{m}\sigma^x\otimes[\gamma_3+\gamma_1\cos(\boldsymbol{k}\cdot\boldsymbol{a}_1)+\gamma_2\cos(\boldsymbol{k}\cdot\boldsymbol{a}_2)] \\
&+\frac{\kappa}{m}\sigma^y\otimes[\gamma_1\sin(\boldsymbol{k}\cdot\boldsymbol{a}_1)+\gamma_2\sin(\boldsymbol{k}\cdot\boldsymbol{a}_2)], \\
\Gamma(\boldsymbol{k})&=\left(i\Omega_0\sigma^y-\frac{b_A+b_B}{2}\sigma_0-\frac{b_A-b_B}{2}\sigma^z\right)\otimes\sigma_0,
\end{split}
\end{equation}
where $m$ denotes the mass, $\kappa$ is the stiffness of the springs, $\eta$ is the control parameter of the prestress, $\gamma_{1,2}=(1-\frac{1}{2}\eta)\sigma_0\pm\frac{\sqrt{3}}{4}\eta\sigma^x+\frac{1}{4}\eta\sigma^z$ and $\gamma_3=(1-\frac{1}{2}\eta)\sigma_0-\frac{1}{2}\eta\sigma^z$, $\Omega_0$ is the rotating angular frequency and $b_A$ ($b_B$) is the friction coefficient of sublattice $A$ ($B$). 
The first Pauli matrix in $\sigma^\mu\otimes\sigma^\nu$ acts on the sublattice degree of freedom while the second one acts on the spatial one. Without the Coriolis force, i.e., $\Omega_0=0$, $H(\boldsymbol{k})$ is chirally symmetric and supports stable exceptional rings, where critical damping dynamics, which is an exponential decay with a polynomial correction (see Sec.~\ref{secspecdec}), can be observed. 
With both the Coriolis force and inversion symmetry, i.e., $\Omega_0\neq0$ and $b_A=b_B$, $H(\boldsymbol{k})$ has a CP symmetry represented by $(\sigma^z\otimes\sigma^x\otimes\sigma_0)\mathcal{K}$ and again exhibits stable exceptional rings. Otherwise, we generally have exceptional points rather than rings.
\\ \\ {\it Non-Hermiticity from dynamical nonreciprocity}

\vspace{3pt}
\noindent
For a large class of rigid mass-spring models, we can employ the so-called equilibrium matrix formalism to analyze the mechanical wave dynamics \cite{TCL15}. By rigid, we mean that the structure at equilibrium is stable against external loads, i.e., external forces applied to the mass points can be balanced by the internal spring tensions. This assumption implies the existence of a dimensionless \emph{equilibrium matrix} $Q$ which relates the tension-induced forces $\bold{f}$ on the mass points to the spring tensions $\bold{t}$:
\begin{equation}
Q\bold{t}=\bold{f}.
\end{equation}
Note that ${\rm Ker}Q$ can be nonempty, since there can be \emph{self-stress states} which have nonzero spring tensions, yet the forces on the mass points cancel out. Another important quantity of rigid mass-spring models is the \emph{compatibility matrix} $R$, which is also dimensionless and relates the spring extensions $\bold{e}$ to the displacement of the mass points $\bold{x}$:
\begin{equation}
R\bold{x}=\bold{e}.
\label{Rxe}
\end{equation}
Again, ${\rm Ker}R$ can be nonempty since there can be zero-frequency floppy modes. In fact, $R$ and $Q$ are not independent. To see this, we apply the principle of virtual work \cite{CRC78} to obtain $\bold{t}^{\rm T}\bold{e}=\bold{f}^{\rm T}\bold{x}=\bold{t}^{\rm T}Q^{\rm T}\bold{x}=\bold{t}^{\rm T}R\bold{x}$, implying
\begin{equation}
R=Q^{\rm T}.
\label{CQT}
\end{equation}
Such a reciprocal relation further enables us to relate the dynamical matrix $D$ to $Q$ through 
\begin{equation}
D=QR=QQ^{\rm T},
\end{equation} 
which is indeed a positive-semidefinite symmetric matrix. Here we have used $m\ddot{\bold{x}}=\bold{f}$, $\bold{t}=-\kappa\bold{e}$ and set $\frac{\kappa}{m}$ to be unity. In a periodic system, we can write $D(\boldsymbol{k})=Q(\boldsymbol{k})Q(\boldsymbol{k})^\dag$.\footnote{After Fourier transform, we obtain $[D(\boldsymbol{k})]_{\alpha\beta}=\sum_{\gamma,\boldsymbol{r}'',\boldsymbol{r}}Q_{\boldsymbol{r}-\boldsymbol{r}'',\alpha\gamma}Q^{\rm T}_{\boldsymbol{r}''-\boldsymbol{r}',\gamma\beta}e^{-i\boldsymbol{k}\cdot(\boldsymbol{r}-\boldsymbol{r}')}=
\sum_{\gamma,\boldsymbol{r}''}Q(\boldsymbol{k})_{\alpha\gamma}Q_{\boldsymbol{r}'-\boldsymbol{r}'',\beta\gamma}e^{i\boldsymbol{k}\cdot(\boldsymbol{r}'-\boldsymbol{r}'')}=\sum_\gamma Q(\boldsymbol{k})_{\alpha\gamma}Q(\boldsymbol{k})^*_{\beta\gamma}=[Q(\boldsymbol{k})Q(\boldsymbol{k})^\dag ]_{\alpha\beta}$.}  Note that $Q(\boldsymbol{k})$  generally differs from $\sqrt{D(\boldsymbol{k})}$ used in the previous subsection by a unitary, which may contain crucial topological information of the system. 
The existence of such an ``ambiguity" in $Q(\boldsymbol{k})$ can already be understood from the fact that different similarity transformations can be applied to Eq.~(\ref{mechmtrx}) to obtain a Schr\"odinger-like equation \cite{SR16}. 
In the present setup, the corresponding Bloch Hamiltonian should be 
\begin{equation}
H(\boldsymbol{k})=\begin{bmatrix} \;\;0\;\; & Q(\boldsymbol{k}) \\ Q(\boldsymbol{k})^\dag & \;\;0\;\; \end{bmatrix},
\end{equation}
which can be considered as a square root of $\sigma_0\otimes D(\boldsymbol{k})$ \cite{KCL14}. This Hamiltonian respects both time-reversal (due to the realness of $Q$) and chiral symmetries and thus belongs to class BDI, which is classified by an integer winding number \cite{SAP08}.

\begin{figure}
\begin{center}
\includegraphics[width=14.5cm]{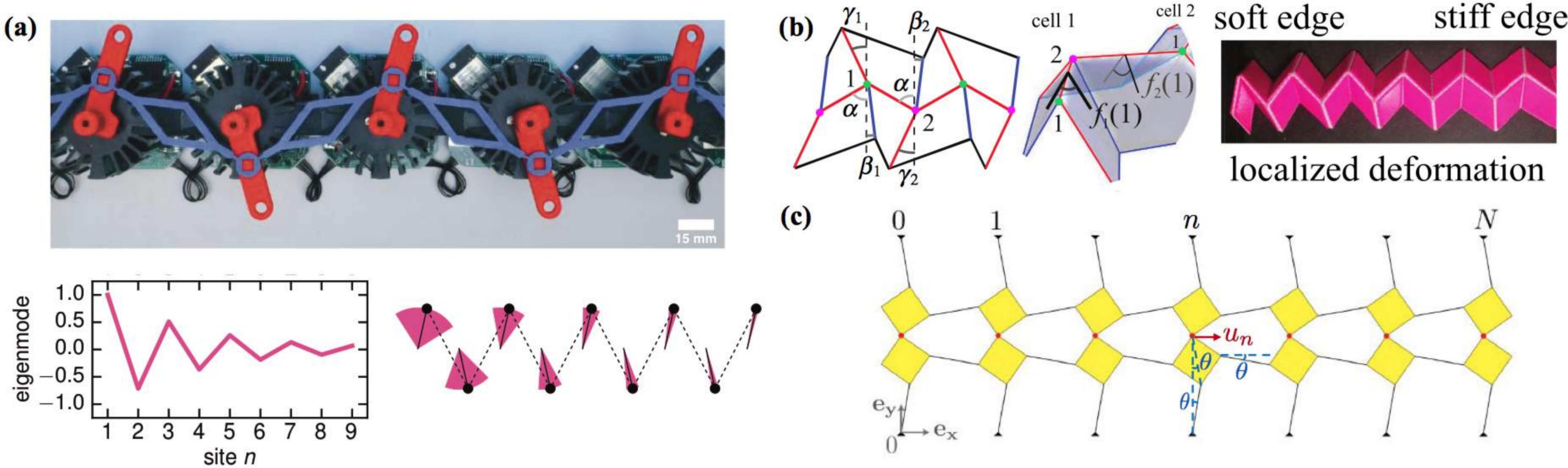}
\end{center}
\caption{(a) Active mechanical metamaterial (upper panel) and its oscillatory eigenmode localized at the left boundary (lower panel). Such an edgemode is protected by non-Hermitian topology (see Eq.~(\ref{wNH})). Adapted from Ref.~\cite{GA19}. (b) Unit cell (left and middle) of a Su-Schrieffer-Heeger-like origami structure and its topological edge mode, i.e., large $f$ angle, localized at the left edge (right). Adapted from Ref.~\cite{CBG16}. Adapted from Ref.~\cite{CBG16}. Copyright \copyright\,   2016 by the American Physical Society. (c) Mechanical metamaterial whose static states exhibits an exponential amplification of the displacement $u_n$. Adapted from the Supplementary information of Ref.~\cite{CC16}.}
\label{fig:3mech}
\end{figure}

Recent developments in 
mechanical metamaterials have made it possible to explicitly break the reciprocity  
by dynamical modulation \cite{WY18,TG19} or active feedback control \cite{BM19,GA19}. The former strategy can be understood from the fact that driving a system effectively inputs energy gain that selectively amplify some modes but almost do not influence the otheres \cite{SDL17}. A careful analysis for time-modulated mechanical systems should be based on the Floquet theory \cite{FC01}. 
We note that non-Hermitian phenomena such as the emergence of exceptional points may effectively occur in such systems \cite{TG19}. 
The latter strategy (feedback) is conceptually more straightforward --- we can break Eq.~(\ref{CQT}) by actively controlling  the dynamical response. This is achievable in active mechanical metamaterials, where the mechanical meta atoms (mass points) are equipped with some controllers that collect the data of the displacements and then impose additional spring tensions according to the data. If the feedback control is linear, we can simply modify the compatibility matrix by
\begin{equation}
R=Q^{\rm T}+R_{\rm fb},
\end{equation}
which becomes $R(\boldsymbol{k})=Q(\boldsymbol{k})^\dag+R_{\rm fb}(\boldsymbol{k})$ for translation-invariant lattice systems with the corresponding dynamical matrix and Hamiltonian given by
\begin{equation}
D(\boldsymbol{k})=Q(\boldsymbol{k})R(\boldsymbol{k}),\;\;\;\;
H(\boldsymbol{k})=\begin{bmatrix} \;\;0\;\; & Q(\boldsymbol{k}) \\ R(\boldsymbol{k}) & \;\;0\;\; \end{bmatrix}.
\end{equation}
This class of non-Hermitian Hamiltonians exhibits the sublattice symmetry and is thus characterized by two winding numbers, one of which is inherited in the Hermitian limit while the other is unique to non-Hermitian systems \cite{ZG18}. The idea of breaking reciprocity by feedback control was executed in Ref.~\cite{GA19} to realized the non-Hermitian SSH model described by
\begin{equation}
Q(k)=-a+be^{-ik},\;\;\;\;R_{\rm fb}(k)=\epsilon(a+be^{-ik}).
\end{equation}
The feedback is performed by further imposing or reducing (so $\epsilon$ can be either positive or negative) the tension of a spring according to the displacement of its two end mass points. The unique non-Hermitian winding number  
\begin{equation}
w_{\rm NH}=\frac{1}{2}\int^{2\pi}_0\frac{dk}{2\pi i}\partial_k\ln\det H(k)
\label{wNH}
\end{equation}
and its associated imbalance in the number of left and right edge modes was observed in this experiment (see Fig.~\ref{fig:3mech}(a)).
\\ \\ {\it Non-Hermitian topology in static mechanical systems}

\vspace{3pt}
\noindent
Previously, we have focused on the wave dynamics in mechanical systems, which can be mapped into a time-dependent Schr\"odinger equation (\ref{mechschro}). If we are only interested in the static state of the system, then it suffices to consider
\begin{equation}
D\bold{x}=\bold{0},
\label{Dx0}
\end{equation}
which may be considered to be an analogy of the time-independent Schr\"odinger equation. The existence of such a static solution (zero mode) can be ensured by the non-trivial topology in $D$, which can be generally asymmetric (non-Hermitian). Even if $D$ is symmetric (Hermitian) but takes the form $D=R^{\rm T}R$ with $R$ being generally asymmetric, as is the case in the previous subsubsection, then Eq.~(\ref{Dx0}) necessarily implies $R\bold{x}=\bold{0}$. In this case, a static solution may again arise from the underlying non-Hermitian topology.

The arguably most easily realizable static non-Hermitian topological phenomenon in practice is the \emph{topological origami} \cite{CBG16}. An ideal origami structure consists of rigid flat polygonal plates, which are connected with each other through their edges. Here by rigid, we mean that the shapes of the plates cannot change at all so that the equilibrium equation given in Eq.~(\ref{Dx0}) is purely geometric. A prototypical example that features non-Hermitian topology is the SSH-like origami structure consisting of tetragons \cite{CBG16}, as shown in Fig.~\ref{fig:3mech}(b). Depending on the crease angles $\alpha$, $\beta$ and $\gamma$, the height of the origami structure can exponentially be amplified along either the left or right direction. This provides a perfect realization of the Hatano-Nelson model \cite{HN96} with a unidirectional hopping amplitude \footnote{This result can be derived from the spherical law of cosines. For the specific origami structure in 
Fig.~\ref{fig:3mech}(b), we have $-\cos(\alpha+\gamma)\cos(\alpha+\beta)+\sin(\alpha+\gamma)\sin(\alpha+\beta)\cos f_1(j)=-\cos(\alpha-\gamma)\cos(\alpha-\beta)+\sin(\alpha-\gamma)\sin(\alpha-\beta)\cos f_2(j)$ and $-\cos(\alpha+\gamma)\cos(\alpha+\beta)+\sin(\alpha+\gamma)\sin(\alpha+\beta)\cos f_2(j)=-\cos(\alpha-\gamma)\cos(\alpha-\beta)+\sin(\alpha-\gamma)\sin(\alpha-\beta)\cos f_1(j+1)$.}
\begin{equation}
\kappa=\left[\frac{\sin(\alpha-\beta)\sin(\alpha-\gamma)}{\sin(\alpha+\beta)\sin(\alpha+\gamma)}\right]^2.
\end{equation}
In other words, the static state is actually an exponentially localized mode at either the left or right edge, with the localization length being $|\ln\kappa|^{-1}$. It is worth mentioning that the particular critical point $\beta=\gamma=0$ with a divergent localization length turns out to be the Miura-ori \cite{GSD13}.

A similar idea was used in Ref.~\cite{CC16} to design mechanical metamaterials with large static nonreciprocity. As shown in Fig.~\ref{fig:3mech}(c), the quasi-1D mechanical metamaterial is characterized by an angle $\theta$. Again, by solely considering the geometric constraints, we can show that the displacement per unit cell is amplified or attenuated exponentially by a factor of
\begin{equation}
g(\theta)=\frac{2+\cos(2\theta)+\sin(2\theta)}{2+\cos(2\theta)-\sin(2\theta)}.
\end{equation}
The critical point lies at $\theta=0$. As for the elastic property of this metamaterial, it is found that the (nonlinear) nonreciprocity susceptibility, which is defined as the coefficient $K$ in $\Delta u\equiv u_{\rm L\to R}+u_{\rm R\to L}=K F_0^2$ with $u_{\rm L\to R}$ ($u_{\rm R\to L}$) being the displacement at the right (left) edge when inputting $F_{\rm L}=F_0$ ($F_{\rm R}=-F_0$), diverges at the critical point \cite{CC16}.

\subsection{Electrical circuits}\label{sec:3ele}
In addition to mechanical systems, electrical circuits provide yet another  platform that enables us to simulate classical wave phenomena at the \emph{macroscopic} scale. Similar to the case of mechanics, studies on electrical circuits date  back to Ohm and Volta, while the recent interest  driven by the experimental progress on simulating novel wave phenomena, especially those related to band topology \cite{NJ15,CWP18,SI18} as well as Landau quantization \cite{ZXX20}. As a general setup, we consider a linear electrical circuit \cite{AVV15} defined on a network consisting of $N$ nodes, some of which are connected to the ground. Suppose that we can inject (alternating) current $I_a$ into each node $a$, where the voltage is $V_a$. Denoting the impedance between two nodes $a$ and $b$ as $Z_{ab}$ and that between $a$ and the ground as $Z_a$, according to Kirchoff's law \cite{YNJ09} (net current at each node must vanish), we have
\begin{equation}
I_a=\frac{V_a}{Z_a}+\sum_{b\neq a}\frac{V_a-V_b}{Z_{ab}}=\sum^N_{b=1} J_{ab} V_b,
\label{IJV}
\end{equation}
where $J$ is an $N\times N$ matrix called the \emph{Laplacian} of the circuit and depends explicitly on the frequency $\omega$ of the input current. We can decompose $J$ into three parts \cite{CHL18}:
\begin{equation}
J=W+D-Y,
\end{equation}  
where $W_{ab}=Z_a^{-1}\delta_{ab}$ is diagonal and depends on how the circuit is grounded, $D_{ab}=\sum_{c\neq a}Z_{ac}^{-1}\delta_{ab}$ is also diagonal and gives the total admittance of each node, and $Y_{ab}=Z_{ab}^{-1}$ ($Y_{aa}\equiv0$) describes the internode admittance. The reciprocal relation $Z_{ab}=Z_{ba}$ holds for usual circuits consisting of only capacitors, inductors and/or resistors, implying that $J=J^{\rm T}$. In particular, $J$ becomes anti-Hermitian in the absence of resistors, and the circuit can be used to simulate closed quantum systems described by a Hermitian Hamiltonian $(i\omega)^{-1}J$. Otherwise, $(i\omega)^{-1}J$ is generally non-Hermitian since the resistors introduce dissipation into the circuit. The transpose symmetry of $J$ can also be broken by integrating nonreciprocal devices such as diodes \cite{EM19}, Hall resistors \cite{HR19} and negative impedance converters with current inversion (INICs) \cite{WKC09}. These device may or may not introduce non-Hermiticity into the effective Hamiltonian $(i\omega)^{-1}J$. In the following, we will separately discuss the cases with and without reciprocity. 
\\ \\ {\it Non-Hermiticity from resistors}

\vspace{3pt}
\noindent
Just like frictions in mechanical systems, integrating resistors into an electrical circuit provides the most natural way of introducing non-Hermiticity. The effects of connecting two nodes and grounding a single node through a resistor are different, although in neither case the reciprocity, i.e., the transpose symmetry $J^{\rm T}=J$ is broken. The former protocol introduces the \emph{same} complex phase into both hopping amplitudes with opposite directions. To see this, we consider a simple situation where we incorporate resistors $R_{ab}$'s ($a\neq b$) through either series or parallel connection into the bonds $ab$'s of an $LC$ circuit with purely imaginary $Z_{ab}$. For parallel connection, the modified hopping amplitudes satisfy $J_{ab}'=Z_{ab}^{-1}+R_{ab}^{-1}=Z_{ba}^{-1}+R_{ba}^{-1}=J_{ba}'$. For series connection, we have $J_{ab}'=(Z_{ab}+R_{ab})^{-1}=J_{ba}'$. Obviously, $J_{ab}'$ becomes complex in both cases, yet the transpose symmetry is kept. The latter protocol introduces a \emph{negative} imaginary on-site potential, which resembles the effect of loss in photonic systems and that of friction in mechanical systems. The negative sign can easily be understood from the fact that resistors always dissipate energies. 
Indeed, incorporating positive $R_a$'s  into an $LC$ circuit always lead to positive real parts in $Z_a$'s, 
so the diagonal components of $(i\omega)^{-1}J$ always have negative imaginary parts.

\begin{figure}
\begin{center}
\includegraphics[width=14.5cm]{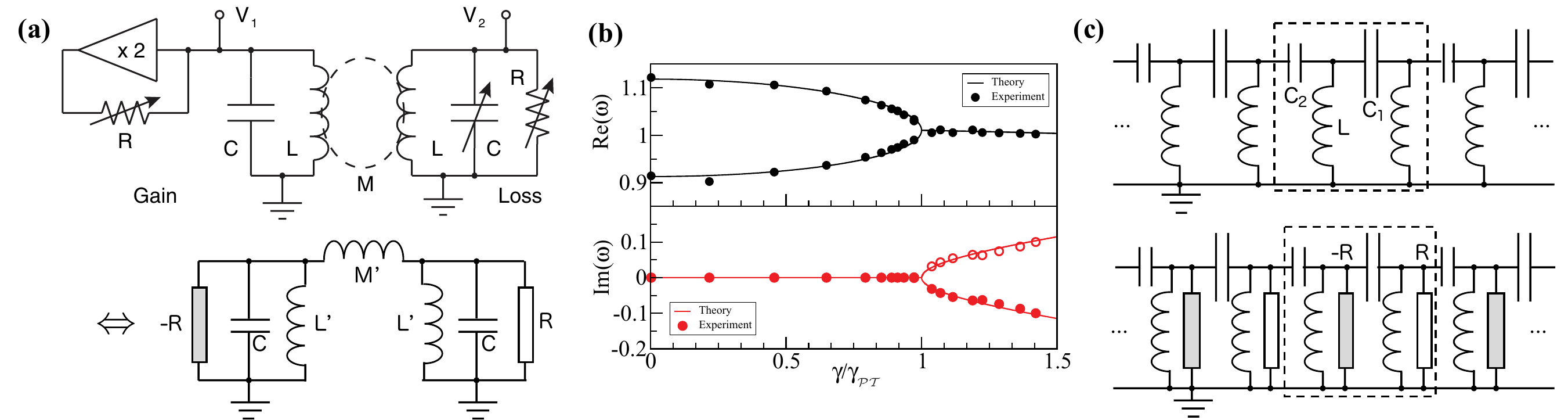}
\end{center}
\caption{(a) PT-symmetric circuit with balanced gain and loss (upper panel) and its equivalent circuit with $L'=L+M$ and $M'=(L^2-M^2)/M$ (lower panel). Here gain is achieved by an effective negative resistor. (b) Real (upper panel) and imaginary (lower panel) parts of the eigenfrequencies $\omega$ as a function of $\gamma=R^{-1}\sqrt{L/C}$. The solid curves (dots) show the theoretical (experimental) results.. Figures (a) and (b) are adapted from Ref.~\cite{SJ2011}. Copyright \copyright\,   2011 by the American Physical Society. (c) Circuit realization of the Su-Schrieffer-Heeger model (upper panel) and its PT-symmetric generalization (lower panel).}
\label{fig:3circuit}
\end{figure}

On the other hand, there does exist electric devices, which necessarily involve some energy resources, that behave like negative resistors. A simple realization can be achieved by a parallel connection of a usual resistor $R$ and a voltage-doubling operational amplifier; then the effective resistance of the combination is $-R$. Such a device was used to simulate PT-symmetry breaking \cite{SJ2011}. As shown in Fig.~\ref{fig:3circuit}(a), by taking the equivalent ``$\Pi$ circuit" of the mutually coupled inductors \cite{CKA99}, we obtain a two-node circuit described by 
\begin{equation}
J=i\sqrt{\frac{C}{L}}\left\{\left[\frac{\omega}{\omega_0}-\frac{\omega_0}{(1-\mu^2)\omega}\right]\sigma_0+\frac{\mu\omega_0}{(1-\mu^2)\omega}\sigma^x+i\gamma\sigma^z\right\},
\label{JRCL}
\end{equation}
where $\omega_0\equiv1/\sqrt{LC}$, $\gamma\equiv R^{-1}\sqrt{L/C}$ and $\mu=M/L$. In the absence of the current injection, a nontrivial oscillation with frequency $\omega$ occurs only if ${\rm det}J=0$, so the possible frequencies are given by
\begin{equation}
\omega=\omega_0\sqrt{\frac{2+\gamma^2(\mu^2-1)\pm\sqrt{4(\mu^2-1)+[2+\gamma^2(\mu^2-1)]^2}}{2(1-\mu^2)}}.
\end{equation}
These two eigenvalues coalesce at the exceptional point
\begin{equation}
\gamma_{\rm PT}=\frac{1}{\sqrt{1-\mu}}-\frac{1}{\sqrt{1+\mu}}.
\end{equation}
As shown in Fig.~\ref{fig:3circuit}(b), the theoretical prediction on the parameter ($\gamma$) dependence of eigenfrequencies perfectly matches the experimental measurement. Finally, we emphasize that although the PT symmetry manifests itself as $\sigma^x\mathcal{K}$ that commutes with $(i\omega)^{-1}J$, i.e.,
\begin{equation}
\sigma^x[(i\omega)^{-1}J]^*\sigma^x=(i\omega)^{-1}J,
\end{equation}
this is not directly relevant to the \emph{dynamical} PT symmetry breaking discussed above. The dynamical PT symmetry should be identified at the level of the full dynamics of both charge and current variables, as in  Eq.~(\ref{mechschro}) for mechanical systems. In such analogy, $i\omega J$ corresponds to the matrix in Eq.~(\ref{mechdet}) for mechanical systems. We will latter return to this point later.

Let us further provide an example of a topological electrical circuit, termed \emph{topolectrical circuit} \cite{CHL18}. The underlying model is a PT-symmetric SSH model (see Eq.~(\ref{PTSSH})), whose photonic realization \cite{WS17} has been discussed in Sec.~\ref{secpop}. The Hermitian limit of this model was proposed in Ref.~\cite{CHL18}, where each node is connected to the adjacent one by two alternating capacitors $C_1$ and $C_2$ and is grounded by an inductor $L$ (see Fig.~\ref{fig:3circuit}(c)). Under the periodic boundary condition, the circuit Laplacian after Fourier transformation reads
\begin{equation}
J_{\rm H}(k)=\left[i\omega(C_1+C_2)+\frac{1}{i\omega L}\right]\sigma_0-i\omega(C_1+C_2\cos k)\sigma^x-i\omega C_2\sin k\sigma^y.
\end{equation}
To introduce gain and loss, we further ground the nodes alternatively by negative and positive resistors through parallel connection, obtaining
\begin{equation}
J_{\rm NH}(k)=J_{\rm H}(k)-\frac{1}{R}\sigma^z.
\end{equation}
One can check that the effective Bloch Hamiltonian $(i\omega)^{-1}J_{\rm NH}(k)$ respects the PT symmetry represented by $\sigma^x\mathcal{K}$. Note that the reciprocity is preserved since 
$J_{\rm NH}(k)^{\rm T}=J_{\rm NH}(-k)$. 

To detect the topological edge mode under the open boundary condition, an experimentally accessible approach is to measure the two-point impedance. For a general circuit described by $J$, the two-point impedance between nodes $a$ and $b$ is given by \cite{CHL18}
\begin{equation}
Z_{ab}=\sum_{j_n\neq0}\frac{(\psi^{\rm R}_{n,a}-\psi^{\rm R}_{n,b})(\psi^{\rm L*}_{n,a}-\psi^{\rm L*}_{n,b})}{j_n},
\end{equation}
where $j_n$ is the $n$th eigenvalue of $J$, $\psi^{\rm L/R}_n$ is the corresponding left/right eigenstate and $\psi^{\rm L/R}_{n,a}$ is the component at node $a$. By tuning the frequency, $j_n$ can be globally shifted by an imaginary value and a divergent impedance peak signals a midgap eigenstate of $J$. This approach has been used to detect the edge mode in the Hermitian SSH circuit \cite{CHL18} and the corner mode in a second-order topolectrical circuit \cite{SI18}. To signal lasing and decaying modes in non-Hermitian circuits, we can further ground each node by identical and tunable resistors through parallel connection, and seek for peaks in the two-point impedance by changing not only the frequency but also the resistance of the auxiliary resistors.
\\ \\ {\it Non-Hermiticity from nonreciprocal devices}

\begin{figure}
\begin{center}
\includegraphics[width=12cm]{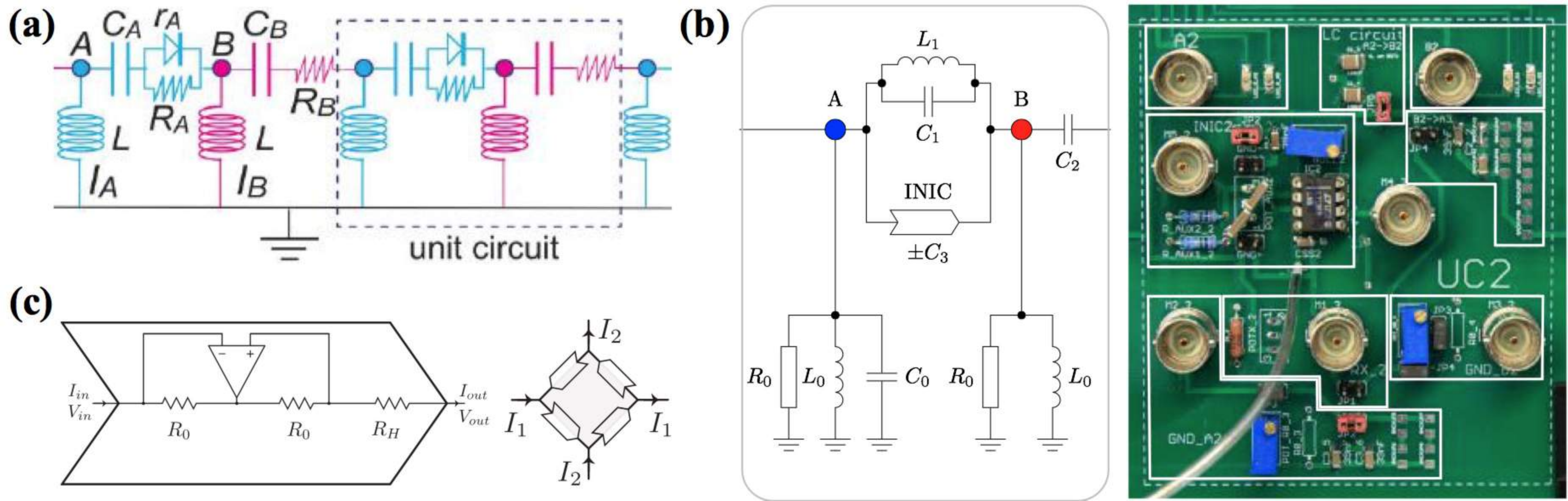}
\end{center}
\caption{(a) A nonreciprocal SSH model with asymmetric hopping amplitudes realized by diodes. Adapted from Ref.~\cite{EM19b}. Copyright \copyright\,   2019 by the American Physical Society. (b) Schematic illustration (left panel) and photo image of a unit-cell circuit of a similar nonreciprocal Su-Schrieffer-Heeger model with asymmetric hopping amplitudes realized by capacitor-like INICs. Adapted from Ref.~\cite{TH19}. (c) Internal structure of a resistor-like INIC (left panel) and a Hall resistor consisting effectively of four INICs (right panel). Adapted from Ref.~\cite{HR19}. Copyright \copyright\,   2019 by the American Physical Society.}
\label{fig:3nrcircuit}
\end{figure}

\vspace{3pt}
\noindent
Let us move on to introduce nonreciprocal electric circuits, for which $J$ generally differs from $J^{\rm T}$. The arguably simplest nonreciprocal device is the \emph{diode}, which, in the ideal case, has zero resistance in one direction but infinite resistance in the opposite direction. Combining such an ideal diode and a reciprocal device with impedance $Z'$ by parallel connection followed by series connection with another reciprocal device with impedance $Z$, we obtain a device, which has impedance $Z$ in one direction but $Z+Z'$ in the opposite direction. Such an idea has been used to realize a non-Hermitian SSH model with asymmetric intra-unit-cell hopping amplitudes (see Fig.~\ref{fig:3nrcircuit}(a)), whose Bloch Hamiltonian takes the form \cite{EM19b}
\begin{equation}
H(k)=\begin{bmatrix} \;\;0\;\; & \;\;t^{\rm L}_A+t_Be^{-ik}\;\; \\ \;\;t^{\rm R}_A+t_Be^{ik}\;\; & \;\;0\;\; \end{bmatrix}.
\label{NHSSH}
\end{equation}
Note that the reciprocal (transpose) symmetry is broken, i.e., $H(k)^{\rm T}\neq H(-k)$ due to $t^{\rm L}_A\neq t^{\rm R}_A$. We mention that Eq.~(\ref{NHSSH}) is a prototypical model that demonstrates the different gap-closing points in non-Hermitian lattices under closed and open boundary conditions and the non-Hermitian skin effect \cite{YS18a}. We will return to these points in Sec.~\ref{sec:bec}.

A more sophisticated nonreciprocal device is the \emph{negative impedance converters with current inversion} (INICs) \cite{WKC09}. As shown in Fig.~\ref{fig:3nrcircuit}(a), such a device involves an operational amplifier and effectively realizes a usual resistor $R$ in one direction but a negative resistor $-R$ in the opposite direction.  
Therefore, $J$ matrix of an INIC reads
\begin{equation}
J=\frac{1}{R}
\begin{bmatrix} 
\;\;-1\;\; & \;\;1\;\; \\ 
\;\;-1\;\; & \;\;1\;\;
\end{bmatrix},
\end{equation}
which explicitly breaks the transpose symmetry. Although a single INIC would introduce imaginary on-site potential, such an effect can be canceled by other INICs. Indeed, a 2D circuit that realizes a Hermitian Chern insulator has been proposed in Ref.~\cite{HT19} by incorporating INICs into an $LC$ circuit. On the other hand, when combined with reciprocal devices, INICs can create asymmetric hopping amplitudes with imbalanced absolute values. For example, if we combine an INIC with resistance $R$ and a usual resistor $R_0$ by parallel connection, then the hopping amplitude in $J$ is $R_0^{-1}+R^{-1}$ in one direction but $R_0^{-1}-R^{-1}$ in the other. This idea has been used to implement topological non-Hermitian quasi-crystals based on Aubry-Andr\'e-Harper models with asymmetric hopping amplitudes \cite{ZQB20}. In a recent experiment, the capacitor variant of INIC, which behaves as a usual capacitor $C$ in one direction but as $-C$ in the opposite direction, has been used to realize a nonreciprocal SSH model \cite{TH19}. The unit cell is shown in Fig.~\ref{fig:3nrcircuit}(b), whose  effective Bloch Hamiltonian also takes the form of Eq.~(\ref{NHSSH}) up to a constant background potential.

Finally, we briefly mention the \emph{Hall resistor}, which, unlike the previous nonreciprocal devices with two nodes, has four nodes for external connections. In this device, the reciprocity is explicitly broken by a magnetic field. In an ideal case, the current in the horizontal (vertical) direction is related to the vertical (horizontal) voltage through \cite{HR19}
\begin{equation}
J=\frac{1}{R_{\rm H}}
\begin{bmatrix} 
\;\;0\;\; & \;\;-1\;\; & \;\;0\;\; & \;\;1\;\; \\ 
\;\;1\;\; & \;\;0\;\; & \;\;-1\;\; & \;\;0\;\;  \\  
\;\;0\;\; & \;\;1\;\; & \;\;0\;\; & \;\;-1\;\; \\
\;\;-1\;\; & \;\;0\;\; & \;\;1\;\; & \;\;0\;\;  
\end{bmatrix},
\end{equation}
where $R_{\rm H}$ is the Hall resistance. This device was used in Ref.~\cite{HR19} for implementation of a Chern circuit on a square lattice. Interestingly, a Hall resistor can effectively be realized by combining four INICs (see Fig.~\ref{fig:3nrcircuit}(c)). A three-node variant consisting of three INICs can be used to build a honeycomb Chern circuit \cite{HR19}. Similar to INICs, ideal Hall resistors themselves do not introduce non-Hermiticity but only phases in the hopping amplitudes in the effective Hamiltonian $(i\omega)^{-1}J$. To make the hopping amplitudes asymmetric, we have to incorporate usual resistors. 
\\ \\ {\it Nonunitary dynamics in non-Hermitian circuits}

\vspace{3pt}
\noindent
So far, we have focused on the simulation of non-Hermitian Hamiltonians based on the \emph{steady-state response} of electrical circuits. This is relevant to the time-independent Schr\"odinger equation and resembles the situation of static mechanical systems discussed previously. While not so widely studied in literature, it is certainly possible to simulate time-dependent Schr\"odinger equation on electrical circuits. In fact, there is a perfect analogy between circuit dynamics and mechanical dynamics, where the voltage and its time derivative (or the current) in the former correspond to the position and the velocity in the latter. 

Let us derive an effective time-dependent Schr\"odinger equation for a general $LCR$ circuit \emph{without} external inputs or outputs. Assuming that all the devices are connected in parallel, we can decompose  
the $J$ matrix in Eq.~(\ref{IJV}) into
\begin{equation}
J=i\omega J_C+J_R+(i\omega)^{-1} J_L,
\end{equation}
where $J_C$, $J_R$ and $J_L$ are attributed to the capacitors, resistors and inductors. Here $J_C$ and $J_L$ are both real, symmetric and positive-semidefinite. There is no general constraint on $J_R$, although it is possible to make it real and skew-symmetric (by, e.g., connecting nodes with ideal INICs) to realize a Hermitian effective Hamiltonian $(i\omega)^{-1}J$. 
After inverse Fourier transforming $i\omega\bold{I}(\omega)=i\omega J\bold{V}(\omega)$ back to the time domain and using the fact that $\dot{\bold{I}}=0$ due to the absence of external connections, we obtain
\begin{equation}
\left(J_C\frac{d^2}{dt^2}-J_R\frac{d}{dt}+J_L\right)\bold{V}=\bold{0}.
\end{equation} 
Provided that $J_C$ is positive definite, the equation of motion for the extended variables $[\tilde{\bold{V}},\dot{\tilde{\bold{V}}}]^{\rm T}$ with $\tilde{\bold{V}}\equiv\sqrt{J_C}\bold{V}$ reads
\begin{equation}
\frac{d}{dt}\begin{bmatrix} \tilde{\bold{V}} \\ \dot{\tilde{\bold{V}}} \end{bmatrix}=
\begin{bmatrix} 
\;\;0\;\; & \;\;I\;\; \\ 
\;\;-J_C^{-\frac{1}{2}}J_LJ_C^{-\frac{1}{2}}\;\; & \;\;J_C^{-\frac{1}{2}}J_RJ_C^{-\frac{1}{2}}\;\;
\end{bmatrix}
\begin{bmatrix} \tilde{\bold{V}} \\ \dot{\tilde{\bold{V}}} \end{bmatrix},
\label{VdV}
\end{equation}
which formally takes the same form of Eq.~(\ref{mechmtrx}) for mechanical systems. We can thus apply a similar technique to transform Eq.~(\ref{VdV}) into a Schr\"odinger equation like Eq.~(\ref{mechschro}), where $D=J_C^{-\frac{1}{2}}J_LJ_C^{-\frac{1}{2}}$ and $\Gamma=J_C^{-\frac{1}{2}}J_RJ_C^{-\frac{1}{2}}$. In particular, the eigenvalue $\omega$ of such an effective Hamiltonian satisfies Eq.~(\ref{mechdet}), which is now explicitly given by
\begin{equation}
\begin{split}
&\det(\omega^2I-i\omega J_C^{-\frac{1}{2}}J_RJ_C^{-\frac{1}{2}}-J_C^{-\frac{1}{2}}J_LJ_C^{-\frac{1}{2}})=0 \\
\Leftrightarrow\;\;&\det(\omega^2J_C-i\omega J_R-J_L)=\det(-i\omega J)=0.
\end{split}
\end{equation}
Now it is clear why we determined the eigenvalues from $\det J=0$ for the PT symmetric two-node circuit in Fig.~\ref{fig:3circuit}(a). We note that such a dynamical treatment was employed in Refs.~\cite{HR19} and \cite{HT19} to obtain the chiral voltage dynamics in Chern circuits.

When a $RCL$ circuit involves series connection, the perfect analogy of mechanical systems generally breaks down. Nevertheless, it is still possible to write an effective Schr\"odinger equation for appropriately chosen variables. For example, the nonunitary dynamics in a translation-invariant 1D non-Hermitian circuit  was studied in Ref.~\cite{EM19c}, which provides some exact solutions for specific choices of parameters. In this circuit, each node is connected to the neighboring ones by a series combination of an inductor $L$ and a resistor $R^L$, and it is further grounded through a parallel combination of a capacitor $C$ and a resistor $R^C$. Choosing the variables to be $\psi=[\bold{V},\sqrt{L/C}\bold{I}]^{\rm T}$, after Fourier transformation into the momentum space, we obtain 
\begin{equation}
i\partial_t\psi_k=H(k)\psi_k,\;\;\;\;
H(k)=\begin{bmatrix}
\;\;-i\frac{R^L}{L}\;\; & \;\;-\frac{i}{\sqrt{LC}}(1-e^{-ik})\;\; \\
\;\;-\frac{i}{\sqrt{LC}}(1-e^{ik})\;\; & -\frac{i}{CR^C}
\end{bmatrix}.
\end{equation}
Such a classical circuit can be used to simulate interference phenomena in quantum walks. When introducing spatial inhomogeneity, we can, for instance, create an SSH model with open boundary and study the edge dynamics. In this case, a large edge voltage or current is found to stay localized in the topological phase, but otherwise quickly diffuses into the bulk in the trivial phase \cite{EM19c}. We can also simulate nonreciprocal dynamics by introducing nonreciprocal devices such as diodes and INICs mentioned above.

\subsection{Biological physics, transport phenomena, and neural networks\label{secbio}}
As the Liouvillean governing classical dynamics is in general non-Hermitian, a variety of physical systems described by the Markovian master equation  can potentially offer an ideal platform to study non-Hermitian physics. More concretely, prominent examples along this direction include stochastic processes such as the asymmetric simple exclusion process (ASEP) \cite{MCT68,MCT69}, which is a paradigmatic model for understanding of biological transport, and localization phenomena in population dynamics and biological networks \cite{MRM72,NDR98}.
Furthermore, dynamical systems with nonlinear interactions are also of interest in terms of non-Hermitian physics. For example, it has recently been argued that the non-Hermiticity in the connectivity matrix of neurons is advantageous in learning dynamics of recurrent neural networks \cite{GS08,KG19}. Below we provide an overview for each of these  topics.    
\subsubsection{Master equation and transport in biological physics\label{secbiomas}}
\begin{figure}
\begin{center}
\includegraphics[width=13cm]{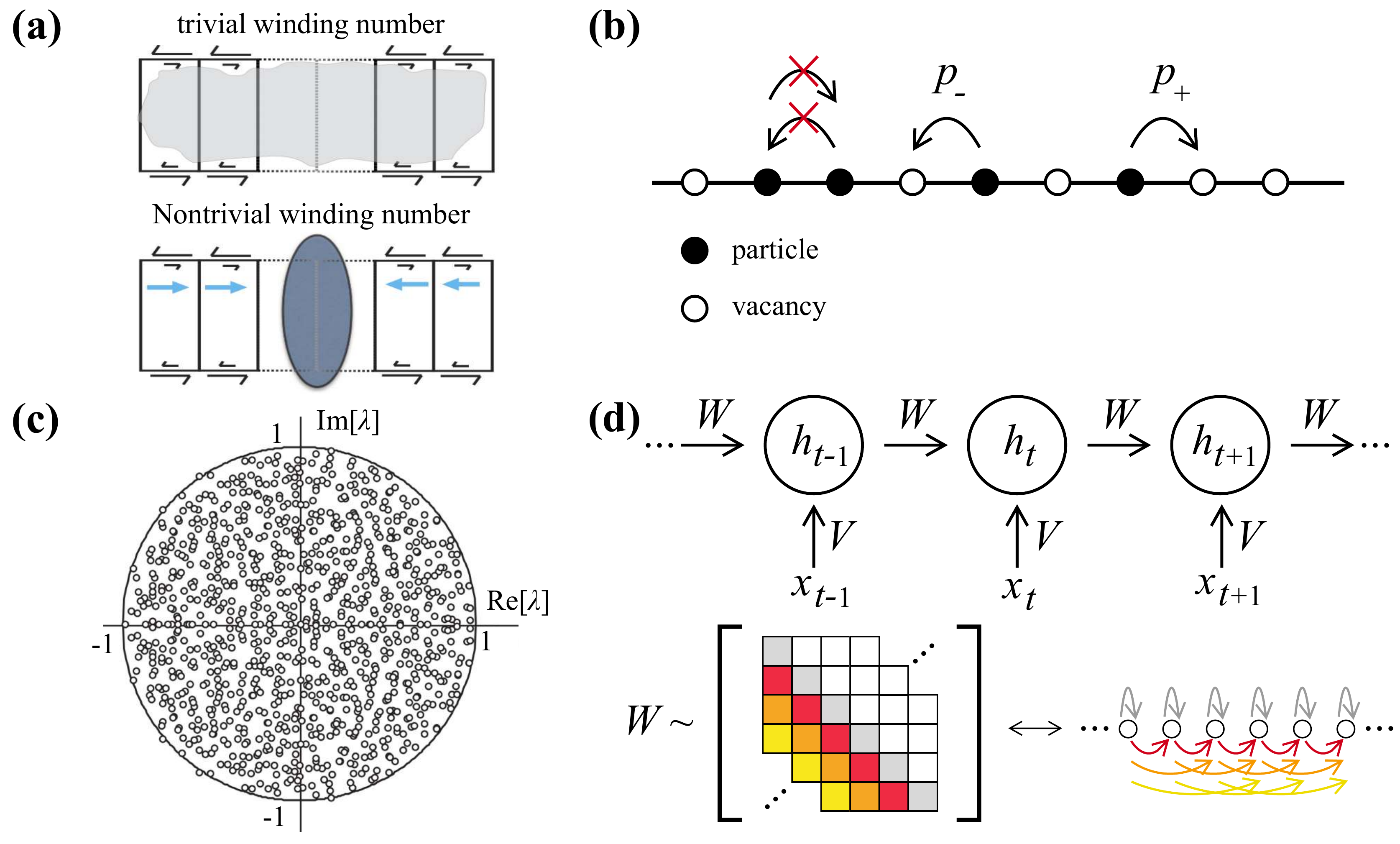}
\end{center}
\caption{ (a) Schematic figure illustrating a localized steady state protected by the nontrivial winding number in the Liouvillean of a biochemical stochastic process. 
Adapted from Ref.~\cite{MA17} licensed under a Creative Commons Attribution 4.0 International License. (b) Schematic illustration of an asymmetric simple exclusion process (ASEP). (c) Eigenvalues of random non-Hermitian $N\times N$ synaptic matrices lying within a unit circle in the complex plane in the limit $N\to\infty$. Adapted from Ref.~\cite{RK06}. Copyright \copyright\,   2006 by the American Physical Society. (d) (Top) Schematic illustration of a recurrent neural network (RNN). (Bottom) Relation between elements of a non-normal connectivity matrix (cf. Eq.~\eqref{sec3nonnormal}) and feed-forward couplings between neurons induced by the nonorthogonality discussed in Ref.~\cite{KG19}. }
\label{fig:3bio}
\end{figure}

Many interesting phenomena in nonequilibrium statistical mechanics and biological physics can often be described by stochastic Markov jump processes between different configurations.  
In general, the \emph{master equation} for Markov processes on discrete (finite) states is given by
\eqn{\label{masterstoch}
\frac{d{\bold{p}}}{dt}=W{\bold{p}},
}
where ${\bold{p}}$ is a vector of the probability distribution and $W$ is the transition matrix (or the Liouvillean) satisfying the condition $\sum_{i}W_{ij}=0$, which ensures the conservation of probability, i.e., $\sum_{i}p_i=1$. If some of state transitions are nonreciprocal, $W$ can be non-Hermitian. If the transition matrix is ergodic, i.e., any two states can be transferred via a sequence of Markov jump processes, the Perron-Frobenius theorem ensures that there exists a unique steady-state solution with zero eigenvalue $W{\bold{p}}^{\rm ss}=0$ \cite{SJ76} while the other eigenvalues have strictly negative real parts corresponding to damping modes. 
We note that, in the infinite-dimensional case, $W$ becomes an operator and its mathematically precise characterization requires a careful analysis based on the functional analysis \cite{SH91}. 
It is worthwhile to mention that a certain type of steady states in finite biochemical networks with nonreciprocal transitions have been interpreted as topological edge modes (or the so-called skin effects) of the non-Hermitian Liouvillean $W$ \cite{MA17,DK2018} (Fig.~\ref{fig:3bio}(a)).  An analogous topological transition in the Liouvillean has been discussed in a different context of the nonanalytic features in full-counting statistics \cite{RJ2013}.

A closely related, another approach to a stochastic process is the Langevin approach, which can be formulated based on the probability density function $p({\bf x},t)$ in continuous space obeying the {\emph{Fokker-Planck equation}} \cite{RH96,GCW04}:
\eqn{
\frac{\partial p({\bf x},t)}{\partial t}&=&\left[-\sum_{i}\frac{\partial}{\partial x_i}\mu_i({\bf x},t)+\sum_{ij}\frac{\partial^2}{\partial x_ix_j}D_{ij}({\bf x},t)\right]p({\bf x},t)\nonumber\\
&\equiv&{\cal L}p({\bf x},t),
}
where $\mu_i$ characterizes the drift motion and $D_{ij}$ is the diffusion constant. The Liouvillean operator $\cal L$ is in general non-Hermitian since the diffusion term is Hermitian while the drift term is anti-Hermitian. A similar type of the Liouvillean operator appears when we discuss a reaction-diffusion system in Sec.~\ref{secbiopop} (cf. Eq.~\eqref{reacdiff}). 

Arguably one of the most paradigmatic Markov jump processes to understand biological transport is the {\emph{asymmetric simple exclusion process}} (ASEP). It describes transport dynamics of particles that are located on a discrete lattice and evolve in continuous time. Hopping particles experience hard-core interaction, i.e., they can randomly jump  to a neighboring site with hopping rates $p_{+}$ in the forward direction and $p_-$ in the backward direction if the target site is empty  (see Fig.~\ref{fig:3bio}(b)).  
The presence of bias in a particular direction ($p_{+}\neq p_-$)  makes the random walk asymmetric and drives the system out of equilibrium, leading to a nonvanishing current. Historically, this model has been first introduced as a prototypical model to understand ribosomes translating dynamics along mRNA \cite{MCT68,MCT69}. In a different context of mathematical physics, the model has appeared in studies of interacting Brownian particles \cite{FS70}. A more general model of stochastic lattice gas problems, which includes the ASEP as a special case, has been introduced in statistical physics \cite{KS84,BD98}. Later, the ASEP has found applications to low-dimensional transport phenomena in a variety of physical systems including ion channels \cite{CT98}, nuclear pore complexes \cite{ZAA07,JTT09}, quantum dots \cite{KT102}, boundary-driven osmosis \cite{CT99} and traffic flow \cite{Evans_1996,DC00}. 
 
The Liouvillean $W$ of the ASEP is in general non-Hermitian due to, for example, the nonequilibrium bias. Because $W$ is real, its eigenvalues and eigenvectors are either real numbers or complex conjugate pairs. For the ASEP on a finite lattice, it has a unique steady-state solution as it satisfies the ergodicity. The integrability of the ASEP, i.e., the presence of an extensive number of integrals of motion, allows one to exactly obtain its spectrum and eigenvectors on the basis of the Bethe ansatz \cite{DD87,GLH92,KD95,ST99}. From a perspective of non-Hermitian physics, a particularly interesting aspect is the mapping of the ASEP to the non-Hermitian quantum spin chain. Without a driving field, the model reduces to the symmetric exclusion process and can be mapped to the (Hermitian) XXZ quantum spin chain. In contrast, the asymmetry breaks the left-right symmetry and violates the Hermiticity in the mapped XXZ spin chain \cite{DD87,GLH92,KD95,dG05}. In the simplest case with the periodic boundary conditions, the non-Hermitian XXZ model after the mapping can be written as
\eqn{
W=\sum_{j=1}^L\left[p_+\sigma_{j}^{+}\sigma_{j+1}^{-}+p_-\sigma_{j}^{-}\sigma_{j+1}^{+}+\frac{p_++p_-}{4}(\sigma_{j}^{z}\sigma_{j+1}^z-1)\right],
}
where the bias (or the nonreciprocal hopping) $p_+\neq p_-$ results in the non-Hermiticity.
It is also possible to map the ASEP to the six vertex model \cite{Kandel_1990,BR08}. These mappings are useful to obtain important  features of $W$ such as the spectral gap \cite{GLH92,KD95} and large deviation functions \cite{DB98,Derrida_1999}.

Other notable studies on biological transport related to non-Hermitian physics include the analyses of photosynthetic phenomena. The efficiency of energy transport in photosynthetic light-harvesting complexes has been studied based on the non-Hermitian model \cite{CG12}. The coupling of a reaction center to the surrounding environment plays the role of energy absorber and induces the non-Hermiticity, which has been argued to influence the energy transfer. On the basis of the similar formalism, the robustness of the energy transfer against disorder and dephasing effects in photosynthesis has also been analyzed \cite{CG14}.

\subsubsection{Random matrices in population evolution and machine learning\label{secbiopop}}
\noindent{\it Population dynamics}

\vspace{3pt}
\noindent
The analysis of non-Hermitian matrices is also useful to understand several fundamental aspects of large complex systems relevant to biological physics. Consider a \emph{nonconservative} dynamical system whose fluctuations around a steady state can be studied by the following linearized equation
\eqn{\label{populationeq}
\frac{dc}{dt}={\cal L}c,
}
where $c\in{\mathbb R}^N$ represents a vector of populations of certain species, and ${\cal L}\in{\mathbb R}^{N\times N}$ governs the linear dynamics and is in general non-Hermitian. A classic result in ecology is the presence of a sharp stable-unstable transition in large complex systems \cite{MRM72}. Suppose that the steady state is stable if couplings between different species are absent. The stability of such ecological systems can be studied based on the eigenspectra of the random matrices,
\eqn{\label{ecomat}
{\cal L}_{\alpha,C}=M_{\alpha,C}-I,
}
where each element of the (non-Hermitian) random matrix $M_{\alpha,C}$ is sampled from a distribution with  zero mean and variance $\alpha$ with probability $0<C\leq1$; otherwise it is set to be zero. While the off-diagonal elements represent couplings between different species, the unit matrix in the right-hand side of Eq.~\eqref{ecomat} denotes the damping rate to the steady state. The system is stable if and only if all the eigenvalues of ${\cal L}_{\alpha,C}$ have negative real parts. It has been shown that the system is almost certainly stable (unstable) if $\alpha<1/\sqrt{NC}$ ($\alpha>1/\sqrt{NC}$) \cite{MRM72}. Since the relative width of the transition regime in $\alpha$ scales as $1/N^{2/3}$, this transition becomes very sharp in a large system with $N\gg 1$, indicating that  the increasing complexity in ecology can induce a sudden stable-to-unstable transition in population dynamics.

Another intriguing aspect of population dynamics in view of non-Hermitian physics is the localization-delocalization transition in disordered systems \cite{NDR98}. This phenomenon has been studied on the basis of the continuum-space variant of Eq.~\eqref{populationeq} with the non-Hermitian random operator, 
\eqn{\label{reacdiff}
{\cal L}_{\bf v}=D\nabla^2-{\bf v}\cdot\nabla+V({\bf x}),
} 
which models the population dynamics in the presence of diffusion $D\nabla^2$ and convective flow $-{\bf v}\cdot\nabla$ together with  random potential $V({\bf x})$. The Liouville operator  ${\cal L}_{\bf v}$ can be obtained from ${\cal L}_{{\bf v}={\bf 0}}$ by performing the {\emph{imaginary}} gauge transformation, $\nabla\to\nabla-{\bf v}/(2D)$. Thus, its lattice discretization corresponds to a higher-dimensional variant of the Hatano-Nelson model \cite{HN96}. While the eigenstates can be spatially localized when ${\bf v}={\bf 0}$ due to disorder, it has been shown that increasing ${\bf v}$ generates unstable delocalized modes whose eigenvalues have positive real parts \cite{NDR98}. Recently, the non-Hermitian localization in disordered systems has also been discussed in the context of ring-shaped biological networks \cite{AA16non,NDR19}. 
\\ \\ {\it Lotka-Volterra equations}

\vspace{3pt}
\noindent
Another paradigmatic model in population dynamics is the {\emph{Lotka-Volterra equations}} \cite{LAJ10,VVJ28}, which describe the evolution of interacting species of organisms relevant to predator-prey ecological processes. In the simplest case, they are defined by a set of the following nonlinear coupled equations:
 \eqn{\label{LVeq}
\frac{dx}{dt}&=&x-xy,\\
\frac{dy}{dt}&=&-y+xy,
 }
where $x(t)$ ($y(t)$) is the population of the prey (predator) species. The equations 
  possess real and positive solutions, which exhibit delayed periodic oscillations. These delayed oscillations can be interpreted as mutual feedback between the predator and prey. It is possible to understand this emergence of periodic real solutions in terms of non-Hermitian physics  \cite{Bender_2007}. To this end, we introduce the parity operation $P$ as the swap operation between predator and prey $(x,y)\to (y,x)$.   
  Then, the Liouvillean operator associated with the present dynamical system, which is defined as $df(x(t),y(t))/dt\equiv -i{\cal L}f$, satisfies the PT symmetry as inferred from the PT invariance of the time-evolution equations. This symmetry implies the real-valuedness of the spectrum of $\cal L$ and thus the existence of periodic bounded solutions of $(x(t),y(t))$. While the symmetry argument alone does not necessarily guarantee the latter, there exists the constant of motion in the Lotka-Volterra equations,
  \eqn{
 x+y-\log(xy)=C,\;\;\;C\in{\mathbb R},
  } 
  which can completely characterize the presence of such periodic bounded solutions. 
Since it is in general highly nontrivial to find the constant of motion in nonlinear dynamical systems, the symmetry argument of the Liouvillean operator as demonstrated above can be potentially useful for qualitative understanding of more complex systems. 
\\ \\ {\it Artificial neural networks and machine learning}

\vspace{3pt}
\noindent
Several fundamental features inherent in non-Hermitian matrices, such as the complex spectra and the nonorthogonality of eigenstates, have also found interesting applications to artificial neural networks and machine learning in recent years. The discrete time evolution of neural networks is given by
\eqn{
h_{t+1}=\phi(Wh_{t}),
}
where $h_t\in{\mathbb R}^{N}$ is a hidden-state vector at time $t\in{\mathbb N}$, $W\in{\mathbb R}^{N\times N}$ characterizes the connectivity between different neurons and $\phi$ is a nonlinear function which is typically chosen to be the sigmoid function or the rectified linear unit function; we omit a bias vector for the sake of simplicity. The understanding of the eigenvalue spectrum of the connectivity matrix $W$, which is in general non-Hermitian, is crucial because it can govern the long-time dynamics of neural networks. If $W$ has eigenvalues whose norms are significantly larger than $1$, training of neural networks can be hampered by divergent gradients at large $t$.  A fundamental result related to this point is the Girko's circle law \cite{GVL85}, which states that the eigenvalues of $N\times N$ random real matrices, whose elements obey a distribution with mean zero and variance $1/N$, lie uniformly within the unit circle in the complex plane in the limit $N\to\infty$.\footnote{While Girko made remarkable progress on generalizing Ginibre's specific result for Gaussian distributions \cite{JG65}, his original proof was found to be flawed \cite{BZD97}. This conjecture was finally proved by Terence Tao and Van H. Vu in 2010 \cite{TT10}.} Later, this statement has been generalized to the case with nonvanishing symmetric correlations \cite{SHJ88}, and to a random synaptic matrix obeying two different distributions representing excitation and inhibition as suitable for realistic neural networks \cite{RK06} (see Fig.~\ref{fig:3bio}(c)). 
Taking the advantage of the restriction of eigenvalues within the unit circle,
 one can use the above types of random matrices as an initial connectivity matrix to prevent neurons from explosions in the long-time regimes. 

When one considers training neural networks in machine learning, an eigenvalue of $W$ whose norm is significantly less than $1$ can also be problematic since it leads to the compressing dynamics in long-time regimes and can cause  vanishing gradients, which are undesirable for realizing an efficient optimization. This problem is considered to be one of the major difficulties in training deep neural networks and recurrent neural networks (RNN). To remedy this, {\it normal} matrices with unit-norm eigenvalues are used as the connectivity matrix $W$ to efficiently train the RNN \cite{AM16,VE17}:
\eqn{
h_{t+1}=\phi(Wh_t+Vx_{t+1}),
}
where $x_t$ is the input vector at time $t$ and $V{\in}{\mathbb R}^{N\times N}$ is a projection matrix (see Fig.~\ref{fig:3bio}(d)). In general, we recall that a complex-valued matrix $M$ is called {\it normal} if and only if $MM^\dagger=M^\dagger M$, or said differently, if and only if there exists a diagonal matrix $D$ and a unitary matrix $U$ such that $M=UDU^\dagger$. Normal matrices completely characterize a set of matrices whose eigenvectors can be chosen to be orthogonal. 
In the present case of real matrices $W$, a normal matrix with unit-norm eigenvalues is equivalent to an orthogonal matrix $WW^{\rm T}=I$. 
 While the RNN based on the orthogonal connectivity matrix has achieved superior performance to the conventional RNN \cite{AM16,VE17}, the orthogonality of eigenstates still restricts the full potential of neural networks to express complex input-output correlations. 
 
Very recently, based on the Schur decomposition, the use of the following {\it non}-normal matrices has been proposed to mitigate this restriction \cite{GS08,KG19}:
\eqn{\label{sec3nonnormal}
W=P(D+\Delta)P^{\rm T},
}
where $P$ is an orthogonal matrix, $D$ is a diagonal matrix whose eigenvalues are set to be 1 at the initial time of learning, and $\Delta$ is restricted to be a strictly lower-triangular matrix. The successive multiplications of the matrix $D+\Delta$ would lead to neither explosion nor compression since $\Delta$ only has a purely feed-forward effect, thus regulating the learning dynamics in long-time regimes. In the limit of $\Delta\to 0$, $W$ reduces to a normal matrix and eigenstates are inevitably orthogonal. In contrast, a nonzero $\Delta$ allows couplings between the orthogonal column vectors of $P$ and this nonorthogonality provides further flexibility in transient dynamics \cite{GS08,GMS09,HG12} that cannot be produced by the RNN with an orthogonal connectivity matrix \cite{AM16,VE17} (see Fig.~\ref{fig:3bio}(d)). This advantage of nontrivial transient dynamics has also been captured by the Fisher-information analysis of the information propagation through the non-normal RNN \cite{GS08}.  As a consequence, the non-normal RNN outperforms the state-of-the-art RNN for certain machine learning tasks \cite{KG19}.

\subsection{Optomechanics and optomagnonics\label{secoptmech}}
Yet another emerging field related to non-Hermitian physics is {\it optomechanics}  \cite{AM14}, which studies hybrid platforms for integrating photonic devices and other degrees of freedom such as phonons in mechanical oscillators. 
With prospect of its applications to quantum optics \cite{HM09,EC11,RM18}, quantum information science \cite{OCAD10,SK11}, and gravitational-wave detection \cite{MDE112}, optomechanics has attracted growing attention in recent years. 
As optomechanical devices are inherently out-of-equilibrium systems due to  external pumping and cavity loss \cite{AM14,VE172}, they can serve as an ideal platform to study rich physics of non-Hermitian systems. We here review several aspects of optomechanical systems in classical regimes from the perspective of non-Hermitian physics.

\begin{figure}
\begin{center}
\includegraphics[width=13cm]{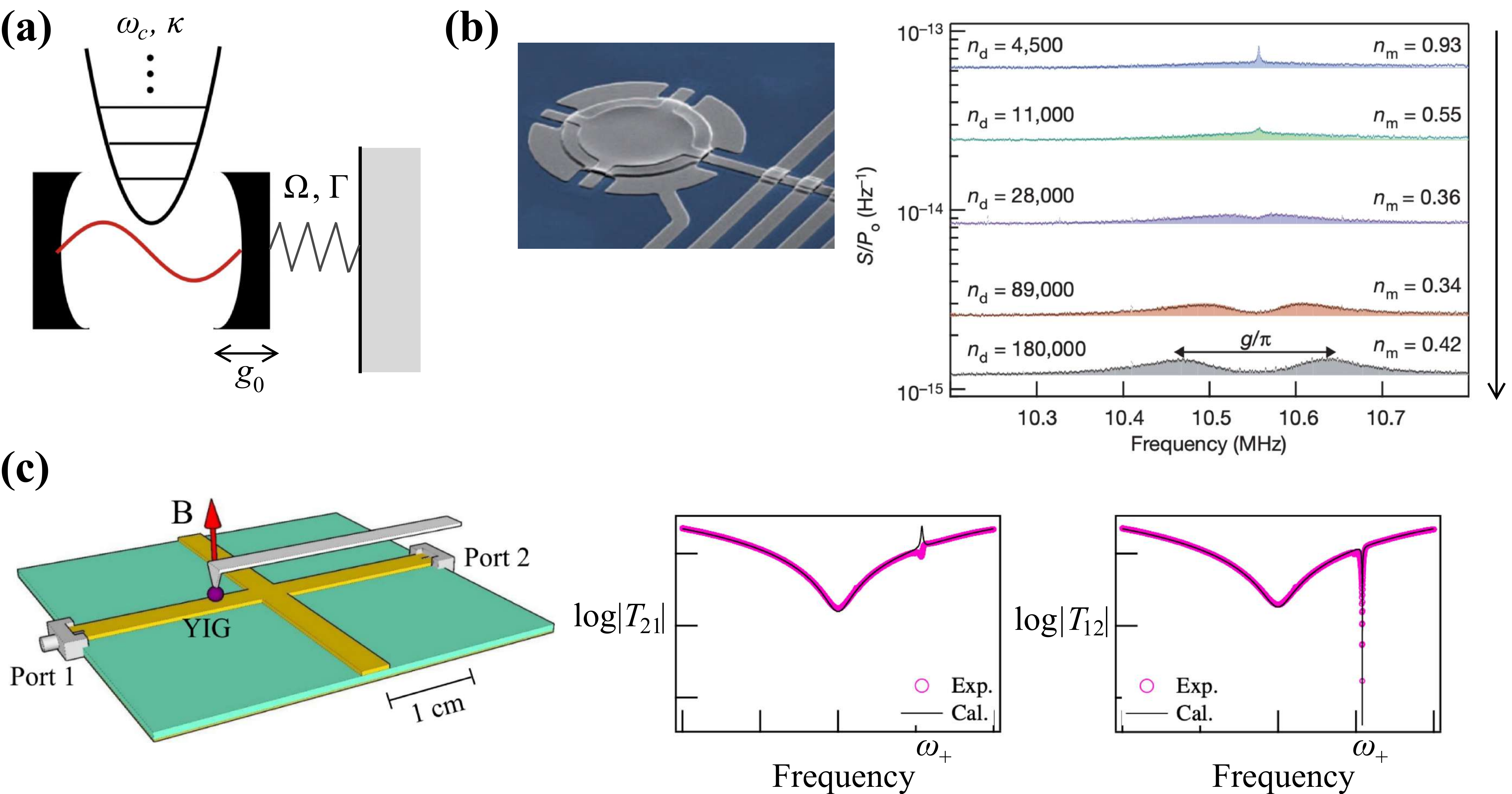}
\end{center}
\caption{(a) Schematic illustration of a prototypical physical setup in optomechanics. An optical cavity with frequency $\omega_{c}$ and decay rate $\kappa$ is coupled to a mechanical mode of the movable mirror with frequency $\Omega$ and damping rate $\Gamma$. The bare coupling strength is characterized by the parameter $g_0$ (see Eqs.~\eqref{sec3optmech1} and \eqref{sec3optmech2}). (b) (Left) A micromechanial membrane coupled to a superconducting microwave resonant circuit. (Right) Experimental results on the normalized sideband noise spectra. From top to bottom panels, the drive power of the cavity is increased and the optomechanical coupling $g$ is enhanced, leading to the emergence of two-resolved peaks in the strong-coupling regime (cf. Eq.~\eqref{lossyoptmechsol}).  Adapted from Ref.~\cite{TJD112}. Copyright \copyright\,   2011 by Springer Nature. (c) Schematic figure illustrating the experimental setup in cavity magnonics (left most) and spectra of transmission coefficients $T_{21}$ ($T_{12}$) from port 1 (2) to 2 (1) (right two panels). Strong asymmetry $T_{21}\neq T_{12}$ (i.e., nonriciprocity) is realized around $\omega=\omega_{+}$ at which the zero-damping condition $|T_{12}(\omega_{+})|=0$ is satisfied.  Adapted from Ref.~\cite{WYP19}. Copyright \copyright\,   2019 by the American Physical Society.  }
\label{fig:3optmech}
\end{figure}

The most prototypical optomechanical system consists of a high-$Q$ optical cavity that has a mechanically displaceable boundary \cite{AM14} (see Fig.~\ref{fig:3optmech}(a)). This system can be modeled as a movable cavity mirror whose excitations can be described based on a mechanical harmonic oscillator with mass $m$ and frequency $\Omega$. Because the radiation pressure of cavity photons acts directly on the mechanical degree of freedom, the optomechanical interaction depends linearly on the mechanical displacement. Specifically, the resulting interaction Hamiltonian is given by $V=-\hbar g_0 (b+b^\dagger)a^\dagger a$, where $b$ ($b^\dagger$) and $a$ ($a^\dagger$) are the  annihilation (creation) operator of phonons and photons, respectively, and $g_0$ is the strength of the optomechanical coupling.
We consider the semiclassical limit of this optomechanical system by assuming sufficiently large photon and phonon numbers such that quantum noise effects can be neglected. The dynamics can then be fully described by classical variables, including a coherent photon amplitude $\alpha(t)=\langle a(t)\rangle$ and  a coherent phonon amplitude $\beta(t)=\langle b(t)\rangle$, which is related to the mechanical displacement as $x(t)=2x_0{\rm Re}[\beta(t)]$ where $x_0\equiv\sqrt{\hbar/(2m\Omega)}$ is zero-point fluctuations.  
The equations of motion for these amplitudes are given by
\eqn{\label{sec3optmech1}
\frac{d\alpha}{dt}&=&-\frac{\kappa}{2}+i\left(\Delta+Gx\right)\alpha,\\
\frac{d\beta}{dt}&=&\left(-i\Omega-\frac{\Gamma}{2}\right)\beta+ig_0\left|\alpha\right|^2,\label{sec3optmech2}
}
where $\kappa$ is the decay rate of cavity photons, $\Delta$ is the detuning frequency of the driving laser from the cavity frequency $\omega_c$, $G=g_0/x_0$ is the optical frequency shift per unit displacement,  and $\Gamma$ is the mechanical damping rate. One can further simplify the equations when linear-order fluctuations around the steady state are taken into account: $\alpha(t)=\overline{\alpha}+\delta\alpha(t)$ and $\beta(t)=\overline{\beta}+\delta\beta(t)$, where $\overline{\alpha}$ and $\overline{\beta}=g_0|\overline{\alpha}|^2/\Omega$ are time-independent photon and phonon mean amplitudes, respectively. Without loss of generality, we assume $\overline{\alpha}\in{\mathbb R}$. Making the standard assumptions $\Delta\sim -\Omega$ (i.e., the rotating wave approximation) and $\Omega\gg\Gamma$, one can arrive at the linearized non-Hermitian equation:
\eqn{\label{lossyoptmech}
\omega
\begin{bmatrix}
\delta\tilde{\alpha}\\
\delta\tilde{\beta}
\end{bmatrix}=
\begin{bmatrix}
\;-\Delta-i\frac{\kappa}{2}\; & \;\;-g\;\; \\
\;-g\; & \;\;\Omega-i\frac{\Gamma}{2}\;\;
\end{bmatrix}
\begin{bmatrix}
\delta\tilde{\alpha}\\
\delta\tilde{\beta}
\end{bmatrix}\equiv \left(H-i\frac{\kappa+\Gamma}{4}I\right)\begin{bmatrix}
\delta\tilde{\alpha}\\
\delta\tilde{\beta}
\end{bmatrix},
}  
where we introduce the Fourier amplitudes, $\delta\tilde{\alpha}(\omega)=\int dt\delta{\alpha}(t)e^{i\omega t}$ and $\delta\tilde{\beta}(\omega)=\int dt\delta{\beta}(t)e^{i\omega t}$, and denote the intermode coupling as $g=g_0\,\overline{\alpha}$. The eigenvalues of the $2\times 2$ non-Hermitian matrix $H$ are given by
\eqn{\label{lossyoptmechsol}
\lambda_{\pm}=\Omega+\frac{\delta}{2}\pm\sqrt{g^2+\left[\frac{\delta+i(\Gamma-\kappa)/2}{2}\right]^2},
}
where $\delta=-\Delta-\Omega$.
We note that $H$ is pseudo-Hermitian if and only if $\delta=0$ (i.e., if the resonance condition is attained) or $\Gamma=\kappa$. In these cases, $\lambda_{\pm}$ are either real or the complex conjugate pair, and the exceptional point (EP) occurs at the spectral transition point. Since any $2\times 2$ pseudo-Hermitian matrix is PT symmetric with an appropriate parity operator (see discussions in {Sec.~\ref{secphqh}}), the present matrix $H$ with $\delta=0$ is also PT symmetric with respect to the parity operator $P=\sigma_x$. 
Physically, the spectral transition triggered by increasing coupling $g$ manifests itself as the splitting of two-well resolved peaks of the same linewidth corresponding to the strongly hybridized optomechanical modes \cite{MF07,DJM08}. Such strong coupling regimes were experimentally reached in a variety of optomechanical systems \cite{GS09,TJD112,TJD11,VE12} (see Fig.~\ref{fig:3optmech}(b)). 

Besides the non-Hermitian system in Eq.~\eqref{lossyoptmech} in effectively lossy (i.e., passive) regimes,  optomechanical systems with balanced gain and loss have also been realized. For instance, a phonon laser accompanying linewidth broadening at an EP \cite{ZJ18} and asymmetric power transfer by encircling an EP were demonstrated \cite{HX2016} (see {Sec.~\ref{secepphys}} for further discussions on the physics of EPs).  
Theoretically, it has been proposed that the enhanced optomechanical coupling can be achieved by coupling an active resonator to a lossy resonator having a mechanical mode, leading to efficient phonon lasing in the vicinity of the EP \cite{HJ14}. The subsequent study showed that the lasing amplitudes of phonons are ultimately saturated by nonlinear effects while oscillations can  still be sustained \cite{DWS16,KVK16}.
An inverted variant of the optomechanically induced transparency (i.e., a transparency dip between two sideband peaks)  has also been predicted \cite{JH15}. 
Coupling an active resonator to a nonlinear and lossy mechanical resonator was proposed to realize nonreciprocal phonon transport \cite{ZJ152}. A higher-order EP \cite{JH17},  low-threshold chaos induced by the dynamical enhancement of nonlinearity \cite{LXY15}, and the analogue of loss-induced lasing in optomechanical systems \cite{LH17}  have also been discussed. 

Finally, it is still largely unexplored how intriguing non-Hermitian phenomena can be transferred to an emerging research area of {\it optomagnonics} while the proof-of-concept experiment of detecting the EP has been reported \cite{ZD17}. 
In optomagnonics, the role of phonons in optomechanics is replaced by magnon excitations  whose dynamics can break the time-reversal symmetry, leading to the Brillouin light scattering \cite{OsA16,OsA18}. In particular, it has recently been observed that the dissipative magnon-photon coupling, whose non-Hermiticity explicitly breaks the time-reversal symmetry, can induce the nonreciprocal microwave transmission in the linear response regime and also trigger the unidirectional invisibility  \cite{WYP19} (see Fig.~\ref{fig:3optmech}(c)).

\subsection{Hydrodynamics\label{sechydro}}
The dynamics of collective motion can often be described by hydrodynamic equations involving a few classical fields. This can be justified if one concerns the length (time) scale that is much longer (slower) than that in dynamics of microscopic constituents. If collective excitations are subject to gain or loss, hydrodynamic equations can be nonconservative and thus non-Hermitian. Even without explicit gain or loss, the linearized hydrodynamic equations around certain steady states can be non-Hermitian due to dynamical (in)stabilities. For instance, historically hydrodynamic instabilities associated with temperature gradients have been studied in the context of fluid mechanics \cite{busse_1970}. Recent realizations of a new type of materials such as fluids with acoustic gain and loss, artificial metamaterials, active matter,  and driven-dissipative exciton-polariton condensates have shed new light on non-Hermitian physics in hydrodynamics as reviewed below.  

\subsubsection{Non-Hermitian acoustics in fluids, metamaterials, and active matter\label{sechydroaco}}
Acoustics is a research area that studies propagation of collective, longitudinal waves through fluids and solid-state materials. Its analogy to the one-body Schr{\"o}dinger equation can be constructed based on the linearized hydrodynamic equation, which describes the dynamics of small pressure fluctuations around their equilibrium values. To this end, we start from the hydrodynamic equation of motion 
\eqn{
\rho\frac{\partial{\bf v}}{\partial t}=-\nabla p,
}
where $\rho$ is the fluid mass density, $\bf v$ is the velocity field and $p$ is the acoustic pressure. The conservation of mass leads to the continuity equation
\eqn{
\frac{\partial p}{\partial t}+\kappa\nabla\cdot {\bf v}=0,
} 
where $\kappa$ characterizes the compressional stiffness of the fluid. Combining these two equations, we arrive at the linear wave equation of acoustic modes
\eqn{
\frac{\partial^2 p}{\partial t^2}=\frac{\kappa}{\rho}\nabla^2 p.
}
The acoustic wave velocity (i.e., the sound velocity) is given by 
$c=\sqrt{\kappa/\rho}$. Comparing this wave equation of sound modes to that of the electromagnetic fields, we notice that the mass density $\rho$ (the bulk modulus $\kappa$) of fluids plays the role of the dielectric constant $\epsilon$ (the inverse of the magnetic permeability $\mu^{-1}$) of electromagnetic materials. It is this analogy that allows one to control sound waves by engineering $\rho$ and $\kappa$ of fluids in a manner similar to what is done for $\epsilon$ and $\mu^{-1}$ in photonic devices as reviewed in Sec.~\ref{secphoto}.

In the conventional solid-state materials, both of $\rho$ and $\kappa$ are restricted to be positive as they are determined from the crystal structure and chemical composition of the media. 
In contrast, in metamaterials, one can realize both positive and negative $\rho$ and $\kappa$ \cite{ZS09,PCM11}, and even make them complex by introducing gain and loss \cite{FR15}.
This can be achieved by, for example, fabricating resonant subwavelength structures.
Along this way, many remarkable acoustic wave phenomena have been realized. 
When either one of the real parts of $\rho$ and $\kappa$ is negative, the phase velocity acquires the imaginary part and can lead to opaque media \cite{LZ00,FN06,YZ08,Lee_2009}. 
When the real parts of both $\rho$ and $\kappa$ are simultaneously negative,  energy can propagate in the direction opposite to that of the wave \cite{LZ12,XY13,LZ13}, which is reminiscent of electromagnetic wave propagation with a negative refractive index \cite{SRA01,SVM07} (cf. Fig.~\ref{fig:3optwave}(c)). With negative $\rho$ and $\kappa$, the reversed Doppler effects \cite{LSH10} and refraction of sounds at negative angles for a wide range of incoming direction \cite{CJ12,GCVM14} have also been observed.
Focusing of sound waves at superresolution (i.e., the spatial resolution below the diffraction limit) was also achieved in a similar setup with the flat lens made of the membrane-based metematerials \cite{PJJ15,KN15}.
 
\begin{figure}
\begin{center}
\includegraphics[width=14.5cm]{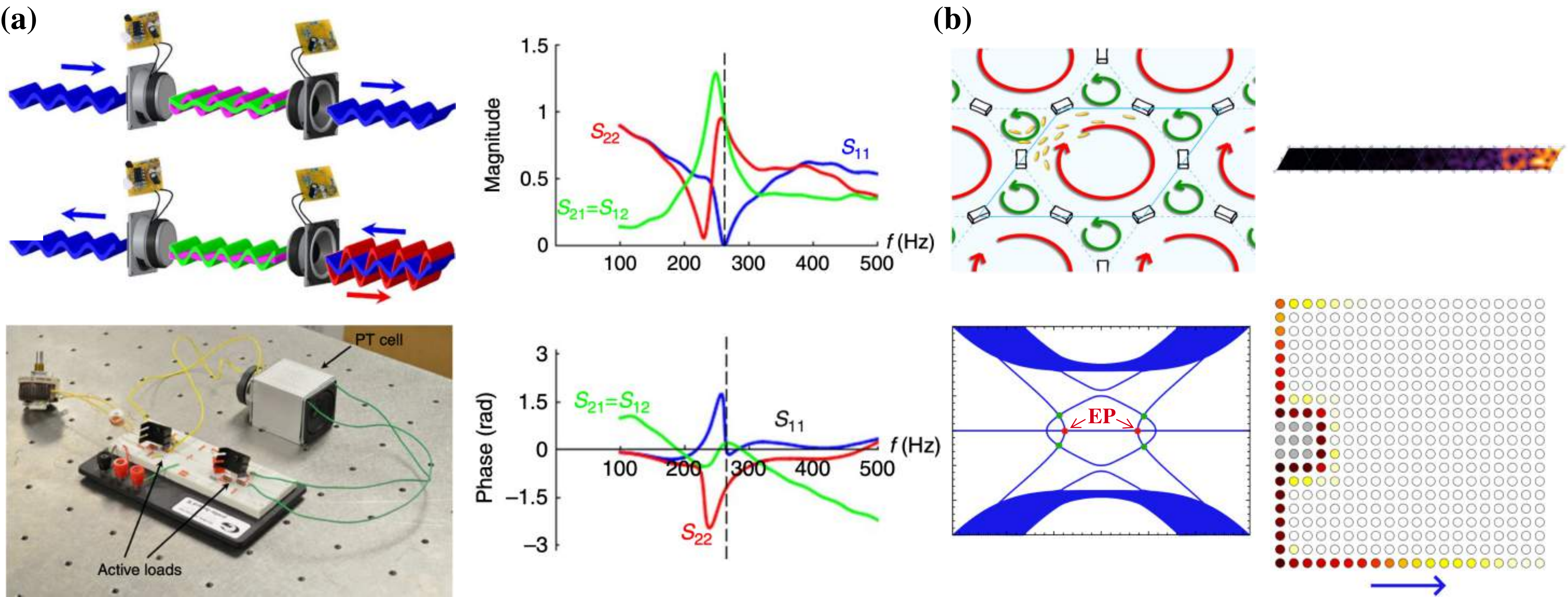}
\end{center}
\caption{ (a) (Top left) Schematic illustration of the PT-symmetric acoustic system realized by a pair of electromechanical resonators. The left (right) resonator is coupled to an absorptive circuit (an acoustic gain element), leading to the unidirectional invisibility of the sound propagation. (Bottom left) Experimental system consisting of acoustic resonators connected with non-Forster electrical circuits. (Right panels) Experimental results on spectra of the magnitudes and the phases of the reflection coefficients $S_{11,22}$ and the transmission coefficients $S_{12,21}$. The unidirectional invisibility is realized at the frequency for which $S_{11}=0$ while the reciprocity $S_{12}=S_{21}$ is always satisfied. Adapted from Ref.~\cite{FR15} licensed under a Creative Commons Attribution 4.0 International License. (b) (Top panels) Proposal for the active matter realization of the quantum anomalous Hall effect (left) and the calculated topological edge mode (right) \cite{SK19}. (Bottom panels) EP-induced gapless edge dispersion in chiral active matter beyond the existing topological classifications (left) and its lasing dynamics, realizing topological insulator laser (right) \cite{SK192}. }
\label{fig:3hydro}
\end{figure}
 
With gain and loss, the parameters $\rho$ and $\kappa$ can acquire significantly nonzero imaginary parts \cite{FR15,AY17}. This can be realized by, e.g., using  the response of piezoelectric materials as acoustic gain \cite{WM14}, where the gain originates from the Cherenkov radiation of phonons generated by electron drift whose velocity is faster than the sound velocity. 
Based on the balanced gain and loss, PT-symmetric acoustic systems have been proposed \cite{ZX14} and subsequently realized in experiments, enabling unconventional acoustic devices with asymmetric responses \cite{SDL15}, the unidirectional invisibility \cite{AY17,CJ16,SC16}, and the enhanced acoustic sensitivity at an EP \cite{FR15} (see Fig.~\ref{fig:3hydro}(a)).
Introducing asymmetric losses in coupled acoustic resonators, the higher-order EP has also been observed \cite{DK162}. 
The idea of creating the spatially homogeneous intensity in non-Hermitian disordered potentials \cite{MKG15,MKG17} has found applications in an acoustically transparent medium that is otherwise an opaque disordered material \cite{RE17,RE18}. Topological edge states associated with the exceptional points of the scattering matrix have also been observed in a lossy acoustic system inspired by the commensurate Aubry-Andr\'e-Harper model \cite{ZW18}.
\\ \\ {\it Active matter}

\vspace{3pt}
\noindent
Another emerging field related to non-Hermitian acoustics is {\it active matter} that studies collective dynamics of self-propelled constituents such as animal flocks \cite{Vicsek1995,Vicsek2012}, biological systems \cite{Brugues2014,Saw2017,Kawaguchi2017,Nishiguchi2018}, catalytic colloids \cite{Palacci2013}, and Janus particles \cite{Jiang2010}. Because of  self-propulsions, the system organizes itself and can reach a nonequilibrium steady state accompanying nonzero spontaneous flows without an external drive or pump.
The collective excitations of active matter can be described by the Toner-Tu equations \cite{Toner1995,Toner1998,Toner2005,Marchetti2013},
\eqn{\label{toner-tu-eq1}
&&\partial_t \rho + \nabla \cdot (\rho \mathbf{v}) = 0,\\
&&\partial_t \mathbf{v} + \lambda (\mathbf{v} \cdot \nabla) \mathbf{v} + \lambda_2 (\nabla \cdot \mathbf{v}) \mathbf{v} + \lambda_3  \nabla |\mathbf{v}|^2 \nonumber\\
&= &(\alpha-\beta|\mathbf{v}|^2)\mathbf{v}-\nabla P +D_B \nabla (\nabla \cdot \mathbf{v}) + D_T \nabla^2 \mathbf{v} + D_2(\mathbf{v} \cdot \nabla)^2 \mathbf{v},
\label{toner-tu-eq2}
}
where $\rho({\bf r},t)$ is the density field of active matter and $\mathbf{v}({\bf r},t)$ is the local average of velocity of self-propelled particles. We here consider the case of two-spatial dimension ${\bf r}=(x,y)^{\rm T}$, which is appropriate for many experimental systems. The first equation~\eqref{toner-tu-eq1} is the continuity equation while the second one~\eqref{toner-tu-eq2} corresponds to the equation of motion, in which  the first term on the right-hand side suggests a preference for a nonzero constant speed $|\mathbf{v}|=\sqrt{\alpha/\beta}$ if $\alpha$ is positive while negative $\alpha$ results in the nonordered state $|\mathbf{v}| = 0$. The terms including $\lambda,\lambda_{2,3}$ indicate the violation of the Galilean invariance due to the self-propulsions and are absent in the (nonactive) conventional fluid. The terms with $D_{B,T,2}$ represent the diffusive effects that are most relevant to the long-distance collective behavior. The coefficients in these equations can be related to  microscopic models \cite{Bertin2006,Peshkov2014,Farrell2012,Suzuki2015}.  The pressure $P$ is assumed to be proportional to $\rho$ as appropriate for an ideal gas. 

The analogy to the Schr{\"o}dinger-like equation can be achieved by linearizing the Toner-Tu equations around a steady-state solution $\rho_{\rm ss}({\bf r})$ and ${\bf v}_{\rm ss}({\bf r})$ \cite{Souslov2018,SK19}. We introduce the sound velocity $c$ as the coefficient in the equation of state $P=c^2\rho/\rho_{\rm ss}$ and the dimensionless fluctuations of density and velocity fields as $\delta\rho({\bf r},t)=[\rho({\bf r},t)-\rho_{\rm ss}({\bf r})]/\rho_{\rm ss}$ and $\delta{\bf v}({\bf r},t)=[{\bf v}({\bf r},t)-{\bf v}_{\rm ss}({\bf r})]/c$. Denoting $\psi=(\delta\rho,\delta v_{x},\delta v_{y})^{\rm T}$ as a vector and assuming $|{\bf v}_{\rm ss}|/c\ll 1$, the linearized equation takes the following form \cite{SK19}:
\eqn{
i\frac{\partial}{\partial t}\psi={ H}\psi,
} 
where the effective Hamiltonian consists of three parts ${ H}=H_0+iH_D+H_\lambda$. The first term is a Hermitian operator $H_0=H_0^\dagger$ and, with the dimensionless steady velocity ${\bf v}_0({\bf r})={\bf v}_{\rm ss}({\bf r})/c$, it can be given by
\eqn{
H_0=
\begin{bmatrix}
\;-i{\bf v}_{0}\cdot\nabla\; & \;\; -i\partial_x\;\; & \;\;-i\partial_y\;\; \\
\;-i\partial_x\; & \;\;-i\lambda{\bf v}_{0}\cdot\nabla\;\; & \;\;0\;\; \\
\;-i\partial_y\; & \;\;0\;\; & \;\;-i\lambda{\bf v}_{0}\cdot\nabla\;\;
\end{bmatrix}.
} 
The diffusive terms lead to the second term $iH_D$ that is anti-Hermitian  $H_D=H_D^\dagger$:
\eqn{
iH_D=i
\begin{bmatrix}
\;\;0\;\; & \;\;0\;\; & \;\;0\;\; \\
\;\;0\;\; & \;\;(D_B+D_T)\partial_x^2\;\; & \;\;D_B\partial_x\partial_y\;\; \\
\;\;0\;\; & \;\;D_B\partial_y\partial_x\;\; & \;\;(D_B+D_T)\partial_y^2\;\;
\end{bmatrix}.
}
We note that the term including $D_2$ can be neglected since it is a higher-order contribution in terms of $|{\bf v}_{\rm ss}|/c$.
Finally, the last term $H_\lambda$ originating from the $\lambda_{2,3}$ terms is intrinsically non-Hermitian and given by
\eqn{
H_\lambda=-i
\begin{bmatrix}
\;\;0\;\; & \;\;0\;\; & \;\;0\;\; \\
\;\;0\;\; & \;\;\lambda_2v_0^x\partial_x+\lambda_3(\partial_x v_0^x+v_0^x\partial_x)\;\; & \;\;\lambda_2v_0^x\partial_y+\lambda_3(\partial_xv_0^y+v_0^y\partial_x)\;\; \\
\;\;0\;\; & \;\;\lambda_2v_0^y\partial_x+\lambda_3(\partial_yv_0^x+v_0^x\partial_y)\;\; & \;\;\lambda_2v_0^y\partial_y+\lambda_3(\partial_yv_0^y+v_0^y\partial_y)\;\;
\end{bmatrix}.
}
Using an analogy to quantum mechanics in this simplest setup of active matter,  topological edge modes reminiscent of the quantum anomalous Hall effect have been predicted \cite{SK19} (see the top panels in Fig.~\ref{fig:3hydro}(b)). It has been pointed out \cite{SK192} that a different variant of active matter, namely, chiral active matter, can exhibit the unique gapless edge mode induced by the EP even when the bulk is topologically trivial (see bottom panels in Fig.~\ref{fig:3hydro}(b)). This new type of a gapless edge mode  violates the bulk-edge correspondence and cannot be categorized by the existing classification tables  of non-Hermitian topological phases (see Sec.~\ref{sec5} for further descriptions). Yet, non-Hermitian physics in active matter is still largely unexplored in both theory and experiment except recent studies \cite{SK192,SC20,SC202}; in particular, the interplay between non-Hermiticity and its unique nonlinear feature has yet to be clarified.

\subsubsection{Exciton polaritons and plasmonics\label{sechydroexc}}
When photons are strongly coupled to a collective excitation of solid-state materials, they form hybridized quasiparticles known as polaritons due originally to Hopfield \cite{HJJ58}. Collective excitations in materials can be  phonons or plasmons, but when they are excitons, i.e., quasiparticles  bound states of electron-hole pairs in semiconductors, the polariton mode is called as {\it exciton polaritons}. 
Experimental realizations of the strong light-matter interactions have enabled one to create exciton polaritons in quantum well structures of microcavities \cite{YY02,KA07}. Exciton polaritons can condense and act as a macroscopic coherent wave \cite{JK06,BT14} while the condensate is intrinsically out of equilibrium since a continuous optical pump is necessary to compensate for the inevitable loss of quasiparticles. 
For instance, a polariton in microcavities typically have a lifetime that is of the  order of several tens of picoseconds. This driven-dissipative system can also feature strong nonlinearity due to inherent quasiparticle interactions, and can provide an alternative platform to study rich physics of non-Hermitian systems in both linear and nonlinear regimes. Below we focus on the mean-field regime, in which exciton polaritons behave as a scalar complex-valued classical field and its dynamics can be described by the Gross-Pitaevskii equation \cite{Gross1961,Pitaevskii61} generalized to including gain and loss.

\begin{figure}
\begin{center}
\includegraphics[width=14.5cm]{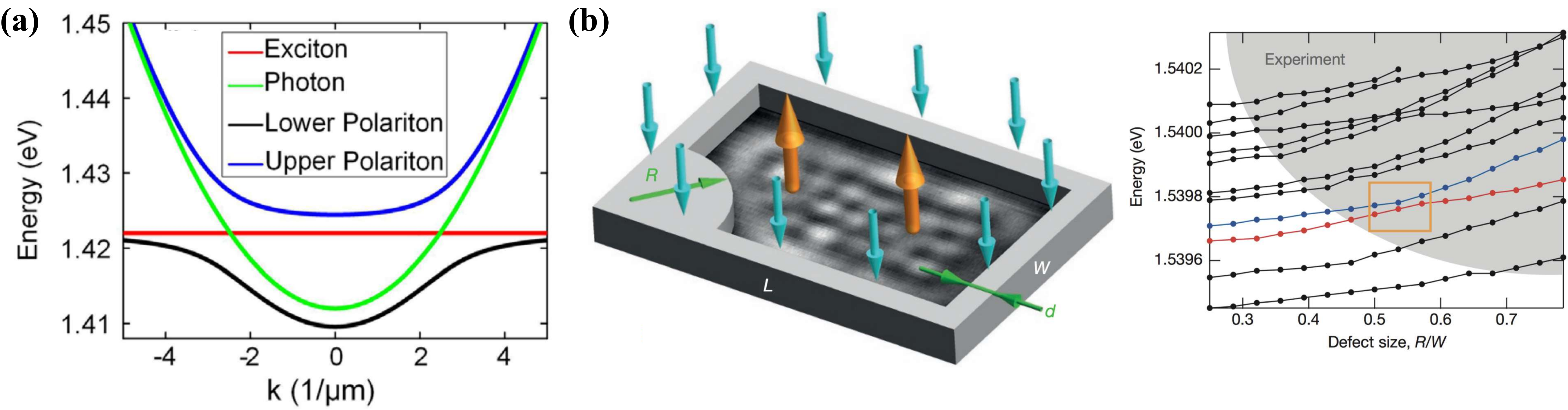}
\end{center}
\caption{(a) Energy dispersions of exciton (red), photon (green), lower  polaritons (black), and upper polaritons (blue curve).   Adapted from Ref.~\cite{Schneider_2016}. Copyright \copyright\,   2016 by IOP Publishing.
(b) Schematic illustration of the experimental setup of the non-Hermitian Sinai's billiard potential in exciton-polariton condensates (left) and experimental results on the measured energy eigenvalues (i.e., the real parts of the complex eigenvalues).
Adapted from Ref.~\cite{GT15}. 
Copyright \copyright\,   2015 by Springer Nature.
}
\label{fig:3excpol}
\end{figure}

Suppose that microcavity photons are coupled to excitons in a quantum well.  Because of the spatial confinement due to the microcavity, the photon excitation in the $z$ direction perpendicular to the system is quantized, resulting in the quadratic in-plane dispersion, $\omega_c+\hbar{\bf k}^2/(2m_c)$, where $\bf k$ is  the two-dimensional wavevector in the $xy$ plane, $\omega_c$ is the cutoff frequency and $m_c$ is the effective mass of photons \cite{DH10,CI13}. The dispersion of excitons can be ignored as its mass is of the order of the electron mass that is several orders of magnitude larger than $m_c$; we thus simply denote the exciton frequency as $\omega_e$. The coupling between microcavity photons and excitons can be induced by their electromagnetic dipole interaction with the Rabi frequency $\Omega_r$. In addition, photons and excitons are subject to decays with loss rates $\gamma_c$ and $\gamma_e$, respectively. The resulting hybridized modes can be obtained from the eigenstates of the following $2\times 2$ non-Hermitian matrix \cite{VS95,KG99}:
\eqn{\label{expol}
\omega
\begin{bmatrix}
\Psi_{c,{\bf k}}\\
\Psi_{e}
\end{bmatrix}=
\begin{bmatrix}
\;\omega_{c}+\frac{\hbar {\bf k}^2}{2m_{c}}-\frac{i\gamma_{c}}{2}\; & \;\;\Omega_{r}\;\; \\
\;\Omega_r\; & \;\;\omega_e-\frac{i\gamma_e}{2}\;\;
\end{bmatrix}
\begin{bmatrix}
\Psi_{c,{\bf k}}\\
\Psi_{e}
\end{bmatrix},
}
where $\Psi_{c,{\bf k}}$ is the mean-field amplitude of cavity photons with in-plane momentum $\bf k$, and $\Psi_e$ is the mean-field amplitude of excitons. This eigenvalue problem is mathematically equivalent to that of Eq.~\eqref{lossyoptmech}. Because the coupling $\Omega_r$ is typically much larger than the other energy scales, there exist two  dispersions with the gap $\sim2\hbar\Omega_r$ corresponding to two different eigenvalues of Eq.~\eqref{expol} (see also Eq.~\eqref{lossyoptmechsol}) (see Fig.~\ref{fig:3excpol}(a)). Polaritons  can be condensed to low momentum states of the lower dispersion, which are called as the lower polaritons (LP). The energy dispersion of the LP can be approximated by
\eqn{
\omega_{{\rm LP},{\bf k}}\simeq\omega_{0}+\frac{\hbar^2{\bf k}^2}{2m}-\frac{i\gamma}{2},
}
where $\omega_0$ is the zero-momentum frequency,  $m$ is the effective mass of the LP and $\gamma=(\gamma_c+\gamma_e)/2$ is its loss rate. 

To compensate for the inherent  loss and sustain the condensation, an external drive is necessary. In many experiments, this has been achieved by incoherent optical pumping that does not retain any coherence to the polariton field. 
The dynamics can then be modeled by the coupling between the coherent polariton field and the incoherent reservoir amplified by the pump field $P({\bf r},t)$.
Because the relaxation rate of the reservoir $\gamma_r$ is typically much larger than the other energy scales, one can adiabatically eliminate the reservoir and arrive at the time-evolution equation including only the coherent polariton field,
\eqn{\label{GGP}
i\frac{\partial}{\partial t}\Psi({\bf r},t)=\left[\omega_{0}-\frac{\hbar\nabla^2}{2m}+V({\bf r})-\frac{i\gamma}{2}+\frac{iP({\bf r},t)}{2}\frac{R}{\gamma_r+R|\Psi|^2}+u|\Psi|^2\right]\Psi({\bf r},t),
}
where $V$ is a potential created by a two-dimensional environment of a quantum well, $R$ is the amplification rate, and $u$ describes the strength of the nonlinear interaction.
This type of equation was originally discussed by Gross \cite{Gross1961} and Pitaevskii \cite{Pitaevskii61} (without gain or loss terms) and applied to laser theory \cite{LWE64}  and  exciton-polariton condensates \cite{WM07}.
In the linear regime with negligible $u|\Psi|^2$ and $R|\Psi|^2$ terms, Eq.~\eqref{GGP} reduces to the non-Hermitian one-body Schr{\"o}dinger equation with the complex potential whose negative and positive imaginary parts are characterized by loss $\gamma$ and gain $P$, respectively.

Recent experimental advances have allowed one to control both real and imaginary parts of the potential via fabricating quantum well structures and also engineering the optical pump field. This ability has found applications to demonstrate the loading of exciton polaritons into a flat band of the  Lieb lattice with gain and loss \cite{BF162}, polariton lasing in the complex potential \cite{ZL15}, and an observation of eigenvalues and topological structures around EPs in chaotic billiard \cite{GT15} (see Fig.~\ref{fig:3excpol}(b) and also Fig.~\ref{fig:2encircle}). 
With the balanced gain and loss, the realization of  the PT-symmetric exciton-polariton condensate in the coupled micropillars has been theoretically proposed \cite{LJY15}. The permanent Rabi oscillation of the PT-symmetric exciton-polariton condensate \cite{CIY16} as well as an emergent EP in the mean-field spectrum \cite{HR192} have also been proposed. The Kardar-Parisi-Zhang scaling in fluctuating finite-temperature dynamics in the one-dimensional case has been predicted in numerical simulations of the Gross-Pitaevskii equation \cite{KM13}. 
 
Quantum-Hall-like topological phenomena have also been predicted in two-dimensional exciton-polariton systems, which can be called \emph{topological exciton-polariton insulators} \cite{RG15,NAV15}. The idea is that the exciton polariton enjoys both properties of photons and electrons, so one can introduce a large spin-orbit coupling, which is usually too small in solid-state materials to support a topological band, on the basis of TE-TM mode splitting. A topological gap can then be opened by a Zeeman splitting induced by a magnetic field. Note that the TE-TM mode splitting alone can lead to the optical spin-Hall effect \cite{KA05}, which is the polaritonic analogue of the electronic spin-Hall effect \cite{HJE99,KYK04} and has been experimentally realized \cite{CL07}. Although the lifetime of an exciton polariton is always finite due to the non-Hermitian nature of the system, this would not significantly alter the topological property as long as the inverse lifetime is much smaller than the topological gap \cite{RG15,NAV15}. Such an idea has recently been implemented with a honeycomb lattice of coupled semiconductor microcavities illuminated by an off-resonant laser, where the chiral motion of edge modes has been observed \cite{SK18}.
\\ \\ {\it Plasmonics}

\vspace{3pt}
\noindent
Another important field concerning the light-matter interaction is {\it plasmonics}, where light is confined into subwavelength-scale structures \cite{TMS13}. Instead of excitons,  
here photons are coupled to plasmons, i.e., collective excitations in metals. Because of inherent dissipation in metals caused by Joule heating, plasmonic systems can potentially serve as another platform for studying non-Hermitian physics. 
The PT-symmetric plasmonic systems have been theoretically proposed by considering the compensation of  the loss by injecting plasmons with stimulated emission of radiation \cite{BDJ03} or by exchanging energies with a gain medium \cite{Benisty:11,Lupu:13}. 
Multi-layered structures with the alternating gain and loss were also suggested to realize unconventional wave properties such as  unidirectional invisibility \cite{AH14,AH142}. 
A more general analysis on the PT-symmetric subwavelength waveguides beyond the paraxial approximation has been performed \cite{Huang:14}. Self-hybridization caused by nonorthogonality between different eigenmodes \cite{HLM18} and exceptional points with \cite{TA19} or without \cite{JHP19} PT symmetry have been observed in plasmonic systems.

\section{Non-Hermitian quantum physics\label{sec4}}
When a quantum system is coupled to a measuring apparatus or a surrounding environment, physics of the original quantum system can strongly be modified, potentially leading to qualitatively new and rich phenomena unseen in isolated, Hermitian systems. Such open quantum systems have  also been of practical importance in view of recent developments  in quantum technologies.  To study this class of quantum systems, several useful descriptions utilizing non-Hermitian operators have been developed, whose validity can be justified under certain conditions to be clarified below. Historically, studies of open quantum systems based on effectively non-Hermitian Hamiltonians date back to the works by Gamow \cite{GG28} and Feshbach \cite{HF58,HF62} in the context of nuclear physics. A theoretical approach developed along this line, known as the Feshbach projection approach (or often as the Cohen-Tannoudji projection technique \cite{CCT68}), has later found its application in mesoscopic physics \cite{AY00}, atomic and molecular physics \cite{SVV08}, and condensed matter physics \cite{DS95,IR09}. 
A common feature of these systems is that a quantum system is 
typically embedded in a macroscopic environment that involves a continuum of delocalized states. On another front, partly motivated by the need of analyzing microscopic quantum systems under continuous measurement, a different theoretical approach to open quantum systems has been developed in quantum optics \cite{MU89,UM90,DJ92,DR92,GCH92,HC93}. Recently, this approach, often referred to as the quantum trajectory approach, has attracted renewed interest especially in the context of quantum many-body physics. 
Together with rapid experimental advances in AMO physics, recent studies have revealed a number of unconventional many-body phenomena. Here we firstly review the main ideas of each of the theoretical approaches above, and then discuss their applications to a variety of physical systems with particular focus on the relevance to non-Hermitian physics.

\subsection{Feshbach projection approach\label{FP}}
The Feshbach projection approach has originally been introduced to study  reactions of heavy nuclei containing dense excitation levels that have long lifetimes \cite{HF58,HF62}. In this case, a quantum system of interest consists of long-lived nuclear states while the environment corresponds to scattering states outside the nucleus. Another paradigmatic examples, for which this approach can be useful, include quantum resonances in mesoscopic systems, and spectral properties of decaying quasiparticles in solids.  As reviewed below, the former has often been discussed in the context of its connection with superradiance \cite{DRH54}. The latter can allow one to interpret several existing features in materials from a simplified non-Hermitian perspective \cite{DS95,IR09}. 

\subsubsection{Non-Hermitian operator}
\label{Sec:NHO}
A salient feature of the Feshbach projection approach is the emergence of a non-Hermitian operator that characterizes an effective time evolution of a subsystem embedded in an environment. The key idea is to project the dynamics of the whole system onto a part of the Hilbert space corresponding to a subsystem, and to derive its effective internal dynamics in a self-consistent manner. The non-Hermiticity in the effective dynamics originates from the unstable nature of some eigenstates in the subsystem due to their couplings to (typically continuum) environmental modes. Eigenstates of the effective non-Hermitian operator can characterize the presence of resonant states, where the imaginary parts of their complex eigenvalues represent the corresponding decay rates. 

We start from the stationary Schr{\"o}dinger equation,
\eqn{
H|\Psi\rangle & = & E|\Psi\rangle,
}
where $H$ is a Hamiltonian acting on the total Hilbert space comprised of system and environmental degrees of freedom,  $E\in\mathbb{R}$ is an energy eigenvalue, and $|\Psi\rangle$ is the corresponding eigenstate. We are interested in effective dynamics in a certain subspace containing $N$ states in the whole Hilbert space. We thus introduce a projector $P$ on this subspace and denote $Q=1-P$ as the projector on the rest of the states.  We have
\eqn{\label{fpanh1}
H_{{\rm eff}}(E)P|\Psi\rangle & = & EP|\Psi\rangle,\\
H_{{\rm eff}}(E) & = & H_{PP}+H_{PQ}\frac{1}{E-H_{QQ}}H_{QP},\label{fpanh}
}
where $H_{XY}$ contains all the matrix elements between the states that belong to the subspaces  $X$ and $Y$. We consider a situation in which $Q$ corresponds to a macroscopic environment and contains a continuum spectrum associated with, e.g., scattering states. We thus have to specify the boundary conditions for dealing with singularities in Eq.~\eqref{fpanh} occurring when the continuum spectrum of $H_{QQ}$ includes an eigenvalue at $E$.  To obtain a solution of physical interest, i.e., a mode decaying in time, we replace $E$ by $E+i\eta$ with $\eta\to+0$ being a vanishingly small positive parameter. We also use the relation
\eqn{
\lim_{\eta\to+0}\frac{1}{E+i\eta-x} & = & {\rm P}\frac{1}{E-x}-i\pi\delta(E-x),
}
where ${\rm P}$ denotes the principal value, and make the spectral decomposition in $Q$
\eqn{
H_{QQ} & = & \sum_{\alpha}\int d\lambda\;\lambda|\phi_{\lambda\alpha}\rangle_{QQ}\langle\phi_{\lambda\alpha}|
}
with $\alpha=1,2,\ldots,C$ labeling a channel of environmental modes. We then arrive at an effective non-Hermitian Hamiltonian
\eqn{\label{fpanonher}
H_{{\rm eff}}(E) & = & H_{PP}+\Delta(E)-\frac{i}{2}\Gamma(E),
}
where $\Delta(E)$ and $\Gamma(E)$ represent an effective energy shift and decay rate, respectively,
\eqn{
\Delta(E) & = & {\rm P}\int d\lambda\frac{A^\dagger(\lambda)A(\lambda)}{E-\lambda},\\
\Gamma(E) & = & 2\pi A^\dagger(E)A(E).\label{fpadec}
}
Here we introduce an operator $A(\lambda)$ characterizing the system-environmental coupling, whose  $C\times N$ matrix elements at energy $\lambda$ are given by
\eqn{
[A(\lambda)]_{\alpha i} & = & _{Q}\langle\phi_{\lambda\alpha}|H_{QP}|\psi_{i}\rangle_P,
}
where $\{|\psi_i\rangle_P\}$ is the orthonormal basis set in the subspace $P$ with $i=1,2,\ldots,N$.  
We note that if a channel does not contain an environmental mode at energy $\lambda$, the corresponding matrix element of $A(\lambda)$ is set to be zero;  the summation in the matrix multiplication of $A^\dagger(E)A(E)$ in Eq.~\eqref{fpadec} is taken only over `open' channels that have environmental modes at a given energy $E$. The factorized structure of $\Gamma(E)$ originates from the unitarity of the dynamics in the entire Hilbert space \cite{DL76}.

The  Green's function for the subsystem can be defined in terms of the effective non-Hermitian operator as \cite{DS95}
\eqn{\label{fpagreen}
G(E) & = & \frac{1}{E-H_{{\rm eff}}(E)}.
}
The scattering matrix is given by
\eqn{
S(E) & = & 1-\frac{i}{\pi}A(E)G(E)A^\dagger(E)=\frac{1-iG_{0}(E)/2}{1+iG_{0}(E)/2},\label{fpasca}
}
where we introduce the propagator for a Hermitian part of the effective operator
\eqn{
G_{0}(E) & = & \frac{1}{E-(H_{PP}+\Delta(E))}.
}
From the last expression in Eq.~\eqref{fpasca}, it is evident that the scattering matrix satisfies the unitarity condition
\eqn{
S(E)S(E)^\dagger=1.
}
The Green's function~\eqref{fpagreen} can be used to analyze resonant modes, which have finite lifetimes, when its argument $E$ is extended to the complex plane. The versatility of this Feshbach projection approach has long been recognized in the studies on   nuclear and atomic physics as well as mesoscopic physics (see discussions below). This approach has also found some applications for understanding quasiparticle spectra in solids  \cite{DS95,IR09,GV11,VK17,SH182,YoT18,ZHG19,MY20}.

\subsubsection{Quantum resonances}\label{Sec:QR}
In physical systems, a pole of the Green function $G(\Lambda)$ at a complex eigenvalue $\Lambda\in{\mathbb C}$ can be detected as a quantum resonant state. When the decay rate $\Gamma$ is energy-independent,  a resonant mode (which we denote as $|\tilde{\Psi}_{r}\rangle$) can be characterized as a mode obeying \cite{SB78}
\eqn{
|\langle\tilde{\Psi}_{r}|e^{-iHt}|\tilde{\Psi}_{r}\rangle|^{2} & = & e^{-\Gamma t},\;\;\;(t>0).\label{fpares2}
}
In this simplified case of a constant $\Gamma$, there is an elegant way to analyze quantum resonances based on complex deformations of the original internal Hamiltonian\footnote{This Hamiltonian should be interpreted as $H_{PP}$ in the notation of the previous subsection; for the sake of simplicity, we here denote it as $H$.} $H$ \cite{NM98}. Because $H$ is Hermitian and possesses real eigenvalues for all the square-integrable wavefunctions, it is evident from Eq.~\eqref{fpares2} that a resonant mode must live outside of the Hilbert space\footnote{To emphasize this point, we introduce the tilde to denote a resonant mode in Eq.~\eqref{fpares2} and distinguish it from a vector in the Hilbert space.}. More specifically, a resonant mode is a nonsquare-integrable function that involves a collection of continuum delocalized modes. 
The main idea of the complex-deformation method is to utilize a similarity transformation ${\cal H}=SHS^{-1}$ and relate a resonant mode $|\tilde{\Psi}_r\rangle$ to a square-integrable eigenfunction $|\Psi_{ r}\rangle$ of the complex-scaled Hamiltonian $\cal H$:
\eqn{
|\Psi_r\rangle&=&S|\tilde{\Psi}_r\rangle,\\
{\cal H}|\Psi_{r}\rangle & = & \left(E_{r}-\frac{i}{2}\Gamma\right)|\Psi_{r}\rangle,\;\;\;E_r,\Gamma\in{\mathbb R}.
}

To gain an insight into this prescription, we consider a simple case of a one-dimensional scattering problem on the semi-infinite space $x\geq0$. In the original physical frame before the similarity transformation, a resonant wavefunction $\tilde{\Psi}_r(x)$ is not square-integrable and has an asymptote
\eqn{
\tilde{\Psi}_{r}(x\to\infty) & \sim e^{iax+bx} & \to\infty,\;a\in\mathbb{R},\;b\geq0.
}
We note that when we consider the time evolution of a resonant mode, it behaves as $\tilde{\Psi}_{r}(x,t)=\tilde{\Psi}_{r}(x)e^{-iE_{r}t/\hbar-\Gamma t/(2\hbar)}$ that still ultimately decays to zero in time for a given $x$. To regulate the function  $\tilde{\Psi}_r(x)$, we can introduce the similarity transformation
\eqn{
S & = & e^{i\theta x\frac{\partial}{\partial x}},\;\;\;\theta\in{\mathbb R},
}
and obtain
\eqn{
\Psi_{r}(x)\equiv S\tilde{\Psi}_{r}(x) & = & \tilde{\Psi}_{r}(xe^{i\theta}).
}
When $\theta$ is set to be a sufficiently large value, the wavefunction $\Psi_r(x)$ in the transformed frame acquires an exponentially convergent factor and one can make it square-integrable. Therefore, the original problem of finding a resonant mode for $H$ reduces to the standard eigenvalue problem for the non-Hermitian operator $\cal H$ with square-integrable wavefunctions. Interestingly, the similarity transformation to a non-Hermitian problem has recently found an application also to the problem of the Jastrow factorization of the many-body wavefunction \cite{CAJ19}.   
\exmp{(Quantum particle on an inverted harmonic potential). As a simple example, consider a single particle subject to an inverted harmonic potential $V(x)=-m\omega^2 x^2/2$. The complex-scaled Hamiltonian is given by
\eqn{
{\cal H}=SHS^{-1} & = & -e^{-2i\theta}\frac{\hbar^{2}}{2m}\frac{\partial^{2}}{\partial x^{2}}-e^{2i\theta}\frac{m\omega^{2}}{2}x^{2}.
}
This operator permits square-integrable eigenfunctions associated with purely imaginary eigenvalues labeled by an integer $n=1,2,\ldots$ as follows:
\eqn{
E_{r}=0, & \;\Gamma_{n}=2\hbar\omega\left(n+\frac{1}{2}\right).
}
In the original frame, this fact shows the presence of discrete resonant modes having decay rates $\Gamma_n$. Further examples of the complex-deformation techniques can be found in an excellent review paper \cite{NM98}.
}
\noindent
{\it Black-hole perturbation}

\vspace{3pt}
\noindent
The physics of quantum resonances above has an intriguing connection with   the problem in a very different realm of physics, namely, black-hole physics \cite{CS83}. In general relativity, a small perturbation of the gravitational field around a black hole can be analyzed by using the following linearized differential equation \cite{CS83}:
\eqn{
\frac{d^{2}\Psi}{dr_{*}^{2}}+(\omega^{2}-V(r_{*}))\Psi & = & 0,\label{fpabh}
}
where $\Psi$ represents an oscillation amplitude, $V$ is a function that can be related to the cosmological constant experienced by a gravitational wave and satisfies $V(|x|\to\infty)=0$, and $r_*$ is the so-called tortoise coordinate. To obtain physically allowed solutions, we must impose appropriate boundary conditions at both the event horizon ($r_*\to-\infty$) and  the spatial infinity ($r_*\to\infty$).  Because nothing could leave the event horizon once it comes in (at least at the classical level), the amplitude must behave as an outgoing solution (or, said differently, an incoming solution into the event horizon)
\eqn{
\Psi(r_{*},t)\sim e^{-i\omega(t+r_{*})} & ,\;\;\;r_{*}\to-\infty.
}
Similarly, at the spatial infinity, it should again behave as an outgoing mode due to the vanishing potential term $V\to 0$
\eqn{
\Psi(r_{*},t)\sim e^{-i\omega(t-r_{*})}, &\;\;\;r_{*}\to\infty.
}
We note that a function satisfying these boundary conditions is not square-integrable and thus lives outside of the Hilbert space. Consequently, such a `radiative' mode can have a complex eigenvalue $\omega\in\mathbb{C}$ with a finite lifetime. This type of resonant modes are known as {\it quasinormal} modes in the context of black-hole physics.  

\exmp{(P{\"o}schl-Teller potential). One of the most  paradigmatic examples supporting quasinormal modes is the so-called P{\"o}schl-Teller model \cite{PG33}:
\eqn{
V(r_{*}) & = & \frac{V_{0}}{\cosh^{2}(kr_{*})},\;\;\;k>0.
}
The exact solutions of quasinormal modes associated with Eq.~\eqref{fpabh} are given in Ref.~\cite{FV84}; they have complex energies
\eqn{
\omega_{n} & = & \pm\sqrt{V_{0}-k^{2}/4}-ik\frac{2n+1}{2},\;\;\;n=0,1,2,\ldots.
} 
Further illustrative examples can be found in Ref.~\cite{Berti_2009}. 
}

\subsubsection{Superradiance}\label{Sec:spr}
One of the primary applications of the Feshbach projection approach is a superradiance phenomenon in mesoscopic and optical systems. The most important feature of this phenomenon is an enhanced emission rate of collective modes compared to a decay rate of an individual mode. Historically, the notion of superradiance was introduced by Dicke \cite{DRH54} in a simple model consisting of an ensemble of two-level systems coupled to a common electromagnetic mode of a cavity. Later, it has been pointed out that the transition into a superradiant phase discussed by Dicke was an artifact due to the fact that the electromagnetic $A^2$ term has been neglected \cite{RK75,NP10}. To circumvent this point and realize a superradiant phase, several proposals based on the use of external pump fields were proposed \cite{DF07,TEG13} and subsequently realized in experiments \cite{MC08,BK10,HC14}. More recently, a route towards realizing a genuine superradiant transition without external driving has  been proposed in thermodynamically large setups \cite{YA20}.

We here briefly review the qualitative physics of this superradiance phenomenon from a perspective of the Feshbach projection approach \cite{CGL09}. We employ the notation in {Sec.~\ref{Sec:NHO}} and focus on the case in which a subsystem with $N$ states is coupled to environmental continuum modes via a single channel, i.e., $C=1$. We also neglect the energy dependence of the operators $\Delta$ and $A$. The effective non-Hermitian Hamiltonian~\eqref{fpanh} is then  given by
\eqn{
[H_{{\rm eff}}]_{jj'} & = & \epsilon_{j}\delta_{jj'}-i\pi A^*_{j}A_{j'},
}  
where we choose the basis in the subspace $P$ in such a manner that the Hermitian part $H_{PP}+\Delta$ is diagonal with real eigenvalues $\{\epsilon_{j}\}$. We note that, in this single-channel case, the system-environment coupling $A$ reduces to a complex-valued vector. We denote a typical level spacing of $\{\epsilon_{j}\}$ by $\delta$ and assume that the coupling strengths take similar values, i.e., $\gamma_{j}\simeq\gamma$ with $\gamma_{j}\equiv 2\pi|A_{j}|^{2}$.  

Suppose that we excite the subsystem at the initial time by, e.g., external driving. We are interested in how the excited resonant modes decay due to their couplings to a common channel of environmental modes. In the weak coupling regime $\delta\gg\gamma$, the perturbation analysis of $H_{\rm eff}$ provides resonance complex energies 
\eqn{
\Lambda_{j} & \simeq & \epsilon_{j}-\frac{i}{2}\gamma_{j}.
}

In the strong coupling regime $\delta\ll\gamma$, the complex eigenvalues can be obtained from the condition
\eqn{\label{fpaeigeq}
1+\frac{i}{2} & \sum_{j} & \frac{\gamma_{j}}{\Lambda-\epsilon_{j}}=0.
}
This condition leads to a single superradiant solution associated with an eigenvalue 
 \eqn{\Lambda_{{\rm sr}} & = & \frac{1}{N}\sum_{j=1}^{N}\epsilon_{j}-\frac{i}{2}\sum_{j=1}^{N}\gamma_{j}.\label{fpasreig}
}
We emphasize that this mode has an enhanced decay rate scaled by the system size 
\eqn{\Gamma_{{\rm sr}}\simeq N\gamma\propto N.} 
Meanwhile, the other $N-1$ solutions have very narrow linewidths 
\eqn{\Gamma_{{\rm nonsr}}\simeq\frac{\delta^{2}}{\gamma}\ll\gamma.\label{fpanonrese}} 
The emergence of these narrow modes can be interpreted as an analogue of the continuous quantum Zeno effect to be discussed later. The state eigenvector $|\psi_{\rm sr}\rangle_P$ of the superradiant solution is almost parallel to $\{A_{{j}}\}$, i.e., $_{P}\langle \psi_{{j}}|\psi_{{\rm sr}}\rangle_P\propto A_{j}$, while the other state vectors are almost orthogonal to the vector  $\{A_{j}\}$. Further contributions to eigenenergies and eigenvectors beyond the leading order can be found in Ref.~\cite{FVV96}.  Generalizations of the present argument to the cases of energy-dependent coupling strengths have also been discussed \cite{VVS92,VA06}.  It is worthwhile to note that Eq.~\eqref{fpaeigeq} has essentially the same mathematical structure as found in the simple case of the Bethe-ansatz equation for solving the central spin problem \cite{BM07,YA19A}. In fact, the original Dicke model has also been exactly solved  by employing a similar technique \cite{TO10}. 

As the coupling strength $\gamma$ is increased, one can thus find a transition from independent unstable modes to a collective superradiant mode with an enhanced decay rate.
Physically, this transition indicates that an ensemble of excited states, which randomly radiate when they are independent, can radiate in a correlated manner when they are coupled to the common environmental mode. Such a transition occurs when the widths of individual adjacent modes grow and begin to overlap so that the environment-mediated couplings become significant. Along this line of arguments, the analogy between a quantum optical superradiance and an enhanced nuclear decay process has been discussed \cite{Rotter_1991},  and further pursued in the subsequent studies \cite{VA05,IR01}. Random matrix analyses have also been applied to understand the fate of superradiance in a quantum chaotic system as detailed below.

\begin{figure}[t]
\begin{center}
\includegraphics[width=13cm]{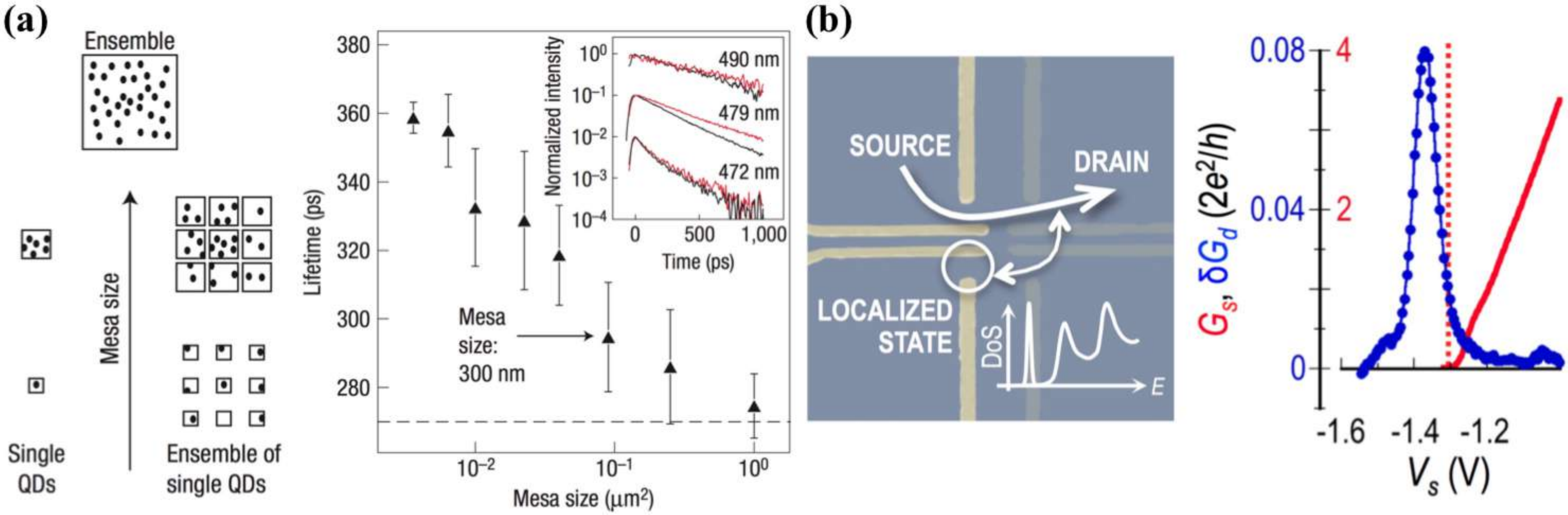}
\end{center}
\caption{(a) Experimental observation of superradiance in quantum dots. The left panel schematically illustrates the setup, where an ensemble of single quantum dots is prepared by gathering mesas of the similar size  arranged in a grid pattern. The right panel shows the measured decay times at different sizes of mesas. Superradiance manifests itself as a decrease in the lifetime against the mesa size. 
Adapted from Ref.~\cite{SM07}. Copyright \copyright\,   2007 by Springer Nature. (b) Experimental observation of the Fano resonance in coupled quantum point contacts. The left panel shows the experimental setup while the right one plots the measured differential conductance. The blue curve represents the detected Fano resonance and the red one shows the associated variation of the conductance. Adapted from Ref.~\cite{FJ14}. Copyright \copyright\,   2014 by ACS Publications.  }
\label{fig:4dot}
\end{figure}

\subsubsection{Physical applications}\label{Sec:4pa}
Owing to its versatility, the Feshbach projection approach has found many applications to several subfields of physics; prominent examples include  atomic and molecular physics,  nuclear physics, and chaotic scattering in mesoscopic physics. Below we summarize several unique features of non-Hermitian quantum physics relevant to these physical systems.
\\ \\ {\it Quantum dots and quantum point contacts}

\vspace{3pt}
\noindent
Quantum dots and quantum point contacts are ideal playgrounds to explore non-Hermitian physics in mesoscopic scales. Both of these systems typically confine a two-dimensional electron gas  into a fabricated narrow region at the interface of two materials in, e.g., GaAs/AlGaAs heterostructure.  
This strong confinement quantizes the energy spectrum. Transmitting electrons experience the resulting discrete spectrum connected with a series of transverse modes. The fabricated structure thus plays the role of a cavity, where quantized energy levels of electrons are confined in the narrow region and coupled to environmental continuum channels.

Several key features expected from the non-Hermitian formalism above have already been observed in these physical setups. 
Superradiance-type behavior was observed in pumped semiconductors as well as in assembled quantum dots by using an intense femtosecond laser pulse \cite{SM07,NG12} (see Fig.~\ref{fig:4dot}(a)). There, the observed strong photon emission is interpreted as a consequence of a correlated radiation from a large number of electron-hole pairs generated by external driving. It has also been pointed out that universal conductance fluctuations reminiscent of cross-section fluctuations in nuclear reactions \cite{TE63} can occur in mesoscopic systems \cite{DS95}; they can be understood from a common perspective of overlapping resonances \cite{HAW90}. The phase lapses of the transmission coefficient acquired by electrons can provide key signatures of the presence of resonant states unique to non-Hermitian physics \cite{MM09}. This phenomenon has been observed in few-electron quantum dots embedded into an Aharanov-Bohm interferometer \cite{YA95,SR97,AKM05}. 
Coupled quantum dots have been used to realize the bound states in the continuum spectrum \cite{GML03}. 
 The environment-mediated coupling between two discrete electron levels has experimentally been measured by using a pair of quantum point contacts, which couple to each other via a common continuum channel \cite{MT03,PVI04,YY09}. 
Coupled quantum point contacts have also been used to study a multichannel resonance \cite{FJ14} originally investigated by Fano \cite{FU61} (see Fig.~\ref{fig:4dot}(b)); a similar analysis has also been applied to transport problems in the tight-binding models \cite{GG15}. 
Quantum point contacts have recently been realized also in ultracold gases  \cite{BJP12,DH15}, and the effect of dissipation has experimentally been studied by employing a non-Hermitian description \cite{LM19,CL19}.  
\\ \\ {\it Random matrix analyses of quantum chaotic scattering and nuclear reactions}

\vspace{3pt}
\noindent
A random matrix analysis, which was originally introduced by Wigner \cite{WEP56}, provides a useful tool to understand statistical aspects of resonant modes in quantum chaotic scattering and nuclear reactions \cite{PCE65}. As resonant states necessarily associate with finite lifetimes, a natural question is how the results from the random matrix theory for Hermitian systems should be modified in the case of open quantum systems. While one may naively expect that the results in the Hermitian limit could still be valid if the widths are narrow and the level overlaps are not significant, recent experiments have demonstrated that this is not necessarily the case \cite{KPE10}.

\begin{figure}[t]
\begin{center}
\includegraphics[width=8cm]{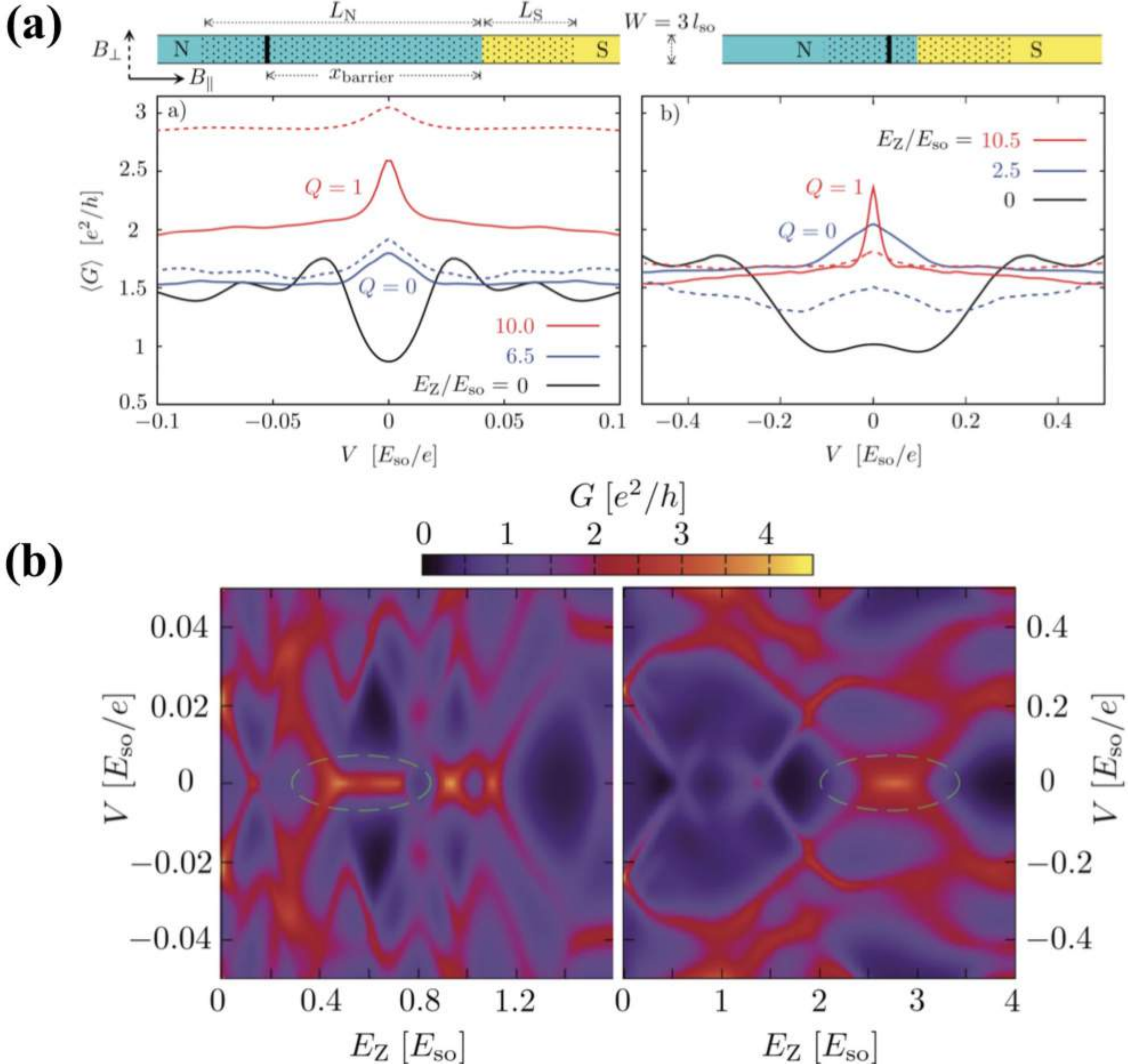}
\end{center}
\caption{(a) Numerical simulations of the disorder-averaged differential conductances plotted against the bias potential in nanowires whose geometries  are shown above each plot.  
(b) The corresponding differential conductance at different bias potentials and magnetic fields in a single disorder realization. The results in the left and right panels are obtained for the geometries shown in (a). As the magnetic field is increased, two resonances coalesce and lead to the emergence of the peak at the zero bias potential (green circles). 
Adapted from Ref.~\cite{Pikulin_2012} licensed under the Creative Commons Attribution 3.0 Unported. 
}
\label{fig:4rmtdot}
\end{figure}

 Studies along this line date back to an old observation made by Ullah \cite{UN69}, who generalized the eigenvalue distribution for the Gaussian orthogonal ensemble (GOE),
\eqn{P(E_{1},E_{2},\ldots,E_{N}) & = & {\cal N}\prod_{i<j}|E_{i}-E_{j}|\exp\left[-\frac{N}{\lambda^{2}}\sum_{i}E_{i}^{2}\right],
}
to the case of single-channel non-Hermitian systems,
\eqn{P(\Lambda_{1},\Lambda_{2},\ldots,\Lambda_{N}) & = & \frac{{\cal N}}{\sqrt{\prod_{i}\Gamma_{i}}}\prod_{i<j}\frac{|\Lambda_{i}-\Lambda_{j}|^{2}}{|\Lambda_{i}-\Lambda_{j}^{*}|}\exp\left[-NF\left(\left\{ \Lambda_{i}\right\} \right)\right],\label{ullah}\\
F\left(\left\{ \Lambda_{i}\right\} \right) & = & \frac{1}{\lambda^{2}}\sum_{i}E_{i}^{2}+\frac{1}{\gamma}\sum_{i}\Gamma_{i}+\frac{1}{2\lambda^{2}}\sum_{i<j}\Gamma_{i}\Gamma_{j},\label{ullahfe}
} 
where  $\cal N$ is a normalization constant, $\lambda$ is a constant independent of the number $N$ of levels, and $E_i$ and $\Gamma_i$ represent real and imaginary parts of complex eigenvalues $\Lambda_i$, respectively. The distribution~\eqref{ullah} should be applicable when (i) the internal chaotic dynamics (governed by $H_{PP}$) satisfies the time-reversal symmetry and can be modeled by the GOE and (ii) a coupling strength of each energy mode to a single decay channel obeys the Gaussian distribution with the variance $\gamma>0$. In the Ullah distribution~\eqref{ullah}, there exist several key features distinct from the GOE statistics in closed systems. Firstly, the quadratic repulsion term $|\Lambda_i-\Lambda_j|^2$ is stronger than the linear one $|E_i-E_j|$ in the GOE while the former is similar to the quadratic repulsion in the Gaussian   unitary ensemble (GUE); this feature can be understood from the emergent violation of the time-reversal symmetry due to environmental couplings. Secondly, the free-energy term (coined by Dyson \cite{DFJ62}) in Eq.~\eqref{ullahfe} contains much richer structures  than that in the closed case. More specifically, the last term in Eq.~\eqref{ullahfe} causes the repulsion of widths and suppresses  the probabilities for (nearly) homogeneous width distributions. In contrast, the probability can significantly be enhanced when a distribution has a one predominantly large width $\Gamma_{\rm sr}$ and $N-1$ smaller homogeneous widths. Physically, this feature indicates that the superradiance discussed in Sec.~\ref{Sec:spr} can survive even when the internal dynamics is chaotic \cite{VVS88,VVS89}. 

Signatures of such superradiance behavior have also numerically been found in the earlier works \cite{MPA68,KP85} done in the context of nuclear physics. 
As a concrete example discussed in these studies, consider the eigenvalue distributions for a one-channel case with the coupling strength $\gamma$. In the Hermitian case $\gamma=0$, the probability at zero spacing is zero.
As $\gamma$ is increased, it  grows from zero, reaches to a maximal value, and converges to zero again after crossing the superradiant transition. The eventual convergence in the large $\gamma$ limit is due to the fact that  complex energies for the residual $N-1$ states (other than the superradiant state) turn out to be repulsive. 
More generally, random matrix analyses as exemplified above have also found many applications to study resonances in chaotic scattering of electron transport in disordered mesoscopic systems \cite{BCWJ97,AY00}. 
For instance, resonant modes can be observed as the enhanced Andreev reflection resulting from quasibound modes close to the Fermi sea \cite{Pikulin_2012} (Fig.~\ref{fig:4rmtdot}). 
The distribution of resonant modes has also been observed in a microresonator setup \cite{KU08}.
It is noteworthy that, in the single-particle semiclassical regime, the random matrix approach can be related \cite{CHL91} to a widely used method based on periodic orbits \cite{GM90} (see also Sec.~\ref{secplight}). 
 \\ \\ {\it Atomic and molecular systems}
 
 \vspace{3pt}
\noindent
A non-Hermitian operator in the Feshbach projection approach is also useful to understand resonances and radiative behavior of individual atoms or molecules. 
For instance, it has been applied to analyze light-induced resonant dissociation of hydrogen ionic molecules and dimers of alkali atoms \cite{SVV08} as well as the onset of the bifurcation in the Bose-Einstein condensate with an attractive $1/r$ interaction \cite{CH08}.
A superradiant phenomenon in atomic systems has been discussed in Ref.~\cite{FVV96}, where the authors argued that the long-range nature of the Coulomb interaction could suppress the onset of the superradiance. Spectroscopic measurements of atomic ensembles have been used to observe the signature of such suppression  by using multi-photon excitations \cite{AM96}, which has theoretically been analyzed in the earlier work \cite{DG90}.
Coupling a Rb atom to confined electromagnetic fields in a cavity, the level structure inherent in the exceptional point has been observed \cite{CY10}. Ferrocene molecules subject to microwave radiation have realized the prototypical model for the Feshbach projection approach, where two specific spins of carbon and hydrogen atoms play the role of a subsystem coupled to the other environmental spins \cite{AGA06,DAD08}. Non-Hermitian degeneracies in the interacting bosonic model, which can be applied to molecular or nuclear physics, have been discussed in terms of their relation to quantum phase transitions \cite{CP072}.

\subsection{Quantum optical approach\label{Sec:qoa}}
When a quantum system is subject to measurement process, the measured system exhibits an unavoidable change, known as the measurement backaction, at the expense of acquiring information about the system.  
To describe such nonunitary quantum dynamics under measurement, theory of continuous quantum measurement has originally been developed in quantum optics \cite{MU89,UM90,DJ92,DR92,GCH92,HC93}, where one typically considers a small quantum system with a few degrees of freedom. This approach, often known as the quantum trajectory approach, provides a consistent and intuitive physical description of nonunitary stochastic dynamics subject to backaction from measurement or a Markovian environment.  Recently, this approach has found applications to a largely different realm of physics, namely, quantum many-body physics,  mainly owing to remarkable experimental developments in AMO systems. In this section, we review a general theory of quantum systems under measurement and discuss its relevance to a plethora of open quantum many-body phenomena with particular focus on non-Hermitian quantum physics.  

\subsubsection{Indirect measurement and quantum trajectory}
\label{sec:qtraj}
\begin{figure}[t]
\begin{center}
\includegraphics[width=12cm]{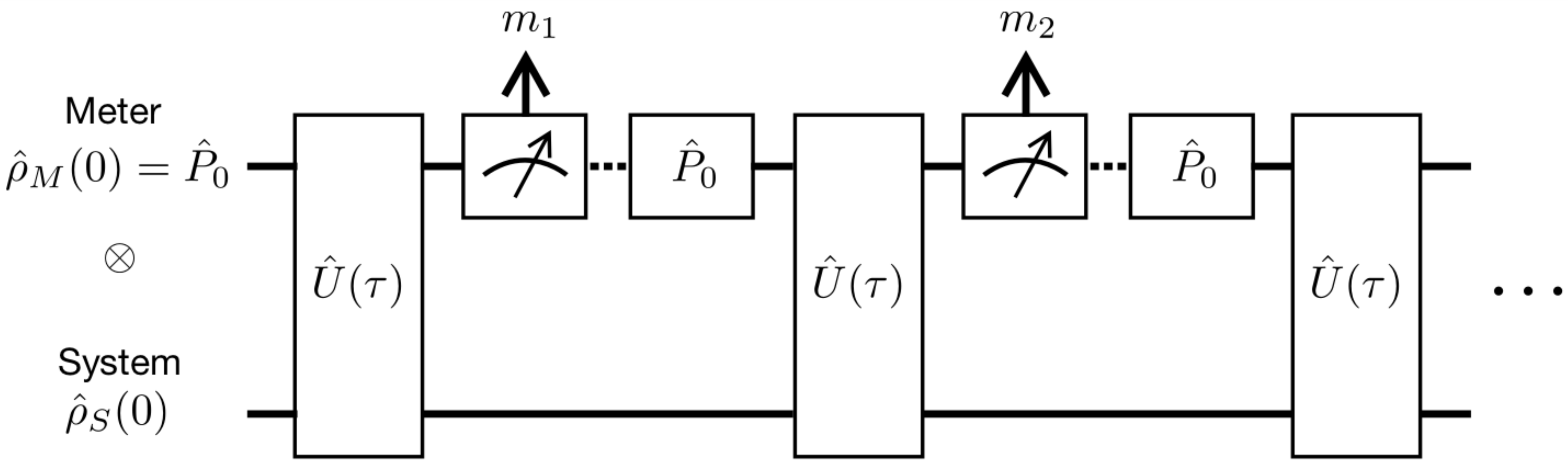}
\end{center}
\caption{Schematic illustration of the model of continuous quantum measurement. The initial state is prepared in the product state between a system and a measuring apparatus (meter). The total state evolves in time via a unitary operator $U(\tau)$ during a short time interval $\tau$. After each evolution, a projection measurement is performed on the meter. The meter  is then reset to the initial state denoted by $P_0$. Repeating the whole process, an observer obtains a sequence of measurement outcomes $\{m_1,m_2,\ldots,\}$.}
\label{fig:4indirectmeas}
\end{figure}

We consider a quantum system and a measuring apparatus (which we call as a meter), whose Hilbert spaces are denoted as $\cal{H}_S$ and $\cal{H}_M$, respectively. Suppose that a quantum system repeatedly interacts with the meter during a short time interval $\tau$ (see Fig.~\ref{fig:4indirectmeas}).  After each interaction, we perform a projection measurement on the meter and reset it to the common state $|\psi_0\rangle_{M}$; the latter process ensures that the meter retains no memory about previous measurement outcomes.  An observer thus obtains the information about the measured system through a sequence of measurement outcomes $\{m_1,m_2,\ldots\}$. We assume that each measurement outcome takes a discrete value from $m_i\in\{0,1,2,\ldots,M\}$ with $i=1,2,\ldots$.

We start from the system-meter product state $ {\rho}(0)= {\rho}_{S}(0)\otimes {P}_{0}$ with $ {P}_{0}=|\psi_{0}\rangle_{MM}\langle\psi_{0}|$ and assume that the dynamics of the total system is governed by a general Hamiltonian
\eqn{\label{hamindirect}
 {H}= {H}_{S}+ {V},\;\; {V}=\gamma\sum_{m=1}^{M} {A}_{m}\otimes {B}_{m}+{\rm H.c.},
}
where $ {H}_{S}$ is a Hamiltonian governing the internal dynamics of the measured system, $\gamma\in{\mathbb R}$ characterizes the strength of the coupling between the system and the meter, and $ {A}_{m}$ is an operator acting on ${\cal H}_{S}$. We assume that $ {B}_{m}$ acts on  ${\cal H}_{M}$ and changes the state of the meter  into the subspace of ${\cal H}_{M}$ corresponding to a measurement outcome $m$, i.e., we impose the relations
\eqn{
 {P}_{m'} {B}_{m}=\delta_{m'm} {B}_{m}\;\; (m'=0,1,\ldots,M;\;m=1,2,\ldots,M),
}
where $ {P}_{m'}$ is a projection measurement on the subspace of ${\cal H}_{M}$ providing a measurement outcome $m'$. A set of $ {P}_{m'}$ satisfies the completeness condition $\sum_{m'=0}^{M} {P}_{m'}= {I}$.
For each indirect measurement process, there are two possibilities, i.e., either  (i) one observes  a change in the state of the meter by obtaining an outcome $m=1,2,\ldots,M$ or  (ii) one observes  no change in the state of the meter from the reset state $|\psi_{0}\rangle_{M}$ and thus obtains the outcome $0$. 

In the first case (i), which is often called as the quantum jump process, the nonunitary mapping ${\cal E}_{m}$ of the system $ {\rho}_{S}$ corresponding to obtaining an outcome $m=1,2,\ldots,M$ is given by
\eqn{\label{nonuni_single}
{\cal E}_{m}( {\rho}_{S}) & ={\rm Tr}_{M}\left[ {P}_{m} {U}(\tau)\left( {\rho}_{S}\otimes {P}_{0}\right) {U}^{\dagger}(\tau) {P}_{m}\right],
}  
where we define $ {U}(t)=e^{-i {H}t}$ and ${\rm Tr}_{M}$ denotes the trace over the meter. This mapping can significantly be simplified in the continuous limit defined by $\gamma\tau\ll 1$ while keeping $\gamma^2\tau$ finite. The leading contribution in this limit is
\eqn{
{\cal E}_{m}( {\rho}_{S})\simeq\tau {L}_{m} {\rho}_{S}(\tau) {L}_{m}^{\dagger},
}
where $ {\rho}_{S}(\tau)= {U}_{S}(\tau) {\rho}_{S} {U}_{S}^{\dagger}(\tau)$ is the  state of the system with the internal dynamics $U_S(\tau)=e^{-iH_S\tau}$  and we introduce the jump operators $ {L}_{m}$ as
\eqn{\label{jump}
 {L}_{m}=\sqrt{\gamma^{2}\tau\langle {B}_{m}^{\dagger} {B}_{m}\rangle_{0}}\;{A}_{m}.
}
 We here denote $\langle\cdots\rangle_{0}$ as an expectation value with respect to $|\psi_{0}\rangle_{M}$.
The probability $p_{m}$ of obtaining an outcome $m=1,2,\ldots,M$ is given by
\eqn{
p_{m}={\rm Tr}_{S}[{\cal E}_{m}( {\rho}_{S})]=\tau{\rm Tr}_{S}[ {L}_{m} {\rho}_{S}(\tau) {L}_{m}^{\dagger}].
} 
In contrast, in the second case (ii) for which one observes no change of the meter state corresponding to the outcome $0$, the nonunitary mapping ${\cal E}_{0}$ can be simplified as
\eqn{
{\cal E}_{0}( {\rho}_{S}) & =&{\rm Tr}_{M}\left[ {P}_{0} {U}(\tau)\left( {\rho}_{S}\otimes {P}_{0}\right) {U}^{\dagger}(\tau) {P}_{0}\right]\\
 & \simeq& {\rho}_{S}(\tau)-\frac{\tau}{2}\left\{ \sum_{m=1}^M {L}_{m}^{\dagger} {L}_{m}, {\rho}_{S}(\tau)\right\},
} 
where   $\{ {O}, {O}'\}\equiv  {O} {O}'+ {O}' {O}$.
The probability $p_{0}$ of  no change being observed is
\eqn{
p_{0}={\rm Tr}_{S}[{\cal E}_{0}( {\rho}_{S})]=1-\tau\sum_{m=1}^M{\rm Tr}_{S}[ {L}_{m} {\rho}_{S}(\tau) {L}_{m}^{\dagger}].
}
We note that the probabilities satisfy the normalization condition $\sum_{m'=0}^{M}p_{m'}=1$.

\exmp{(Two-level system under photon counting).\label{tlsqt} 
As the simplest example, consider a two-level system coupled to a single bosonic mode, whose system-meter coupling can be described by the Jaynes-Cumming (JC) interaction \cite{ETJ63}
\eqn{
V_{{\rm JC}}=\gamma\left(\sigma^{-}a^{\dagger}+\sigma^{+}a\right),
}
where $a$ ($a^\dagger$) is the bosonic annihilation (creation) operator of, for example, a photon in the radiation mode. Suppose that we are in a small photon-number regime, in which at most  a single photon is detected, i.e., the reset state corresponds to the vacuum $|\psi_0\rangle_M=|0\rangle_M$. In this case, there exists only a single jump operator corresponding to the spontaneous emission of a photon,   
\eqn{
L_{-}=\sqrt{\Gamma}\sigma^{-},
} 
where $\Gamma$ characterizes the decay rate. 
}

\exmp{\label{qgmexmp}(Many-body system measured by photon scattering). 
Consider ultracold atoms trapped in optical lattice. Suppose that we use an off-resonant light to probe atoms  via (dispersive) light scattering. Under certain conditions, the light-matter interaction can effectively be modeled as (see Refs.~\cite{JJ95,YA15} for the derivation from microscopic Hamiltonians),
\eqn{V=\gamma\sum_{m}\left( {E}_{m}^{(-)}+ {E}_{m}^{(+)}\right)n_{m},}
where $n_m$ is an occupation-number operator of atoms at site $m$, and ${E}_{m}^{(+)}$ (${E}_{m}^{(-)}$) is the positive (negative) frequency component of the scattered electric field, which physically corresponds to the annihilation (creation) operator of photon modes scattered by an atom at site $m$. Suppose that photodetectors are prepared around each site and we work in a low-photon regime as in the previous example. Then, the jump operators labeled by site index $m$ can be obtained as 
\eqn{ {L}_{m}\propto\sqrt{\langle {E}_{m}^{(+)} {E}_{m}^{(-)}\rangle_{0}}\; {n}_{m}.
}
The corresponding jump process is a detection of an atom positioned at site $m$ and thus physically describes a site-resolved position measurement on a  many-body system.
While a more realistic treatment relevant to quantum gas microscopy \cite{BWS09} is possible \cite{YA15}, the conclusions are essentially the same as presented here.
}

\vspace{5pt}

\noindent{\it Quantum trajectory method}

\vspace{3pt}
\noindent
The dynamics of a quantum system under measurement formulated above can be described on the basis of a stochastic time-evolution equation. Suppose that we consider the evolution during a time interval $dt=N\tau$, which is much shorter than typical time scale of the internal dynamics $H_S$ but still contains a large number of repetitive interactions with the meter, i.e., $N\gg 1$. Because the probabilities of observing a change of the meter scale as $p_{m=1,\ldots,M}\sim(\gamma\tau)^2\ll 1$, one can assume that such a jump event occurs at most once during each interval $dt$. Under these conditions, the nonunitary dynamical mapping of the system during the interval $dt$ can be described as 
\eqn{
\Phi_{m=1,\ldots,M}( {\rho}_{S}) & =&\sum_{i=1}^{N}\left({\cal E}_{0}^{N-i}\circ{\cal E}_{m}\circ{\cal E}_{0}^{i-1}\right)( {\rho}_{S})= {M}_{m} {\rho}_{S} {M}_{m}^{\dagger}+O(dt^{2}),\\
\Phi_{0}( {\rho}_{S}) & =&{\cal E}_{0}^{N}( {\rho}_{S})=M_0 {\rho}_{S}M_0^{\dagger}+O(dt^{2}).
}
Here, we introduce the measurement operators as 
\eqn{
 {M}_{m}&=& {L}_{m}\sqrt{dt}\;\;\;\;\;(m=1,2,\ldots,M),\label{conmeass}\\
 {M}_{0}&=&1-i {H}_{{\rm eff}}dt,\label{conmeas}
}
where $H_{\rm eff}$ is an effective non-Hermitian Hamiltonian 
\eqn{\label{qoptnh}
 {H}_{{\rm eff}}= {H}_{S}-\frac{i}{2}\sum_{m=1}^{M} {L}_{m}^{\dagger} {L}_{m}.
}
An operator $ {M}_{m}$ acts on a quantum state if the jump event with outcome $m$ is observed, which occurs with a probability ${\rm Tr}_S[M_m\rho_S M_m^\dagger]$. 
The operator $ {M}_{0}$ acts on the state if no jumps are observed during the time interval $[t,t+dt]$, and this no-count process occurs with the probability ${\rm Tr}_S[M_0\rho_S M_0^\dagger]$. 
We note that the measurement operators satisfy the normalization condition $\sum_{m'=0}^{M} {M}_{m'}^{\dagger} {M}_{m'}=1$ aside from negligible contributions on the order of $O(dt^2)$.

For the sake of simplicity, suppose that the initial state is pure and thus the state $|\psi\rangle_{S}$ remains so in the course of time evolution. 
Due to the probabilistic nature of quantum measurement, an appropriate way to formulate the dynamics under measurement is to use a stochastic differential time-evolution equation \cite{GAP14}. To this end, we have to introduce a discrete random variable $dN_{m}$ with $m=1,\ldots,M$, whose  mean value is given by
\eqn{
{\mathbb{E}}[dN_{m}]=\langle {M}_{m}^{\dagger} {M}_{m}\rangle_{S}=\langle {L}_{m}^{\dagger} {L}_{m}\rangle_{S}\;dt,
}
where ${\mathbb{E}}[\cdot]$ represents the ensemble average over the stochastic process and $\langle\cdots\rangle_{S}$ denotes an expectation value with respect to a quantum state $|\psi\rangle_{S}$ of the system. 
The random variables $dN_m$ satisfy the following stochastic calculus:
\eqn{\label{stN}
dN_{m}dN_{n}=\delta_{mn}dN_{m}.
}
We remark that the stochastic process $dN_{m}$ is not a simple Poisson process because its intensity depends on a stochastic vector $|\psi\rangle_{S}$; 
it is rather known as a {\it marked point process} in the field of  stochastic analysis \cite{BA12}.
The stochastic change of a quantum state $|\psi\rangle_S$ during the time interval $[t,t+dt]$ can now be obtained as
\eqn{\label{stochastic}
|\psi\rangle_{S}\to|\psi\rangle_{S}+d|\psi\rangle_{S}\!=\!\!\left(1\!-\!\sum_{m=1}^{M}{\mathbb{E}}[dN_{m}]\right)\!\frac{ {M}_{0}|\psi\rangle_{S}}{\sqrt{\langle {M}_{0}^{\dagger} {M}_{0}\rangle_{S}}}\!+\!\sum_{m=1}^{M}\!dN_{m}\frac{ {M}_{m}|\psi\rangle_{S}}{\sqrt{\langle {M}_{m}^{\dagger} {M}_{m}\rangle_{S}}}.
}
Physically, the first term on the rightmost side describes the no-count process occurring with the probability $1-\sum_{m=1}^{M}{\mathbb{E}}[dN_{m}]$ while the second term describes the detection of a jump process $m$ occurring with a probability ${\mathbb{E}}[dN_{m}]$. We note that the denominator in each term is introduced to ensure the normalization of the state vector. 
From Eqs.~(\ref{conmeas}) and (\ref{conmeass}), we can rewrite Eq.~(\ref{stochastic}) as
\eqn{\label{conmeas2}
d|\psi\rangle_{S}\!=\!\left(1\!-\!i {H}_{{\rm eff}}\!+\!\frac{1}{2}\sum_{m=1}^{M}\langle {L}_{m}^{\dagger} {L}_{m}\rangle_{S}\right)dt|\psi\rangle_{S}\!+\!\sum_{m=1}^{M}\left(\frac{ {L}_{m}|\psi\rangle_{S}}{\sqrt{\langle {L}_{m}^{\dagger} {L}_{m}\rangle_{S}}}-|\psi\rangle_{S}\right)dN_{m}.
}
The first term on the right-hand side describes the non-Hermitian time evolution, in which the factor $\sum_{m=1}^M\langle {L}_{m}^{\dagger} {L}_{m}\rangle_{S}/2$ plays the role of maintaining the normalization of the state vector. In the second term, when the  jump event $m$ is detected, an operator $ {L}_{m}$ acts on the quantum state and causes its discontinuous change. A specific realization of this stochastic differential equation is often referred to as the {\it quantum trajectory} \cite{HC93}. 

Taking the ensemble average over all the possible trajectories, one reproduces the master equation in the Gorini-Kossakowski-Sudarshan-Lindblad form \cite{GV76,LG76}. To see this explicitly, let us rewrite Eq.~(\ref{conmeas2}) using the density matrix $ {\rho}_{S}=|\psi\rangle_{S}{}_{S}\langle\psi|$: 
\eqn{\label{conmeas3}
d {\rho}_{S}=-i\left( {H}_{{\rm eff}} {\rho}_{S}- {\rho}_{S} {H}_{{\rm eff}}^{\dagger}\right)dt+\sum_{m=1}^{M}\langle {L}_{m}^{\dagger} {L}_{m}\rangle_{S} {\rho}_{S}dt+\sum_{m=1}^{M}\left(\frac{ {L}_{m} {\rho}_{S} {L}_{m}^{\dagger}}{\langle {L}_{m}^{\dagger} {L}_{m}\rangle_{S}}- {\rho}_{S}\right)dN_{m},\nonumber\\
}
where we take the leading order of $O(dt)$ and use the stochastic calculus~(\ref{stN}). We note that this equation remains valid for a generic density matrix $ {\rho}_{S}$ that is not necessarily pure.
Introducing the ensemble-averaged density matrix ${\mathbb{E}[{\rho}_{S}]}$ and taking the average of Eq.~(\ref{conmeas3}), one arrives at the master equation  \cite{GV76,LG76}:
\eqn{\label{lindbladchap2}
\frac{d\,{\mathbb{E}[{\rho}_{S}]}}{dt}=-i\left( {H}_{{\rm eff}}{\mathbb{E}[{\rho}_{S}]}-{\mathbb{E}[{\rho}_{S}]} {H}_{{\rm eff}}^{\dagger}\right)+\sum_{m=1}^{M} {L}_{m}{\mathbb{E}[{\rho}_{S}]} {L}_{m}^{\dagger}.
}

The master equation~\eqref{lindbladchap2} describes the dynamics when no information about measurement outcomes is available to an observer. It  can also be applied to analyze dissipative dynamics of a quantum system coupled to a Markovian environment. The latter can be justified provided that the following three conditions are satisfied. Firstly, the strength of the system-environment coupling must be sufficiently weak in such a way that the Born approximation is valid. In other words, if the total system has no system-environmental correlations at the initial time, an evolved state at a later time must also remain so. Secondly, the environmental correlation time $\tau_{\rm env}$ must be much shorter than the relaxation time scale $\tau_{\rm rel}$ of the whole dynamics including both the system and the environment; $\tau_{\rm rel}$  can typically be estimated as $\tau_{\rm rel}\propto\Delta/\mu^2$ with $\Delta$ being typical level spacing in the internal dynamics of the system and $\mu$ being the strength of the system-environment coupling. This condition is known as the Markov approximation. The final assumption is the  condition $1/\Delta\ll\tau_{\rm rel}$, which can ensure that the jump operators $L_m$ in the master equation are time-independent.  These three conditions are sufficient to trace out the environmental degrees of freedom and to justify the use of Eq.~\eqref{lindbladchap2} for a time scale satisfying $1/\Delta\ll dt\ll\tau_{\rm rel}$. In practice, however, we note that there certainly exist several cases in which the master equation (or the Redfield-type master equation) can still remain useful even when some of these conditions  are not strictly satisfied \cite{PA16,FD19,Cattaneo_2019}. 

We mention that the effective non-Hermitian Hamiltonians $H_{\rm eff}$ found in both the quantum optical approach (cf. Eq.~\eqref{qoptnh}) and the Feshbach projection approach (cf. Eq.~\eqref{fpanonher}) share several common features  and thus play the similar roles. First, both of them have the factorized anti-Hermitian terms that are positive definite; this feature has its root in the unitarity of the whole theory including all the system and environmental degrees of freedom \cite{DL76}. Second, when the Markov approximation is valid, it is well known that the non-Hermitian operator in the Feshbach projection approach can also be used to analyze the subsystem dynamics during a certain time interval in a  manner similar to the quantum optical approach \cite{JGM04,ZR60}. Finally, we note that a paradigmatic setup in non-Hermitian physics is a many-body system subject to a decay process such as   unstable nuclei undergoing spontaneous emission or ultracold atoms experiencing loss of particles (see Secs.~\ref{Sec:4pa} and \ref{Sec:PS}). In both of these systems described by the Feshbach projection approach and the quantum trajectory approach, respectively, the spectra of the non-Hermitian operators  $H_{\rm eff}$ contain rich information underlying in lossy many-body physics (cf. Sec.~\ref{Sec:CDC}). In fact, the non-Hermitian Hamiltonian in the trajectory approach can in general have the complete information about the Liouvillean spectrum of a lossy master equation (i.e., without gain) \cite{BHJ93,TJM14}.
Despite these resemblances, we mention that concrete expressions of the operators $L_m$ or $A(\lambda)$ in each of two approaches are in general different even if the resulting anti-Hermitian terms in $H_{\rm eff}$ could be identified; those expressions must be determined by starting from microscopic Hamiltonians for each physical system.

\exmp{(Two-level system). 
Consider Example~\ref{tlsqt}, in which the coupling to the radiation mode causes  the spontaneous decay of a two-level system.  The effective non-Hermitian Hamiltonian is
\eqn{H_{\rm eff}=H_S-\frac{i\Gamma}{2}\sigma^+\sigma^-=H_S-\frac{i\Gamma}{2}P_{e},
}
where $H_S$ describes the internal dynamics of the two-level system, and $P_e=(1+\sigma^z)/2$ is the projection on the excited state. During the no-count process, we emphasize that an observer actually gains the information about the system that no spontaneous emission has been detected. This information acquisition indicates that the system is more likely to be in the ground state than in the excited state, which is captured by the measurement backaction represented by the anti-Hermitian term in $H_{\rm eff}$.  Meanwhile, when the jump process (described by $L_-=\sqrt{\Gamma}\sigma^-$) is detected, an observer acquires the information that the system is projected onto the ground state. If all the measurement outcomes are averaged out or discarded, the time evolution reduces to the master equation (cf. Eq.~\eqref{lindbladchap2}). When the system is further subject to an external drive, this master equation is called the optical Bloch equation \cite{CCT98}.
}

\begin{figure}[t]
\begin{center}
\includegraphics[width=12cm]{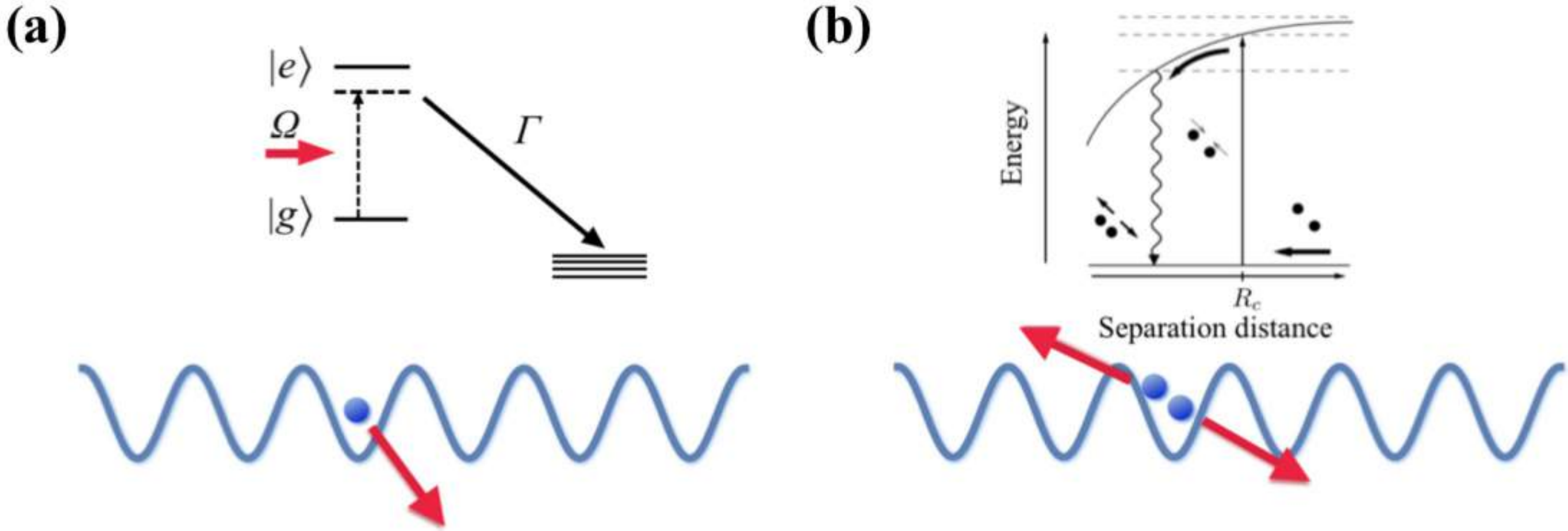}
\end{center}
\caption{(a) Schematic illustration of a one-body loss process. 
A particle is excited from the ground state $|g\rangle$ to an excited state $|e\rangle$ by a near-resonant laser with frequency $\Omega$. The excited state immediately decays to the continuum with the decay rate $\Gamma$ other outside of the ground state manifold. One can thus use a laser beam nearly resonant to the transition between the ground and excited states to realize a one-body loss of particles. After adiabatically eliminating the excited state, this effect can be taken into account by the quadratic non-Hermitian terms in Eqs.~\eqref{contloss} and \eqref{latticeloss}. (b) Schematic illustration of a two-body loss process. A red-detuned laser beam induces the excitation along the attractive molecular potential of two particles. The resulting energy gain, which is accumulated during the lifetime of the excited state, is converted into the kinetic energy, leading to an inelastic collision and the subsequent two-body loss process. This effect leads to the non-Hermitian interaction terms in Eqs.~\eqref{contloss} and \eqref{latticeloss}. }
\label{fig:4loss}
\end{figure}

\exmp{(Many-body system subject to losses). 
Consider a quantum many-body system subject to one-body and two-body loss processes. In general, the effective Hamiltonian can be obtained as
\eqn{\label{contloss}
H_{{\rm eff}}=\int d{\bf x}\left\{ \Psi^{\dagger}({\bf x})\left[-\frac{\hbar^{2}\nabla^{2}}{2m}+V_{\rm r}({\bf x})-iV_{\rm i}({\bf x})\right]\Psi({\bf x})+\frac{g-i\gamma}{2}\Psi^{\dagger}({\bf x})\Psi^{\dagger}({\bf x})\Psi({\bf x})\Psi({\bf x})\right\}, \nonumber\\
}
where $\Psi$ ($\Psi^\dagger$) is the annihilation (creation) operator of either a fermion or boson; for the sake of simplicity, we omit spin degrees of freedom. A one-body loss process can be taken into account by an imaginary potential, $-iV_{\rm i}({\bf x})$, while the two-body loss effectively changes an interaction parameter to a complex value $g-i\gamma$. Physically, the one-body loss can be realized by a near-resonant optical potential (Fig.~\ref{fig:4loss}(a)) while the two-body loss can be induced by light-assisted inelastic collisions (Fig.~\ref{fig:4loss}(b)) or internal atomic or molecular reactions. In ultracold atoms, a many-body system is typically prepared in a periodic optical potential, allowing one to describe it on the basis of a Hubbard-type lattice model \cite{GEM08,YA16crit}. When this tight-binding treatment is valid, we can rewrite Eq.~\eqref{contloss} as
\eqn{\label{latticeloss}
H_{{\rm eff}}=\sum_{mn}\left(J_{mn}^{\rm r}-iJ_{mn}^{\rm i}\right)\left(b_{m}^{\dagger}b_{n}+{\rm H.c.}\right)+\frac{U_{\rm r}-iU_{\rm i}}{2}\sum_{m}b_{m}^{\dagger}b_{m}^{\dagger}b_{m}b_{m},
} 
where $b_m$ ($b^\dagger_m$) is the annihilation (creation) operator of a particle at site $m$.  The sum in the first term is taken over the on-site contributions ($n=m$) as well as the nearest-neighbor hoppings ($|n-m|=1$). We note that a one-body loss process in general leads to (possibly inhomogeneous) imaginary on-site/hopping terms through the spatial dependence of $V_{\rm i}({\bf x})$ (see Example~\ref{ptband} and Fig.~\ref{fig:2epchiral1}(a) for an illustrative example).
}

\vspace{5pt}

\noindent{\it Diffusive limit}

\begin{figure}[t]
\begin{center}
\includegraphics[width=14.5cm]{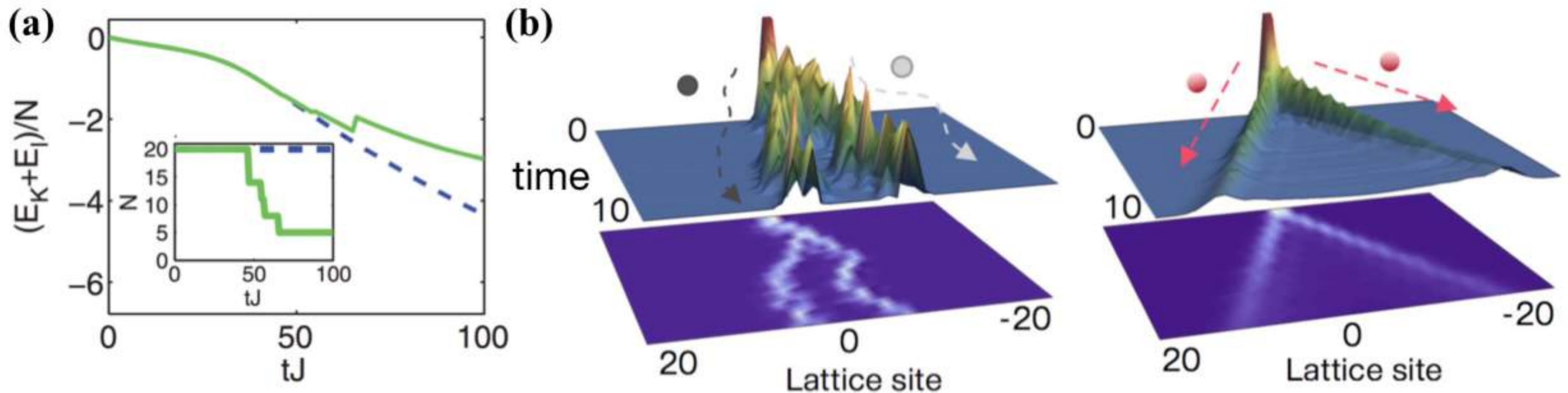}
\end{center}
\caption{Quantum trajectory dynamics associated with (a) discrete and (b) continuous stochastic processes. (a) Single trajectory dynamics under three-body loss of particles. The main (inset) panel plots the time evolution of the sum of kinetic and interaction energies per particle (the number of particles). The blue dashed curve corresponds to the non-Hermitian dynamics with no loss events  and the green solid curve shows the trajectory dynamics associated with several jump events.
Adapted from Ref.~\cite{DAJ09}. Copyright \copyright\,  2009 by the American Physical Society.
 (b) Single trajectory dynamics in the diffusive limit of a continuous position measurement. The left (right) panel shows the dynamics of distinguishable particles (fermions), indicating uncorrelated diffusive behavior (anti-correlated ballistic behavior) \cite{YA17}.   }
\label{fig:4traj}
\end{figure}

\vspace{3pt}
\noindent
We have so far discussed a measurement process in which an operator $ {L}_{m}$ induces a discontinuous change of a quantum state. It is worthwhile to mention that there is another type of continuous measurement associated with a diffusive stochastic process. To this end, we assume that measurement backaction induced by an operator $ {L}_{m}$ is weak in the sense that it satisfies
\eqn{
 {L}_{m}=\sqrt{\Gamma}(1-\epsilon {a}_{m}),
}
where $\Gamma$ characterizes the detection rate of jump events  and $\epsilon\ll 1$ is a small dimensionless parameter.
We also assume that detections of jump events are so frequent that the expectation value $\delta N_m$ of the number of jump-$m$ events being observed during a time interval $\delta t$ is sufficiently large. From the central-limit theorem, it can  be approximated as
\eqn{
\delta N_{m}  & \simeq&\Gamma(1-\epsilon\langle {a}_{m}+ {a}_{m}^{\dagger}\rangle_{S})\delta t+\sqrt{\Gamma}\left(1-\frac{\epsilon}{2}\langle {a}_{m}+ {a}_{m}^{\dagger}\rangle_{S}\right)\delta W_{m},
}
where $\delta W_{m}\in{\cal N}(0,\delta t)$ is a random variable obeying the normal distribution with the zero mean and the variance $\delta t$. We then take the diffusive measurement limit \cite{WHM10}, i.e., $\epsilon\to 0$ and $\Gamma\to\infty$ while $\Gamma \epsilon^2$ is kept constant.  The resulting stochastic  differential equation is
\eqn{\label{diffusiveTE}
d {\rho}_{S}&=&\left[-i[ {H}_{S}, {\rho}_{S}]-\frac{1}{2}\sum_{m=1}^{M}\left(\left\{  {l}_{m}^{\dagger} {l}_{m}, {\rho}_{S}\right\} -2 {l}_{m} {\rho}_{S} {l}_{m}^{\dagger}\right)\right]dt\nonumber\\
&&+\sum_{m=1}^{M}\left[\left( {l}_{m}-\langle {l}_{m}\rangle_{S}\right) {\rho}_{S}+ {\rho}_{S}\left( {l}_{m}^{\dagger}-\langle {l}_{m}^{\dagger}\rangle_{S}\right)\right]dW_{m},
}
where we introduce operators $ {l}_{m}=\sqrt{\Gamma \epsilon^2}\,{a}_{m}$ as well as the Wiener stochastic processes that satisfies
\eqn{
{\mathbb{E}}[dW_m]=0,\;\;dW_m dW_n=\delta_{m n}dt.
}
When a quantum state is pure, the  stochastic time-evolution equation~(\ref{diffusiveTE}) can be rewritten as
\eqn{
d|\psi\rangle_{S}&=&\left[1-i {H}_{S}-\frac{1}{2}\sum_{m=1}^{M}\left( {l}_{m}^{\dagger} {l}_{m}- {l}_{m}\langle {l}_{m}+ {l}_{m}^{\dagger}\rangle_{S}+\frac{1}{4}\langle {l}_{m}+ {l}_{m}^{\dagger}\rangle_{S}^{2}\right)\right]dt|\psi\rangle_{S}\nonumber\\
&&+\sum_{m=1}^{M}\left( {l}_{m}-\frac{1}{2}\langle {l}_{m}+ {l}_{m}^{\dagger}\rangle_{S}\right)dW_{m}|\psi\rangle_{S}.
}
The stochastic Schr{\"o}dinger equation with the diffusive process \cite{NG92} was originally applied to analyze nonunitary dynamics of few-body systems such as a cavity under homodyne measurement {\cite{MJC87,MGJ93}} as well as a quantum particle under continuous position measurement \cite{KJ06,BA13}; recently, it has been extended to many-particle regimes in the context of minimally destructive observation \cite{YA17}. Typical examples of single trajectory dynamics under continuous observation are shown in Fig.~\ref{fig:4traj}.
\\ \\ {\it Classical limit}

\vspace{3pt}
\noindent
Some of physical phenomena in open quantum systems have complete analogues in  classical systems. This is the case especially when neither quantum entanglement nor quantum correlations play nontrivial roles in determining physical properties. Since we have reviewed a number of classical aspects of non-Hermitian physics in other sections (e.g., Secs.~\ref{sec3} and \ref{sec5}), here we will mainly focus on genuinely quantum aspects of non-Hermitian physics, which are distinct from classical wave phenomena frequently observed in, e.g., optics. 
Before proceeding further, however, we mention two specific cases in which open-system dynamics formulated here becomes equivalent to a classical problem. 

The first (rather trivial) case is when a system is subject to a one-body loss (i.e., a jump operator is proportional to the annihilation operator of a particle) and the system consists only of a single particle \cite{DS15A}. In this case, the quantum jump term (the last term in Eq.~(\ref{conmeas3})) is irrelevant to dynamics (aside from the wavefunction normalization) because a quantum state simply changes to the particle vacuum after the jump process {\cite{GZ17}}. The residual no-count process exactly corresponds to  a non-Hermitian Schr{\"o}dinger wave equation also discussed in classical optics (see, e.g., Eq.~\eqref{tmmode}).   The second case is when a bosonic many-particle system is subject to loss processes and a certain mode is macroscopically occupied such that a quantum state is well approximated by a coherent state (i.e., an eigenstate of the annihilation operator). In this case, the jump term does not alter the quantum state and the problem reduces to the mean-field problem described by the complex Gross-Pitaevskii equation. Physically, this case is relevant to, e.g., mean-field physics of dissipative atomic Bose-Einstein condensates \cite{WP09,BG13,LR16}, driven-dissipative exciton-polariton condensates (see {Sec.~\ref{sechydroexc}}), or superconducting wires \cite{RJ07,NMC12}.

\subsubsection{Role of conditional dynamics}
\label{Sec:RCD}
We make several important remarks on the role of dynamics conditioned on measurement outcomes, which are often overlooked in literature. 
Such {\it conditional} dynamics is in general distinct from the {\it unconditional} dynamics obeying the master equation; the latter has been well explored over the last couple of decades especially in the context of driven-dissipative systems, and we refer to excellent review papers \cite{DH10,MM12,RH13,CI13,Sieberer_2016,WH19} for developments in that direction. In contrast, as detailed in Sec.~\ref{Sec:QMBP}, conditional dynamics can potentially exhibit new and rich physics with many open questions. Importantly, it has recently become relevant to a number of physical setups thanks to remarkable developments in measurement techniques of AMO systems at the level of single quanta.

First of all, we recall that an appropriate theoretical description of open quantum dynamics changes depending on how much information about measurement outcomes is available to an observer. For instance, if we can access the complete information about measurement outcomes, i.e., all the times and types of quantum jumps, the dynamics is described by a single realization of the quantum trajectory:
\eqn{
|\psi_{{\rm traj}}\rangle_{S}=\prod_{k=1}^{n}\left[ {{\cal U}}_{{\rm eff}}(\Delta t_{k}) {L}_{m_{k}}\right] {{\cal U}}_{{\rm eff}}(t_{1})|\psi_{0}\rangle_{S}/{\|\cdot\|},\label{trajD}
}
where $0\!<\!t_1\!<\!\cdots\!\!<\!t_n\!<\!t$ are the times of quantum jumps whose types are $\{m_1,m_2,\ldots,m_n\}$, $\Delta t_{k}=t_{k+1}-t_{k}$ is the time difference with $t_{n+1}\equiv t$, and $ {\cal U}_{\rm eff}(t)=e^{-i {H}_{\rm eff}t}$ is the non-Hermitian evolution operator.
Meanwhile, even when we cannot access such complete information,  we may have an ability to access partial information about measurement outcomes. As mentioned later, one natural situation is that we know the total number of quantum jumps occurred during a certain time interval, but not their types and occurrence times. In such a case, the quantum state is described by the {\it full-counting} density matrix conditioned on the number $n$ of quantum jumps that have occurred during the time interval $[0,t]$: 
\eqn{
 {\rho}^{(n)}(t)&\propto&\sum_{\alpha\in{\cal D}_{n}}|\psi_{\alpha}\rangle_{S}{}_{S}\langle\psi_{\alpha}|\nonumber\\
 &=&\!\sum_{\{m_{k}\}_{k=1}^{n}}\!\int_{0}^{t}dt_{n}\cdots\int_{0}^{t_{2}}dt_{1}\nonumber\\
&&\;\;\;\;\;\;\prod_{k=1}^{n}\left[ {\cal U}_{\rm eff}(\Delta t_{k}) {L}_{m_{k}}\right] {\cal U}_{\rm eff}(t_{1}) {\rho}_{S}(0) {\cal U}_{\rm eff}^{\dagger}(t_{1})\prod_{k=1}^{n}\left[ {L}_{m_{k}}^{\dagger} {\cal U}^{\dagger}_{\rm eff} (\Delta t_{k})\right],
\label{fcs2}
}
where the ensemble average is taken over the subspace ${\cal D}_{n}$ spanned by all the trajectories having $n$ quantum jumps during $[0,t]$. In other words, this can be interpreted as a coarse-grained dynamics of pure quantum trajectories~(\ref{trajD}), where the number of jumps is known while the information about the times and types of jumps are averaged out \cite{YA18}. The no-jump process $ {\rho}^{(0)}(t)$ is the simplest  example of this class of dynamics.  

\begin{figure}[t]
\begin{center}
\includegraphics[width=13.5cm]{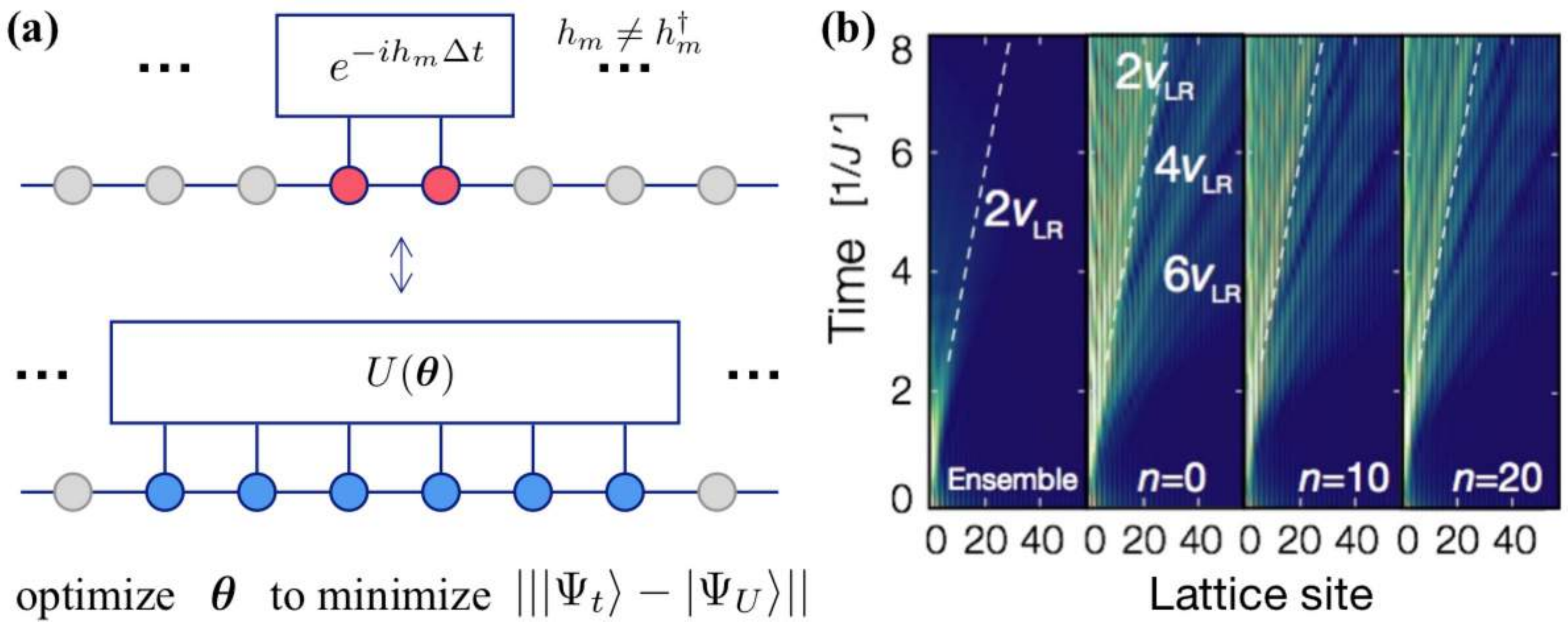}
\end{center}
\caption{(a)  Schematic  illustration of the mapping between local non-Hermitian  evolution and nonlocal unitary evolution. A normalized state $|\Psi_t\rangle$ evolved by a local non-Hermitian Hamiltonian can be approximated by a quantum state $|\Psi_U\rangle$ obtained by unitary operators, which locally act on $\tilde{N}$ qubits. The divergence of $\tilde{N}$ in the thermodynamic limit (cf. Eq.~\eqref{nonlocthm}) indicates the  nonlocality inherent in the dynamics conditioned on measurement outcomes. In practice, this fact may provide a way to approximately simulate the conditional dynamics by variational optimizing parameters in unitary operators \cite{Moll_2018}. (b) Spatiotemporal dynamics of correlations in the unconditional dynamics (the  leftmost panel) and the conditional dynamics for different numbers $n$ of quantum jumps (other panels). The inherent nonlocality manifests itself as the  supersonic propagation beyond the Lieb-Robinson bound, which is indicated by the white dashed lines.  
 Adapted from Ref.~\cite{YA18}.
 }
\label{fig:4nonloc}
\end{figure}

We then emphasize that such conditional dynamics should in general be considered in a separate manner from the unconditional, master-equation dynamics. The most fundamental distinctive feature is that the locality constraint can be mitigated in conditional dynamics while this is not the case in the unconditional one \cite{PD10,BT12,KM14b}. This fact was first pointed out in Ref.~\cite{YA18}, and has recently been discussed in unitary-circuit dynamics under repeated projective measurements \cite{LY20}.
The authors in Ref.~\cite{MM20} point out the similar feature in the following theorem.
\begin{theorem}[Nonlocality in conditional dynamics]\label{nonloctheoremnum} Consider a $k$-local non-Hermitian Hamiltonian $H_{\rm eff}=\sum_{m=1}^{N}h_m$ acting on qubits on a $d$-dimensional lattice with the operator norm ${\|h_m\|}\leq 1$. Let ${|\Psi_t\rangle={e^{-iH_{\rm eff}t}|\Psi_0\rangle}/{{\|\cdot\|}}}$ be a normalized state evolved by $H_{\rm eff}$ starting from the product state $|\Psi_0\rangle$. 
 Let $|\Psi_U\rangle=U_{NT}\cdots U_{2}U_{1}|\Psi_0\rangle$ be a quantum state obtained by a sequence of $NT$ unitary operators,  each of which locally acts on $\tilde{N}$ qubits. Let $C$ be an upper bound on the correlation length of $|\Psi_t\rangle$. Then, for any $\epsilon>0$, one can approximate $|\Psi_t\rangle$ by $|\Psi_U\rangle$ as ${\| |\Psi_t\rangle-|\Psi_U\rangle\|}\leq \epsilon$ with  unitary operators having the support 
 \eqn{\tilde{N}=k(2C)^d\ln^d\left(2\sqrt{2}NT\epsilon^{-1}\right),\label{nonlocthm}}
 which diverges in the thermodynamic limit $N\to\infty$.
\end{theorem}
\noindent The  statement should readily be extended to generic conditional trajectory dynamics including both non-Hermitian evolution and jump processes. The theorem indicates that, even if the underlying physical system is locally interacting and measurement acts on local regions, the resulting conditional dynamics can in general not be reproduced by local unitary operators, thus leading to the violation of the locality at least in the conventional sense (Fig.~\ref{fig:4nonloc}(a)). 
This fact makes a sharp contrast with the (unconditional) master-equation dynamics, which satisfies the locality constraint as long as the Liovillean consists of operators acting on local regions \cite{PD10,BT12,KM14b}. 
Thus, the conditional dynamics can potentially exhibit rich unconventional phenomena \cite{DAJ09,LTE14,KS14,WAC15,WAC16,KW17,YA18,YA18therm,JJS18,MK18,LY18,CA19,SB19,YF20,MN19}, part of which is fundamentally inaccessible at the level of the master equation (see Fig.~\ref{fig:4nonloc}(b) for an illustrative example and Sec.~\ref{Sec:CDC} for further details).

Despite these remarks on distinctions between conditional and the unconditional dynamics, we mention that previous studies have often found the non-Hermitian description, a particular type of conditional dynamics, indeed useful to understand physics underlying in unconditional dynamics and, in certain cases, even to provide quantitatively accurate predictions \cite{DAJ09,DAJ14,KW16,LM19,LC19,LDA19}. 
More specifically, for the purpose of understanding dissipative dynamics obeying the master equation, the non-Hermitian description can particularly be useful when the following two conditions are met: (i) phenomena of our interest occur in a transient regime rather than the steady state of the master equation;
(ii) an exact analysis of the master equation, such as the exact numerical diagonalization of the Liouvillean or the use of the Bethe ansatz techniques \cite{Prosen_2015}, is infeasible. 
Regarding the condition (i), we note that a variety of theoretical tools, such as the variational method, are readily available if one is only interested in the steady state (i.e., the zero mode of the Liouvillean)  \cite{WH19}. The condition (ii) naturally arises from the fact that the investigation of transient phenomena in principle requires the full spectra and eigenstates of the Liouvillean, which in general makes an exact analysis  challenging. The non-Hermitian description offers an effective theoretical tool to capture key features of such transient phenomena.

Importantly, physical situations satisfying these conditions are fairly ubiquitous in interacting AMO many-body systems that can be naturally subject to spontaneous decay or loss processes \cite{NS08,RMS09,YB13,ZB14,EU15,PYS15,TT17}. In these lossy systems, the analysis of the Liouvillean of the master equation can significantly be simplified to that of the non-Hermitian operator $H_{\rm eff}$ as the Liouvillean possesses the tridiagonal matrix structure and thus shares the same spectral feature with that of $H_{\rm eff}$ \cite{BHJ93,TJM14}. For instance, the PT-symmetry breaking in $H_{\rm eff}$ can lead to the qualitatively different behavior of the loss rate in the unconditional dynamics \cite{YT20}.   Several comparative analyses have also been performed in Refs.~\cite{Faisal_1981,Graefe_2010,KGZ14,ABJ20,CCL20}, where it has been found that the non-Hermitian description can correctly capture qualitative physics and even quantitatively agree  with the master-equation results in certain regimes. A possible extension of the non-Hermitian description to long-time regimes has been proposed in Refs.~\cite{EA18,Varguet_2019,CCL20}.
In a recent experiment \cite{LM19,CL19}, the non-Hermitian analysis has also been found to be useful for understanding dissipative transport dynamics in many-body ultracold atoms subject to loss. Overall, the non-Hermitian description can provide a simple and intuitive way to understand essential physics behind lossy dissipative dynamics in transient regimes, whose analysis must typically resort to numerical methods that are, in many-body regimes, demanding and often less insightful (see Sec.~\ref{Sec:QMBP} for further details). 

\subsection{Quantum many-body physics}
\label{Sec:QMBP}
A phenomenon of great interest in quantum physics is the collective behavior that never appears in the individual constituents, but can emerge only when a large number of constituents interact with each other \cite{APW72}. 
Recently, the quantum optical approach outlined in Sec.~\ref{Sec:qoa} has attracted renewed interest mainly owing to its relevance to such {\it many-body} regimes. Below we review several fundamental aspects of open quantum many-body phenomena with a focus on non-Hermitian physics as well as conditional trajectory dynamics, and outline a variety of physical setups that can be used to explore  them.  We also  review several attempts to go beyond the quantum optical approach, aiming for advancing our understanding of open systems in non-Markovian regimes, in which the master-equation formalism becomes inapplicable. 

\subsubsection{Criticality, dynamics, and chaos\label{Sec:CDC}}
Quantum critical behavior is arguably one of the most prominent collective phenomena emerging from strong correlations between many quantum degrees of freedom. 
Studies of  quantum criticality in nonunitary many-body systems date back to Fisher's work \cite{MEF78} on $i\phi^3$-field theory, which corresponds to the effective field theory of the Yang-Lee edge singularity (cf. Eq.~\eqref{iphi3}). Later, Cardy analyzed this setup in the case of the one-dimensional system by considering the {${\cal M}(5,2)$} conformal field theory  \cite{JC85}; the subsequent and recent developments along this direction can be found in {Sec.~\ref{nonunitary_cft}} and Refs.~\cite{SAM19,SAM192,AJ20,AJ202,AJ19}. 

More generally, a broad range of low-energy behavior in one-dimensional quantum systems can be described by the sine-Gordon model, whose critical phase corresponds to the Tomonaga-Luttinger liquid (TLL). In Hermitian systems, the critical behavior in the TLL is solely characterized by the single parameter $K$ known as the TLL parameter \cite{TG03}:
\eqn{C_{\theta}(r)\propto\left(\frac{1}{r}\right)^{\frac{1}{2K}},\;\;\;C_{\phi}(r)=-\frac{K}{2\pi^{2}r^{2}}+{\rm const.}\times\frac{\cos\left(2\pi\rho_{0}r\right)}{r^{2K}},
}
where $C_\theta$ ($C_\phi$) is the one-particle (density-density) correlation function, and $\rho_0$ is the mean particle density. 
 There are essentially two possible ways to extend this paradigmatic critical behavior to non-Hermitian regimes. The first way is to add the imaginary quadratic term $\propto-i\int dx(\partial_x\phi)^2$, which effectively changes the  TLL parameter $K$ to be a complex value. This change induces the bifurcation of the critical exponents between $C_\theta$ and $C_\phi$ \cite{YA16crit}. The other way is to add the relevant imaginary potential $\propto -i \int dx\cos(2\phi+\delta)$, where $\delta$ is the phase difference from the real potential in the sine-Gordon model. When $\delta\neq \pm\pi/2$,  this term again makes $K$ complex along the RG flows, and merely leads to effects similar to the ones induced by the imaginary quadratic perturbation.   The influences unique to the imaginary potential perturbation can thus be identified in the case of $\delta=\pm\pi/2$, at which $K$ remains to be real along RG flows. The resulting field theory is the generalized sine-Gordon model considered in Eq.~\eqref{sG} \cite{CMB05,YA17nc}. We recall that these types of the anti-Hermitian terms can physically originate from the underlying one-body and two-body loss processes (cf. Eqs.~\eqref{contloss} and \eqref{latticeloss}). 

\begin{figure}[t]
\begin{center}
\includegraphics[width=11.5cm]{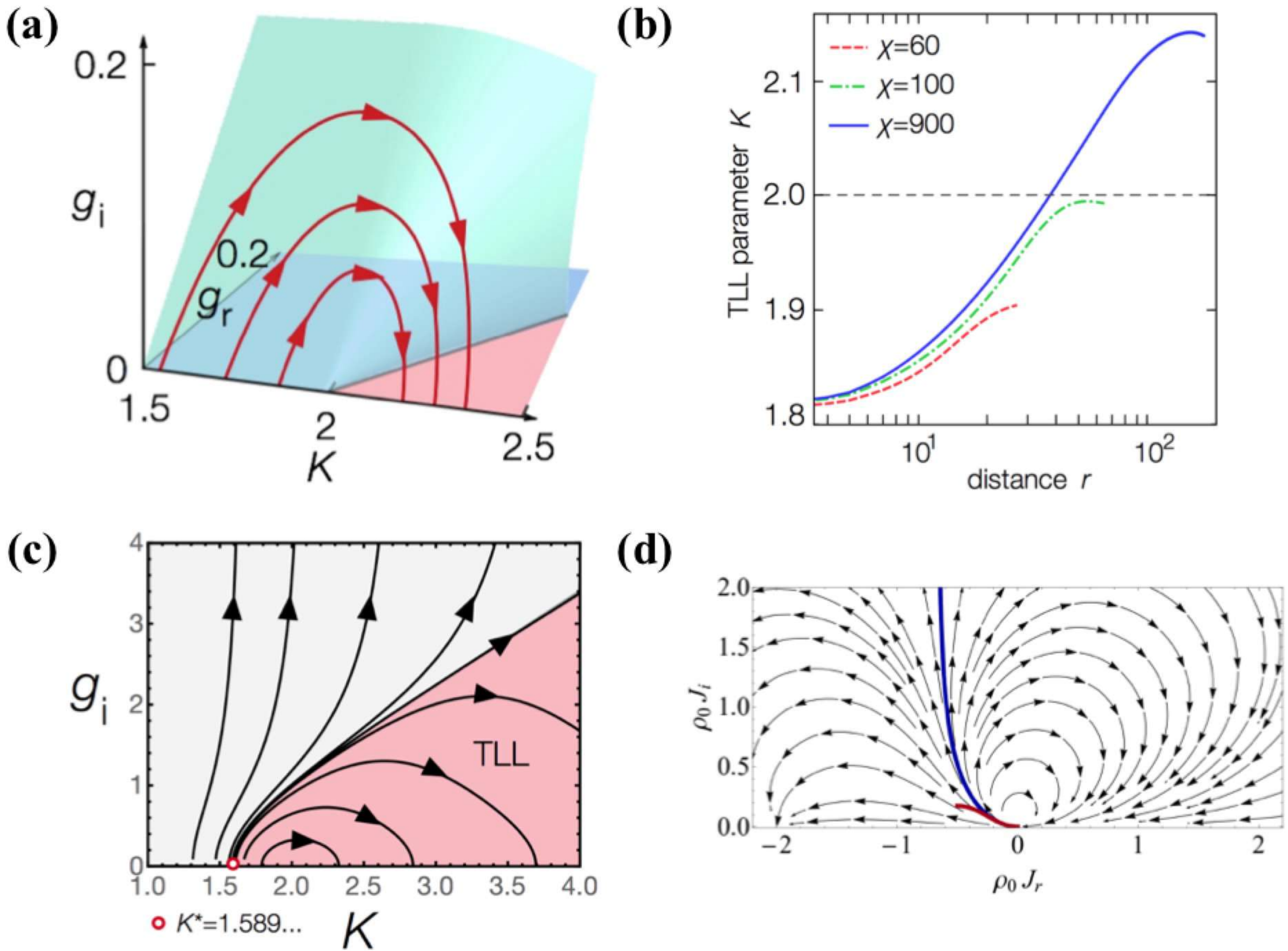}
\end{center}
\caption{(a) Semicircular RG flows in the generalized sine-Gordon model, where $g_i$ represents the strength of the imaginary potential and $K$ is the TLL parameter. This RG flow violates Zamolodchikov's $c$-theorem and is thus unique to non-Hermitian regimes. (b) Numerical verification of the enhancement of the TLL parameter $K$ at a large distance predicted in (a). The results are obtained by the iTEBD calculations and $\chi$ represents the bond dimension of the matrix product states. The panels (a) and (b) are adapted from  Ref.~\cite{YA17nc}. (c) RG flows in the nonperturbative regimes of the same model. In the vicinity of $K\sim2$, the RG flows reproduce the semicircular flows found in the perturbative results in (a). Below $K<K^*$, the flows no longer return to the TLL fixed line and the coupling strength $g_i$ diverges. Adapted from Ref.~\cite{YAbook}. (d) Circular RG flows in the generalized Kondo model, where $J_i$ and $J_r$ represent imaginary and real parts of the Kondo coupling strength. Adapted from Ref.~\cite{NM18}.
}
\label{fig:4crit}
\end{figure}

One of the prominent features in the generalized sine-Gordon model is the emergence of the unconventional RG fixed points unique to non-Hermitian regimes \cite{YA17nc}, which belong to the so-called Liouville CFT \cite{NS90} (cf. Eq.~\eqref{livcft}). Indeed, it has been shown that the three-point correlation functions at these fixed points  exhibit unconventional critical behavior different from the unitary theories \cite{IY16}.
The other crucial feature is the emergence of the anomalous RG flows (Fig.~\ref{fig:4crit}(a)), which violates the fundamental theorem in the conformal field theory, known as Zamolodchikov's $c$-theorem \cite{ZAB86}, 
\eqn{\frac{dc}{dl}<0,\label{ctheorem}}
where $c$ is the central charge of the theory (cf. Eq.~\eqref{virasoroa}) and $l$ indicates the logarithmic RG scale. 
Equation~\eqref{ctheorem} indicates the irreversibility  along RG flows, namely,  the number of effective degrees of freedom must monotonically decrease along the coarse-graining procedures. The $c$-theorem has recently been extended to a certain class of the nonunitary field theory in the PT-unbroken regime \cite{OAC17}.

The semicircular RG flows found in the generalized sine-Gordon model (Fig.~\ref{fig:4crit}(a)) indicate that the system that initially resides in the critical phase with $K<2$ will, after winding of the flow, end up again in the critical phase but with the enhanced TLL parameter $K>2$.  Since both the initial and final critical phases correspond to the TLL phase (associated with the same value of $c$), this RG flow violates the $c$-theorem and thus has no counterparts in Hermitian regimes. We note that this unconventional RG flow has been pointed out in the earlier work by Fendley, Saleur, and Zamolodchikov \cite{FP93} in the particular case of a purely imaginary coupling ($\alpha_{\rm r}=0$). A first concrete evidence of these  RG predictions have recently been provided in Ref.~\cite{YA17nc} (Fig.~\ref{fig:4crit}(b)). There, the predicted enhancement of the TLL parameter has been verified based on the infinite time-evolving decimation (iTEBD) {\cite{GV07}} analysis of the spin-chain model\footnote{We note that this model can be realized as a special case of Eq.~\eqref{latticeloss}, including the position-dependent one-body loss. To see this, one can use the mapping $\sigma^-\to b$ and $\sigma^z\to b^\dagger b-1/2$ with $b$ ($b^\dagger$) being the annihilation  (creation) operator of hard-core bosons ($b^{\dagger 2}=0$). A concrete experimental realization in ultracold atoms can be found in Ref.~\cite{YA17nc}.}, 
\eqn{H_{{\rm eff}}\!=\!\sum_{m=1}^{N}\left[\!-\!\left(J^{\rm r}\!+\!\left(-1\right)^{m}iJ^{\rm i}\right)\left(\sigma_{m}^{x}\sigma_{m+1}^{x}\!+\!\sigma_{m}^{y}\sigma_{m+1}^{y}\right)\!+\!\Delta\sigma_{m}^{z}\sigma_{m+1}^{z}\!+\!\left(-1\right)^{m}h_{s}\sigma_{m}^{z}\right]\!,}
whose effective theory corresponds to the generalized sine-Gordon model. We note that the signature of the enhanced superfluidity is already evident at a modest scale, e.g., $10$-$20$ sites as shown in Fig.~\ref{fig:4crit}(b).
While the semicircular RG flows can easily go beyond the range of validity of the perturbative treatments, the functional RG analysis has also been given in Ref.~\cite{YAbook}. This analysis further validates the unconventional RG flows in perturbative regimes and also indicates a possible transition into a gapped phase in nonperturbative regimes  (Fig.~\ref{fig:4crit}(c)). 

From a broader perspective, a non-Hermitian TLL  has also been discussed by Affleck \emph{et al.} \cite{IA04} in the context of  vortex pinnings originally discussed by Hatano and Nelson for a single-particle lattice model \cite{HN96}. The analogous winding RG flows have also been discussed in the non-Hermitian Kondo models \cite{LAS18,NM18,Yoshi20}  (Fig.~\ref{fig:4crit}(d)), where the Bethe-ansatz techniques have been found to remain useful. The related techniques have been applied to solve Richardson and Gaudin's central spin model \cite{MG76,RWR63} in non-Hermitian regimes \cite{RDA18}. Unconventional superfluidity \cite{GA18,YK19} as well as topological phases and excitations \cite{SJP16,ZQB16,SN19,YT192,GC19} in non-Hermitian regimes have also been discussed.
While some of the  phenomena reviewed here have once been considered to be of purely academic interest, they are now ready to be explored in actual physical systems thanks to rich experimental toolkits available in AMO systems (see Sec.~\ref{Sec:PS}).

\begin{figure}[t]
\begin{center}
\includegraphics[width=13cm]{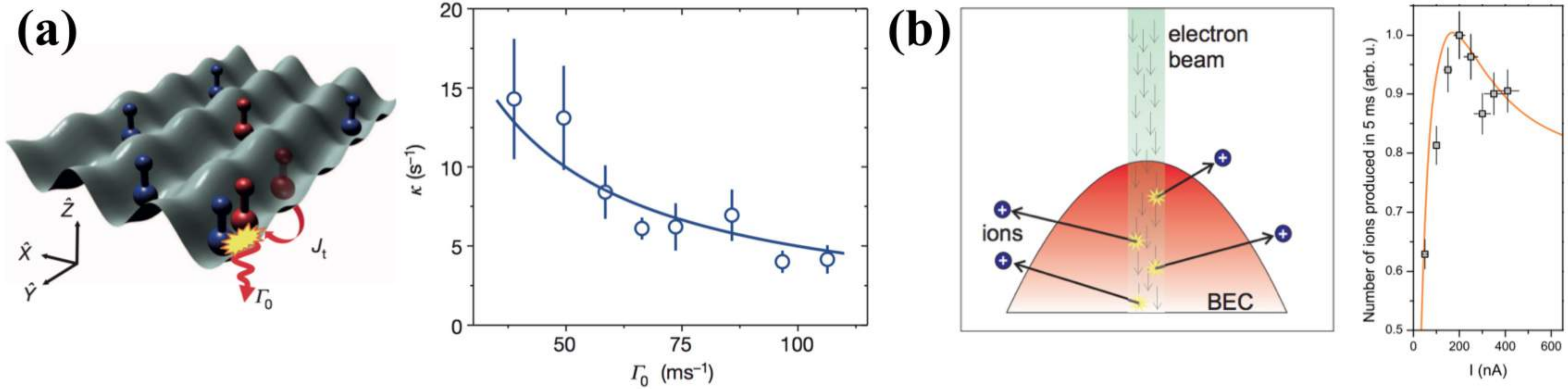}
\end{center}
\caption{(a) The left panel shows the schematic illustration of a two-body loss process induced by inelastic collisions between two molecules. The right panel shows experimental results of the loss rate against the rate of the inelastic collision $\Gamma_0$. The suppression of the loss rate at a large $\Gamma_0$ with $1/\Gamma_0$ dependence (indicated by the blue solid curve) is consistent with the continuous quantum Zeno effect. Adapted from Ref.~\cite{YB13}. Copyright \copyright\,   2013 by Springer Nature. 
(b) The left panel shows the schematic illustration of a one-body loss process induced by the electron beam that locally ionizes atoms. The right panel shows experimental results of the number of ions (i.e., the number of particles lost from an optical trap) against the intensity $I$ of the electron beam. The solid curve is obtained from the numerical simulations. The suppression of the number of lost particles for large $I$ is qualitatively consistent with the continuous quantum Zeno effect.  Adapted from Ref.~\cite{BG13}. Copyright \copyright\,  2013 by the American Physical Society.}
\label{fig:4exp}
\end{figure}

The nonunitary perturbation can also qualitatively alter underlying transient dynamical phenomena. A prototypical phenomenon is the so-called continuous quantum Zeno effect, where the coupling strength $\Gamma$ of the system to  measuring apparatus or environment is so large that broadening of linewidths beyond an energy scale $\Delta$ in the internal dynamics effectively suppresses the state-transition rate (i.e., the rate of jumps being observed). Similar to the analysis done in the Feshbach projection approach (cf. discussions below Eq.~\eqref{fpasreig}), this suppression can be best understood by analyzing the spectrum of the non-Hermitian effective Hamiltonian. Namely, the continuous quantum Zeno effect {\cite{BA00,ZP14,BM20,ZG20}} in general manifests itself as the suppressed imaginary parts of its complex eigenvalues, scaling as $\Delta^2/\Gamma$ in the limit of $\Gamma\to\infty$ (see Fig.~\ref{fig:4exp} for experimental observations). This feature is common both in few- and many-body systems while the latter can lead to rich phenomena such as enhanced particle correlations \cite{DAJ09,JJGR09,DSG09}, the shift of the quantum critical point \cite{YA16crit,TT17}, the engineering of non-Abelian gauge fields \cite{KS14}, multiband effects due to lattice confinement \cite{ZB14,YB13}, and (transiently) stabilizing otherwise unstable quantum states \cite{PB07,DAJ09,DS10,MOROZ2010491,ZZ20,NM202,NM20,BB20,PL20}. Regarding other dynamical aspects, the unconventional Kibble-Zurek mechanism is proposed {\cite{YS17,DB192,PX20}}, and insights from exceptional points in disordered non-Hermitian many-body systems are argued to be useful for understanding transient destabilization of the many-body localization due to environments \cite{LDJ19}. 

\exmp{(Continuous quantum Zeno effect in the dissipative Bose-Hubbard model). Consider ultracold bosons trapped in an optical lattice subject to two-body loss \cite{YA16crit,TT17}. The system is effectively described by the extended Bose-Hubbard Hamiltonian (cf. Eq.~\eqref{latticeloss})
\eqn{\label{bh2loss}
H_{{\rm eff}}=-J\sum_{m=1}^N\left(b_{m}^{\dagger}b_{m+1}+{\rm H.c.}\right)+\frac{U_{\rm r}-iU_{\rm i}}{2}\sum_{m=1}^N n_{m}\left(n_{m}-1\right),
}
where $b_m$ ($b_m^\dagger$) is the bosonic annihilation (creation) operator at site $m$, and $n_m=b_m^\dagger b_m$. An imaginary interaction parameter $-iU_{\rm i}$ results from an inelastic two-body collision. While this model is not exactly solvable, its asymptotically exact spectrum can be obtained by the strong-coupling-expansion analysis \cite{MF11} in the limit of $U_{{\rm r},{\rm i}}\gg J$. For instance, the decay rate $\Gamma_{\rm Mott}$ of the Mott-insulator state at filling $\rho=1$  (i.e., the imaginary part of the corresponding complex eigenvalue in Eq.~\eqref{bh2loss}) is given by \cite{YA16crit}
\eqn{\frac{\Gamma_{{\rm Mott}}}{N}=\frac{3J^{2}U_{\rm i}}{4\left(U_{\rm r}^{2}+U_{\rm i}^{2}\right)}.
}
The continuous quantum Zeno effect manifests itself as the suppression of the loss rate that scales as $\Gamma_{\rm Mott}\propto J^2/U_{\rm i}$ in the limit of $U_{\rm i}\to\infty$. We recall that this suppression and its scaling are generic features also found in the Feshbach projection approach (cf. Eq.~\eqref{fpanonrese}). Meanwhile, the real part of an eigenvalue represents an effective energy of each eigenstate, whose value is shifted by the enhanced interaction parameter $U_{\rm r}\to |U_{\rm r}-iU_{\rm i}|$, leading to a shift in the quantum critical point in favor of the Mott insulator phase \cite{YA16crit}. Since this is a lossy system, these spectral features exactly share  the same information with the Liouvillean for the dissipative dynamics \cite{BHJ93,TJM14}. From a broader perspective, the Hubbard-type non-Hermitian Hamiltonians have been analyzed in the earlier works \cite{LRA98,FT98} in the context of vortex pinning or the dielectric breakdown (see also Refs.~\cite{OT10,PA20,TK19}).
}

\begin{figure}[t]
\begin{center}
\includegraphics[width=14.5cm]{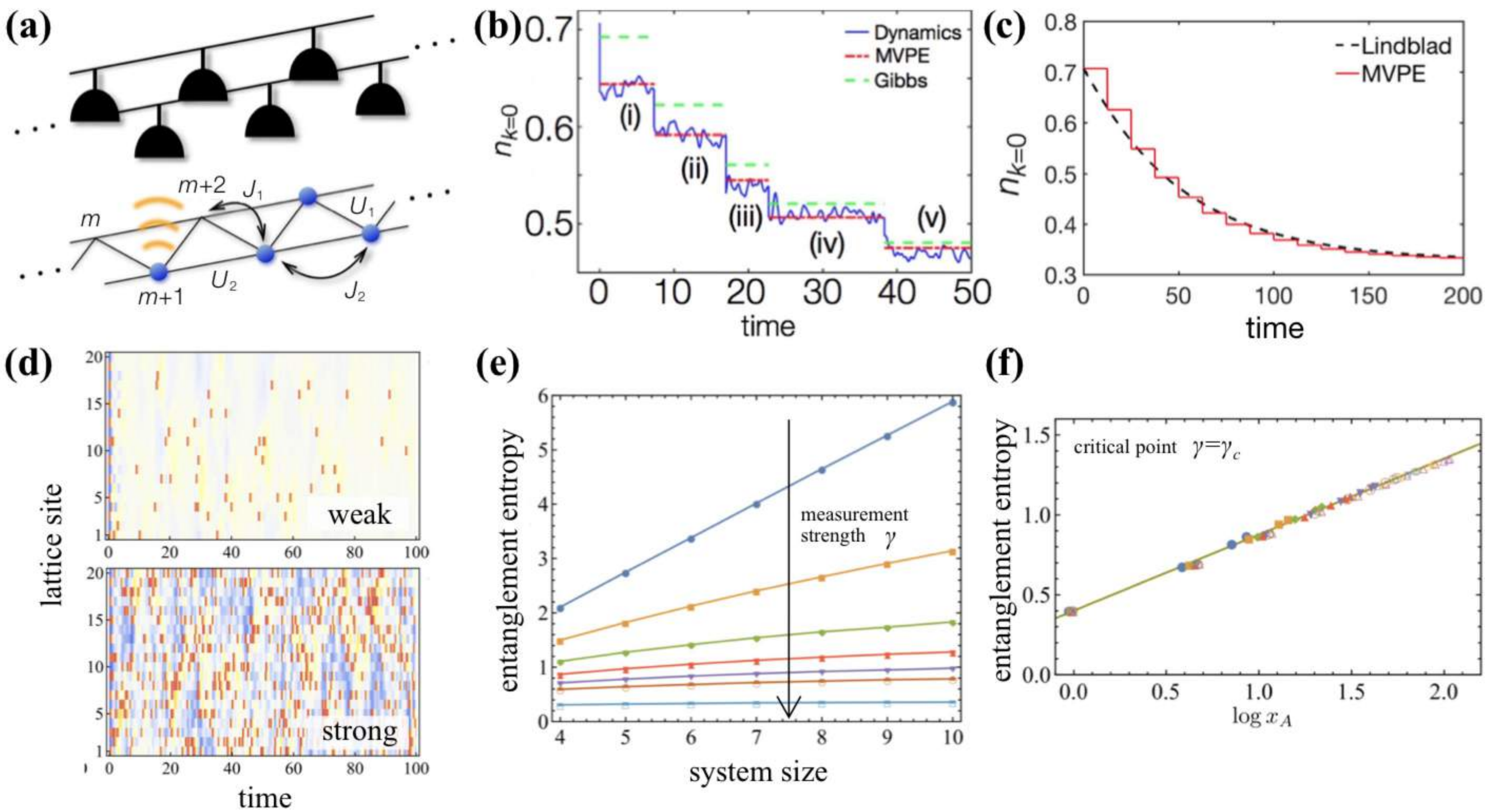}
\end{center}
\caption{(a) Schematic illustration of a nonintegrable quantum many-body system subject to continuous measurement of local occupation number via light scattering. (b) The blue curve plots numerical results on typical trajectory dynamics of the occupation number at zero momentum. The green dashed lines show the predictions from the Gibbs ensemble at fitted temperatures. The red horizontal lines represent the predicted values obtained from the efficient approach termed as the matrix-vector product ensemble (MVPE). (c) Comparison between the unconditional dynamics obtained from the master equation (black dashed curve) and the single trajectory result obtained from the MVPE (red lines). (d) Spatiotemporal trajectory dynamics of occupation number at each lattice site in the setup illustrated in (a). Top and bottom panels correspond to dynamics under weak and strong measurement strengths, respectively. (e) The corresponding half-chain entanglement entropy at different measurement strengths $\gamma$. As $\gamma$ is increased, the entanglement entropy $S_{\rm EE}$ exhibits the qualitative change from the volume-law scaling $S_{\rm EE}\propto L$ to the area-law scaling $S_{\rm EE}\propto L^0$. (f) At the critical point $\gamma=\gamma_c$ between the two phases found in (e), the entanglement entropy is consistent with the logarithmic law $S_{\rm EE}\sim(c/3)\log L$ (cf. Eq.~\eqref{EEcc}) with the extracted central charge being $c\simeq 1.33$. Adapted from Refs.~\cite{YA18therm,YF20}.}
\label{fig:4heating}
\end{figure}

Another fundamental phenomenon in open many-body systems is  heating and the subsequent thermalization \cite{LDA16,GF10,PH10,SJ14,YA18therm,ST20,BA20}. This phenomenon is  ubiquitous in many realistic setups; for instance, when a jump operator is Hermitian (or more generally normal), the steady state of the master equation should be the infinite-temperature state\footnote{This fact can be inferred from the facts that the master equation~\eqref{lindbladchap2} permits the solution $\rho\propto 1$ (identity matrix) when $L_j$ is Hermitian (or normal) for all $j=1,2,\ldots,M$ and that, in most cases, this is the unique solution due to the Perron-Frobenius theorem (see discussions below Eq.~\eqref{masterstoch}).}. As this steady state is featureless, interesting and nontrivial phenomena can basically occur in transient heating regimes.  A prototypical model to study such phenomena is an open many-body system subject to dephasing, or said differently, continuous position measurement, whose jump operator is given by $L_m\propto b^\dagger_m b_m$ (Fig.~\ref{fig:4heating}(a)). As the jump term conserves the particle number, the non-Hermitian analysis does not provide much insights in this case; thus, it is crucial to analyze single-trajectory dynamics accompanying quantum jumps \cite{PH10,SJ14,YA18therm}. 

To solve such a many-body problem,  one usually has to resort to a numerical technique such as tDMRG, which, however, becomes increasingly challenging at longer times due to high computational cost. Recently, it has been pointed out that, for generic many-body systems obeying the eigenstate thermalization hypothesis (ETH), the  significant simplifications can be made possible. Specifically, it has been shown that the master-equation dynamics can efficiently be approximated by considering only a (typical) single trajectory \cite{YA18therm}. This finding allows one to replace the diagonalization of a $D^2\times D^2$ Liouvillean ($D$ being the dimension of the Hilbert space) by that of a $D\times D$ operator with {\it no} need of taking the ensemble average over trajectories, thus enabling the efficient simulation of open many-body systems (Fig.~\ref{fig:4heating}(b,c)). From a broader perspective, a single trajectory dynamics has also been gaining attentions owing to its possible relevance to studies on noisy unitary circuits \cite{HP16,BL17,NA17,MK18,LY18,CA19,SB19}. An intriguing possibility of the measurement-induced transition into a critical state at the trajectory level has recently been pursued in realistic open many-body systems in steady-state regimes \cite{XC19,TQ20,YF20} and in unstable regimes \cite{GS20} (see Fig.~\ref{fig:4heating}(d-f) and its figure caption for further details).

\exmp{(Heating and measurement-induced criticality in chaotic trajectory dynamics). Consider a chaotic (i.e., nonintegrable) many-body system under continuous  measurement. Let us consider the following Hamiltonian \cite{KH14} and the jump operator
\eqn{
H&=&\!-\!\sum_{m}\left(J_{1}b_{m}^{\dagger}b_{m+1}\!+\!J_{2}b_{m}^{\dagger}b_{m+2}+{\rm H.c.}\right)+\sum_{m}\left(U_{1}n_{m}n_{m+1}+U_{2}n_{m}n_{m+2}\right),\\
L_m&=&\sqrt{\gamma}n_m.
}
This model physically corresponds to the site-resolved position measurement realized in quantum gas microscopy (see Fig.~\ref{fig:4heating}(a) and Example~\ref{qgmexmp}). 
Its single-trajectory dynamics is given by {Eq.~(\ref{trajD})}, 
where $H_{\rm eff}=H-(i\gamma/2)\sum_m n_m^2$. In the weakly perturbed regime, where the measurement strength $\gamma$ is smaller than any energy scales, it has been shown that $|\psi_{\rm traj}\rangle$ exhibits a sharp energy distribution and (almost always) thermalizes itself as a consequence of  the ETH  \cite{YA18therm}. Meanwhile, as $\gamma$ is increased, intriguing signatures of the eigenstate transition and the associated criticality at the transition point have been found \cite{YF20}. Interestingly,  these features are qualitatively consistent with predictions from the random circuit dynamics \cite{LY19}, while the quantitative discrepancies indicate that  measurement-induced criticality in quantum trajectory can exhibit conformal criticality different from the conventional universality classes (see Fig.~\ref{fig:4heating}(d-f) and Sec.~\ref{nonunitary_cft}). We emphasize that such transition and the associated criticality are accessible only when we go beyond the unconditional, master-equation dynamics.
}

Finally, going beyond AMO systems, open-system treatments have also been applied to understand chaotic behavior in quantum chromodynamics (QCD). This is because the Euclidean Dirac operator at nonzero baryon chemical potential is intrinsically non-Hermitian, while it is anti-Hermitian in the absence of the chemical potential. In the weak non-(anti-)Hermiticity regime, a rigorous mathematical approach based on the method of orthogonal polynomials has been applied to analyze the spectral density and correlations in complex spectra of non-Hermitian random matrices \cite{RMTbook}. Similar techniques have found applications to the effective Lagrangian with the imaginary gauge potential \cite{EKB97,EKB972}. These examples share the common feature that two-point functions  can be factorized into a bosonic and fermionic part, which has its root in the Toda lattice equation \cite{KS04}. 
Further discussions on non-Hermitian random matrix theories can be found in an excellent review paper \cite{Fyodorov_2003}. In the strong non-(anti-)Hermiticity regime of the Dirac operator, the spectral feature is similar to that of the Ginibre ensemble \cite{JG65}, where the elliptic law holds true; several features beyond these results have recently been reported in Refs.~\cite{AG18,AG19,LACT19,DS19,CT19,WK20}.
\exmp{(Eigenvalue statistics of chiral non-Hermitian random matrices in QCD).  
Consider a non-Hermitian random matrix $H-i\gamma\Gamma$, where $H$ and $\Gamma$ are taken independently from the Gaussian unitary ensemble with the same variance, and $\gamma$ is the dimensionless parameter controlling the strength of the non-Hermiticity in the model. This class of random matrices is of physical importance because of its relevance to  QCD, in particular, to the spectrum of the Dirac operator at nonzero chemical potential. One of the well-known key results in this context is the complex eigenvalue statistics \cite{FYV97}
\eqn{
P\left(\Lambda_{1},\Lambda_{2},\ldots,\Lambda_{N}\right)={\cal N}\prod_{i<j}|\Lambda_{j}-\Lambda_{k}|^{2}\!\exp\left[\frac{-N}{1-\tau^{2}}\sum_{j=1}^{N}\left[|\Lambda_{j}|^{2}\!-\!\frac{\tau}{2}\left(\Lambda_{j}^{2}\!+\!\Lambda_{j}^{*2}\right)\right]\right],
}
where $\tau=(1-\gamma^2)/(1+\gamma^2)$. In the strong non-Hermiticity case, i.e., $\lim_{N\to\infty}(1-\tau)>0$,\footnote{Here we note that $\tau$ in general depends on $N$. This condition is to be contrasted with the weak non-Hermiticity regime $0<\lim_{N\to\infty}N(1-\tau)<\infty$, which implies the presence of a scaling relation $\tau\sim 1-C/N$ for a large $N$, where $C>0$ is a constant independent of $N$.} this distribution becomes equivalent to the Ginibre's formula \cite{JG65} while in the Hermitian limit $\tau\to 1$ it reduces to the conventional Wigner-Dyson statistics.
}

\subsubsection{Physical systems}
\label{Sec:PS}
Owing to remarkable experimental developments, 
rich open many-body phenomena reviewed in Sec.~\ref{Sec:CDC} have become relevant to a variety of physical systems. We here briefly outline these developments with a particular focus on advances in AMO many-body systems.

A quantum many-body system can be coupled to an external world in such way that its decay processes such as losses of particles are engineered in a highly controllable manner. The loss of particles can be described on the basis of the quantum optical approach, provided that kinetic energies acquired by particles during loss processes are much larger than energy scales of optical potentials so that particles can immediately escape from the trap before influencing the internal dynamics. This condition has been very well met in typical AMO systems.  
 Different types of loss processes have been realized depending on the number of particles lost at each quantum jump event. Firstly, one-body loss process has been realized by using near-resonant probe light  \cite{KSJ98,MKO99,AT03,RS05,YA17nc,LHP17}, as well as by recoil energies due to light-induced transitions \cite{BWS09,RB16} and ionization due to electron beam \cite{GT08,WP09,BG13,LR16}.
 These processes allow one to implement \cite{BP11,WD11,VI14,YT20}  the quadratic anti-Hermitian terms in the effective Hamiltonian (cf. the first terms in Eqs.~\eqref{latticeloss} and \eqref{contloss}). 
 
 Secondly,  the control of two-body loss processes in many-body systems has firstly been demonstrated in Ref.~\cite{NS08}. There, weakly bound molecules have been prepared in optical potentials, and energies injected by inelastic collisions between two particles are large enough to cause loss of particles from the trap. Two-body loss processes have also been realized by chemical reactions between ground-state KRb molecules \cite{YB13} and light-assisted inelastic collisions \cite{TT17}. 
 On another front, there has recently been significant interest in many-body systems associated with Rydberg excitations, including possibilities to realize programmable quantum simulators \cite{Barredo1021,Endres1024}, unconventional nonequilibrium states \cite{LTE12,BH17,YA19L}, and crystalline equilibrium states \cite{SP12}. This class of systems naturally accompany decaying processes due to finite lifetimes of Rydberg states. In particular, an exotic two-body loss, which occurs at two atomic positions separated from each other by a certain distance, has been proposed in Refs.~\cite{AC12,EB14}. In general, two-body loss processes lead to anti-Hermitian interaction terms in the effective Hamiltonian \cite{NS08,JJGR09,YB13,ZB14,TT17} (cf. the second terms in Eqs.~\eqref{latticeloss} and \eqref{contloss}).  Finally, three-body loss processes have also been realized in experiments  \cite{KT06,EU15}, which can effectively induce the anti-Hermitian three-body interaction \cite{PB07,RA07,DAJ09,DS10,CYC11}. As mentioned before, in these lossy systems, the non-Hermitian analysis can be useful to understand essential features of underlying physics in their unconditional dissipative dynamics (as well as in conditional ones). 

 Remarkable developments in measurement techniques of AMO systems at the level of single quanta also provide new possibilities to study conditional {\it many-body} dynamics beyond the unconditional one, as originally pioneered in small quantum systems \cite{HS132}. To be concrete, we here explain one possible procedure to access conditional dynamics in lossy systems.
 First of all, we note that an initial many-body state of trapped atoms can be prepared at the single-particle resolution  \cite{IR15}. Then, suppose that the system  evolves in time under loss processes. After a certain time interval, the total number of particles in the final state can be measured at the single-atom resolution by quantum gas microscopy  \cite{PPM152}. Experimentally, the fidelity of measuring the atom number has already reached almost unit fidelity (99.5\% according to Ref.  \cite{SJF10}). Comparing this measured atom number with  the initial value, one can classify trajectories according to the number of jumps that have occurred during the time evolution. To describe such conditional dynamics, one should use the full-counting density matrix introduced in Eq.~\eqref{fcs2} rather than the unconditional density matrix. The exact non-Hermitian evolution corresponds to the no-count case, while key signatures of a non-Hermitian system can be transferred to other realizations accompanying quantum jumps (at least) when conditional dynamics is concerned  \cite{YA18}. 
 We note that this type of selective procedures of certain realizations have routinely  been performed in ultracold atomic experiments in rather different contexts such as 
the simulation of an effectively lower-temperature state (see, e.g., Refs.~\cite{EM11,FT15,IR15,PM16f}). It is noteworthy that the similar idea of studying conditional dynamics has been realized in a number of quantum optical setups \cite{TJS16,LX17,NaM19,Quiroz-Juarez:19,CW20,SCW20}.

The analysis of conditional trajectory dynamics accompanying quantum jumps  becomes particularly important when the jump process conserves the particle number (as in, for example, dephasing). Such a situation has also been routinely realized in AMO many-body setups, where dephasing can be introduced by dispersive light scattering \cite{PYS14,RB16,LHP17}. While most of the earlier studies have so far analyzed the dynamics at the level of the master equation,  conditional trajectory dynamics should readily be explored by combining such light scattering techniques with the optical setup for high spatially resolved imaging as realized in quantum gas microscopy. This possibility has been theoretically pursued in Ref.~\cite{YA15} and experimental efforts towards this direction have recently been reported in Refs.~\cite{PYS15,AA16,SS192,MM19}. Further developments of the related techniques will allow one to study a variety of measurement-induced quantum phenomena, including trajectory-level heating as well as the measurement-induced criticality as reviewed in Sec.~\ref{Sec:CDC}.

From a broader perspective, the quantum optical approach has also found applications to a broad range of systems beyond ultracold gases. Examples include trapped ions \cite{JE03,MM12}, atom-cavity systems \cite{BK10,BF13,MG16,Dogra1496,RCEI18}, optomechanical systems \cite{AM14}, quantum transport problems \cite{KM11,LM19,CL19,PT14},  and photosynthetic complexes \cite{MM08}. On another front, noisy intermediate-scale quantum (NISQ) technology may provide an entirely different way to simulate conditional dynamics of open many-body systems. Specifically, one may optimize parameters in the unitary gates to best approximate (normalized) non-Hermitian many-body dynamics (see Fig.~\ref{fig:4nonloc}(a)); this approach has already found applications in certain quantum chemistry problems \cite{Moll_2018,MM20}. 
The logarithmic scaling found in Theorem~\ref{nonloctheoremnum} indicates that, for a modest system size, it should be feasible to use the near-term quantum computers to simulate conditional nonunitary dynamics, especially non-Hermitian evolution, via the available unitary-circuit architectures \cite{Moll_2018}.  

\subsubsection{Beyond the Markovian regimes}\label{Sec:BMR}
There have been remarkable progresses  in analyzing open many-body systems beyond the realm of the Markovian regimes considered above. Studies on a possible extension of the quantum optical approach to a non-Markovian case can be traced to the early work \cite{IA94}, where the finite environmental correlation time $\tau_{\rm env}>0$ has been taken into account by introducing auxiliary bosonic modes. When the non-Markovian dynamics is local in time,  the master equation can still be useful in which  the jump operators $ {L}_{m}(t)$ and the associated coefficients $\gamma_{m}(t)$ become in general time-dependent \cite{BHP07}. Yet, possible negativity of a coefficient $\gamma_{m}(t)$  makes it challenging to apply the standard quantum optical approach to non-Markovian dynamics. Interestingly, the subsequent study \cite{PJ08} has extended the quantum trajectory approach to a non-Markovian case by introducing a ``backward" jump operator $ {D}_{\alpha\to\alpha'}^{m}$ for $\gamma_{m}(t)<0$, where $\alpha$ and $\alpha'$ represent a source state  and a target state, respectively, and the operator acts on the state as $|\psi_{\alpha'}(t+\delta t)\rangle= {D}^{m}_{\alpha\to\alpha'}(t)|\psi_{\alpha}(t)\rangle$ such that $|\psi_\alpha (t)\rangle= {L}_{m}(t)|\psi_{\alpha'}(t)\rangle/{\| \cdot \|}$. Several alternative approaches have also been proposed, including the expanded Hilbert space \cite{dVI17}, and the explicit inclusion of a part of environmental degrees of freedom \cite{RT16,Dorda_2017}.

\begin{figure}[t]
\begin{center}
\includegraphics[width=14.5cm]{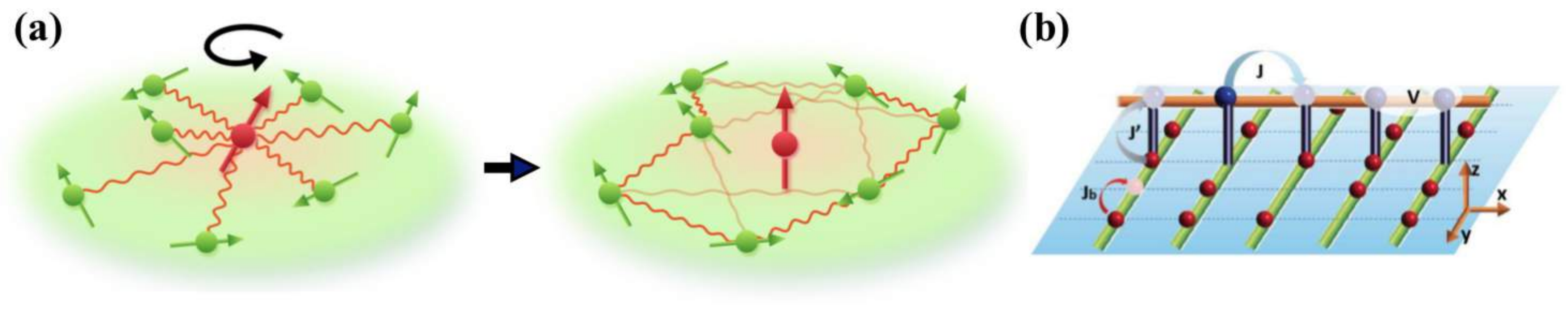}
\end{center}
\caption{Open quantum systems beyond the Markovian regimes. (a) Schematic illustrations of a quantum spin coupled to a generic many-body non-Markovian environment and the unitary transformation disentangling the spin from the environment degrees of freedom. Adapted from Ref.~\cite{YA18B}.   
(b) Schematic illustration of a lattice model subject to local non-Markovian dissipations, where environmental effects are modeled by independent  chains (green lines), each of which plays the role of a reservoir. Adapted from Ref.~\cite{YZ18}. Copyright \copyright\, 2018 by the American Physical Society. }
\label{fig:4nonmarkov}
\end{figure}

Going beyond AMO systems, it is also ubiquitous in nature (especially in solid state physics) that a quantum system is strongly correlated with an external environment such that the master-equation description can be violated \cite{UW99}. 
In this case, one has to explicitly take into account both system and environmental degrees of freedom. This line of studies have been initiated by the pioneering works by Caldeira and Leggett on a two-level impurity system coupled to bosonic bath modes (known as the spin-boson model) \cite{CAO81} as well as by Schmidt \cite{SA83} and Guinea, Hakim, and Muramatsu  \cite{GF85} on a mobile particle coupled to a bosonized environment. Subsequently, the equivalence between the spin-boson model and the Kondo model (in the infinite-bandwidth limit) was pointed out by Leggett et al. in Ref.~\cite{LAJ87}. The ensuing studies have developed a number of theoretical approaches to reveal the equilibrium properties of such a paradigmatic class of open quantum systems; prominent examples include the numerical renormalization group (NRG) \cite{WKG75,BR03,BCA10} and the Bethe-ansatz techniques  \cite{NK81,AN83,SP89}.  There have also been significant efforts towards unveiling nonequilibrium dynamics, including the real-time Monte Carlo  calculations \cite{STL08,WP09d,SM09,WP10,CG13}, the coherent-state expansion \cite{BS14,FS15,BCZ17}, the time-dependent NRG \cite{AFB05,LB14}, and the time evolving decimation \cite{NM15,DB17}. Despite these remarkable progresses, revealing the long-time many-body dynamics has remained a major challenge. To resolve this problem, an efficient variational approach based on the impurity-decoupling unitary transformation has recently been developed for both fermionic and bosonic non-Markovian environments \cite{YA18L,YA18B,YA19L,YA19A} (Fig.~\ref{fig:4nonmarkov}(a)).  Numerical approaches based on the quantum Monte Carlo techniques \cite{CZ14,YZ18} and the matrix-product states \cite{dVA15,TC20} have also been proposed to analyze a quantum system coupled to multiple bosonic non-Markovian baths  (Fig.~\ref{fig:4nonmarkov}(b)).

\subsection{Quadratic problems}\label{Sec:QP}
Many of phenomena in open many-particle problems that are quadratic in terms of creation/annihilation operators can find their counterparts in classical non-Hermitian systems reviewed in Secs.~\ref{sec3} and \ref{sec5}. Therefore,  we here  briefly review several aspects in a class of quadratic problems that are important in view of non-Hermitian {\it quantum} physics. A general description of non-Hermitian quadratic problems is given in Appendix~\ref{app2} for both fermions and bosons.

First of all, we recall that the master equation can formally be mapped to a non-Hermitian system defined on a doubled Hilbert space {(usually called the \emph{Liouville space} \cite{EM09})}, where a density matrix of the original open system is vectorized. This idea has once been discussed in the context of a single-particle quantum master equation \cite{FWR90}. In other early works \cite{MMA97,OG04}, this direction  has been pursued for the purpose of analyzing the classical master equation (cf. Eq.~\eqref{populationeq}) of a reaction-diffusion process by the quadratic nonunitary quantum field theory. More recently, the vectorization method has found applications to the quantum master equations \cite{PT12,PA19,HJ20}. For a specific class of problems in quadratic fermions, this vectorization idea permits the so-called {\it third} quantization approach \cite{PT08}, which has been used to obtain solvable non-Hermitian lattice models \cite{SN19,SN19b}. Importantly, based on the fermionic Gaussian states, a classification of steady states in driven-dissipative quadratic fermions has been given by constructing the non-Hermitian zero mode of the vectorized master equation \cite{CEB2013} (see also Ref.~\cite{LS20}). A possible protocol to probe topology there has also been proposed \cite{CEB2018}.

Non-Hermitian solvable models have also been discussed in a number of other contexts. For instance, the early work by Baxter has shown that a certain class of non-Hermitian PT-symmetric spin chains (known as the clock chains with $\mathbb{Z}_n$ symmetry) can be mapped to models of quadratic parafermions \cite{BRJ892,RJB89} (see the review paper~\cite{Fendley_2014} for further developments). In the context of open quantum systems, the exact dynamics of quadratic models under measurement have been discussed in Refs.~\cite{YA18,DB19}, where the violation of the Lieb-Robinson bound in lossy conditional dynamics has been found in many-particle regimes (cf. Fig.~\ref{fig:4nonloc}(b)).
This finding is a direct manifestation of the nonlocal nature in dynamics conditioned on measurement outcomes (cf. {Sec.~\ref{Sec:RCD}}), and is absent at the level of the master equation. The emergent supersonic propagation can be best understood from the analysis of the underlying non-Hermitian Hamiltonian~\eqref{pt2S}, while this feature is maintained in the presence of  quantum jumps as long as  conditional dynamics is of interest \cite{YA18}. 
Meanwhile, topological aspects of non-Hermitian solvable fermionic models have also been investigated in Refs.~\cite{ZQB16,KM17,JH182,LE20}; we will revisit further developments related to band topology in Sec.~\ref{sec5}.

Finally, in the context of the Bose-Einstein condensate, it is well known that weakly excited states on top of the macroscopically occupied bosonic mode can be analyzed by non-Hermitian quadratic bosonic operators \cite{BECbook}. Specifically,  linearizing the Gross-Pitaevskii equation around the equilibrium condensate, one can reduce the problem of determining excitation energies to the diagonalization of the non-Hermitian matrix in the basis of the Bogoliubov quasiparticles (see Appendix~\ref{app2}). The presence of positive imaginary parts in the spectrum indicates parametric amplification and the subsequent dynamical instability \cite{CC01,BNR14}. Such a non-Hermitian aspect of Bogoliubov excitations has recently been revisited in the context of band topology \cite{OT20}. Lattice models that demonstrate the same essential physics, such as the bosonic Kitaev chain \cite{MA18}, have been proposed to be realizable in optomechanical arrays and superconducting circuits subject to parametric drivings \cite{AN19}. Yet, a generic non-Hermitian quadratic problem of single-mode boson  has been analyzed in detail in the earlier work \cite{SMS04}.

\subsection{Nonunitary conformal field theory}
\label{nonunitary_cft}
Historically, non-Hermitian or nonunitary quantum field theory appeared naturally in the studies of $(1+1)$D {conformal field theories} (CFTs) \cite{DFP12}, where there is an {infinite} number of independent conformal symmetries \cite{AAB84}. These theories are widely used to describe {critical} 1D quantum systems, as well as their 2D statistical mechanical counterparts. Owing to the high symmetry, CFTs are captured by only a few universal parameters such as central charges and conformal dimensions. Interestingly, it turns out that many exactly solvable CFTs are nonunitary as we review below. 
\\ \\ {\it Minimal models}

\vspace{3pt}
\noindent
On the 2D complex plane, local conformal transformations consist of all the analytic (holomorphic) functions $f(z)=\sum_{n\in\mathbb{Z}}c_nz^n$ and their infinitesimal versions $z^{n+1}\partial_z$'s ($n\in\mathbb{Z}$) form the Witt algebra. When the theory is quantized, the symmetry algebra is center extended to the \emph{Virasoro algebra} \cite{VMA70}, which also has an infinite number of generators $L_n$'s satisfying
\begin{equation}\label{virasoroa}
[L_m,L_n]=(m-n)L_{m+n}+\frac{c}{12}\delta_{n,-m}m(m^2-1),
\end{equation}
where $c$ is the \emph{central charge} that commutes all $L_n$'s. The state space of any CFT\footnote{Rigorously speaking, the full conformal-symmetry algebra consists of two copies of Virasoro algebra, which correspond to both holomorphic and anti-holomorphic infinitesimal transformations. What we discuss here is actually the {diagonal CFT, whose representations of the two Virasoro algebras coincide with each other}.} is specified by a representation of the Virasoro algebra, and the simplest class with finite-dimensional representations is called the \emph{minimal models}. In analogy with the fact that a finite-dimensional angular momentum ($\mathfrak{su}(2)$ algebra)  representation requires the spin number to be an integer or half-integer, minimal models require $c$ to take specific discrete values. Moreover, similar to the spin projection quantum number in the angular momentum theory, in any minimal model there exists a highest-weight state $|h\rangle$, which is created by a primary field and satisfies
\begin{equation}
L_n|h\rangle=0\;\forall n\in\mathbb{Z}^+,\;\;\;\;L_0|h\rangle=h|h\rangle,
\end{equation}
where $h$ is the \emph{conformal dimension}. It can be shown that the central charge $c$ and conformal dimension $h$ of minimal models $\mathcal{M}(p,q)$ can only take the following values \cite{FD84,AAB84}:
\begin{equation}
c=1-\frac{6(p-q)^2}{pq},\;\;\;\;
h_{r,s}=\frac{(pr-qs)^2-(p-q)^2}{4pq},
\label{ch}
\end{equation}
where $p>q\ge2$ are two coprime ($(p,q)=1$) integers and $1\le r<q$, $1\le s<p$. The finiteness of the representation dimension does not necessarily mean that the theory is unitary. In fact, by examining whether the Gram matrix $M_{jj'}=\langle j|j'\rangle$ of a basis $|j\rangle$'s is positive-semidefinite, one can find that most of the minimal models are \emph{nonunitary} except for those with $p=q+1$. These minimal models cover all the unitary CFTs with $c<1$, while CFTs with $c\ge1$ are all unitary \cite{FD84}. A prototypical example of a unitary minimal model is the critical Ising model, which corresponds to $\mathcal{M}(4,3)$ with $c=1/2$ \cite{FD84}. In contrast, a prototypical example of a nonunitary minimal model is the one for the Yang-Lee edge singularity \cite{MEF78}, whose effective theory has the action
{\begin{equation}
S=\int d^2\boldsymbol{x}\left[\frac{1}{2}(\nabla\phi(\boldsymbol{x}))^2
+\frac{1}{3}ig\phi(\boldsymbol{x})^3\right]\label{iphi3}
\end{equation}}
and corresponds to $\mathcal{M}(5,2)$ with $c=-22/5$ \cite{CJL85}.

It is well-known that the conformal dimension determines the algebraic (power-law) decay of spatiotemporal correlation functions of primary fields \cite{AAB84}. Meanwhile, the central charge plays a role under a {macroscopic} length cutoff $l$, which is the system or subsystem size. For example, the vacuum (Casimir) energy is proportional to $\propto cl^{-1}$ \cite{BHWJ86}. In light of the development of quantum information, it has become clear that $c$ of a unitary CFT manifests itself in the entanglement entropy $S_{\rm EE}$ of a subsystem \cite{PC04}:
\eqn{S_{\rm EE}\sim\frac{c}{3}\ln l,\label{EEcc}
}
where ``$\sim$" means that the rhs of this symbol shows the leading term for large $l$. 
This result stays formally valid for nonunitary minimal models, for which the entanglement $n$-R\'enyi entropy is given by \cite{DB15}
\begin{equation}
S_n\equiv\frac{1}{1-n}\ln{\rm Tr}[\rho_l^n]\sim\frac{c_{\rm eff}(n+1)}{6n}\ln l.
\label{Sn}
\end{equation}
Here $\rho_l$ is the reduced density operator of a length-$l$ subsystem. However, the central charge is now replaced by an effective one:
\begin{equation}
c_{\rm eff}=c-24h_{\min}=1-\frac{6}{pq},
\end{equation} 
where $h_{\min}=[1-(p-q)^2]/(4pq)$ is the smallest conformal dimension given in Eq.~(\ref{ch}). We mention that the entanglement entropy is well-defined because the left and right eigenstates coincide for nonunitary CFTs. While the subleading contributions to the R\'enyi entropy in general depend on the boundary conditions \cite{JC10}, they can also exhibit universal features characterized by the conformal dimension \cite{OK15}. 
Recently, there are attempts to generalize entanglement entropy to non-Hermitian systems  on the basis of the quantum group \cite{KC072,CR2017}. In this case, the left and right eigenstates differ in general and so do their entanglement entropies. Finally, we note that a non-Hermitian free-fermion realization (PT-symmetric SSH model at the PT-symmetry breaking critical point) of a $c=-2$ nonunitary CFT has been reported in Ref.~\cite{PYC19}. While this is not a minimal model, Eq.~(\ref{Sn}) turns out to be valid with $c_{\rm eff}=c=-2$ ($p=2$ and $q=1$), provided that the reduced density operator is defined in terms of both left and right eigenstates and is thus non-Hermitian.
\\ \\ {\it Topological quantum field theory}

\vspace{3pt}
\noindent
As mentioned in the beginning of this section, CFTs emerge in critical quantum (or statistical) systems, which often require parameter fine-tunings in order to reach the critical point. However, for a large class of gapped (2+1)D systems whose low-energy properties are captured by {topological quantum field theories} (TQFTs) \cite{AMF88}, the criticality of their edge theories is topologically ensured and CFTs naturally emerge without parameter fine-tunings \cite{GM89,EF04}. 
For example, the $\nu=5/2$ fractional quantum Hall effect (FQHE), whose ground state is given by the Moore-Read Pfaffian state, can be described by an Ising-type edge theory $\mathcal{M}(4,3)$ with $c=1/2$ \cite{GM1991}. A Levin-Wen model \cite{LMA05} with Fibonacci anyon excitations has an edge theory captured by the minimal model $\mathcal{M}(5,4)$ with $c=7/10$ \cite{FA2007,CG09,EA11}.

Interestingly, some  many-body wave functions proposed for describing factional quantum Hall effects turn out to correspond to nonunitary CFT edge theories. Examples include the Haldane-Rezayi state as a candidate for $\nu=5/2$ FQHE, which corresponds to a nonunitary CFT with $c=-2$ \cite{FDMH1988,VG1997}, and the Gaffnian state for $\nu=2/5$ FQHE, which corresponds to a nonunitary minimal model $\mathcal{M}(5,3)$ with $c=-3/5$ \cite{SHS2007}. Some nonunitary CFTs/TQFTs can be related to their unitary counterparts through \emph{Galois conjugations} \cite{EA2011}. 
For example, the Yang-Lee CFT/TQFT can be obtained from the Fibonacci CFT/TQFT through $t=e^{\frac{2\pi i}{5}}\to e^{\frac{4\pi i}{5}}$, which is a parameter in the $F$-symbol that serves as a building block of the Levin-Wen model \cite{LMA05}. In the Hamiltonian picture, it is well known that the complex conjugation, which is a special Galois conjugation $e^{\frac{2\pi i}{4}}\to e^{\frac{6\pi i}{4}}$,  
preserves the Hermiticity. However, general Galois conjugations, which leave the real spectrum unchanged, typically destroy the Hermiticity. 
Remarkably, the left/right ground eigenstate may have {no} physical realizations, i.e., it is impossible to construct a Hermitian parent Hamiltonian (of which the state of interest is the ground state) that is gapped and local \cite{MHF2012}. Recently, these novel many-body wave functions have been constructed in terms of tensor network states \cite{LL20}.
\\ \\ {\it Liouville conformal field theory}

\vspace{3pt}
\noindent
An important class of CFTs beyond minimal models is the \emph{Liouville} CFT  \cite{SN1990}:
\begin{equation}
S=\int d^2\boldsymbol{x}\left[\frac{1}{2}(\nabla\phi(\boldsymbol{x}))^2+\mu e^{2b\phi(\boldsymbol{x})}\right],\label{livcft}
\end{equation}
where $\mu\in{\mathbb R}$ is the so-called cosmological constant and $b$ is a complex constant that determines the central charge:
\begin{equation}
c=1+6Q^2,\;\;\;\;Q=b+b^{-1}.
\end{equation}
This CFT can be considered as a special case of the generalized sine-Gordon model~\eqref{sG} \cite{CMB05,YA17nc}. In fact, it is irrational, i.e., the state space cannot be decomposed into a finite number of irreducible representations of the Virasoro algebra. Yet, this CFT is solvable, in the sense that the analytical expressions of the three-point correlation functions have been obtained \cite{HD92,AZ96}. A Liouville CFT becomes nonunitary whenever $b$ is not real (otherwise, $c\ge 25$), especially for a purely imaginary $b$ such that $c\le 1$. Nevertheless, these nonunitary Liouville CFTs still correspond to some statistical mechanical models such as the Potts model \cite{MP13}, and consistent results of the three-point functions have been reported \cite{IY16}.

\subsection{Non-Hermitian analysis of Hermitian systems}
\label{Sec:NHH}
The knowledge on non-Hermitian systems may in turn help us gain new insights into Hermitian systems. 
This line of thoughts have appeared in many different subjects since 1950s, such as the Yang-Lee phase transition theory in statistical physics \cite{LTD52}, analytic properties of Wannier functions in condensed matter physics \cite{KW59}, and even the proof of the KAM theorem in classical mechanics \cite{AVI63}; in all of these applications, the analytic continuation  plays a central role. We here review these classical results as well as some recent progress in the context of nonequilibrium quantum dynamics \cite{EJ15}. We also review another relatively recent topic --- matrix product states \cite{MF92,perez07mps,FV08}, which are widely used in dealing with 1D Hermitian many-body problems, but are actually closely related to non-Hermitian analyses.
\\ \\ {\it Yang-Lee zeros and Fisher zeros} 

\vspace{3pt}
\noindent
In the seminal work~\cite{LTD52}, Lee and Yang considered the statistical mechanics of a general Ising model defined on a lattice $\Lambda$. In the presence of a homogeneous magnetic field, the Hamiltonian reads 
\begin{equation}
H=-\sum_{j,j'\in\Lambda}J_{jj'}\sigma^z_j\sigma^z_{j'}-h\sum_{j\in\Lambda}\sigma^z_j.
\end{equation}
Denoting the volume of the lattice as $|\Lambda|$ (i.e., the total number of spins), we can formally write down the partition function as
\begin{equation}
Z\equiv{\rm Tr}e^{-\beta H}=e^{-|\Lambda|\beta h}\mathfrak{p}(e^{2\beta h}),
\end{equation}
where $\mathfrak{p}(z)$ is a polynomial with degree $|\Lambda|$. Provided that all the couplings are ferromagnetic, i.e., $J_{jj'}\ge0$ $\forall j,j'\in\Lambda$, it was found that all the zeros of $\mathfrak{p}(z)$, dubbed \emph{Yang-Lee zeros} (or Lee-Yang zeros), are located on the unit circle in the complex $z$ plane. This implies $Z=0$ only if the magnetic field is {purely imaginary}, in which case $H$ is non-Hermitian. This result has been generalized to higher (even continuous) spins and inhomogeneous magnetic fields \cite{NCM74,LEH81}. The distribution of the Yang-Lee zeros contains crucial information about the thermodynamic properties of the original (Hermitian) Ising model. In particular, a necessary condition for observing a phase transition in the absence of magnetic fields is that some Yang-Lee zeros should approach $1$ on the real axis in the thermodynamic limit $|\Lambda|\to\infty$ \cite{LTD52}. Recently, Yang-Lee zeros have been observed in an NMR-based interferometric experiment \cite{WBB12,PX15}. The main idea is that an imaginary magnetic field multiplied by a real inverse temperature can be considered as a real field multiplied by an imaginary inverse temperature, which then becomes equivalent to a time development. It is also worth mentioning the Yang-Lee theory  has a dynamical counterpart in the context of full counting statistics \cite{FC13}, where time serves as the system size and the thermodynamic limit is the long-time limit; such dynamical Yang-Lee zeros have experimentally been observed \cite{BK17}. 

Following the proposal of Yang-Lee zeros, Fisher suggested that one can consider the complex extension of the inverse temperature, which is applicable to all the Hamiltonians including the Ising models \cite{FME65}. According to the Weierstrass factorization theorem, the generalized partition function can generally be written as
\begin{equation}
Z(z)\equiv{\rm Tr}e^{-zH}=e^{g(z)}\prod_n\left(1-\frac{z}{z_n}\right),
\end{equation}
where $g(z)$ is an entire function and $z_n$'s are called \emph{Fisher zeros}. Recently, there have been growing interest in Fisher zeros in the context of quench dynamics \cite{MH18}. Precisely speaking, to analyze quench dynamics in isolated quantum systems, the definition of Fisher zeros should be modified to be the zeros of $\langle\Psi_0|e^{-zH}|\Psi_0\rangle$, where $|\Psi_0\rangle$ is the initial state and $H$ is the postquench Hamiltonian. In analogy to the free energy density in statistical mechanics, we can define a {rate function} for such a quench process as
\begin{equation}
l(t)\equiv-\lim_{|\Lambda|\to\infty}|\Lambda|^{-1}\ln|\langle\Psi_0|e^{-iHt}|\Psi_0\rangle|^2.
\label{lt}
\end{equation}
If some Fisher zeros approach the {imaginary} axis in the thermodynamic limit, then $l(t)$ would exhibit certain singularity, such as the divergence of the time derivative, at some critical times. This phenomenon is often called a ``{dynamical quantum phase transition}" \cite{HM13}, in the sense that it is a quantum dynamical analogy of the non-analytical behaviors of free energies at thermal phase transitions. Owing to experimental developments, dynamical quantum phase transitions have been observed in various AMO platforms such as trapped ions \cite{JP17}, ultracold atoms \cite{NF18}, photonic quantum walks \cite{WK19}, and superconducting qubits \cite{GXY19}. We also note that dynamical quantum phase transitions have been studied for non-Hermitian lattice systems \cite{ZL18c}
\\ \\ {\it Wannier exponential decay and Lieb-Robinson bound}

\vspace{3pt}
\noindent
Analytic continuation also finds its applications to band theory. Such an application dates back to the seminal work by Kohn \cite{KW59}, who proves in a continuous setup that a Wannier function in 1D decays exponentially in space. A closely related result is that the many-particle state with one or several bands being filled has exponentially decaying spatial correlations \cite{CJD64}, since the spatial correlations are determined by the off-diagonal entries of the projector $P$ onto the bands and $P=\sum_{x\in\mathbb{Z}}\sum_{\alpha:{\rm occ.}}|W_{x\alpha}\rangle\langle W_{x\alpha}|$, where $W_{x\alpha}$ is the Wannier function of band $\alpha$ centered at site $x$.\footnote{Assuming $\|\langle x+r|W_{x\alpha}\rangle\|\le ce^{-\kappa |r|}$ $\forall \alpha$ with $c,\kappa>0$, we have $\|\langle x_1| P|x_2\rangle \| \le \sum_{x\in\mathbb{Z}}\sum_{\alpha\in{\rm occ.}}c^2e^{-\kappa(|x-x_1|+|x-x_2|)}=n_{\rm occ}c^2[|x_1-x_2|+\tanh (2\kappa)]e^{-\kappa|x_1-x_2|}$, where $n_{\rm occ}$ is the number of occupied bands. Here we use $\|\cdot\|$ instead of $|\cdot|$ because of the existence of internal states.} 
Below we review the lattice version of the latter result. 

\begin{figure}[!t]
\begin{center}
\includegraphics[width=12cm]{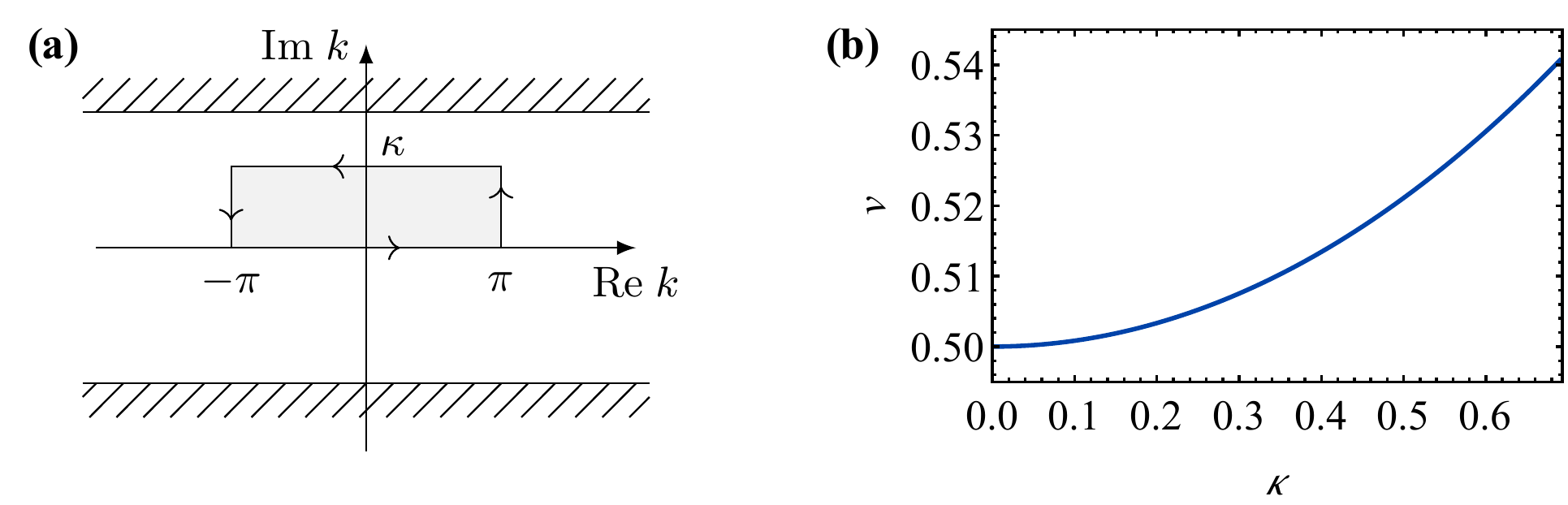}
\end{center}
\caption{(a) Contour in the complex wave-number plane that does not include any pole. Accordingly, for an analytic integrand $f(k)$ satisfying $f(k)=f(k+2\pi)$, we have $\int^{\pi+i\kappa}_{-\pi+i\kappa} dkf(k)=\int^\pi_{-\pi} dkf(k)$. (b) $\kappa$ dependence of the Lieb-Robinson velocity $v$ in Eq.~(\ref{vkappa}) for an SSH model (\ref{SSH}) with $J_1=1$ and $J_2=0.5$. The observation that $v$ increases monotonically with $\kappa$ is actually a universal property for any analytical Bloch Hamiltonian. Adapted from Refs.~\cite{ZG2019}.}
\label{fig:LR}
\end{figure}

We consider a 1D Bloch Hamiltonian $H(k)$ which is analytic on $\{k:|{\rm Im}k|<\kappa_0\}$ and satisfies $H(k)=H(k+2\pi)$. The former condition actually implies long-range hoppings are exponentially suppressed. When the wave number $k$ is extended to be complex, $H(k)$ generally becomes non-Hermitian and the band dispersions become complex. Suppose that there is a set of bands separated from the rest for $k\in\mathbb{R}$, 
these bands should stay isolated for a sufficiently small ${\rm Im}k$. This observation allows us to conclude that the projector 
\begin{equation}
P(k)=\oint_C \frac{dz}{2\pi i} \frac{1}{zI-H(k)},
\end{equation}
where $C$ encircles the bands of interest, is analytical on $\{k:|{\rm Im}k|<\kappa_1\}$ for some $\kappa_1\le \kappa_0$. Accordingly, we can shift the contour in the integral expression of the correlation function $\langle x| P | x'\rangle=\int^\pi_{-\pi}\frac{dk}{2\pi} P(k)e^{ik(x-x')}$ to $\int^{\pi\pm i\kappa}_{-\pi\pm i\kappa}$ with $\kappa<\kappa_1$ (see Fig.~\ref{fig:LR}(a)), obtaining 
\begin{equation}
\begin{split}
\langle x| P | x'\rangle&=\int^\pi_{-\pi}\frac{dk}{2\pi} P(k+i{\rm sgn}(x-x')\kappa)e^{ik(x-x')-\kappa|x-x'|} \\
\Rightarrow\;\;\;\;\|\langle x| P | x'\rangle\|&\le e^{-\kappa|x-x'|}\int^\pi_{-\pi}\frac{dk}{2\pi} \|P(k+i
\kappa)\|,
\end{split}
\end{equation}
where we have used $P(k+i\kappa)=P(k-i\kappa)^\dag \Rightarrow \|P(k+i\kappa)\|=\|P(k-i\kappa)\|$. We note that the generalization to higher dimensions is {not} straightforward. In fact, exponentially decaying Wannier functions in more than 1D require the Chern number to vanish \cite{BC2007}. Otherwise, both the Wannier functions and the correlation functions obey a {power-law} decay \cite{WTB13}.

Recently, the idea of analytic continuation of the wave number has been applied to {tighten} the 
Lieb-Robinson bound on the correlation propagation in free-fermion systems \cite{ZG2019}. Consider a quantum quench in a 1D free-fermion lattice, with the initial state being the ground state of a gapped Bloch Hamiltonian $H_0(k)$, while the postquench Bloch Hamiltonian $H(k)$ is not necessarily gapped. Denoting $P_0(k)$ as the ground-state projector of $H_0(k)$, we know that the time-evolved state at $t$ consists of all the single-particle modes in $P(t)\equiv e^{-iH(k)t}P_0(k)e^{iH(k)t}$ \cite{GZ18}. Suppose that $P_0(k)$ and $H(k)$ are both analytical on $\{k:|{\rm Im}k|<\kappa_0\}$, for some $\kappa<\kappa_0$ such that $H(k\pm i\kappa)$ stays diagonalizable, we can upper bound the correlation function at time $t$ by \cite{ZG2019}
\begin{equation}
\|\langle x| P(t) | x'\rangle\|\le C e^{-\kappa (|x-x'|-vt)},
\end{equation}
where $C$ is a time-independent constant depending on both $P_0(k)$ and $H(k)$, and the Lieb-Robinson velocity $v$ is given by
\begin{equation}
v=\kappa^{-1}\max_{k\in[-\pi,\pi],\alpha,\beta}{\rm Im}[\epsilon_\alpha(k+i\kappa)-\epsilon_\beta(k+i\kappa)],
\end{equation}
where $\epsilon_\alpha(k)$ is the dispersion of band $\alpha$. We can prove that $v$ increases monotonically with $\kappa$, so that its minimal value gives the maximal {relative} group velocity:
\begin{equation}
v=\max_{k\in[-\pi,\pi],\alpha,\beta}\left[\frac{d\epsilon_\alpha(k)}{dk}-\frac{d\epsilon_\beta(k)}{dk}\right].
\label{vkappa}
\end{equation} 
This idea has recently been generalized to interacting systems by considering analytically continued single-particle Green's functions \cite{ZW19}. These results are important both technically and conceptually --- they not only provide the tightest Lieb-Robinson bound for each system \cite{ZW19}, but also give a rigorous justification of the quasiparticle picture for quench dynamics\footnote{We remark, however, that the Lieb-Robinson bound can in general be violated in nonunitary dynamics conditioned on measurement outcomes, including non-Hermitian evolutions, where a crucial difference is that the state vector should be normalized during the quench dynamics  \cite{YA18,DB19} (see Sec.~\ref{Sec:RCD} and Fig.~\ref{fig:4nonloc}(b)).} \cite{EA15}.
\\ \\ {\it Matrix-product states}

\vspace{3pt}
\noindent
In the previous subsection, we have seen that a non-Hermitian analysis on band theory can help us understand the exponential decay of Wannier and correlation functions, and even the Lieb-Robinson bound. 
Complementary to such a momentum-space analysis, here we discuss a class of 1D quantum many-body states, called \emph{matrix-product states} \cite{MF92,perez07mps,FV08}, whose correlation properties can straightforwardly be understood from a non-Hermitian analysis in real space. These states are known to faithfully represent ground states of gapped and local 1D (Hermitian) Hamiltonians \cite{VF06}, a fact considered to underpin the efficiency of the DMRG algorithm \cite{US11}. 

For simplicity, we focus on bosonic (spin) translational-invariant MPSs. The basic properties are shared by the inhomogeneous MPSs, and the generalization to fermions (parafermions) is possible by introducing a $\mathbb{Z}_2$-graded ($\mathbb{Z}_n$-graded) algebra structure \cite{BN17,XWT17}. An MPS with bond dimension $D$ and length $L$ is generated by a set of $D\times D$ matrices $M_m$'s through
\begin{equation}
|\Psi\rangle=\sum_{\{m_j\}^L_{j=1}}{\rm Tr}[M_{m_1}M_{m_2}...M_{m_L}]|m_1m_2...m_L\rangle,
\label{MPS}
\end{equation}
where $|m_j\rangle$ denotes a local state at the $j$th site ($j=1,2,..,L$). Such a many-body wave function has a bounded entanglement entropy $\ln D$ for an arbitrarily large subsystem, and thus obeys the entanglement area law \cite{EJ10}. Provided that the MPS (\ref{MPS}) is normalized in the thermodynamic limit $L\to\infty$, we can perform a gauge transformation $M_m\to VM_m V^{-1}$ ($V\in\mathbb{C}^{D\times D}$ is invertible) such that 
\begin{equation}
\mathcal{E}(\;\cdot\;)\equiv\sum_m M_m\;\cdot\;M_m^\dag
\label{MPSchan}
\end{equation}
is a {unital channel}, i.e., $\sum_m M_mM_m^\dag =\mathbb{1}$ ($\mathbb{1}$ is the identity operator on the virtual Hilbert space $\mathbb{C}^D$). The spectrum of $\mathcal{E}$, which coincides with that of the dual quantum channel $\mathcal{E}^\dag$, is known to be the same as that of the non-Hermitian {transfer matrix} $\sum_mM_m\otimes M^*_m$ on the Liouville space, and is located within the unit circle in the complex plane \cite{WMM08}. For a {normal} MPS, whose associated unital channel (\ref{MPSchan}) has a {unique} fixed point $\mathbb{1}$, the spatial correlation in the thermodynamic limit can be upper bounded by (see Appendix~\ref{app3})
\begin{equation}
|\langle\Psi|O_XO_Y|\Psi\rangle - \langle\Psi|O_X|\Psi\rangle\langle\Psi|O_Y|\Psi\rangle|
\le\sqrt{D} \|O_X\|\|O_Y\|\|\mathcal{E}^l-\mathcal{E}^\infty\|,
\label{MPScor}
\end{equation}
where $O_{X/Y}$ is supported on $X/Y\subset\mathbb{Z}$, $l={\rm dist}(X,Y)-1$ and the superoperator norm $\|\mathcal{L}\|\equiv\max_{\|O\|_2=1}\|\mathcal{L}(O)\|_2$ ($\|O\|_2\equiv\sqrt{{\rm Tr}[O^\dag O]}$ is the Schatten-2 norm) is nothing but the operator norm on the Liouville space. According to Eq.~(\ref{rhonorm}), $\forall \epsilon>0$ there exists a constant $c_\epsilon$ such that $\|\mathcal{E}^l-\mathcal{E}^\infty\|=\|(\mathcal{E}-\mathcal{E}^\infty)^l\|\le c_\epsilon (\mu+\epsilon)^l$, where $\mu<1$ is 
the second largest value in the normed spectrum of $\mathcal{E}$ \cite{MF92}. This result links the exponential decay in spatial correlations in MPSs to the exponential convergence to the steady state of a stroboscopic open-quantum dynamics generated by $\mathcal{E}^\dag$. Note that the generalization to continuous quantum fields has been achieved \cite{VF10}. In this continuous limit, we can associate with an MPS an adjoint master equation (cf. Eq.~\eqref{lindbladchap2}), which is the differential version of a unital channel.

Due also to the non-Hermitian nature of the transfer matrix, especially the possible existence of nontrivial Jordan blocks (see Sec.~\ref{secspecdec}), there can be polynomial corrections to the exponentially decaying correlations in MPSs \cite{SO15}. In other words, the lhs of Eq.~(\ref{MPScor}) may not be upper bounded by $c\mu^l$ $\forall l$ with $c$ being a constant. We may readily construct such a state from a classical Markov chain specified by a transition-probability matrix $[T]_{ba}=p_{a\to b}$, which satisfies $p_{a\to b}\ge0$ and $\sum_b p_{a\to b}=1$ $\forall a$ \cite{GM18,GI19}. The MPS is generated by $M_{a\to b}=\sqrt{p_{a\to b}}|a)(b|$\footnote{Here $|\cdot)$ refers to a state in the virtual Hilbert space, which should be distinguished from a physical state denoted by $|\cdot\rangle$.} and is normal whenever the classical steady state $\bold{\pi}$ determined by $T\bold{\pi}=\bold{\pi}$ is unique and satisfies $\pi_a>0$ $\forall a$. The associated unital channel is given by
\begin{equation}
\mathcal{E}(\;\cdot\;)=\sum_{a,b}p_{a\to b} |a)(b|\cdot|b)(a|,
\end{equation}
whose spectrum (nonzero part) coincides with that of $T$. Choosing $O_X=|a\to b\rangle\langle a\to b|$ and $O_Y=|a'\to b'\rangle\langle a'\to b'|$ that are separated by $l$ sites, we have  
\begin{equation}
\langle\Psi|O_XO_Y|\Psi\rangle - \langle\Psi|O_X|\Psi\rangle\langle\Psi|O_Y|\Psi\rangle=\pi_a p_{a\to b}[T^l-T^\infty]_{a'b}p_{a'\to b'},
\label{Tcor}
\end{equation}
where $T^\infty=\pi\bold{1}^{\rm T}$ ($\bold{1}\equiv[1,1,...,1]^{\rm T}$), and therefore a polynomial correction appears as long as $T$ is non-diagonalizable. A minimal construction  
can be achieved by a three-state Markov chain specified by $p_{1\to2}=(2-\sqrt{3})p$, $p_{2\to3}=2p$, $p_{3\to1}=(2+\sqrt{3})p$ ($p\le 2-\sqrt{3}$) and $p_{a\to b}=0$ for any other $a\neq b$. One can check that the spectrum of the unital channel consists of $1$, $1-3p$ and $0$, so the exponential part of a correlation function should be $(1-3p)^l$. In addition, since $T$ has a $2\times2$ Jordan block, there is generally a prefactor linear in $l$. For example, when $a\to b=a'\to b'=2\to3$, Eq.~(\ref{Tcor}) reads $(\frac{2p}{9})^2(\frac{3p}{3p-1}l-1)(1-3p)^l$. Note that here the bond dimension is $D=3$, so such an MPS gives a counterexample of Lemma~22 in Ref.~\cite{FGSLB15}.

\section{Band topology in non-Hermitian systems\label{sec5}}
Interesting topological aspects of non-Hermitian systems have been pointed out in the context of exceptional points as mentioned in Sec.~\ref{seceptopo} \cite{DC01,WDH12}.
They arise 
from the topological structure of the Riemann surfaces around the exceptional points \cite{HWD01}.
In this section, we provide yet another topological structure of 
non-Hermitian systems --- 
\emph{band topology}.

\subsection{Brief review of band topology in Hermitian systems}\label{Sec:5rev}
Before discussing non-Hermitian band topology, it is worthwhile to briefly review the band topology in Hermitian systems. Understanding these well-established concepts would be helpful for considering appropriate non-Hermitian generalizations. Moreover, later we will explicitly apply the results obtained for classifying Hermitian systems. 

\subsubsection{Definition of band topology}\label{Sec:5def}
Consider a general Hermitian free-fermion system on a $d$D lattice $\Lambda$ with translation invariance and under the periodic boundary condition (PBC). For simplicity, we focus on particle-number conserving systems, while the case without particle-number conservation will later be illustrated by specific models. On a single-particle level, we write down the Hamiltonian as
\begin{equation}
H=\sum_{\boldsymbol{r},a,\boldsymbol{r}',a'} H_{\boldsymbol{r}a,\boldsymbol{r}'a'}|\boldsymbol{r}a\rangle\langle \boldsymbol{r}'a'|,
\end{equation} 
where $|\boldsymbol{r}a\rangle$ is the single-particle state localized at site $\boldsymbol{r}\in\Lambda$ with internal state $a$, which includes spins, obitals, sublattices, etc. We assume $H_{\boldsymbol{r}a,\boldsymbol{r}'a'}= H^*_{\boldsymbol{r}'a',\boldsymbol{r}a}=H_{\boldsymbol{r}-\boldsymbol{r}',aa'}$, so that we can Fourier transform it to obtain
$H=\sum_{\boldsymbol{k},a,a'} [H(\boldsymbol{k})]_{a,a'}|\boldsymbol{k}a\rangle\langle\boldsymbol{k}a'|$,
where 
\begin{equation}
[H(\boldsymbol{k})]_{a,a'}=\sum_{\boldsymbol{r}-\boldsymbol{r}'}H_{\boldsymbol{r}-\boldsymbol{r}',aa'}e^{-i\boldsymbol{k}\cdot(\boldsymbol{r}-\boldsymbol{r}')}
\end{equation}
is the Bloch Hamiltonian and $|\boldsymbol{k}a\rangle\equiv |\Lambda|^{-1/2}\sum_{\boldsymbol{r}}e^{i\boldsymbol{k}\cdot\boldsymbol{r}}|\boldsymbol{r}a\rangle$ is the quasi-momentum eigenstate with wave number $\boldsymbol{k}$ and internal state $a$. One can check from the Hermiticity of $H_{\boldsymbol{r}a,\boldsymbol{r}'a'}$ that $H(\boldsymbol{k})^\dag=H(\boldsymbol{k})$. The real eigenvalues $\epsilon_\alpha(\boldsymbol{k})$'s of $H(\boldsymbol{k})$ and the corresponding eigenstates $|u_{\boldsymbol{k}\alpha}\rangle$'s are called band dispersions and Bloch wave functions, respectively \cite{NWA76}.

Provided that $H_{\boldsymbol{r}a,\boldsymbol{r}'a'}$ only contains short-range hoppings\footnote{This is a typical situation in literature but is only a sufficient condition. We may loose the condition to be no slower than a power-law decaying long-range hopping $H_{\boldsymbol{r}a,\boldsymbol{r}'a'}\sim |\boldsymbol{r}-\boldsymbol{r}'|^{-\alpha}$ with $\alpha>d$, where $d$ is the spatial dimension.} and $a$ takes $n$ different values, the corresponding Bloch Hamiltonian $H(\boldsymbol{k})$ with $n$ bands is a \emph{continuous map} from the 
Brillouin zone $T^d$ ($d$D torus) to an $n\times n$ Hermitian-matrix space. 
Roughly speaking, band topology concerns the question about whether two such continuous maps, say $H_0(\boldsymbol{k})$ and $H_1(\boldsymbol{k})$, can continuously be deformed into each other under certain conditions. That is, whether there exists an 
interpolation $H_\lambda(\boldsymbol{k})$ with $\lambda\in[0,1]$, i.e., a continuous map from $T^d\times[0,1]$ to the matrix space of interest \cite{AK09}. We basically focus on the classification of insulators (or superconductors), for which we should prescribe a Fermi energy $E_{\rm F}$. In this situation, we only consider \emph{gapped} 
$H(\boldsymbol{k})$'s whose \emph{real} spectra are separated by $E_{\rm F}$ into (at least) two disjoint parts $\forall\boldsymbol{k}\in T^d$. More generally, we may impose certain symmetries to $H(\boldsymbol{k})$, which further constrain the matrix space.

Precisely speaking,  the above definition is called \emph{homotopy} equivalence, which is conceptually simple but practically hard to handle \cite{KR15}. In fact, there is a weaker version of equivalence, that is, 
$H_0(\boldsymbol{k})$ and $H_1(\boldsymbol{k})$ are said to be \emph{stably} equivalent if a continuous interpolation is possible between $H_0(\boldsymbol{k})\oplus H'(\boldsymbol{k})$ and $H_1(\boldsymbol{k})\oplus H'(\boldsymbol{k})$ for some gapped $H'(\boldsymbol{k})$ that respects the same symmetry constraint as $H_0(\boldsymbol{k})$. Physically, such a definition takes into account possible additional bands that are projected out from a more complete effective theory of interest, and this fact explains why the equivalence is stable. The classification for stable equivalence is well captured by \emph{$K$-theory} \cite{MK08}, which will briefly be reviewed in Sec.~\ref{ptAZ}. Note that the homotopy classification is more refined compared to that based on $K$-theory; in fact,  a homotopically nontrivial band topology may become trivial in the sense of $K$-theory \cite{KR15}. A well-known example is a two-band Hopf insulator \cite{MJE08}, which features 
a nontrivial Hopf linking number but becomes trivial in the presence of additional bands.

\subsubsection{Prototypical systems}\label{ProSys}
It is instructive to first look at some concrete models of topological free-fermion systems. The arguably most important example is quantum Hall or Chern (anomalous Hall) insulators in 2D \cite{KKv80}, which are characterized by integer (first) Chern numbers without the need of symmetry protection \cite{TDJ82}. Denoting the wave-number-resolved projector onto the Fermi sea as $P_<(\boldsymbol{k})\equiv\sum_{\epsilon_\alpha(\boldsymbol{k})<E_{\rm F}}|u_{\boldsymbol{k}\alpha}\rangle\langle u_{\boldsymbol{k}\alpha}|$, the Chern number is given by \cite{CCK16}
\begin{equation}
{\rm Ch}=i\int_{T^2}\frac{d^2\boldsymbol{k}}{2\pi }{\rm Tr}[P_<(\boldsymbol{k})[\partial_{k_x}P_<(\boldsymbol{k}),\partial_{k_y}P_<(\boldsymbol{k})]]\in\mathbb{Z}.
\label{Chgen}
\end{equation}
If there is only a single band below the Fermi surface, the integrand in the above formula reduces to the Berry curvature $B_{xy}(\boldsymbol{k})\equiv i(\langle\partial_{k_x}u_{\boldsymbol{k}}|\partial_{k_y}u_{\boldsymbol{k}}\rangle-\langle\partial_{k_y}u_{\boldsymbol{k}}|\partial_{k_x}u_{\boldsymbol{k}}\rangle)$, 
which can actually be applied to any isolated (not necessarily the ground) band \cite{XD10}. In fact, a Chern insulator can already be realized in a two-band model \cite{BAB13}:
\begin{equation}
H(k_x,k_y)=\sin k_x\sigma^x+\sin k_y\sigma^y+(M-\cos k_x-\cos k_y)\sigma^z.
\label{ChH}
\end{equation} 
 We can check from Eq.~(\ref{Chgen}) that the Chern number of the above model is $1$ ($-1$) when $0<M<2$ ($-2<M<0$) and vanishes when $|M|>2$. 
 
We note that a necessary condition to have a nontrivial Chern number is the explicit breaking of the time-reversal symmetry (TRS) $\mathcal{T}$. Otherwise, the integrand in Eq.~(\ref{Chgen}) is anti-symmetric over the Brillouin zone and thus the Chern number vanishes identically. Nevertheless, if the TRS is anti-involutory, i.e., $\mathcal{T}^2=-1$, we instead have a well-defined $\mathbb{Z}_2$ invariant given by \cite{KCL05,FL06}
\begin{equation}
\nu=\prod_{\Gamma}\frac{{\rm Pf} w(\Gamma)}{\sqrt{\det w(\Gamma)}}\in\{-1,1\}.
\label{Z2ind}
\end{equation}
Here $[w(\boldsymbol{k})]_{\alpha\beta}=\langle u_{-\boldsymbol{k}\alpha}|\mathcal{T}|u_{\boldsymbol{k}\beta}\rangle$ is a unitary matrix that satisfies $w(\boldsymbol{k})^{\rm T}=-w(-\boldsymbol{k})$ and thus becomes anti-symmetric at TRS-invariant high-symmetry points $\Gamma$'s (with each component being either $0$ or $\pi$), ${\rm Pf}$ is the Pfaffian and the branch of $\sqrt{\det w(\boldsymbol{k})}$ should be chosen such that it changes smoothly over the Brillouin zone. This formula (\ref{Z2ind}) applies also to 3D topological insulators \cite{FL07} and can greatly be simplified in the presence of an additional inversion symmetry into 
$\nu=\prod_{\Gamma}\prod_{j:{\rm occ.}}\xi_{2j}(\Gamma)$,
where $\xi_{2j}(\Gamma)$ is the parity of the $j$th Kramers pair in the occupied band at a high symmetry point \cite{FL07b}. A minimal realization of a 2D topological insulator with TRS is the four-band Bernevig-Hughes-Zhang model described by \cite{BBA06}
\begin{equation}\label{bhzhermitian}
H(\boldsymbol{k})=\begin{bmatrix} h(\boldsymbol{k}) & 0 \\ 0 & h^*(-\boldsymbol{k}) \end{bmatrix},
\end{equation}
where $h(\boldsymbol{k})$ is given in Eq.~(\ref{ChH}). This model exhibits both a TRS $\mathcal{T}=i(\sigma^y\otimes\sigma_0)\mathcal{K}$ and an inversion symmetry $P=\sigma_0\otimes\sigma^z$, and we can therefore apply the simplified formula 
to obtain $\nu=-1$ (nontrivial) for $0<|M|<2$ and $\nu=1$ for $|M|>2$, which correspond to a nontrivial and trivial Chern number of $h(\boldsymbol{k})$, respectively.

In fact, topological phases appear already in 1D, but only in the presence of symmetries. For example, if the Hamiltonian has a \emph{chiral symmetry} (CS) $\Gamma=\sigma^z\otimes I$, i.e., $\{H(k),\Gamma\}=0$ for $\forall k$, we can define two complementary projectors $\frac{1}{2}(1\pm\Gamma)$ under which $H(k)$ is block-off-diagonal. Denoting one of the block-off-diagonal parts as $q(k)\equiv\frac{1}{4}(1+\Gamma)H(k)(1-\Gamma)$, provided that $H(k)$ is gapped, we can define a \emph{winding number} as \cite{CCK16}
\begin{equation}
w=i\int^{2\pi}_0\frac{dk}{2\pi} \partial_k\ln\det q(k).
\label{wn}
\end{equation}
Such a topological number stays well-defined in the presence of an involutory TRS or particle-hole symmetry (PHS). A quintessential example is the SSH model \cite{SWP79}:
\begin{equation}
H(k)=-(J_1+J_2\cos k)\sigma^x-J_2\sin k\sigma^y,\;\;\;\;J_{1,2}\in\mathbb{R}^+.
\label{SSH}
\end{equation} 
We can check from Eq.~(\ref{wn}) that $w=1$ for $J_2>J_1$ and $w=0$ for $J_2<J_1$. If there is only an involutory PHS, which is inherent in superconductors on the mean-field level, we no longer have the block-off-diagonal structure, but can still define a $\mathbb{Z}_2$ number as \cite{AYK01}
\begin{equation}
\nu=\prod_{\Gamma=0,\pi}{\rm sgn}[{\rm Pf}(A(\Gamma))],
\label{DZ2}
\end{equation}
where $A(\Gamma)$ is an anti-symmetric real matrix related to $iH(\Gamma)$ through a basis transformation such that under the new basis (which always exists \cite{WE60}) the PHS operator is simply represented by $\mathcal{K}$. A prototypical model characterized by this $\mathbb{Z}_2$ number is the Kitaev chain \cite{AYK01}:
\begin{equation}
H=-\sum_j\left[\frac{J}{2}(c_{j+1}^\dag c_j+{\rm H.c.})+\mu\left(c_j^\dag c_j-\frac{1}{2}\right)+(\Delta c^\dag_{j+1}c^\dag_j+{\rm H.c.})\right],
\end{equation}
which describes 1D $p$-wave superconductor. With $\Delta\in\mathbb{R}$ assumed, its Bogoliubov-de Gennes (BdG) Hamiltonian is given by (see Appendix~\ref{app2} for a general derivation)
\begin{equation}
H_{\rm BdG}(k)=-\frac{1}{2}(J\cos k+\mu)\sigma^z-\frac{1}{2}\Delta\sin k\sigma^y,
\end{equation}
According to Eq.~(\ref{DZ2}), whenever $\Delta\neq0$ such that the Hamiltonian is gapped, we have $\nu=-1$ for $|J|>\mu$ and $\nu=1$ for $|J|<\mu$.

Similar to the 1D case, there is no stable\footnote{By this, we mean that we do not consider Hopf insulators \cite{MJE08}, which are not stable against additional bands.} gapped topological phase in 3D without symmetry protection. However, there exists topologically nontrivial \emph{gapless} phases, represented by Weyl semimetals \cite{WX11}. The topological characterization is ultimately related to that for gapped phases \cite{SM13}. For the case of Weyl semimetals, which do not need any symmetry protection, we can consider a closed 2D surface surrounding a Weyl point and calculate the Chern number using Eq.~(\ref{Chgen}), where $T^2$ should be replaced by the surface. The Chern number does not depend on the choice of the surface (as long as it does not include additional Weyl points) and is an intrinsic character of the topological charge of the Weyl point. As a simple illustration, we consider a slightly modified version of Eq.~(\ref{ChH}) by adding $-\cos k_z$ to the coefficient of $\sigma^z$ \cite{ANP18}:
\begin{equation}
H(k_x,k_y,k_z)=\sin k_x\sigma^x+\sin k_y\sigma^y+(M-\cos k_x-\cos k_y-\cos k_z)\sigma^z.
\label{WeylH}
\end{equation}
We can check that the system stays gapless for $|M|\le3$ and has two Weyl points located at $\boldsymbol{k}^{\pm}_{\rm W}=(0,0,\pm\arccos (M-2))$ for $1<M<3$, around which the dispersion can be expanded as
\begin{equation}
H(\boldsymbol{k}=\boldsymbol{k}^{\pm}_{\rm W}+\delta\boldsymbol{k})\simeq \delta k_x\sigma^x+\delta k_y\sigma^y\pm\sqrt{1-(M-2)^2}\delta k_z\sigma^z.
\end{equation}
Fixing $k_z$ in Eq.~(\ref{WeylH}), we find that the 2D Bloch Hamiltonian on the $k_x-k_y$ plane undergoes phase transitions across the above two Weyl points. Precisely speaking, the Chern number is $1$ for $|k_z|<\arccos (M-2)$ but vanishes otherwise, implying that the two Weyl points possess opposite topological charges.  
We recall that a single Weyl point can never be realized in a local and static Hermitian lattice system at equilibrium, as a result of the torus topology of the Brillouin zone. This no-go theorem is well-known as the \emph{Nielsen-Ninomiya theorem} \cite{HBN81a,HBN81b}. A violation of a 1D version of the Nielsen-Ninomiya theorem has also been reported in the PT-symmetric lattice model \cite{Chernodub_2017}.

\subsubsection{Periodic table for Altland-Zirnbauer classes}
\label{ptAZ}
We have so far reviewed several  free-fermion systems with or without symmetries and their topological invariants.  
Here, we briefly review the systematic classification of gapped Hermitian free-fermion systems belonging to the \emph{Altland-Zirnbauer} (AZ) classes \cite{AA97,SAP08,AK09}. There are in total 10 AZ classes, including 2 complex ones and 8 real ones. The former includes class A, which does not require any symmetry at all, and class AIII, which possesses a single CS. At the level of Bloch (or BdG) Hamiltonians, a CS $\Gamma$ is a unitary matrix that satisfies
\begin{equation}
\Gamma H(\boldsymbol{k}) \Gamma^{-1}=-H(\boldsymbol{k}),\;\;\;\;\Gamma^2=1. 
\label{CS}
\end{equation}
A real AZ class includes at least one anti-unitary symmetry. For a single TRS $\mathcal{T}$, we have
\begin{equation}
\mathcal{T}H(\boldsymbol{k}) \mathcal{T}^{-1}=H(-\boldsymbol{k}),\;\;\;\;\mathcal{T}^2=\pm1, 
\label{TRS}
\end{equation}
where $\mathcal{T}^2=1$ ($\mathcal{T}^2=-1$) correspond to class AI (AII). For a single PHS $\mathcal{C}$, we have
\begin{equation}
\mathcal{C}H(\boldsymbol{k}) \mathcal{C}^{-1}=-H(-\boldsymbol{k}),\;\;\;\;\mathcal{C}^2=\pm1, 
\label{PHS}
\end{equation}
where $\mathcal{C}^2=1$ ($\mathcal{C}^2=-1$) correspond to class D (C). The remaining 4 real classes involve both TRS and PHS, which \emph{commute} with each other and their combination gives a CS. For $\mathcal{C}^2=1$ ($\mathcal{C}^2=-1$), $\mathcal{T}^2=1$ and $\mathcal{T}^2=-1$ correspond to classes BDI (CI) and DIII (CII), respectively. 

To classify gapped Bloch Hamiltonians, there are at least two complementary methods --- one starts from a gapless Dirac Hamiltonian and analyzes how the gap is opened by an additional mass term \cite{SAP08}, the other directly counts the number of topologically different flat-band representatives of gapped Hamiltonians \cite{AK09}. Here we adopt the latter, which is conceptually more straightforward. Setting $E_{\rm F}=0$ without loss of generality, we can continuously deformed $H(\boldsymbol{k})$  into \cite{AK09}
\begin{equation}
H_{\rm flat}(\boldsymbol{k})=1-2P_<(\boldsymbol{k}),
\label{Hflat} 
\end{equation}
while maintaining all the fundamental symmetries in Eqs.~(\ref{CS}), (\ref{TRS}) and (\ref{PHS}), if any. We start from the zero-dimensional case, i.e., the classification of Hamiltonians without $\boldsymbol{k}$ dependence. There is a unified way to work out the classification for all the AZ classes on the basis of the \emph{Clifford algebra}. A complex (real) Clifford algebra ${\rm C}\ell_p(\mathbb{C})$ (${\rm C}\ell_{p,q}(\mathbb{R})$) is a set of $p$ involutory (anti-involutory) elements $\{e_j\}^p_{j=1}$ (and $q$ involutory elements $\{e_j\}^{p+q}_{j=p+1}$) which \emph{anti-commute} with each other. It is always possible to relate a flattened Hamiltonian in a complex (real) AZ class to an extension (i.e., adding an element) of a complex (real) Clifford algebra. For example, a flattened Hamiltonian in class AIII corresponds to an extension ${\rm C}\ell_1(\mathbb{C})=\{\Gamma\}\to {\rm C}\ell_2(\mathbb{C})=\{\Gamma,H\}$ and that in class BDI corresponds to ${\rm C}\ell_{1,2}(\mathbb{R})=\{\mathcal{C},i\mathcal{C},i\mathcal{T}\mathcal{C}\}\to {\rm C}\ell_{1,3}(\mathbb{R})=\{\mathcal{C},i\mathcal{C},i\mathcal{T}\mathcal{C},H\}$. These Clifford algebras satisfy several nice properties such as ${\rm C}\ell_p(\mathbb{C})\simeq{\rm C}\ell_{p+2}(\mathbb{C})$, ${\rm C}\ell_{p+2,q}(\mathbb{R})\simeq{\rm C}\ell_{q,p}(\mathbb{R})$, ${\rm C}\ell_{p,q}(\mathbb{R})\simeq{\rm C}\ell_{p+1,q+1}(\mathbb{R})$ and ${\rm C}\ell_{p,q}(\mathbb{R})\simeq{\rm C}\ell_{p,q+8}(\mathbb{R})\simeq{\rm C}\ell_{p+8,q}(\mathbb{R})$.\footnote{Here ``$\simeq$" means the \emph{Morita equivalence}, which is the  isomorphsim, up to tensoring ${\rm M}_n(\mathbb{C})$ or ${\rm M}_n(\mathbb{R})$, i.e., up to $n\times n$ matrix space with complex or real entries. For example, ${\rm C}\ell_{p+2}(\mathbb{C})={\rm M}_2(\mathbb{C})\otimes{\rm C}\ell_p(\mathbb{C})$ and ${\rm C}\ell_{p+n,q+n}(\mathbb{R})={\rm M}_{2^n}(\mathbb{R})\otimes{\rm C}\ell_{p,q}(\mathbb{R})$, where we temporarily use the symbol ``$=$" for the usual isomorphism.} By identifying the Clifford-algebra extensions for all the AZ classes, we can determine the classifications, which are \emph{$K$-groups} $K_{\mathbb{F}}(s;d)$ with $d=0$, as the disjoint sectors (zeroth homotopy group) of the classifying spaces: 
\begin{equation}
K_{\mathbb{C}}(s;0)=\pi_0({\mathscr C}_s),\;\;\;\;
K_{\mathbb{R}}(s;0)=\pi_0({\mathscr R}_s),
\end{equation}
where ${\mathscr C}_s:{\rm C}\ell_s(\mathbb{C})\to {\rm C}\ell_{s+1}(\mathbb{C})$ and $\mathscr {R}_s: {\rm C}\ell_{0,s}(\mathbb{R})\to {\rm C}\ell_{0,s+1}(\mathbb{R})$ and the correspondence between $s$ and AZ classes is listed in Table~\ref{table2}. 

\begin{table*}[tbp]
\caption{Periodic table for Hermitian topological insulators and superconductors \cite{SAP08,AK09}. Here the first two (remaining eight) rows correspond to complex (real) AZ classes, whose classifying spaces are given by $\mathscr{C}_s$'s ($\mathscr{R}_s$'s).}
\begin{center}
\begin{tabular}{ccccccccccccc}
\hline\hline
AZ class & TRS & PHS & CS & \;$K$-group\; & $d=0$ & \;1\; & \;2\; & \;3\; & \;4\; & \;5\; & \;6\; & \;7\; \\
\hline
A & 0 & 0 & 0 & $K_{\mathbb{C}}(0;d)$  & $\mathbb{Z}$ & 0 & $\mathbb{Z}$ & 0 & $\mathbb{Z}$ & 0 & $\mathbb{Z}$ & 0 \\
AIII &0 & 0 & 1 & $K_{\mathbb{C}}(1;d)$ & 0 & $\mathbb{Z}$ & 0 & $\mathbb{Z}$ & 0 & $\mathbb{Z}$ & 0 & $\mathbb{Z}$ \\
\hline
AI & $+$ & 0 & 0 & $K_{\mathbb{R}}(0;d)$ & $\mathbb{Z}$  & 0 & 0 & 0 & $2\mathbb{Z}$ & 0 & $\mathbb{Z}_2$ & $\mathbb{Z}_2$ \\
BDI & $+$ & $+$ & 1 & $K_{\mathbb{R}}(1;d)$ & $\mathbb{Z}_2$ & $\mathbb{Z}$ & 0 & 0 &0 & $2\mathbb{Z}$ & 0 & $\mathbb{Z}_2$ \\
D & 0 & $+$ & 0 & $K_{\mathbb{R}}(2;d)$ & $\mathbb{Z}_2$ & $\mathbb{Z}_2$ & $\mathbb{Z}$ & 0 & 0 & 0 & $2\mathbb{Z}$ & 0 \\
DIII & $-$ & $+$ & 1 & $K_{\mathbb{R}}(3;d)$ & 0 & $\mathbb{Z}_2$ & $\mathbb{Z}_2$ & $\mathbb{Z}$ & 0 & 0 & 0 & $2\mathbb{Z}$ \\
AII & $-$ & 0 & 0 & $K_{\mathbb{R}}(4;d)$ & $2\mathbb{Z}$ & 0 & $\mathbb{Z}_2$ & $\mathbb{Z}_2$ & $\mathbb{Z}$ & 0 & 0 & 0 \\
CII & $-$ & $-$ & 1 & $K_{\mathbb{R}}(5;d)$ & 0 & $2\mathbb{Z}$ & 0 & $\mathbb{Z}_2$ & $\mathbb{Z}_2$ & $\mathbb{Z}$ & 0 & 0 \\
C & 0 & $-$ & 0 & $K_{\mathbb{R}}(6;d)$ & 0 & 0 & $2\mathbb{Z}$ & $0$ & $\mathbb{Z}_2$ & $\mathbb{Z}_2$ & $\mathbb{Z}$ & 0 \\
CI & $+$ & $-$ & 1 & $K_{\mathbb{R}}(7;d)$ & 0 & 0 & 0 & $2\mathbb{Z}$ & 0 & $\mathbb{Z}_2$ & $\mathbb{Z}_2$ & $\mathbb{Z}$ \\
\hline\hline
\end{tabular}
\end{center}
\label{table2}
\end{table*}

The classification in higher dimensions can easily be obtained from the identity
\begin{equation}
K_{\mathbb{F}}(s;d)=K_{\mathbb{F}}(s+1;d+1), 
\label{KFsd}
\end{equation}
which can be proved by establishing an explicit isomorphism between these two $K$-groups \cite{TJCY10}. Since the symmetry-class label $s$ shifts by 1, this isomorphism should change a chiral symmetric class into a class without CS and vice versa. In terms of representative flattened Hamiltonians, these two types of maps are given by
\begin{equation}
\begin{split}
H_{\rm nc}(\boldsymbol{k},\theta)=\cos\theta H_{\rm c}(\boldsymbol{k})&+\sin\theta \Gamma, \\
H_{\rm c}(\boldsymbol{k},\theta)=\cos\theta [\sigma^z\otimes H_{\rm nc}(\boldsymbol{k})]&+\sin\theta (\sigma^\mu\otimes I),
\end{split}
\label{Kiso}
\end{equation}
where $\theta$ is an additional momentum variable and $\mu=x$ or $y$ depends on the symmetry class. For example, when mapping from class AI ($s=0$), we should choose $\mu=y$ so that $\mathcal{T}H_{\rm c}(\boldsymbol{k},\theta)\mathcal{T}^{-1}=H_{\rm c}(-\boldsymbol{k},-\theta)$ is satisfied. Moreover, the emergent PHS $\mathcal{C}=\sigma^x\otimes\mathcal{T}$ satisfies $\mathcal{C}^2=1$ so the image is indeed within class BDI ($s=1$). These two maps (\ref{Kiso}) are  $K$-group homomorphisms, since they are linear under direct sums of Hamiltonians. To prove that the homomorphism is actually an isomorphism, we should prove that any flattened Hamiltonian can continuously be deformed into the forms shown on the right-hand side of Eq.~(\ref{Kiso}), which is achievable, though rather technical, with the help of Morse theory \cite{TJCY10}. Thanks to Eq.~(\ref{KFsd}), we can fulfill the entire periodic table from the leftmost ($d=0$) column.

Finally, we mention that the above classification of gapped free-fermion phases can readily be applied to classify Fermi surfaces in gapless systems. Denoting the spatial dimension and the Fermi-surface dimension as $d_{\rm BZ}$ and $q$, respectively, the classification is again given by $K_{\mathbb{F}}(s;d)$ but with $d=d_{\rm BZ}-q-1$ \cite{SM13}, which is the dimension of the loop or surface that encircles or encompasses the Fermi surface. For example, a Weyl point is a $q=0$-dimensional Fermi surface in class A with $\mathbb{F}=\mathbb{C}$ and $s=0$ in $d_{\rm BZ}=3$ dimensions; therefore its classification reads $K_{\mathbb{C}}(0;2)=\mathbb{Z}$, consistent with the Chern number character discussed in the previous subsection.

\subsection{Complex energy gaps}\label{ceg}
We recall that a crucial constraint on topological insulators and superconductors is that the Bloch or BdG Hamiltonian should be \emph{gapped}. 
To define topological phases for non-Hermitian systems, the first question we must address is how to generalize the notion of gap. One of the most remarkable differences between Hermitian and non-Hermitian Hamiltonians is that the former spectra are always real while the latter are generally complex. From a mathematical point of view, we cannot define an order relation on the complex plane \cite{LVA79}. Therefore, in general, we cannot interpret `phases' of non-Hermitian systems as equilibrium states of matter.\footnote{We emphasize that this is a statement for a \emph{general} situation but does not rule out the possibility of introducing well-defined ground states in several setups, such as non-Hermitian systems with real spectra or long-lived states.} 
Accordingly, the notion of gap introduced above for Hermitian systems cannot be generalized to non-Hermitian cases in a straightforward manner.

\begin{figure}[!t]
\begin{center}
\includegraphics[width=14.5cm]{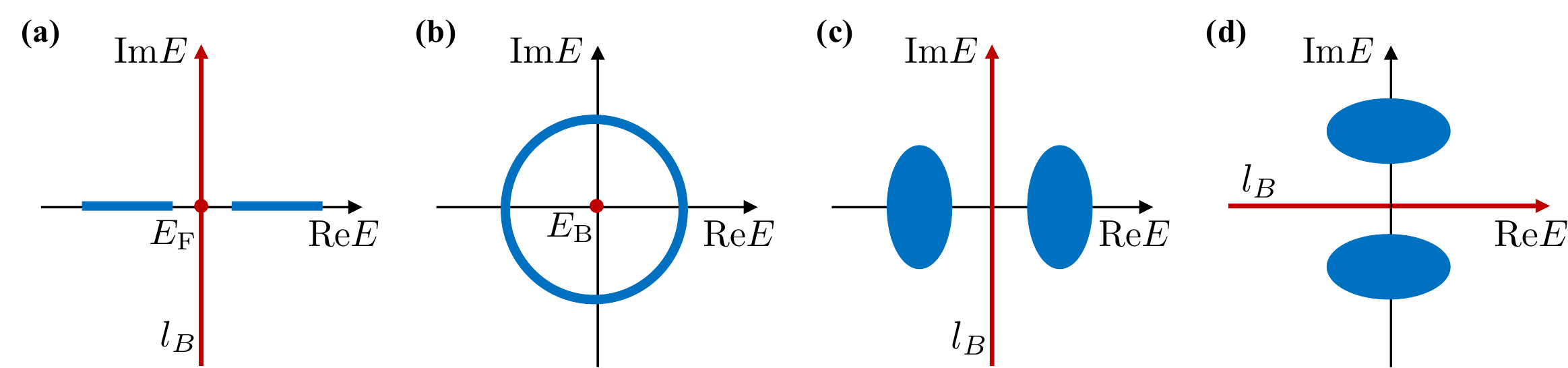}
\end{center}
\caption{(a) Real spectrum of a gapped Hermitian system. The lower and upper bands are separated by both the Fermi energy $E_{\rm F}$ and the imaginary energy axis. (b) Complex spectrum of a non-Hermitian system forming a loop. The system is point-gapped at $E_{\rm B}$, but has no line gap. (c) Complex spectrum of a real- and (d) an imaginary-line-gapped non-Hermitian system. 
In all the figures, the eigenenergies are marked in blue, while the base (Fermi) energy points $E_{\rm B}$ and lines $l_{\rm B}$ are marked in red.}
\label{fig:gap}
\end{figure}

Nevertheless, there are at least two possible ways of defining complex energy gaps; one is unique to non-Hermitian regimes while the other turns out to be essentially equivalent to that of Hermitian systems. To see this, we recall that a Hermitian Bloch (BdG) Hamiltonian $H(\boldsymbol{k})$ is said to be gapped with respect to the Fermi energy $E_{\rm F}$ if $E_{\rm F}$ is not touched by the spectrum of $H(\boldsymbol{k})$ for all $\boldsymbol{k}\in T^d$ (see Fig.~\ref{fig:gap}(a)). Removing the constraint of Hermiticity, we obtain the following notion of a \emph{point gap} \cite{ZG18,ZH19}: A non-Hermitian Bloch Hamiltonian $H(\boldsymbol{k})$ is said to have a point gap if there exists a base energy $E_{\rm B}$, which is generally complex, such that
\begin{equation}
\det[H(\boldsymbol{k})-E_{\rm B}]\neq 0,\;\;\;\;\forall\boldsymbol{k}\in T^d.
\label{HEB}
\end{equation}

Alternatively, the property of a Hermitian Bloch Hamiltonian $H(\boldsymbol{k})$ being gapped can be understood as the fact that the spectrum is separated by $E_{\rm F}$ on the real energy axis. Without Hermiticity, the energy spectrum of the entire Hamiltonian $\bigoplus_{\boldsymbol{k}}H(\boldsymbol{k})$ generally forms areas and thus can be separated by a 1D curve. This gives rise to the notion of a \emph{line gap}: 
$H(\boldsymbol{k})$ is said to have a line gap if there exists a base line $l_{\rm B}$ on the complex energy plane such that  Eq.~(\ref{HEB}) is valid for $\forall E_{\rm B}\in l_{\rm B}$, and the eigenvalues distribute on both sides of $l_{\rm B}$ \cite{KK19c}. Clearly, a line-gapped system is always point-gapped. However, the converse is not true since even a single band system, whose spectrum can never be separated by a line, may have a point gap (see Fig.~\ref{fig:gap}(b)), as is exemplified by the Hatano-Nelson model with a connected elliptical spectrum \cite{HN96}. 

More generally, topological structures inherent in point-gapped bands are intrinsically unique to non-Hermitian systems, while this is not the case in line-gapped bands. Indeed, the topology of line-gapped bands is {\it essentially Hermitian} in the sense that one can show the following theorem:
\begin{theorem}\label{linedeform}
Any line-gapped non-Hermitian bands can continuously be deformed to either Hermitian or anti-Hermitian ones while keeping the line gap and all the  relevant symmetries. 
\end{theorem}
\noindent
Note that an anti-Hermitian band is always equivalent to the Hermitian one after simply multiplying the entire spectrum by the imaginary unit $i$. 
The proof of the above statement can be found in Appendix~\ref{app4}, which is, to our knowledge, not provided in literature, but given by this review article for the first time\footnote{The ``proof" argued in Ref.~\cite{KK19c} was incomplete due to the crucial reasons detailed in Sec.~\ref{Sec:NHPT}.}.

We require the choices of base points/lines to be consistent with the symmetries, if any. 
For a point-gapped $H(\boldsymbol{k})$, the shifted Bloch Hamiltonian $H(\boldsymbol{k})-E_{\rm B}$ should maintain all the symmetries of $H(\boldsymbol{k})$. For example, $E_{\rm B}$ should be real (zero) if $H(\boldsymbol{k})$ has a TRS (CS). Note that this requirement appears already in Hermitian systems, such as $E_{\rm F}=0$ for superconductors described by BdG Hamiltonians with PHS \cite{FS09}. In the following, we will  set $E_{\rm B}=0$ otherwise stated  \cite{ZG18}. As for a base line, we require it to respect the spectral constraint on $H(\boldsymbol{k})$ imposed by the symmetries. For example, a TRS (CS) enforces the spectrum to be symmetric with respect to the real axis (origin), so the base line should be either exactly the real axis or perpendicular to it (go through the origin). Two natural choices of base lines are the real and imaginary axes, as will be adopted hereafter. 
The line gap is said to be \emph{real} (\emph{imaginary}) if the base line is chosen to be the imaginary (real) axis (see Figs.~\ref{fig:gap}(c) and (d)).

Finally, we mention some other notions of gaps. The arguably most important one is {band spearation} \cite{SH18}, which means that the dispersions of two non-Hermitian bands $\alpha$ and $\beta$ do not cross each other, i.e., $\epsilon_\alpha(\boldsymbol{k})\neq \epsilon_\beta(\boldsymbol{k})$ for $\forall\boldsymbol{k}\in T^d$. In particular, band $\alpha$ is said to be \emph{isolated} if it is separated from any band $\beta\neq\alpha$. The notion of band separation and isolation allows us to discuss the properties of individual non-Hermitian bands, but has the disadvantage that it is not applicable to disordered systems, in contrast to point and line gaps \cite{ZL19}. In addition, as a natural generalization of point and line gaps, one may consider the spectral constraint that the eigenvalues cannot lie on a closed (Jordan) curve on the complex plane \cite{BDS20} . 

\subsection{Prototypical examples and topological invariants}\label{Sec:5peti}
There had recently been many  
works concerning concrete non-Hermitian topological models \cite{RMS09,HYC11,HS13,LR14,MS15,CY15,PSJ16,XY17,LD17,LTE16,NX18,YC18,LS18,ZL18,LC18,YC18}. Typically, these systems are built through introduction of non-Hermiticity into well-known Hermitian topological models, some of which have been reviewed in Sec.~\ref{ProSys}. In the following, we discuss two complementary cases --- introduction of non-Hermiticity  into gapped and gapless Hermitian systems.  
Here, we only focus on the \emph{bulk} topological properties  
because the boundary physics turns out to be very subtle and complicated, and thus will be discussed separately in Sec.~\ref{sec:bec}. 
\\ \\ {\it Non-Hermitian models built from gapped Hermitian systems}

\vspace{3pt}
\noindent
If we start from a gapped Hermitian Hamiltonian, according to the spectral stability of Hermitian matrices (Theorem~\ref{HNH}), we know that a line gap (and thus a point gap) should stay open as long as the introduced non-Hermiticity is weak enough. Such non-Hermitian systems are  \emph{essentially Hermitian} in the sense that their topological invariants can straightforwardly be determined from their Hermitian counterparts. In fact,  we can prove (Theorem~\ref{linedeform}) that any line-gapped non-Hermitian Bloch Hamiltonian can continuously be deformed into a Hermitian or anti-Hermitian Hamiltonian while keeping the line gap and all the symmetries (see Sec.~\ref{Sec:NHPT} and Appendix~\ref{app4} for further details). 

Let us first consider the generalization of the Chern number (\ref{Chgen}) to non-Hermitian systems. As a na\"ive guess, we may simply replace the Hermitian projector onto the Fermi sea $P_<(\boldsymbol{k})$ by a non-Hermitian $P(\boldsymbol{k})$ ($P(\boldsymbol{k})^2=P(\boldsymbol{k})$ but generally $P(\boldsymbol{k})^\dag\neq P(\boldsymbol{k})$) onto an isolated energy cluster on the complex plane, such as all the bands with ${\rm Re}\epsilon_\alpha(\boldsymbol{k})<0$ in a real-line-gapped $H(\boldsymbol{k})$. This natural generalization turns out indeed to be a topological invariant. To see this, we note that a continuous deformation of $H(\boldsymbol{k})$ implies that of $P(\boldsymbol{k})$ due to 
\begin{equation}
P(\boldsymbol{k})=\oint_C \frac{dz}{2\pi i}\frac{1}{zI-H(\boldsymbol{k})}.
\label{Blochproj}
\end{equation}
A continuous deformation of $P(\boldsymbol{k})$ can be shown to be equivalent to perform a similarity transformation by an invertible operator $V(\boldsymbol{k})$, which is smooth in $\boldsymbol{k}$, provided that $P(\boldsymbol{k})$ is.\footnote{The construction of $V(\boldsymbol{k})$ is not unique since it can be replaced by $V(\boldsymbol{k})W(\boldsymbol{k})$, where $W(\boldsymbol{k})$ is invertible and continuous in $\boldsymbol{k}$ and satisfies $[W(\boldsymbol{k}),P(\boldsymbol{k})]=0$. However, there is a natural construction in terms of a given continuous path $P_\lambda(\boldsymbol{k})$ ($\lambda\in[0,1]$): $V(\boldsymbol{k})=\hat{{\rm \Lambda}}e^{\int^1_0d\lambda[1-2P_\lambda(\boldsymbol{k})]\partial_\lambda P_\lambda(\boldsymbol{k})}$, where $\hat{{\rm \Lambda}}$ indicates the ordering of $\lambda$, such that $P_1(\boldsymbol{k})=V(\boldsymbol{k})P_0(\boldsymbol{k})V(\boldsymbol{k})^{-1}$ and the smoothness in $\boldsymbol{k}$ follows that in $P_\lambda(\boldsymbol{k})$.} We can then check that the Chern number (\ref{Chgen}) calculated in terms of $V(\boldsymbol{k})P(\boldsymbol{k})V(\boldsymbol{k})^{-1}$ is exactly the same as $P(\boldsymbol{k})$. In particular, if $H(\boldsymbol{k})$ is built from a Hermitian Chern insulator by introducing a sufficiently weak non-Hermiticity such that the bands of interest do not touch the remaining ones, the Chern number of these bands is exactly the same as that for the Hermitian model.

\begin{figure}[!t]
\begin{center}
\includegraphics[width=14.5cm]{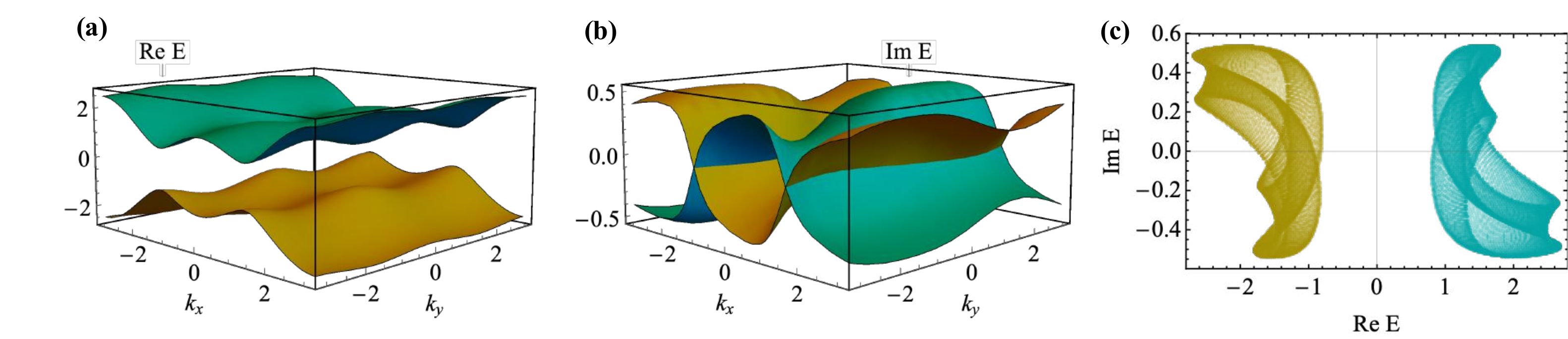}
\end{center}
\caption{(a) Real and (b) imaginary parts of the complex energy dispersions of the two-band non-Hermitian Chern insulator given in Eq.~(\ref{NHCh}) with $t=1$, $\kappa_x=0.2$, $\kappa_y=0.3$, $\delta=0.4$ and $m=-0.5$. (c) Real-line gapped energy spectrum of (a) and (b) on the complex plane. 
Such a non-Hermitian Chern insulator is created through introduction of non-Hermiticity into a Hermitian one with Chern number ${\rm Ch}=-1$.}
\label{fig:NHCh}
\end{figure}

It is worth mentioning the special case of a single band, which means $P(\boldsymbol{k})=| u^{\rm R}_{\boldsymbol{k}}\rangle\langle u^{\rm L}_{\boldsymbol{k}}|$ is of rank $1$. We can easily construct such a model by, for example, introducing on-site non-Hermitian spin couplings and gain/loss in Eq.~(\ref{ChH}) to obtain \cite{SH18}
\begin{equation}
H(k_x,k_y)=(t\sin k_x+i\kappa_x)\sigma^x+(t\sin k_y+i\kappa_y)\sigma^y+(t\cos k_x+t\cos k_y+m+i\delta)\sigma^z.
\label{NHCh}
\end{equation}
As shown in Fig.~\ref{fig:NHCh}, while the dispersions become complex and form areas, the two single bands stay separated (and the real line gap stays open) for sufficiently small $\kappa_x$, $\kappa_y$ and $\delta$. In this case, we can continuously deform $P(\boldsymbol{k})$ into $P(\boldsymbol{k})P(\boldsymbol{k})^\dag /{\rm Tr}[P(\boldsymbol{k})P(\boldsymbol{k})^\dag]=| u^{\rm R}_{\boldsymbol{k}}\rangle\langle u^{\rm R}_{\boldsymbol{k}}|$, $P(\boldsymbol{k})^\dag P(\boldsymbol{k})/{\rm Tr}[P(\boldsymbol{k})^\dag P(\boldsymbol{k})]=| u^{\rm L}_{\boldsymbol{k}}\rangle\langle u^{\rm L}_{\boldsymbol{k}}|$ or $P(\boldsymbol{k})^\dag=| u^{\rm L}_{\boldsymbol{k}}\rangle\langle u^{\rm R}_{\boldsymbol{k}}|$ through $O_\lambda (\boldsymbol{k})P(\boldsymbol{k})/{\rm Tr}[O_\lambda (\boldsymbol{k})P(\boldsymbol{k})]$, $P(\boldsymbol{k})O_\lambda (\boldsymbol{k})/{\rm Tr}[P(\boldsymbol{k})O_\lambda (\boldsymbol{k})]$ or $O_\lambda (\boldsymbol{k})P(\boldsymbol{k})O_\lambda (\boldsymbol{k})/{\rm Tr}[O_\lambda (\boldsymbol{k})P(\boldsymbol{k})O_\lambda (\boldsymbol{k})]$, where $O_\lambda (\boldsymbol{k})=(1-\lambda)I+\lambda P(\boldsymbol{k})^\dag$. This means the four types of Berry curvatures in terms of left and/or right eigenstates:
\begin{equation}
B^{\rm XY}_{xy}(\boldsymbol{k})\equiv i(\langle\partial_{k_x}u^{\rm X}_{\boldsymbol{k}}|\partial_{k_y}u^{\rm Y}_{\boldsymbol{k}}\rangle-\langle\partial_{k_y}u^{\rm Y}_{\boldsymbol{k}}|\partial_{k_x}u^{\rm X}_{\boldsymbol{k}}\rangle),\;\;\;\;{\rm X,Y=L,R}.
\label{BXY}
\end{equation} 
These Berry curvatures differ from each other locally, but they give rise to the same Chern number upon the integration over the 
Brillouin zone. Note that $|u^X_{\boldsymbol{k}}\rangle$ should be normalized in Eq.~(\ref{BXY}) if $X=Y$, and otherwise $\langle u^{\rm L}_{\boldsymbol{k}}| u^{\rm R}_{\boldsymbol{k}}\rangle=1$ is required. Meanwhile, 
only $B^{\rm RR}_{xy}$ manifests itself in the wave-packet dynamics since the physical state is the right eigenstate and a non-Hermitian Hamiltonian is the generator acting on the right state vector (cf. Eqs.~\eqref{fpanh1} and \eqref{qoptnh} in Sec.~\ref{sec4}).

Another widely studied model in the literature is the non-Hermitian SSH model \cite{LS18}. Although there are various non-Hermitian generalizations, we may roughly divide them into two classes, one of which is obtained by introducing on-site gain and loss \cite{RMS09}, while the other  by introducing asymmetry in the hopping amplitudes \cite{LTE16}. 
A prototypical model in the former class is the PT-symmetric SSH model (see also Eq.~(\ref{PTSSH}) in Sec.~\ref{sec3}):
\begin{equation}
H(k)=-(J_1+J_2\cos k)\sigma^x-J_2\sin k\sigma^y-i\gamma\sigma^z,\;\;\;\;\gamma\in\mathbb{R},
\label{PTSSH2}
\end{equation}
which differs from Eq.~(\ref{SSH}) by a purely imaginary $\sigma^z$ term and can be regarded as the imaginary-potential counterpart of the Rice-Mele model \cite{RMJ82}. See Fig.~\ref{fig:3band}(a) for its typical band dispersions. While the sublattice symmetry is lost, this model~(\ref{PTSSH2}) exhibits a chiral symmetry: 
\begin{equation}
H(k)=-\sigma^zH(k)^\dag \sigma^z.
\end{equation}
Assuming $\gamma<|J_1-J_2|$ so that a real line gap is open, we can characterize the system by a winding number given in Eq.~(\ref{wn}), where $q(k)$ reads
\begin{equation}
q(k)=-\frac{J_1+J_2e^{-ik}}{\sqrt{J_1^2+J_2^2-\gamma^2+2J_1J_2\cos k}}.
\end{equation}
We can check that, as expected, the winding number 
stays the same as the Hermitian case ($\gamma=0$). In general, provided that there is a real line gap, we can calculate the winding number using Eq.~(\ref{wn}) with $q(k)$ determined by (see Appendix~\ref{app4})
\begin{equation}
q(k)=\frac{1}{4}(I+\Gamma)[I-P(k)-P(k)^\dag](I-\Gamma),
\end{equation}
where $P(k)$ is given in Eq.~(\ref{Blochproj}) with $C$ being an arbitrary closed contour that encircles all the eigenvalues with negative real parts.  
It is worth mentioning that although the model (\ref{PTSSH2}) also respects the PT symmetry represented by $\sigma^x\mathcal{K}$, this symmetry can only protect a $\mathbb{Z}_2$ topology for both point and line gaps \cite{ZG18}, and does not pin the edge mode at the zero energy. More recently, a non-Hermitian extension of the Bernevig-Hughes-Zhang model~\eqref{bhzhermitian} has also been proposed in Ref.~\cite{SK192}.

\begin{figure}[!t]
\begin{center}
\includegraphics[width=14.5cm]{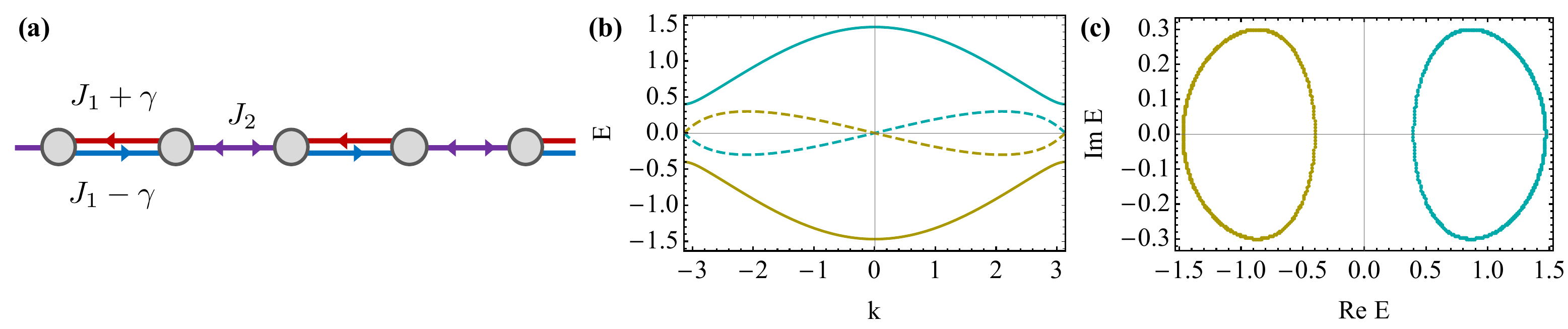}
\end{center}
\caption{(a) Schematic illustration of the non-Hermitian SSH model given in Eq.~(\ref{asyhSSH}). 
(b) Real (solid) and imaginary (dashed) parts of the complex band dispersions. 
(c) Full spectrum on the complex-energy plane. Here, the parameters are chosen to be $J_1=0.5$, $J_2=1$ and $\gamma=0.3$. Such a non-Hermitian SSH model is constructed through introduction of non-Hermiticity into a Hermitian one with winding number $w=1$.}
\label{fig:NHSSH}
\end{figure}

We then turn to the case of asymmetric hopping amplitudes. A prototypical model in this case is given by \cite{LTE16,YS18a}
\begin{equation}
H(k)=-(J_1+J_2\cos k)\sigma^x-(J_2\sin k+i\gamma)\sigma^y,
\label{asyhSSH}
\end{equation}
which differs from Eq.~(\ref{SSH}) by a purely imaginary shift in the $\sigma^y$ component, implying an asymmetry in the intra-unit-cell hopping amplitudes. See Fig.~\ref{fig:NHSSH} for a schematic illustration of the model and its typical band dispersions. Since the sublattice symmetry is respected, we can easily identify $q(k)$ from the off-block-diagonal component of $H(k)$ as
\begin{equation}
q(k)=-J_1-\gamma-J_2e^{-ik}.
\end{equation}
However, due to the absence of Hermiticity, the other off-block-diagonal component is not necessarily the Hermitian conjugate of $q(k)$, and thus the corresponding winding number may not be the opposite, provided that the non-Hermiticity is strong enough. Such a situation actually gives an example of genuine non-Hermitian topological phases, on which we focus in the following. 
\begin{figure}[!t]
\begin{center}
\includegraphics[width=14.5cm]{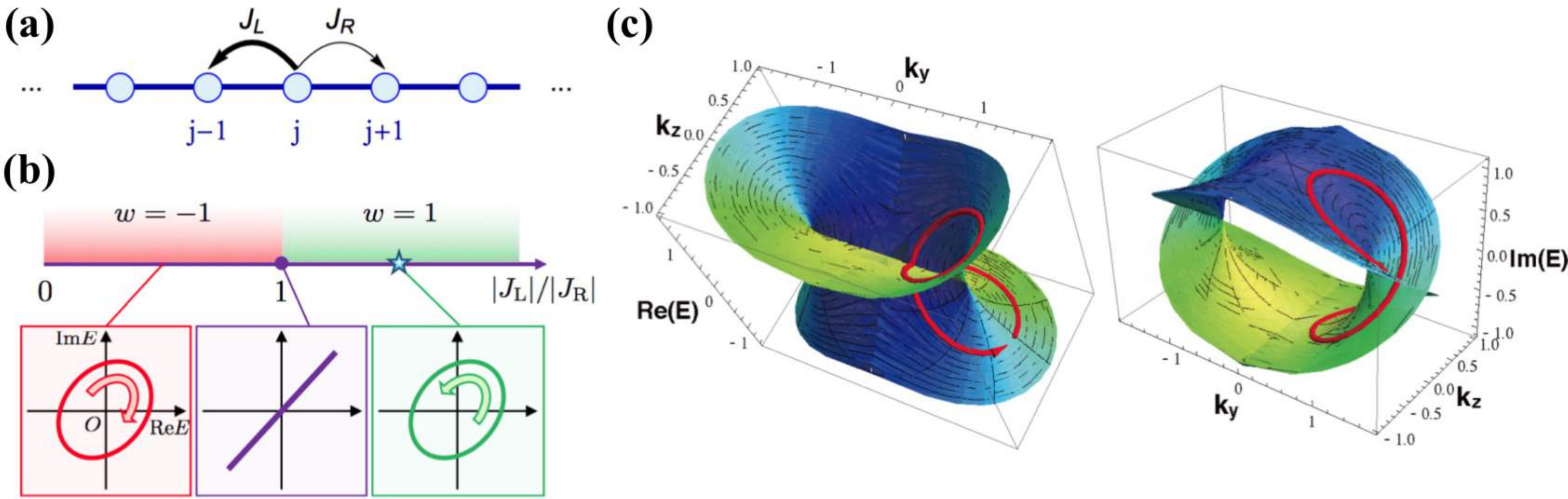}
\end{center}
\caption{(a) Schematic illustration of the Hatano-Nelson model (\ref{RSHN}), which is a 1D lattice with asymmetric leftward hopping $J_{\rm L}$ and rightward hopping $J_{\rm R}$. (b) Phase diagram of the Hatano-Nelson model. The bottom panels indicate that different phases are distinguished by the spectral winding number (\ref{NHwn}). Adapted from Ref.~\cite{ZG18}. (c) Weyl exceptional ring. Real (left) and imaginary parts (right) of the complex band as functions of $k_y$ and $k_z$ at $k_x=0$ (see also Fig.~\ref{fig:3band}(c)). The spectral winding number can be defined by the line integral along the red solid curve. Adapted from Ref.~\cite{XY17}. Copyright \copyright\,    2017 by the American Physical Society. }
\label{fig:ptgtopo}
\end{figure}
\\ \\ {\it Non-Hermitian models built from gapless Hermitian systems}

\vspace{3pt}
\noindent
One of the most important differences between non-Hermitian and Hermitian systems is that the energy spectrum of the former is generally complex while that of the latter is always real. This remarkable difference leads to not only a richer gap structure (see Sec.~\ref{ceg}), but also a unique topology related solely to the complex spectrum. That is, the spectrum can form a loop that encircles a base point $E_{\rm B}$, and such a loop cannot be removed while keeping the system point-gapped at $E_{\rm B}$. In 1D, such a loop can quantitatively be characterized by a winding number of the spectrum \cite{ZG18}: 
\begin{equation}
w=\int^\pi_{-\pi}\frac{dk}{2\pi i} \frac{d}{dk}\ln\det[H(k)-E_{\rm B}].
\label{NHwn}
\end{equation} 
This is a spectral winding number because  
$\det[H(k)-E_{\rm B}]=\prod_\alpha [\epsilon_\alpha(k)-E_{\rm B}]$ and thus 
\begin{equation}
w=\sum_\alpha\int^\pi_{-\pi}\frac{dk}{2\pi} \frac{d}{dk}{\rm arg}[\epsilon_\alpha(k)-E_{\rm B}].
\end{equation}
Obviously, the winding number $w$ of a Hermitian system gapped at $E_{\rm B}$ always vanishes, implying that a non-Hermitian system with $w\neq0$ is not near any gapped Hermitian system. Nevertheless, we can realize a nontrivial spectral winding by perturbing a gapless Hermitian system. To illustrate this point, we consider the Hatano-Nelson model \cite{HN96} (cf. Fig.~\ref{fig:ptgtopo}(a))
\begin{equation}
H=\sum_j(J_{\rm L}|j-1\rangle\langle j|+J_{\rm R}|j\rangle\langle j-1|),
\label{RSHN}
\end{equation}
whose Bloch Hamiltonian is given by
\begin{equation}
H(k)=J_{\rm L}e^{ik}+J_{\rm R}e^{-ik}.
\label{KHN}
\end{equation}
In the Hermitian limit $J_{\rm L}=J_{\rm R}^*$, the system is gapless at $E_{\rm B}=0$, while a spectral winding $w=1$ ($w=-1$) appears immediately for $|J_{\rm L}|>|J_{\rm R}|$ ($|J_{\rm L}|<|J_{\rm R}|$) (see Fig.~\ref{fig:ptgtopo}(b)). Of course, it is possible to achieve $w\neq0$ by introducing sufficiently strong non-Hermiticity (necessarily across a gapless point) into a gapped Hermitian model. For example,  
the non-Hermitian SSH model 
in Eq.~(\ref{asyhSSH}) has a nontrivial \emph{total} winding number when $|J_1-J_2|<|\gamma|<J_1+J_2$, provided that $J_{1,2}\in\mathbb{R}^+$. Here we emphasize ``total" because Eq.~(\ref{asyhSSH}) has a sublattice symmetry and thus has two independent off-block-diagonal parts, each of which is associated with a winding number \cite{ZG18}.

\begin{figure}[!t]
\begin{center}
\includegraphics[width=14.5cm]{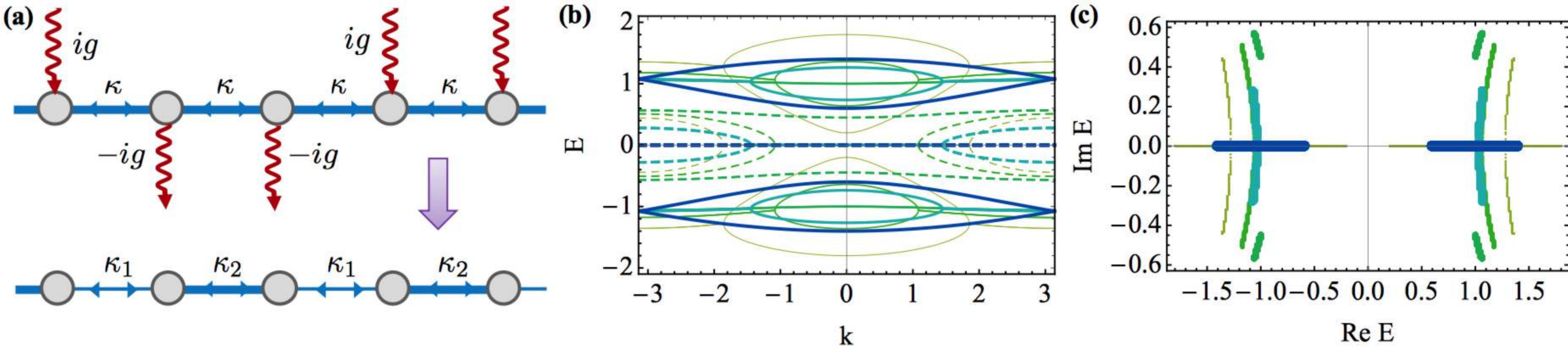}
\end{center}
\caption{(a) Schematic illustration of the four-band model given in Eq.~(\ref{4bm}) (upper panel) and the Hermitian SSH model (lower panel). The former can continuously be deformed into the latter while keeping the real line gap and the chiral symmetry. (b) Snapshots of the complex band dispersions, whose real and imaginary parts are indicated by real and dashed curved, and (c) the full energy spectra during the continuous deformation. Thicker curves correspond to later stages in the deformation. Here, the parameters are chosen to be $\kappa_2=\kappa=1$, $\kappa_1=0.4$ and $g=0.6$.}
\label{fig:Htopo}
\end{figure}

Another well-known example unique to non-Hermitian systems is the Weyl exceptional ring arising from a perturbed Weyl point, which has recently been realized in a photonic lattice (see Sec.~\ref{secphoto}) \cite{AC19}. One of the simplest constructions is given by \cite{XY17}  
\begin{equation}
H(\boldsymbol{k})\simeq k_x\sigma^x+k_y\sigma^x+(k_z+i\gamma)\sigma^z,
\label{EWR2}
\end{equation}
whose corresponding exceptional ring is 
$k_x^2+k_y^2=\gamma^2$ within the $k_x$-$k_y$ plane (i.e., $k_z=0$);  see also Fig.~\ref{fig:3band}(c). Of course, this is just an asymptotic form and a lattice version can be obtained by replacing $k_\mu$ ($\mu=x,y,z$) by $\sin k_\mu$. 
While Eq.~(\ref{EWR2})  is a \emph{gapless} non-Hermitian system, one can analyze it by simply looking at the lower-dimensional gapped topology  \cite{SM13}. Similar to a Weyl point in a Hermitian system (see Sec.~\ref{ProSys}), a Weyl exceptional ring is characterized by the Chern number (\ref{NHCh}) defined on a closed surface, such as a sphere, that encompasses the ring. In addition, a Weyl exceptional ring is characterized by the spectral winding number (\ref{NHwn}) defined on a loop that encircles the ring (cf. Fig.~\ref{fig:ptgtopo}(c)). These two topological invariants have different origins ---  they come from the line-gap and point-gap topology, respectively. We also note that some specific non-Hermitian perturbations to Weyl points may lead to exceptional Hopf links \cite{YZ19,CJ19}.

Introduction of non-Hermiticity into a gapless Hermitian system may not always yield a genuine non-Hermitian topological phase. This can be illustrated by a four-band SSH model with on-site gain and loss (see Fig.~\ref{fig:Htopo}(a)) \cite{TK18}:
\begin{equation}
H(k)=\begin{bmatrix} 
\;\;ig\;\; & \;\;\kappa_1\;\; & \;\;0\;\; & \;\;\kappa_2e^{-ik}\;\;  \\
\;\;\kappa_1\;\; & \;\;-ig\;\; & \;\;\kappa_2\;\; & \;\;0\;\; \\
\;\;0\;\; & \;\;\kappa_2\;\; & \;\;-ig\;\; & \;\;\kappa_1\;\; \\
\;\;\kappa_2e^{ik}\;\; & \;\;0\;\; & \;\;\kappa_1\;\; & \;\;ig\;\; \\
\end{bmatrix}.
\label{4bm}
\end{equation}
This Bloch Hamiltonian respects a chiral symmetry $(\sigma_0\otimes\sigma^z)H(k)(\sigma_0\otimes\sigma^z)=-H(k)^\dag$. One can analytically obtain the dispersion relations as
\begin{equation}
\epsilon(k)=\pm\sqrt{\kappa_1^2+\kappa_2^2-g^2\pm2\kappa_2\sqrt{\kappa_1^2\cos^2\frac{k}{2}-g^2}}.
\end{equation}
When $g=0$, the model reduces to the 
Hermitian SSH model (\ref{SSH}) and is gapless at $\kappa_1=\kappa_2=\kappa$. Apart 
from such a gapless point, 
the system is driven into the topological phase with $\kappa_1<\kappa_2$ by a sufficiently small $g$. To see this, we only have to show that a continuous deformation  $(\kappa_1,\kappa_2,g)=(\kappa,\kappa,g)\to(\kappa-\delta\kappa,\kappa,0)$ (with $\delta\kappa>0$ but not too large) does not close the real line gap. For example, we can change $\kappa_1=\kappa$ into $\kappa-\delta\kappa$ while keeping $g$ unchanged, followed by reducing $g$ to zero with fixed $\kappa_1$ and $\kappa_2$. See Figs.~\ref{fig:Htopo}(b) and (c) for an illustration.

\begin{figure}[!t]
\begin{center}
\includegraphics[width=12cm]{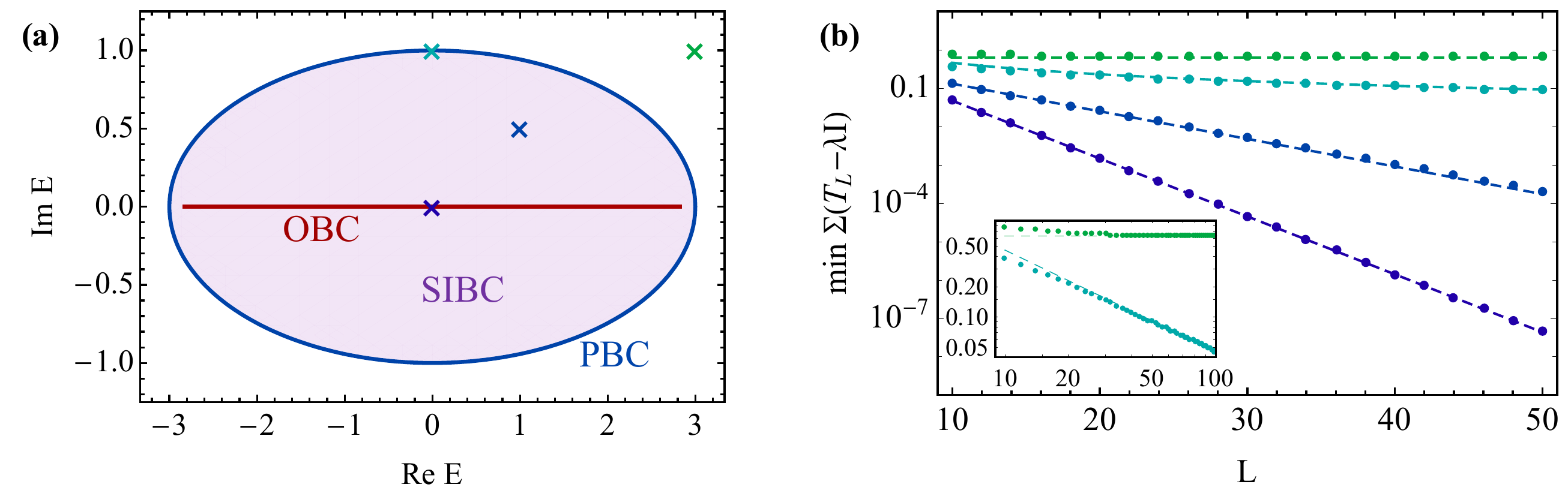}
\end{center}
\caption{(a) Spectrum of the Hatano-Nelson model (\ref{RSHN}) with $J_{\rm L}=2$ and $J_{\rm R}=1$ under the periodic boundary conditions (PBC) (blue), the semi-infinite boundary condition (SIBC) (light purple), and the open boundary conditions (OBC) (red). (b) Minimal singular value of $T_L-\lambda I$ for four base energies $\lambda$ (marked as ``$\times$" in (a)) plotted against the system size $L$. We only plot the data points with even $L$. For odd $L$, one can show that the minimal singular value of $T_L$ is exactly zero from its vanishing determinant. The dots refer to the actual data and the dashed lines are fitted to them. For $\lambda=0$ and $1+0.5i$ in the SIBC spectrum, $\min\Sigma(T_L-\lambda I)$ decays exponentially with respect to $L$ (see Eq.~(\ref{mTLS})). For $\lambda=i$ in the PBC spectrum and $\lambda=3+i$ outside the SIBC spectrum, $\min\Sigma(T_L-\lambda I)$ decays as $L^{-1}$ and saturates to a constant value in the large-$L$ limit, respectively (see the inset for a double-log plot).} 
\label{fig:HN}
\end{figure}

\subsection{Bulk-edge correspondence}
\label{sec:bec}
In Sec.~\ref{Sec:5peti}, we have introduced several prototypical non-Hermitian models with nontrivial bulk topology. As is well-known for Hermitian topological systems, a nontrivial bulk topology usually implies the emergence of edge states under the open boundary condition (OBC). This statement is usually called \emph{bulk-edge correspondence} \cite{HY93}, which has been rigorously proved for Hermitian systems respecting various symmetries with the help of index theorems \cite{EP16}. It is thus natural to ask whether or not we have a similar bulk-edge correspondence for non-Hermitian topological systems. The answer to this question has been found to be apparently \emph{negative} in many numerical studies concerning the role of boundary conditions \cite{LTE16,YX18,YS18a}. 

As the simplest example 
we recall the single-band Hatano-Nelson model 
(\ref{KHN}), which has an elliptic spectrum under  PBC. While each point within the ellipse is associated with a nontrivial winding number, we can straightforwardly check that the spectrum under OBC collapses into an interval $[-2\sqrt{J_{\rm L}J_{\rm R}},2\sqrt{J_{\rm L}J_{\rm R}}]$ on the real axis (see Fig.~\ref{fig:HN}(a)) and there does not seem to be a special topological mode. Meanwhile, all the eigenstates under OBC might be considered as edge modes since they are exponentially localized at one side of the lattice simply due to asymmetric hoppings, a phenomenon often called the \emph{skin effect} \cite{YS18a}.

\begin{figure}[!t]
\begin{center}
\includegraphics[width=12cm]{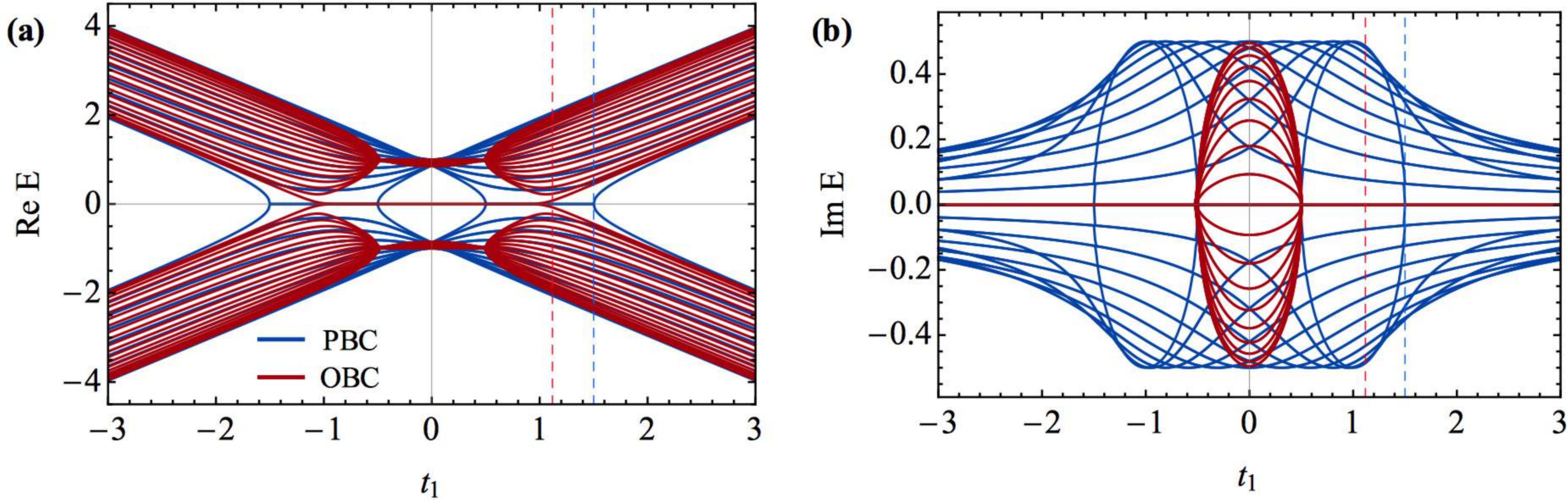}
\end{center}
\caption{(a) Real and (b) imaginary parts of the full energy spectrum of the non-Hermitian SSH model (\ref{SLSNHH}) under the PBC (blue) and the OBC (red) for $t_2=1$, $\gamma=0.5$, and $L=20$. The difference between the two boundary conditions  is small for $t_1\gg\gamma$ (i.e., small non-Hermiticity), but becomes significant when $t_1$ is comparable to $\gamma$.  The blue and red dashed lines indicate the PBC gap-closing point $t_1=t_2+\gamma$ and the OBC gap-closing point $t_1=\sqrt{t_2^2+\gamma^2}$, respectively.}
\label{fig:POBC}
\end{figure}

Another minimal example demonstrating the violation of the bulk-boundary correspondence is a two-band 
non-Hermitian SSH model with asymmetric intra-unit-cell hopping amplitudes \cite{LTE16,YS18a,JL19}. The real-space Hamiltonians is
\begin{equation}
H=\sum_j[(t_1-\gamma)|jB\rangle\langle jA|+(t_1+\gamma)|jA\rangle\langle jB|+t_2(|j+1,A\rangle\langle jB|+{\rm H.c.})],
\label{SLSNHH}
\end{equation}
where $A$ and $B$ label the two sublattices. The corresponding Bloch Hamiltonian under the PBC is already given in Eq.~(\ref{asyhSSH}) and reads $H(k)=(t_1+t_2\cos k)\sigma^x+(t_2\sin k+i\gamma)\sigma^y$ in the new notations. This Hamiltonian has a sublattice symmetry 
\begin{equation}
H=-SHS^\dag,\;\;\;\;S=\sum_j(|jA\rangle\langle jA|-|jB\rangle\langle jB|), 
\end{equation}
and thus its spectrum is inversion-symmetric with respect to the origin. We can determine the bulk gap-closing point from $\det H(k)=0$ for some $k\in[-\pi,\pi]$. Assuming $t_{1,2},\gamma\in\mathbb{R}$ for simplicity, this condition is obtained to be $|t_1|=|\gamma\pm t_2|$. Unlike the Hatano-Nelson model, we {do} have zero-energy edge modes in the OBC spectrum for a certain range of parameters. However, the critical points corresponding to the emergence/disappearance of edge modes are given by $|t_1|=\sqrt{\gamma^2\pm t_2^2}$ (where ``$-$" is taken only for $|\gamma|>|t_2|$) \cite{YS18a,KFK18}, mismatching the bulk gap-closing points (see Fig.~\ref{fig:POBC}). Such a discrepancy implies that the characterization of the edge modes by the bulk topological invariants is not appropriate. We note that Eq.~(\ref{SLSNHH}) also exhibits the non-Hermitian skin effect, i.e., all the OBC eigenstates  
are trivially localized at one boundary due to asymmetric hoppings.

In the following, we review recent efforts on restoring the bulk-edge correspondence in non-Hermitian topological systems. We shall discuss see two complementary approaches --- one modifies the definition of edge modes, the other modifies the topological invariants. Both approaches have (dis)advantages,  and we should apply either or both of them depending on the concrete situation.
\\ \\ {\it Pseudo-spectrum approach}

\vspace{3pt}
\noindent
In Sec.~\ref{sec2}, we have seen that the spectrum of a large Jordan block, which can be considered as the Hamiltonian of a Hatano-Nelson model with unidirectional hoppings, is extremely sensitive to perturbations. To quantitatively study the sensitivity, Reichel and Trefethen suggested to consider the \emph{pseudo-spectrum} and \emph{pseudo-eigenstates}  \cite{Reichel1992}. In their original paper, they focused on general non-Hermitian \emph{Toeplitz matrices}, \emph{Toeplitz operators} and \emph{Laurent operators}. In the language of physics, these are nothing but the Hamiltonians of translation-invariant single-band 1D lattices under the OBC, semi-infinite boundary condition (SIBC) and PBC.  The following theorem is proved in Ref.~\cite{Reichel1992}:
\begin{theorem}\label{Toeplitz}
Given $\{a_j\in\mathbb{C}\}_{j\in\mathbb{Z}}$ with $\sum_{j\in\mathbb{Z}}|a_j|<\infty$\footnote{While $f'(z)$ in Eq.~(\ref{LamLT}) might diverge at some $z$ under this condition, this does affect the well-definedness of the integral since it simply counts the phase winding of $f(z)-\lambda$, which is well-defined (i.e., bounded) along the path.} specifying a series of Toeplitz matrices $[T_L]_{jj'}=a_{j-j'}$ with $j,j'\in\{0,1,...,L-1\}$ and $L\in\mathbb{Z}^+$, a Toeplitz operator $[T]_{jj'}=a_{j-j'}$ with $j,j'\in\mathbb{N}$ and a Laurent operator $[L]_{jj'}=a_{j-j'}$ with $j,j'\in\mathbb{Z}$, we have
\begin{equation}
\begin{split}
\Lambda(L)&=\{f(z):|z|=1,z\in\mathbb{C}\},\\
\Lambda(T)&=\Lambda(L)\cup\{\lambda\in\mathbb{C}\backslash\Lambda(L):w(\lambda)\neq0\},\;\;\;\;w(\lambda)\equiv\oint_{|z|=1}\frac{dz}{2\pi i}\frac{f'(z)}{f(z)-\lambda}, \\
\Lambda_\epsilon(T)&=\lim_{L\to\infty}\Lambda_\epsilon(T_L)\;\;\Rightarrow\;\;\Lambda(T)=\lim_{\epsilon\to0}\lim_{L\to\infty}\Lambda_\epsilon(T_L),
\end{split}
\label{LamLT}
\end{equation}
where $\Lambda$ is the spectrum without taking into account the degeneracy, $\Lambda_\epsilon$ is the $\epsilon$-pseudo-spectrum defined in Eq.~(\ref{Lamep}) and $f(z)=\sum_{j\in\mathbb{Z}}a_jz^j$ is called the \emph{symbol} of the Toeplitz matrix.
\end{theorem}
The above result was rediscovered in Ref.~\cite{ZG18} in a physical context, and is reinterpreted as a novel bulk-edge correspondence with a \emph{continuum} of (pseudo-)edge modes under the SIBC (OBC). Indeed, comparing $\Lambda(L)$ (PBC) and $\Lambda(T)$ (SIBC), we find that the latter 
covers the entire interior of the former, 
wherever the spectral winding number $w(\lambda)$, whose equivalent version with $z$ replaced by $e^{ik}$ already appears in Eq.~(\ref{NHwn}), does not vanish (see Fig.~\ref{fig:HN}(a)). In fact, we can further show that $w(\lambda)>0$ ($w(\lambda)<0$) indicates the existence of left (right) edge modes under the right (left) SIBC and $|w(\lambda)|$ gives the degeneracy at energy $\lambda$. Moreover, under the OBC, a normalized pseudo-edge mode $\bold{u}$ 
at $E_{\rm B}=\lambda$ with $w(\lambda)\neq0$ only has an exponentially small error (see Theorem 3.2 in Ref.~\cite{Reichel1992} for a more precise statement):
\begin{equation}
\|(T_L-\lambda I)\bold{u}\|\sim e^{-O(L)}.
\label{mTLS}
\end{equation}
See Fig.~\ref{fig:HN}(b) for an illustration based on the Hatano-Nelson model. Physically, this fact implies that a wave packet initialized as a pseudo-edge mode $\bold{u}$ 
evolves like an eigenstate with energy $\lambda$, i.e., $e^{-iT_Lt}\bold{u}\simeq e^{-i\lambda t}\bold{u}$, until some time scale proportional to $L$ 
\cite{ZG18}.

We argue that the pseudo-spectrum approach is applicable to general non-Hermitian lattice models in arbitrary dimensions and symmetry classes. The argument is based on the one-to-one correspondence between non-Hermitian Hamiltonians and chiral symmetric Hermitian Hamiltonians, as is already shown in Eq.~(\ref{HM}). For our present purpose, we modify Eq.~(\ref{HM}) into 
\begin{equation}
H_{\rm H}=\begin{bmatrix} \;0\; & \;\;H-E_{\rm B}I\;\; \\ \;H^\dag-E_{\rm B}^*I\; & \;\;0\;\; \end{bmatrix},
\end{equation}
so that $H$ being point gapped at $E_{\rm B}$ is equivalent to $H_{\rm H}$ being gapped at $E_{\rm F}=0$. Thanks to this correspondence, a topologically nontrivial non-Hermitian $H$ implies that $H_{\rm H}$ is also topologically nontrivial as a chiral symmetric Hamiltonian and thus zero-energy edge modes should emerge under the OBC. Precisely speaking, the energy is exactly zero only in the thermodynamic limit, while it stays finite but exponentially small in terms of the system size. Recalling that the positive half of the spectrum of $H_{\rm H}$ is nothing but the singular value spectrum of $H-E_{\rm B}$, we conclude that $H$ has an exact edge mode (a pseudo-edge mode) at $E_{\rm B}$ under the SIBC (OBC). We emphasize again that this argument does not rely on specific spatial dimensions or symmetries, so that it is of course possible to have pseudo-edge modes characterized by a topological invariant other than the winding number \cite{ON19}. Within the framework of this pseudo-spectrum approach, it has been shown that a pseudoedge mode (or a skin mode) can be related to the point-gap topology \cite{ZG18,KZ19}. For instance, one can show that \cite{NO20}
\begin{theorem}\label{OPSI}
The spectrum of the Hamiltonian of a 1D non-Hermitian lattice system under the OBC is always contained in the spectrum under the SIBC, or equivalently, encircled by the spectrum under the PBC. Moreover, for an OBC mode with energy $E$, it is an exponentially localized skin mode whenever $w(E)\neq0$. 
\end{theorem}
\begin{figure}[!t]
\begin{center}
\includegraphics[width=12cm]{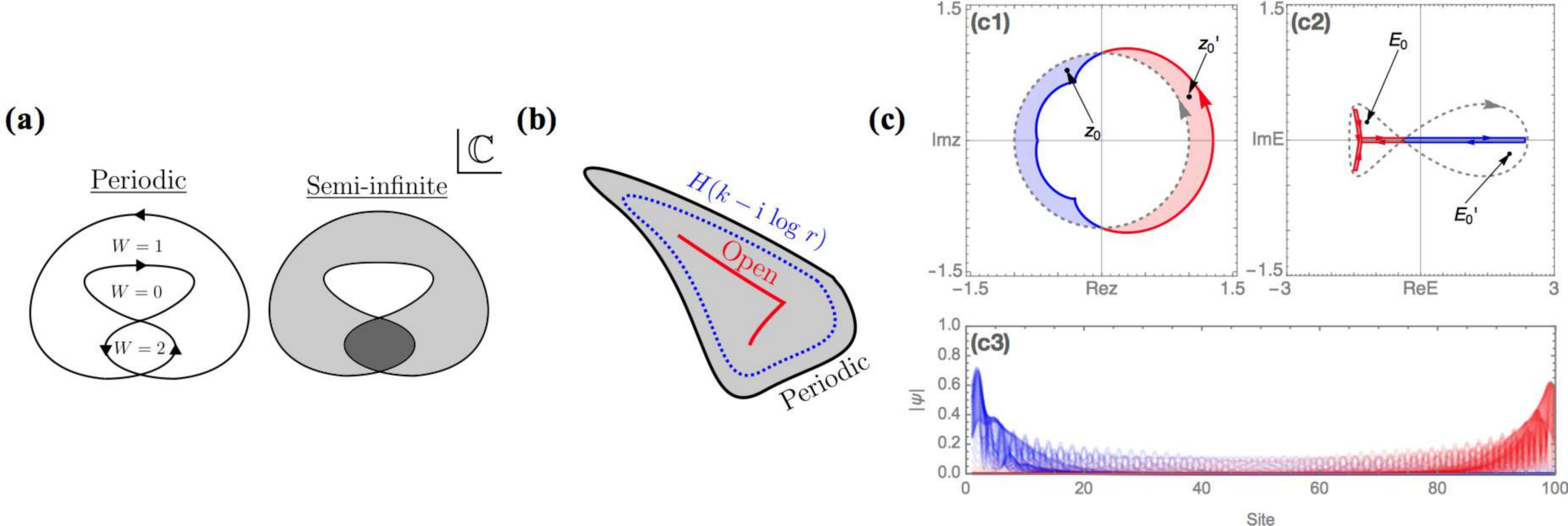}
\end{center}
\caption{(a) Schematic illustration of the fact that the SIBC spectrum (gray region) is the interior of the PBC spectrum (black curve) with nonzero winding numbers. (b) The OBC spectrum (red curve) is inside the SIBC spectrum, and is thus encircled by the PBC spectrum. Moreover, it is also encircled by the spectrum under an arbitrary twisted boundary condition by an imaginary flux (dashed blue curve). (c) (c1) Original (gray) and generalized (blue and red) Brillouin zones, (c2) spectra under the PBC (gray) and OBC (blue and red), and (c3) all the eigen-wavefunctions of Eq.~(\ref{fish}) with $J_{\rm L}=J_{\rm R}=1$ and $J_2=0.4$  under the OBC. Here, $z_0$ and $z'_0$ are two randomly chosen points encompassed by the original and generalized Brillouin zones, and the corresponding energies $E_0$ and $E_0'$ fall in the SIBC spectrum. The majority of the eigenstates are localized at either of the boundaries (depending on the sign of the winding number), while those with eigenenergies near the point that connects the ``fish tail" and ``body" are extended. Adapted from Refs.~\cite{NO20} and \cite{KZ19}. Copyright \copyright\,  2020 by the American Physical Society.}
\label{fig:SKE}
\end{figure}
\noindent See Figs.~\ref{fig:SKE}(a) and (b) for a schematic illustration. The essence of this result has already been evident in the Hatano-Nelson model (\ref{RSHN}) as discussed in Ref.~\cite{ZG18}, whose OBC spectrum is a line segment within the elliptical PBC spectrum (see Fig.~\ref{fig:HN}). Another example is a single-band system studied in Ref.~\cite{KZ19} (see Fig.~\ref{fig:SKE}(c)) whose Hamiltonian is given by
\begin{equation}
H=\sum_j(J_{\rm L}|j-1\rangle\langle j| +J_{\rm R}|j\rangle\langle j-1|+J_2 |j+1\rangle\langle j-1|).
\label{fish}
\end{equation}
The PBC and OBC spectra just look like a ``fish" and a ``fishbone", respectively. Accordingly, all the pseudo-eigenstates at the OBC spectrum, including the exact eigenstates under the OBC, should be exponentially localized edge modes.

Moreover, {symmetry-protected} skin effects have been proposed in Ref.~\cite{NO20}. A prototypical example is the following two-band model in 1D:
\begin{equation}
H(k)=2t\cos k\sigma_0+2\Delta\sin k\sigma^x+2i g\sin k \sigma^z,
\end{equation} 
which is constructed from two copies of Hatano-Nelson models with one of them transposed, and thus exhibits a so-called TRS$^\dag$ (cf. Eq.~(\ref{NHAZdag}) below):
\begin{equation}
\sigma^y H(k)^{\rm T}\sigma^y=H(-k).
\label{syT}
\end{equation}
We can check from Eq.~(\ref{syT}) that the winding number (\ref{NHwn}) always vanishes regardless of $E_{\rm B}$. Nevertheless, all the eigenstates under the OBC are found to be localized, but in a pairwise manner at both left and right edges due to the Kramers degeneracy. This turns out to be a \emph{symmetry-protected} skin effect since the OBC eigenstates can be delocalized by adding symmetry-breaking perturbations. Moreover, this skin effect is characterized by a $\mathbb{Z}_2$ point-gap topological number (see class AII$^\dag$ in Table~\ref{table4}) and can again be destroyed by coupling two copies of the system while keeping the symmetry. 
While these subtle topological aspects may warrant further studies from a mathematical point of view, at the present time these so-called {\it non-Hermitian skin effects} appear to be rather trivial from a perspective of physics, which are already well known in the context of convective flows in hydrodynamics or stochastic processes/complex networks (see, e.g., Ref.~\cite{MA17} and Sec.~\ref{secbio}). 

The pseudo-spectrum approach is compatible with the bulk-edge correspondence for point-gap topology. However, its obvious disadvantage is that it does not provide any \emph{quantitative} information about the exact eigenvalues and eigenstates under the OBC. Indeed, Theorem~\ref{Toeplitz} concerns $\lim_{\epsilon\to0}\lim_{L\to\infty}\Lambda_\epsilon(H_{\rm OBC})$ while Theorem~\ref{OPSI} only gives a \emph{qualitative} statement on $\lim_{L\to\infty}\lim_{\epsilon\to0}\Lambda_\epsilon(H_{\rm OBC})$. To  determine the spectrum of physical importance 
together with the corresponding eigenstates, we need a complementary approach, which is explained as follows.
\\ \\ {\it Non-Bloch-wave approach}

\vspace{3pt}
\noindent
Since any actual physical systems are finite and have the corresponding boundaries,  
 the bulk topological number (obtained under PBC) is generally no longer a faithful indicator of the presence or absence of edge modes in non-Hermitian systems, as has been exemplified in the beginning of this section. To  restore the genuine bulk-edge correspondence, we have to figure out the appropriate topological invariants under the OBC. There have been several efforts along this line, one of which is often known as the \emph{non-Bloch-wave approach} \cite{YS18a,YS18b,KY19}. This approach has been developed  for arbitrary dimensional systems with the OBC in only one direction \cite{KY19}, and for 
the non-Hermitian Chern insulators with the OBC in two directions \cite{YS18b}. For simplicity, we 
focus on 1D systems hereafter. 

The main idea of this approach is that, to reproduce the spectrum under the OBC, we have to take into account the non-Bloch-wave nature of the eigenstates by deforming the Brillouin zone $\beta=e^{ik}\in\mathbb{C}$ from the unit circle to a model-specific contour $C_\beta$, which is called the \emph{generalized Brillouin zone}. Accordingly, the appropriate topological invariant for characterizing the edge modes should be defined with respect to the generalized Brillouin zone. To illustrate how to determine $C_\beta$, let us take the non-Hermitian SSH model (\ref{SLSNHH}) as an example. Our ansatz of a non-Bloch right eigenstate under the OBC with eigenenergy $E$ takes the form \cite{YS18a}
\begin{equation}
|\psi\rangle=\sum^2_{\mu=1}\sum^L_{j=1}\beta_\mu^j(c_{\mu A}|jA\rangle+c_{\mu B}|jB\rangle).
\end{equation}
Here $L$ is the number of unit cells and $\beta_\mu$'s ($\mu=1,2$) satisfy
\begin{equation}
\det[H(\beta)-E]=(t_1-\gamma+t_2\beta)(t_1+\gamma+t_2\beta^{-1})-E^2=0,
\label{BKRS}
\end{equation}
where $H(\beta)$ is obtained from the Bloch Hamiltonian $H(k)$ by replacing $e^{ik}$ with $\beta$. In addition, the coefficients $\bold{c}_\mu\equiv [c_{\mu A},c_{\mu B}]^{\rm T}$'s should satisfy 
\begin{equation}
[H(\beta)-E]\bold{c}_\mu=\bold{0}\;\;\;\;\Rightarrow\;\;\;\; c_{\mu A}=\frac{E}{t_1-\gamma+t_2\beta_\mu}c_{\mu B}.
\label{BKRC}
\end{equation}
These two equations (\ref{BKRS}) and (\ref{BKRC}) ensure $H|\psi\rangle=E|\psi\rangle$ in the bulk, i.e., $\langle j\alpha|H|\psi\rangle=E\langle j\alpha|\psi\rangle$ for $j=2,3,...,L-1$ and $\alpha=A,B$. To make $|\psi\rangle$ an exact eigenstate, we only have to impose the boundary conditions:
\begin{equation}
c_{1B}+c_{2B}=0,\;\;\;\;\beta_1^{L+1}c_{1A}+\beta_2^{L+1}c_{2A}=0.
\label{BDRC}
\end{equation}
Combining Eq.~(\ref{BDRC}) with Eq.~(\ref{BKRC}), we obtain $\beta_1^{L+1}(t_1-\gamma+t_2\beta_2)=\beta_2^{L+1}(t_1-\gamma+t_2\beta_1)$, 
which implies $|\beta_1|=|\beta_2|$ in the thermodynamic limit $L\to\infty$. Recalling that $\beta_\mu$'s satisfy Eq.~(\ref{BKRS}), we must have $\beta_1\beta_2=\frac{t_1-\gamma}{t_1+\gamma}$ and thus $|\beta_1|=|\beta_2|=\sqrt{\frac{|t_1-\gamma|}{|t_1+\gamma|}}$. Therefore, the generalized Brillouin zone $C_\beta$ is a circle with radius $r=\sqrt{\frac{|t_1-\gamma|}{|t_1+\gamma|}}$. We can check that the gap closing point of $H(\beta)$ over $C_\beta$ exactly gives $|t_1|=\sqrt{\gamma^2\pm t_2^2}$. Moreover, the winding number can be calculated as
\begin{equation}
w=i\int_{C_\beta}\frac{d\beta}{2\pi}\partial_\beta\ln\det q(\beta),
\label{GBZwn}
\end{equation}
which is a natural generalization of Eq.~(\ref{wn}). For the current model (\ref{SLSNHH}), we have $q(\beta)=t_1+\gamma+t_2\beta^{-1}$, implying $w=1$ for $\sqrt{\gamma^2-t_2^2}<|t_1|<\sqrt{\gamma^2+t_2^2}$ if $|\gamma|>|t_2|$ and for $|t_1|<\sqrt{\gamma^2+t_2^2}$ if $|\gamma|<|t_2|$ and $w=0$ otherwise.

\begin{figure}[!t]
\begin{center}
\includegraphics[width=12cm]{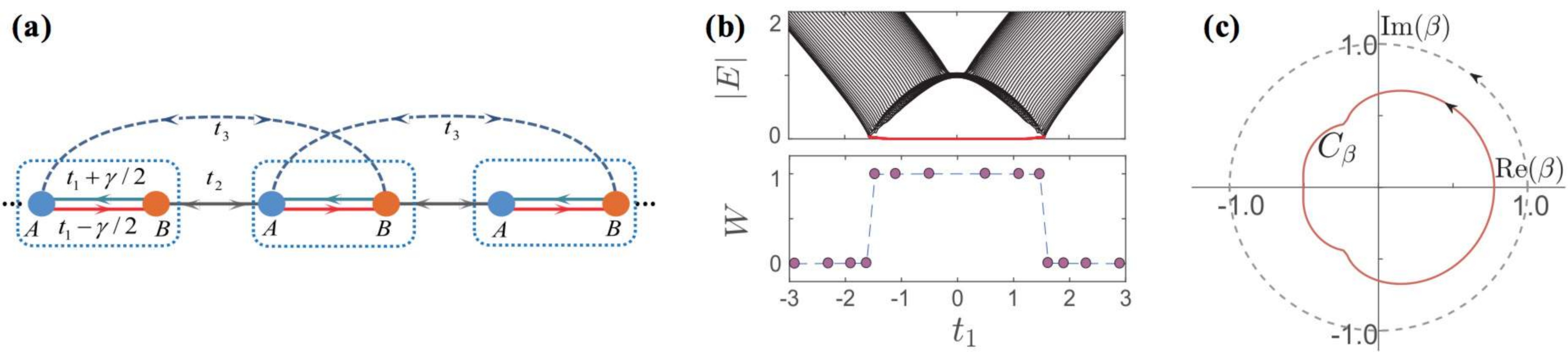}
\end{center}
\caption{(a) Schematic illustration of the sublattice-symmetric model given in Eq.~(\ref{SSHt3}). (b) Absolute values of the OBC spectrum (upper panel) and the winding numbers (\ref{GBZwn}) defined on the generalized Brillouin zones (lower panel). (c) Original (gray) and generalized (red) Brillouin zone for $t_1=1.1$. In both (b) and (c), the parameters are chosen to be $t_2=1$, $\gamma=2/3$ and $t_3=0.2$. Adapted from Ref.~\cite{YS18a}. Copyright \copyright\,  2018 by the American Physical Society.}
\label{fig:GBZ}
\end{figure}

So far we have analyzed a specific model whose generalized Brillouin zone $C_\beta$  turns out to be a circle. In general, however, $C_\beta$ is not necessarily a circle, as has already been exemplified by a single-band model in Eq.~(\ref{fish}) (see Fig.~\ref{fig:SKE}(c)). Here, we provide another example studied in Ref.~\cite{YS18a}. As illustrated in Fig.~\ref{fig:GBZ}(a), the Hamiltonian reads
\begin{equation}
H=H_{\rm NHSSH}+t_3\sum_j(|j+1,B\rangle\langle jA|+{\rm H.c.}),
\label{SSHt3}
\end{equation}
where $H_{\rm NHSSH}$ is given  
in Eq.~(\ref{SLSNHH}). Following a similar procedure presented above, we can determine $C_\beta$ 
by solving the OBC eigenvalue problem and then calculate the winding number using Eq.~(\ref{GBZwn}). As  shown in Fig.~\ref{fig:GBZ}(b), the winding number defined on $C_\beta$,  
which is not a circle (see Fig.~\ref{fig:GBZ}(c)), indeed gives a faithful characterization for the edge modes. 

For general 1D models, a universal criterion for determining $C_\beta$ has been proposed  to be $|\beta_M|=|\beta_{M+1}|$ \cite{TR19,KY19}, where $|\beta_1|\le|\beta_2|\le...\le|\beta_{2M-1}|\le|\beta_{2M}|$ are $2M$ zeros of the polynomial $\det[H(\beta)-E]$ and $M=DR$ with $R$ and $D$ being the hopping range and the number of internal states per unit cell, respectively. To see this, denoting the internal states as $s_1,s_2,...,s_D$, we first 
write down a general eigenstate ansatz 
with energy $E$ as $|\psi\rangle=\sum^{2M}_{\mu=1}\sum^L_{j=1}\sum^D_{\alpha=1}c_\mu \beta^j_\mu u_{\mu s_\alpha}|js_\alpha\rangle$, 
where $\bold{u}_\mu\equiv[u_{\mu s_1},u_{\mu s_2},...,u_{\mu s_D}]^{\rm T}$ is a normalized Bloch eigenvector determined from $[H(\beta_\mu)-E]\bold{u}_\mu=\bold{0}$. Imposing the OBC based simply on a cutoff (although the result holds for more general OBC \cite{KY19}), we obtain the following boundary condition for the coefficients $c_\mu$'s:
\begin{equation}
\begin{bmatrix} 
\;\beta_1^{-(R-1)}\bold{u}_1\; & \;\beta_2^{-(R-1)}\bold{u}_2\; & \;\cdots\; & \;\beta_{2M}^{-(R-1)}\bold{u}_{2M}\; \\
\;\beta_1^{-(R-2)}\bold{u}_1\; & \;\beta_2^{-(R-2)}\bold{u}_2\; & \;\cdots\; & \;\beta_{2M}^{-(R-2)}\bold{u}_{2M}\; \\
\;\vdots\; & \;\vdots\; & \;\ddots\; & \;\vdots\; \\
\;\bold{u}_1\; & \;\bold{u}_2\; & \;\cdots\; & \;\bold{u}_{2M}\; \\
\;\beta_1^{L+1}\bold{u}_1\; & \;\beta_2^{L+1}\bold{u}_2\; & \;\cdots\; & \;\beta_{2M}^{L+1}\bold{u}_{2M}\; \\
\;\beta_1^{L+2}\bold{u}_1\; & \;\beta_2^{L+2}\bold{u}_2\; & \;\cdots\; & \;\beta_{2M}^{L+2}\bold{u}_{2M}\; \\
\;\vdots\; & \;\vdots\; & \;\ddots\; & \;\vdots\; \\
\;\beta_1^{L+R}\bold{u}_1\; & \;\beta_2^{L+R}\bold{u}_2\; & \;\cdots\; & \;\beta_{2M}^{L+R}\bold{u}_{2M}\; \\
\end{bmatrix}
\begin{bmatrix}
c_1 \\ c_2 \\ \vdots \\ c_{2M}
\end{bmatrix}
=\bold{0}.
\label{bcfc}
\end{equation}
To ensure a nontrivial solution of $c_\mu$'s, the determinant of the $2M\times 2M$ matrix\footnote{Since $\bold{u}_\mu$ has $D$ elements and there are totally $2R$ $\bold{u}_\mu$'s, the total number of rows is given by $2RD=2M$.} in Eq.~(\ref{bcfc}) must vanish. Recalling the assumption $|\beta_1|\le|\beta_2|\le ...\le|\beta_{2M-1}|\le|\beta_{2M}|$, we know that, for a sufficiently large $L$, the leading and sub-leading terms in the determinant should contain the factors $(\beta_{M+1}\beta_{M+2}...\beta_{2M})^L$ and $(\beta_{M}\beta_{M+2}...\beta_{2M})^L$, respectively. Suppose that $|\beta_M|<|\beta_{M+1}|$, then the leading term will overwhelm all the other terms in the thermodynamic limit and the determinant cannot be zero, leading to a contradiction. Therefore, we necessarily have $|\beta_M|=|\beta_{M+1}|$. Note that for a Hermitian system $H(\beta)$ satisfies $H(\beta)^\dag=H(1/\beta^*)$ and the spectrum under the OBC is real. This implies the solutions of $\det[H(\beta)-E]=0$ come into pairs $(\beta,1/\beta^*)$, and in particular $|\beta_M|=|1/\beta_M^*|\Rightarrow |\beta_M|=1$, implying that $C_\beta$ is the usual Brillouin zone (unit circle) \cite{KY19}. While the above procedure for determining $C_\beta$ is applicable to general 1D non-Hermitian systems, there are still many open problems, such as how to deal with 
higher dimensions \cite{YS18b}, and whether $C_\beta$ always forms a loop and what is the physical origin of the cusps in $C_\beta$ \cite{KY19}.

Finally, we briefly mention several complementary methods in  literature which are essentially along the same line of taking into account the non-Bloch-wave nature of eigenstates. Reference~\cite{KFK18} proposes the notion of the \emph{biorthogonal bulk-edge correspondence}, which suggests that the appropriate topological number, such as the polarization \cite{VD93}, should be defined in terms of both right and left eigenstates. This idea was successfully applied to deal with the model in Eq.~(\ref{SLSNHH}) to predict the correct phase boundary, and has later been incorporated in a systematic \emph{transfer-matrix formalism} \cite{KFK19}. This formalism, which was originally developed for Hermitian systems \cite{DV16}, can provide a solution to the eigenvalue problem under the OBC and the PBC. Moreover, some crucial information can directly be read out from the transfer matrix. For example, the zero determinant of the transfer matrix signals the existence of exceptional points.
 More recently, a modified non-Bloch-wave approach is proposed in Ref.~\cite{KII19}, which suggests that it is sufficient to consider \emph{circular} $C_\beta$ with various imaginary wave numbers $r=\log|\beta|$, but the phase diagram should be defined in the extended parameter space with the $r$ axis and that the OBC corresponds to a slice in this space. These related methods may conceptually  deepen and/or technically improve the non-Bloch-wave approach discussed above.

\subsection{Topological classifications}\label{Sec:5topcl}
We here review possible topological classifications of non-Hermitian Bloch Hamiltonians. A non-Hermitian periodic table has firstly been proposed in Ref.~\cite{ZG18} and later found a number of possible generalizations \cite{ZH19,LCH19,KK19c,LCH19b}. Technically, we focus on  classifications based on $K$-theory, which is the standard approach used for  Hermitian free fermions (see Sec.~\ref{ptAZ}), while there is some recent progress on the homotopy classifications \cite{RK18,GJ19,ZL19,CCW19}. A  classification not only incorporates known non-Hermitian topological invariants  into a common framework, but also can potentially guide the search for new topological invariants \cite{AG192}. Nevertheless, we emphasize that these classifications are not given under the OBC (that are relevant to actual physical setups), but for Bloch Hamiltonians that are under the PBC. Thus, unlike Hermitian systems, a classification often fails to correctly predict the presence of topological edge states because the breakdown of the bulk-boundary correspondence  is rather common in non-Hermitian systems as discussed above. In this respect, physical significance of non-Hermitian periodic tables can be challenged when robust  gapless edge modes can emerge from an alternative mechanism unique to non-Hermitian regimes \cite{SK192}.

\subsubsection{Bernard-LeClair classes}\label{Sec:5BL}
The symmetries in non-Hermitian systems are richer than those in Hermitian systems \cite{DB02,EK11,SM12,RJDH19}. For example, the CS $H=-\Gamma H^\dag \Gamma^\dag$ differs generally from the \emph{sublattice symmetry} (SLS) $H=-S H S^\dag$ in non-Hermitian systems, whereas  they are equivalent in the Hermitian case. In particular, the fundamental symmetry classification, i.e., the non-Hermitian version of the Altland-Zirnbauer classification \cite{AA97}, is the \emph{Bernard-LeClair} (BL) classification  \cite{DB01}. The symmetry constraint on a BL class can generally be written as
\begin{equation}
H=\epsilon_X U_X\mathcal{X}(H)U^\dag_X,
\label{BLsym}
\end{equation}
where $\epsilon_X\in\{\pm1\}$, $U_X$ is unitary and $\mathcal{X}(\cdot)$ can be the identity, Hermitian conjugation, transpose and complex conjugation, corresponding to $X=P$, $Q$, $C$ and $K$ symmetries. As a matrix in the fundamental symmetry classes, $H$ cannot possess any unitary symmetry. After iterating Eq.~(\ref{BLsym}) twice, we find $[U_X^2,H]=0$ for $X=P,Q$ and $[U_XU_X^*,H]=0$ for $X=C,K$, implying that $U_X^2$ or $U_XU_X^*$ should be proportional to the identity matrix. For $X=P,Q$, we can always choose an appropriate  ${\rm U}(1)$ phase so that $U_X^2=I$. For $X=C,K$, we have $U_XU_X^*=\pm I$.\footnote{We can assume $U_XU_X^*=e^{i\phi}I\Leftrightarrow U_X^*U_X=e^{-i\phi}I$ from the unitarity. Then $U_X^*=e^{i\phi}U^\dag=e^{-i\phi}U^\dag$, leading to $e^{i\phi}=\pm1$.} In the context of band topology, it is more convenient to use the Bloch-Hamiltonian version: 
\begin{equation}
\begin{split}
&P\;{\rm symmetry}\;\;H(\boldsymbol{k})=-U_PH(\boldsymbol{k}) U_P^\dag,\;\;\;\;\;\;\;\;\;\;\;\;\;\;\;\;\;\;\;\;U^2_P=I, \\
&Q\;{\rm symmetry}\;\;H(\boldsymbol{k})=\epsilon_QU_QH(\boldsymbol{k})^\dag U_Q^\dag,\;\;\;\;\;\;\;\;\;\;\;\;\;\;\;\;\;U^2_Q=I, \\
&C\;{\rm symmetry}\;\;H(\boldsymbol{k})=\epsilon_CU_CH(-\boldsymbol{k})^{\rm T} U_C^\dag,\;\;\;\;\;\;\;\;U_CU_C^*=\eta_CI, \\
&K\;{\rm symmetry}\;\;H(\boldsymbol{k})=\epsilon_KU_KH(-\boldsymbol{k})^*U_K^\dag,\;\;\;\;\;\;U_KU_K^*=\eta_KI. 
\end{split}
\label{PQCK}
\end{equation}
Note that the wave number is reversed for $C$ and $K$ symmetries, since transpose and complex conjugate correspond to spatial inversion and time reversal, respectively. We will discuss in Sec.~\ref{Sec:NHHC} their relation to PHS and TRS. 
The previously mentioned SLS corresponds to $Q$ symmetry with $\epsilon_Q=-1$ and that with $\epsilon_Q=1$ implies pseudo Hermiticity (see Sec.~\ref{secphqh}).

Similarly to the fact that the coexistence of two in the CS, PHS and TRS implies the other in an AZ class, here the coexistence of two symmetries in $Q$, $C$ and $K$ implies the other. Moreover, when two or more symmetries coexist, we have 
\begin{equation}
\begin{split}
U_P=\epsilon_{PQ}U_QU_PU_Q^\dag,\;\;\;\;\;\;\;\;&U_C=\epsilon_{PC}U_PU_CU_P^{\rm T}, \\
U_K=\epsilon_{PK}U_PU_KU_P^{\rm T},\;\;\;\;\;\;\;\;&U_C=\epsilon_{QC}U_QU_CU_Q^{\rm T}, 
\end{split}
\label{epXY}
\end{equation}
due again to the absence of unitary symmetries. Here $\epsilon_{XY}=\pm1$ as can be seen by 
an iteration of each of Eq.~\eqref{epXY} together with $U_Q^2=U_P^2=I$. We emphasize that $\epsilon_X$, $\epsilon_{XY}$ and $\eta_X$ are generally \emph{not} independent. For example, once we know $\epsilon_Q$, $\epsilon_C$ and $\epsilon_{QC}$, we have $\epsilon_K=\epsilon_Q\epsilon_C$, $U_K=U_QU_C$ and $\eta_K=\epsilon_{QC}\eta_C$. Also, some apparently different combinations are equivalent. For example, when $P$ and $C$ coexist, $\epsilon_C$ can be reversed if we redefine $U_C$ by $U_PU_C$ such that $\eta_C$ becomes $\epsilon_{PC}\eta_C$, i.e., $(\epsilon_C,\eta_{PC},\eta_C)$ becomes equivalent to $(-\epsilon_C,\eta_{PC},\eta_{PC}\eta_C)$. In addition, since the band topology of $H(\boldsymbol{k})$ does not change upon being multiplied by $i$, which reverse $\epsilon_Q$ and $\epsilon_K$, some BL classes are unified for the point gap. Such a Wick rotation also establishes a map between real- and imaginary-line-gapped systems, but it does \emph{not} reduce symmetry classes for a given type (either real or imaginary) of line gaps.  After  taking these factors into a simple counting procedure (see Appendix~\ref{Sec:countBL}), we can obtain $38$ classes for the point gap and $54$ classes for a given type of line gaps.

\subsubsection{Periodic tables}
\label{Sec:NHPT}
We introduce a method of systematically classifying both point-gapped and line-gapped non-Hermitian Bloch Hamiltonians. For the former, we set $E_{\rm B}=0$ without loss of generality, so that point-gapped simply means 
invertible. This property allows a unique \emph{polar decomposition} \cite{ZG18}:
\begin{equation}
H(\boldsymbol{k})=U(\boldsymbol{k})P(\boldsymbol{k}),
\end{equation}
where $P(\boldsymbol{k})=\sqrt{H(\boldsymbol{k})^\dag H(\boldsymbol{k})}$ is a positive-definite Hermitian operator and $U(\boldsymbol{k})=H(\boldsymbol{k})P(\boldsymbol{k})^{-1}$ is unitary, both of which are continuous in $\boldsymbol{k}$. This is a natural generalization of band flattening \cite{AK09} since when $H(\boldsymbol{k})$ is Hermitian the unitaried Hamiltonian is nothing but the flattened Hamiltonian.\footnote{If a matrix $H$ is Hermitian and has a unique polar decomposition $H=UP$, we have $H=H^\dag=PU^\dag=U^\dag(UPU^\dag)$. From the uniqueness, we have $U=U^\dag$ and thus $U^2=UU^\dag=I$, implying that the spectrum of $U$ consists of $\pm1$.} Using the uniqueness of the decomposition, we can show that $U(\boldsymbol{k})$ shares all the BL symmetries as $H(\boldsymbol{k})$ \cite{ZH19}. This is because 
\begin{equation}
\begin{split}
H(\boldsymbol{k})&=U(\boldsymbol{k})P(\boldsymbol{k})=\epsilon_XU_X\mathcal{X}[H(s_X\boldsymbol{k})]U_X^\dag \\
&=\epsilon_X U_X\mathcal{X}[U(s_X\boldsymbol{k})]U_X^\dag U_X\mathcal{X}[P(s_X\boldsymbol{k})]U_X^\dag 
\end{split}
\end{equation}
if $\mathcal{X}$ is the identity or complex conjugate and
\begin{equation}
U(\boldsymbol{k})P(\boldsymbol{k})=\epsilon_X U_X\mathcal{X}[U(s_X\boldsymbol{k})]U_X^\dag U_X\mathcal{X}[U(s_X\boldsymbol{k})P(s_X\boldsymbol{k})U(s_X\boldsymbol{k})^\dag]U_X^\dag 
\end{equation}
if $\mathcal{X}$ is the transpose or Hermitian conjugate, implying
\begin{equation}
U(\boldsymbol{k})=\epsilon_X U_X\mathcal{X}[U(s_X\boldsymbol{k})]U_X^\dag
\end{equation} 
for both cases. Therefore, a simple interpolation $H_\lambda(\boldsymbol{k})=(1-\lambda) H(\boldsymbol{k})+\lambda U(\boldsymbol{k})$ ($\lambda\in[0,1]$) gives a continuous symmetry-preserving path, which  also preserves the point gap since $H_\lambda(\boldsymbol{k})=\lambda U(\boldsymbol{k})[(1-\lambda)P(\boldsymbol{k})+\lambda I]$ is a product of a unitary operator and a positive-definite operator. 

\begin{table*}[tbp]
\caption{Topological classifications of point-gapped non-Hermitian Bloch Hamiltonians with BL symmetries. Here we always fix $\epsilon_Q=-1$ and $\epsilon_K=1$ by Wick rotation. If $C$ and $P$ symmetries coexist, we also fix $\epsilon_C=-1$ by redefining $U_C$ as $U_PU_C$. The parameters in the brackets can be reversed by simultaneously Wick rotating the Hamiltonian and redefining the symmetry unitary.}
\begin{center}
\begin{tabular}{ccc}
\hline\hline
BL sym. & $\epsilon_X$, $\eta_X$, $\epsilon_{XY}$ & $K$-group \\ 
\hline
None & N/A  & $K_{\mathbb{C}}(1;d)$  \\ 
$P$ & N/A & $K_{\mathbb{C}}(1;d)\oplus K_{\mathbb{C}}(1;d)$ \\ 
$Q$ & N/A & $K_{\mathbb{C}}(0;d)$ \\ 
$C_{++}$ & $\epsilon_C=1,\eta_C=1$ & $K_{\mathbb{R}}(7;d)$ \\ $C_{+-}$ & $\epsilon_C=1,\eta_C=-1$ & $K_{\mathbb{R}}(3;d)$ \\ 
$C_{-+}$ & $\epsilon_C=-1,\eta_C=1$ & $K_{\mathbb{R}}(3;d)$ \\ 
$C_{--}$ & $\epsilon_C=-1,\eta_C=-1$ & $K_{\mathbb{R}}(7;d)$ \\
$K_+$ & 
$\eta_K=1$ & $K_{\mathbb{R}}(1;d)$ \\
$K_-$ & 
$\eta_K=-1$ & $K_{\mathbb{R}}(5;d)$ \\
$PQ^+$ & 
$\epsilon_{PQ}=1$ & $K_{\mathbb{C}}(1;d)$ \\
$PQ^-$ & 
$\epsilon_{PQ}=-1$ & $K_{\mathbb{C}}(0;d)\oplus K_{\mathbb{C}}(0;d)$ \\
$PC^+_+$ & 
$\eta_C=1,\epsilon_{PC}=1$ & $K_\mathbb{C}(1;d)$ \\
$PC^+_-$ & 
$\eta_C=-1,\epsilon_{PC}=1$ & $K_\mathbb{C}(1;d)$ \\
$PC^-_+$ & 
$\eta_C=1,\epsilon_{PC}=-1$ & $K_\mathbb{R}(3;d)\oplus K_\mathbb{R}(3;d)$\\
$PC^-_-$ & 
$\eta_C=-1,\epsilon_{PC}=-1$ & $K_\mathbb{R}(7;d)\oplus K_\mathbb{R}(7;d)$\\
$PK^+_+$ & 
$\eta_K=1,\epsilon_{PK}=1$ & $K_\mathbb{R}(1;d)\oplus K_\mathbb{R}(1;d)$\\
$PK^+_-$ & 
$\eta_K=-1,\epsilon_{PK}=1$ & $K_\mathbb{R}(5;d)\oplus K_\mathbb{R}(5;d)$\\
$PK^-$ & 
($\eta_K=1$) $\epsilon_{PK}=-1$ & $K_\mathbb{C}(1;d)$ \\
$QC^+_{++}$ & 
$\epsilon_C=1,\eta_C=1,\epsilon_{QC}=1$ & $K_\mathbb{R}(0;d)$ \\
$QC^+_{+-}$ & 
$\epsilon_C=1,\eta_C=-1,\epsilon_{QC}=1$ & $K_\mathbb{R}(4;d)$ \\
$QC^+_{-+}$ & 
$\epsilon_C=-1,\eta_C=1,\epsilon_{QC}=1$ & $K_\mathbb{R}(2;d)$ \\
$QC^+_{--}$ & 
$\epsilon_C=-1,\eta_C=-1,\epsilon_{QC}=1$ & $K_\mathbb{R}(6;d)$ \\
$QC^-_{++}$ & 
$\epsilon_C=1,\eta_C=1,\epsilon_{QC}=-1$ & $K_\mathbb{R}(6;d)$ \\
$QC^-_{+-}$ & 
$\epsilon_C=1,\eta_C=-1,\epsilon_{QC}=-1$ & $K_\mathbb{R}(2;d)$ \\
$QC^-_{-+}$ & 
$\epsilon_C=-1,\eta_C=1,\epsilon_{QC}=-1$ & $K_\mathbb{R}(4;d)$ \\
$QC^-_{--}$ & 
$\epsilon_C=-1,\eta_C=-1,\epsilon_{QC}=-1$ & $K_\mathbb{R}(0;d)$ \\
$PQC^{+++}_+$ & 
$\eta_C=1,\epsilon_{PQ}=1,\epsilon_{PC}=1,\epsilon_{QC}=1$ & $K_\mathbb{R}(1;d)$ \\
$PQC^{+++}_-$ & 
$\eta_C=-1,\epsilon_{PQ}=1,\epsilon_{PC}=1,\epsilon_{QC}=1$ & $K_\mathbb{R}(5;d)$ \\
$PQC^{++-}_+$ & 
$\eta_C=1,\epsilon_{PQ}=1,\epsilon_{PC}=1,\epsilon_{QC}=-1$ & $K_\mathbb{R}(5;d)$ \\
$PQC^{++-}_-$ & 
$\eta_C=-1,\epsilon_{PQ}=1,\epsilon_{PC}=1,\epsilon_{QC}=-1$ & $K_\mathbb{R}(1;d)$ \\
$PQC^{+-}_+$ & 
$\eta_C=1,\epsilon_{PQ}=1,\epsilon_{PC}=-1$ ($\epsilon_{QC}=1$) & $K_\mathbb{R}(3;d)$ \\
$PQC^{+-}_-$ & 
$\eta_C=-1,\epsilon_{PQ}=1,\epsilon_{PC}=-1$ ($\epsilon_{QC}=1$) & $K_\mathbb{R}(7;d)$ \\
$PQC^{-+}_+$ & 
$\eta_C=1,\epsilon_{PQ}=-1,\epsilon_{PC}=1$ ($\epsilon_{QC}=1$) & $K_\mathbb{C}(0;d)$ \\
$PQC^{-+}_-$ & 
$\eta_C=-1,\epsilon_{PQ}=-1,\epsilon_{PC}=1$ ($\epsilon_{QC}=1$) & $K_\mathbb{C}(0;d)$  \\
$PQC^{--+}_+$ & 
$\eta_C=1,\epsilon_{PQ}=-1,\epsilon_{PC}=-1,\epsilon_{QC}=1$ & $K_\mathbb{R}(2;d)\oplus K_\mathbb{R}(2;d)$ \\
$PQC^{--+}_-$ & 
$\eta_C=-1,\epsilon_{PQ}=-1,\epsilon_{PC}=-1,\epsilon_{QC}=1$ & $K_\mathbb{R}(6;d)\oplus K_\mathbb{R}(6;d)$ \\
$PQC^{---}_+$ & 
$\eta_C=1,\epsilon_{PQ}=-1,\epsilon_{PC}=-1,\epsilon_{QC}=-1$ & $K_\mathbb{R}(4;d)\oplus K_\mathbb{R}(4;d)$ \\
$PQC^{---}_-$ & 
$\eta_C=-1,\epsilon_{PQ}=-1,\epsilon_{PC}=-1,\epsilon_{QC}=-1$ & $K_\mathbb{R}(0;d)\oplus K_\mathbb{R}(0;d)$ \\
\hline\hline
\end{tabular}
\end{center}
\label{table3}
\end{table*}

So far, we have shown that the classification of point-gapped (invertible) non-Hermitian Bloch Hamiltonians is equivalent to that of Bloch unitaries with the same symmetries. 
The latter has been well studied in the context of Floquet topological phases \cite{HF20} and quantum walks \cite{KT12}. Indeed, there is a well-developed technique called \emph{Hermitianization} --- we can construct a Hermitian Bloch Hamiltonian from $U(\boldsymbol{k})$ as \cite{HF17}
\begin{equation}
H_U(\boldsymbol{k})=\begin{bmatrix} \;0\; & \;\;U(\boldsymbol{k})\;\; \\ \;U(\boldsymbol{k})^\dag\; & \;\;0\;\; \end{bmatrix},
\label{HUk}
\end{equation}
which is given in Eq.~(\ref{HM}). Such a Hermitianized Hamiltonian is already flattened ($H_U(\boldsymbol{k})^2=\sigma_0\otimes I$) and has a chiral symmetry $\Sigma=\sigma^z\otimes I$ ($\{H_U(\boldsymbol{k}),\Sigma\}=0$). Moreover, all the BL symmetries can be promoted to certain AZ symmetries of $H_U(\boldsymbol{k})$, which may commute or anti-commute with $\Sigma$. To be concrete, we can take $U'_X=\sigma^x\otimes U_X$ with $\{U'_X,\Sigma\}=0$ for $X=Q,C$ and $U'_X=\sigma_0\otimes U_X$ with $[U'_X,\Sigma]=0$ for $X=P,K$, 
such that $H_U(\boldsymbol{k})=\epsilon_XU'_XH_U(\boldsymbol{k})(U'_X)^\dag$ and $(U'_X)^2=\sigma_0\otimes I$ for $X=P,Q$ and $H_U(\boldsymbol{k})=\epsilon_XU'_XH_U(-\boldsymbol{k})^*(U'_X)^\dag$ and $U'_X(U'_X)^*=\eta_X\sigma_0\otimes I$ for $X=C,K$. We can also check that the commutation and anti-commutation relations between $U'_P$, $U'_Q$, $U'_C\mathcal{K}$ and $U'_K\mathcal{K}$ are simply determined by $\epsilon_{XY}$ in Eq.~(\ref{epXY}). Therefore, we have eventually transformed the problem into the classification of Hermitian Bloch Hamiltonians with mutually (anti-)commutating two-fold (anti-)unitary (anti-)symmetries, which can be solved using the standard method introduced in Sec.~\ref{ptAZ}.

As some simple illustrations, let us consider the classification for the BL classes with at most one symmetry. If there is no symmetry, then due to the inherent chiral symmetry $\Sigma$ of $H_U(\boldsymbol{k})$, the relevant BL class is the same as class AIII and is given by $K_{\mathbb{C}}(1;d)$. In the following, we consider the case with a single symmetry $X$. For $X=P$, we have $[U'_P,\Sigma]=0$ and thus $U'_P\Sigma$ 
commutes with both $\Sigma$ and $H_U(\boldsymbol{k})$, so the classification is given by $K_{\mathbb{C}}(1;d)\oplus K_{\mathbb{C}}(1;d)$. For $X=Q$, we have $\epsilon_Q=-1$ (otherwise Wick rotating the Bloch Hamiltonian) and $\{U'_Q,\Sigma\}=\{U'_Q,H_U\}=0$, so the classification is $K_{\mathbb{C}}(2;d)=K_{\mathbb{C}}(0;d)$, which is the same as class A. For $X=C$, we have $\{U'_C\mathcal{K},\Sigma\}=0$, $U'_CH_U(\boldsymbol{k})(U'_C)^\dag=\epsilon_CH_U(-\boldsymbol{k})$ and $U'_C(U'_C)^*=\eta_C\sigma_0\otimes I$. Therefore, if $\epsilon_C\eta_C=-1$, the classification is $K_{\mathbb{R}}(3;d)$, which is the same as class DIII; 
otherwise, $\epsilon_C\eta_C=1$, the classification is $K_{\mathbb{R}}(7;d)$, which is the same as class CI. 
For $X=K$, we have $\epsilon_K=1$ (otherwise Wick rotating $H(\boldsymbol{k})$) and $\{U'_K\mathcal{K},\Sigma\}=0$, so the relevant class is the same as class BDI (CII) and is given by $K_{\mathbb{R}}(1;d)$ ($K_{\mathbb{R}}(5;d)$) if $\eta_K=1$ ($\eta_K=-1$). These results are summarized in the first $8$ rows in Table~\ref{table3}. See Appendix~\ref{Sec:pgBL} for more detailed derivations of the remaining results.

A classification of line-gapped systems follows from a straightforward application of the results that are well studied for Hermitian Hamiltonians (see Appendix~\ref{Sec:lgBL}).
This is because any line-gapped non-Hermitian bands can continuously be deformed to either Hermitian or anti-Hermitian ones (Theorem~\ref{linedeform}). While this reduction to a Hermitian band was claimed to be ``proved" in Ref.~\cite{KK19c}, the ``proof" is incomplete because it only applies to diagonalizable $H(\boldsymbol{k})$ 
and does not explain how to handle many important non-Hermitian bands, such as the ones having exceptional points\footnote{Regarding the case with exceptional points, it is stated in Ref.~\cite{KK19c} that ``they can be pair-annihilated without closing a line gap" without any justifications on how this can be achieved, especially under various symmetry constraints.}. 
Here we provide the complete, yet much simpler, constructive proof of the feasibility of continuous Hermitianization (see Appendix~\ref{app4} for details). The main idea is that we can first continuously flatten 
$H(\boldsymbol{k})$ into $H_1(\boldsymbol{k})=P_+(\boldsymbol{k})-P_-(\boldsymbol{k})$ ($H_1(\boldsymbol{k})=i[P_+(\boldsymbol{k})-P_-(\boldsymbol{k})]$) through simple interpolation $H_\lambda(\boldsymbol{k})=(1-\lambda) H(\boldsymbol{k}) + \lambda H_1(\boldsymbol{k})$ while keeping the real (imaginary) line gap and all the symmetries, where $P_\pm(\boldsymbol{k})$ is the \emph{non-Hermitian} projector onto the generalized eigenspace of $H(\boldsymbol{k})$ with eigenvalues whose real  (imaginary) parts are positive for $P_{+}(\boldsymbol{k})$ and negative for $P_{-}(\boldsymbol{k})$. Then we can prove that $H_1(\boldsymbol{k})$ can be deformed into its (anti-)Hermitian part, i.e., $\frac{1}{2}[H_1(\boldsymbol{k})+H_1(\boldsymbol{k})^\dag]$ ($\frac{1}{2}[H_1(\boldsymbol{k})-H_1(\boldsymbol{k})^\dag]$), again through a simple interpolation. Such a proof circumvents the apparent difficulty encountered in Ref.~\cite{KK19c}, which arises from the possible absence of a global section in $R(\boldsymbol{k})$ used to diagonalize $H(\boldsymbol{k})$ (i.e., $H(\boldsymbol{k})=R(\boldsymbol{k})\Lambda(\boldsymbol{k})R(\boldsymbol{k})^{-1}$ with $\Lambda(\boldsymbol{k})$ being diagonal).

Finally, we mention that it is possible to systematically determine the intrinsic non-Hermitian (i.e., point-gapped)  topological phases by removing the overlap with the line-gapped ones. 
Technically speaking, the map from a set of point- or line-gapped Hamiltonians (with symmetries) to the corresponding $K$-group on a fixed base manifold is a functor. Since a line-gapped Hamiltonian is always point-gapped, we have a natural inclusion of the former into the latter, and such an inclusion induces a homomorphism between the $K$-groups. The complement of the image of such a group homomorphism then gives genuine non-Hermitian (i.e., point-gapped)  topological phases.

\subsubsection{Non-Hermitian-Hermitian correspondence}
\label{Sec:NHHC}
There is a special BL subclass which can be considered as a natural generalization of AZ classes. Such a subclass is specified by recognizing $Q$ symmetry, $C$ symmetry with $\epsilon_C=-1$ and $K$ symmetry with $\epsilon_K=1$ as CS, PHS and TRS, respectively:
\begin{equation}
\begin{split}
{\rm CS:}\;\;\;\; &U_QH(\boldsymbol{k})^\dag U_Q^\dag=-H(\boldsymbol{k}), \\
{\rm PHS:}\;\;\;\; &U_CH(\boldsymbol{k})^{\rm T}U_C^\dag=-H(-\boldsymbol{k}),  \\
{\rm TRS:}\;\;\;\; &U_KH(\boldsymbol{k})^*U_K^\dag=H(-\boldsymbol{k}). \\
\end{split}
\label{NHAZ}
\end{equation}
Note that while the TRS is exactly the same as that for Hermitian systems, the PHS transposes the Hamiltonian instead of taking the complex conjugate. As a result, the CS combined from the TRS and PHS involves a Hermitian conjugate. Nevertheless, Eq.~(\ref{NHAZ}) reduces to Eqs.~(\ref{CS}), (\ref{TRS}) and (\ref{PHS}) for Hermitian Hamiltonians, which satisfy $H(\boldsymbol{k})^{\rm T}=H(\boldsymbol{k})^*$. We can also single out another subclass called AZ$^\dag$, for which the CS stays the same but  PHS and TRS respectively become PHS$^\dagger$ and TRS$^\dagger$ as follows:
\begin{equation}
\begin{split}
{\rm PHS^\dag:}\;\;\;\; &U_CH(\boldsymbol{k})^*U_C^\dag=-H(-\boldsymbol{k}),  \\
{\rm TRS^\dag:}\;\;\;\; &U_KH(\boldsymbol{k})^{\rm T}U_K^\dag=H(-\boldsymbol{k}). \\
\end{split}
\label{NHAZdag}
\end{equation} 
Again, these two symmetries reduce to the conventional PHS and TRS for a Hermitian $H(\boldsymbol{k})$.

\begin{table*}[tbp]
\caption{Periodic table for AZ$^\dag$ (AZ) class non-Hermitian systems with point gaps \cite{JYL19}, which is part of Table~\ref{table3}. Compared with the Hermitian counterpart in Table~\ref{table2}, the $K$-group is shifted from $K_{\mathbb{F}}(s;d)$ to $K_{\mathbb{F}}(s-1;d)$ ($K_{\mathbb{F}}(s+1;d)$).}
\begin{center}
\begin{tabular}{ccccccccccc}
\hline\hline
AZ$^\dag$ (AZ) class & BL symmetries & \;$K$-group\; & $d=0$ & \;1\; & \;2\; & \;3\; & \;4\; & \;5\; & \;6\; & \;7\; \\
\hline
A$^\dag$ (A) & None & $K_{\mathbb{C}}(1;d)$ & 0 & $\mathbb{Z}$ & 0 & $\mathbb{Z}$ & 0 & $\mathbb{Z}$ & 0 & $\mathbb{Z}$ \\
AIII$^\dag$ (AIII) & $Q$ & $K_{\mathbb{C}}(0;d)$  & $\mathbb{Z}$ & 0 & $\mathbb{Z}$ & 0 & $\mathbb{Z}$ & 0 & $\mathbb{Z}$ & 0 \\
\hline
AI$^\dag$ (C) & $C_{++}$ ($C_{--}$) & $K_{\mathbb{R}}(7;d)$ & 0 & 0 & 0 & $2\mathbb{Z}$ & 0 & $\mathbb{Z}_2$ & $\mathbb{Z}_2$ & $\mathbb{Z}$ \\
BDI$^\dag$ (CI) & $QC^+_{++}$ ($QC^-_{--}$) & $K_{\mathbb{R}}(0;d)$ & $\mathbb{Z}$  & 0 & 0 & 0 & $2\mathbb{Z}$ & 0 & $\mathbb{Z}_2$ & $\mathbb{Z}_2$ \\
D$^\dag$ (AI) & $K_+$ & $K_{\mathbb{R}}(1;d)$ & $\mathbb{Z}_2$ & $\mathbb{Z}$ & 0 & 0 &0 & $2\mathbb{Z}$ & 0 & $\mathbb{Z}_2$ \\
DIII$^\dag$ (BDI) & $QC^-_{+-}$ ($QC^+_{-+}$) & $K_{\mathbb{R}}(2;d)$ & $\mathbb{Z}_2$ & $\mathbb{Z}_2$ & $\mathbb{Z}$ & 0 & 0 & 0 & $2\mathbb{Z}$ & 0 \\
AII$^\dag$ (D) & $C_{+-}$ ($C_{-+}$) & $K_{\mathbb{R}}(3;d)$ & 0 & $\mathbb{Z}_2$ & $\mathbb{Z}_2$ & $\mathbb{Z}$ & 0 & 0 & 0 & $2\mathbb{Z}$ \\
CII$^\dag$ (DIII) & $QC^+_{+-}$ ($QC^-_{-+}$) & $K_{\mathbb{R}}(4;d)$ & $2\mathbb{Z}$ & 0 & $\mathbb{Z}_2$ & $\mathbb{Z}_2$ & $\mathbb{Z}$ & 0 & 0 & 0 \\
C$^\dag$ (AII) & $K_-$ & $K_{\mathbb{R}}(5;d)$ & 0 & $2\mathbb{Z}$ & 0 & $\mathbb{Z}_2$ & $\mathbb{Z}_2$ & $\mathbb{Z}$ & 0 & 0 \\
CI$^\dag$ (CII) & $QC^-_{++}$ ($QC^+_{--}$) & $K_{\mathbb{R}}(6;d)$ & 0 & 0 & $2\mathbb{Z}$ & $0$ & $\mathbb{Z}_2$ & $\mathbb{Z}_2$ & $\mathbb{Z}$ & 0 \\
\hline\hline
\end{tabular}
\end{center}
\label{table4}
\end{table*}

The classifications for pointed-gapped non-Hermitian Hamiltonians in the AZ and AZ$^\dag$ classes exhibit the periodicity  not only in the spatial dimension $d$ but also in the symmetry class $s$. In fact, compared with the Hermitian AZ classes (see Table~\ref{table2}), $s$ in the $K$-group is shifted to $s+1$ ($s-1$) for the non-Hermitian AZ (AZ$^\dag$) classes, as can be understood from an additional $\Sigma$ ($i\Sigma$) in the Clifford-algebra (see Appendix~\ref{Sec:pgBL}). Recalling Eq.~(\ref{KFsd}), we know that the classification of $d$D non-Hermitian AZ$^\dag$ classes coincides with that of  $(d+1)$D Hermitian AZ classes. Such a \emph{non-Hermitian-Hermitian correspondence} was unveiled in Ref.~\cite{JYL19}, which further suggests that the anomalous edge state of the latter is realized in the long-time bulk dynamics in the former. 

Let us explain the correspondence in further detail. It is instructive to first look at the simplest example --- non-Hermitian class A ($=$A$^\dag$) in 1D. In this case, the corresponding Hermitian anomalous edge state is the chiral edge mode in quantum Hall or Chern insulators, which are 2D systems in class A. This correspondence is quite intuitive since the prototypical model on the non-Hermitian side is the Hatano-Nelson model with asymmetric hopping amplitudes \cite{HN96}, which should feature certain kind of chirality. Indeed, it is known that the Hatano-Nelson model exhibits unidirectional transport which is robust against disorder \cite{LS15}. Quantitatively, we can describe a chiral edge mode by a Dirac (Weyl) Hamiltonian $H_{\rm edge}(k)=vk$. This Hamiltonian is anomalous in the sense that it cannot be realized in the bulk of a 1D Hermitian system \cite{HBN81b}. In a non-Hermitian system, however, such a Hamiltonian becomes realizable in the long-time limit, where the modes with the largest imaginary parts dominate. For example, we can construct $H(k)=v\sin k+i\gamma\cos k$ with $\gamma>0$, so that only the modes near $k=0$ survive after a long time. 
Such a Bloch Hamiltonian corresponds to the Hatano-Nelson model with $J_{\rm L}=\frac{i}{2}(\gamma+v)$ and $J_{\rm R}=\frac{i}{2}(\gamma-v)$, whose winding number $w={\rm sgn} v$ is nontrivial. We note that a recent work analyzes a 3D non-Hermitian model that corresponds to the 4D quantum Hall effect \cite{FKK20}. 

Let us finally consider a general AZ class $s$ in $(d+1)$D. It suffices to consider the cases with unit topological numbers, since then an arbitrary topological number can be generated by stacking. Without loss of generality, 
we can describe the anomalous edge states by 
\begin{equation}
H_{\rm edge}(\boldsymbol{k})=\sum^d_{j=1}k_j\Gamma_j,\;\;\;\;\{\Gamma_j,\Gamma_{j'}\}=2\delta_{jj'}I,
\label{Hedge}
\end{equation}
where $\Gamma_j$'s are all Hermitian and satisfy $U_K\Gamma_j^* U_K^\dag=-\Gamma_j$ or/and $U_C\Gamma_j^{\rm T} U_C^\dag=\Gamma_j$, if the system has PHS or/and  TRS. The corresponding non-Hermitian Bloch Hamiltonian can then be constructed as
\begin{equation}
H(\boldsymbol{k})=\sum^d_{j=1}\sin k_j\Gamma_j-i\gamma\left(m-\sum^d_{j=1}\cos k_j\right)I,
\label{NHBH}
\end{equation}
which satisfies $U_KH(\boldsymbol{k})^*U_K^\dag=-H(-\boldsymbol{k})$ and $U_CH(\boldsymbol{k})^{\rm T}U_C^\dag=H(-\boldsymbol{k})$ and thus belongs to the AZ$^\dag$ class $s^\dag$. Setting $m\in(d-2,d)$, we find that only the wave numbers near $\boldsymbol{k}=\bold{0}$ survive in the long-time limit and give rise to the desired anomalous edge (\ref{Hedge}). Moreover, such a non-Hermitian Bloch Hamiltonian (\ref{NHBH}) is indeed characterized by a unit topological number, since $H_U(\boldsymbol{k})=\sum^d_{j=1}\sin k_j\sigma^x\otimes\Gamma_j+\gamma\left(m-\sum^d_{j=1}\cos k_j\right)\sigma^y\otimes I$ (see Eq.~(\ref{HUk})) is trivial when $m>d$ and undergoes a sign inversion in the mass term at $\boldsymbol{k}=\bold{0}$ when $m$ decreases below $d$. 

\section{Miscellaneous subjects\label{sec6}}
In light of the accelerating growth of non-Hermitian physics, there have emerged a large variety of interdisciplinary subjects on top of what we have discussed in the previous sections. While it is impossible to cover all of them, here we would like to pick up a few representatives, including nonreciprocal transport, non-Hermitian quantum dynamics and thermodynamics, and further topics on non-Hermitian topology.

\subsection{Nonreciprocal transport\label{sec:6np}}
One of the most notable features in non-Hermitian waves is their (possibly) nonreciprocal responses as mentioned in Secs.~\ref{secepphys} and \ref{secplight}. While p-n junctions and Faraday isolators are  nonreciprocal devices of commercial success, there is still significant interest in achieving nonreciprocal transport by alternative mechanisms and in other devices. Here we review recent developments in this direction.

To understand nonreciprocal phenomena, it is instructive to start from the simplest setup, i.e.,  a one-dimensional scattering problem of a single quantum particle of mass $m$ governed by a Hamiltonian
\eqn{
H  =  H_{0}+V,
}
where $H_0=p^2/(2m)$ describes an asymptotic free evolution, and $V$ is a general perturbation that can include both of the position $x$ and momentum $p$ operators and can be non-Hermitian. Let $|\psi_{{\rm in}}\rangle$ ($|\psi_{{\rm out}}\rangle$) be an asymptotic incoming (outgoing) plane-wave solution evolving under $H_{0}$. The scattering process between these two asymptotes, $|\psi_{{\rm out}}\rangle=S|\psi_{{\rm in}}\rangle$,
can fully be characterized by the scattering operator $S$ defined
by
\eqn{
S  =  \Omega_{-}^{\dagger}\Omega_{+},\;\;\;\Omega_{+}  \equiv  \lim_{t\to\infty}e^{-iHt/\hbar}e^{iH_{0}t/\hbar},\;\;\;\Omega_{-} \equiv  \lim_{t\to\infty}e^{iH^{\dagger}t/\hbar}e^{-iH_{0}t/\hbar}.
}
Since we consider a one-dimensional case, there are only two channels and the asymptotic states are characterized by the momentum eigenvalue $p$, resulting in the input-output relation in the momentum basis
\eqn{
\left[\begin{array}{c}
\langle p|\psi_{{\rm out}}\rangle\\
\langle-p|\psi_{{\rm out}}\rangle
\end{array}\right] =  S(p)\left[\begin{array}{c}
\langle p|\psi_{{\rm in}}\rangle\\
\langle-p|\psi_{{\rm in}}\rangle
\end{array}\right]\equiv\left[\begin{array}{cc}
T_{l}(p) & R_{r}(p)\\
R_{l}(p) & T_{r}(p)
\end{array}\right]\left[\begin{array}{c}
\langle p|\psi_{{\rm in}}\rangle\\
\langle-p|\psi_{{\rm in}}\rangle
\end{array}\right],\;\;\;p>0.
}

In a Hermitian case, $H=H^\dagger$, the unitarity of the scattering operator, $S^\dagger S=1$, leads to the equal transmission/reflection probabilities 
\eqn{
\left|T_{l}\right|^{2}=\left|T_{r}\right|^{2}, \;\;\; \left|T_{a}\right|^{2}=1-\left|R_{a}\right|^{2} \;\;\; (a=l,r).
}
In addition, if the time-reversal symmetry is satisfied, i.e., ${\cal T}H{\cal T}^{-1}=H$, we obtain the equal transmission/reflection coefficients 
\eqn{\label{nrtrs}
S(p)=\sigma^{x}S^{\rm T}(p)\sigma^{x},\;\;S(p)=S^{\rm T}(p)\Longleftrightarrow  T_{l}=T_{r}, \; R_{l}=R_{r}.
} 
The conditions~\eqref{nrtrs} also follow from the parity symmetry ${\cal P}H{\cal P}^{-1}=H$. With channels more than two, the unitarity condition is relaxed and it alone no longer ensures the equal transmission/reflection probabilities. However, the presence of the time-reversal or parity symmetry still leads to the equal transmission/reflection coefficients (cf. Sec.~\ref{secplight}). Thus, to realize nonreciprocal linear scattering in Hermitian systems, one must violate both of the time-reversal and parity symmetry. This can typically be achieved by, e.g., the Faraday effect in magneto-optical systems.

\begin{table*}[b]
\caption{\label{tablenonrec} Summary of the possibilities of nonreciprocity in linear two-channel scattering problems for transmission coefficients $T_{l,r}$ and reflection coefficients $R_{l,r}$. In Hermitian systems, the unitarity of the scattering matrix ensures the reciprocal responses. The parity symmetry, ${\cal P}H{\cal P}^{-1}=H$, leads to the reciprocal responses even in non-Hermitian systems. With the broken parity symmetry, non-Hermitian systems can exhibit  nonreciprocal reflection $|R_l|\neq|R_r|$ owing to the nonunitarity of the scattering matrix. In addition, if transposition symmetry is broken, ${\cal T}H{\cal T}^{-1}\neq H^\dagger$, transmission can be nonreciprocal $|T_l|\neq|T_r|$. }
\footnotesize
\begin{tabular}{@{}  >{\centering\arraybackslash} m{6.1cm}@{}  >{\centering\arraybackslash} m{4.2cm} @{}  >{\centering\arraybackslash} m{4.2cm}}
\midrule \midrule
Systems and symmetries & Nonreciprocal linear transmission \newline $|T_l|\neq|T_r|$&Nonreciprocal linear reflection \newline $|R_l|\neq|R_r|$ \\ \midrule 
Hermitian & \xmark & \xmark
\\
Non-Hermitian, parity broken, transposition unbroken & \xmark &\cmark
\\
Non-Hermitian, parity broken, transposition broken & \cmark &\cmark
\\
\midrule \midrule
\end{tabular}
\end{table*}

A similar reciprocity relation can hold even in non-Hermitian systems. To see this, we note that the non-Hermiticity allows us to introduce a variant of  time-reversal symmetry (denoted as TRS$^\dagger$ in Sec.~\ref{sec5}); we shall refer to it as transposition symmetry in this section,  
\eqn{\label{6trsdag}
{\cal T}H{\cal T}^{-1}  = H^{\dagger}\Longleftrightarrow\langle x|V|x'\rangle=\langle x'|V|x\rangle,
}
which results in the reciprocal transmission coefficients  
\eqn{
S(p)=\sigma^{x}S^{\rm T}(p)\sigma^{x}\Longleftrightarrow  T_{l}=T_{r}.
}
Note that it reduces to time-reversal symmetry for the Hermitian case. 
Remarkably, the condition~\eqref{6trsdag} is satisfied for any complex potentials $V(x)\neq V^*(x)$, showing that merely introducing complex potentials via gain/loss does not lead to nonreciprocal transmissions in linear regimes \cite{AZ01,JGM04}. In contrast, owing to the nonunitarity of the scattering operator, reflection coefficients can still be nonreciprocal. In the extreme case, this leads to the unidirectional and reflectionless transport with $R_l=0$ and $R_r\neq 0$, which occurs at the exceptional point of the scattering matrix (cf. Sec.~\ref{secepphys}). In fact, the related arguments on the reflectionless transport can be generalized to a wider class of non-Hermitian systems \cite{Longhi:15,HSAR15} (see also Refs.~\cite{PL96,MK05} for earlier suggestions). 
We note, however, that the parity symmetry leads to the equal transmission/reflection coefficients in the same manner as in Hermitian systems. In Table~\ref{tablenonrec}, we summarize possibilities of realizing nonreciprocal responses in linear two-channel scattering setups.

\begin{figure}[!t]
\begin{center}
\includegraphics[width=14cm]{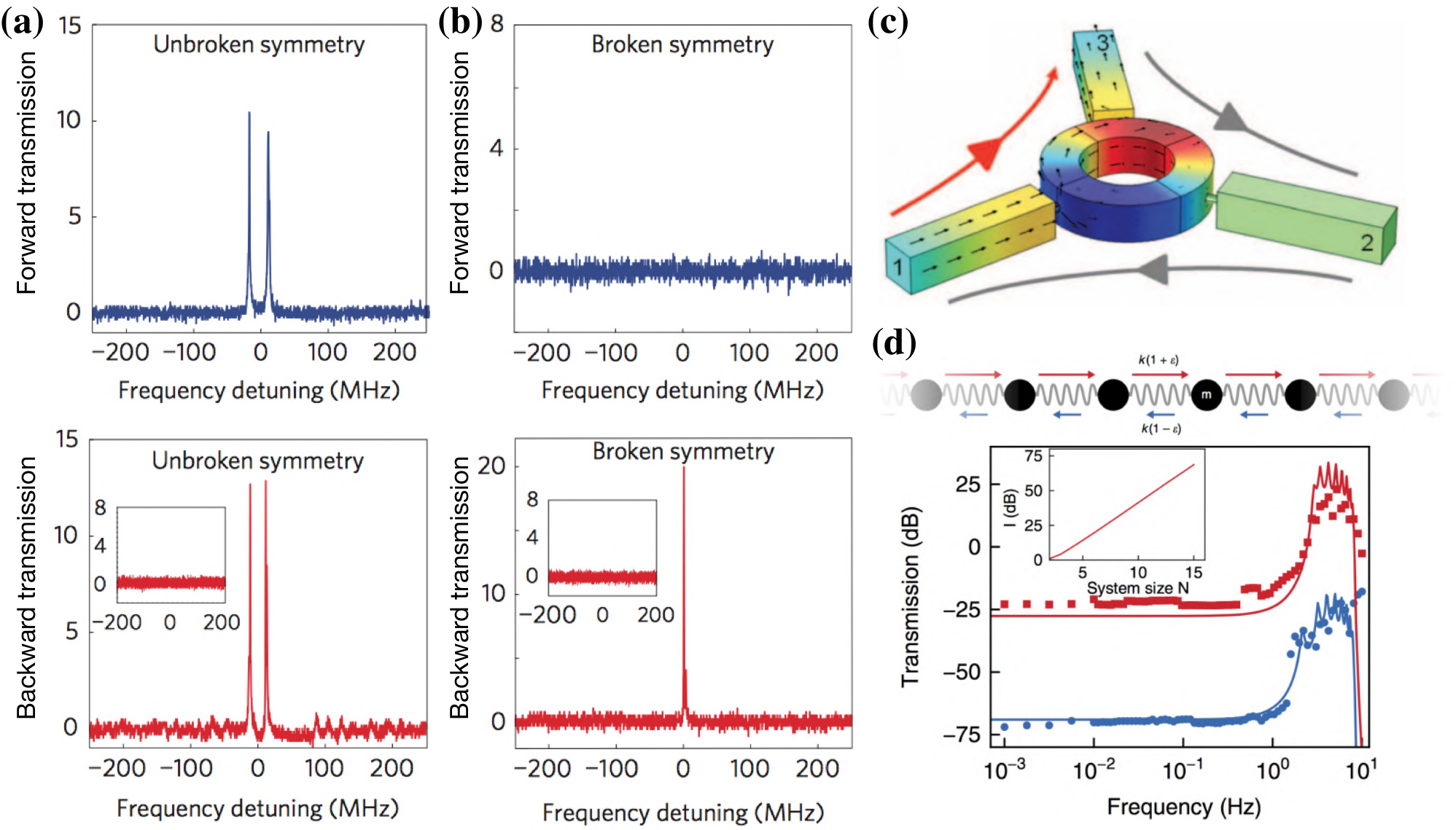}
\end{center}
\caption{Nonreciprocal transmissions in linear and nonlinear regimes. (a,b)  Nonlinear nonreciprocal transmission in PT-symmetric non-Hermitian resonators. Transmission coefficients in the forward and backward directions are plotted in top and bottom panels, respectively. Because of the scalar nature of gain/loss terms, the system respects the transposition symmetry~\eqref{6trsdag}, leading to the reciprocal transmission in a linear regime as shown in (a) (cf. the second line in Table~\ref{tablenonrec}). With the broken PT symmetry, the lasing behavior and the resulting nonlinearity can lead to the nonreciprocal transmission as shown in (b). Adapted from Ref.~\cite{BP14}. Copyright \copyright\,  2014 by Springer Nature.  (c) Linear nonreciprocal transmission of sound waves through a circulator. While the unitarity of the scattering matrix is satisfied, the broken time-reversal symmetry together with the three-channel setup can lead to nonreciprocal transmission in a linear regime. Adapted from Ref.~\cite{Fleury516}. Copyright \copyright\,  2014 by American Association for the Advancement of Science.  
(d) Linear nonreciprocal transmission in a non-Hermitian robotic metamaterial. Left-to-right and right-to-left transmission coefficients are plotted as red and blue curves, respectively (bottom panel). Nonreciprocal hopping (top panel) violates Hermiticity, parity and time-reversal symmetries simultaneously, leading to two-channel nonreciprocal transmission in a linear regime (cf. the last line in Table~\ref{tablenonrec} and the caption therein). Adapted from Ref.~\cite{BM19} licensed under a Creative Commons Attribution 4.0 International License.}
\label{fig:6nonrecip}
\end{figure}

To attain the nonreciprocal transmission $|T_l|\neq|T_r|$, one must go beyond the constraints discussed above. There are several strategies for this purpose. 
Firstly, one can employ nonlinear effects to break the reciprocity. For instance, lasing behavior in the broken PT symmetry enabled one to attain the strong nonlinear response at lower power in comparison with the PT unbroken regime, thus leading to the large nonreciprocal light transmission in the broken phase \cite{RH10,CL14,BP14} (see Fig.~\ref{fig:6nonrecip}(a,b)). Nonlinearity-induced nonreciprocity was also realized in a single active microcavity \cite{JX16}.  A related phenomenon was theoretically explored in a setup of Fano resonances in a photonic circuit \cite{Nazari:14}, optical meta-atom \cite{LS152}, low-threshold phonon diodes \cite{ZJ152}, and many-particle dynamics \cite{YA18}.  
At the fundamental level, we note that nonlinearity alone can lead to nonreciprocal transport without introducing any non-Hermiticity provided that the parity symmetry is broken \cite{LB09,LF143,CC16,TY182}. Yet, non-Hermiticity can still be of practical importance because it allows one to compensate losses inherent in strongly nonlinear regimes, thus potentially leading to stable nonreciprocal transmission.

Secondly, one can break transposition symmetry~\eqref{6trsdag} to realize nonreciprocal {\it linear} transmission through introducing time-dependent operations \cite{HY15,Shaltout:15} or implementing (an analogue of) a magnetic field  \cite{RP97}. The former has recently been realized in mechanical metamaterials \cite{WY18,TG19}, metasurfaces \cite{TG19}, and  acoustic resonators \cite{WQ15}. While the latter is a common technique  utilized in a Faraday isolator, it finds interesting applications to other classical devices, such as a three-channel acoustic resonator with  circulating medium \cite{Fleury516} (see Fig.~\ref{fig:6nonrecip}(c)). These two approaches remain useful independent of whether or not a system is Hermitian. Meanwhile, several mechanisms unique to non-Hermitian regimes have been introduced to attain linear nonreciprocal phenomena. One common way is to implement nonreciprocal hopping (or said differently, an imaginary gauge potential) \cite{HN96}, which can naturally lead to zero-bias spontaneous current as inferred from the analogy to convective flow in the hydrodynamic limit (cf. Eq.~\eqref{reacdiff}). From a fundamental perspective, this phenomenon is nothing more than trivial nonzero  flow associated with broken detailed balanced conditions \cite{US12b,JJH15} (cf. Sec.~\ref{secbiomas}). Nevertheless, it has recently found applications to mechanical metamaterials \cite{BM19,GA19} (see Fig.~\ref{fig:6nonrecip}(d) and Sec.~\ref{sec:mech}) and electrical circuits \cite{ZQB20} (see Sec.~\ref{sec:3ele}). 
Other mechanisms inherent in non-Hermitian systems include nonreciprocal dissipative magnon-photon couplings \cite{WYP19}, two-channel setups with the broken unitarity via circulations \cite{MS18,ZH20}, and wavefunction collapses via continuous measurement \cite{BP19}.

\subsection{Speed limits, shortcuts to adiabaticity, 
and quantum thermodynamics\label{sec:6sl}}
In light of the rapid development of non-Hermitian physics, 
many fundamental results in quantum mechanics, such as the uncertainty relation \cite{WH49}, the adiabatic theorem \cite{BM28}, and quantum fluctuation relations \cite{CM11}, have been reexamined in the absence of Hermiticity or unitarity. Here we provide several examples of particular recent interest.
\\ \\ {\it Speed limits}

\vspace{3pt}
\noindent
Uncertainty relations lie at the heart of quantum mechanics \cite{WH49}. In particular, the uncertainty relation between time and energy \cite{AY61} sets the fundamental limit on how fast a quantum state can evolve, which is called the \emph{quantum speed limit} \cite{SD17}. In its original formulation, the quantum speed limit $\tau_{\rm QSL}$ refers to a lower bound on the time it takes for a quantum state $|\psi_0\rangle$ to evolve into an orthogonal one under a Hermitian Hamiltonian $H$. Such a bound was found by Mandelstam and Tamm to be \cite{LM45}
\begin{equation}
\tau_{\rm QSL}=\frac{\pi}{2\Delta H},\;\;\;\;\Delta H\equiv\sqrt{\langle\psi_0| H^2|\psi_0\rangle-\langle\psi_0|H|\psi_0\rangle^2},
\label{MTbound}
\end{equation}
where we have set $\hbar=1$. While this result can be derived from the time-energy uncertainty relation by choosing the observable to be $|\psi_0\rangle\langle\psi_0|$, it is worthwhile to mention an alternative derivation from a geometric viewpoint \cite{AJ90}. That is, whenever we can define a distance $\mathcal{L}$ between two quantum states in the Hilbert space, we can obtain a quantum speed limit to be
\begin{equation}
\mathcal{L}(|\psi_0\rangle,|\psi_\tau\rangle)\le \int^\tau_0 \mathcal{L}(|\psi_{t+dt}\rangle,|\psi_t\rangle)\;\;\;\;\Leftrightarrow\;\;\;\;\tau\ge\tau_{\rm QSL}= \frac{\mathcal{L}(|\psi_0\rangle,|\psi_\tau\rangle)}{\frac{1}{\tau}\int^\tau_0 \mathcal{L}(|\psi_{t+dt}\rangle,|\psi_t\rangle)},
\label{Ltau}
\end{equation}
where $|\psi_t\rangle=e^{-iHt}|\psi_0\rangle$ is the time-evolved state. For example, if we apply the statistical distance $\mathcal{L}(|\phi\rangle,|\psi\rangle)=\arccos |\langle\phi|\psi\rangle|$ \cite{WWK81}, we obtain 
\begin{equation}
\mathcal{L}(|\psi_{t+dt}\rangle,|\psi_t\rangle)=\sqrt{\langle d\psi_t|[1-|\psi_t\rangle\langle\psi_t|]|d\psi_t\rangle}=\Delta H dt,
\label{dcalL}
\end{equation}
which implies Eq.~(\ref{MTbound}) for $\langle\psi_0|\psi_\tau\rangle=0$ since $\mathcal{L}(|\psi_\tau\rangle,|\psi_0\rangle)=\frac{\pi}{2}$ in this case. The above proof can readily be generalized to time-dependent Hamiltonians by replacing $\Delta H$ with $\overline{\Delta H}\equiv\frac{1}{\tau}\int^\tau_0dt\Delta H(t)$, where $\Delta H(t)\equiv \sqrt{\langle\psi_t| H(t)^2|\psi_t\rangle-\langle\psi_t| H(t)|\psi_t\rangle^2}$ is the instantaneous energy variance \cite{SD13}. Another well-known bound is found by Margolus and Levitin \cite{LBL98}:
\begin{equation}
\tau_{\rm QSL}=\frac{\pi}{2(\langle H\rangle_0-E_{\rm g})},\;\;\;\;\langle H\rangle_0\equiv\langle\psi_0|H|\psi_0\rangle,
\end{equation}  
where $E_{\rm g}$ is the ground-state energy of $H$. The time-dependent generalization of the Margolus-Levitin bound stays an open question since it is only achieved under specific conditions \cite{SS19}.   This is to be contrasted with the time-dependent generalization of the Mandelstam-Tamm bound, which holds true for arbitrary driven quantum systems.

So far we have discussed the quantum speed limits for Hermitian Hamiltonians. It is natural to ask whether, and if yes, how the quantum speed limits would be modified for non-Hermitian Hamiltonians. To address this question, Bender and Brody studied a minimal example of a PT-symmetric two-level system described by \cite{BCM09}
\begin{equation}
H=-(r\cos\theta \sigma_0+ir\sin\theta \sigma^z+s\sigma^x),\;\;\;\;r,s,\theta\in\mathbb{R}^+.
\label{HPTQSL}
\end{equation} 
Starting from $|\psi_0\rangle=|1\rangle$ (eigenstate of $\sigma^z=|1\rangle\langle 1|-|0\rangle\langle 0|$ with eigenvalue $1$), one can straightforwardly calculate the time-evolved state to be
\begin{equation}
|\tilde\psi_t\rangle\equiv e^{-iHt}|\psi_0\rangle=\frac{e^{itr\cos\theta}}{\cos\alpha}[\cos(\omega t+\alpha)|1\rangle+i\sin(\omega t)|0\rangle],
\end{equation}
where the tilde in $|\tilde\psi_t\rangle$ indicates that the state is not normalized, $\alpha=\arcsin(r\sin\theta/s)$ and $\omega=\sqrt{s^2-r^2\sin^2\theta}$. Superficially, it seems that it may take an arbitrarily small time $\tau=(\frac{\pi}{2}-\alpha)/\omega$ for the state to become orthogonal to $|1\rangle$ when $\alpha$ approaches $\frac{\pi}{2}$, which is actually an exceptional point. However, this turns out not to be the case since $\omega$ will necessarily approach $0$ when $\alpha$ appraoches $\frac{\pi}{2}$. By taking the limit $s\to r\sin\theta$, we obtain $\tau=s^{-1}$, which is finite. In fact, there is a natural generalization of the Mandelstam-Tamm bound (\ref{MTbound}) for non-Hermitian Hamiltonians:
\begin{equation}
\tau_{\rm QSL}=\frac{\pi}{2\overline{\Delta H}},\;\;\;\;\overline{\Delta H}=\frac{1}{\tau}\int^\tau_0dt\sqrt{\langle\psi_t| H^\dag H|\psi_t\rangle-|\langle\psi_t| H|\psi_t\rangle|^2},
\label{NHQSL}
\end{equation}
where $|\psi_t\rangle=e^{-iHt}|\psi_0\rangle/\|e^{-iHt}|\psi_0\rangle\|$ is the normalized time-evolved state. This result can be derived by substituting $|d\psi_t\rangle=-i(H+\frac{1}{2}\langle\psi_t|H^\dag -H|\psi_t\rangle)|\psi_t\rangle dt$ into Eq.~(\ref{dcalL}) followed by using Eq.~(\ref{Ltau}) \cite{RU12}. Interestingly, the PT-symmetric Hamiltonian (\ref{HPTQSL}) at the exceptional point $s=r\sin\theta$ gives an example of the saturation of the non-Hermitian quantum speed limit in Eq.~(\ref{NHQSL}): In this case, we have $|\psi_t\rangle\propto (1-st)|1\rangle+ist|0\rangle$ and $\overline{\Delta H}=s^2\int^{s^{-1}}_{0}dt[(1-st)^2+s^2t^2]^{-1}=\frac{\pi}{2}s$, so that $\tau_{\rm QSL}=s^{-1}=\tau$. We note that the non-Hermitian Mandelstam-Tamm bound can readily be generalized to time-dependent cases, while the non-Hermitian Margolus-Levitin bound has only been derived under specific constraints \cite{SS19}. 
By utilizing the statistical distance for mixed states \cite{BSL94}, we can also derive the quantum speed limits for dissipative systems \cite{DS13,PDP16,KF19}.

It is clear from Eq.~(\ref{Ltau}) that the quantum speed limit is rooted in the underlying geometric structure of the state space. 
Indeed, we have shown that the quantum speed limit exists also in non-Hermitian systems. From this point of view, one may argue that the quantum speed limit does \emph{not} necessarily rely on quantumness \cite{SB18,OM18}. In this sense, a more appropriate name should simply be the speed limit. In particular, speed limits appear in classical stochastic processes (cf. Sec.~\ref{secbio}) governed by generally non-Hermitian (even not pesudo-Hermitian, if the detailed balance condition breaks down) generators \cite{SN18}, where the state space consists of distribution functions, for which we can define various statistical distances \cite{TMC06}. Remarkably, some speed limits in this context have clear thermodynamic interpretations --- the speed of evolution is limited by the amount of entropy production \cite{SN18,AD18}. It is also worth mentioning some closely related recent developments on  
\emph{thermodynamic uncertainty relations}, which concern the trade-off relationship between entropy productions and current fluctuations \cite{BAC15,GTR16,HY19,HJM20,KL19}. 
\\ \\ {\it Shortcuts to adiabaticity}

\vspace{3pt}
\noindent
\emph{Shortcuts to adiabaticity} is a protocol with considerable recent interest for controlling quantum dynamics \cite{GOD19}. As the name indicates, this protocol refers to a finite-time realization of a quantum adiabatic evolution by a time-dependent Hamiltonian $H(t)$, which otherwise requires an infinite time according to the quantum adiabatic theorem \cite{BM28}. This is achieved by adding a well-designed \emph{counter-diabatic} time-dependent Hamiltonian $H_{\rm CD}(t)$ to $H(t)$ so that the dynamics governed by the entire Hamiltonian is transitionless, in the sense that starting from the $n$th eigenstate $|n_0\rangle$ of $H(t=0)$, the time-evolved state is exactly the instantaneous $n$th eigenstate $|n_t\rangle$ of $H(t)$ \cite{MVB09}. 

Let us first derive a general expression of $H_{\rm CD}(t)$ for Hermitian systems. We first apply the instantaneous spectral decomposition $H(t)=\sum_n\epsilon_n(t)|n_t\rangle\langle n_t|$. The gauge-invariant (under $|n_t\rangle\to e^{i\theta_n(t)}|n_t\rangle$) adiabatic evolution unitary is thus given by
\begin{equation}
U(t)=\sum_n e^{-i\phi_n(t)} |n_t\rangle\langle n_0|,\;\;\;\;\
\phi_n(t)=\int^t_0dt'[\epsilon_n(t')-i\langle n_{t'}|\partial_{t'} n_{t'}\rangle],
\end{equation}
where $\phi_n(t)$ consists of both dynamical and geometric phases \cite{BMV84}. Accordingly, $H_{\rm CD}(t)$ can be determined from $i\partial_t U(t)=[H(t)+H_{\rm CD}(t)] U(t)\Leftrightarrow H(t)+H_{\rm CD}(t)=i  [\partial_t U(t)] U(t)^\dag$. The result turns out to be
\begin{equation}
H_{\rm CD}(t)=i\sum_n(|\partial_t n_t\rangle \langle n_t| - |n_t\rangle\langle n_t|\partial_t n_t\rangle\langle n_t|),
\label{HCD}
\end{equation}
which is again gauge-invariant and Hermitian.\footnote{To confirm the Hermiticity, we only have to use the fact that $\langle n_t|\partial_t n_t\rangle$ is purely imaginary and $\partial_t I=\partial_t(\sum_n|n_t\rangle\langle n_t|)=\sum_n|\partial_t n_t\rangle\langle n_t|+\sum_n|n_t\rangle\langle \partial_t n_t|=0$ ($I$: identity).}

\begin{figure}[!t]
\begin{center}
\includegraphics[width=14cm]{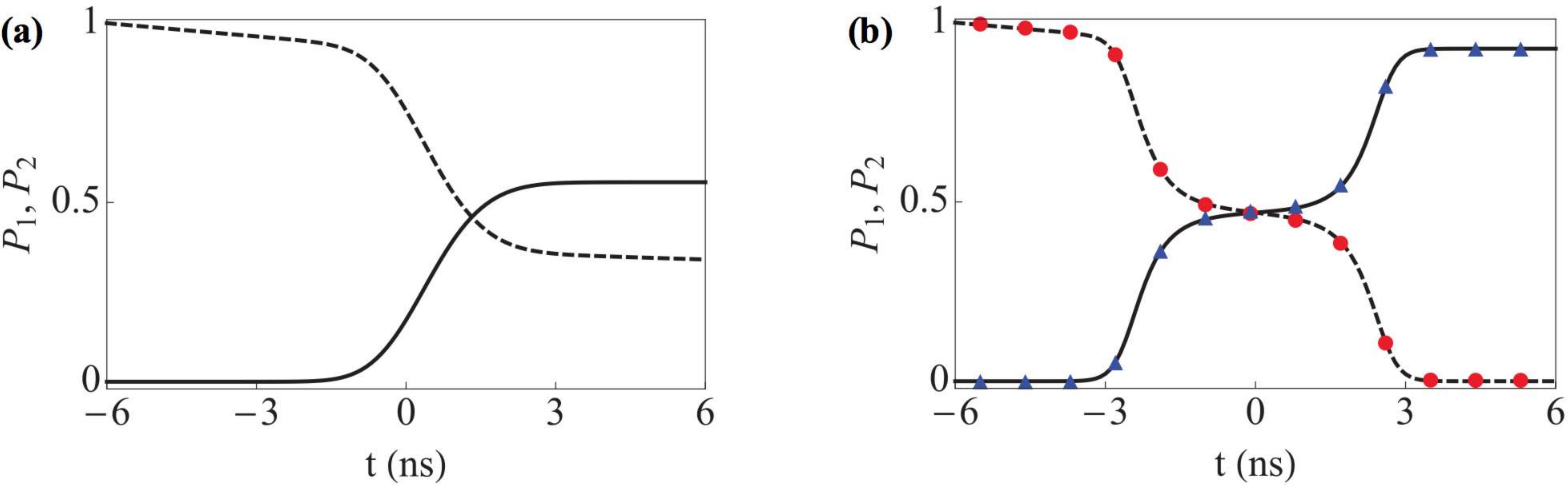}
\end{center}
\caption{Time evolution of the population of the ground state $P_1$ (solid curve) and that of the excited state $P_2$ (dashed curve) governed by the non-Hermitian Hamiltonian in Eq.~(\ref{TLH}) (a) without and (b) with the counter-diabatic Hmailtonian in Eq.~(\ref{TLHCD}). Note that $P_1+P_2<1$ (except for the initial time) due to the lossy nature of the dynamics. Here the driving protocol is chosen to be $\Omega(t)=\Omega_0 e^{-At^2}$, $\Delta(t)=-2Bt$ and $\Gamma(t)=\Gamma$, where $\Omega_0=2\pi\times 0.1\;{\rm GHz}$, $A=(2\pi)^2\times0.01\;{\rm GHz}^2$, $B=(2\pi)^2\times 0.00025\;{\rm GHz}^2$ and $\Gamma=2\pi\times 2\;{\rm MHz}$. Adapted from Ref.~\cite{IS11}. Copyright \copyright\,  2011 by the American Physical Society.}
\label{fig:STA}
\end{figure}

To generalize the above result to non-Hermitian systems, we only have to pay attention to the fact that there are both left and right eigenstates, which generally differ from each other. Having this in mind, we can straightforwardly modify Eq.~(\ref{HCD}) into 
\begin{equation}
H_{\rm CD}(t)=i\sum_n(|\partial_t n^{\rm R}_t\rangle \langle n^{\rm L}_t| - |n^{\rm R}_t\rangle\langle n^{\rm L}_t|\partial_t n^{\rm R}_t\rangle\langle n^{\rm L}_t|),
\label{NHHCD}
\end{equation}
which is gauge-invariant under $|n^{\rm R}_t\rangle\to c_n(t)|n^{\rm R}_t\rangle$ and $|n^{\rm L}_t\rangle\to {c^*_n(t)}^{-1}|n^{\rm L}_t\rangle$ ($c_n(t)\in\mathbb{C}\backslash\{0\}$). As a minimal example, we consider a two-level atom with time-dependent driving $\Omega(t)$, detuning $\Delta(t)$ and excited-state decaying $\Gamma(t)$ \cite{IS11,TBT13}:
\begin{equation}
H(t)=-\frac{1}{2}\Delta(t)\sigma^z+\frac{1}{2}\Omega(t)\sigma^x+\frac{i}{4}\Gamma(t)(\sigma^z-\sigma_0).
\label{TLH}
\end{equation}
The counter-diabatic Hamiltonian turns out to be
\begin{equation}
H_{\rm CD}(t)=\frac{1}{2}C(t)\sigma^y,\;\;\;\;C(t)=\frac{\Omega(t)[\dot\Delta(t)-i\dot\Gamma(t)/2]-\dot\Omega(t)[\Delta(t)-i\Gamma(t)/2]}{[\Delta(t)-i\Gamma(t)/2]^2+\Omega(t)^2},
\label{TLHCD}
\end{equation}
which is generally non-Hermitian as well. In Fig.~\ref{fig:STA}, the dynamics with and without the counter-diabatic driving are compared for a specific protocol. Note that $C(t)$ diverges at an exceptional point $\Gamma(t)=2\Omega(t)$ and $\Delta(t)=0$. Moreover, as already mentioned in Sec.~\ref{seceptopo}, when a closed parameter trajectory encircles an exceptional point, an eigenstate will flip into another rather than return to itself \cite{RU11,MVB11}, in stark contrast to the Hermitian case. 

Similarly to the speed limits, shortcuts to adiabaticity applies also to classical systems. For example, as discussed in Sec.~\ref{sec:mech}, the dynamics of a classical mechanical system can be rewritten into a non-Hermitian Schr\"odinger equation, allowing us to apply Eq.~(\ref{NHHCD}). This idea was ultilized in Ref.~\cite{IS11} to determine the shortcut-to-adiabaticity protocol for a classical harmonic oscillator with a time-dependent frequency. Another application of considerable interest is to classical stochastic systems (cf. Sec.~\ref{secbio}). In particular, the notion of shortcuts to isothermality has been proposed for fast preparing thermal equilibrium states in Langevin systems with time-dependent potentials \cite{LG17}. Some more recent studies focus on shortcuts to stochastic near-adiabatic pumping \cite{FK20,TK20}. Here the target states are not the instantaneous steady states but instead the first-order near-adiabatic solution, which is known to give rise to a geometric contribution to the pumped current \cite{RJ10,ST11},
\\ \\ {\it Quantum thermodynamics}

\vspace{3pt}
\noindent
\emph{Quantum thermodynamics} is an emergent field that aims at extending the conventional thermodynamics for macroscopic classical systems to microscopic quantum systems \cite{JG09,SV16,JG16}. A fundamental question in this field is how to identify basic thermodynamic quantities, such as work, heat and entropy production, in the framework of quantum mechanics \cite{KR13}. While there exists several different formalisms to tackle this question \cite{BFGSL13,JO13,PS14,LM15,MPM15,RN15,LDA16}, one important direction is based on the \emph{quantum fluctuation relations} (or quantum fluctuation theorems) \cite{EM09,CM11,Funoreview}. They are \emph{exact} identities concerning fluctuations of thermodynamic quantities that are valid even far from equilibrium and can be regarded as a refined second law of thermodynamics \cite{JC11,US12b}. In the simplest setup, we consider a closed driven system described by a time-dependent Hamiltonian $H(t)=\sum_n \epsilon_n(t)|n_t\rangle\langle n_t|$, so there is only work but no heat. Starting from a diagonal ensemble $\rho_0=\sum_n p_n|n_0\rangle\langle n_0|$, the quantum work distribution at time $t=\tau$ is given by \cite{HT00,JK00}
\begin{equation}
P(W)=\sum_{m,n}|\langle m_\tau |U(\tau)|n_0\rangle|^2p_n\delta(W-\epsilon_m(\tau)+\epsilon_n(0)),
\label{PW} 
\end{equation}
where $U(\tau)=\hat{{\rm T}}e^{-i\int^\tau_0 dtH(t)}$ ($\hat{{\rm T}}$: time ordering) is the time-evolution operator. One can check that, provided that the initial state $\rho_0=e^{-\beta H(0)}/Z(0)$ is at thermal equilibrium ($\beta$: inverse temperature; $Z(t)\equiv {\rm Tr}[e^{-\beta H(t)}]$: partition function), such a work distribution (\ref{PW}) validates the Crooks fluctuation theorem \cite{CGE99} and the Jarzynski equality \cite{JC97}: 
\begin{equation}
P(W)e^{-\beta (W-\Delta F)}=\bar P(-W)\;\;\;\Rightarrow\;\;\;
\langle e^{-\beta(W-\Delta F)}\rangle \equiv\int dW P(W)e^{-\beta (W-\Delta F)}=1.
\label{JECFT}
\end{equation}
Here $\bar P(W)$ is the quantum work distribution for the time-reversal process and $\Delta F=-\beta \ln [Z(\tau)/Z(0)]$ is the free-energy difference. Moreover, the quantum work distribution is physically accessible through the two-time measurement protocol \cite{TP07} and has indeed been measured in experiments \cite{SA15}.

\begin{figure}[!t]
\begin{center}
\includegraphics[width=14cm]{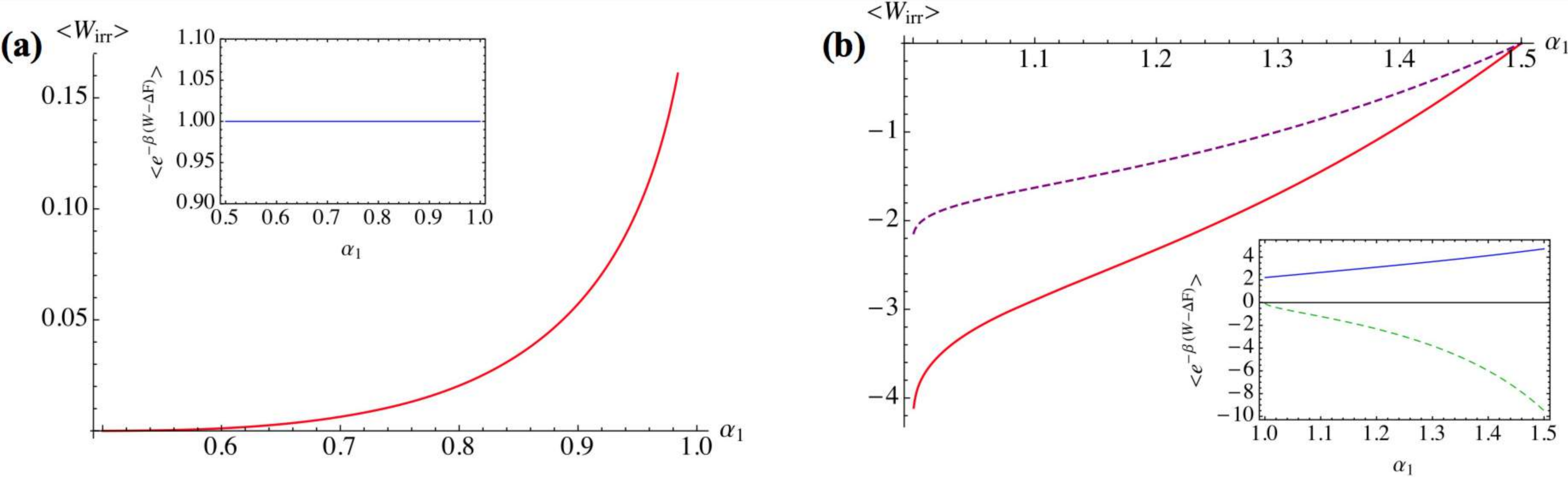}
\end{center}
\caption{Mean of the irreversible work $\langle W_{\rm irr}\rangle\equiv\langle W\rangle-\Delta F$ (main panel) and its exponentiated mean $\langle e^{-\beta (W-\Delta F)}\rangle$ (inset) for the driven PT-symmetric two-level system in Eq.~(\ref{Halpha}) with (a) $\alpha_0=1/2$ and $\alpha_1<1$ in the PT-unbroken regime and (b) $\alpha_0=3/2$ and $\alpha_1>1$ in the PT-broken regime. The other parameters are fixed as $\kappa=1$, $\beta=1$ and $\tau=1$.  Adapted from Ref.~\cite{DSSA15}. Copyright \copyright\,  2015 by the American Physical Society.}
\label{fig:QWD}
\end{figure}

If we na\"ively apply Eq.~(\ref{PW}) to driven non-Hermitian systems, then we will not obtain a normalized distribution function, let alone the validity of the fluctuation theorems. Nevertheless, if we confine ourselves to pseudo-Hermitian time-dependent Hamiltonians with real spectra, a formal generalization is achievable \cite{SD16}. In this case, we have (cf. Theorem~\ref{realspectrum})
\begin{equation}
H(t)^\dag =\eta(t) H(t) \eta(t)^{-1},
\end{equation}   
where $\eta(t)$ is a positive-definite Hermitian operator. Remarkably, the instantaneous right eigenstates $|n_t^{\rm R}\rangle$'s of $H(t)$ can be made orthonormal to each other with respect to the metric $\eta(t)$:
\begin{equation}\label{sec7etaexp}
\langle m_t^{\rm R}|\eta(t)|n_t^{\rm R}\rangle=\delta_{mn}.
\end{equation}
Provided that the dynamics preserves the modified unitarity~\eqref{sec7etaexp} defined with respect to $\eta(t)$, the Sch\"odinger equation is given by \cite{JG13}
\begin{equation}
i\partial_t|\psi_t\rangle=\left[H(t)-\frac{i}{2}\eta(t)^{-1}\partial_t\eta(t)\right]|\psi_t\rangle.
\label{nhtd}
\end{equation}
It is found that the fluctuation theorems (\ref{JECFT}) can be restored if the dynamical evolution follows Eq.~(\ref{nhtd}) and the transition probabilities are evaluated with respect to the metric $\eta(\tau)$, i.e., $|\langle m_\tau |U(\tau)|n_0\rangle|^2\to|\langle m^{\rm R}_\tau |\eta(\tau)U(\tau)|n^{\rm R}_0\rangle|^2$. This is particularly the case for PT-symmetric systems in the PT-unbroken regime \cite{DSSA15,WBB18}, as can be illustrated in a simple driven two-level system \cite{DSSA15}:
\begin{equation}
H(\alpha_t)=\kappa(i\alpha_t\sigma^z-\sigma^x),\;\;\;\;\alpha_t=\alpha_0+(\alpha_1-\alpha_0)t/\tau,
\label{Halpha}
\end{equation}
where $\kappa,\alpha_{0,1}\in\mathbb{R}$. The metric operator in the PT-unbroken regime ($\alpha_t<1$) can be chosen to be $\eta(\alpha_t)=(\sigma_0 - \alpha_t \sigma^y)/\sqrt{1-\alpha_t^2}$, leading to $\eta(\alpha_t)^{-1}\partial_t\eta(\alpha_t)=\dot{\alpha}_t\sigma^y/(\alpha_t^2-1)$. As shown in Fig.~\ref{fig:QWD}, the Jarzynski equality is indeed valid in the PT-unbroken regime. If we na\"ively apply the same gauge potential $\eta(\alpha_t)^{-1}\partial_t\eta(\alpha_t)=\dot{\alpha}_t\sigma^y/(\alpha_t^2-1)$ in the Schr\"odinger equation (\ref{nhtd}) to the PT-broken regime ($\alpha_t>1$), we will find that the Jarzynski equality generally breaks down (see Fig.~\ref{fig:QWD}(b)). Interestingly, when the parameter approaches the exceptional point, the mean work diverges (see Figs.~\ref{fig:QWD}(a) and (b)) and thus signals a thermodynamic singularity for the PT-symmetry breaking transition.

In a more realistic setup, we should consider a driven open quantum system, whose energy change should consist of both work and heat \cite{KR13}. 
The two-time measurement protocol has been generalized in this context --- we should jointly perform the energy measurement on both the system and the heat bath (environment) at the initial and final times, and the total energy difference and that of the heat bath are identified as work and heat, respectively \cite{CM09}. If the system and the heat bath are both at thermal equilibrium at the initial time, one can check that both the Jarzynski and the Crooks fluctuation theorem are valid. Moreover, in the weak-coupling limit, the dynamics of the reduced density operator of the system can be well described by a Lindblad equation under reasonable approximations (see discussions below Eq.~\eqref{lindbladchap2} and also Ref.~\cite{TA12}). It is found that the basic quantum thermodynamic quantities can consistently be defined along individual quantum trajectories \cite{HJM12,HFWJ13,GZ16,MG18}, where the dynamics is governed by a non-Hermitian Hamiltonian between quantum jumps (cf. Sec.~\ref{sec:qtraj}). Here by consistently we mean that the quantum-trajectory-based work/heat distributions can be reproduced from the two-point measurement protocol and also validate the fluctuation relations \cite{SM14,LF16}.  

\subsection{Miscellaneous topics on non-Hermitian topological systems\label{sec:6top}}
The family of topological phases has undergone rapid growth in the past decades. There have appeared several large branches such as (quasi-)disordered topological phases \cite{YEK16}, Floquet topological phases \cite{HF20}, and more recent higher-order topological phases \cite{REG19} and non-Hermitian topological phases (cf. Sec.~\ref{sec5}). These branches are not independent, but rather interconnected with each other, leading to more complicated novel topological phases including Flouqet Anderson insulators \cite{TP16}, higher-order Floquet insulators \cite{HH20} and higher-order topological quasicrystals \cite{CR20}. Here, we discuss how the branch of non-Hermitian topological systems intertwines with the others.
\\ \\ {\it Disordered and quasi-disordered non-Hermitian topological phases}

\vspace{3pt}
\noindent
Let us first discuss the interplay between non-Hermitian topology and disorder. In fact, the original Hatano-Nelson model, which is a prototypical non-Hermitian topological model, has both asymmetric hopping amplitudes and an on-site random potential \cite{HN96}:
\begin{equation}
H=\sum^L_{j=1} (J_{\rm R}|j+1\rangle\langle j| + J_{\rm L}|j\rangle\langle j+1|+V_j|j\rangle\langle j|).
\label{disHNH}
\end{equation}  
\begin{figure}[!t]
\begin{center}
\includegraphics[width=14cm]{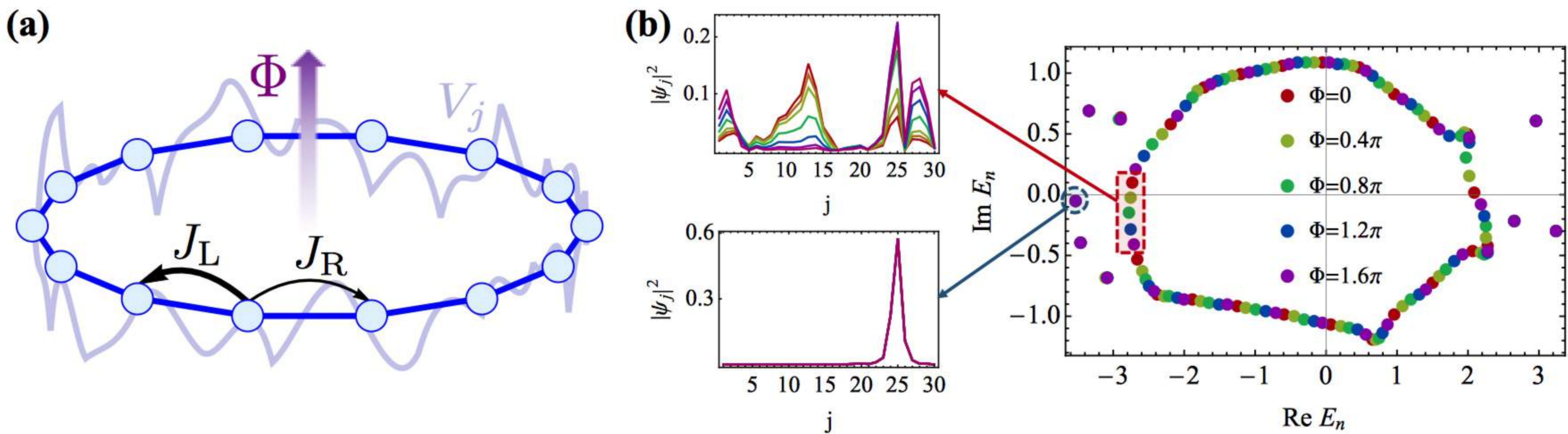}
\end{center}
\caption{(a) Schematic illustration of the twisted boundary-condition induced by a flux in the disordered Hatano-Nelson model (\ref{disHNH}). (b) Spectral evolution when the flux increases from $0$ to $2\pi$ and the typical profiles of an extended (left upper panel) and a localized (left lower panel) eigen-wave function. The eigenvalues of the extended modes encircles the origin, giving rise to a nontrivial winding number. Here the random potential is complex with disorder strength $W=2.5$ and the other parameters read $J_{\rm L}=2$, $J_{\rm R}=1$ and $L=30$. Adapted from Ref.~\cite{ZG18}.}
\label{fig:flux}
\end{figure}
\noindent Here we adopt the periodic boundary condition, since otherwise all the eigenstates will be localized due to the non-Hermitian skin effect (cf. Sec.~\ref{sec:bec}). To characterize the topological property of this disordered model (\ref{disHNH}), we can apply the \emph{twisted boundary condition} induced by a magnetic flux $\Phi$ (see Fig.~\ref{fig:flux}(a)), which is a well-established technique used to determine the Chern number of a disordered quantum Hall insulator \cite{NQ85}:
\begin{equation}
H(\Phi)=\sum^L_{j=1} (J_{\rm R}e^{-i\Phi/L}|j+1\rangle\langle j| + J_{\rm L}e^{i\Phi/L}|j\rangle\langle j+1|+V_j|j\rangle\langle j|).
\label{HPhi}
\end{equation}  
One can check that increasing $\Phi$ by $2\pi$ is equivalent to performing a large gauge transformation, so $\det H(\Phi)$ is a periodic function of $\Phi$ with periodicity $2\pi$. This fact enables us to define the following winding number \cite{ZG18}:
\begin{equation}
w=\int^{2\pi}_0 \frac{d\Phi}{2\pi i}\partial_\Phi\ln\det H(\Phi),
\label{fluxwn}
\end{equation}
which can be shown to reduce to Eq.~(\ref{wn}) in the absence of disorder. See Fig.~\ref{fig:flux}(b) for an example with a nontrivial winding number. 

Meanwhile, we can quantify the degree of localization of the eigen-wave functions by the \emph{inverse participation ratio} (IPR) \cite{HN98}:
\begin{equation}
{\rm IPR}=\frac{\sum^L_{j=1} \rho_j^2}{(\sum^L_{j=1}\rho_j)^2},\;\;\;\;\rho_j=|\langle j|\psi^{\rm R}\rangle\langle\psi^{\rm L}| j\rangle|,
\end{equation} 
where $|\psi^{\rm R}\rangle$ is a right eigenstate of $H$ and $|\psi^{\rm L}\rangle$ is the corresponding left eigenstate. 
When the eigen-wave function is localized (delocalized), the IPR is of the order of one (vanishes) in the thermodynamic limit. More precisely speaking, assuming $\rho_j$ follows an algebraic decay $\propto|j-j_0|^{-\alpha}$, we have
\begin{equation}
{\rm IPR}\sim\left\{\begin{array}{ll} 
\mathcal{O}(L^{-1}), & \;\;\;\;0\le\alpha<\frac{1}{2}; \\  
\mathcal{O}(L^{-1}\ln L), & \;\;\;\;\alpha=\frac{1}{2}; \\ 
\mathcal{O}(L^{2(\alpha-1)}), & \;\;\;\;\frac{1}{2}<\alpha<1; \\ 
\mathcal{O}(\ln^{-2} L), & \;\;\;\;\alpha=1; \\ 
\mathcal{O}(1), & \;\;\;\;\alpha>1. 
\end{array}\right.
\label{IPRalpha}
\end{equation}
In particular, a vanishing IPR only requires $\rho_j$ to decay no faster than $|j-j_0|^{-1}$, and thus not necessarily delocalize as a Bloch wave ($\alpha=0$).

\begin{figure}[!t]
\begin{center}
\includegraphics[width=14.5cm]{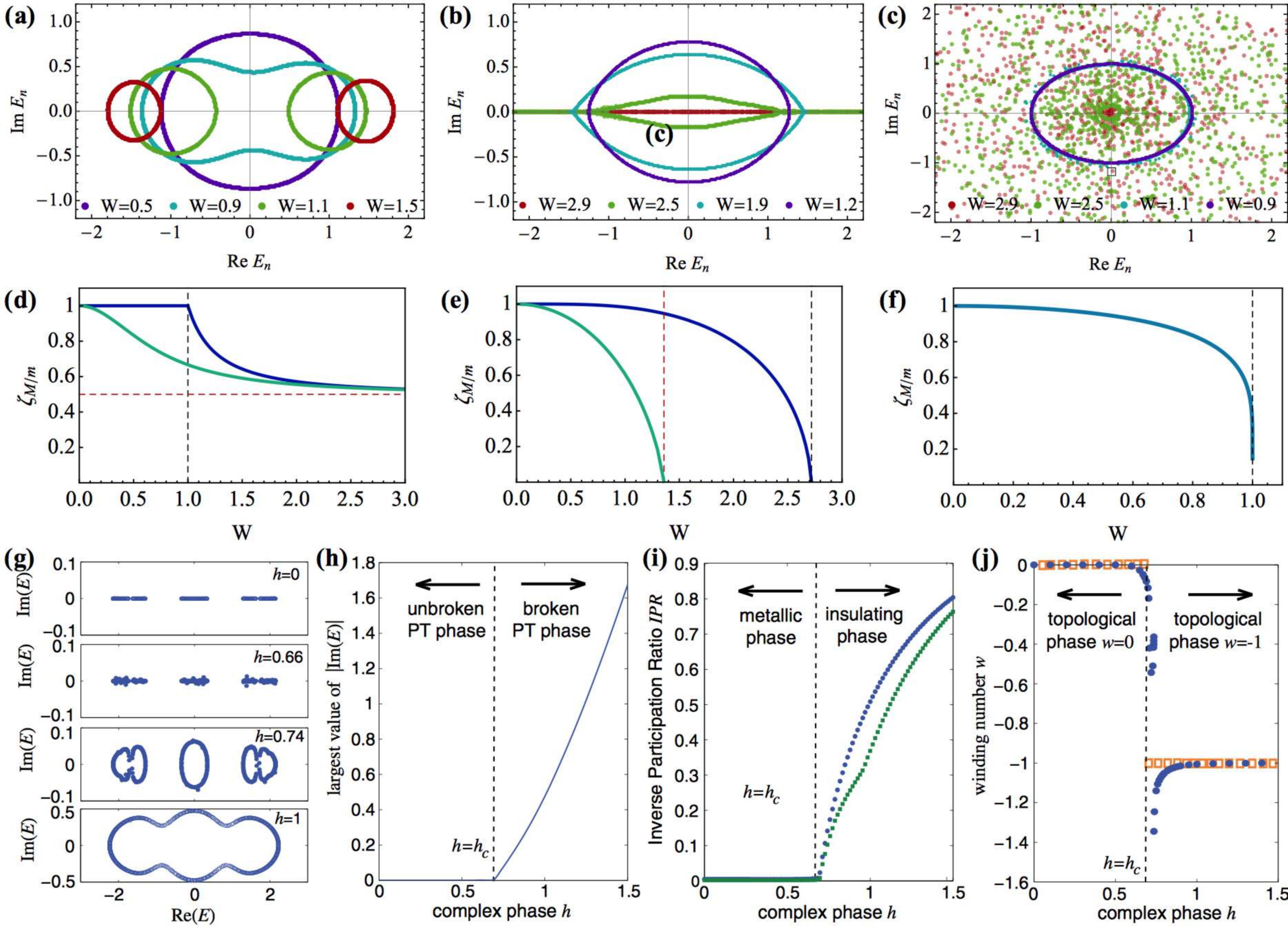}
\end{center}
\caption{Spectra of the Hatano-Nelson models (\ref{disHNH}) with unidirectional hopping ($J_{\rm L}=1$, $J_{\rm R}=0$) and (a) real binary, (b) real uniform and (c) complex on-site random potentials under different disorder strength. (a) is adapted from Ref.~\cite{ZG18}. Copyright \copyright\, 2018 by the American Physical Society. The corresponding localization characters $\zeta\equiv ({\rm IPR}\times L)^{-1}$'s are given in (d), (e) and (f), respectively. Here the thermodynamic limit is taken and the largest (smallest) values $\zeta_M$'s ($\zeta_m$'s) are plotted in blue (green) curves. (g) Spectra and (h) the maximal imaginary eigenenergy of the non-Hermitian quasicrystal given in Eq.~(\ref{Htheta}) for different $h$, which controls the degree of non-Hermiticity. (i) and (j) show the corresponding inverse participation ratio (maximum in blue and minimum in green) and topological winding number. (g)-(j) are adapted from Ref.~\cite{LS19}. Copyright \copyright\, 2019 by the American Physical Society.}
\label{fig:IPR}
\end{figure}

Several analytical results have been obtained in Ref.~\cite{ZG18} for the limit of unidirectional hopping, such as $J_{\rm L}=J$ and $J_{\rm R}=0$. The winding number is found to be $0$ for $|J|<J_{\rm c}$ and $1$ for $|J|>J_{\rm c}$, respectively, where $J_{\rm c}$ is determined by $\ln J_{\rm c}=\overline{\ln |V_j|}\equiv L^{-1}\sum^L_{j=1}\ln |V_j|$. In the thermodynamic limit, according to the central limit theorem, $\ln J_{\rm c}$ should be given by $\int dV P(V)\ln |V|$ with $P(V)$ being the distribution function of the on-site random potential. For example, as was studied in the original paper by Hatano and Nelson \cite{HN96}, if $V_j$ is real and follows a uniform distribution over $[-W,W]$ ($W\in\mathbb{R}^+$), we obtain $J_{\rm c}=W/e$; if $V_j$ follows a Lorentz distribution with linewidth $W$, then $J_{\rm c}=W$; the same critical value appears for a binary disorder $V_j=\pm W$. While Ref.~\cite{ZG18} only provides numerical results on the IPR of the eigenstates, here we point out that the IPR has the following analytical expression 
in the unidirectional-hopping limit:
\begin{equation}
{\rm IPR}=\frac{\sum^L_{j=1}|E-V_j|^{-2}}{(\sum^L_{j=1}|E-V_j|^{-1})^2},
\label{IPRana}
\end{equation} 
where $E$ is an eigenenergy that satisfies $\prod^L_{j=1}(E-V_j)=J^L$. For a typical realization with a sufficiently large $L$, we may replace the numerator and denominator by $L\int dV P(V)|E-V|^{-2}$ and $(L\int dV P(V)|E-V|^{-1})^2$, respectively. Applying Eq.~(\ref{IPRana}) to the case of real uniform disorder, we find that the maximal IPR vanishes as $\propto L^{-1}$ for $W<e|J|/2$. For $e|J|/2<W<e|J|$, the maximal IPR is of $\mathcal{O}(1)$ while the minimal IPR still vanishes as $\propto L^{-1}$ (see Fig.~\ref{fig:IPR}(e)). It is worthwhile to mention that the IPR scales as $\propto \ln^{-2} L$ at the mobility edge, corresponding to an algebraic decaying with unit power (cf. Eq.~(\ref{IPRalpha})). For $W>e|J|$, even the minimal IPR is of $\mathcal{O}(1)$ so all the eigenstates are localized. In this case, the localization transition coincides with the topological transition. This fact allows us to interpret the emergence of Anderson localization in the Hatano-Nelson model to be a manifestation of the unique non-Hermitian topology \cite{ZG18}. Note that Anderson localization does not occur in 1D Hermitian systems \cite{AE79}.
\begin{table*}[t]
\caption{\label{table:IPR} {Scalings of the minimal and maximal IPRs for different types of disorder. See also Figs.~\ref{fig:IPR}(d)-(f).}}
\footnotesize
\begin{tabular}{c|cccc}
\hline\hline
\;\;\;\;\;Type of disorder\;\;\;\;\; & \;\;\;\;\;\;$W<J$\;\;\;\;\;\; &  \multicolumn{1}{|c|}{\;\;\;\;\;\;\;$J<W<eJ/2$\;\;\;\;\;\;\;} & \;\;\;\;\;\;\;$eJ/2< W < eJ$\;\;\;\;\;\;\; &  \multicolumn{1}{|c}{ \;\;\;\;\;\;$W> eJ$\;\;\;\;\;\; }\\ 
\hline
Binary & $L^{-1}$,\;$\mathcal{O}(L^{-1})$ & \multicolumn{3}{|c}{
$\mathcal{O}(L^{-1})$} \\
\hline
Real &  \multicolumn{2}{c|}{
$\mathcal{O}(L^{-1})$} & $\mathcal{O}(L^{-1})$,\;$\mathcal{O}(1)$  &  \multicolumn{1}{|c}{ 
$\mathcal{O}(1)$} \\
\hline
Complex & $\mathcal{O}(L^{-1})$ & \multicolumn{2}{|c}{
$\mathcal{O}(L^{-1}\ln L)$,\;$\mathcal{O}(1)$} &  \multicolumn{1}{|c}{$\mathcal{O}(1)$} \\
\hline\hline
\end{tabular}
\end{table*}

On the other hand, applying Eq.~(\ref{IPRana}) to a binary disorder with equal numbers of $\pm W$, we find that even the maximal IPR is smaller than $2L^{-1}$ no matter how large $W$ is (see Fig.~\ref{fig:IPR}(d)). This fact implies that all the eigenstates are always delocalized, and thus the topological transition in the spectrum is \emph{not} necessarily accompanied by the localization transition. A more complicated situation is when $V_j$ becomes complex, such as $P(V=re^{i\phi})=\theta (W-r)/(2\pi rW)$ corresponding also to $J_{\rm c}=W/e$. Unlike the real-disorder case, the eigenenergies of the localized states, which appear when $W>|J|$, cover a 2D \emph{area} instead of 1D segments and all the delocalized states form the mobility edge (see Fig.~\ref{fig:IPR}(c)). The IPRs of these delocalized states scale as $L^{-1}\ln L$ instead of $L^{-1}$, corresponding to an algebraic decay with power $1/2$ (cf Eq.~(\ref{IPRalpha})). {All the scaling behaviors of the IPRs are summarized in Table~\ref{table:IPR}.}
It is already clear from these simple examples that the correspondence between topological and localization transitions is rather complicated and not universal, but substantially depends on a specific choice of disorder. It is naturally expected that a possible interplay between topology and disorder can further be enriched by symmetries \cite{KK20}.

In addition to the genuine disorder discussed above, there are considerable recent efforts on exploring the physics of \emph{quasi-disordered} non-Hermitian systems. The arguably simplest model is the Hatano-Nelson model (\ref{disHNH}) with a quasi-periodic potential $V_j=2\Delta\cos(2\pi\beta j)$ with $\beta$ being an irrational number. This model reduces to the incommensurate Aubry-Andr\'e-Harper model in the Hermitian limit $J_{\rm L}=J_{\rm R}^*$ \cite{SA80}, and it is known that all the eigenstates undergo a delocalization transition when $|\Delta|$ increases across $|J_{\rm L}|=|J_{\rm R}|$ \cite{GT91}. When $J_{\rm L}\neq J_{\rm R}^*$, it is found in Ref.~\cite{JH19} that the transition point becomes $|\Delta|=\max\{J_{\rm R},J_{\rm L}\}$ and is associated with the change of the winding number defined in Eq.~(\ref{fluxwn}). While the coincidence between topological and localization transition is reminiscent of the Hatano-Nelson model with uniform disorder, there is a qualitative difference in that the fraction of localization modes increases gradually from $0$ to $1$ in the Hatano-Nelson model but undergoes a sudden change for the quasi-periodic potential. Similar phenomenon was observed in a dual model with symmetric hopping but a complex quasi-periodic potential \cite{LS19}:  
\begin{equation}
H(\theta)=\sum^L_{j=1} (J|j+1\rangle\langle j| + J|j\rangle\langle j+1|+V\cos (2\pi\beta j + ih+ \theta/L )|j\rangle\langle j|),
\label{Htheta}
\end{equation}
where $J,V,h,\theta\in\mathbb{R}$ and $\theta$ plays a role similar to $\Phi$ in Eq.~(\ref{HPhi}). By dual we mean that there exists a unitary transformation that changes Eq.~(\ref{Htheta}) into Eq.~(\ref{HPhi}) with $\Phi=\theta$, $V_j=2J\cos (2\pi \beta j)$, $J_{\rm R}=Ve^h/2$ and $J_{\rm L}=Ve^{-h}/2$.  The critical point is therefore determined by $2|J|=|V|e^{|h|}$ and the winding number can also be defined as Eq.~(\ref{fluxwn}) with $\Phi$ replaced by $\theta$. On the other hand, in contrast to all of the previous models, this dual model has a trivial (nontrivial) winding number in the delocalized (localized) phase. 

In fact, there are several other non-Hermitian generalizations of the Aubry-Andr\'e-Harper model \cite{CY14,HAK16,ZQB20}. In particular, Zeng \emph{et al.} studied an off-diagonal non-Hermitian Aubry-Andr\'e model with asymmetric quasi-periodic hopping amplitudes \cite{ZQB20}:
\begin{equation}
H=J\sum^L_{j=1}[(1-\gamma+\lambda_j)|j+1\rangle\langle j|+(1+\gamma+\lambda_j)|j\rangle\langle j+1|],\;\;\;\;\lambda_j=i\lambda\cos(2\pi\beta j+\delta),
\label{ODAA}
\end{equation}
where $J,\gamma,\lambda$, and $\delta$ are real parameters. In certain regimes, such a model is found to exhibit not only 1D point-gap non-Hermitian topology, but also nontrivial 2D line-gap non-Hermitian topology since Eq.~(\ref{ODAA}) can be considered as a non-Hermitian Chern insulator if we regard $\delta$ as an additional wave number. This work demonstrates that the manifestation of higher-dimensional topology in lower-dimensional quasicrystals, a well-known phenomenon in Hermitian systems \cite{KYE12,LLJ12,GS13,PE15}, appears also in non-Hermitian systems, at least for line-gap topology. It would be interesting to consider the possibility of realizing genuine non-Hermitian topology (i.e., point-gap topology) in lower-dimensional non-Hermitian quasicrystals.

Finally, we mention that sometimes disorder may play a positive role in stabilizing the band topology. In particular, if an insulator is trivial in the clean limit and is driven into a topological phase due to an appropriate amount of disorder, then the disorder-induced topological phase is called a \emph{topological Anderson insulator} \cite{LJ09,GCW09,JH09,GHM10}. The essential physics underlying this phenomenon is that disorder can effectively renormalize the parameters, as can be understood from the Green's function \cite{GCW09,JH09,GHM10}, and thus induce a topological transition. This idea has recently been applied to non-Hermitian sublattice-symmetric topological insulators in 1D \cite{XWL19,DWZ20} 
and non-Hermitian Chern insulators in 2D \cite{LZT20}
with the help of a real-space formula for computing the topological invariants \cite{SF19b}. 
\\ \\ {\it Topological phenomena in non-unitary dynamics} 

\vspace{3pt}
\noindent
The main focus of recent studies on Hermitian topological systems has   
shifted to various nonequilibrium regimes, such as periodic driving \cite{OT09,KT10b,KT10,JL11} and quench dynamics \cite{CMD15,BJC16,WC17,LL18,GZ18,McM18,LZ18}. This tendency also spreads its impact on non-Hermitian topological systems. In particular, there is considerable recent interest in topological (discrete) \emph{non-unitary quantum walks}, which are described by Bloch time-evolution operators $U(\boldsymbol{k})$'s that are generally not unitary, i.e., $U(\boldsymbol{k})^{-1}\neq U(\boldsymbol{k})^\dag$. Na\"ively, one may think that the topology of $U(\boldsymbol{k})$ simply coincides with that of non-Hermitian Bloch Hamiltonians since $U(\boldsymbol{k})$ is also non-Hermitian in general. However, there is a crucial difference in the \emph{symmetry constraint}. If a (non-unitary) quantum walk respects a symmetry, it means the effective (non-Hermitian) Hamiltonian $H_{\rm eff}(\boldsymbol{k})$ determined from $U(\boldsymbol{k})=e^{-iH_{\rm eff}(\boldsymbol{k})}$ rather than $U(\boldsymbol{k})$ itself satisfies the symmetry. For example, a $P$-symmetry (see Eq.~(\ref{PQCK})) implies $U(\boldsymbol{k})=U_PU(\boldsymbol{k})^{-1}U_P^\dag$, while a $C$-symmetry implies $U(\boldsymbol{k})=U_C[U(\boldsymbol{k})^{\epsilon_C}]^{\rm T}U_C^\dag$. We emphasize that $H_{\rm eff}(\boldsymbol{k})$ is not unique and may not even be continuous in $\boldsymbol{k}$ (implying non-locality in real space \cite{GD12}). On the other hand, the classification of invertible $U(\boldsymbol{k})$'s can still be carried out by using the unitarization technique (cf. Sec.~\ref{Sec:NHPT}), so the problem should be equivalent to that for unitary quantum walks without any spectral constraint. The latter problem has partially been solved in Ref.~\cite{HS19} for AZ classes.

\begin{figure}[!t]
\begin{center}
\includegraphics[width=14.5cm]{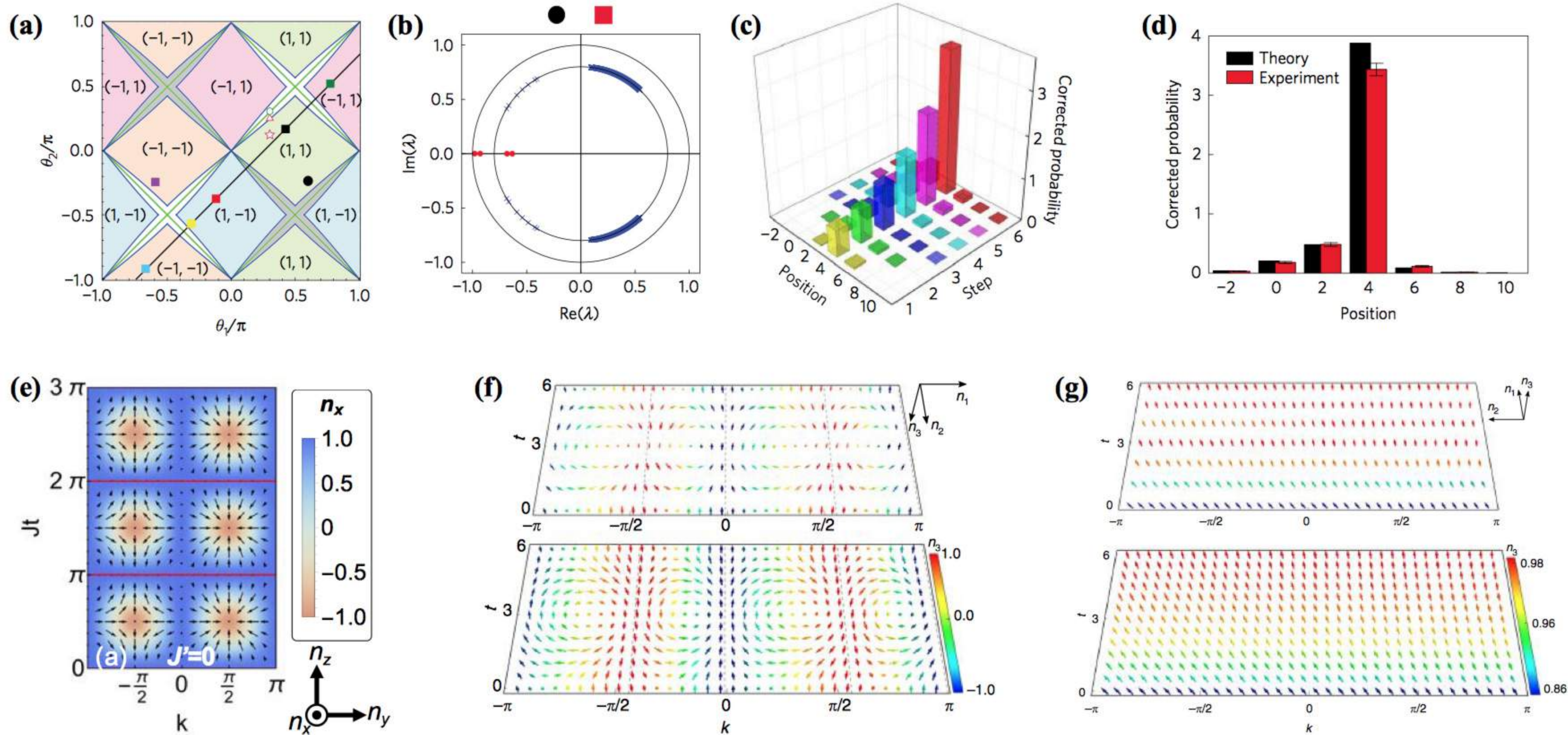}
\end{center}
\caption{(a) Phase diagram of the non-unitary PT-symmetric quantum walk governed by Eq.~(\ref{PTUk}). (b) Spectrum of an inhomogeneous quantum walk with different $(\theta_1,\theta_2)$ for the left and right half spaces. The parameters are indicated by the black circle and red square in (a). (c) Corrected (by introducing a background gain to make the lossy system PT-symmetric) time evolution and (d) the final profile of a quantum walk starting from the interface where $(\theta_1,\theta_2)$ suddenly changes. Adapted from Ref.~\cite{LX17}. Copyright \copyright\, 2017 by Springer Nature. (e) Skyrmion spin texture in the momentum-time space generated by a flat-band quench in the SSH model from a trival phase to a topological phase. Adapted from Ref.~\cite{GZ18}. Copyright \copyright\, 2018 by the American Physical Society. Observed (upper) and simulated (lower) spin texture in PT-symmetric quench dynamics implemented by photonic quantum walks in the (f) PT-unbroken and (g) PT-broken regimes. Momentum-time skyrmions appear (disappear) in the former (latter) case. Adapted from Ref.~\cite{KW19} licensed under a Creative Commons Attribution 4.0 International License.}
\label{fig:dyn}
\end{figure}

The minimal setup of a quantum walk is on a 1D lattice with two internal states $s=\uparrow,\downarrow$ per site. The one-step time-evolution operator is built from some (unitary) coin operators $C=\mathbb{I}\otimes S$ acting only on the internal states ($\mathbb{I}$: spatial identity) and some spin-dependent translation operators $T_s=\mathbb{I}\otimes| \bar s\rangle\langle \bar s|+\sum_j|j+1\rangle\langle j|\otimes|s\rangle\langle s|$ ($\bar s= \uparrow/\downarrow$ if $s=\downarrow/\uparrow$). These operators can be block-decomposed into $C=\sum_k |k\rangle\langle k|\otimes C(k)$ and $T_s=\sum_k |k\rangle\langle k|\otimes T_s(k)$ with
$C(k)=S$ and 
$T_s(k)=| \bar s\rangle\langle \bar s|+e^{-ik}|s\rangle\langle s|$.
The non-unitarity can arise from the operator $G$, which may involve spin-dependent gain or/and loss. Let us discuss two examples, both of which have been realized in experiments. The first example is a PT-symmetric quantum walk governed by \cite{DK16,LX17}
\begin{equation}
U(k)=R(\theta_1/2) G^{-1}T(k)R(\theta_2)T(k)G R(\theta_1/2),
\label{UPT}
\end{equation}
where $T(k)=T_\downarrow(k)^\dag T_\uparrow(k)=e^{-ik\sigma^z}$, 
$G=e^{-\gamma\sigma^z}\sigma^z$ ($\gamma\in\mathbb{R}$) and $R(\theta)=e^{-i\theta\sigma^y}$. Here the PT symmetry is represented as $\sigma^z\mathcal{K}$ and thus $U(k)$ satisfies
\begin{equation}
U(k)=\sigma^z[U(k)^{-1}]^*\sigma^z.
\label{PTUk}
\end{equation}
Such a symmetry enforces the spectrum of $U(k)$ to be located on the unit circle in the complex plane for a sufficiently small $\gamma$.\footnote{Precisely speaking, what is realized in the \emph{loss-only} photonic quantum walk experiment \cite{LX17} is a passive PT symmetry, in the sense that Eq.~(\ref{PTUk}) is valid only after a background loss in $U(k)$ is removed. The experimental data  shown in Figs.~\ref{fig:dyn}(c) and (d) are corrected in this manner.} On the other hand, when $U(k)$ is in the topological regime, its edge modes always break the PT symmetry and exhibit both lasing and decaying (see Figs.~\ref{fig:dyn}(b)-(d)). This property resembles very much the Hamiltonian counterpart in Eq.~(\ref{PTSSH}) \cite{WS17}. However, a crucial difference is that here the edge modes may appear near either/both $1$ or $-1$, corresponding to a \emph{pair} of winding numbers (see Fig.~\ref{fig:dyn}(a)) \cite{DK16}. These winding numbers can be detected through measurement of the average displacement of the walker \cite{ZX17}, a method developed originally for continuous non-unitary quantum walks \cite{RMS09,ZJM15}. 

The second example is a sublattice-symmetric non-unitary quantum walk governed by \cite{LX20} 
\begin{equation}
U(k)=R(\theta_1/2) T_\downarrow(k)^\dag R(\theta_2/2) G R(\theta_2/2)  T_\uparrow(k) R(\theta_1/2),
\end{equation}
where $R(\theta)$ and $G$ follow the same definitions as those in Eq.~(\ref{UPT}). The sublattice symmetry is represented by $\sigma_x$ and thus $U(k)$ satisfies
\begin{equation}
U(k) = \sigma^x U(k)^{-1}\sigma^x. 
\end{equation}
In the unitary limit, such a quantum walk is known to be characterized by a pair of winding numbers, provided that the spectrum of $U(k)$ is gapped at both $\pm1$ \cite{AJK13}. For a general non-unitary $U(k)$, it is experimentally demonstrated that the appropriate winding numbers that exactly 
characterize the edge modes should be defined in terms of the generalized Brillouin zone \cite{LX20}. This is one of the earliest experiments that demonstrates the non-Hermitian bulk-edge correspondence (cf. Sec.~\ref{sec:bec}).

A closely related setup to non-unitary quantum walks is non-Hermitian Floquet topological phases \cite{CY15b,ZL18b,BH20b,ZL20,ZX20,HW20}, which are described by time-periodic non-Hermitian Hamiltonians $H(t)=H(t+T)\neq H(t)^\dag$ with $T$ being the driving period. We can straightforwardly determine the one-period time-evolution (Floquet) operator from $U_{\rm F}=\hat T e^{-i\int_0^T dt H(t)}$ as well as the effective Hamiltonian $H_{\rm eff}$ from $U_{\rm F}=e^{-iH_{\rm eff}T}$. A remarkable feature of periodic driving is that it can effectively generate long-range hopping in $H_{\rm eff}$, which can, in principle, lead to arbitrarily large topological numbers \cite{ZL18b,ZL20}.  
It should be emphasized that, in contrast to (non-unitary) quantum walks, the topology of (non-Hermitian) Floquet topological phases is \emph{not} determined completely by $U_{\rm F}$ or $H_{\rm eff}$. Instead, an explicit time-dependence in $H(t)$ should be taken into account. A well-known example in Hermitian Floquet systems is the anomalous chiral Floquet phase \cite{RMS13,PHC16,HF17b}, whose Floquet unitary in the bulk can trivially be the identity (i.e., no dynamics at all), yet there exist robust chiral edge modes. 
The non-Hermitian extensions of these intrinsic Floquet topological phases seem to be largely unexplored. 
Interestingly, a unitary anomalous chiral Floquet topological phase is suggested to exhibit certain emergent non-Hermiticity in 
a segment cut from its boundary \cite{BH20}.

If the time-evolution operator is simply generated by a time-dependent Hamiltonian, such a setup is usually called a quantum quench \cite{ZL18c} or continuous quantum walk \cite{RMS09}. In Hermitian systems, the topology underlying quench dynamics has been studied in Refs.~\cite{CMD15,BJC16,WC17,LL18,GZ18,McM18,LZ18}. 
In particular, Refs.~\cite{LL18} and \cite{GZ18} unveiled the dynamical Chern number underlying 
 $(1+1)$D quench dynamics in Hermitian two-band systems, which is associated with skyrmion spin textures in the momentum-time space (see Fig.~\ref{fig:dyn}(e)). 
A similar phenomenon is predicted \cite{XQ19} and experimentally verified \cite{KW19} in some PT-symmetric non-unitary quench dynamics in the PT-unbroken regime (see Fig.~\ref{fig:dyn}(f)). It is also found that these momentum-time skyrmions will be destroyed by PT-symmetry breaking  (see Fig.~\ref{fig:dyn}(g)) \cite{XQ19,KW19}. 
\\ \\ {\it Higher-order non-Hermitian topological phases}

\vspace{3pt}
\noindent
\emph{Higher-order topological phases} have been attracting increasing interest from the community of topological materials in recent years \cite{BWA17,BWA17b,SZ17,LJ17,SF18,KE18b}. Generally speaking, an $n$th-order topological insulator (superconductor) in $d$ dimensions ($d\ge n$) exhibit $(d-n)$D topologically protected edge states \cite{LJ17,KE18b}. In particular, conventional topological insulators (superconductors) can be considered as first-order topological phases. The minimal example beyond conventional ones is the second-order topological insulator in 2D, whose edge states are 0D corner states. In the seminal paper by Benalcazar \emph{et al.} \cite{BWA17}, a minimal construction in a four-band Bloch Hamiltonian in class BDI is proposed:
\begin{equation}
H(k_x,k_y)=(t+\lambda\cos k_x)\Gamma_4-\lambda\sin k_x\Gamma_3-(t+\lambda\cos k_y)\Gamma_2-\lambda\sin k_y\Gamma_1, 
\label{SOH}
\end{equation} 
where $t,\lambda\in\mathbb{R}$ and $\Gamma_1=\sigma^y\otimes\sigma^x$, $\Gamma_2=\sigma^y\otimes\sigma^y$, $\Gamma_3=\sigma^y\otimes\sigma^z$ and $\Gamma_4=\sigma^x\otimes\sigma_0$ anti-commute with each other. The TRS and PHS are represented as $\mathcal{K}$ and $\Gamma_0\mathcal{K}$, respectively, whose combination gives the CS (or sublattice symmetry) $\Gamma_0=\sigma^z\otimes\sigma_0$. The corner states can intuitively be understood as conventional edges modes appearing at the interfaces of 1D topological insulators (in class BDI) with different topological (winding) numbers, which are realized at the 1D boundaries of the system. This recipe can straightforwardly be generalized to higher orders and dimensions as well as other symmetry classes \cite{LJ17,KE18b}. There have been a number of experimental realizations of higher-order topological phases in natural materials \cite{FSZW18,SNK19} and metamaterials \cite{MSG18,CWP18,SI18,HX18,XN18,AEH19,JHJ19}.

\begin{figure}[!t]
\begin{center}
\includegraphics[width=14.5cm]{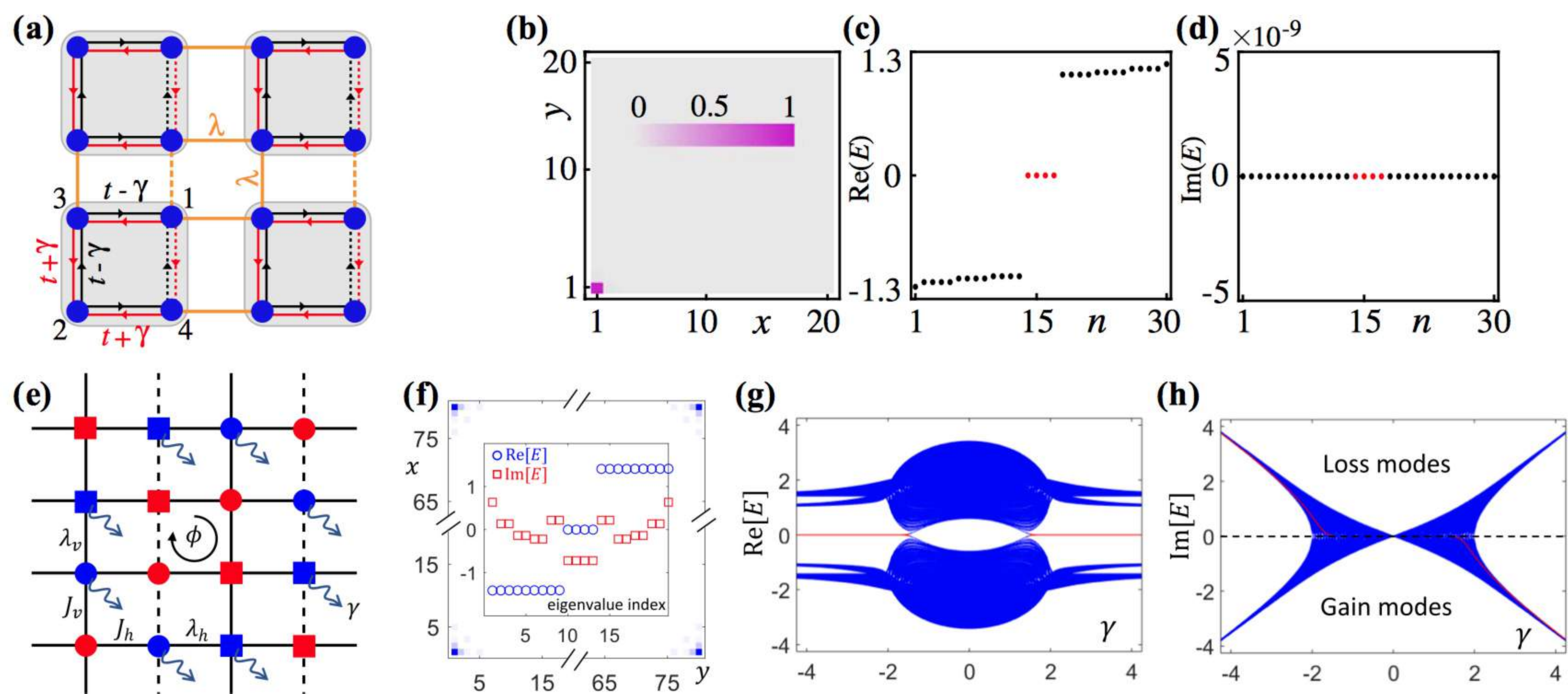}
\end{center}
\caption{(a) Schematic illustration of a 2D second-order non-Hermitian topological insulator (\ref{NHSOH}) with asymmetric hopping amplitudes in both directions. (b) Profiles of the corner states, which are concentrated in the left-lower corner. (c) Real and (d) imaginary parts of the eigenvalues near zero, where those corresponding to the corner states are marked in red. In (b), (c) and (d), the parameters in Eq.~(\ref{NHSOH}) are chosen to be $t=0.6$, $\lambda=1.5$ and $\gamma=0.4$. (e) Schematic illustration of a 2D second-order non-Hermitian topological insulator with balanced gain (red sites) and loss (blue sites). (f) Profiles of the corner states (main panel), which are distributed over all the corners, and their eigenvalues compared with nearby ones (inset). Here the strength of gain and loss is chosen to be $\gamma=2$. (g) Real and (h) imaginary parts of the full energy spectra for different $\gamma$. In (f), (g) and (h), the parameters in (a) are chosen to be $J_h=J_v=\sqrt{2}$ and $\lambda_\nu=\lambda_h=1$. Adapted from Refs.~\cite{LT19} and \cite{LXW19}. Copyright \copyright\,  2019 by the American Physical Society.}
\label{fig:HOTI}
\end{figure}

There have also been some recent efforts on merging non-Hermitian and higher-order topological phases \cite{LT19,LL19,EE19,EM19a,EM19b,ZZ19,LXW19}. For example, Liu \emph{et al.} studied an asymmetric-hopping generalization of the prototypical model in Eq.~(\ref{SOH}) (see Fig.~\ref{fig:HOTI}(a)) \cite{LT19}:
\begin{equation}
H(k_x,k_y)=(t+\lambda\cos k_x)\Gamma_4-(\lambda\sin k_x+i\gamma)\Gamma_3+(t+\lambda\cos k_y)\Gamma_2+(\lambda\sin k_y+i\gamma)\Gamma_1,
\label{NHSOH}
\end{equation}
where $\gamma\in\mathbb{R}$ and all the other notations follow those in Eq.~(\ref{SOH}). It is found that, due to the interplay between non-Hermitian skin effect and higher-order topology, the corner states will concentrate on one of the four corners (see Fig.~\ref{fig:HOTI}(b)). A similar phenomenon was observed in other lattice models \cite{LL19,EE19,EM19a,EM19b} and for the hinge modes in 3D non-Hermitian second-order topological insulators \cite{LT19}. 
Opposite situations with only gain and loss have also been considered for 2D second-order topological insulators \cite{ZZ19,LXW19}, where there is no skin effect and the corner states are distributed over all the four corners (see Fig.~\ref{fig:HOTI}(f)). It is found in Ref.~\cite{ZZ19} that, in the presence of PT symmetry, the energy degeneracy of the corner states can be lifted due to PT symmetry breaking. It is found in Ref.~\cite{LXW19} that starting from a trivial gapped Hermitian system with sublattice symmetry, we can drive it into a second-order topological phase solely by introducing sufficiently strong gain and loss (see Figs.~\ref{fig:HOTI}(g) and (h)). The model used to demonstrate this phenomenon has 16 bands (see Fig.~\ref{fig:HOTI}(e) for an illustration of a single unit cell) and can be regarded as the higher-order generalization of the four-band model given in Eq.~(\ref{4bm}).

\section{Summary and outlook\label{sec7}}
Non-Hermitian physics has become an interdisciplinary area of research at the intersection of many different subfields of science. The present review aims to fill a gap between different areas and stimulate further interest in studies of novel physical phenomena beyond Hermitian regimes, thus cultivating the fertile ground of the field of non-Hermitian physics. 

In Sec.~\ref{sec2} we reviewed key results/concepts of non-Hermitian linear algebra, including the Jordan normal form, biorthogonal eigenstates, exceptional points, pseudo-Hermiticity, and parity-time symmetry. We provided a number of illustrative examples to relate these central aspects with various unconventional physical phenomena, such as unidirectional invisibility, enhanced sensitivity, chirality/topological aspect of exceptional points, quantum criticality and many-body exceptional points, and spectral sensibility against boundary conditions. In Sec.~\ref{sec3}, we explained how one can utilize the formal equivalence between the one-body Schr{\"o}dinger equation and a linearized wave equation for the purpose of simulating non-Hermitian wave physics in various classical systems, ranging from photonics, mechanics, electrical circuits, acoustics, active matter, optomechanics, to neural networks.  We reviewed how the non-Hermiticity in these systems can manifest itself as rich wave phenomena such as coherent perfect absorption, single-mode lasing, topological energy transfer, nonreciprocal linear transmission,  robust biological transport, and efficient learning of deep neural networks.

In Sec.~\ref{sec4}, we explained in detail how non-Hermitian operators emerge as an effective description of open quantum systems on the basis of the Feshbach projection approach and the quantum trajectory approach. 
We clarified under what conditions these approaches can be justified and applied to understand essential physics underlying  various open quantum systems. We then reviewed a number of prominent physical phenomena occurring in such quantum regimes, including quantum resonances, superradiance, continuous quantum Zeno effect, quantum critical phenomena, the Dirac spectra in quantum chromodynamics, and nonunitary conformal invariance. We also discussed how they can be observed in various experimental systems realized in the fields of atomic, molecular and optical physics, mesoscopic physics, and nuclear physics. In Sec.~\ref{sec5}, we introduced possible extensions of band topology to complex spectra of non-Hermitian systems and presented their classifications by providing the constructive proof that is, to our knowledge, not provided in literature, but given in this review article for the first time. We also reviewed a number of instructive examples as well  as several efforts on restoring the bulk-edge correspondence by, e.g., modifying the definition of edge modes or introducing modified topological invariants appropriate for open boundary conditions. Finally, in Sec.~\ref{sec6} we reviewed a number of interdisciplinary subjects, such as nonreciprocal transport, speed limits, shortcuts to adiabacity, quantum thermodynamics, higher-order topology, and nonunitary quantum walk.

Despite much progress made toward elucidating the role of non-Hermiticity in both classical and quantum physics, there are still many open questions to be addressed. 
After a couple of decades of effort, it seems to us at this moment that non-Hermitian wave phenomena in linear classical systems have been well explored at least from a fundamental point of views. One possible exception in this direction is the subtle nature of topology in complex bands, whose complete understanding is still elusive. In particular, the understanding on the bulk-edge correspondence in non-Hermitian topological systems is far from complete. As mentioned in Sec.~\ref{sec:bec}, the generalized Brillouin zone approach has shown its great success in dealing with 1D systems under open boundary conditions, but a systematic generalization to higher dimensions has yet to be achieved. The pseudo-spectrum approach stays largely unexplored, due partially to its complicated mathematical nature. As mentioned in Sec.~\ref{sec:6top}, the family of Hermitian topological systems is still growing, and thus we expect that there is plenty of room for further studies of novel non-Hermitian topological phenomena.

It is natural and promising to investigate a role of nonlinear effects in non-Hermitian waves. In this regard, besides photonic or mechanical systems, a particularly intriguing and largely unexplored direction is to study nonlinear and non-Hermitian phenomena in intrinsically out-of-equilibrium setups, such as active matter and biophysical systems. In fact, nonlinear dynamics can lead to many interesting dynamical phenomena, such as chaos \cite{OE90}, bifurcations \cite{RMM76}, and synchronizations \cite{AJA05}, which typically associate with dissipations,  indicating that the underlying dynamical generator should be non-Hermitian. Further developments and applications of these fundamental aspects will not only advance the subject itself, but also may provide crucial insights into some urgent problems represented by explaining the success of machine learning based on (nonlinear) deep neural networks \cite{LZ20}. The interplay between nonlinearity and non-Hermitian topology would be yet another topic of interest.

On another front, in quantum regimes there are still many unanswered questions even at the level of few-body systems or noninteracting many-particle systems. For instance, it is important to elucidate how our understanding of a number of peculiar wave phenomena found in photonic systems with gain/loss should be modified in genuine quantum regimes. Also, while (solvable) non-Hermitian quadratic problems have recently been subjected to extensive research, their nonequilibrium aspects are largely unexplored and have just come under studies as briefly reviewed in Sec.~\ref{Sec:QP}. It is worthwhile to further explore the idea of utilizing nonunitary quadratic quantum field theories to analyze classical stochastic processes. The idea of using non-Hermitian point of view to extract new insights into Hermitian systems, which dates back to the seminal works by Yang and Lee in the 1950s \cite{LTD52}, also deserves further attentions. We expect that this idea could shed new light on long-lasting problems in non-Hermitian physics and bring about fruitful cross-fertilization between non-Hermitian physics and statistical physics. A remarkable recent achievement is the proof of the equivalence between the Lieb-Robinson velocity and the maximal relative group velocity for noninteracting fermion systems \cite{ZG2019}, as reviewed in Sec.~\ref{Sec:NHH}.

Studies of open quantum systems will become even more interesting when combined with many-body interactions, with great potential for further developments. While in Sec.~\ref{Sec:QMBP} we reviewed recent developments on several many-body aspects such as quantum criticality, phase transitions, and thermalization, these fundamental problems have just begun to attract attention. Much remains unknown also for entanglement growth in fluctuating quantum trajectories, especially for its possible relevance to studies on random unitary circuits. The impact of dissipation on the so-called Floquet time crystals \cite{KV16,EDV16} has not yet been fully explored \cite{GZHR18,GFM19,BZ19}. In light of rapid developments of quantum technologies, it will increasingly be an important direction to extend the frameworks of open quantum systems to non-Markovian regimes; several recent efforts were reviewed in Sec.~\ref{Sec:BMR}.
 Another interesting subject is topological phases in interacting non-Hermitian many-body systems as briefly discussed in Sec.~\ref{nonunitary_cft}. Therein, we mentioned that some 2D many-body states, including candidates for describing fractional quantum Hall effects, cannot be the ground states of any gapped Hermitian Hamiltonians with local interactions. Yet, they can still be the ground states of certain non-Hermitian local Hamiltonians with real spectra, whose edge theories are described by some nonunitary CFTs. It is then natural to ask the following questions: Which many-body states fall into a class unique to non-Hermitian Hamiltonians and how can one classify them? How will the classification be altered by dimensions and symmetries? A recent work seems to suggest a rather negative answer for 1D \cite{WX19}, while in higher dimensions we do not know the complete answer even in the absence of symmetries.

These future studies lie at the intersection of many research areas, including AMO physics, photonics, mechanics, quantum optics, condensed matter physics, biophysics, and statistical physics, and will not only enrich the scope of the field, but also may open a door for  potential applications. Thus, exploring non-Hermitian physics in quantum and classical systems should provide a research avenue for searching qualitatively new phenomena, many of which are yet to be unveiled.

\section*{Acknowledgments}
We are grateful to Yohei Fuji, Takeshi Fukuhara, Hosho Katsura, Kyogo Kawaguchi, Flore K. Kunst, Naoto Nagaosa, Masaya Nakagawa, Shuta Nakajima, Daiki Nishiguchi, Takahiro Sagawa, Kazuki Sone, Kazumasa A. Takeuchi, and Hidenori Tanaka for fruitful discussions. 
Y.A. acknowledges support from the Japan Society for the Promotion of Science through Grant No.~JP16J03613, and Harvard University for hospitality. Z.G. was supported by MEXT. We acknowledge support through KAKENHI Grant No.~JP18H01145 and a Grant-in-Aid for Scientific Research on Innovative Areas ``Topological Materials
Science" (KAKENHI Grant No.~JP15H05855) from the Japan Society for the Promotion of Science.  

\begin{appendices}
\section{Details on the Jordan normal form and the proofs}\label{app1}
\begin{figure}[b]
\begin{center}
\includegraphics[width=8cm]{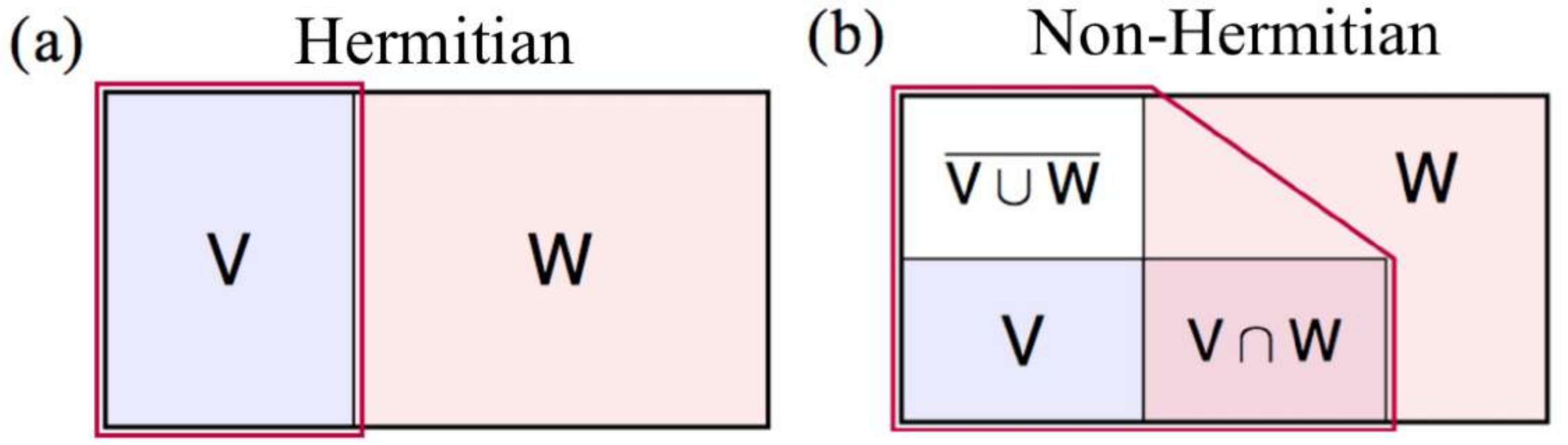}
\end{center}
\caption{Given an eigenvalue $\lambda$ of an $n\times n$ matrix $M$ with geometric multiplicity $m^{\rm g}$ and algebraic multiplicity $m^{\rm a}$, we can define the kernel $\textsf{V}\equiv{\rm Ker}(M-\lambda I)\subseteq\tilde{\textsf{V}}\equiv{\rm Ker}(M-\lambda I)^{m^{\rm a}-m^{\rm g}+1}$ and the image $\textsf{W}\equiv{\rm Im}(M-\lambda I)$ of mapping $M-\lambda I$, where $I$ is the $n\times n$ identity matrix (see Eqs.~(\ref{VM}), (\ref{WM}) and (\ref{tVM})). (a) If $M$ is Hermitian, then $\textsf{V}$ (blue rectangle) coincides with $\tilde{\textsf{V}}$ (magenta polygon) and is complementary to $\textsf{W}$ (red rectangle). (b) If $M$ is non-Hermitian, then in general $\textsf{V}\neq \tilde{\textsf{V}}\neq\overline{\textsf{W}}$. Both $\textsf{V}\cap \textsf{W}$ (purple rectangle) and $\overline{\textsf{V}\cup \textsf{W}}$ (white rectangle) can be non-empty and always share the same dimension. Moreover, $\tilde{\textsf{V}}$ contains $\overline{\textsf{V}\cup \textsf{W}}$ and may have a nonzero overlap with $\textsf{W}\backslash \textsf{V}$. }
\label{fig:app1}
\end{figure}

We first sketch the crucial steps in the derivation of the Jordan normal form in Theorem~\ref{Thm:JNF} and relegate further technical details later. We also provide the proofs for other statements in Sec.~\ref{sec2}.   
\\
\\
{\it Proof of Theorem~\ref{Thm:JNF}}

\vspace{3pt}
\noindent
First of all, we elucidate the origin of the possible difference between $m^{\rm g}_j$ and $m^{\rm a}_j$, which motivates the introduction of  \emph{generalized eigenvectors} and \emph{Jordan chains} \cite{CDM00}. This line of thinking eventually gives rise to the notion of the \emph{Jordan normal form} (sometimes called the \emph{Jordan canonical form} \cite{DH09}) introduced in Sec.~\ref{sec2}. 

To understand how $m^{\rm g}_j$ is equal to or smaller than $m^{\rm a}_j$, we first introduce
\begin{equation}
\textsf{W}_M(\lambda_j)\equiv{\rm Im}(M-\lambda_jI)\equiv\{\bold{u}:\bold{u}=(M-\lambda_jI)\bold{v},\bold{v}\in\mathbb{C}^n\},
\label{WM}
\end{equation}
which is again an invariant subspace of $M$. If $\textsf{V}_M(\lambda_j)\cap \textsf{W}_M(\lambda_j)=\{\bold{0}\}$, as is (but not necessarily) the case of a Hermitian matrix, we must have $m^{\rm g}_j=m^{\rm a}_j$ (see Fig.~\ref{fig:app1}(a)).  In general, $\textsf{V}_M(\lambda_j)\cap \textsf{W}_M(\lambda_j)$ can be a nontrivial invariant subspace. Moreover, its dimension can be shown to be the same as that of $\overline{\textsf{V}_M(\lambda_j)\cup \textsf{W}_M(\lambda_j)}$ (see Fig.~\ref{fig:app1}(b)). 
Such a coincidence of the dimensions is ultimately related to a close correspondence between the vectors in these two subspaces. In fact, any vector $\bold{v}$ in $\textsf{V}_M(\lambda_j)\cap \textsf{W}_M(\lambda_j)$ can be related to a vector $\bold{u}$ (not unique in general) in $\overline{\textsf{V}_M(\lambda_j)\cup \textsf{W}_M(\lambda_j)}$ through
\begin{equation}
\bold{v}=(M-\lambda_j)^p\bold{u},
\label{vpu}
\end{equation}
where $p$ is a positive integer equal to or less than $m^{\rm a}_j-m^{\rm g}_j$. Conversely, given any $\bold{u}\in \overline{\textsf{V}_M(\lambda_j)\cup \textsf{W}_M(\lambda_j)}$ we can specify a vector $\bold{v}\in \textsf{V}_M(\lambda_j)\cap \textsf{W}_M(\lambda_j)$ through Eq.~(\ref{vpu}). This relation (\ref{vpu}) can easily be examined for Example~\ref{minimalnond}, where 
\begin{equation}
\begin{split}
&\textsf{V}_M(\lambda)=\textsf{W}_M(\lambda)=\textsf{V}_M(\lambda)\cap\textsf{W}_M(\lambda)=\{[c,0]^{\rm T}:c\in\mathbb{C}\}, \\
&\overline{\textsf{V}_M(\lambda)\cup\textsf{W}_M(\lambda)}=\{[0,c]^{\rm T}:c\in\mathbb{C}\},
\end{split}
\end{equation}
and we have $[1,0]^{\rm T}=(M-\lambda)[0,1]^{\rm T}$. Moreover, Eq.~(\ref{vpu}) implies that $\bold{u}$ satisfies
\begin{equation}
(M-\lambda_j)^{p+1}\bold{u}=\bold{0},\;\;\;\;(M-\lambda_j)^p\bold{u}\neq\bold{0}.
\label{gev}
\end{equation}
Such a property defines the concept of \emph{generalized eigenvectors} with rank $p+1$, which reduce to the conventional eigenvectors for $p=0$. All the generalized eigenvecors span the \emph{generalized eigenspace}
\begin{equation}
\tilde{\textsf{V}}_M(\lambda_j)\equiv{\rm Ker}(M-\lambda_j I)^{m^{\rm a}_j-m^{\rm g}_j+1},
\label{tVM}
\end{equation}
whose dimension turns out to be $m^{\rm a}_j$. It is now clear that the difference between $m^{\rm a}_j$ and $m^{\rm g}_j$ arises from the existence of generalized eigenvectors.

Let us move on to analyze the action of $M$ on $\tilde{\textsf{V}}_M(\lambda_j)$. We consider a set of linearly independent vectors $\bold{v}_1,\bold{v}_2,...,\bold{v}_s$ that constitute a basis of $\textsf{V}_M(\lambda_j)\cap \textsf{W}_M(\lambda_j)$. Then we can find $\bold{u}_1,\bold{u}_2,...,\bold{u}_s$ in $\overline{\textsf{V}_M(\lambda_j)\cup \textsf{W}_M(\lambda_j)}$ according to Eq.~(\ref{vpu}). It turns out that $\bold{u}_1,\bold{u}_2,...,\bold{u}_s$ are also linearly independent, so they are actually a basis of $\overline{\textsf{V}_M(\lambda_j)\cup \textsf{W}_M(\lambda_j)}$. Denoting the rank of $\bold{u}_r$ as $p_r+1$ ($r=1,2,...,s$), the following series of eigenvectors
\begin{equation}
\begin{split}
&\bold{u}^{(0)}_r\equiv\bold{u}_r,\;\;
\bold{u}^{(1)}_r\equiv(M-\lambda_j I)\bold{u}_r,\;\;
\bold{u}^{(2)}_r\equiv(M-\lambda_j I)^2\bold{u}_r,\;\;... \\
&\bold{u}^{(p_r)}_r\equiv(M-\lambda_j I)^{p_r}\bold{u}_r=\bold{v}_r
\end{split}
\end{equation}
form a length-$(p_r+1)$ \emph{Jordan chain} with head $\bold{u}_r$ and tail $\bold{v}_r$, on which $M$ acts like
\begin{equation}
M[\bold{u}^{(p_r)}_r,\bold{u}^{(p_r-1)}_r,...,\bold{u}^{(0)}_r] =[\bold{u}^{(p_r)}_r,\bold{u}^{(p_r-1)}_r,...,\bold{u}^{(0)}_r] J_{p_r+1}(\lambda_j), 
\end{equation}
where $J_{p_r+1}(\lambda_j)$ is the size-$(p_r+1)$ \emph{Jordan block} with eigenvalue $\lambda_j$. 
If we regard the vectors in $\textsf{V}_M(\lambda_j)\backslash\textsf{W}_M(\lambda_j)$ as length-$1$ Jordan chains with identical heads and tails, it can be shown that all the Jordan chains with eigenvalue $\lambda_j$ are linearly independent and form a complete basis of $\tilde{\textsf{V}}_M(\lambda_j)$. Combining this result with the fact that the entire linear space is a direct sum of $\tilde{\textsf{V}}_M(\lambda_j)$'s, we know that there exists a basis in which $M$ is a direct sum of Jordan blocks, which is called the \emph{Jordan normal form} of $M$ (see Theorem~\ref{Thm:JNF}).

We now provide further technical details of the proof of Theorem~\ref{Thm:JNF}. We start from explaining why $\textsf{V}_M(\lambda_j)\cap \textsf{W}_M(\lambda_j)=\{\bold{0}\}$ implies $m^{\rm g}_j=m^{\rm a}_j$. This is because, otherwise suppose that $m^{\rm g}_j<m^{\rm a}_j$, the characteristic polynomial of the restriction of $M$ on $\textsf{W}_M(\lambda_j)$ will contain a factor $\lambda-\lambda_j$, implying the existence of some $\bold{u}\in \textsf{W}_M(\lambda_j)$ such that $M\bold{u}=\lambda_j\bold{u}$. This contradicts the assumption $\textsf{V}_M(\lambda_j)\cap \textsf{W}_M(\lambda_j)=\{\bold{0}\}$ since by definition $\bold{u}\in\textsf{V}_M(\lambda_j)$.

We turn to explain why 
\begin{equation}
{\rm dim}\textsf{V}_M(\lambda_j)\cap\textsf{W}_M(\lambda_j)={\rm dim}\overline{\textsf{V}_M(\lambda_j)\cap\textsf{W}_M(\lambda_j)}. 
\label{dimeq}
\end{equation}
According to the definitions of $\textsf{V}_M(\lambda_j)$ and $\textsf{W}_M(\lambda_j)$, we know from the \emph{rank-nullity theorem} that
\begin{equation}
{\rm dim}\textsf{V}_M(\lambda_j)+{\rm dim}\textsf{W}_M(\lambda_j)=n.
\label{RNthm}
\end{equation}
On the other hand, the dimension of the union of two linear spaces always satisfies
\begin{equation}
{\rm dim}\textsf{V}_M(\lambda_j)\cup \textsf{W}_M(\lambda_j)={\rm dim}\textsf{V}_M(\lambda_j)+{\rm dim}\textsf{W}_M(\lambda_j)-{\rm dim}\textsf{V}_M(\lambda_j)\cap \textsf{W}_M(\lambda_j).
\label{dimunion}
\end{equation} 
Combining Eqs.~(\ref{RNthm}) and (\ref{dimunion}), we obtain Eq.~(\ref{dimeq}).

Let us proceed by presenting several properties of the generalized eigenvectors. We first note that a generalized eigenvector $\bold{u}$ with an arbitrary rank $p+1$ must be defined with respect to eigenvalues, since $\bold{v}=(M - \lambda I)^p\bold{u}$ is a usual eigenvector. Moreover, a generalized eigenvector $\bold{u}$ with a given eigenvalue $\lambda$ cannot be a generalized eigenvector (including usual one) with a different eigenvalue $\lambda'$. Otherwise, suppose that $(M - \lambda' I)^{q+1}\bold{u}=\bold{0}$, we have the following contradiction:
\begin{equation}
\bold{0}=(M - \lambda I)^p(M - \lambda' I)^{q+1}\bold{u}=(M - \lambda' I)^{q+1}\bold{v}=(\lambda-\lambda')^{q+1}\bold{v}\neq\bold{0}.
\end{equation}
This result implies that the invariant subspaces consisting of all the generalized eigenvectors with different eigenvalues, which can be denoted as $\tilde{\textsf{V}}_M(\lambda)={\rm Ker}(M-\lambda I)^n$, have zero overlap with each other. 
To see why it is sufficient to write ${\rm Ker}(M-\lambda I)^n$, we only have to show the absence of a generalized eigenvector with rank large than $n$. Suppose that such a generalized eigenvector $\bold{u}$ with rank $n'>n$ exists, then $\{(M-\lambda I)^k\bold{u}\}^n_{k=0}$ are $n+1$ linearly independent vectors, contradicting the fact that the entire linear space is $n$-dimensional. The linear independence can be confirmed by solving 
\begin{equation}
\sum^n_{k=0} c_k(M-\lambda I)^k\bold{u}=\bold{0}
\end{equation}
By multiplying $(M-\lambda I)^{n'-1}$, we obtain $c_0(M-\lambda I)^{n'-1}\bold{u}=\bold{0}$ so that $c_0=0$. Then by multiplying $(M-\lambda I)^{n'-2}$ we will obtain $c_1=0$. Repeating this process, we will end up with $c_0=c_1=...=c_n=0$. 
Later we will see that actually it is already safe to use $\tilde{\textsf{V}}_M(\lambda)={\rm Ker}(M-\lambda I)^{m^{\rm a}-m^{\rm g}+1}$.

With the above basic properties of generalized eigenvectors in mind, we are ready to prove
\begin{theorem}
Given an $n\times n$ matrix $M$ with different eigenvalues $\lambda_j$'s and corresponding algebraic multiplicities $m^{\rm a}_j$'s, we have 
\begin{equation}
{\rm dim}\tilde{\textsf{V}}_M(\lambda_j)=m^{\rm a}_j, 
\label{dimVM}
\end{equation}
and thus $\mathbb{C}^n=\bigoplus_j \tilde{\textsf{V}}_M(\lambda_j)$.
\label{dimtildeVj}
\end{theorem}
\emph{Proof.---} First, from the fact that $\tilde{\textsf{V}}_M(\lambda_j)$'s are disjoint subspaces of $\mathbb{C}^n$, we have
\begin{equation}
\sum_j{\rm dim}\tilde{\textsf{V}}_M(\lambda_j)\le n.
\label{sumVM}
\end{equation}
On the other hand, suppose that $\tilde{\textsf{V}}_M(\lambda_j)<m^{\rm a}_j$, then the characteristic polynomial of $M$ restricted on $\tilde{\textsf{V}}_M(\lambda_j)$ has at most $(m^{\rm a}_j-1)$ multiplicities of $(\lambda-\lambda_j)$ factor. This implies that there is at least one eigenvector of $\lambda_j$ lying outside $\tilde{\textsf{V}}_M(\lambda_j)$, contradicting with the definition of $\tilde{\textsf{V}}_M(\lambda_j)$. Therefore, we must have
\begin{equation}
{\rm dim}\tilde{\textsf{V}}_M(\lambda_j)\ge m^{\rm a}_j,\;\;\;\;\forall j.
\label{VMma}
\end{equation}
Combining Eqs.~(\ref{sumVM}) and (\ref{VMma}) with the identity $\sum_jm^{\rm a}_j=n$, we obtain Eq.~(\ref{dimVM}). Moreover, it can be inferred from the disjointness of $\tilde{\textsf{V}}_M(\lambda_j)$'s and $\sum_j{\rm dim}\tilde{\textsf{V}}_M(\lambda_j)=n$ that the entire linear space $\mathbb{C}^n$ is a direct sum of $\tilde{\textsf{V}}_M(\lambda_j)$'s. $\square$

To prove that all the Jordan chains are linearly independent, we need the following theorem
\begin{theorem}
Suppose $\lambda$ is an eigenvalue of $M$ and $\bold{u}_j$'s ($j=1,2,...,m$) are a set of generalized eigenvectors of ranks $(p_j+1)$'s with $0\le p_1\le p_2\le...\le p_m$. Defining $\bold{u}^{(k_j)}_j\equiv (M-\lambda I)^{k_j}\bold{u}_j$ with $0\le k_j \le p_j$, then $\bold{u}^{(k_j)}_j$'s are $m+\sum^m_{j=1}{p_j}$ linearly independent vectors if and only if $\bold{u}^{(p_j)}_j$'s are $m$ linearly independent vectors.
\label{lind}
\end{theorem}
\emph{Proof.---} We first note that ``only if" is trivial since $\bold{u}^{(p_j)}_j$'s are a subset of $\bold{u}^{(k_j)}_j$'s. To prove ``if", we directly consider the possible solution to
\begin{equation}
\sum^m_{j=1}\sum^{p_j}_{k_j=0} c_{j,k_j} \bold{u}^{(k_j)}_j = \bold{0}.
\label{cjkj}
\end{equation}
Multiplying $(M-\lambda I)^{p_m}$ to both sides gives
\begin{equation}
\bold{0}=(M-\lambda I)^{p_m}\sum^m_{j=1}\sum^{p_j}_{k_j=0} c_{j,k_j} \bold{u}^{(k_j)}_j = \sum_{j:p_j=p_m} c_{j,0} \bold{u}^{(p_j)}_j.
\end{equation}
Since $\bold{u}^{(p_j)}_j$'s are by assumption linearly independent, we must have $c_{j,0}=0$ for those $j$'s satisfying $p_j=p_m$. 

Suppose that we already know $c_{j,0}=c_{j,1}=...=c_{j,p_j-p_m+(r-1)}=0$ for those $j$'s satisfying $p_j\ge p_m-(r-1)$, which is true at least for $r=1$. Whenever $r\le p_m$, we can multiply $(M-\lambda I)^{p_m-r}$ to Eq.~(\ref{cjkj}) to obtain
\begin{equation}
\bold{0}=(M-\lambda I)^{p_m-r}\sum^m_{j=1}\sum^{p_j}_{k_j=\max\{0,p_j-p_m+r\}} c_{j,k_j} \bold{u}^{(k_j)}_j = \sum_{j:p_j\ge p_m-r} c_{j,p_j-p_m+r} \bold{u}^{(p_j)}_j.
\end{equation}
Again, due to the linear independence of $\bold{u}^{(p_j)}_j$, we have $c_{j,0}=c_{j,1}=...=c_{j,p_j-p_m+r}=0$ for those $j$'s satisfying $p_j\ge p_m-r$. In particular, when taking $r=p_m$, we know that all the coefficients in Eq.~(\ref{cjkj}) should vanish. $\square$ 

To understand Eq.~(\ref{vpu}), we only have to note that $\bold{v}\in \textsf{W}_M(\lambda_j)$ and thus by definition $\bold{v}=(M-\lambda_jI)\bold{u}^{(1)}$ for some $\bold{u}^{(1)}\in\overline{\textsf{V}_M(\lambda_j)}$. If $\bold{u}^{(1)}\in\overline{\textsf{W}_M(\lambda_j)}$, then $\bold{u}=\bold{u}^{(1)}$ and $p=1$. Otherwise, $\bold{u}^{(1)}\in \textsf{W}_M(\lambda_j)$ and we can find $\bold{u}^{(2)}$ such that $\bold{u}^{(1)}=(M-\lambda_j I)\bold{u}^{(2)}$ and carry out a similar analysis for $\bold{u}^{(2)}$. Such an iteration should stop after at most $m^{\rm a}_j-m^{\rm g}_j$ steps. Otherwise, according to Theorem~\ref{lind}, the Jordan chain ended at $\bold{v}$ with length larger than $m^{\rm a}_j-m^{\rm g}_j+1$ together with other linearly independent vectors in $\textsf{V}_M(\lambda_j)$ would span a subspace of $\tilde{\textsf{V}}_M(\lambda_j)$ with dimension larger than $m^{\rm a}_j$, leading to a contradiction with Theorems~\ref{dimtildeVj}. 

Now it is time to show that all the Jordan chains (including those with length $1$) with eigenvalue $\lambda$ span the generalized-eigenvector subspace $\tilde{\textsf{V}}_M(\lambda)$. First, by definition, all the Jordan chains should span a subspace of $\tilde{\textsf{V}}_M(\lambda)$. Second, for any generalized (including usual) eigenvector $\bold{w}$, it should belong to one of $\textsf{V}_M(\lambda)$, $\overline{\textsf{V}_M(\lambda)\cup \textsf{W}_M(\lambda)}$ and $\textsf{W}_M(\lambda)\backslash \textsf{V}_M(\lambda)$. For the former two cases, $\bold{w}$ can be expressed as a linear combination of the ends or heads of the Jordan chains. For the last case, we can find some $\bold{u}\in \overline{\textsf{V}_M(\lambda)\cup \textsf{W}_M(\lambda)}$ and $k\in\mathbb{Z}^+$ such that $\bold{w}=(M-\lambda)^k\bold{u}$, so $\bold{w}$ is linear combination of some middle vectors of the Jordan chains, with the coefficients following those in the decomposition of $\bold{u}$ into the heads of the Jordan chains.

It is useful to introduce a general formula for calculating the number of Jordan blocks with size $r$ and eigenvalue $\lambda$: 
\begin{equation}
n_M(\lambda,r)=2{\rm dim\;Ker}(M-\lambda I)^r-{\rm dim\;Ker}(M-\lambda I)^{r+1}-{\rm dim\;Ker}(M-\lambda I)^{r-1}.
\label{nlr}
\end{equation}
To derive this result, it is sufficient to show that, given an eigenvalue $\lambda$, the number of Jordan
with size no less than $r$ is given by
\begin{equation}
N_M(\lambda,r)={\rm dim\;Ker}(M-\lambda I)^r-{\rm dim\;Ker}(M-\lambda I)^{r-1},
\label{Nlr}
\end{equation}
which is related to Eq.~(\ref{nlr}) through $n_M(\lambda,r)=N_M(\lambda,r)-N_M(\lambda,r+1)$. It is rather straightforward to examine Eq.~(\ref{Nlr}) from the following property of a Jordan normal block:
\begin{equation}
{\rm dim\;Ker}(J_R(\lambda)-\lambda I)^r=\min\{R,r\},
\end{equation} 
which implies
\begin{equation}
{\rm dim\;Ker}(J_R(\lambda)-\lambda I)^r-{\rm dim\;Ker}(J_R(\lambda)-\lambda I)^{r-1}=\left\{\begin{array}{ll} 1,\;\; & \;\;R\ge r \\ 0,\;\; & \;\;R<r \end{array}\right..
\label{JRr}
\end{equation}
We obtain Eq.~(\ref{Nlr}) by summing Eq.~(\ref{JRr}) up with respect to all the Jordan blocks in $M$ with eigenvalue $\lambda$.
\\
\\
{\it Proof of Eq.~\eqref{fM}}

\vspace{3pt}
\noindent
The formula~\eqref{fM} for a function of a matrix using the Jordan normal form can be derived by
\begin{equation}
\begin{split}
f(M)&\equiv\sum^\infty_{l=0}c_l M^l=\sum^\infty_{l=0}c_l\left[\sum^J_{j=1}\sum^{m^{\rm g}_j}_{\alpha=1} (\lambda_jP_{j\alpha}+N_{j\alpha})\right]^l \\
&=\sum^\infty_{l=0}\sum^J_{j=1}\sum^{m^{\rm g}_j}_{\alpha=1}c_l (\lambda_j P_{j\alpha}+N_{j\alpha})^l=\sum^J_{j=1}\sum^{m^{\rm g}_j}_{\alpha=1}\sum^\infty_{l=0}\sum^l_{p=0}\frac{l!c_l}{p!(l-p)!} \lambda^{l-p}_j P_{j\alpha}^{l-p}N_{j\alpha}^p \\
&=\sum^\infty_{l=0}\sum^J_{j=1}\sum^{m^{\rm g}_j}_{\alpha=1}\left[c_l\lambda_j^l P_{j\alpha}+\sum^l_{p=1}\frac{l!c_l}{p!(l-p)!} \lambda_j^{l-p}N_{j\alpha}^p\right],
\end{split}
\label{fMpre}
\end{equation}
where we have used Eqs.~(\ref{onp}) and (\ref{pn}). Using Eq.~\eqref{nip} and the fact that the $p$th derivative of $f(z)$ is given by
\begin{equation}
f^{(p)}(z)=\sum^\infty_{l=p} \frac{l!c_l}{(l-p)!}z^{l-p},
\end{equation}
we can simplify Eq.~(\ref{fMpre}) into the formula~\eqref{fM} in the main text. 
\\
\\
{\it Proof of Eq.~\eqref{SVD} for a noninvertible matrix}

\vspace{3pt}
\noindent
The existence of the singular value decomposition~\eqref{SVD} when $M$ is not invertible, i.e., $m^{\rm g}_0\equiv{\rm dim\;Ker}M={\rm dim\;Ker}M^\dag M\ge1$, can be proven as follows.  We first refine Eq.~(\ref{VMMV}) into  
\begin{equation}
VM^\dag MV^\dag=\begin{bmatrix} V_1^\dag \\ V_2^\dag \end{bmatrix} M^\dag M \begin{bmatrix} V_1 & V_2\end{bmatrix} = \begin{bmatrix} \;V_1^\dag M^\dag M V_1\; & \;V_1^\dag M^\dag M V_2\; \\ \;V_2^\dag M^\dag M V_1\; & \;V_2^\dag M^\dag M V_2\; \end{bmatrix} = \begin{bmatrix} \;\Sigma_1^2\; & \;0\; \\ \;\;0\;\; & \;\;0\;\; \end{bmatrix}=\Sigma^2,
\end{equation}
where the dimension of $\Sigma_1$, which is positive definite, is $n-m^{\rm g}_0$. Obviously, we have $MV_2=0$ due to $V_2^\dag M^\dag M V_2=0$. Also, we can introduce an $n\times(n-m^{\rm g}_0)$ matrix
\begin{equation}
W_1=MV_1\Sigma_1^{-1},
\end{equation}
which satisfies $W_1^\dag W_1=I_1$ and is thus an \emph{isometry}. This means that the columns of $W_1$ form a set of orthonormal vectors and can be extended to be complete, i.e., we can find an $n\times m^{\rm g}_0$ matrix such that $W=[W_1\;\;W_2]$ is a unitary matrix. Recalling that $MV_2=0$, we have
\begin{equation}
W\Sigma=\begin{bmatrix} W_1 & W_2\end{bmatrix} \begin{bmatrix} \;\Sigma_1\; & \;0\; \\ \;\;0\;\; & \;\;0\;\; \end{bmatrix}=\begin{bmatrix} W_1\Sigma_1 & 0\end{bmatrix}=M\begin{bmatrix} V_1 & V_2\end{bmatrix}=MV,
\end{equation}
which gives rise to Eq.~(\ref{SVD}) if we multiply $V^\dag$ from the right. 
\\
\\
{\it Proof of Eq.~\eqref{rhonorm}}

\vspace{3pt}
\noindent
To show the right identity of Eq.~\eqref{rhonorm}, we first note that $\|M^k\|\ge \rho(M^k)=\rho(M)^k$ according to the left inequality in Eq.~\eqref{rhonorm}. On the other hand, for any $\epsilon>0$, we can show on the basis of the Jordan normal form (especially by using Eq.~(\ref{fM})) that 
\begin{equation}
\lim_{k\to\infty} \left[\frac{M}{\rho(M)+\epsilon}\right]^k=0, 
\end{equation}
implying the existence of $k_\epsilon$ such that $\|M^k\|< (\rho(M)+\epsilon)^k$ for $\forall k>k_\epsilon$. Combining these two aspects, we know that for $\forall\epsilon>0$, there exists an integer $k_\epsilon$ such that for $\forall k>k_\epsilon$
\begin{equation}
\rho(M)\le\|M^k\|^{\frac{1}{k}}\le\rho(M)+\epsilon\;\;\;\;\Rightarrow\;\;\;\;
|\|M^k\|^{\frac{1}{k}}-\rho(M)|\le\epsilon.
\end{equation}
This is nothing but the definition of $\rho(M)=\lim_{k\to\infty}\|M^k\|^{\frac{1}{k}}$.
\\
\\
{\it Proof of Eq.~\eqref{Weylpert}}

\vspace{3pt}
\noindent
The result~\eqref{Weylpert} in the main text can  be derived from the \emph{min-max principle} (or \emph{Courant-Fischer theorem} \cite{CDM00}), which states that the $j$th largest eigenvalue $\lambda_j$ of a Hermitian matrix $M$ is given by
\begin{equation}
\lambda_j=\max_{\textsf{V}\subseteq\mathbb{C}^n:{\rm dim}\textsf{V}=j}\min_{\bold{v}\in \textsf{V}:\|\bold{v}\|=1}\bold{v}^\dag M\bold{v}.
\end{equation}
Denoting $\textsf{V}^{(j)}_{M+E}$ as the $j$-dimensional subspace within which 
\begin{equation}
\lambda'_j=\min_{\bold{v}\in \textsf{V}^{(j)}_{M+E}:\|\bold{v}\|=1}\bold{v}^\dag(M+E)\bold{v}, 
\end{equation}
we have
\begin{equation}
\begin{split}
\lambda'_j&=\min_{\bold{v}\in \textsf{V}^{(j)}_{M+E}:\|\bold{v}\|=1}\bold{v}^\dag M\bold{v}+\min_{\bold{v}\in \textsf{V}^{(j)}_{M+E}:\|\bold{v}\|=1}\bold{v}^\dag E\bold{v} \\
&\le \max_{\textsf{V}\subseteq\mathbb{C}^n:{\rm dim}\textsf{V}=j}\min_{\bold{v}\in \textsf{V}}\bold{v}^\dag M\bold{v}+\max_{\bold{v}\in \mathbb{C}^n:\|\bold{v}\|=1}\bold{v}^\dag E\bold{v} \\
&=\lambda_j+\|E\|.
\end{split}
\label{lpl}
\end{equation}
Regarding $M=(M+E)-E$, we can obtain 
\begin{equation}
\lambda_j\le\lambda'_j+\|E\| 
\label{llp}
\end{equation}
following a similar procedure. Combining Eqs.~(\ref{lpl}) and (\ref{llp}), we obtain  Eq.~(\ref{Weylpert}). 
\\
\\
{\it Proof of Theorem~\eqref{NHSS}}

\vspace{3pt}
\noindent
Theorem~\ref{NHSS} can be proven as follows. We only have to consider those $\lambda'_j$'s satisfying $\min_{j'}|\lambda'_j-\lambda_{j'}|>0$, i.e., $\lambda'_j\notin\Lambda(M)$. Since $\lambda'_j$ is an eigenvalue of $M+E$, there exists a vector $\bold{v}_j\neq\bold{0}$ such that 
\begin{equation}
(M+E)\bold{v}_j=\lambda'_j\bold{v}_j.
\end{equation}
Defining $\bold{u}_j\equiv V^{-1}\bold{v}_j\neq\bold{0}$ and using the decomposition $M=VDV^{-1}$, we obtain 
\begin{equation}
(D+V^{-1}EV)\bold{u}_j=\lambda'_j\bold{u}_j.
\end{equation}
Recalling the assumption $\lambda'_j\notin\Lambda(M)$, we can rewrite the above equation into
\begin{equation}
\bold{u}_j=(\lambda'_jI-D)^{-1}V^{-1}EV\bold{u}_j.
\label{umuD}
\end{equation}
Taking the norm for both sides of Eq.~(\ref{umuD}) gives
\begin{equation}
\begin{split}
\|\bold{u}_j\|&=\|(\lambda'_jI-D)^{-1}V^{-1}EV\bold{u}_j\| \\
&\le \|(\lambda'_jI-D)^{-1}\|\|V^{-1}\|\|E\|\|V\|\|\bold{u}_j\| \\
&=(\min_{j'}|\lambda'_j-\lambda_{j'}|)^{-1}{\rm cond}(V)\|E\|\|\bold{u}_j\|,
\end{split}
\end{equation}
which  leads to
\begin{equation}
\min_{j'}|\lambda'_j-\lambda_{j'}|\le {\rm cond}(V)\|E\|.
\end{equation}
Since this inequality holds for $\forall j$ satisfying $\min_{j'}|\lambda'_j-\lambda_{j'}|>0$ and obviously also for those with $\min_{j'}|\lambda'_j-\lambda_{j'}|=0$, we end up with Eq.~(\ref{SpeS}) in the main text. 
\\
\\
{\it Proof of Eq.~\eqref{rjrjp}}

\vspace{3pt}
\noindent
The relation~(\ref{rjrjp}) can be derived by substituting 
\begin{equation}
\lambda_j-\lambda_j^*=\langle r_j|M-M^\dag|r_j\rangle,\;\;\;\;
\langle r_{j'}| r_j\rangle(\lambda_j-\lambda_{j'}^*)=\langle r_{j'}|M-M^\dag|r_j\rangle
\end{equation}
into Eq.~(\ref{rjrjp}), which gives an equivalent inequality
\begin{equation}
|\langle r_{j'}|i(M-M^\dag)|r_j\rangle|^2\le|\langle r_j|i(M-M^\dag)|r_j\rangle||\langle r_{j'}|i(M-M^\dag)|r_{j'}\rangle|.
\label{rjpmrj}
\end{equation}
If $i(M-M^\dag)$ ($i(M^\dag -M)$) is positive semi-definite, we can introduce $|\phi_j\rangle=\sqrt{i(M-M^\dag)}|r_j\rangle$ ($|\phi_j\rangle=\sqrt{i(M^\dag-M)}|r_j\rangle$) to simplify the above inequality (\ref{rjpmrj}) into $|\langle\phi_{j'}|\phi_j\rangle|^2\le|\langle\phi_{j'}|\phi_{j'}\rangle||\langle\phi_j|\phi_j\rangle|$, which is nothing but the Cauchy-Schwarz inequality.

\section{General description of quadratic Hamiltonians}\label{app2}
We provide a description of a general quadratic problem for both fermionic and bosonic systems, and discuss their excitation spectra in terms of the Bogoliubov-de Gennes (BdG) Hamiltonians. 
\subsection*{Fermions}
We start from the most general form of a non-Hermitian fermionic Hamiltonian that is quadratic in terms of $N$ different fermion modes: 
\begin{equation}
H=\sum^N_{j,j'=1}\left(J_{jj'}c_j^\dag c_{j'}+\frac{1}{2}\Delta^+_{jj'}c_j^\dag c_{j'}^\dag+\frac{1}{2}\Delta^-_{jj'}c_jc_{j'}\right)-\frac{1}{2}{\rm Tr} J,
\label{ffH}
\end{equation}
where $J$ and $\Delta^\pm$ are all $N\times N$ matrices. Without loss of generality, we assume the pairing matrices $\Delta^\pm$ to be anti-symmetrized, i.e., $(\Delta^\pm)^{\rm T}=-\Delta^\pm$, since $\{c_j,c_{j'}\}=\{c_j^\dag,c_{j'}^\dag\}=0$. With this assumption, there is no other redundant degrees of freedom in Eq.~(\ref{ffH}). The reason that we introduce a constant term $-\frac{1}{2}{\rm Tr}J$ is to impose an inherent PHS, 
as will be clear later.

To simplify the notation, we introduce $\bold{c}=[c_1,c_2,...,c_N]^{\rm T}$, so that Eq.~(\ref{ffH}) can be rewritten into the Nambu representation:
\begin{equation}
H=\frac{1}{2}\begin{bmatrix} \;\bold{c}^\dag\; & \;\bold{c}\; \end{bmatrix} 
\begin{bmatrix} \;J\; & \;\Delta^+\; \\ \;\Delta^-\; & \;-J^{\rm T}\; \end{bmatrix}
\begin{bmatrix} \;\bold{c}\; \\ \;\bold{c}^\dag\; \end{bmatrix},
\label{namb}
\end{equation}
where we have used $\{c_j^\dag,c_{j'}\}=\delta_{jj'}$. Alternatively, we can introduce $2N$ Majorana operators related to $\bold{c}$ and $\bold{c}^\dag$ via 
\begin{equation}
\begin{bmatrix} \;\boldsymbol{\gamma}_{\rm o}\; \\ \;\boldsymbol{\gamma}_{\rm e}\; \end{bmatrix} = 
\begin{bmatrix} \;1\; & \;1\; \\ \;i\; & \;-i\; \end{bmatrix}
\begin{bmatrix} \;\bold{c}\; \\ \;\bold{c}^\dag\; \end{bmatrix},
\label{gamc}
\end{equation}
where $\boldsymbol{\gamma}_{\rm o}=[\gamma_1,\gamma_3,...,\gamma_{2N-1}]^{\rm T}$ and $\boldsymbol{\gamma}_{\rm e}=[\gamma_2,\gamma_4,...,\gamma_{2N}]^{\rm T}$, with $\gamma_j$'s satisfying $\gamma_j^\dag=\gamma_j$ for $\forall j=1,2,...,2N$ and $\{\gamma_j,\gamma_{j'}\}=2\delta_{jj'}$ for $\forall j,j'$. Combining Eqs~(\ref{gamc}) and (\ref{namb}), we obtain the following Majorana representation:
\begin{equation}
H=\frac{1}{8}\begin{bmatrix} \;\boldsymbol{\gamma}_{\rm o}\; & \;\boldsymbol{\gamma}_{\rm e}\; \end{bmatrix} 
\begin{bmatrix} \;J-J^{\rm T}+\Delta^++\Delta^-\; & \;i(\Delta^+-\Delta^--J-J^{\rm T})\; \\ \;i(J+J^{\rm T}+\Delta^+-\Delta^-)\; & \;J-J^{\rm T}-\Delta^+-\Delta^-\; \end{bmatrix}
\begin{bmatrix} \;\boldsymbol{\gamma}_{\rm o}\; \\ \;\boldsymbol{\gamma}_{\rm e}\; \end{bmatrix},
\label{Majrep}
\end{equation}
Denoting the matrix in Eq.~(\ref{Majrep}) as $2iA$, we find that
\begin{equation}
A^{\rm T}=-A.
\end{equation}
Conversely, starting from 
\begin{equation}
H=\frac{i}{4}\begin{bmatrix} \;\boldsymbol{\gamma}_{\rm o}\; & \;\boldsymbol{\gamma}_{\rm e}\; \end{bmatrix} A \begin{bmatrix} \;\boldsymbol{\gamma}_{\rm o}\; \\ \;\boldsymbol{\gamma}_{\rm e}\; \end{bmatrix},\;\;\;\;A=\begin{bmatrix} \;A_{\rm o}\; & \;B\;\; \\ \;-B^{\rm T}\; & \;A_{\rm e}\;\; \end{bmatrix},
\label{iA}
\end{equation}
with $A^{\rm T}_{\rm o,e}=-A_{\rm o,e}$, we can specify the matrices in Eq.~(\ref{Majrep}) as $J=\frac{i}{2}(A_{\rm o}+A_{\rm e})-\frac{1}{2}(B+B^{\rm T})$ and $\Delta^\pm=\frac{i}{2}(A_{\rm o}-A_{\rm e})\pm\frac{1}{2}(B-B^{\rm T})$. Equation~(\ref{iA}) is arguably the most compact form for representing a free-fermion Hamiltonian and has been demonstrated to be helpful for proving some analytical results \cite{FL10}. The necessary and sufficient condition for $H$ being Hermitian is $A^\dag=-A$, implying $A^*=A$ according to $A^{\rm T}=-A$. 

Let us move on to explain how the PHS manifests itself in the Nambu representation (\ref{namb}), which is also known as the BdG representation. First, we should emphasize that the PHS operator, although transforming $c_j$'s into $c_j^\dag$'s, is \emph{unitary} on the many-body level \cite{CCK16}. This implies the existence of a unitary matrix $U_{\rm C}$ such that
\begin{equation}
\mathcal{C}\bold{c}\mathcal{C}^{-1}=U_{\rm C}\bold{c}^\dag\;\;\;\;\Leftrightarrow\;\;\;\;
\mathcal{C}\bold{c}^\dag\mathcal{C}^{-1}=U_{\rm C}^*\bold{c}.
\end{equation}
Therefore, $[\mathcal{C},H]=0$ gives
\begin{equation}
\begin{bmatrix} \;0\; & \;U_{\rm C}^{\rm T}\; \\ \;U_{\rm C}^\dag\; & \;0\;  \end{bmatrix}  
\begin{bmatrix} \;J\; & \;\Delta^+\; \\ \;\Delta^-\; & \;-J^{\rm T}\; \end{bmatrix}
\begin{bmatrix} \;0\; & \;U_{\rm C}\; \\ \;U_{\rm C}^*\; & \;0\;  \end{bmatrix}
=\begin{bmatrix} \;J\; & \;\Delta^+\; \\ \;\Delta^-\; & \;-J^{\rm T}\; \end{bmatrix},
\end{equation}
which turns out to be equivalent to
\begin{equation}
\begin{bmatrix}  \;U_{\rm C}^{\rm T}\; & \;0\; \\  \;0\; & \;U_{\rm C}^\dag\; \end{bmatrix}  
\begin{bmatrix} \;J\; & \;\Delta^+\; \\ \;\Delta^-\; & \;-J^{\rm T}\; \end{bmatrix}^{\rm T}
\begin{bmatrix}  \;U_{\rm C}^*\; & \;0\; \\  \;0\; & \;U_{\rm C}\; \end{bmatrix} 
=-\begin{bmatrix} \;J\; & \;\Delta^+\; \\ \;\Delta^-\; & \;-J^{\rm T}\; \end{bmatrix},
\label{PHST}
\end{equation}
where we have used $(\Delta^\pm)^{\rm T}=-\Delta^\pm$. In particular, transpose is equivalent to complex conjugate for a Hermitian Hamiltonian with $J^\dag=J$ and $\Delta^+=(\Delta^-)^\dag$, so the representation of $\mathcal{C}$ becomes effectively anti-unitary on the single-particle level. We note that Eq.~(\ref{PHST}) is especially advantageous for particle-number conserving systems, for which the matrix becomes block-diagonal and $\mathcal{C}$ can be represented $U_C^{\rm T}\mathcal{K}$ acting only on the particle sector.

Finally, let us figure out the explicit BdG Hamiltonian for Eq.~(\ref{ffH}) on a translation-invariant lattice. In this case, we have $j=(\boldsymbol{r},a)$, where $\boldsymbol{r}$ is a lattice site and $a$ denotes an internal state, and the translation invariance implies $J_{\boldsymbol{r}a,\boldsymbol{r}'a'}=J_{\boldsymbol{r}-\boldsymbol{r}',aa'}$ and $\Delta^\pm_{\boldsymbol{r}a,\boldsymbol{r}'a'}=\Delta^\pm_{\boldsymbol{r}-\boldsymbol{r}',aa'}$. After Fourier transform, we can rewrite Eq.~(\ref{ffH}) into 
\begin{equation}
\begin{split}
H=&\sum_{\boldsymbol{k},a,a'}\left\{[J(\boldsymbol{k})]_{aa'}c_{\boldsymbol{k}a}^\dag c_{\boldsymbol{k}a'}
+\frac{1}{2}[\Delta^+(\boldsymbol{k})]_{aa'}c_{\boldsymbol{k}a}^\dag c_{-\boldsymbol{k}a'}^\dag+
\frac{1}{2}[\Delta^-(\boldsymbol{k})]_{aa'}c_{-\boldsymbol{k}a} c_{\boldsymbol{k}a'}\right\} \\
&-\frac{1}{2}\sum_{\boldsymbol{k}}{\rm Tr}[J(\boldsymbol{k})],
\end{split}
\end{equation}
where $c_{\boldsymbol{k}a}=N^{-1/2}\sum_{\boldsymbol{r}}e^{-i\boldsymbol{k}\cdot\boldsymbol{r}}c_{\boldsymbol{r}a}$, $[J(\boldsymbol{k})]_{aa'}=\sum_{\boldsymbol{r}-\boldsymbol{r}'}e^{-i\boldsymbol{k}\cdot(\boldsymbol{r}-\boldsymbol{r}')}J_{\boldsymbol{r}-\boldsymbol{r}',aa'}$ and $[\Delta^\pm(\boldsymbol{k})]_{aa'}=\sum_{\boldsymbol{r}-\boldsymbol{r}'}e^{-i\boldsymbol{k}\cdot(\boldsymbol{r}-\boldsymbol{r}')}\Delta^\pm_{\boldsymbol{r}-\boldsymbol{r}',aa'}$. Note that $\Delta^\pm(\boldsymbol{k})^{\rm T}=-\Delta^\pm(-\boldsymbol{k})$ due to $\Delta_{\boldsymbol{r}-\boldsymbol{r}',aa'}=-\Delta_{\boldsymbol{r}'-\boldsymbol{r},a'a}$. Applying the equivalence between Eqs.~(\ref{ffH}) and (\ref{namb}) for each $\boldsymbol{k}$, we obtain
\begin{equation}
H=\sum_{\boldsymbol{k}} \begin{bmatrix} \;\bold{c}_{\boldsymbol{k}}^\dag\; & \;\bold{c}_{-\boldsymbol{k}}\; \end{bmatrix} H_{\rm BdG}(\boldsymbol{k})\begin{bmatrix} \;\bold{c}_{\boldsymbol{k}}\; \\ \;\bold{c}^\dag_{-\boldsymbol{k}}\; \end{bmatrix},
\end{equation}
where the BdG Hamiltonian is given by
\begin{equation}
H_{\rm BdG}(\boldsymbol{k})=\frac{1}{2}\begin{bmatrix} \;J(\boldsymbol{k})\; & \;\Delta^+(\boldsymbol{k})\; \\ \;\Delta^-(\boldsymbol{k})\; & \;-J(-\boldsymbol{k})^{\rm T}\;  \end{bmatrix}.
\end{equation}
Using $\Delta^\pm(\boldsymbol{k})^{\rm T}=-\Delta^\pm(-\boldsymbol{k})$, we can check that
\begin{equation}
U_{\rm C}H_{\rm BdG}(\boldsymbol{k})^{\rm T}U_{\rm C}=-H_{\rm BdG}(-\boldsymbol{k}),\;\;\;\;U_{\rm C}=\sigma^x\otimes I.
\end{equation}
In the Hermitian limit, we have $J(\boldsymbol{k})=J(\boldsymbol{k})^\dag$, $\Delta^+(\boldsymbol{k})=[\Delta^-(\boldsymbol{k})]^\dag$ and thus $H_{\rm BdG}(\boldsymbol{k})^\dag=H_{\rm BdG}(\boldsymbol{k})$, leading to an effectively anti-unitary PHS:
\begin{equation}
\mathcal{C}H_{\rm BdG}(\boldsymbol{k})\mathcal{C}^{-1}=-H_{\rm BdG}(-\boldsymbol{k}),\;\;\;\;
\mathcal{C}=(\sigma^x\otimes I)\mathcal{K}.
\end{equation}
A single-particle  excitation spectrum can be obtained by diagonalizing $H_{\rm BdG}$ via a unitary matrix.

\subsection*{Bosons}
Using the same notation and procedure in the fermionic case above, one can arrive at a general quadratic Hamiltonian for bosons as
\begin{equation}
H=\sum_{\boldsymbol{k}} \begin{bmatrix} \;\bold{a}_{\boldsymbol{k}}^\dag\; & \;\bold{a}_{-\boldsymbol{k}}\; \end{bmatrix} H(\boldsymbol{k})\begin{bmatrix} \;\bold{a}_{\boldsymbol{k}}\; \\ \;\bold{a}^\dag_{-\boldsymbol{k}}\; \end{bmatrix},
\end{equation}
where $H(\boldsymbol{k})$ is given by
\begin{equation}
H(\boldsymbol{k})=\frac{1}{2}\begin{bmatrix} \;J(\boldsymbol{k})\; & \;\Delta^+(\boldsymbol{k})\; \\ \;\Delta^-(\boldsymbol{k})\; & \;J(-\boldsymbol{k})^{\rm T}\;  \end{bmatrix}.
\end{equation}
Because of the commutation relations of bosonic operators, the off-diagonal matrices now satisfy $\Delta^{\pm}({\boldsymbol k})^{\rm T}=\Delta^{\pm}(-\boldsymbol{k})$. The most important distinctive feature of bosonic systems is that one must use a \emph{paraunitary} matrix $T({\boldsymbol k})$ \cite{JHPC78,MR13} (instead of a unitary matrix in fermionic systems) to obtain a single-particle excitation spectrum:
\eqn{
\begin{bmatrix} \;\bold{a}_{\boldsymbol{k}}\; \\ \;\bold{a}^\dag_{-\boldsymbol{k}}\; \end{bmatrix}=T({\boldsymbol k})\begin{bmatrix} \;\bold{b}_{\boldsymbol{k}}\; \\ \;\bold{b}^\dag_{-\boldsymbol{k}}\; \end{bmatrix},\;\;\;T({\boldsymbol k})^\dag(\sigma^z\otimes I)T({\boldsymbol k})=(\sigma^z\otimes I).
}
This constraint comes from the commutation relations for the transformed bosonic operators $\bold{b}_{\boldsymbol{k}}$ and $\bold{b}_{\boldsymbol{k}}^\dagger$. 
The BdG Hamiltonian is thus given by
\eqn{
H&=&\sum_{\boldsymbol{k}} \begin{bmatrix} \;\bold{b}_{\boldsymbol{k}}^\dag\; & \;-\bold{b}_{-\boldsymbol{k}}\; \end{bmatrix} T({\boldsymbol k})^{-1}H_{\rm BdG}(\boldsymbol{k})T({\boldsymbol k})\begin{bmatrix} \;\bold{b}_{\boldsymbol{k}}\; \\ \;\bold{b}^\dag_{-\boldsymbol{k}}\; \end{bmatrix},\\
H_{\rm BdG}(\boldsymbol{k})&=&\frac{1}{2}\begin{bmatrix} \;J(\boldsymbol{k})\; & \;\Delta^+(\boldsymbol{k})\; \\ \;-\Delta^-(\boldsymbol{k})\; & \;-J(-\boldsymbol{k})^{\rm T}\;  \end{bmatrix}.
}
In the same manner as in the fermionic case, in the Hermitian limit, we have $J(\boldsymbol{k})=J(\boldsymbol{k})^\dag$, $\Delta^+(\boldsymbol{k})=[\Delta^-(\boldsymbol{k})]^\dag$ and thus $H_{\rm BdG}$ satisfies the PHS:
\begin{equation}
\mathcal{C}H_{\rm BdG}(\boldsymbol{k})\mathcal{C}^{-1}=-H_{\rm BdG}(-\boldsymbol{k}),\;\;\;\;
\mathcal{C}=(\sigma^x\otimes I)\mathcal{K},
\end{equation}
leading to the excitation spectrum 
\eqn{
T({\boldsymbol k})^{-1}H_{\rm BdG}(\boldsymbol{k})T({\boldsymbol k})&=&\frac{1}{2}\begin{bmatrix} \;\bold{\epsilon}({\boldsymbol k})\; & \;0\; \\ \;0\; & \;-\bold{\epsilon}^*(-{\boldsymbol k})\;  \end{bmatrix}.
}
In this case, we emphasize that, even if the original Hamiltonian is Hermitian,  excitation energies $\bold{\epsilon}({\boldsymbol k})$ can be complex.
In particular, $H_{\rm BdG}$ is $\eta$-pseudo-Hermitian with $\eta=\sigma^z\otimes I$ \cite{OT20} and thus has either entirely real or complex conjugate pairs of eigenvalues (see Eq.~\eqref{etapseudo} and Theorem~\ref{pseudoh}). 
When the system is stable and only contains real-valued eigenvalues, the physical elementary excitations correspond to the positive components $\bold{\epsilon}({\boldsymbol k})>0$ \cite{BECbook}.  Physically, the emergence of complex-valued modes indicates the instability of the underlying bosonic condensate, which is the starting point (i.e., the vacuum state) of the present mean-field description.

\section{Bound on correlations in matrix-product states}\label{app3}
Without loss of generality, we assume $O_X$ and $O_Y$, which are \emph{not} necessarily Hermitian, to be on-site. Otherwise, we can block multiple adjacent sites into one and redefine $|m_1m_2...m_{|X|/|Y|}\rangle$ to be $|m\rangle$. In terms of the MPS expression (\ref{MPS}), the lhs of Eq.~(\ref{MPScor}) can be rewritten into 
\begin{equation}
|{\rm Tr}[\Lambda\mathcal{E}_X(\mathcal{E}^l-\mathcal{E}^\infty)\mathcal{E}_Y(\mathbb{1})]|=|{\rm Tr}[\mathcal{E}_X^\dag(\Lambda)^\dag(\mathcal{E}^l-\mathcal{E}^\infty)\mathcal{E}_Y(\mathbb{1})]|, 
\end{equation}
where $\Lambda=\Lambda^\dag>0$ with ${\rm Tr}\Lambda=1$ is the unique fixed point of the dual quantum channel $\mathcal{E}^\dag(\;\cdot\;)=\sum_m M_m^\dag\;\cdot\;M_m$ and 
\begin{equation}
\mathcal{E}_X^\dag(\;\cdot\;)=\sum_{m,m'} \langle m|O_X^\dag|m'\rangle M_m^\dag\;\cdot\;M_{m'},\;\;\;\;
\mathcal{E}_Y(\;\cdot\;)=\sum_{m,m'} \langle m'|O_Y|m\rangle M_m\;\cdot\;M_{m'}^\dag.
\end{equation}
In the following, we prove $\|\mathcal{E}_X^\dag (\Lambda)\|_2\le\|O_X\|$ and $\|\mathcal{E}_Y(\mathbb{1})\|_2\le\sqrt{D}\|O_Y\|$, which imply Eq.~(\ref{MPScor}) according to the definition of the superoperator norm.

Recalling that $\|O\|_2\equiv\sqrt{{\rm Tr}[O^\dag O]}$, we first write down the explicit expression of
\begin{equation}
\begin{split}
\|\mathcal{E}_X^\dag (\Lambda)\|_2^2&=
\sum_{m,m',n,n'}\langle m'|O_X|m\rangle\langle n|O_X^\dag|n'\rangle {\rm Tr}[M_{m'}^\dag\Lambda M_m M_n^\dag\Lambda M_{n'}] \\
&=\sum_{m,m',n,n'}\langle m'n'|O_X\otimes O_X^*|mn\rangle{\rm Tr}[(\sqrt{\Lambda}M_{m'}M_{n'}^\dag\sqrt{\Lambda})^\dag\sqrt{\Lambda}M_mM_n^\dag\sqrt{\Lambda}] \\
&={\rm Tr}[(O_X\otimes O_X^*)\rho_\Lambda],
\end{split}
\end{equation}
where $\rho_\Lambda=\sum_{m,m',n,n'}{\rm Tr}[(\sqrt{\Lambda}M_{m'}M_{n'}^\dag\sqrt{\Lambda})^\dag\sqrt{\Lambda}M_mM_n^\dag\sqrt{\Lambda}]|mn\rangle\langle m'n'|= \rho_\Lambda^\dag\ge0$ and ${\rm Tr}\rho_\Lambda=\sum_{m,n}{\rm Tr}[M_m^\dag\Lambda M_m M_n^\dag\Lambda M_n]={\rm Tr}[\mathcal{E}^\dag(\Lambda)^2]={\rm Tr}[\Lambda^2]\le 1$. Using the H\"older inequality $|{\rm Tr}[A^\dag B]|\le \|A\|\|B\|_1$ ($\|B\|_1\equiv{\rm Tr}[\sqrt{B^\dag B}]$ is the Schatten-$1$ norm) \cite{BB11}, we obtain
\begin{equation}
\|\mathcal{E}_X^\dag (\Lambda)\|_2^2\le \|O_X\otimes O_X^*\| \|\rho_\Lambda\|_1=\|O_X\|^2{\rm Tr}[\Lambda^2]\le \|O_X\|^2.
\end{equation}
So far we have proved $\|\mathcal{E}_X^\dag (\Lambda)\|_2\le\|O_X\|$. Similarly, we can express $\|\mathcal{E}_Y(\mathbb{1})\|_2^2$ as
\begin{equation}
\begin{split}
\|\mathcal{E}_Y(\mathbb{1})\|_2^2&=\sum_{m,m',n,n'}\langle m|O_Y^\dag|m'\rangle\langle n'|O_Y|n\rangle {\rm Tr}[M_{m'} M_m^\dag M_n M_{n'}^\dag] \\
&=\sum_{m,m',n,n'}\langle m'n'|O_Y^*\otimes O_Y|mn\rangle {\rm Tr}[(M_{m'}^\dag M_{n'})^\dag  M_m^\dag M_n ] \\
&={\rm Tr}[(O_Y^*\otimes O_Y)\rho_{\mathbb{1}}],
\end{split}
\end{equation}
where $\rho_{\mathbb{1}}=\sum_{m,m',n,n'}{\rm Tr}[(M_{m'}^\dag M_{n'})^\dag  M_m^\dag M_n]|mn\rangle \langle m'n'|=\rho_{\mathbb{1}}^\dag\ge0$ and ${\rm Tr}\rho_{\mathbb{1}}=\sum_{m,n}{\rm Tr}[M_mM_m^\dag M_n M_n^\dag]={\rm Tr}[\mathcal{E}(\mathbb{1})^2]={\rm Tr}\mathbb{1}=D$. Therefore, we have
\begin{equation}
\|\mathcal{E}_Y(\mathbb{1})\|_2^2\le \|O_Y^*\otimes O_Y\| \|\rho_{\mathbb{1}}\|_1= D\|O_Y\|^2,
\end{equation}
which completes the proof of $\|\mathcal{E}_Y(\mathbb{1})\|_2\le\sqrt{D}\|O_Y\|$.

\section{Continuous Hermitianization of line-gapped Bloch Hamiltonians}\label{app4}
We prove that an arbitrary non-Hermitian Bloch Hamiltonian with a real line gap can continuously be Hermitianized while keeping the line gap and all the BL symmetries, if any, and thus can be further flattened by unitarization. Once we can prove this, it follows that an imaginarily line-gapped Bloch Hamiltonian can continuously be anti-Hermitianized, since it can be mapped into a really line-gapped Bloch Hamiltonian by simply multiplying $i$. Our proof consists of two steps -- one is continuous non-Hermitian flattening, and the other is continuous Hermitianization.

The continuous path of non-Hermitian flattening is constructed as
\begin{equation}
H_{\lambda}(\boldsymbol{k})=(1-\lambda)H(\boldsymbol{k})+\lambda\left(\oint_{C_+}-\oint_{C_-}\right)\frac{dz}{2\pi i}\frac{1}{zI-H(\boldsymbol{k})},\;\;\;\;\lambda\in[0,1],
\label{NHflat}
\end{equation}
where $C_+\equiv\{z:|z|=r,{\rm Re}z>0\}\cup\{z:{\rm Re}z=0,|{\rm Im}z|\le r\}$ and $C_-\equiv-C_+$ with $r>\max_{\boldsymbol{k}}\|H(\boldsymbol{k})\|$, so the spectrum of $H(\boldsymbol{k})$ is not touched by $C_\pm$ (we recall that $H(\boldsymbol{k})$ has a real line gap). Obviously, $H_{\lambda}(\boldsymbol{k})$ is continuous in both $\lambda$ and $\boldsymbol{k}$. At the end of the path, we have
\begin{equation}
H_1(\boldsymbol{k})=P_+(\boldsymbol{k})-P_-(\boldsymbol{k}),
\label{H1Ppm}
\end{equation}
which is involutory since $P_\pm(\boldsymbol{k})$ is the (generally non-Hermitian) projector onto the eigenspaces whose eigenvalues have positive/negative real parts. Moreover, we note that each eigenvalue of $H_{\lambda}(\boldsymbol{k})$ takes the form of
\begin{equation}
\epsilon_\lambda(\boldsymbol{k})=(1-\lambda)\epsilon(\boldsymbol{k})+\lambda{\rm sgn}{\rm Re}\epsilon(\boldsymbol{k}),
\end{equation}
where $\epsilon(\boldsymbol{k})$ is the eigenvalue of $H(\boldsymbol{k})$. Therefore, the sign of the real part of each eigenvalue stays unchanged, and thus there is always a real line gap.

The remaining problem is whether the path given in Eq.~(\ref{NHflat}) keeps all the symmetries. To examine this, we first write down a general BL into the following compact form:
\begin{equation}
H(\boldsymbol{k})=\epsilon_XU_X\mathcal{X}[H(s_X\boldsymbol{k})]U_X^\dag,
\end{equation}
where $\epsilon_X,s_X\in\{\pm1\}$, $U_X$ is a unitary and $\mathcal{X}$ refers to either identity ($X=P$), transpose ($X=C$), complex conjugate ($X=K$) or Hermitian conjugate ($X=Q$). Note that $C_\pm$ has no overlap with the spectrum of $H(\boldsymbol{k})$ if and only if it has no overlap with that of $\mathcal{X}[H(\boldsymbol{k})]$, since $\|H(\boldsymbol{k})\|=\|\mathcal{X}[H(\boldsymbol{k})]\|$. Therefore, for $\forall z\in C_\pm$, we can take the inverse of both $zI-H(\boldsymbol{k})$ and $zI-\mathcal{X}[H(\boldsymbol{k})]$ to obtain
\begin{equation}
\frac{1}{zI-H(\boldsymbol{k})}=\epsilon_X U_X\frac{1}{\epsilon_X zI-\mathcal{X}[H(s_X\boldsymbol{k})]} U_X^\dag.
\label{BLinv}
\end{equation}
Moreover, we can show that for all types of $\mathcal{X}$ 
\begin{equation}
\mathcal{X}\left[\oint_{C_\pm}\frac{dz}{2\pi i}\frac{1}{zI-H(\boldsymbol{k})}\right]=\oint_{C_\pm}\frac{dz}{2\pi i}\frac{1}{zI-\mathcal{X}[H(\boldsymbol{k})]}.
\label{Xint}
\end{equation}
This is obvious for identity and transpose, so we only have to prove the case of complex conjugate, whose combination with transpose gives the Hermitian conjugate. This property can directly be shown:
\begin{equation}
\left[\oint_{C_\pm}\frac{dz}{2\pi i}\frac{1}{zI-H(\boldsymbol{k})}\right]^*=-\oint_{C^*_\pm}\frac{dz^*}{2\pi i}\frac{1}{z^*I-H(\boldsymbol{k})^*}=\oint_{C_\pm}\frac{dz}{2\pi i}\frac{1}{zI-H(\boldsymbol{k})^*},
\end{equation} 
where we have used the fact that $C^*_\pm$ stays the same as $C_\pm$ except for that the direction of the contour integral is reversed. Combining Eqs.~(\ref{BLinv}) and (\ref{Xint}), we obtain
\begin{equation}
\begin{split}
P_+(\boldsymbol{k})&=\oint_{C_+}\frac{dz}{2\pi i}\frac{1}{zI-H(\boldsymbol{k})}=\oint_{C_+}\frac{d(\epsilon_Xz)}{2\pi i}U_X\frac{1}{\epsilon_XzI-\mathcal{X}[H(s_X\boldsymbol{k})]}U_X^\dag \\
&=U_X\mathcal{X}\left[\oint_{C_+}\frac{d(\epsilon_Xz)}{2\pi i}\frac{1}{\epsilon_XzI-H(s_X\boldsymbol{k})}\right]U_X^\dag \\
&=U_X\mathcal{X}\left[\oint_{C_{\epsilon_X}}\frac{dz}{2\pi i}\frac{1}{zI-H(s_X\boldsymbol{k})}\right]U_X^\dag=U_X\mathcal{X}[P_{\epsilon_X}(s_X\boldsymbol{k})]U_X^\dag.
\end{split}
\label{PpUX}
\end{equation}
Since $P_\pm(\boldsymbol{k})$ are complementary to each other, i.e., $P_+(\boldsymbol{k})+P_-(\boldsymbol{k})=I$, we have $P_\pm(\boldsymbol{k})=\frac{1}{2}[I\pm H_1(\boldsymbol{k})]$ according to Eq.~(\ref{H1Ppm}). Applying this relation to Eq.~(\ref{PpUX}), we finally obtain
\begin{equation}
H_1(\boldsymbol{k})=\epsilon_XU_X \mathcal{X}[H_1(s_X\boldsymbol{k})]U_X^\dag,
\end{equation}
implying that $H_\lambda(\boldsymbol{k})$ as a real linear combination of $H(\boldsymbol{k})$ and $H_1(\boldsymbol{k})$ respects all the BL symmetries.

We move on to prove that $H_1(\boldsymbol{k})$ can continuously be Hermitianized while keeping all the BL symmetries as well as the real line gap. Given a non-Hermitian Bloch Hamiltonian $H_1(\boldsymbol{k})$, we can identify two Hermitian Hamiltonians in the Hermitian-anti-Hermitian decomposition $H_1(\boldsymbol{k})=h_+(\boldsymbol{k})+ih_-(\boldsymbol{k})$:
\begin{equation}
h_+(\boldsymbol{k})\equiv \frac{1}{2}[H_1(\boldsymbol{k})+H_1(\boldsymbol{k})^\dag],\;\;\;\;
h_-(\boldsymbol{k})\equiv \frac{1}{2i}[H_1(\boldsymbol{k})-H_1(\boldsymbol{k})^\dag].
\end{equation}
Obviously, $h_+(\boldsymbol{k})$ and $h_-(\boldsymbol{k})$ are both continuous in $\boldsymbol{k}$ and the former satisfies all the BL symmetries of $H_1(\boldsymbol{k})$ due to that it is a real linear combination of $H_1(\boldsymbol{k})$ and $H^\dag_1(\boldsymbol{k})$. Furthermore, since $H_1(\boldsymbol{k})$ is involutory, we have
$[h_+(\boldsymbol{k})+ih_-(\boldsymbol{k})]^2=h_+(\boldsymbol{k})^2-h_-(\boldsymbol{k})^2+i\{h_+(\boldsymbol{k}),h_-(\boldsymbol{k})\}=I$,
which implies
\begin{equation}
h_+(\boldsymbol{k})^2=I+h_-(\boldsymbol{k})^2,\;\;\;\;\{h_+(\boldsymbol{k}),h_-(\boldsymbol{k})\}=0.
\label{hpm}
\end{equation}
Now let us consider the following symmetry-preserving continuous path:
\begin{equation}
\tilde H_\lambda(\boldsymbol{k})=(1-\lambda)H_1(\boldsymbol{k})+\lambda h_+(\boldsymbol{k}),\;\;\;\;
\lambda\in[0,1].
\end{equation}
Using Eq.~(\ref{hpm}), $\tilde H_\lambda(\boldsymbol{k})^2$ can be obtained to be
\begin{equation}
\tilde H_\lambda(\boldsymbol{k})^2=I+[1-(1-\lambda)^2]h_-(\boldsymbol{k})^2\ge I,
\end{equation}
which means all the eigenvalues of $\tilde H_\lambda(\boldsymbol{k})^2$ are real and larger than $1$. This in turn implies that all the eigenvalues of $\tilde H_\lambda(\boldsymbol{k})$ are real and of absolute values no less than $1$, and thus a line gap stays open along the path.  We thus have completed the proof by constructing an explicit symmetry- and gap-preserving continuous path of Hermitianization.

\section{Topological classifications of the Bernard-LeClair classes}\label{app5}
In this appendix, we explain in detail the number of genuinely different BL classes for both point and line gaps. We also complete the classifications of all these BL classes. 

\subsection{Counting the Bernard-LeClair classes}\label{Sec:countBL}
In this appendix, we provide a detailed analysis on the number of genuinely different BL classes for both point and line gaps. We first consider the BL classes without $P$ symmetry, some of which reduce to the AZ classes in the Hermitian limit. The simplest case is no symmetry, which has a single class for both point and line gaps. We then consider the case of a single symmetry $X\neq P$. If $X=Q$, we have $1$ class for point gap while $2$ classes for line gap, since $\epsilon_Q=\pm1$ but it can be reversed by Wick rotation in the former case. If $X=C$, we have $4$ classes for both point and line gaps since $\epsilon_C=\pm1$, $\eta_C=\pm1$ and Wick rotation cannot reverse $\epsilon_C$. If $X=K$, we have $2$ classes for point gap while $4$ classes for line gap, since $\epsilon_K=\pm1$, $\eta_K=\pm1$ and the former can be reversed by Wick rotation. So far, when there is at most a single symmetry $X\neq P$, we have obtained $8$ and $11$ classes for point and line gaps, respectively.

We move on to analyze the case without $P$ symmetry but with two of the remaining symmetries. Recalling that the coexistence of any two of $Q$, $C$ and $K$ symmetries implies the other, we only have consider, for example, $Q$ and $C$ symmetries. Since there are $4$ binary parameters $\epsilon_Q$, $\epsilon_C$, $\eta_C$ and $\epsilon_{QC}$, we have $16$ classes for line gap. As for point gap, we have $8$ classes since Wick rotation only reverses $\epsilon_Q$ while leaves the other parameters unchanged.

Now let us focus on the BL classes with $P$ symmetry. If there is only a single $P$ symmetry, we have $1$ class for both point and line gaps. We then consider the case of a single additional symmetry $X\neq P$. If $X=Q$, we have $2$ classes for both point and line gaps since $\epsilon_Q=\pm1$, $\epsilon_{PQ}=\pm1$ and the former can be reserved by redefining $U_Q$ as $\sqrt{\epsilon_{PQ}}U_PU_Q$. If $X=C$, we have $4$ classes for both point and line gaps since there is an equivalence between $(\epsilon_C,\eta_C,\epsilon_{PC})$ and $(-\epsilon_C,\epsilon_{PC}\eta_C,\epsilon_{PC})$. If $X=K$, we have $3$ classes for point gap and $4$ classes for line gap. This is because, similar to $X=C$, there is an equivalence between $(\epsilon_K,\eta_K,\epsilon_{PK})$ and $(-\epsilon_K,\epsilon_{PK}\eta_K,\epsilon_{PK})$, and Wick rotation further unifies $(\epsilon_K,\eta_K,\epsilon_{PK})$ with $(-\epsilon_K,\eta_K,\epsilon_{PK})$ and thus $(\epsilon_K,\epsilon_{PK}\eta_K,\epsilon_{PK})$ for point gap. Therefore, when there is at most a single symmetry on top of $P$, we have $10$ and $11$ classes for point and line gaps, respectively.

Finally, we turn to the case with all types of symmetries. Recalling that only two of $Q$, $C$ and $K$ symmetries are independent, it suffices to consider the interplay between $P$, $Q$ and $C$ symmetries. Following the previous analysis, we can always fix $\epsilon_Q$ and $\epsilon_C$ to be $-1$ by otherwise redefining $U_Q$ or/and $U_C$ as $\sqrt{\epsilon_{PQ}}U_PU_Q$ or/and $U_PU_C$. Such a gauge transformation unifies $(\epsilon_Q,\epsilon_C,\eta_C,\epsilon_{PQ},\epsilon_{PC},\epsilon_{QC})$ with $(-\epsilon_Q,\epsilon_C,\eta_C,\epsilon_{PQ},\epsilon_{PC},\epsilon_{PC}\epsilon_{PQ}\epsilon_{QC})$ and $(\epsilon_Q,-\epsilon_C,\epsilon_{PC}\eta_C,\epsilon_{PQ},\epsilon_{PC},\epsilon_{PQ}\epsilon_{QC})$. Since there are $4$ remaining binary parameters $(\eta_C,\epsilon_{PQ},\epsilon_{PC},\epsilon_{QC})$, we have $16$ classes for line gap. As for point gap, $\epsilon_Q$ can be reversed by Wick rotation, which leaves all the other parameters unchanged. Therefore, $(\eta_C,\epsilon_{PQ},\epsilon_{PC},\epsilon_{QC})$ and $(\eta_C,\epsilon_{PQ},\epsilon_{PC},\epsilon_{PC}\epsilon_{PQ}\epsilon_{QC})$ are unified, leading to $12$ genuinely different classes.

In total, there are $8+8+10+12=38$ BL classes for point gap and $11+16+11+16=54$ BL classes for line gap, respectively.


\subsection{Classifications for real and imaginary line gaps}\label{Sec:lgBL}
As has been proven in Appendix~\ref{app4} of the present review, any non-Hermitian Bloch Hamiltonian with a real (imaginary) line gap can continuously be (anti-)Hermitianized and flattened while keeping all the symmetries. 
As a result, if the line gap is real (imaginary), $P$ symmetry is unified with $Q$ symmetry with $\epsilon_Q=1$ ($\epsilon_Q=-1$) and $C$ symmetry with $\epsilon_C=\pm1$ is unified with $K$ symmetry with $\epsilon_K=\pm1$ ($\epsilon_K=\mp1$). Moreover, by performing Wick rotation, we can map a real-line-gapped system into an imaginary-line-gapped one with $\epsilon_X$ reversed for $X=Q,K$. Having these in mind, we can figure out the entire periodic table by only considering some BL classes for real line gap, which we focus on in the following.

\begin{table*}[tbp]
\caption{Topological classifications of real-line-gapped non-Hermitian Bloch Hamiltonians with BL symmetries.}
\begin{center}
\small
\begin{tabular}{ccc}
\hline\hline
BL sym. & $\epsilon_X$, $\eta_X$, $\epsilon_{XY}$ & $K$-group \\ 
\hline
None & N/A  & $K_{\mathbb{C}}(0;d)$  \\ 
$P$ & N/A & $K_{\mathbb{C}}(1;d)$ \\ 
$Q_+$ & $\epsilon_Q=1$ & $K_{\mathbb{C}}(0;d)\oplus K_{\mathbb{C}}(0;d)$ \\ 
$Q_-$ & $\epsilon_Q=-1$ & $K_{\mathbb{C}}(1;d)$ \\ 
$C_{++}$ & $\epsilon_C=1,\eta_C=1$ & $K_{\mathbb{R}}(0;d)$ \\ 
$C_{+-}$ & $\epsilon_C=1,\eta_C=-1$ & $K_{\mathbb{R}}(4;d)$ \\ 
$C_{-+}$ & $\epsilon_C=-1,\eta_C=1$ & $K_{\mathbb{R}}(2;d)$ \\ 
$C_{--}$ & $\epsilon_C=-1,\eta_C=-1$ & $K_{\mathbb{R}}(6;d)$ \\
$K_{++}$ & $\epsilon_K=1,\eta_K=1$ & $K_{\mathbb{R}}(0;d)$ \\
$K_{+-}$ & $\epsilon_K=1,\eta_K=-1$ & $K_{\mathbb{R}}(4;d)$ \\
$K_{-+}$ & $\epsilon_K=-1,\eta_K=1$ & $K_{\mathbb{R}}(2;d)$ \\
$K_{--}$ & $\epsilon_K=-1,\eta_K=-1$ & $K_{\mathbb{R}}(6;d)$ \\
$PQ^+$ & 
$\epsilon_{PQ}=1$ & $K_{\mathbb{C}}(1;d)\oplus K_{\mathbb{C}}(1;d)$ \\
$PQ^-$ & 
$\epsilon_{PQ}=-1$ & $K_{\mathbb{C}}(0;d)$ \\
$PC^+_+$ & 
$\eta_C=1,\epsilon_{PC}=1$ & $K_{\mathbb{R}}(1;d)$ \\
$PC^+_-$ & 
$\eta_C=-1,\epsilon_{PC}=1$ & $K_{\mathbb{R}}(5;d)$ \\
$PC^-_+$ & 
$\eta_C=1,\epsilon_{PC}=-1$ & $K_{\mathbb{R}}(3;d)$ \\
$PC^-_-$ & 
$\eta_C=-1,\epsilon_{PC}=-1$ & $K_{\mathbb{R}}(7;d)$ \\
$PK^+_+$ & 
$\eta_K=1,\epsilon_{PK}=1$ & $K_{\mathbb{R}}(1;d)$ \\
$PK^+_-$ & 
$\eta_K=-1,\epsilon_{PK}=1$ & $K_{\mathbb{R}}(5;d)$ \\
$PK^-_+$ & 
$\eta_K=1,\epsilon_{PK}=-1$ & $K_{\mathbb{R}}(7;d)$ \\
$PK^-_-$ & 
$\eta_K=-1,\epsilon_{PK}=-1$ & $K_{\mathbb{R}}(3;d)$ \\
$QC^+_{+++}$ & $\epsilon_Q=1,\epsilon_C=1,\eta_C=1,\epsilon_{QC}=1$ & $K_{\mathbb{R}}(0;d)\oplus K_{\mathbb{R}}(0;d)$ \\
$QC^+_{++-}$ & $\epsilon_Q=1,\epsilon_C=1,\eta_C=-1,\epsilon_{QC}=1$ & $K_{\mathbb{R}}(4;d)\oplus K_{\mathbb{R}}(4;d)$ \\
$QC^+_{+-+}$ & $\epsilon_Q=1,\epsilon_C=-1,\eta_C=1,\epsilon_{QC}=1$ & $K_{\mathbb{R}}(2;d)\oplus K_{\mathbb{R}}(2;d)$ \\
$QC^+_{+--}$ & $\epsilon_Q=1,\epsilon_C=-1,\eta_C=-1,\epsilon_{QC}=1$ & $K_{\mathbb{R}}(6;d)\oplus K_{\mathbb{R}}(6;d)$ \\
$QC^-_{+++}$ & 
$\epsilon_Q=1,\epsilon_C=1,\eta_C=1,\epsilon_{QC}=-1$ & $K_{\mathbb{C}}(0;d)$ \\
$QC^-_{++-}$ & 
$\epsilon_Q=1,\epsilon_C=1,\eta_C=-1,\epsilon_{QC}=-1$ & $K_{\mathbb{C}}(0;d)$ \\
$QC^-_{+-+}$ & 
$\epsilon_Q=1,\epsilon_C=-1,\eta_C=1,\epsilon_{QC}=-1$ & $K_{\mathbb{C}}(0;d)$ \\
$QC^-_{+--}$ & 
$\epsilon_Q=1,\epsilon_C=-1,\eta_C=-1,\epsilon_{QC}=-1$ & $K_{\mathbb{C}}(0;d)$ \\
$QC^+_{-++}$ & 
$\epsilon_Q=-1,\epsilon_C=1,\eta_C=1,\epsilon_{QC}=1$ & $K_{\mathbb{R}}(1;d)$ \\
$QC^+_{-+-}$ & 
$\epsilon_Q=-1,\epsilon_C=1,\eta_C=-1,\epsilon_{QC}=1$ & $K_{\mathbb{R}}(5;d)$ \\
$QC^+_{--+}$ & 
$\epsilon_Q=-1,\epsilon_C=-1,\eta_C=1,\epsilon_{QC}=1$ & $K_{\mathbb{R}}(1;d)$ \\
$QC^+_{---}$ & 
$\epsilon_Q=-1,\epsilon_C=-1,\eta_C=-1,\epsilon_{QC}=1$ & $K_{\mathbb{R}}(5;d)$ \\
$QC^-_{-++}$ & 
$\epsilon_Q=-1,\epsilon_C=1,\eta_C=1,\epsilon_{QC}=-1$ & $K_{\mathbb{R}}(7;d)$ \\
$QC^-_{-+-}$ & 
$\epsilon_Q=-1,\epsilon_C=1,\eta_C=-1,\epsilon_{QC}=-1$ & $K_{\mathbb{R}}(3;d)$ \\
$QC^-_{--+}$ & 
$\epsilon_Q=-1,\epsilon_C=-1,\eta_C=1,\epsilon_{QC}=-1$ & $K_{\mathbb{R}}(3;d)$ \\
$QC^-_{---}$ & 
$\epsilon_Q=-1,\epsilon_C=-1,\eta_C=-1,\epsilon_{QC}=-1$ & $K_{\mathbb{R}}(7;d)$ \\
$PQC^{+++}_+$ & 
$\eta_C=1,\epsilon_{PQ}=1,\epsilon_{PC}=1,\epsilon_{QC}=1$ & $K_{\mathbb{R}}(1;d)\oplus K_{\mathbb{R}}(1;d)$ \\
$PQC^{+++}_-$ & 
$\eta_C=-1,\epsilon_{PQ}=1,\epsilon_{PC}=1,\epsilon_{QC}=1$ & $K_{\mathbb{R}}(5;d)\oplus K_{\mathbb{R}}(5;d)$ \\
$PQC^{++-}_+$ & 
$\eta_C=1,\epsilon_{PQ}=1,\epsilon_{PC}=1,\epsilon_{QC}=-1$ & $K_{\mathbb{C}}(1;d)$ \\
$PQC^{++-}_-$ & 
$\eta_C=-1,\epsilon_{PQ}=1,\epsilon_{PC}=1,\epsilon_{QC}=-1$ & $K_{\mathbb{C}}(1;d)$ \\
$PQC^{+-+}_+$ & 
$\eta_C=1,\epsilon_{PQ}=1,\epsilon_{PC}=-1,\epsilon_{QC}=1$ & $K_{\mathbb{C}}(1;d)$ \\
$PQC^{+-+}_-$ & 
$\eta_C=-1,\epsilon_{PQ}=1,\epsilon_{PC}=-1,\epsilon_{QC}=1$ & $K_{\mathbb{C}}(1;d)$ \\
$PQC^{+--}_+$ & 
$\eta_C=1,\epsilon_{PQ}=1,\epsilon_{PC}=-1,\epsilon_{QC}=-1$ & $K_{\mathbb{R}}(3;d)\oplus K_{\mathbb{R}}(3;d)$ \\
$PQC^{+--}_-$ & 
$\eta_C=-1,\epsilon_{PQ}=1,\epsilon_{PC}=-1,\epsilon_{QC}=-1$ & $K_{\mathbb{R}}(7;d)\oplus K_{\mathbb{R}}(7;d)$ \\
$PQC^{-++}_+$ & 
$\eta_C=1,\epsilon_{PQ}=-1,\epsilon_{PC}=1,\epsilon_{QC}=1$ & $K_{\mathbb{R}}(0;d)$ \\
$PQC^{-++}_-$ & 
$\eta_C=-1,\epsilon_{PQ}=-1,\epsilon_{PC}=1,\epsilon_{QC}=1$ & $K_{\mathbb{R}}(4;d)$ \\
$PQC^{-+-}_+$ & 
$\eta_C=1,\epsilon_{PQ}=-1,\epsilon_{PC}=1,\epsilon_{QC}=-1$ & $K_{\mathbb{R}}(2;d)$ \\
$PQC^{-+-}_-$ & 
$\eta_C=-1,\epsilon_{PQ}=-1,\epsilon_{PC}=1,\epsilon_{QC}=-1$ & $K_{\mathbb{R}}(6;d)$ \\
$PQC^{--+}_+$ & 
$\eta_C=1,\epsilon_{PQ}=-1,\epsilon_{PC}=-1,\epsilon_{QC}=1$ & $K_{\mathbb{R}}(2;d)$ \\
$PQC^{--+}_-$ & 
$\eta_C=-1,\epsilon_{PQ}=-1,\epsilon_{PC}=-1,\epsilon_{QC}=1$ & $K_{\mathbb{R}}(6;d)$ \\
$PQC^{---}_+$ & 
$\eta_C=1,\epsilon_{PQ}=-1,\epsilon_{PC}=-1,\epsilon_{QC}=-1$ & $K_{\mathbb{R}}(4;d)$ \\
$PQC^{---}_-$ & 
$\eta_C=-1,\epsilon_{PQ}=-1,\epsilon_{PC}=-1,\epsilon_{QC}=-1$ & $K_{\mathbb{R}}(0;d)$ \\
\hline\hline
\end{tabular}
\end{center}
\label{tableA1}
\end{table*}

\begin{table*}[tbp]
\caption{Topological classifications of imaginary-line-gapped non-Hermitian Bloch Hamiltonians with BL symmetries.}
\begin{center}
\small
\begin{tabular}{ccc}
\hline\hline
BL sym. & $\epsilon_X$, $\eta_X$, $\epsilon_{XY}$ & $K$-group \\ 
\hline
None & N/A  & $K_{\mathbb{C}}(0;d)$  \\ 
$P$ & N/A & $K_{\mathbb{C}}(1;d)$ \\ 
$Q_+$ & $\epsilon_Q=1$ & $K_{\mathbb{C}}(1;d)$ \\ 
$Q_-$ & $\epsilon_Q=-1$ & $K_{\mathbb{C}}(0;d)\oplus K_{\mathbb{C}}(0;d)$ \\ 
$C_{++}$ & $\epsilon_C=1,\eta_C=1$ & $K_{\mathbb{R}}(2;d)$ \\ 
$C_{+-}$ & $\epsilon_C=1,\eta_C=-1$ & $K_{\mathbb{R}}(6;d)$ \\ 
$C_{-+}$ & $\epsilon_C=-1,\eta_C=1$ & $K_{\mathbb{R}}(0;d)$ \\ 
$C_{--}$ & $\epsilon_C=-1,\eta_C=-1$ & $K_{\mathbb{R}}(4;d)$ \\
$K_{++}$ & $\epsilon_K=1,\eta_K=1$ & $K_{\mathbb{R}}(0;d)$ \\
$K_{+-}$ & $\epsilon_K=1,\eta_K=-1$ & $K_{\mathbb{R}}(4;d)$ \\
$K_{-+}$ & $\epsilon_K=-1,\eta_K=1$ & $K_{\mathbb{R}}(2;d)$ \\
$K_{--}$ & $\epsilon_K=-1,\eta_K=-1$ & $K_{\mathbb{R}}(6;d)$ \\
$PQ^+$ & 
$\epsilon_{PQ}=1$ & $K_{\mathbb{C}}(1;d)\oplus K_{\mathbb{C}}(1;d)$ \\
$PQ^-$ & 
$\epsilon_{PQ}=-1$ & $K_{\mathbb{C}}(0;d)$ \\
$PC^+_+$ & 
$\eta_C=1,\epsilon_{PC}=1$ & $K_{\mathbb{R}}(1;d)$ \\
$PC^+_-$ & 
$\eta_C=-1,\epsilon_{PC}=1$ & $K_{\mathbb{R}}(5;d)$ \\
$PC^-_+$ & 
$\eta_C=1,\epsilon_{PC}=-1$ & $K_{\mathbb{R}}(3;d)$ \\
$PC^-_-$ & 
$\eta_C=-1,\epsilon_{PC}=-1$ & $K_{\mathbb{R}}(7;d)$ \\
$PK^+_+$ & 
$\eta_K=1,\epsilon_{PK}=1$ & $K_{\mathbb{R}}(1;d)$ \\
$PK^+_-$ & 
$\eta_K=-1,\epsilon_{PK}=1$ & $K_{\mathbb{R}}(5;d)$ \\
$PK^-_+$ & 
$\eta_K=1,\epsilon_{PK}=-1$ & $K_{\mathbb{R}}(3;d)$ \\
$PK^-_-$ & 
$\eta_K=-1,\epsilon_{PK}=-1$ & $K_{\mathbb{R}}(7;d)$ \\
$QC^+_{+++}$ & $\epsilon_Q=1,\epsilon_C=1,\eta_C=1,\epsilon_{QC}=1$ & $K_{\mathbb{R}}(1;d)$ \\
$QC^+_{++-}$ & $\epsilon_Q=1,\epsilon_C=1,\eta_C=-1,\epsilon_{QC}=1$ & $K_{\mathbb{R}}(5;d)$ \\
$QC^+_{+-+}$ & $\epsilon_Q=1,\epsilon_C=-1,\eta_C=1,\epsilon_{QC}=1$ & $K_{\mathbb{R}}(1;d)$ \\
$QC^+_{+--}$ & $\epsilon_Q=1,\epsilon_C=-1,\eta_C=-1,\epsilon_{QC}=1$ & $K_{\mathbb{R}}(5;d)$ \\
$QC^-_{+++}$ & $\epsilon_Q=1,\epsilon_C=1,\eta_C=1,\epsilon_{QC}=-1$ & $K_{\mathbb{R}}(7;d)$ \\
$QC^-_{++-}$ & $\epsilon_Q=1,\epsilon_C=1,\eta_C=-1,\epsilon_{QC}=-1$ & $K_{\mathbb{R}}(3;d)$ \\
$QC^-_{+-+}$ & $\epsilon_Q=1,\epsilon_C=-1,\eta_C=1,\epsilon_{QC}=-1$ & $K_{\mathbb{R}}(3;d)$ \\
$QC^-_{+--}$ & $\epsilon_Q=1,\epsilon_C=-1,\eta_C=-1,\epsilon_{QC}=-1$ & $K_{\mathbb{R}}(7;d)$ \\
$QC^+_{-++}$ & $\epsilon_Q=-1,\epsilon_C=1,\eta_C=1,\epsilon_{QC}=1$ & $K_{\mathbb{R}}(0;d)\oplus K_{\mathbb{R}}(0;d)$ \\
$QC^+_{-+-}$ & $\epsilon_Q=-1,\epsilon_C=1,\eta_C=-1,\epsilon_{QC}=1$ & $K_{\mathbb{R}}(4;d)\oplus K_{\mathbb{R}}(4;d)$ \\
$QC^+_{--+}$ & $\epsilon_Q=-1,\epsilon_C=-1,\eta_C=1,\epsilon_{QC}=1$ & $K_{\mathbb{R}}(2;d)\oplus K_{\mathbb{R}}(2;d)$  \\
$QC^+_{---}$ & $\epsilon_Q=-1,\epsilon_C=-1,\eta_C=-1,\epsilon_{QC}=1$ & $K_{\mathbb{R}}(6;d)\oplus K_{\mathbb{R}}(6;d)$ \\
$QC^-_{-++}$ & $\epsilon_Q=-1,\epsilon_C=1,\eta_C=1,\epsilon_{QC}=-1$ & $K_{\mathbb{C}}(0;d)$ \\
$QC^-_{-+-}$ & $\epsilon_Q=-1,\epsilon_C=1,\eta_C=-1,\epsilon_{QC}=-1$ & $K_{\mathbb{C}}(0;d)$ \\
$QC^-_{--+}$ & $\epsilon_Q=-1,\epsilon_C=-1,\eta_C=1,\epsilon_{QC}=-1$ & $K_{\mathbb{C}}(0;d)$ \\
$QC^-_{---}$ & $\epsilon_Q=-1,\epsilon_C=-1,\eta_C=-1,\epsilon_{QC}=-1$ & $K_{\mathbb{C}}(0;d)$ \\
$PQC^{+++}_+$ & 
$\eta_C=1,\epsilon_{PQ}=1,\epsilon_{PC}=1,\epsilon_{QC}=1$ & $K_{\mathbb{R}}(1;d)\oplus K_{\mathbb{R}}(1;d)$ \\
$PQC^{+++}_-$ & 
$\eta_C=-1,\epsilon_{PQ}=1,\epsilon_{PC}=1,\epsilon_{QC}=1$ & $K_{\mathbb{R}}(5;d)\oplus K_{\mathbb{R}}(5;d)$ \\
$PQC^{++-}_+$ & 
$\eta_C=1,\epsilon_{PQ}=1,\epsilon_{PC}=1,\epsilon_{QC}=-1$ & $K_{\mathbb{C}}(1;d)$ \\
$PQC^{++-}_-$ & 
$\eta_C=-1,\epsilon_{PQ}=1,\epsilon_{PC}=1,\epsilon_{QC}=-1$ & $K_{\mathbb{C}}(1;d)$ \\
$PQC^{+-+}_+$ & 
$\eta_C=1,\epsilon_{PQ}=1,\epsilon_{PC}=-1,\epsilon_{QC}=1$ & $K_{\mathbb{R}}(3;d)\oplus K_{\mathbb{R}}(3;d)$ \\
$PQC^{+-+}_-$ & 
$\eta_C=-1,\epsilon_{PQ}=1,\epsilon_{PC}=-1,\epsilon_{QC}=1$ & $K_{\mathbb{R}}(7;d)\oplus K_{\mathbb{R}}(7;d)$ \\
$PQC^{+--}_+$ & 
$\eta_C=1,\epsilon_{PQ}=1,\epsilon_{PC}=-1,\epsilon_{QC}=-1$ & $K_{\mathbb{C}}(1;d)$ \\
$PQC^{+--}_-$ & 
$\eta_C=-1,\epsilon_{PQ}=1,\epsilon_{PC}=-1,\epsilon_{QC}=-1$ & $K_{\mathbb{C}}(1;d)$ \\
$PQC^{-++}_+$ & 
$\eta_C=1,\epsilon_{PQ}=-1,\epsilon_{PC}=1,\epsilon_{QC}=1$ & $K_{\mathbb{R}}(2;d)$ \\
$PQC^{-++}_-$ & 
$\eta_C=-1,\epsilon_{PQ}=-1,\epsilon_{PC}=1,\epsilon_{QC}=1$ & $K_{\mathbb{R}}(6;d)$ \\
$PQC^{-+-}_+$ & 
$\eta_C=1,\epsilon_{PQ}=-1,\epsilon_{PC}=1,\epsilon_{QC}=-1$ & $K_{\mathbb{R}}(0;d)$ \\
$PQC^{-+-}_-$ & 
$\eta_C=-1,\epsilon_{PQ}=-1,\epsilon_{PC}=1,\epsilon_{QC}=-1$ & $K_{\mathbb{R}}(4;d)$ \\
$PQC^{--+}_+$ & 
$\eta_C=1,\epsilon_{PQ}=-1,\epsilon_{PC}=-1,\epsilon_{QC}=1$ & $K_{\mathbb{R}}(2;d)$ \\
$PQC^{--+}_-$ & 
$\eta_C=-1,\epsilon_{PQ}=-1,\epsilon_{PC}=-1,\epsilon_{QC}=1$ & $K_{\mathbb{R}}(6;d)$ \\
$PQC^{---}_+$ & 
$\eta_C=1,\epsilon_{PQ}=-1,\epsilon_{PC}=-1,\epsilon_{QC}=-1$ & $K_{\mathbb{R}}(4;d)$ \\
$PQC^{---}_-$ & 
$\eta_C=-1,\epsilon_{PQ}=-1,\epsilon_{PC}=-1,\epsilon_{QC}=-1$ & $K_{\mathbb{R}}(0;d)$ \\
\hline\hline
\end{tabular}
\end{center}
\label{tableA2}
\end{table*}

If there is no symmetry, the classification is the same as class A and is given by $K_{\mathbb{C}}(0;d)$. We then consider the case with a single symmetry $X$. If $X=P$, the classification is the same as class AIII and is given by $K_{\mathbb{C}}(1;d)$. According to the equivalence discussion above, the same classification applies to $X=Q$ with $\epsilon_Q=-1$. If $\epsilon_Q=1$, $U_Q$ commutes with $H$ and the classification is given by $K_{\mathbb{C}}(0;d)\oplus K_{\mathbb{C}}(0;d)$. The classification for $X=C,K$ is exactly the same, so it suffices to consider the former. Note that $C$ with $\epsilon_C=1$ ($\epsilon_C=-1$) symmetry reduces to TRS (PHS) for a Hermitian Bloch Hamiltonian, the classification should be the same as class AI (D) if $\eta_C=1$ and class AII (C) if $\eta_C=-1$, with the corresponding $K$-group given by $K_{\mathbb{R}}(0;d)$ ($K_{\mathbb{R}}(2;d)$) and $K_{\mathbb{R}}(4;d)$ ($K_{\mathbb{R}}(6;d)$). So far we have filled the first $12$ rows.

We move on the case with two symmetries. If one is $P$ symmetry, then $\epsilon_{X=Q,C,K}$ can be fixed and the remaining two parameters are $\eta_X$ and $\epsilon_{PX}$. For $X=Q$, we actually only have a single parameter $\epsilon_{PQ}$. If $\epsilon_{PQ}=-1$, then $U_Q$ anti-commutes with both $U_P$ and $H(\boldsymbol{k})$ and the $K$-group is shifted to $K_{\mathbb{C}}(2;d)=K_{\mathbb{C}}(0;d)$. If $\epsilon_{PQ}=1$, then we can construct $U_PU_Q$ which commutes with both $U_P$ and $H(\boldsymbol{k})$ and the $K$-group is doubled into $K_{\mathbb{C}}(1;d)\oplus K_{\mathbb{C}}(1;d)$. For $X=C$, we fix $\epsilon_C=-1$ so that $U_C\mathcal{K}=\mathcal{C}$ is a PHS and $U_P=\Gamma$ is a CS. If $\epsilon_{PC}=1$ ($\epsilon_{PC}=-1$), we have $\Gamma=\mathcal{T}\mathcal{C}$ ($\Gamma=i\mathcal{T}\mathcal{C}$) and $\mathcal{T}^2=\mathcal{C}^2=\eta_C$ ($\mathcal{T}^2=-\mathcal{C}^2=-\eta_C$), so the classification is the same as class BDI (DIII) if $\eta_C=1$ and class CII (CI) if $\eta_C=-1$, whose $K$-groups read $K_{\mathbb{R}}(1;d)$ ($K_{\mathbb{R}}(3;d)$) and $K_{\mathbb{R}}(5;d)$ ($K_{\mathbb{R}}(7;d)$), respectively. Exactly the same results apply to $X=K$ if $\epsilon_K$ is fixed to be $1$. Otherwise, as is the case in Table~\ref{tableA1}, the classifications for $\eta_K=\pm1$ and $\epsilon_{PK}=-1$ should be exchanged. So far, we have filled another $10$ rows.

We continue to consider the case with two symmetries, but neither of which is $P$. Recalling that $Q$, $C$ and $K$ symmetries are not independent, it suffices to focus on the combination of $Q$ and $C$ symmetries. If $\epsilon_Q=1$ and $\epsilon_{QC}=1$ ($\epsilon_{QC}=-1$), $U_Q$ ($iU_Q$ ) commutes with $i$, $U_C\mathcal{K}$ and $H$, 
so the $K$-group is doubled (complexified) compared with that without $Q$ symmetry (see $4$-$8$ rows in Table~\ref{tableA1}). If $\epsilon_Q=-1$, then the BL class reduces to an AZ class with CS, as can be determined from $15$-$22$ rows. So far, another $16$ rows are filled.

We finally turn to the case with all the BL symmetries. To determine the $K$-group, we can consider how the Clifford-algebra extension without $P$ symmetry, which corresponds to a chiral symmetric AZ class, is modified by the additional $P$ symmetry. If $\epsilon_{PQ}=1$, we can construct either $\sqrt{\epsilon_{PC}\epsilon_{QC}}U_PU_Q$ which commutes with $i$, $U_{\rm C}\mathcal{K}$, $U_Q$ and $H(\boldsymbol{k})$. Therefore, the $K$-group is doubled (complexified) if $\epsilon_{PC}\epsilon_{QC}=1$ ($\epsilon_{PC}\epsilon_{QC}=-1$). Otherwise, $\epsilon_{PQ}=-1$ so $U_P$ commutes with either $\mathcal{C}=U_C\mathcal{K}$ (if $\epsilon_{PC}=1$) or $\mathcal{T}=\eta_C\sqrt{\epsilon_{QC}}U_QU_C\mathcal{K}$ (if $\epsilon_{PC}=-1$) and anti-commutes with the other.  Therefore, we can construct the original Clifford-algebra extension as $\{\mathcal{C},i\mathcal{C},i\mathcal{T}\mathcal{C}\}\to\{\mathcal{C},i\mathcal{C},i\mathcal{T}\mathcal{C},H\}$ and add $iU_P$ ($U_P$) if $\epsilon_{PC}=1$ ($\epsilon_{PC}=-1$), 
leading to a shift $K_{\mathbb{R}}(s;d)\to K_{\mathbb{R}}(s-1;d)$ ($K_{\mathbb{R}}(s;d)\to K_{\mathbb{R}}(s+1;d)$). Using these properties, we have filled the remaining $16$ rows.

\subsection{Classifications for point gap}\label{Sec:pgBL}
In Sec.~\ref{Sec:NHPT} we have figured out the classifications for those BL classes with at most one symmetry. Here we consider the remaining BL classes with at least two symmetries. We first consider the combination of $P$ symmetry with another symmetry $X=Q,C,K$. For $X=Q$, there are two cases with $\epsilon_{PQ}=1$ and $\epsilon_{PQ}=-1$. In the former case, we can construct a complex Clifford-algebra extension as ${\rm C}\ell_3(\mathbb{C})=\{\Sigma,U_Q',i\Sigma U'_QU'_P\}\to{\rm C}\ell_4(\mathbb{C})=\{\Sigma,U_Q',i\Sigma U'_QU'_P,H_U\}$ for the Hermitianized Hamiltonian, so the $K$-group is given by $K_{\mathbb{C}}(3;d)=K_{\mathbb{C}}(1;d)$. In the latter case, we can find an element $\Sigma U'_P$ which commutes with $\Sigma$, $U'_Q$ and $H_U$, so the $K$-group should be $K_{\mathbb{C}}(2;d)\oplus K_{\mathbb{C}}(2;d)=K_{\mathbb{C}}(0;d)\oplus K_{\mathbb{C}}(0;d)$. For $X=C$, there are four cases with $\eta_C=\pm1$ and $\epsilon_{PC}=\pm1$. If $\eta_C=1$ and $\epsilon_{PC}=1$, we have $U'_C(U'_C)^*=1$ and $[U'_C,U'_P]=0$, so the symmetry class of $H_U(\boldsymbol{k})$ with $\Sigma$ excluded is BDI. Moreover, since $\{U'_C\mathcal{K},\Sigma\}=0$, we can construct $i U'_P\Gamma$ which is anti-involutory and commutes with all the elements, so the $K$-group is turned to be complex and is thus given $K_\mathbb{C}(1;d)$. If $\eta_C=-1$ and $\epsilon_{PC}=1$, we can apply exactly the same technique and the $K$-group is turned from $K_\mathbb{R}(5;d)$ into $K_\mathbb{C}(5;d)=K_\mathbb{C}(1;d)$. If $\epsilon_{PC}=-1$, we can still apply a similar technique but now $U'_P\Sigma$ instead of $i U'_P\Gamma$ should be added into the Clifford algebra corresponding to class DIII (CI), if $\eta_C=1$ ($\eta_C=-1$). In this case, the $K$-group is simply doubled as $K_\mathbb{R}(3;d)\oplus K_\mathbb{R}(3;d)$ ($K_\mathbb{R}(7;d)\oplus K_\mathbb{R}(7;d)$). For $X=K$, there are three cases, two of which are $\eta_K=\pm1$ and $\epsilon_{PK}=1$ and the remaining one is $\epsilon_{PK}=-1$. If $\epsilon_{PK}=1$, we can construct $U'_P\Sigma$ which is involutory and commutes with all the other elements, leading to a doubled $K$-group $K_\mathbb{R}(1;d)\oplus K_\mathbb{R}(1;d)$ ($K_\mathbb{R}(5;d)\oplus K_\mathbb{R}(5;d)$) for $\eta_K=1$ ($\eta_K=-1$). If $\epsilon_{PK}=1$, we can construct $iU'_P\Sigma$ which is anti-involutory and commutes with all the other elements, leading to a complexified $K$-group $K_\mathbb{C}(7;d)=K_\mathbb{C}(1;d)$. Things become much easier if neither of the two symmetries is $P$. If $\epsilon_C=-1$ ($\epsilon_C=1$), $\Sigma$ ($i\Sigma$) anti-commutes with $U'_Q$, $U'_C\mathcal{K}$ and $H$, so the $K$-group is simply shifted from $K_{\mathbb{R}}(s;d)$ ($s$ is odd due to the CS) to $K_{\mathbb{R}}(s+1;d)$ ($K_{\mathbb{R}}(s-1;d)$).

Now we turn to the case with all the BL symmetries. We first consider the Clifford-algebra extension for $H_U$ without $U'_P$. Denoting $\mathcal{C}=U'_C\mathcal{K}$ and $\mathcal{T}=\sqrt{\epsilon_{QC}}U'_QU'_C\mathcal{K}$, the extension can be constructed as $\{\mathcal{C},i\mathcal{C},i\mathcal{T}\mathcal{C},\Sigma\}\to\{\mathcal{C},i\mathcal{C},i\mathcal{T}\mathcal{C},\Sigma,H_U\}$, with the corresponding $K$-group given by $K_{\mathbb{R}}(1+2\eta_C-\epsilon_{QC};d)$. If $\epsilon_{PQ}=-1$, we can add either $\sqrt{-\epsilon_{PC}}U'_P\Sigma$ which commutes with all the elements, leading to a doubling ($\epsilon_{PC}=-1$) or complexification ($\epsilon_{PC}=1$). Otherwise, $\epsilon_{PQ}=1$ and we can add $\sqrt{\epsilon_{PC}\epsilon_{QC}}U'_PU'_Q\Sigma$ which anti-commutes with all the elements, leading to a shift from $K_{\mathbb{R}}(s;d)$ to $K_{\mathbb{R}}(s-\epsilon_{PC}\epsilon_{QC};d)$, where $s=1+2\eta_C-\epsilon_{QC}$. One can check that the result is indeed gauge invariant under $\epsilon_{QC}\to\epsilon_{PQ}\epsilon_{PC}\epsilon_{QC}$.

\end{appendices}

\bibliographystyle{tADP}
\bibliography{./reference.bib}
\end{document}